\documentclass[preprint,trackchanges]{aastex63}

\hypersetup{linkcolor=magenta,citecolor=cyan,filecolor=yellow,urlcolor=blue}
\usepackage{indentfirst}
\usepackage{comment}
\usepackage{multirow} 
\usepackage{booktabs}
\usepackage{bigints}
\usepackage{mathrsfs,amsmath} 
\received{2020 December 4}
\revised{2021 February 18}
\accepted{2021 February 19}
\published{2021 June 7}

\shorttitle{Bayesian Time-Resolved Spectroscopy of Multipulse GRBs}
\shortauthors{Li et al.}

\begin{document}

\title{Bayesian Time-Resolved Spectroscopy of Multipulse GRBs:\\ Variations of Emission Properties amongst Pulses}

\author[0000-0002-1343-3089]{Liang Li}
\affiliation{ICRANet, Piazza della Repubblica 10, I-65122 Pescara, Italy}
\affiliation{INAF -- Osservatorio Astronomico d'Abruzzo, Via M. Maggini snc, I-64100, Teramo, Italy}
\affiliation{ICRA, Dipartimento di Fisica, Sapienza Università di Roma, P.le Aldo Moro 5, I–00185 Rome, Italy}
\affiliation{Department of Physics, KTH Royal Institute of Technology, and the Oskar Klein Centre for Cosmoparticle Physics, SE-10691 Stockholm, Sweden}
\affiliation{Department of Physics, Stockholm University, AlbaNova, SE-10691 Stockholm, Sweden}

\author{Felix Ryde}
\affil{Department of Physics, KTH Royal Institute of Technology, and the Oskar Klein Centre for Cosmoparticle Physics, 10691 Stockholm, Sweden}

\author{Asaf Pe{\textquoteright}er}
\affiliation{Department of Physics, Bar-Ilan University, Ramat-Gan 52900, Israel}

\author{Hoi-Fung Yu}
\affiliation{Faculty of Science, The University of Hong Kong, Pokfulam, Hong Kong}

\author{Zeynep Acuner}
\affil{Department of Physics, KTH Royal Institute of Technology, and the Oskar Klein Centre for Cosmoparticle Physics, 10691 Stockholm, Sweden}

\correspondingauthor{Liang Li}
\email{liang.li@icranet.org}

\begin{abstract}
Gamma-ray bursts (GRBs) are highly variable and exhibit strong spectral evolution. In particular, the emission properties vary from pulse to pulse in multipulse bursts. Here we present a time-resolved Bayesian spectral analysis of a compilation of GRB pulses observed by the {\it Fermi}/Gamma-ray Burst Monitor. The pulses are selected to have at least four timebins with a high statistical significance, which ensures that the spectral fits are well determined and that spectral correlations can be established. The sample consists of 39 bursts, 117 pulses, and 1228 spectra. We confirm the general trend that pulses become softer over time, with mainly the low-energy power-law index $\alpha$ becoming smaller. A few exceptions to this trend exist, with the hardest pulse occurring at late times. The first pulse in a burst is clearly different from the later pulses; three-fourths of them violate the synchrotron line of death, while around half of them significantly prefer photospheric emission. These fractions decrease for subsequent pulses. We also find that in two-thirds of the pulses, the spectral parameters ($\alpha$ and peak energy) track the light-curve variations. This is a larger fraction compared to what is found in previous samples. In conclusion, emission compatible with the GRB photosphere is typically found close to the trigger time, while the chance of detecting synchrotron emission is greatest at late times. This allows for the coexistence of emission mechanisms at late times.

\end{abstract}


\section{Introduction} \label{sec:intro}

Gamma-ray burst (GRB) emission exhibits significant variability and spectral evolution. The prompt emission light curves typically have irregular, multipulse temporal profiles, in some cases having a complex structure \citep[e.g.,][]{Norris1996, Norris2005}. The emission spectrum typically changes from having both large spectral peak energies and  hard spectral slopes below the peak (so-called hard spectra) to having {lower peak energies and softer spectral slopes (soft spectra)}. Such variations occur both within individual pulse structures  and as an overall trend during the burst \citep{Mazets1982, Ford1995, Crider1997}. 

Within the fireball model of GRBs, there are two main emission sources for the prompt gamma rays. The first is radiation from where the jet becomes transparent, namely, the photosphere region \citep[e.g.,][]{Goodman1986, Rees2005, Peer2006a}.  The second is radiation from a region at larger distance from the progenitor, where the kinetic energy of the jet is dissipated and radiated away in the form of synchrotron emission \citep[e.g,][]{Piran1993,  Sari1998, Lloyd2000a}. The time scale for the gamma-ray emission, relative to the launching of the jet, depends on the distances to the different emission sites, $r_\gamma$, and the Lorentz factor of the jet, $\Gamma$, according to $t_\gamma = r_\gamma/(c \Gamma^2)$, where $c$ is the speed of light. For the typical outflow parameters, the photosphere is expected to occur at $r_\gamma^{\rm ph} \sim 3 \times 10^{12}$ cm, while the synchrotron emission is expected at larger radii, for instance, $r_\gamma^{\rm IS} \sim 10^{13}$ cm for internal shocks and $r_\gamma^{\rm ES} \sim 10^{17}$ cm for external shocks \cite[e.g, ][]{Rees1998}. Therefore, a central engine activity will first be observed by its thermal emission and be followed by synchrotron emission, with a delay of a few seconds or even hundreds of seconds \citep{Rees1994}. In such a case, one should expect to observe an initial photospheric emission episode followed by synchrotron emission activity.

Alternatively, variations of the jet property, such as the entropy and magnetization \citep[e.g., ][]{Rees1994, Beloborodov2013, Uhm2014, Zhang2018, Li2019a, Li2020, Li2021} can cause the variations in the observed emission. For instance, the interaction between the jet and the surrounding material as the jet passes though the progenitor star will cause a variable mixing and thereby a change in the entropy of the flow \citep{Lazzati2009a, Lopez2014, Ito2015}. This directly affects the properties of the observed emission, since the entropy  determines the position of the photosphere, $r_{\rm ph}$, relative to the  saturation radius, $r_{\rm s}$ \citep[e.g., ][]{Meszaros2000}.  If $r_{\rm ph}/r_{\rm s}  \lesssim 1$, a photosphere will dominate the emission,\footnote{A typical result from an analysis of the photospheric spectra is that the initial radius of the jet is $r_0 \sim \rm{few} \times 10^9$ cm \citep[e.g., ][]{Peer2007, Ryde2009, Iyyani2015}. This is also expected theoretically due to interaction between the jet and the progenitor surrounding it, which prevents the jet from accelerating while it is within the star \citep[e.g., ][]{Thompson2006, Gottlieb2019}. A consequence of this fact is that the saturation radius of the jet $r_{\rm s} = \Gamma r_0 \sim r_{\rm ph}$.} while in the opposite cases, $r_{\rm ph}/r_{\rm s} \gg 1$, the photosphere will be very weak, and only nonthermal emission, such as synchrotron, will be expected. Furthermore, in the case of photospheric emission, the entropy variation will allow for a varying amount of broadening of the spectral shape due to subphotospheric dissipation. The reason is that strong dissipation is not expected below the saturation radius. Therefore, significant spectral broadening of the photospheric spectrum is only expected when $r_{\rm ph}/r_{\rm s} \gtrsim 1$. In this alternative scenario, the evolution of the spectral properties of consecutive pulses is interpreted as variations of properties in the jet.  

The two main emission sources can also coexist, giving rise to multicomponent emission spectra \citep[e.g.][]{Ryde2005, Ryde2009, Guiriec2011, ajello2020}.  The observed evolution in spectral shape might then be driven by the relative change in the contribution of the two emission sources in superposition.

In all cases, observed trends and variations among pulses carry important clues to the physics of GRBs. The multipulse nature indicates a continuous ejection from the progenitor, implying that (i) it is not destroyed immediately, (ii) the variation in flux between pulses represents the variation of the available energy, and (iii) the time difference between the pulses represents some characteristic internal time, possibly the fallback time of the material, or accretion into the surface of the newly formed neutron star until it reaches a critical mass and reexplodes. In this paper, we therefore analyze a sample of multipulse bursts in order to investigate whether there is any variation in spectral characteristics depending on the sequel position among pulses within a burst. In particular, we examine the spectral shape, correlations between spectral parameters, and compatibility with photospheric emission. Our sample is obtained from the first 11 yr of the {\it Fermi} Gamma-ray Burst Monitor (GBM) mission, consisting of 117 well-separated pulses and 1228 spectra from 39 bursts.

The paper is organized as follows. The methods are presented in Section 2. The detailed observational properties are summarized in Section 3. In Section 4, we present an assessment of the compatibility with emission models. The discussion and summary are presented in Sections 5 and Section 6, respectively. Throughout the paper, the standard $\Lambda$-CDM cosmology with the parameters $H_{0}= 67.4$ ${\rm km s^{-1}}$ ${\rm Mpc^{-1}}$, $\Omega_{M}=0.315$, and $\Omega_{\Lambda}=0.685$ are adopted \citep{PlanckCollaboration2018}. 

\section{Methodology} \label{sec:data}

\subsection{Initial Burst Selection}

We use data obtained by the {\it Fermi Gamma-ray Space Telescope}, which was launched in 2008, and carries two instruments: the GBM \citep{Meegan2009} and the Large Area Telescope \citep[LAT;][]{Atwood2009}. Together, they cover an energy range from a few keV to a few hundred GeV. The GBM harbors 14 detectors, of which 12 are sodium iodide (NaI; 8 keV-1 MeV) and two are bismuth germanate (BGO; 200 keV-40 MeV) scintillators. By 2019 June, {\it Fermi} had completed 11 yr of operation, and at least 2388 GRBs had been observed. 

Among these bursts, we want to identify the ones that have at least two individual emission episodes, which can be assumed to be independent from each other. Since there is no theoretical prediction of the shape and form of individual emission episodes, we choose to make a general definition for our selection, which does not limit the sample to smooth pulses only. Indeed, numerical simulations of jet emission propagating through the progenitor star show that not only smooth pulses are expected but also more complex morphologies \citep{Lazzati2009a, Lopez2014}. This choice also avoids omitting too many bursts. Following \cite{Yu2019}, we thus allow subdominant variations on top of the main pulse structures. In addition, we also include emission activities with many smaller spikes but whose heights are limited by an approximate pulse-shaped envelope. The pulse-shaped envelope indicates that such emission could be connected, in spite of its large variability. Examples of the light curves that are selected for the sample are shown in \S \ref{sec:classification} and in the Appendix.

We first visually inspected the Time-Tagged Events (TTE) light curves from each of the 2388 GRBs obtained from NASA/HEASARC\footnote{\url{https://heasarc.gsfc.nasa.gov/W3Browse/fermi/fermigbrst.html}}. We identified more than 120 bursts that had at least two clear emission episodes. The light curves exhibit diverse temporal properties. Some have a precursor-like pulse followed by a main pulse, while others exhibit a main pulse followed by a small pulse, and yet others consist of several pulses with similar strength. Many of the pulses are well separated by quiescent intervals, while others have a slight temporal overlap. 

\subsection{Detector, Source, and Background Selections}

We use the standard selection criteria adopted in GBM catalogs \citep{Goldstein2012, Gruber2014, Yu2016}.  Following the common practice, we choose the triggered detectors that have a viewing angle of less than 60 degrees \citep{Goldstein2012}.  Typically, one to three  NaI and one BGO detector are selected. The period of GRB emission that is considered is in most cases somewhat longer than the $T_{90}$ reported in the HEASARC database. This is done so that all relevant features in the light curve can be incorporated. In order to determine the background emission, we select one interval located tens of seconds before the triggered time and one interval located tens of seconds after the emission has ended. We fit the background photon count level with a polynomial function, which typically has an order lower than 4. The optimal order is determined through a likelihood ratio test. The polynomial is applied to fit all the 128 energy channels and then interpolated into the pulse interval to yield the background photon count estimate. In a few bursts, there are several pulses that are separated by long quiescent intervals. In these cases, we select three background intervals to better constrain the background levels (e.g., GRB 140810782). The spectral energy range is set from 10 to 900 keV for the NaI detectors and 300 keV to 30 MeV for the BGO detectors. In order to avoid the K-edge at 33.17 keV,  we ignore the range 30 -- 40 keV.

\subsection{Light-curve Binning}

In order to follow and study spectral evolution in individual bursts, detailed time-resolved spectroscopy is needed \citep[see, e.g.,][]{Crider1997, Kaneko2006, Goldstein2012}. The question is then how to divide the light curve into time bins, since this choice might affect the results. Foremost, it is important to minimize the amount of variation of the emission during a time bin, since such variations will obscure the intrinsic spectral shape. A method to account for this is to identify Bayesian blocks (BBlocks) in the light curve (\citealt{Scargle2013}). Therefore, we apply the BBlock method with a false-alarm probability $p_{0}$=0.01 to the TTE light curve of the brightest NaI detector \citep[see ][for further details]{Li2019a,Yu2019}.  This binning is then used for the other detectors as well. 

The BBlock method will create time bins that have varied signal-to-noise ratios. Therefore, some time bins will not have enough signal for a fit to be reliable. A suitable measure of the signal-to-noise ratio is the statistical significance $S$\footnote{The statistical significance, $S$, is suitable for data with Poisson sources with Gaussian backgrounds, which is the case for GBM data \citep{Vianello2018}.}. \citet{Dereli-Begue2020} demonstrated that to fully determine the spectral shape, a value of $S=15$--$20$ is needed. In particular, this ensures that the model parameters converge  properly \citep[see also, e.g., ][]{Vianello2018, Li2019a, Li2019b, Ryde2019}. We therefore follow this recommendation and only select the BBlock time bins that have $S$ $\geq$ 20. 

This specific selection thus provides time bins during which (i) there is no significant spectral evolution and (ii) there is highly significant data. Such spectra are required in order to make firm inferences on the intrinsic emission spectrum. However, a consequence of this selection is that intervals with rapid variations will be dismissed. This is often the case for the rise phase of pulses, which could carry particular information about the emission process (see further discussion in \S \ref{sec:tracking}). A possibility of incorporating these dismissed intervals would be perform joint fits of many intervals at the same time by assuming a prescription of the spectral evolution. This would increase the statistical significance while maintaining the temporal resolution. However, since further assumptions need to be made on the spectral evolution \citep[e.g., ][]{Ryde1999}, such studies are deferred to other publications.

\subsection{Sample (Pulse) Definition}

The purpose of this study is to relate the emission properties between pulses within a burst. We therefore only consider bursts that have multiple distinguishable pulses. One of the properties that we will relate is the spectral evolution during the individual pulses. Hence, in order to include a pulse in the study, it needs to have at least four time bins with $S$ $\geq$20 ($N_{(\rm S \geq20)} \geq 4$). This allows one to determine the spectral evolution and the correlation between spectral parameters. Our final selection criterion is, therefore, that the burst should have at least two such pulses. This criterion strongly reduces the sample size. However, the pulses that pass the criterion will provide well-determined spectra and evolution characteristics. This final selection yields a sample of 39 bursts. From these, we obtain 117 pulses and 1228 time-resolved spectra, of which 103 pulses have $N_{(\rm S \geq20)} \geq 4$), consisting of 944 spectra.

The properties of our sample are summarized in Table \ref{table:global}, and includes the {\it Fermi}-GBM ID (column 1), redshift (column 2), and observed duration $t_{90}$ (column 3), together with the used detectors (column 4), and the selected source and background intervals (columns, 5 and 6), the number of total (column 7) and the used ($S \geq$20, parentheses in column 7) BBlock time bins, and the number of total (column 8) and the used (parentheses in column 8) pulses for each burst. The detector in parentheses is the brightest one, used for the BBlocks and background determinations.

Inevitably, there will be some overlap between many of the pulses. During such periods, emission from both pulses will be present, and the observed spectrum will be a superposition of two spectra. Depending on their mutual strengths, a fit with a single spectral component might give misleading results. We therefore group the pulses in the 39 bursts into three subsamples.

\begin{itemize}
\item {\it Gold}: No overlap, and the pulses are completely independent. There are 13 such pulses, which are listed in Table \ref{table:Pulse} (column 5). 
\item {\it Silver}: Slight overlap. There are 90 such pulses (column 5 in Table \ref{table:Pulse}).
\item {\it Bronze}:  Pulses that have less than four time bins $S \geq$ 20. There are 14 such pulses (column 5 in Table \ref{table:Pulse}). Note that these pulses are not used in our final statistical analysis. 
\end{itemize}

\subsection{Spectral Fitting}

The physical model of the GRB spectra is not yet firmly established. Likewise, the possible regimes of the suggested emission models are not either clarified. Therefore, empirical models are typically used in the analysis. This offers the flexibility to explore many different models and, at the same time, to explore the models without limiting the certain spectral regimes. The typically used functions are the Band function, which is a broken power-law function, and the cutoff power-law (CPL) function \citep{Band1993,Gruber2014}.  Another advantage of these empirical models is that they are used in all catalogs mapping the behavior and characteristics of observed bursts.

However, the empirical models do not necessarily correspond to the underlying physical model \citep[e.g.,][]{Lloyd-Ronning2002, Burgess2014, Acuner2019}. Therefore, the empirical parameter values that are inferred from the data cannot be directly translated into physical model parameters. However, the way the physical model parameters map onto the Band parameters has been established for a few emission models \citep[e.g., ][ see also \S \ref{sec:EM}]{Lloyd2000a, Acuner2019}. Moreover, \citet{Acuner2020} showed that the values of, e.g., $\alpha$ can be used to decisively distinguish between various types of emission models, as long as the data have a high signal strength (as is the case in the present study). 

We will therefore make use of these empirical functions. The Band function is defined by the low-energy power-law index ($\alpha$) and the high-energy index ($\beta$), which are connected by a smoothly around the break energy ($E_{\rm 0}$).

The photon number spectrum is defined as
\begin{eqnarray}
f_{\rm BAND}(E)=A \left\{ \begin{array}{ll}
(\frac{E}{E_{\rm piv}})^{\alpha} \rm exp (-\frac{{\it E}}{{\it E_{0}}}), & E \le (\alpha-\beta)E_{0}  \\
\lbrack\frac{(\alpha-\beta)E_{0}}{E_{\rm piv}}\rbrack^{(\alpha-\beta)} \rm exp(\beta-\alpha)(\frac{{\it E}}{{\it E_{\rm piv}}})^{\beta}, & E\ge (\alpha-\beta)E_{0}\\
\end{array} \right.
\label{eq:Band} 
\end{eqnarray}
The energy of the spectral peak of the $\nu F_{\nu}$ spectrum (assuming $\beta<-2$) is in units of keV,
\begin{equation}
E_{\rm p}=(2+\alpha)E_{0},
\end{equation}
and $A$ is the normalization factor at 100 keV in units of ph cm$^{-2}$keV$^{-1}$s$^{-1}$, $E_{\rm piv}$ is the pivot energy fixed at 100 keV, and $\alpha$ and $\beta$ are the low- and high-energy power-law photon spectral indices, respectively. 

The CPL function is given by 
\begin{equation}
f_{\rm CPL}(E) =A \left(\frac{E}{E_{\rm piv}}\right)^{\alpha}\rm exp(-\frac{\it E}{\it E_{c}})
\label{CPL}
\end{equation}
Note that the CPL model approaches the Band model as $\beta$ tends to -$\infty$. The peak energy $E_{\rm p}$ of the $\nu F_\nu$ spectrum is related to the $E_{c}$ through $E_{\rm p}$=(2+$\alpha$)$E_{\rm c}$. 

The analysis in this work is performed with the Multi-Mission Maximum Likelihood Framework (3ML; \citealt{Vianello2015}) and we use a Bayesian approach, using the Markov Chain Monte Carlo (MCMC) method \citep[e.g.,][]{Foreman-Mackey2013}. We use the typical spectral parameters obtained from the previous {\it Fermi}-GBM catalog as the prior information of the Bayesian inference.

For the Band model,
\begin{eqnarray}
\left\{ \begin{array}{ll}
A \sim \log  \mathcal{N} (\mu=0,\sigma=2) & \rm cm^{-2}keV^{-1}s^{-1}\\
\alpha \sim \mathcal{N} (\mu=-1,\sigma=0.5) \\
\beta \sim \mathcal{N} (\mu=-2,\sigma=0.5) \\
E_{\rm p} \sim \log \mathcal{N} (\mu=2,\sigma=1) & \rm keV\\
\end{array} \right.
\label{eq:priorBand}
\end{eqnarray}
and for CPL model,
\begin{eqnarray}
\left\{ \begin{array}{ll}
A \sim \log  \mathcal{N} (\mu=0,\sigma=2) & \rm cm^{-2}keV^{-1}s^{-1}\\
\alpha \sim \mathcal{N} (\mu=-1,\sigma=0.5) \\
E_{\rm c} \sim \log \mathcal{N} (\mu=2,\sigma=1) & \rm keV\\
\end{array} \right.
\label{eq:priorCPL}
\end{eqnarray}
A posterior distribution is obtained from the prior distribution and the likelihood that combines the model and the observed data. The best model parameters are estimated from the posterior probability distribution obtained by MCMC sampling. When using MCMC sampling, in order to obtain the steady-state chains of the parameter distributions, the sampling needs to reach a certain number of times; therefore, the first part of the sample that has not reached the steady-state distribution is discarded. Therefore, for each parameter estimation, we take 10000 MCMC samples and discard the initial 20\%. The fitted parameters are estimated by the Maximum A posteriori Probability (MAP) with uncertainties at 1$\sigma$ (68\%) Bayesian credible level and are based the last 80\% of the MCMC samples. We also provide all of the analysis results of each time-resolved spectrum, which includes the best parameter value estimates, covariance matrices, and the statistical information criteria. They can be retrieved at doi: \url{https://zenodo.org/record/4746267}.

\subsection{High-energy Power Law, $\beta$, and the Preferred Model Selection}\label{sec:bestmodel}

For a majority of time-resolved burst spectra, the CPL model is a sufficient model \citep[e.g.][]{Kaneko2006, Goldstein2012, Burgess2019, Yu2019}. However, in some cases, a high-energy power law significantly improves the fit, indicating a significant flux contribution beyond the spectral peak. In the case of synchrotron emission, the high-energy power law provides information about the particle acceleration and the nature of the shocks, while in the case of photospheric emission, it provides information about the energy dissipation in the emitting region.

In order to determine whether a high-energy power law gives significant improvement in our analysis, we compare the information criteria of the Band and the CPL fits. \citet{Acuner2020} showed that the information criteria capture everything important in the fits and that the difference in information criteria can be used in the model comparison. In particular, they showed that a significance of 99\% of preferring one model over the other is found for a difference in log evidence greater than 5 (their Eq. 6) and that this corresponds approximately to a difference in information criteria of 10 (further discussion can be found in \citealt{Acuner2019a}\footnote{Z. Acuner, ‘Statistical Investigations of the Emission Processes in Gamma-ray Bursts’, PhD dissertation, KTH Royal Institute of Technology, Stockholm, 2019.}). In this work, we therefore adopt the information criterion to compare models, and, following the literature in the field \citep[e.g., ][]{Greiner2016, Yu2019}, we particularly use the deviance information criterion (DIC; \citealt{Spiegelhalter2002, Moreno2013}), which is defined as DIC=-2log[$p$(data$\mid\hat{\theta}$)]+2$p_{\rm DIC}$, where $\hat{\theta}$ is the posterior mean of the parameters, and $p_{\rm DIC}$ is a term to penalize the more complex model for overfitting \citep{Gelman2014}. We consequently accept the Band function as the preferred model if the difference between the Band's DIC and the CPL's DIC $\Delta$DIC = DIC$_{\rm Band} - $DIC$_{\rm CPL} > -10$. 

For consistency, one needs to use the same empirical model throughout the whole pulse in order to avoid artificial fluctuation due to change of spectral model \citep[see][]{Yu2019}. Consequently, if one time bin has $ \Delta$DIC$ <-10$ then we use the Band function throughout the pulse. Otherwise, we use the simpler CPL model. Therefore, the pulses that are fitted by a Band function have at least one time bin in which a high-energy power law is significantly detected. The model used for each pulse is listed in column 6 in Table \ref{table:Pulse}.

Among all of the individual time bins, we find that in only 29 \% ({274/944}) is $\Delta$DIC$<-10$, i.e., the Band function is preferred. For these time bins, a high-energy power law is required by the data \citep[see also ][]{Gruber2014, Yu2019}. {For the sequence of pulses, we find the corresponding fractions to be 36\% (122/338) for the first pulse $P_1$, 28\% (101/361) for $P_2$, 22\% (34/156) for $P_3$, 15\% (7/46) for $P_4$, 15\% (4/26) for $P_5$, and 29\% (5/17) for $P_6$.} There is only a weak trend, with the first pulse $P_{1}$ having the largest fraction.

Turning over to the pulses, we find that {66\% (77/117)} of them have at least one time bin with $\Delta$DIC$<-10$, therefore the Band function is used for the entire pulse. For the sequence of pulses, we find that the corresponding fractions are 77\% (30/39) for $P_1$, 69\% (27/39) for $P_2$, 63\% (15/24) for $P_3$, 22\% (2/9) for $P_4$, 50\% (2/4) for $P_5$, and 50\% (1/2) for $P_6$. Again, there is only a weak trend, with the first pulse $P_{1}$ having a larger fraction.

In addition to the $\Delta$DIC criterion, we also examined the $p_{\rm DIC}$ values following \citet{Yu2019}. In some instances, the values were found to be anomalously large (hundreds of thousands), which typically indicates that a local rather than a global minimum of the likelihood function is found. In these cases, we reran the Bayesian fits with new initial values to ensure proper convergence to the global minimum.
  
\section{Observational Properties among Pulses}\label{sec:observational}

We now investigate how the emission properties vary among the pulses within bursts. We compare the group consisting of the first pulse in every burst ($P_1$) with the group containing the second pulse in each burst ($P_2$), and so on. The number of spectra in each pulse group decreases, with the first three groups having above 100 spectra each (338, 361, and 156) and the last three only having a few tens (46, 26, and 17). 
 
\begin{figure*}
\begin{center}
\includegraphics[angle=0,scale=0.70]{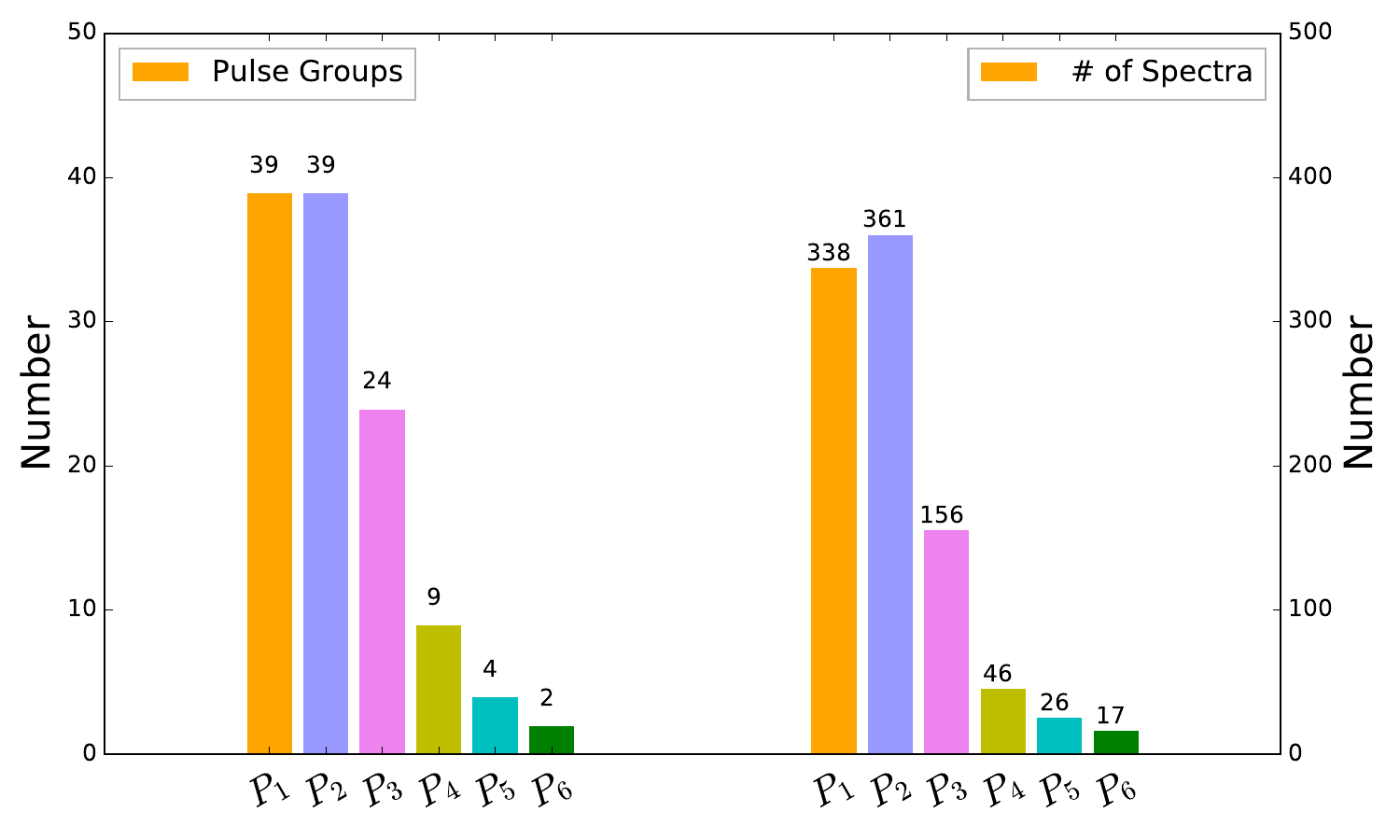}
\end{center}
\caption{{Histograms of the numbers of pulses (left) and spectra (right) in each pulse grouping.}}\label{fig:histogram_Pulses}
\end{figure*}

The identification and binning of the analyzed pulses, as well as their spectral properties, are listed in Table \ref{table:Pulse}. The table includes the {\it Fermi}-GBM ID (column 1), the time-series pulse (Column 2), the pulsewise source intervals (column 3), the total and used ($S \geq$20) time bins (column 4) by using the BBlocks across the source intervals for each pulse, the identified grade of the pulse (column 5), the best spectral model (column 6), the preferred physical model (column 7), the pulsewise evolution of the spectral parameters (column 8 for $E_{\rm p}$ and column 9 for $\alpha$), and the type of parameter relations with the Spearman’s rank coefficient, $r$, in the parentheses: $F$-$\alpha$ (column 10), $F$-$E_{\rm p}$ (column 11), and $\alpha$-$E_{\rm p}$ (column 12). Figure \ref{fig:histogram_Pulses} shows the histograms of the number of pulses in each pulse group (left side) and the corresponding number of spectra (right side). In the following section, we only consider the best-fit model for each pulse according to \S \ref{sec:bestmodel}.

\subsection{Parameter Distributions}\label{sec:distribution}

Figure \ref{fig:distrubution_best_model} shows the parameter distributions for every group of pulses, including $\alpha$, $E_{\rm p}$, and the $K$-corrected energy flux $F$ (erg cm$^{-2}$ s$^{-1}$), over the range 1-10$^{4}$ keV, as well as the duration, $\Delta t$, of the pulses. In the following figures, the data points are colored orange ($P_{1}$), blue-magenta ($P_{2}$), violet ($P_{3}$), yellow ($P_{4}$ or $P_{4+5+6}$)\footnote{Note that, in some cases, we combine the data for the three last pulse groups $(P_{4+5+6})$, since the parameter distributions are typically similar and an increased sample size improves the fits.}, cyan ($P_{5}$), and green ($P_{6}$). The best Gaussian fit for each distribution is presented, and the corresponding average values and standard deviations are presented in Table \ref{table:distribution}. The average value of the full sample, including all pulses, $\alpha$=-0.84$\pm$0.35 and $E_{\rm p}$=$\rm log_{10}$(214)$\pm$0.42. These values are in agreement with previous catalogs \citep[e.g.,][]{Kaneko2006}.

In order to assess whether the distributions change between the different pulse groups, we use the Kolmogorov-Smirnov (K-S) test. This test determines the chance probability, $P$, that two distributions are sampled from populations with identical distributions. For our purposes, we use $P < 10^{-2}$ to ensure that the distributions are truly different. In Table \ref{tab:KStest} we present the $P$-values for all the $\alpha$ and $E_{\rm p}$ distributions for the pulse groups.

These results indicate that the groups of pulses could be divided into two categories with two different behaviors. The first category, with $P_1$ and $P_2$, is different from the rest of the pulse groups. In the first group of two pulses ($P_1$ and $P_2$), the peak energy $E_{\rm p}$ does not change, while the spectral slope $\alpha$ exhibits a clear softening. Compared to the second category, with the later pulse groups ($P_3$--$P_6$), there is a change, with both $\alpha$ and $E_{\rm p}$ decreasing. However, within this second category, the distributions are not significantly different. Based on this analysis alone, it can, therefore, be argued that these two categories represent different types of spectral characteristics and, therefore, possibly different emissions.

The comparison between the fluxes and pulse durations is shown in the bottom panels of Figure \ref{fig:distrubution_best_model}. The fluxes have a similar pattern as $\alpha$. There is a steady decrease, except for the last two groups, which have a similar distribution. Finally, the pulse duration distributions are comparable, and the first three groups have increasing average values. 

Another way to compare the variation of emission properties in between pulses is to study the maximal value of the parameters of $\alpha$ and $E_{\rm p}$ in each pulse and how they vary. Note that $\alpha_{\rm max}$ and $E_{\rm p, max}$ do not necessarily need to be at the same time bin. In the upper panels in Figure \ref{fig:max}, we present $\alpha_{\rm max} = \alpha_{\rm max}(t)$ and $E_{\rm p, max} = E_{\rm p, max}(t)$ versus time. Here, the trend of softening is again revealed for $\alpha_{\rm max}$. Later pulses typically have smaller $\alpha_{\rm max}$. There is also a trend that $\alpha_{\rm max}$ for the first pulse is softer the further it is delayed from the trigger. For $E_{\rm p, max}$, however, there is no trend, and the distinction between the two categories from the K-S test is not as apparent. This means that, when it comes to $E_{\rm p, max} $ and $\alpha_{\rm max}$, it is mainly the spectral shape that changes and not the location of the peak energy when we compare the six pulse groups.

In the lower panels in Figure \ref{fig:max}, the dependency of $\alpha_{\rm max}$ and $E_{\rm p, max}$ on pulse duration $\Delta t$ is shown.  The colored star correspond to the average values. For the first three groups, the increase in average pulse duration is correlated with the decrease in the average values of $\alpha_{\rm max}$ and $E_{\rm p, max}$ (see Fig. \ref{fig:distrubution_best_model}). Again, the last group with $P_{(4+5+6})$ differs from the trend by having a shorter (average) pulse duration. However, instrumental effects might play a role here. The fluxes of later pulses are lower; hence, the full duration of the pulses might not be apparent above the background noise level.

\subsection{Spectral Evolution}\label{sec:classification}

Early investigations of GRB emission identified a significant correlation between spectral properties, such as a relation between the intensity and the shape of the spectrum \citep{Wheaton1973}, and a correlation between the intensity and the spectral peak \citep{Golenetskii1983}. In this section, we classify the evolution of $\alpha$ and $E_{\rm p}$ relative to the count light curves, as commonly done in the literature \citep[e.g., ][]{Kargatis1994, Ford1995}. In the \S \ref{sec:SC} we will quantify this further by investigating the actual correlations between the parameters.

Examples of the evolution of the spectral parameters $E_{\rm p}$, and $\alpha$, and the $\nu F_{\nu}$ flux are provided in the upper panels of Figure \ref{fig:example}. Here the parameter evolutions are overlaid on the count light curve. In the Appendix, we further provide the corresponding figures for all bursts (Figure \ref{fig:Ep_Best}, \ref{fig:Alpha_Best}, \ref{fig:Flux_Best}) as well as the figures for all bursts for the evolution of the spectral parameter $\beta$. Following the traditional classification, we use the notation that the parameter value is denoted as `hard' when referring to large values of both $\alpha$ and $E_{\rm p}$, as opposed to soft values (low values of $\alpha$ and $E_{\rm p}$).  We categorize the evolution in the two main groups: those with a hard-to-soft ({\it h.}-t.-{\it s.}) pattern, that is, the parameters decrease independent of the rise and decay of the pulse, and those with a flux-tracking ({\it f.}-{\it t.}) pattern, that is, the parameters are correlated with the rise and decay of the flux with or without a time lag. In a handful of cases, other patterns are also identified. The classification of the pulses is given in Table \ref{table:Pulse} (columns 8-9).
 
For the $E_{\rm p}$ evolution (see Figure \ref{fig:Ep_Best} and Column 8 in Table \ref{table:Pulse}), we find that the hard-to-soft and flux-tracking patterns are the two dominant patterns in our multipulse sample. We find that about two-thirds (63/103 = 61\%) of the pulses show a flux-tracking pattern and about one-third (32/103 = 31\%) of the pulses exhibit a hard-to-soft pattern. Other evolution patterns are rarely observed. For instance, we only identify two cases showing the hard-to-soft-to-hard pattern (e.g., the $P_{4}$ in GRB 101014175), one case displaying the soft-to-hard evolution (the $P_{1}$ in GRB 170207906), and three cases exhibiting a hard-to-soft followed by a flux-tracking evolution within a pulse (e.g., the $P_{1}$ in GRB 091127976). Here we note that previous investigations found that about two-thirds of cases have a hard-to-soft behavior, while a smaller fraction has a flux-tracking behavior \citep[e.g.,][]{Ford1995, Crider1997, Lu2012, Yu2019}. This fact is at odds with the observation in our sample. This is further discussed in Section \ref{sec:PreviousWorksComparison}.

For the $\alpha$ evolution (see Figure \ref{fig:Alpha_Best} and column 9 in Table \ref{table:Pulse}), we find, similarly, that the hard-to-soft and the flux-tracking patterns dominate. The flux-tracking pattern accounts for 65\% (67/103) of the pulses, while the hard-to-soft pattern accounts for 25\% (26/103) of the pulses. These two patterns thus account for 90\% of the pulses. Among the rest of the pulses, we find that three cases have a soft-to-hard evolution (e.g., the $P_{1}$ in GRB 110625881), and one case has a hard-to-soft-to-hard evolution (the $P_{4}$ in GRB 101014175). Moreover, we also identify two cases showing a ``flat'' (or weak rise) behavior throughout the pulse (e.g., the $P_{1}$ in GRB 120129580), and two cases have no clear trend at all (e.g., $P_{1}$ in GRB 120728434), which is not found in the $E_{\rm p}$ evolution.

We note that the $E_{\rm p}$ and $\alpha$ evolutionary patterns during a single pulse are not necessarily the same. In roughly half of the pulses (51\% = 53/103), the $E_{\rm p}$ and $\alpha$ evolution are classified to have the same pattern, while 49\% (50/103) of the pulses do not have the same pattern.

Finally, the patterns of the spectral evolution for $E_{\rm p}(t)$ and $\alpha$(t) can vary from pulse to pulse within a burst. We find that in only about half of the pulses, the patterns of the $E_{\rm p}(t)$ evolution between two adjacent pulses are the same (55\% = 35/64), and the rest 45\% (29/64) are the inconsistent cases. Similarly, for $\alpha(t)$, we also find that for about half of the pulses, the spectral evolution among different pulses shares a similar pattern, accounting for 58\% (37/64) of the pulses, while the inconsistent cases account for 42\% (23/64) of the pulses. However, there is no significant variation in the fraction of pulses that are classified as having a tracking behavior between the six pulse groups, which all have around a two-thirds fraction (Table \ref{tab:SepctralPatterns} and Figure \ref{fig:histogram}) 

\subsection{Parameter Correlations}\label{sec:SC}

We now turn to investigating the correlation between the following parameter pairs: ($\log F$, $\alpha$), ($\log F$, $\log E_{\rm p}$), and ($\alpha$, $\log E_{\rm p}$). Examples of these relations are provided in the lower panels of Figure \ref{fig:example}, where functional fits are made. The leftmost panel shows the fit of $F = F_0 \, e^{k \alpha}$, where $k \sim 3.37 \pm 0.49$ \citep[a typical value; ][]{Ryde2019}, the middle panel shows the power law between $F$ and $E_{\rm p}$ with index $1.50 \pm 0.15$ \citep[a typical value; ][]{Borgonovo2001}, and finally the rightmost panel shows a fit to $\alpha = k_2 \ln (E_{\rm p}/E_0) + \alpha_0$, where $k_2 = -2.01 \pm 0.20$. In the Appendix, we further provide the corresponding figures of these correlations for the full sample (Figs. \ref{fig:FluxAlpha_Best}, \ref{fig:FluxEp_Best}, \ref{fig:EpAlpha_Best}).

We visually inspect the correlations and classify them according to the scheme in \citet{Yu2019}, who classified them into three behaviors. The first behavior is a monotonic relation,  defining  Type 1. It can be divided into three categories: type 1p, monotonic positive correlation; type 1n, monotonic negative correlation; and type 1f, flat relation. The second behavior has two piecewise monotonic relations combined at a break point. This behavior is defined as type 2, and is divided into two subcategories, either a concave (type 2p) or a convex (type 2n) function. No clear trend is classified as type 3. The classification is given in Table \ref{table:Pulse}.

The strengths of the correlations are given by the Spearman's rank $r$. Strong correlations have $r>0.7$, and weak correlations have $r<0.4$. For both the $F$-$\alpha$ and $F$-$E_{\rm p}$ relations, around half of the pulses have strong correlations (47/103 and 54/103), and a quarter have weak correlations (27/103 and 26/103). In contrast, the $\alpha$-$E_{\rm p}$ relations have the reverse properties: strong correlations in 21/103 and weak correlations in 57/103. 

Among the pulses with strong correlations, we find that for the $F$-$\alpha$ relation, the vast majority has a positive monotonic relation (1p; 38/47), only a few have a negative relation (1n; 6/47). Similarly, for the $F$-$E_{\rm p}$ relation, a vast majority have a positive power law (1p; 45/51). Finally, for the $\alpha$-$E_{\rm p}$-relation, a negative and positive relations are equally common (1n; 9/21) and (1p; 8/21).

The three relations  $F$-$\alpha$, $F$-$E_{\rm p}$, and $E_{\rm p}$-$\alpha$ typically do not have strong correlations at the same time. However, in a few cases, they do: 13 out of the 103 pulses have $r>0.7$ or $r<-0.7$ for all correlations at the same time. An example of a spectral evolution with simultaneous strong correlations is the single-pulse burst GRB 131231A, where all three relations have monotonic positive correlations \citep{Li2019}. In the following discussions, we summarize the correlations of all of the pulses independent of their $r$ values. 

\subsubsection{Individual $F$-$\alpha$ Relation}

For each individual burst, the $F$-$\alpha$ plot is shown in Figure \ref{fig:FluxAlpha_Best}, and the identified type, as well as its Spearman’s coefficient $r$ is summarized in column 10 in Table \ref{table:Pulse}. The statistical results are presented in the lower left panel in Figure \ref{fig:histogram}.

We find that the dominant $F$-$\alpha$ relation is a monotonic correlation in the log-linear plots (type 1) accounting for 87 pulses (84\%). Of these, 69 are of type 1p, accounting for 67\%, and 16 are of type 1n, accounting for 15\%. Among the rest, 13\% (13/103) are of type 3. 

To identify any change in spectral properties among the pulses, we compare the frequency of the dominating types 1p and 1n in each pulse group. The proportion of the frequencies of these types are all high. For $P_1$, the proportion of type 1p versus type 1n is 19/8, and for the following pulses, it is 28/6 ($P_2$), 14/1 ($P_3$), 4/1 ($P_4$), 3/0 ($P_5$), and 1/0 ($P_6$).  No apparent variation in the relative frequency is  identified, apart from the tendency that the negative $F$-$\alpha$ relations are proportionally more common in $P_1$. An example is given by GRB150330, which has three well-separated pulses; the first pulse is a clear type 1n, while the following two pulses have positive relations (type 1p), also studied in detail by \citet{Li2019a}. The change in correlation pattern corresponds to a change in the range of $\alpha$ values for the pulses. \citet{Li2019a} therefore suggested that a change in jet properties should account for both of these properties.

We also note that there are only three pulses that are classified as type 2 (relation with a break), and all of these are type 2p and all are identified in the first pulse (GRB 081009, GRB 100719, and GRB 100826). Finally, we find that in 25\% (14/57) of cases, the types change between adjacent pulses.

\subsubsection{Individual $F$-$E_{\rm p}$ Relation}

The classification of the $F$-$E_{\rm p}$ relation in Figure \ref{fig:FluxEp_Best}, is shown in column 11 of Table \ref{table:Pulse}, and the lower middle panel of Figure \ref{fig:histogram}. We find that a monotonic correlation in the log-log plots is again the most common type in the $F$-$E_{\rm p}$ relation, accounting for 75 (72\%) of the pulses, of which 74 (71\%) are type 1p (similar to the fraction for the $F$-$\alpha$ relation), and only one case ($P_{3}$ in GRB 090131) is identified as type 1n (a much lower fraction compared to the $F$-$\alpha$ relation), and no type 1f pulse is found. Among the rest, 18\% (19/103) of the pulses have no clear trend (type 3) and 8\% (8/103) have a broken power-law behavior (type 2p). 

We again compare the frequency of the dominating types, which in this case are type 1p versus type 2p. The proportions are all very high, without any significant variation amongst the groups:  {20/6 ($P_{1}$), 33/1 ($P_{2}$), 14/1 ($P_{3}$), 4/0 ($P_{4}$), 2/0 ($P_{5}$), and 1/0 ($P_{6}$)}. There is a slight tendency, though, for type 2p to be proportionally more prevalent in $P_1$. We note that the $F$-$E_{\rm p}$ relation does not change between two adjacent pulses for a majority of cases.

\subsubsection{Individual $\alpha$-$E_{\rm p}$ Relation}

For the $\alpha$-$E_{\rm p}$ relation (see Figure \ref{fig:EpAlpha_Best} and column 12 in Table \ref{table:Pulse}), we find that pulses with a monotonic relation (type 1) dominate (72\% = 74/103). However, compared to the $F$-$\alpha$ and $F$-$E_{\rm p}$ relations, the number of type 1p and type 1n are more even: 32\% = 33/103 and 36\% = 37/103, respectively. Type 2 pulses account for 20\% (21/103) of cases, of which 15 are type 2p and six are type 2n. Type 3 has 8\% (8/103). 

The proportion between the frequencies of the two dominating types, type 1p and type 1n, for the six pulse groups are 2/15 ($P_{1}$), 18/11 ($P_{2}$), 8/7 ($P_{3}$), 3/2 ($P_{4}$), 2/1 ($P_{5}$), and 0/1 ($P_{6}$). Apart from $P_1$, there is no apparent variation between the pulse groups. However, for $P_1$, there is a disparity; only 44\% (17/39) are of Type 1, compared 72\% for all pulses. Out of these 17, type 1n clearly dominates, in contrast to all other pulses, where type 1p is slightly dominat instead. We note that, in contrast to the other correlations, most of the $\alpha$-$E_{\rm p}$ correlations are weak (\S \ref{sec:SC}). To confirm this tendency, we therefore particularly investigate only the strong correlations (21 pulses with $r>0.7$). Of these, there are seven $P_1$, and they all are of type 1n. Ten cases are $P_2$, and of these, three are of type 1n and five of type 1p. We therefore conclude that for the $\alpha$-$E_{\rm p}$ relation, the first pulses have (i) a smaller fraction of type 1 and (ii) of these type 1n have a higher fraction compared to the later pulses. Finally, we find that 46\% (26/57) of the $\alpha$-$E_{\rm p}$ relations change between two adjacent pulses.

\section{Assessment of the Compatibility with Emission Models}\label{sec:EM}

In both synchrotron and photosphere models, there are many different types of spectra that can be produced, mainly depending on the location of the emission site and the flow properties. The spectral shape from a synchrotron emitting source depends on the relation between the radiative  cooling time and other timescales, related to heating and adiabatic expansion \citep{Lloyd2000a,Tavani2000}. The steepest low-energy power law that is allowed (for isotropically distributed pitch angles) is $\alpha =-2/3$ for slow-cooled emission (SCS), while steeper slopes down to $\alpha =-3/2$ are expected for (marginally) fast-cooling emission. Similarly, the properties of the emission from the photosphere depend on the emission site and the amount of dissipation that occurs in the vicinity of the photosphere. If there is no energy dissipation, the spectrum is expected to be slightly broader than a Planck function, namely the nondissipative photosphere \citep[NDP;][]{Beloborodov2011, Acuner2018, Meng2019}. Typically, though, dissipation is expected around the photosphere, which thus causes the spectral shape to broaden further \citep{Giannios2005, Rees2005, Peer2006, Vurm2013}.

A few attempts have been made to fit particular cases of these models to the data directly \citep[e.g., ][]{Lloyd-Ronning2002, Ryde2004, Ryde2005, Ahlgren2015, Ryde2017, Oganesian2019, Acuner2020, Burgess2020}. However, such studies are limited by the range of models that are used and by the limited samples that can be studied due to the computationally costly procedures. Alternatively, synthetic data from a certain physical model can be produced by using the detector response. The synthetic data can then be fitted with empirical models, accounting for the limitations of the typically adopted analysis methods. Such a procedure identifies the parameter space of the empirical model that corresponds to that particular physical model. 

The relation between $\alpha$ and $E_{\rm p}$ that such investigations yield for slow-cooled synchrotron (SCS) and  the NDP are shown by the green and pink lines in the upper panel of Figure \ref{fig:max_dis}. These lines are reproduced from Figure 4 of \citet{Burgess2015} and Figure 3 of \citet{Acuner2019}. Using these lines, general assessments of the emission process can be made. For instance, if we assume that the same emission mechanisms operate throughout the pulse, a single data point above the SCS line indicates that the pulse has a higher probability of being of a photospheric origin \citep{Acuner2019, Dereli-Begue2020}.

\subsection{Spectral Shape}

In the upper panel of Figure \ref{fig:max_dis} we plot the values of $\alpha_{\rm max}$ and the corresponding value of $E_{\rm p}$, with one data point from every pulse. Comparing with the limiting lines for SCS and the NDP, we find that 67\% (79/103) of all the pulses have at least one data point above the SCS line and 21\% (22/103) of all the pulses have a data point above the NDP line.

The number of pulses that are above the NDP and SCS lines for the sequence of pulses is shown in the lower panel of Figure \ref{fig:max_dis}, together with the corresponding fractions. The first two pulses in a burst have a larger fraction above the SCS line. There is a clear decrease in later pulses. There are, however, a couple of bursts in which the hardest spectrum occurs at late times. This fact is reflected in the large fraction for $P_5$ in Figure \ref{fig:max_dis}. Examples of these bursts are GRB 121225 and GRB 140810, where the largest $\alpha$ occurs at around 50s after the trigger. 

As mentioned above, the spectra with $(E_{\rm p}, \alpha_{\rm max})$ values lying above the SCS line can be expected to have a higher probability of having a photospheric origin. However, how big this probability is can only be answered by a model comparison of the physical models, for instance, through Bayesian evidence. In any case, for spectra close to the SCS line, a  model comparison will be inconclusive, since both models can produce similar spectra (over the observed energy range). \citet{Acuner2020} investigated this point quantitatively and found that a photospheric preference can be claimed, with great confidence, only for spectra with $\alpha \sim > -0.5$, as long as the data have a high significance. To illustrate this point, we perform the Bayesian model comparison for two example time bins in GRB150330 (at 1.4 and 137.0 s). We calculate the Bayesian evidence for the NDP spectrum, $Z_{\rm NDP}$, and the evidence for the the slow-cooled synchrotron spectrum, $Z_{\rm SCS}$. As shown in \citet{Acuner2020}, a log-evidence difference of $\ln Z_{NDP}/\ln Z_{SCS} \gtrsim 2$  indicates that the NDP spectrum is preferred, while if the ratio is less than -2, the SCS spectrum is the preferred model. For the 1.4 s time bin, $\alpha = -0.24$, and  we find that $\ln Z_{NDP}/\ln Z_{SCS} = 33.6$, strongly favoring a photospheric origin. Correspondingly, the 137.0 s time bin, which has $\alpha = -0.9$, we find that $\ln Z_{NDP}/\ln Z_{SCS} < -78.9$, strongly favoring an SCS origin (in the comparison between these two specific models). This shows again that the $\alpha$ value can be used to make an approximate model comparison in order to identify the preferred model, which is sufficient for the present study. A full model comparison investigation based on Bayesian evidence is beyond the scope of this paper.

We therefore identify spectra that have $\alpha_{\rm max}> -0.5$ according to the \citet{Acuner2020} criterion. These spectra are denoted by $Ph$ in the Table \ref{table:Pulse} and shown by the blue bars in the lower panel of Figure \ref{fig:max_dis}. The fraction of these spectra steadily decreases for subsequent pulses. For the first pulse in a burst, nearly half of all pulses (18/39)  pass this very stringent criterion. The fraction decreases by approximately a half for every following pulse.

\subsection{Spectral Evolution and Correlations}

We now examine the spectral evolution and correlations for the pulses that are compatible with photospheric emission ($Ph$ in Table \ref{table:Pulse}). The purpose is to assess whether they have a different characteristics of their spectral properties.

For the photospheric pulses, the majority are classified as flux-tracking or hard-to-soft for both $E_{\rm p}$ and $\alpha$. For the $E_{\rm p}$-evolution, the tracking pattern is found in about half, or 47\% (20/43), which is only somewhat lower than in the full sample. However, for the $\alpha$ evolution flux tracking is still dominant with 63\% (27/43), similar to the full sample. 

Turning over to the parameter correlations, we find that {47\% (20/43), 51\% (22/43), and 26\% (11/43) have a strong correlation ($r > 0.7$) for the $F$-$\alpha$ relations, the $F$-$E_{\rm p}$ relation, and the $\alpha$-$E_{\rm p}$ relation, respectively}. These fractions are similar to the ones for the full sample, which indicates that the strength of the correlation does not depend on whether the pulses are photospheric or not.

We again consider the two main types of correlations for each pair of parameters. For the $F$-$\alpha$ relation, the two main types in the full sample were positive (1p) and negative (1n) monotonic correlations. For the photospheric pulses, these fractions are 60\% (26/43) type 1p and 21\% (9/43) type 1n. For the $F$-$E_{\rm p}$-relation, the two main patterns are type 1p and convex relation (type 2p) and the photospheric pulses have 60\% (26/43) type 1p and 16\% (7/43) type 2p. Finally, the $\alpha$-$E_{\rm p}$ relation has 21\% (9/43) type 1p and 35\% (15/43) type 1n. Compared to the full sample (\S \ref{sec:observational}) these fractions are not significantly different.  

\section{Discussion}

\subsection{Comparison with a Single-pulse Sample}\label{sec:SinglePulseComparison}

The bursts in our sample were specifically selected, since they have multiple pulses. Here we address the question of whether or not the spectral properties of the pulses in our sample are similar to those of the single-pulse bursts. To do this, we compare the results in \citet{Yu2019}, who studied a  sample of 37 single-pulse bursts.  

In Figure \ref{fig:comparison} we compare the distributions of duration $T_{90}$, peak flux, fluence, and the time-integrated spectral shape ($E_{\rm p}$, $\alpha$, and $\beta$ from the Band function fit). Our sample is shown by magenta lines, and the \citet{Yu2019} sample is shown by the green lines. All of these parameters are collected from the {\it Fermi}-GBM catalog. Both the peak flux (1024 ms timescale) and the fluence are taken over the 10-1000 keV range. The average values and standard deviations (1$\sigma$ error) are presented in Table \ref{tab:Comparison}. Both the peak flux and the fluence are approximately double for the multipulse bursts. Also, the $\alpha$ distribution is shifted to softer values for the multipulse bursts. On the other hand, the $E_{\rm p}$ and $\beta$ values are similar between the samples. The corresponding probabilities from K-S tests are 0.02, $<10^{-2}$, $<10^{-2}$, 0.86, $<10^{-2}$, and 0.19 for $t_{90}$, peak flux, fluence, $E_{\rm p}$, $\alpha$, and $\beta$, respectively.

In Figure \ref{fig:distribution_Band_CPL_time_resolved} we compare the distributions of the time-resolved spectral parameters. In order to make a consistent comparison, we compare the results from using the CPL function throughout for both samples. This choice is based on the fact that in the \citet{Yu2019} sample, the CPL model is the best model. Likewise, we find that in only 29\% of the time-resolved bins is the Band function a better choice (\S \ref{sec:bestmodel}). We find that the average distribution of $\alpha$ is softer and the flux is larger for the multipulse sample. However, the $E_{\rm p}$ distributions are similar. These results are similar to the time-integrated spectra comparison. If we instead particularly identify the distribution of $P_1$, we find that $\alpha$ distributions have similar peak values. On the other hand, the flux of $P_1$ is double the corresponding flux for the single-pulse bursts.

We also compare the frequency of the parameter relations between the samples. Most proportions between the patterns are similar. However, for the $\alpha$-$E_{\rm p}$ relation, there is a notable difference. In our sample, the proportions are (72\%) 74/103 for type 1 and (20\%) 21/103 for type 2, while for the single-pulse sample, the proportions are (32\%) 12/38 for type 1 and (45\%) 17/38 for type 2. There is still a similarity in that the two pattern types are frequent in both samples; there is no strong dominance. However, in the multipulse sample the type 1 pattern is relatively more frequent.

\subsection{On the Dominance of the Flux-tracking Pattern}\label{sec:PreviousWorksComparison}\label{sec:tracking}

We find that the largest fraction of pulses in our sample follow a tracking pattern, which is at odds with earlier investigations that instead found that the “hard-to-soft” evolution accounts for about two-thirds and the “flux-tracking” evolution only accounts for one-third of the observations \citep[e.g.,][]{Ford1995, Lu2012, Basak2013, Yu2019}.

\citet{Lu2012} argued that if there is a significant overlap between pulses, the classified pattern might be affected. For instance, an apparently flux-tracking pulse might be a consequence of two overlapping hard-to-soft pulses. In our sample, many pulses are slightly overlapping, which means that in the transition period between the pulses, there are contributions from both pulses. Therefore, this might affect our results. To investigate this point, we particularly study the {“gold” sample (see column 5 in Table \ref{table:Pulse})}. These nine bursts\footnote{GRB 081009, GRB 101014, GRB 140416, GRB 140508, GRB 150118, GRB 150330, GRB 151231, GRB 160802, and GRB171120.} have 22 pulses that are clearly separated by periods of no emission. Among these, the flux-tracking pattern accounts for 59\% (13/22) of the pulses, while the hart-to-soft pattern only accounts for 36\% (8/22) of the pulses. These fractions are similar to the ones from the full sample (see \S \ref{sec:classification}). We conclude, therefore, that our sample is not greatly affected by the overlap between pulses.

Another possible reason for the difference is the binning methods used. We use a combination of BBlocks and significance and only consider the most significant time bins. The advantage of our method is that the substantial variations in the light curve are captured and the spectral fits are reliable. On the other hand, we might miss some of the rising phases of the pulses. Including a larger fraction of the pulse duration by using less significant data ($S<20$), could change the apparent pattern of evolution. The spectral evolution of the added periods might indeed be different, but more seriously, the spectral fits are less reliable and could therefore give misleading results.  

A further uncertainty is the inherent problem in this type of classification. A subjective decision needs to be made on where a pulse starts and ends. This can affect the classification. Another point is the fact that the tracking behavior, in many cases, involves a timelag, both positive and negative. If this lag is large compared to the pulse size, the pulse could be classified as a hard-to-soft ({or soft-to-hard}) pulse instead. An example is the $E_{\rm p}$ pattern for GRB120328, where the tracking pattern depends on one data point during the rise phase. The same issue happens for $\alpha$ in other cases\footnote{GRB 110301, GRB 130606, GRB 131014, GRB 140213, GRB 141222, GRB 160422, and GRB 160802.}. These uncertainties can affect the classification of different samples. However, these cases are not sufficient to explain the whole discrepancy.  

Our results thus indicate that the flux-tracking pattern is the prevalent pattern for $\alpha$ and $E_{\rm p}$, at least around the peak of the pulse, where the significance is the largest. This is also consistent with the global $F$-$E_{\rm p}$ relation in the middle panels in Figure \ref{fig:globalcorrelation}. 

\subsection{Implication for the Radiation Process}

\begin{figure*}
\includegraphics[angle=0,scale=0.45]{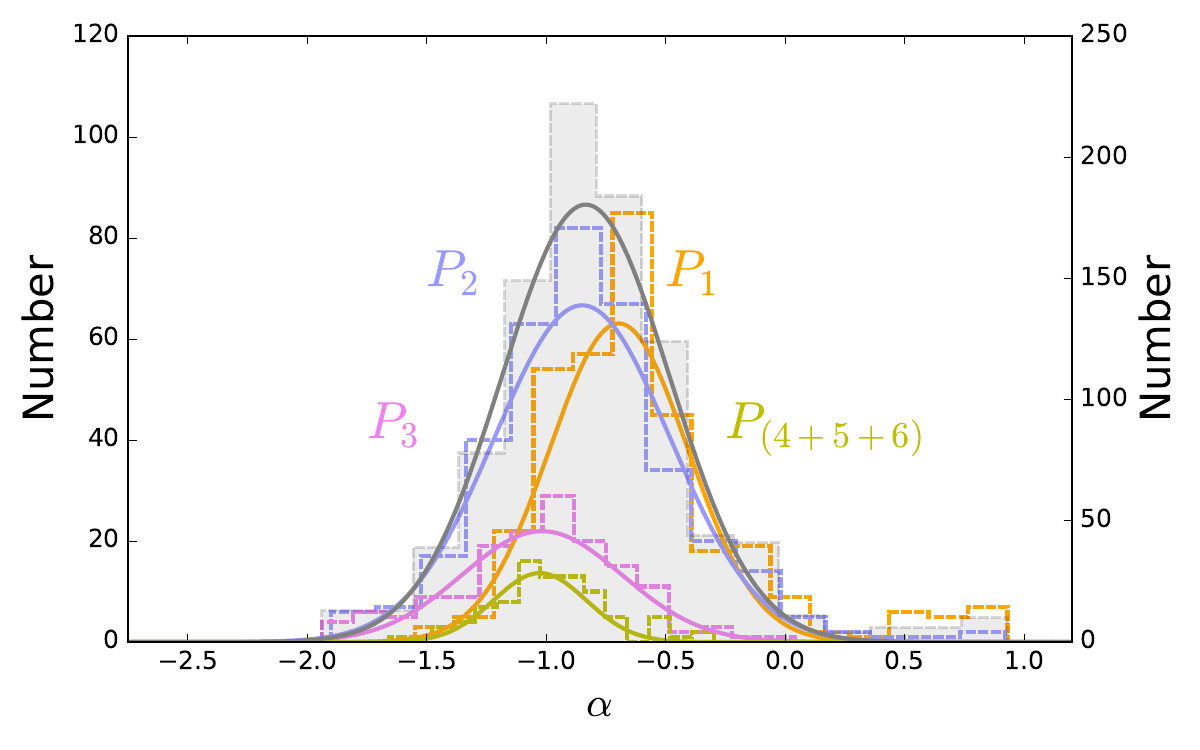}
\includegraphics[angle=0,scale=0.45]{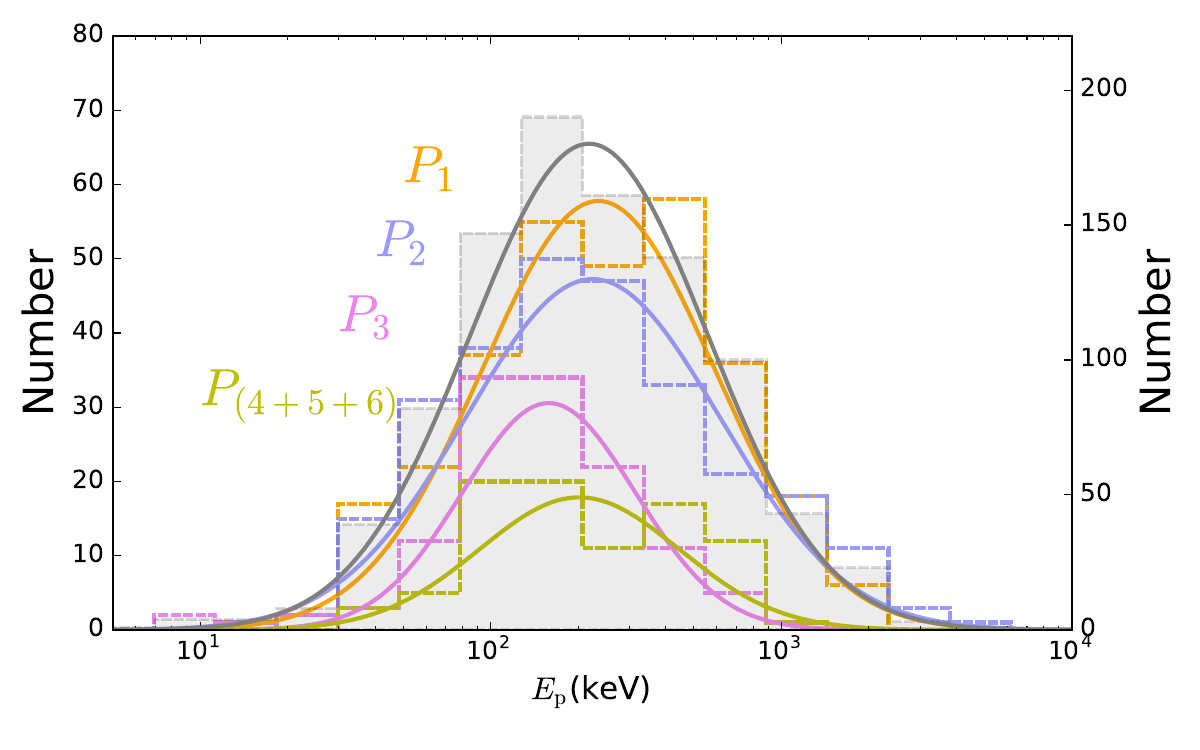}
\includegraphics[angle=0,scale=0.45]{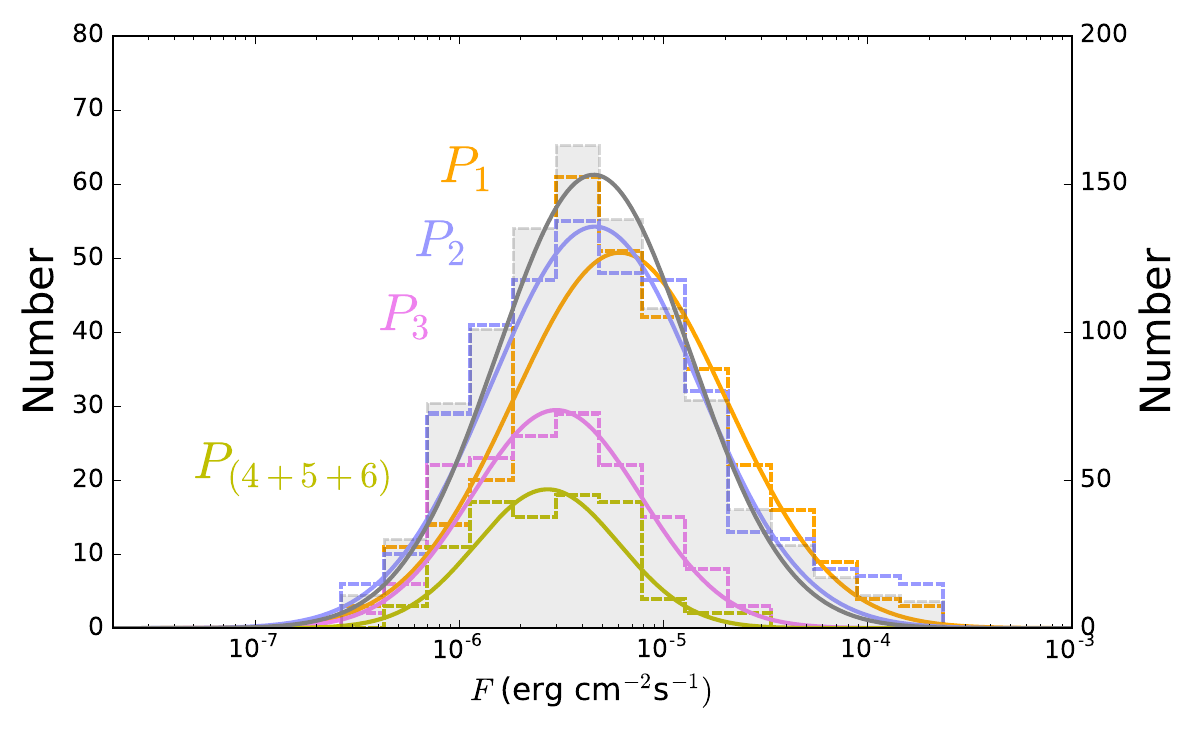}
\includegraphics[angle=0,scale=0.45]{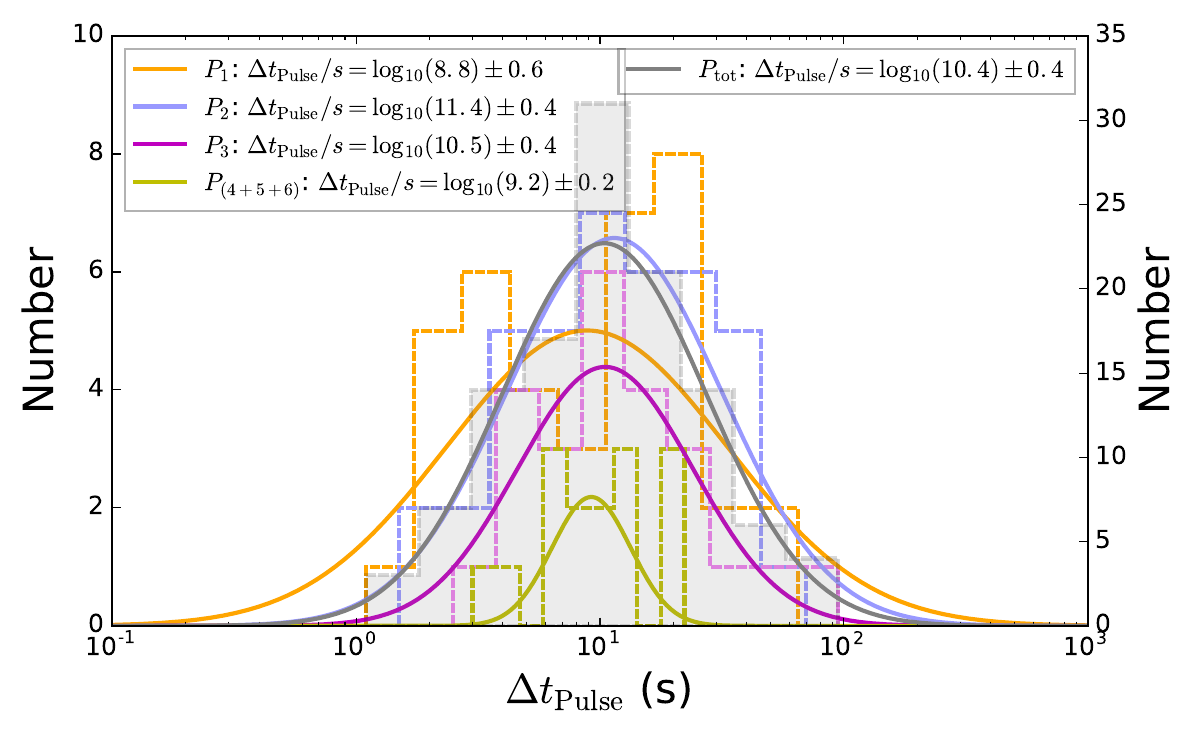}
\caption{Distributions of $\alpha$, $E_{\rm p}$, and energy flux, as well as the duration ($t_{90}$) for each pulse group. The distributions correspond to $P_{1}$ (orange), $P_{2}$ (turquoise), $P_{3}$ (violet), $P_{4}$ or $P_{(4+5+6)}$ (yellow), $P_{5}$ (cyan), and $P_{6}$ (pink). The black line is for the full sample. The y-axes on the left are for the  pulse groups, while the ones on the  right are for the full sample.}\label{fig:distrubution_best_model}
\end{figure*}

\begin{figure*}
\includegraphics[angle=0,scale=0.45]{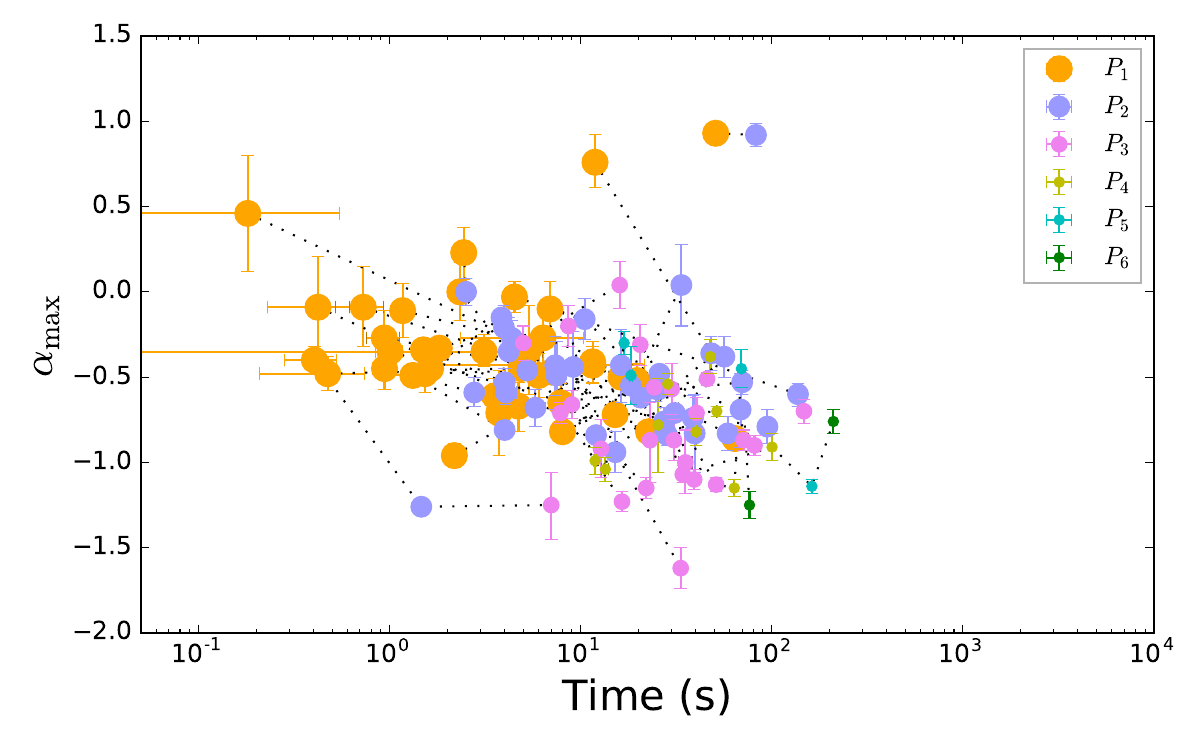}
\includegraphics[angle=0,scale=0.45]{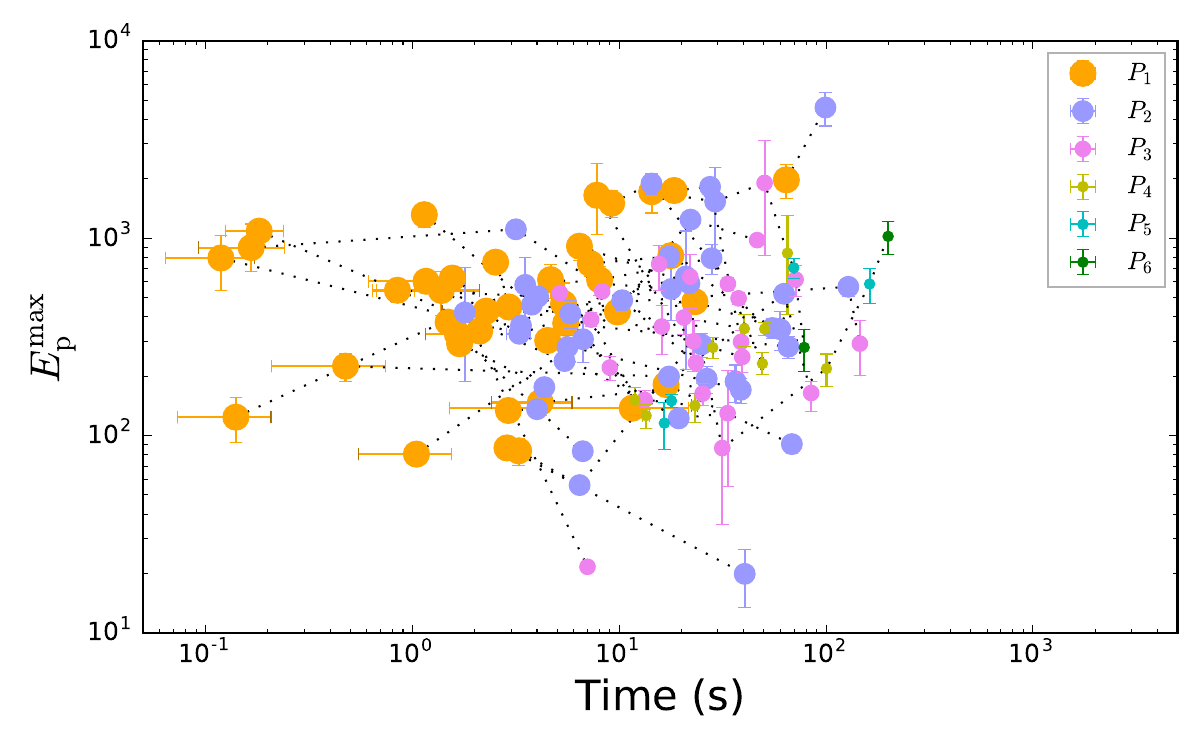}
\includegraphics[angle=0,scale=0.45]{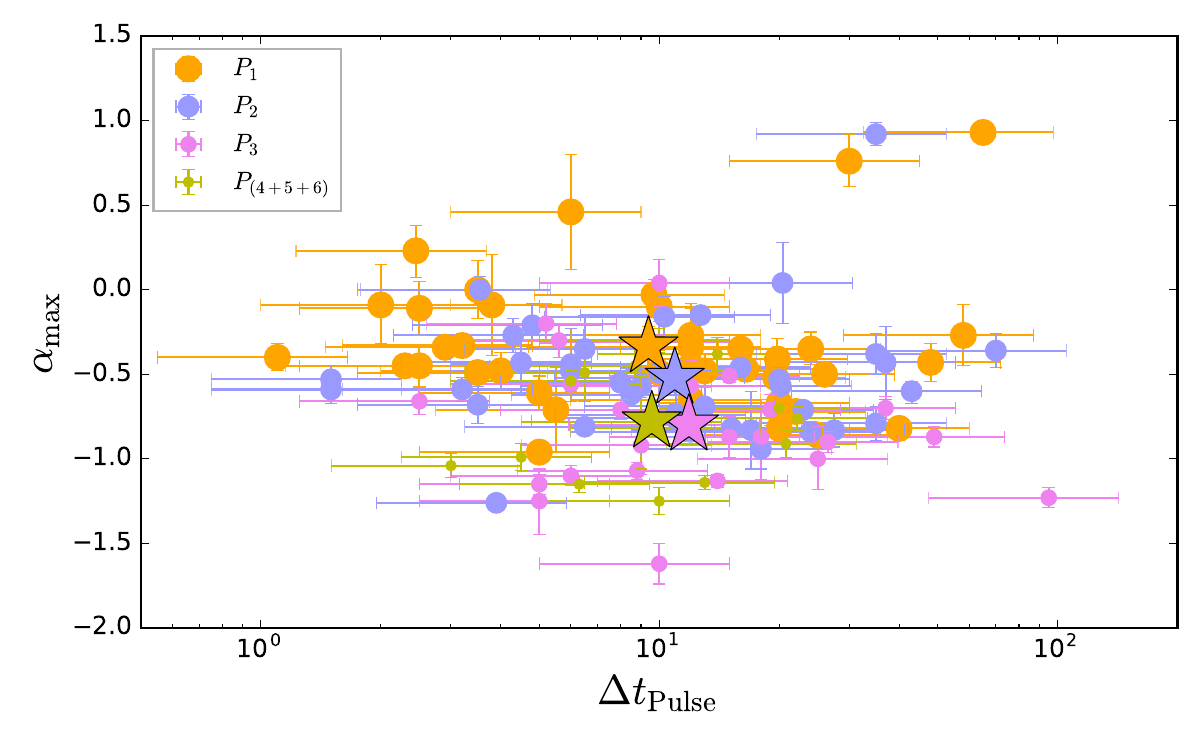}
\includegraphics[angle=0,scale=0.45]{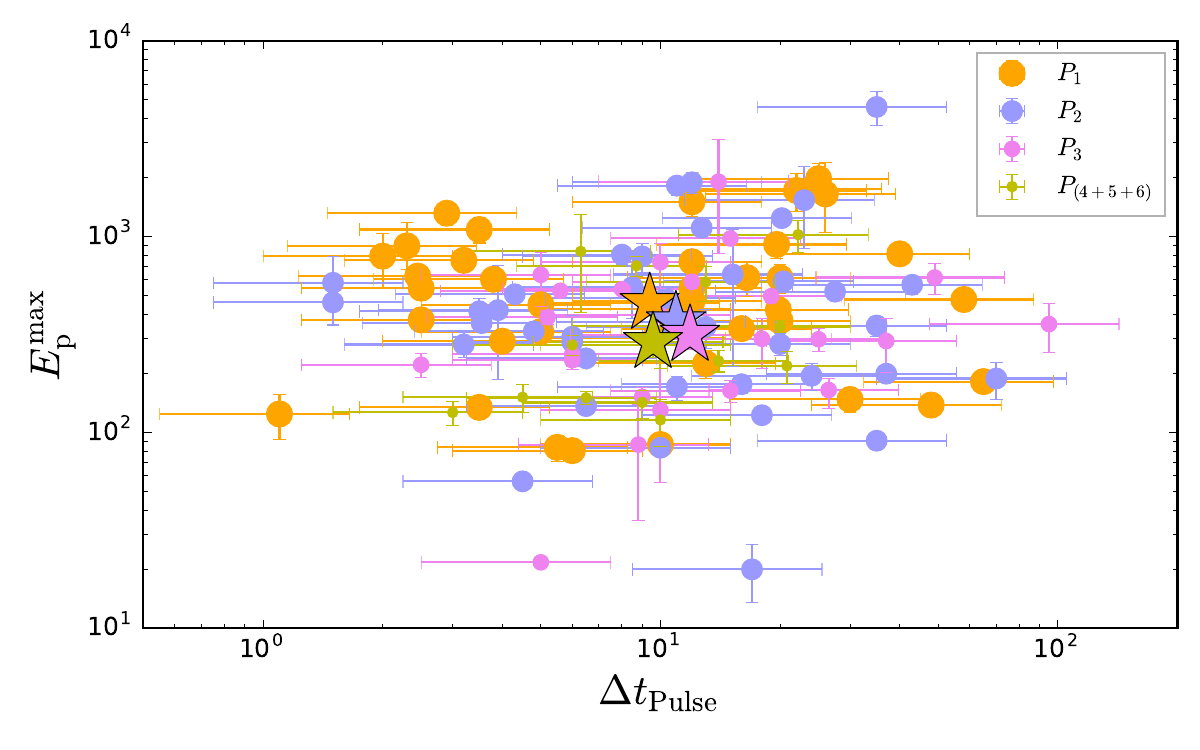}
\caption{Upper panels: evolution of $\alpha_{\rm max}$ (left panel) and $E_{\rm p, max}$ (right panel). Pulses from each individual burst are connected by dashed lines. Color notation is in the same as in Figure \ref{fig:distrubution_best_model}; lower panels: $\alpha_{\rm max}$ (left panel) and $E_{\rm p, max}$ (right panel) vs. pulse duration. The corresponding mean value of each individual pulse group is indicated by colored stars.}\label{fig:max}
\end{figure*}

\begin{figure*}
\includegraphics[angle=0,scale=0.3]{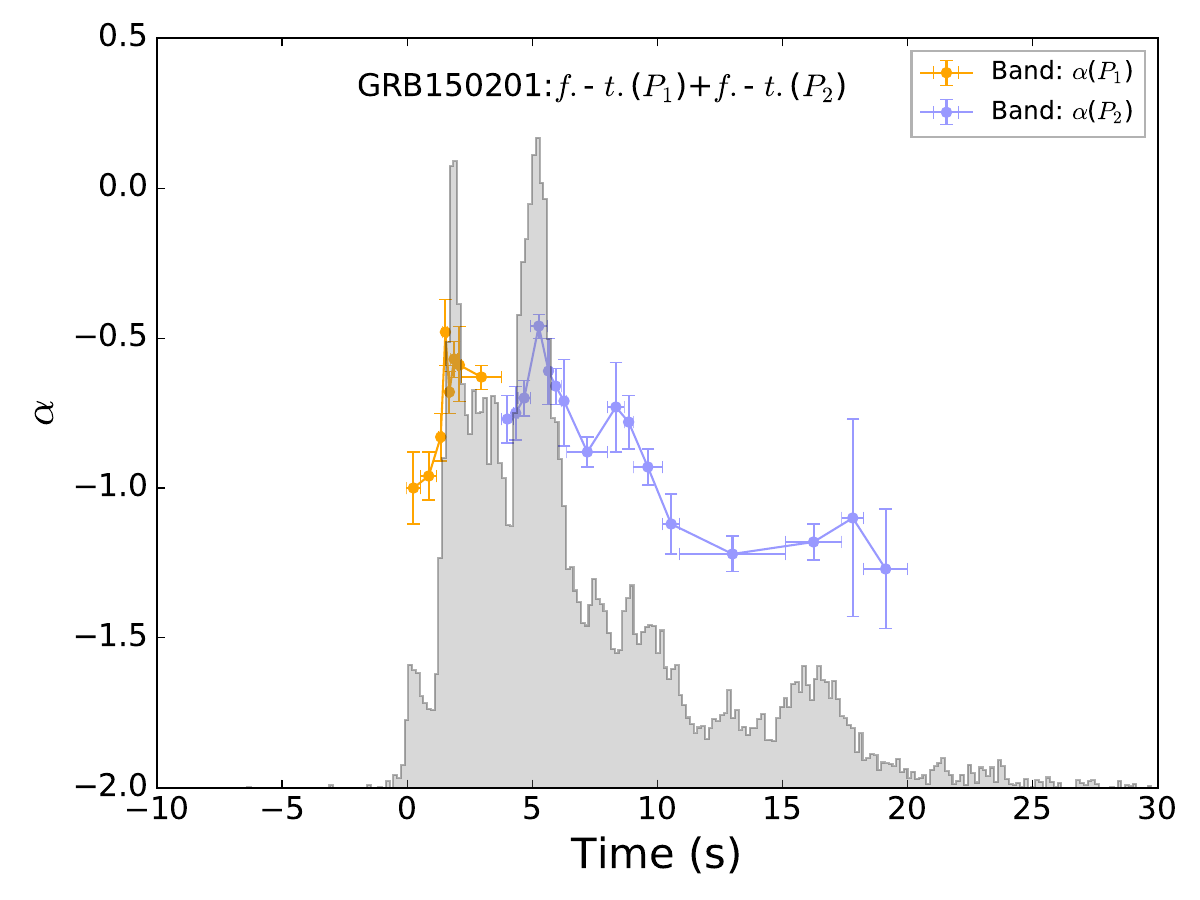}
\includegraphics[angle=0,scale=0.3]{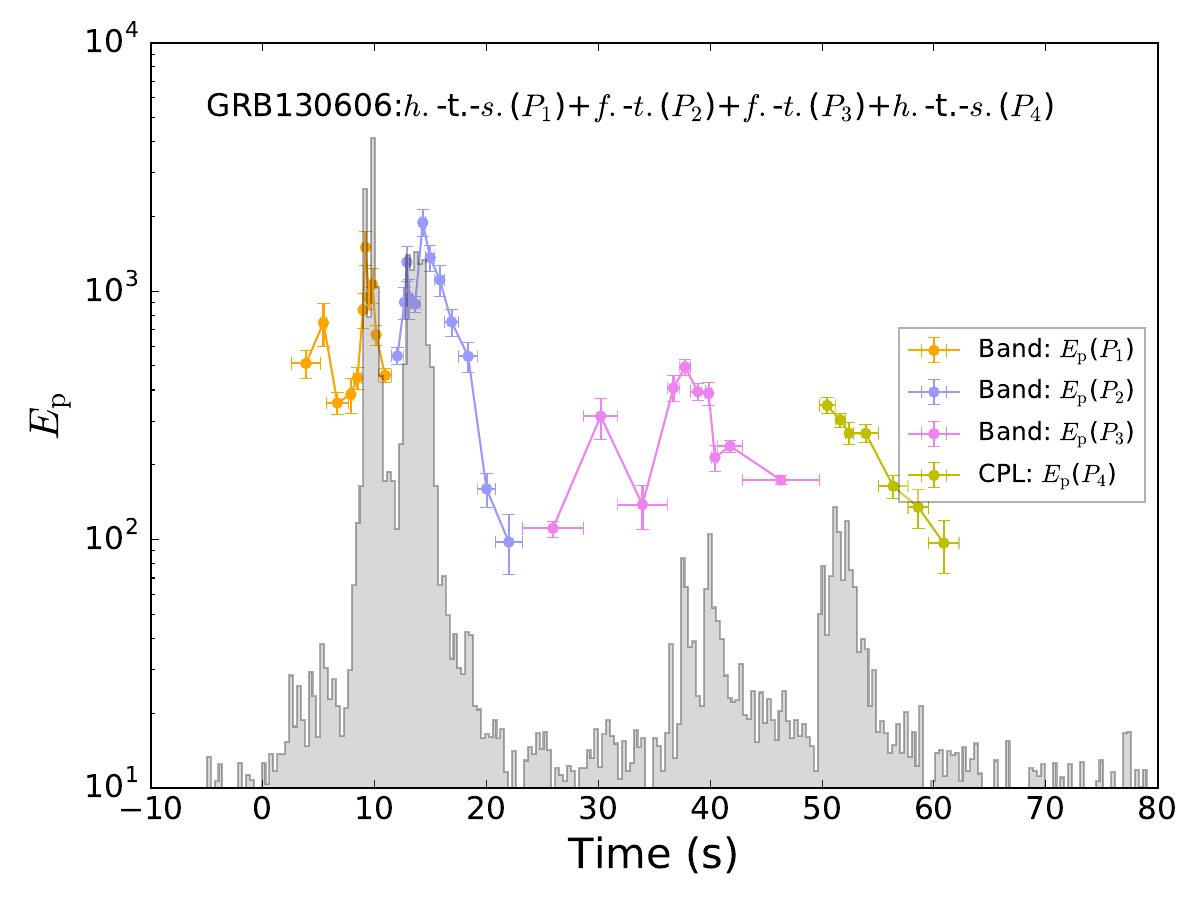}
\includegraphics[angle=0,scale=0.3]{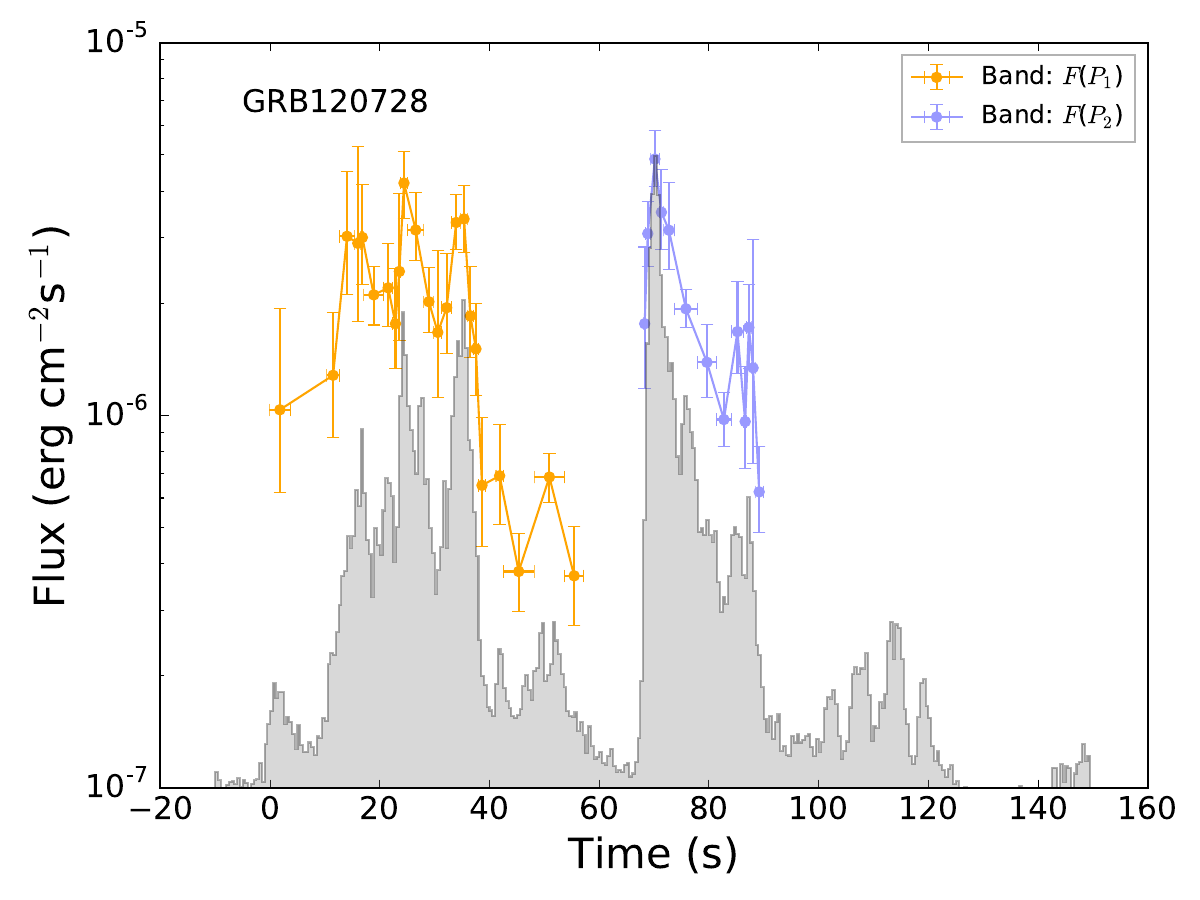}
\includegraphics[angle=0,scale=0.3]{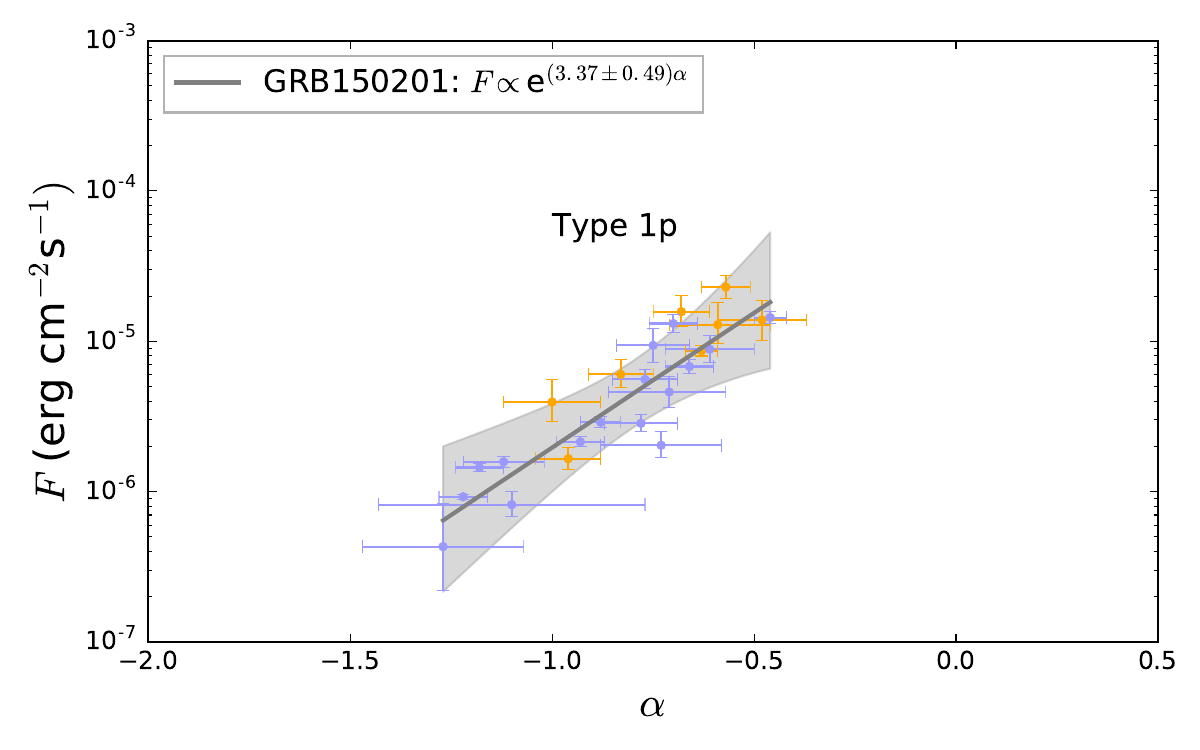}
\includegraphics[angle=0,scale=0.3]{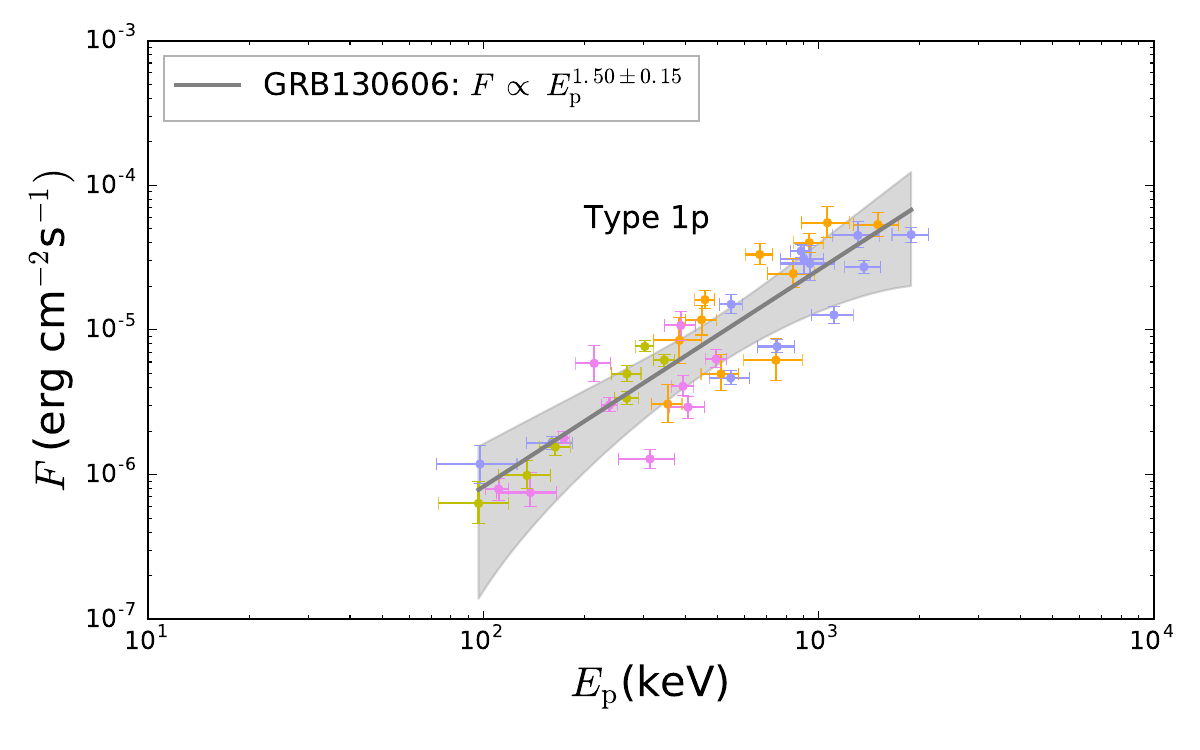}	
\includegraphics[angle=0,scale=0.3]{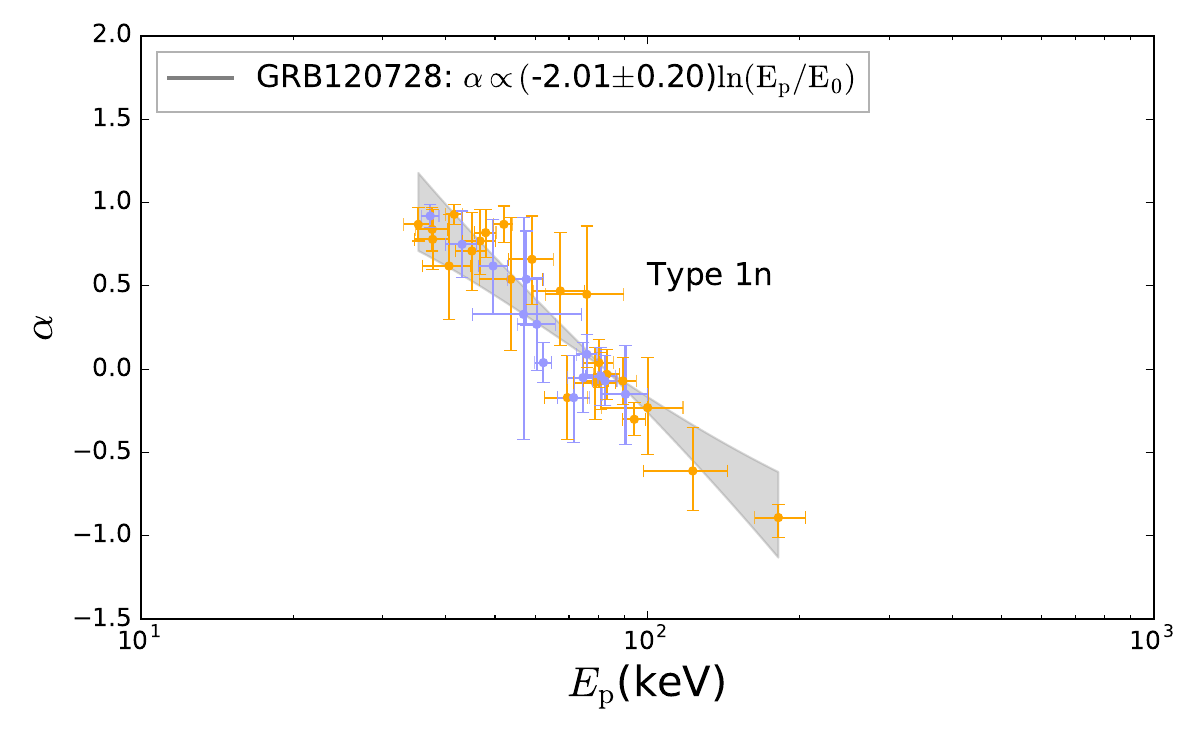}
\caption{Upper panels: examples of the parameter evolution of the $E_{\rm p}$, $\alpha$, and the $\nu F_{\nu}$ flux. The count light curves are shown (in arbitrary units) by the gray histograms. Lower panels: examples of the parameter relations of the $F-\alpha$, $F-E_{\rm p}$, and $\alpha-E_{\rm p}$, as well as the best-fit relations with the 2$\sigma$ error region.}\label{fig:example}
\end{figure*}

\begin{figure*}
\includegraphics[angle=0,scale=0.45]{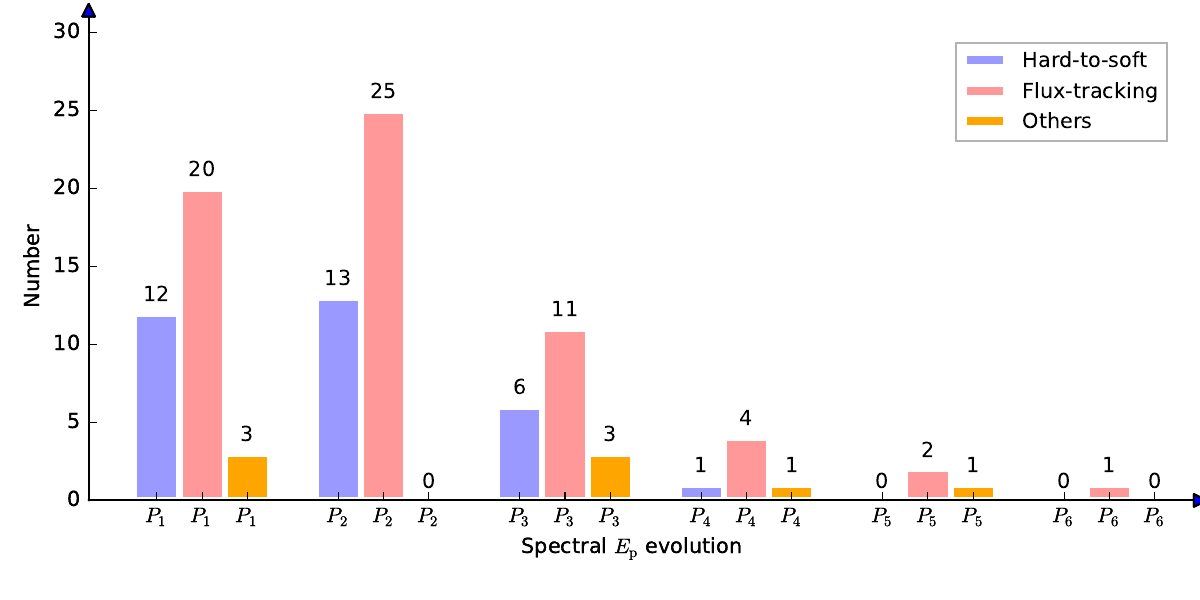}
\includegraphics[angle=0,scale=0.45]{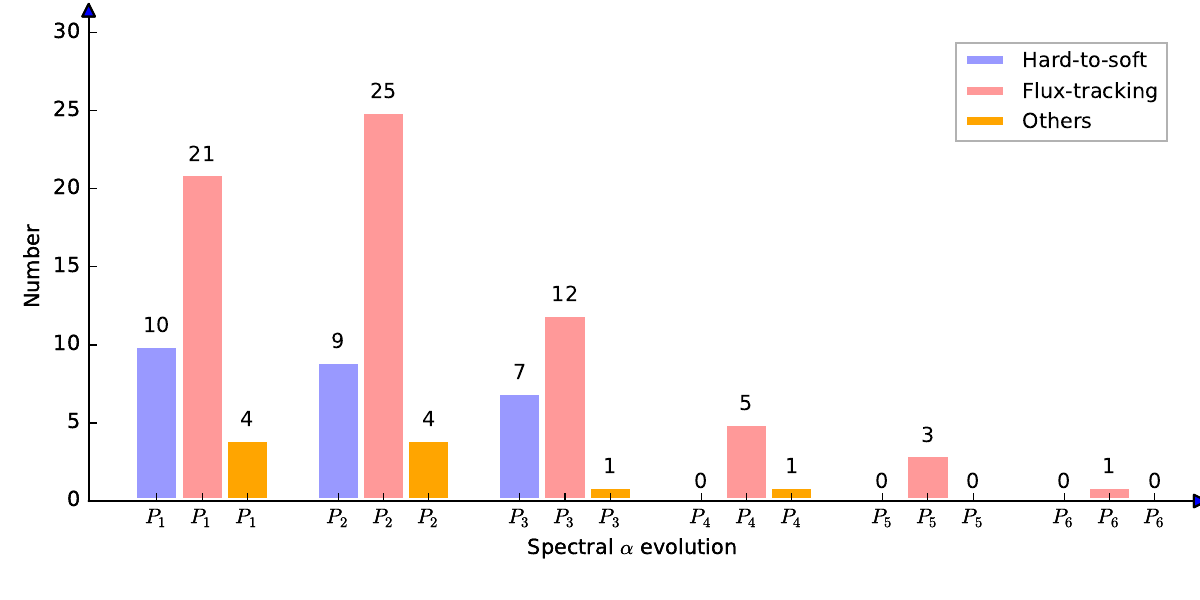}
\includegraphics[angle=0,scale=0.40]{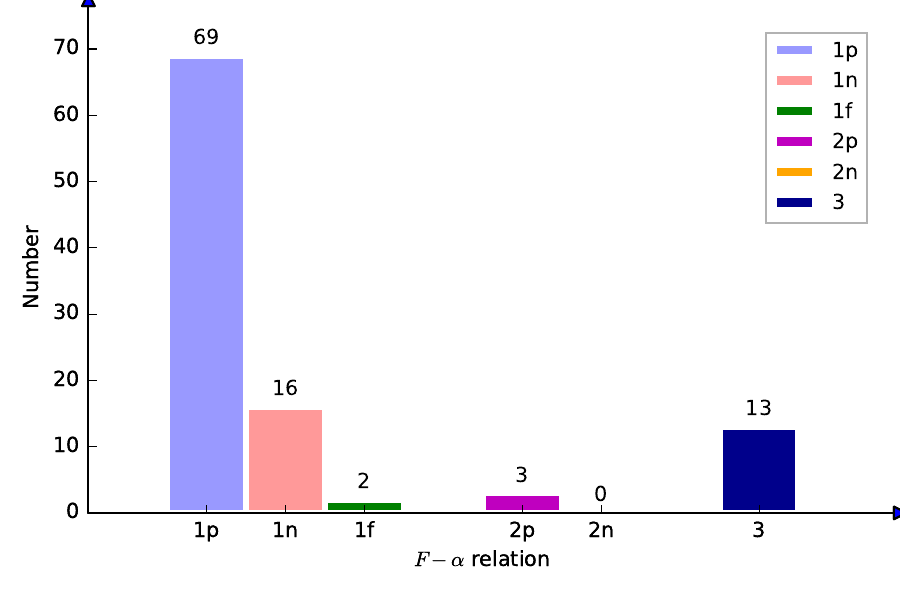}
\includegraphics[angle=0,scale=0.40]{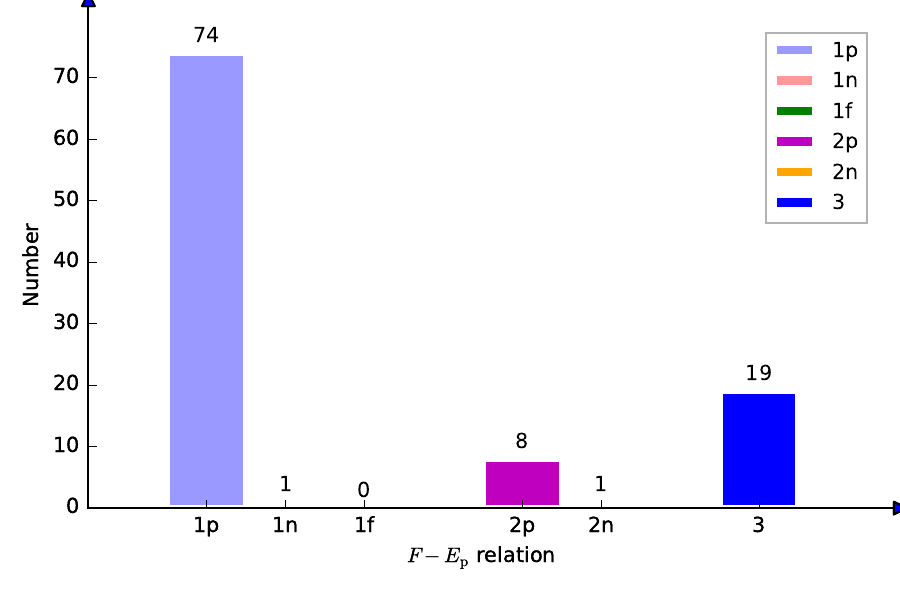}
\includegraphics[angle=0,scale=0.40]{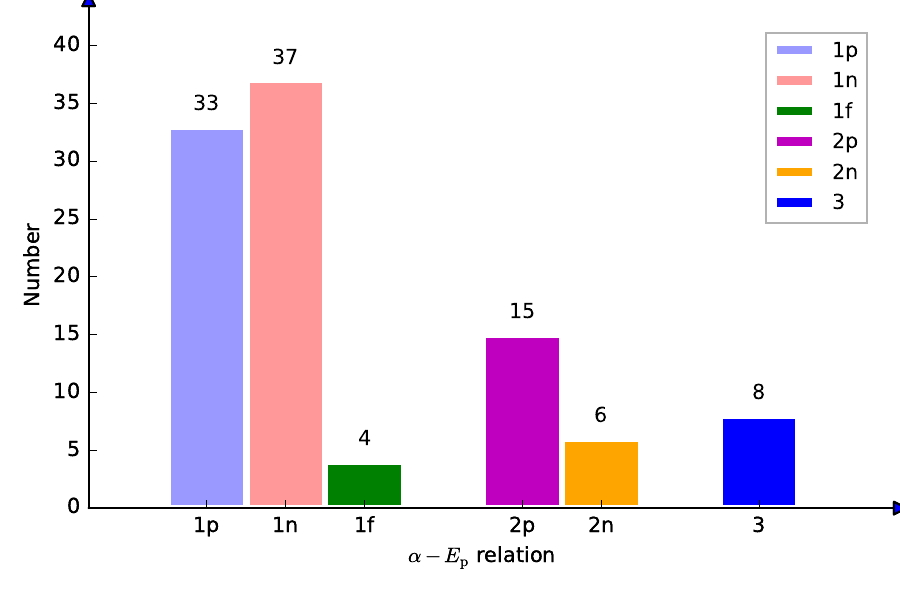}
\caption{Histograms of the spectral evolution and parameter relations. Upper left panel: $E_{\rm p}$ evolution. Upper right panel: $\alpha$ evolution. For each pulse, the histograms show the number of hart-to-soft and flux-tracking patterns, as well as some particular cases, denoted as ``others''.
Lower panels: classification of parameter relations in the entire sample. The histograms display the number pulses in the three main categories, as well as the subcategories for categories 1 and 2, as defined in the text. Lower left panel: $F$-$\alpha$ relation. Lower middle panel: $F$-$E_{\rm p}$ relation. Lower right panel: $\alpha$-$E_{\rm p}$ relation.}\label{fig:histogram}
\end{figure*}

Based on the analysis of the change in $\alpha$ and $E_{\rm p}$ (Section \ref{sec:distribution} and Fig. \ref{fig:distrubution_best_model}) alone, it was suggested that there are two categories of the pulse groups. The first one, consisting of $P_1$ and $P_2$, has a constant $E_{\rm p}$, while $\alpha$ softens. Within the second category, there is not much change, but they are all different from the first two pulses. Moreover, we found that the initial pulses had different spectral properties (frequency of different parameter correlation) as a group (\S \ref{sec:SC}). These distinctions lead to the possibility that they are due to different emission mechanisms. 

In the simplest internal shock model, the late pulses ($P_{i}$ with $i>1$) occur above the photosphere and hence must have a synchrotron origin. We do not see such a clear distinction, indicating that at least the most simplest version of the internal shock model does not represent what we see.

For the photospheric scenario, the $\alpha$ value reflects the energy dissipation and photon production below the photosphere \citep[e.g., ][]{Peer2006, Vurm2016}. The $E_{\rm p}$ instead is mainly set by properties at high optical depths. Therefore, if the first two pulses are photospheric, the fact that  $E_{\rm p}$ is similar while $\alpha$ varies could be due to similar properties close to the central engine but a varying amount of turbulence in the flow causing the dissipation. The variability and pulse structures of the photospheric emission have been reproduced by numerical simulations of a jet passing through the progenitor surrounding. For instance, \citet{Lopez2014} showed that even with a steady central engine, a light curve with a pulse structure and large variability arises. The main cause is the Rayleigh-Taylor instabilities that arise in the contact between the layers of jet and the surrounding progenitor material. This leads to a variable amount of mixing between the layers and thereby a variable baryon load of the jet, which has a direct influence on the radiative efficiency and spectral shape \citep[see, e.g., ][]{Rees2005, Gottlieb2019}. 

On the other hand, the pulse groups, $P_3$--$P_6$, all have $\alpha \sim -1$ (Fig. \ref{fig:distrubution_best_model}). This could be explained by synchrotron emission from electrons that are marginally fast-cooled \citep{Daigne2011, Yu2015, Geng2018}. It can thus be argued that the first two pulses are typically photospheric, while the rest are due to synchrotron emission. If this interpretation is correct, one would expect that the two categories would have distinct and different spectral properties. There are several points in the analysis above that, therefore, do not support this interpretation. First, the values of $E_{\rm p, max}$ are similar for all of the pulse groups (Fig. \ref{fig:max}); there is no clear distinction between the pulse groups. Second, there is no clear distinction between pulse groups when it comes to the spectral evolution (\S \ref{sec:classification}) and types of correlations (apart from the initial pulses; \S \ref{sec:SC}). Third, the average pulse durations are also similar in all pulse groups. All of these points indicate that there is no drastic change in emission pattern between the pulses (apart from $\alpha$).

On the contrary, the properties of the four last pulse groups can also be interpreted as photospheric emission. Previous studies have shown that pulses that, beyond any doubt, are photospheric do all have significant spectral evolution \citep[e.g., ][]{Ryde2019}. In particular, at the end of such pulses, $\alpha$ values down to -1 and below are common, which is interpreted as the result of subphotospheric dissipation with varying jet properties \citep[e.g., ][]{Ryde2010, Ryde2011}. Consequently, pulses with $\alpha \sim -1$ could be the result of dissipative photospheres throughout the pulse. Indeed, the observation that the last four pulse groups all peak at around $\alpha \sim -1$ is in line with the theoretical expectations of fully dissipative photospheres, i.e., flows where there is no strong limitation, not on the soft photon production deep in the flow or the amount of dissipation in the parts of the flow where the spectrum is formed \citep{Beloborodov2010, Vurm2013}. 

Even though this line of argument makes the subphotospheric dissipation model appealing for most of the pulses, some admixture of synchrotron pulses is expected. The most prominent example of this is GRB190114C, which, apart from a very hard spectral component, showed a clear afterglow emission component already during the prompt phase \citep{ajello2020}. This proved that a synchrotron component was present during the prompt phase \citep[see also][]{Axelsson2012, Iyyani2013}. Further early examples of such a suggestion are given in \citet{Ryde2005, Ryde2009, Guiriec2011} and more recently in \citet{Wang2019, Burgess2019, Wang2019b}.

Many of these synchrotron components are observed at late times. Combined with the fact that the fraction of (certain) photospheric pulses decreases with pulse number (Fig. \ref{fig:max_dis}), emission compatible with synchrotron therefore appears to prevail mainly at the end of the prompt phases. Different possibilities exist for synchrotron emission to arise toward the end of the prompt phase.  One example is GRB 150330, for which \citet{Li2019a} suggested that the jet composition changes from a baryon-dominated flow during the first pulse to a Poynting-flux dominated flow during the rest of the burst, producing synchrotron emission \citep[see also][]{Zhang2018a}. Alternatively, during the early phases of the afterglow, shocks due to interaction between the jet and slower-moving material ahead of the jet will also produce efficient synchrotron emission \citep[e.g., ][]{Duffell2015}. 

However, from the GBM data alone, it is not possible to make a firm conclusion as to whether a soft pulse is due to synchrotron emission or due to photospheric emission subject to dissipation below the photosphere, since they are indistinguishable in many cases \citep[e.g., ][]{Acuner2020}. Simultaneous data from other wavelengths can be useful in some cases be useful \citep[e.g., ][]{Ravasio2018, Ahlgren2019, ajello2020} or polarization measurements \citep[e.g., ][]{Sharma2019}.

\section{Summary}

In this paper, we performed time-resolved spectroscopy on a sample of multipulse GRBs observed by {\it Fermi}-GBM during the first 11 yr of its mission. We investigated  the variation in emission properties between the pulses in light of the prediction of emission models. Our sample consists of 39 bursts, from which have 117 distinct pulse structures, entailing 1228 time-resolved spectra. All of the spectra have a very high statistical significance of $S \geq 20$. This ensures that the spectral fits are well determined and that the low-energy power-law index $\alpha$ can be used to discriminate between spectra that are compatible with the photosphere or with synchrotron emission. The emission properties we studied include the spectral shape and the correlations between spectral shape parameters. 

For the sample as a whole, we found that flux-tracking evolution is more common than the hard-to-soft evolution, independent of the $E_{\rm p}$ or $\alpha$ evolution, differing from previous findings. We also found that a positive correlation is most common for the $F$-$\alpha$ and $F$-$E_{\rm p}$ relations. In contrast, for the $\alpha$-$E_{\rm p}$ relation, both the positive and negative correlations are equally common. In addition, we compared our sample to that of the single-pulse sample of \citet{Yu2019}. We found that the peak flux is significantly larger and the average $\alpha$ value is softer in our sample, while peak energies are similar. On the other hand, we found that the average $\alpha$ value of the initial pulses of our multi-pulse sample is similar to that of the single-pulse sample.

Specifically, we searched for signatures of any characteristic variation in the emission properties between pulses that might reveal different underlying emission processes. We find that the characteristics of the pulses remain astonishingly similar. It is mainly the low-energy power-law index, $\alpha$ that has a significant softening (gets smaller). In addition, we find that, on average, the first pulse in each burst behaves slightly differently than consecutive pulses when it comes to correlations between spectral shape parameters.

\begin{figure*}
\centering
\includegraphics[angle=0,scale=0.63]{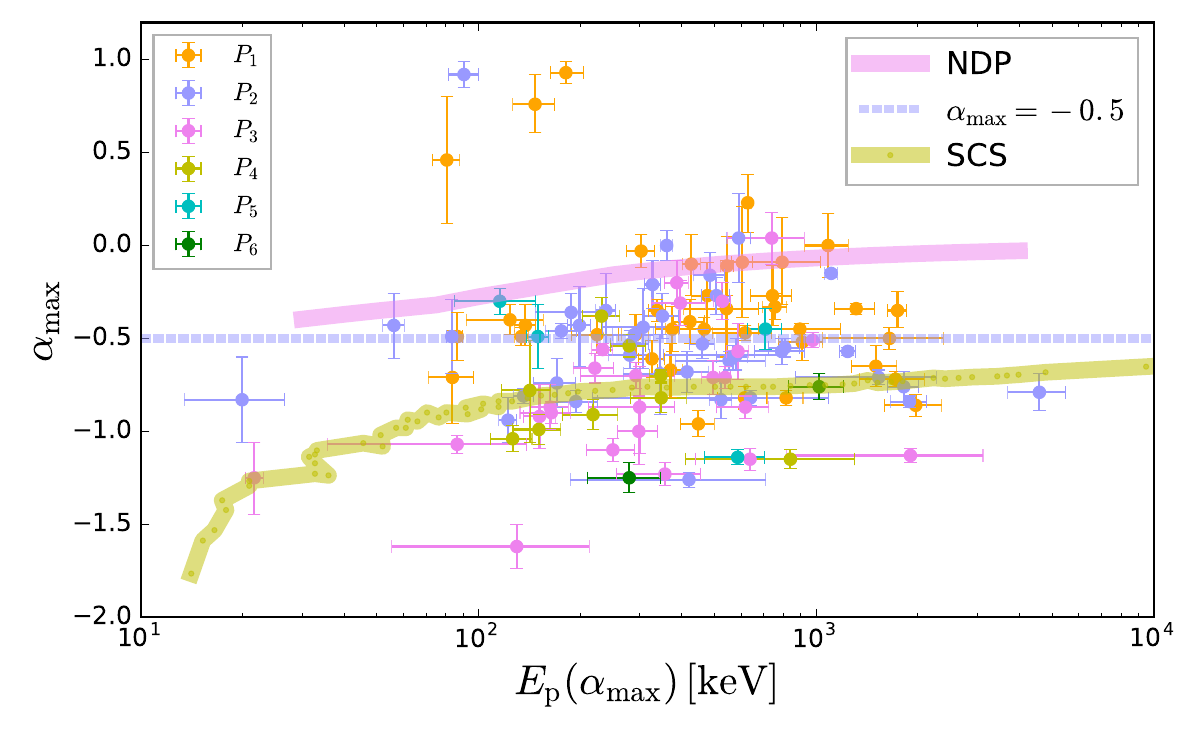}
\includegraphics[angle=0,scale=0.60]{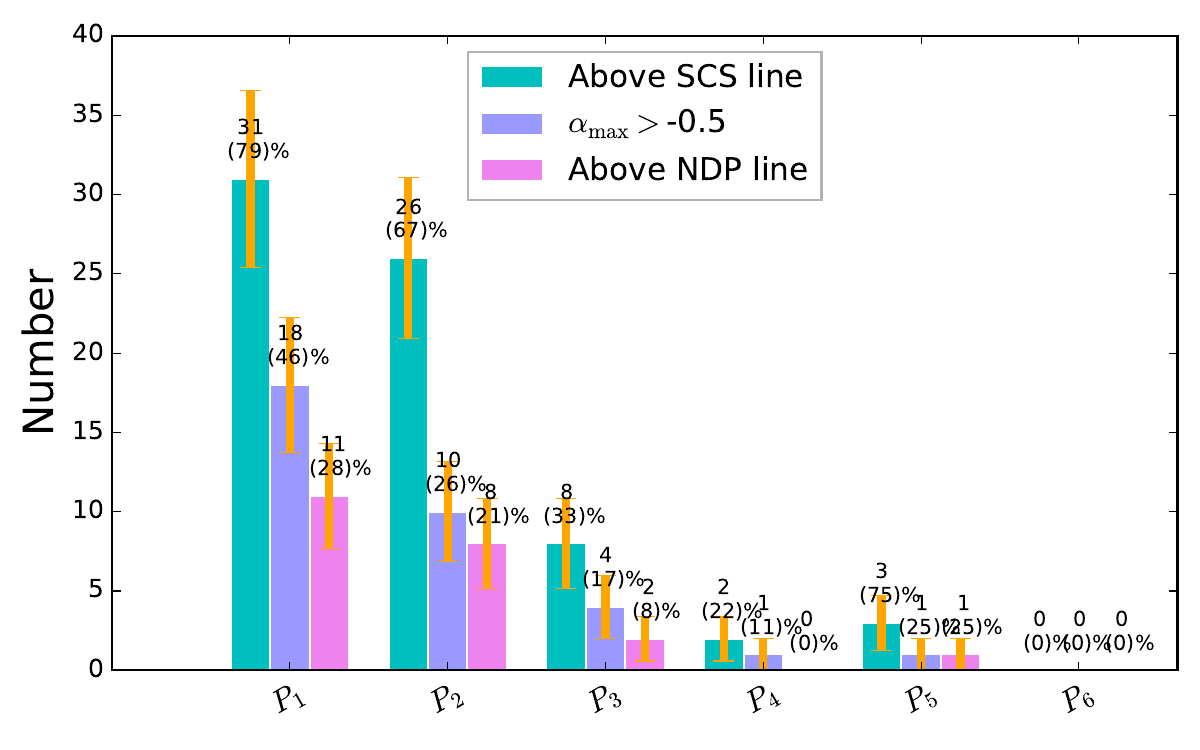}
\caption{Compatibility with photospheric emission models. Upper panel: $\alpha_{\rm max}$ vs. the corresponding $E_{\rm p}$ with the same color notation as in Fig. \ref{fig:distrubution_best_model}. The pink line is the limiting line for the NDP and the green line is the limiting line for SCS. Lower panel: The number and fraction of $\alpha_{\rm max}$ bins that lie above the SCS line (green), the number and fraction of bins with $\alpha_{\rm max}> -0.5$ (blue), and the number and fraction of bins that lie above the NDP line (pink).}\label{fig:max_dis}
\end{figure*}

\begin{figure*}
\includegraphics[angle=0,scale=0.5]{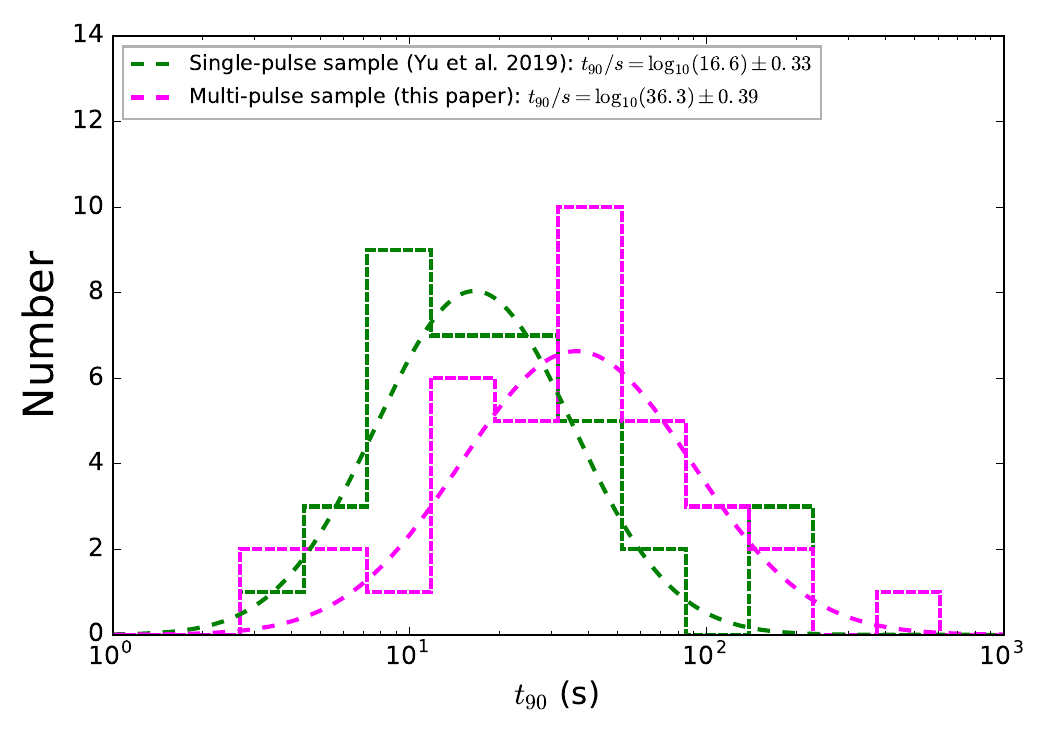}
\includegraphics[angle=0,scale=0.5]{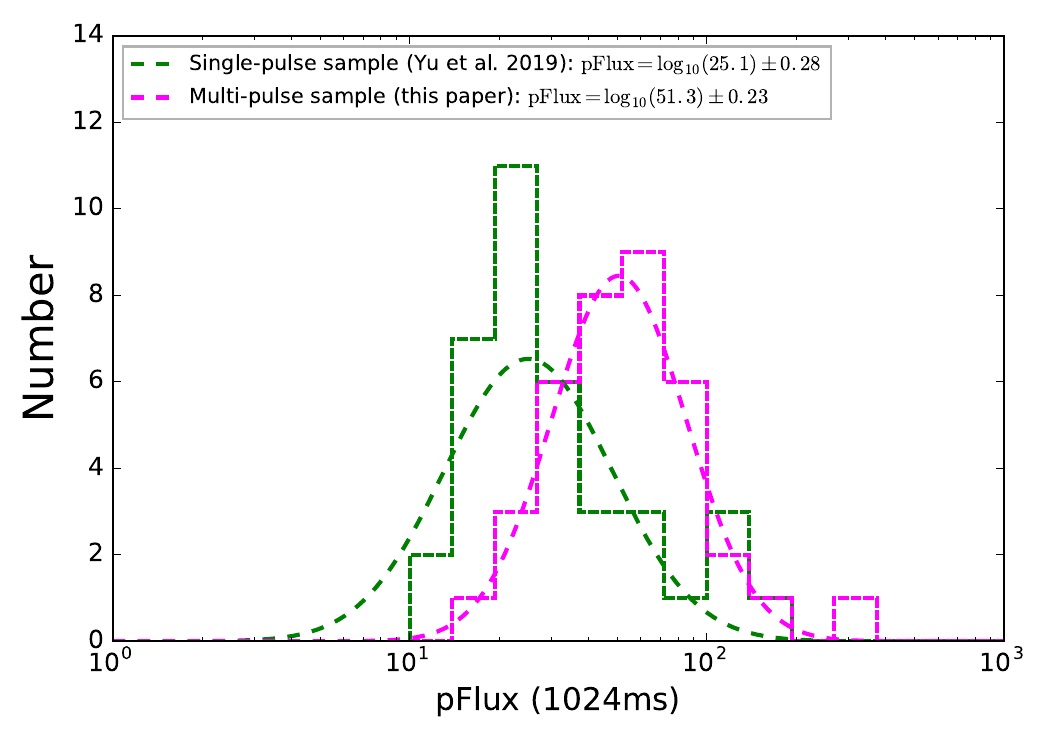}
\includegraphics[angle=0,scale=0.5]{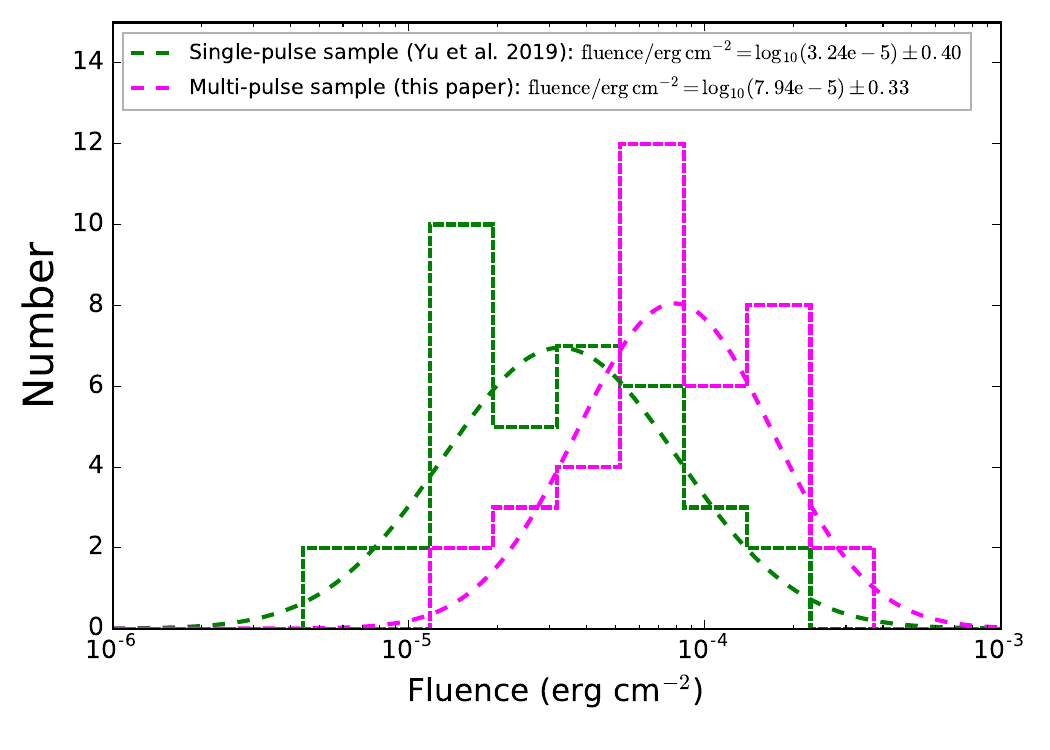}
\includegraphics[angle=0,scale=0.5]{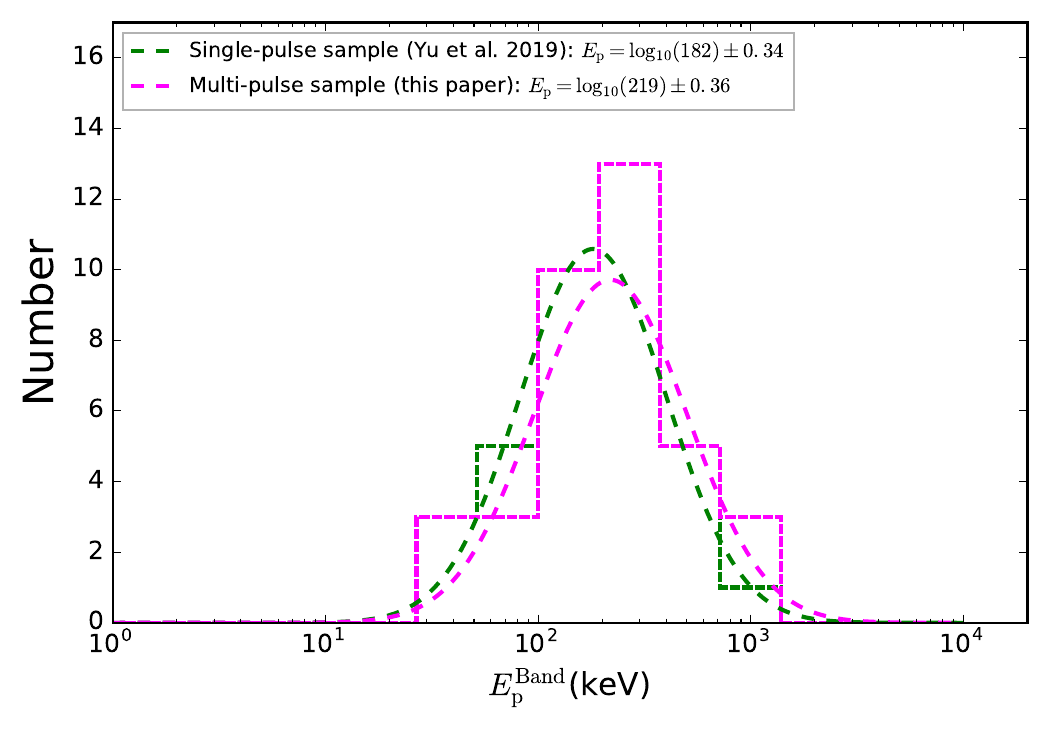}
\includegraphics[angle=0,scale=0.5]{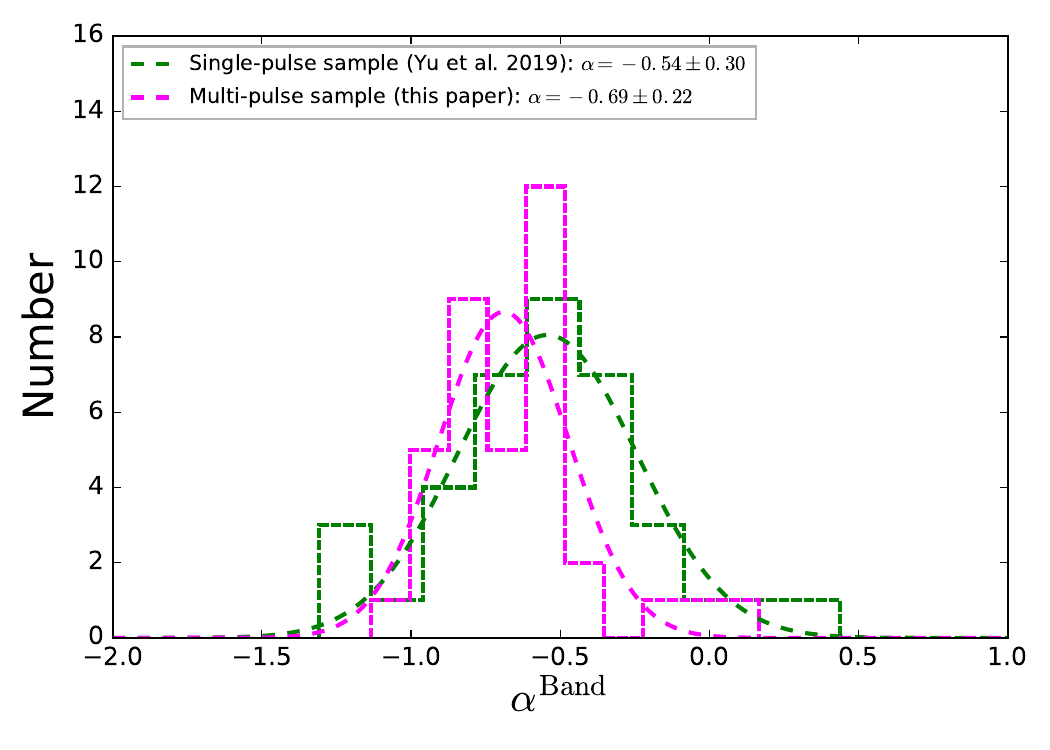}
\includegraphics[angle=0,scale=0.5]{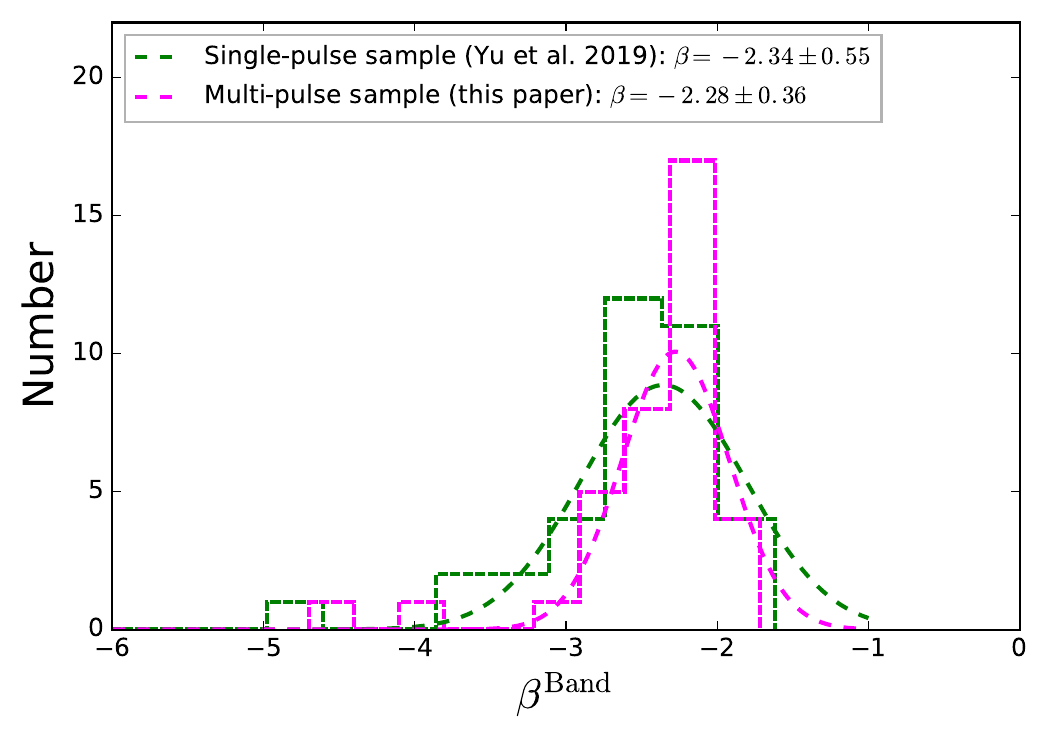}
\caption{Parameter distributions of the duration $T_{\rm 90}$, peak flux, fluence, $E_{\rm p}$, $\alpha$, and $\beta$ from time-integrated spectra. The sample in this paper is shown by the magenta curves, while the single-pulse sample in \citet{Yu2019} is shown by the green curves. }\label{fig:comparison}
\end{figure*}

\begin{figure*}
\includegraphics[angle=0,scale=0.45]{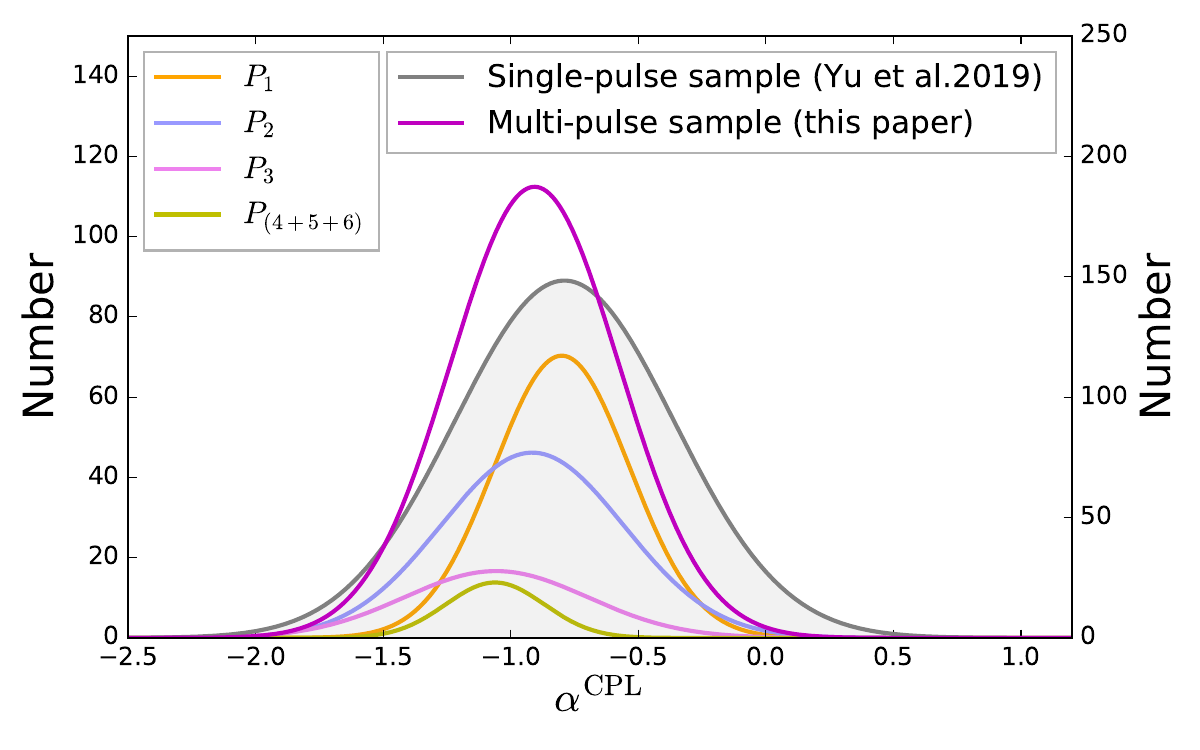}
\includegraphics[angle=0,scale=0.45]{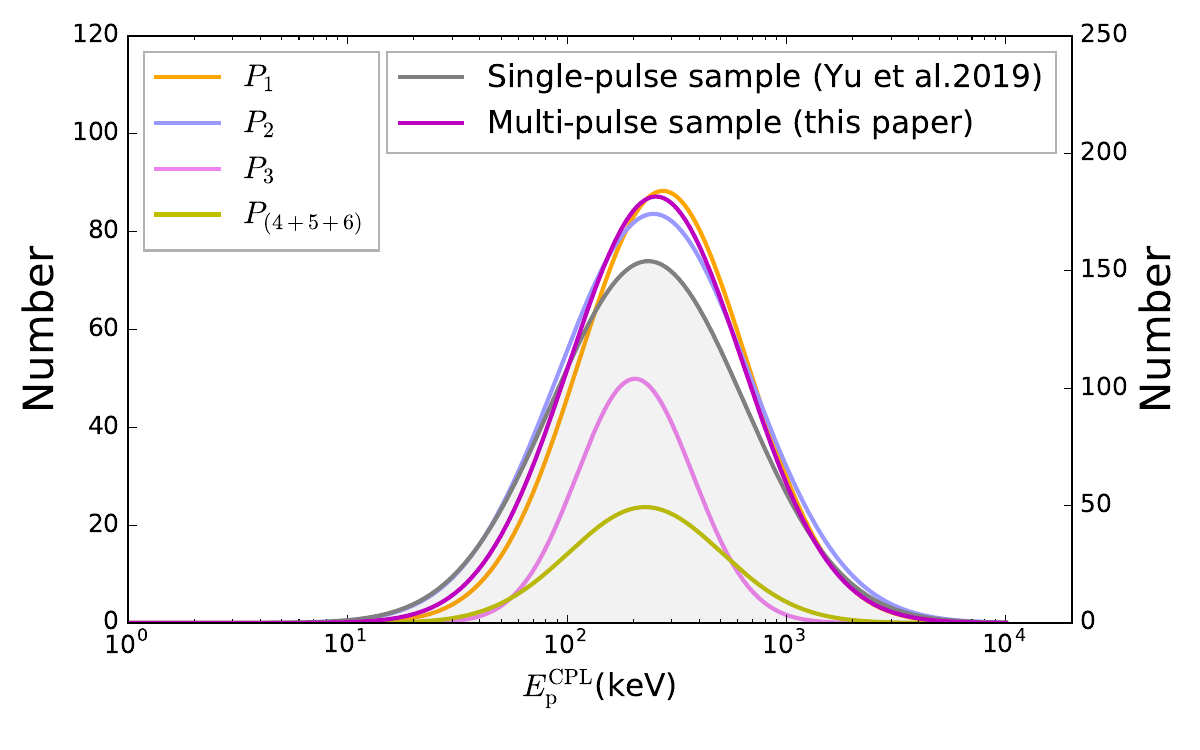}
\begin{center}
\includegraphics[angle=0,scale=0.45]{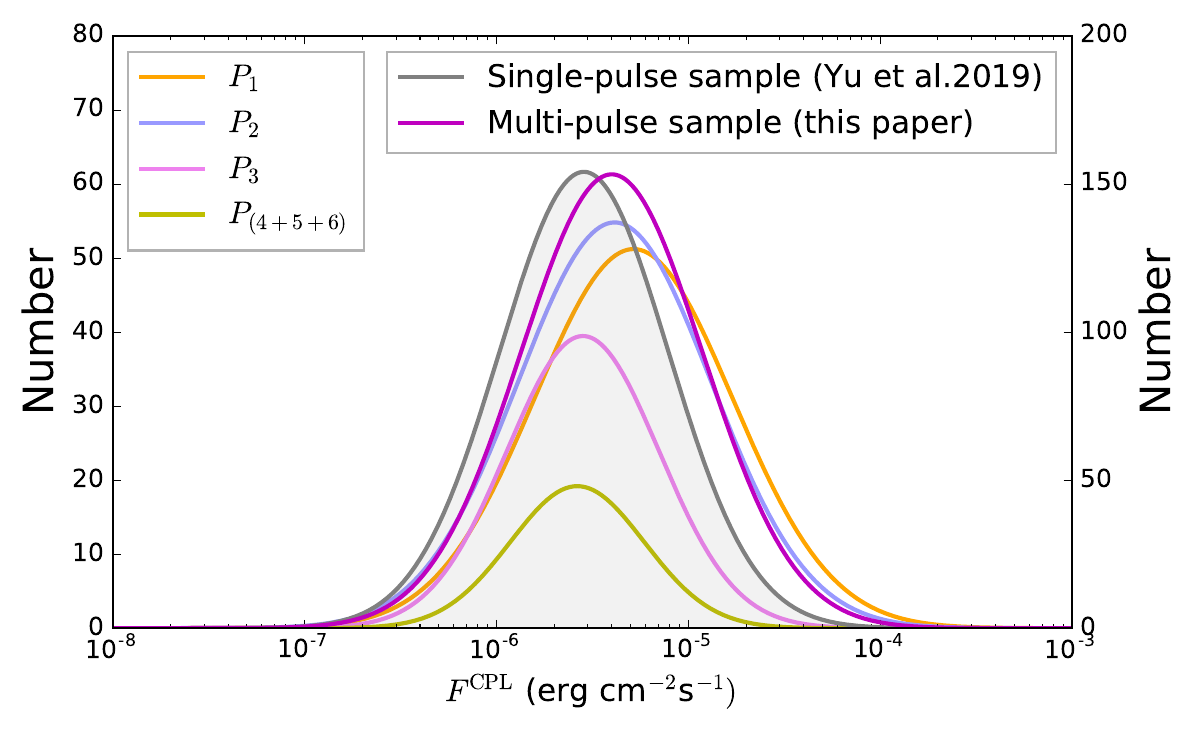}   
\end{center}
\caption{Same as Figure \ref{fig:comparison}, but the analysis is based on the time-resolved spectral results. The y-axes on the left are for the pulse groups, while the ones on the right are for the full samples. The color notation is the same as in Fig. \ref{fig:distrubution_best_model} for the pulse group samples. The black line is for the full sample in \cite{Yu2019} while the purple line is for the full sample in this paper.}\label{fig:distribution_Band_CPL_time_resolved}
\end{figure*}

We further assessed the compatibility of the data with emission models for each individual pulse. Assuming that the same emission mechanism operates during a pulse, any single time bin that violates the synchrotron limit indicates that a photospheric origin is more probable (Fig. \ref{fig:max_dis}). We also used the criterion that $\alpha_{\rm max} > -0.5$ which identifies pulses that significantly prefer photospheric emission \citep{Acuner2020}. The first pulse in a burst is clearly different from the later pulses: three-fourths of them violate the synchrotron emission, and half of them prefer photospheric emission. These fractions decrease rapidly for subsequent pulses. 

We argue that in many cases, synchrotron emission in later pulses might contribute to these trends. However, the similarity between pulses, averaging over the whole sample, points to a photospheric origin of most pulses, albeit with greatly varying dissipation properties

We conclude that photospheric emission can be found at any time during the burst duration; however, it is more common in the early phase. In order to make a general statement of the emission mechanism in a GRB, the spectral softening of $\alpha$ between pulses is a property that needs to be considered. In particular, the analysis of individual pulses will be influenced  by their occurrence relative to the trigger time. The chance to detect the photosphere is largest among the first few pulses, while synchrotron emission is mainly found at late times. This also allows for the coexistence of emissions at late times.

\acknowledgments

We would like to thank Drs. Magnus Axelsson, Damien B\'egu\'e, H\"usne Dereli-B\'egu\'e, and Yu Wang for useful discussions. This research is supported by the Swedish National Space Agency and  made use of the High Energy Astrophysics Science Archive Research Center (HEASARC) Online Service at the NASA/Goddard Space Flight Center (GSFC). In particular, we thank the GBM team for providing the tools and data that were used in this research. F.R. is supported by the G\"oran Gustafsson Foundation for Research in Natural Sciences and Medicine and the Swedish Research Council (Vetenskapsr{\aa}det), while A.P. acknowledges support from the EU via the ERC grant O.M.J. Part of this work was performed during Dr. Liang Li's visits with Professors Anzhong Wang, Qiang Wu, and Tao Zhu at the United Center for Gravitational Wave Physics (UCGWP), and Professors Yefei Yuan and Yifu Cai at the University of Science and Technology of China (USTC), China.

\vspace{10mm}
\facilities{{\it Fermi}/GBM}
\software{3ML\citep{Vianello2015}}
\bibliography{MyReferences.bib}

\clearpage

\clearpage

\clearpage

\clearpage

\clearpage

\clearpage

\clearpage

\clearpage

\clearpage
\begin{deluxetable}{cccccccc}
\tablewidth{0pt}
\tabletypesize{\scriptsize}
\tablecaption{Global Properties of the Sample}
%\tablenum{1}
\tablehead{
\colhead{GRB}
&\colhead{$z$}
&\colhead{$T_{90}$}
&\colhead{Detectors}
&\colhead{$\Delta T_{\rm src}$}
&\colhead{$[\Delta T_{\rm (bkg,1)},\Delta T_{\rm (bkg,2)}]$}
&\colhead{Spectra ($N_{S\geq20}$)}
&\colhead{Pulse ($N_{(S\geq20)}\geq 4$)}\\
&
&(s)
&
&(s)
&(s)
&(Number)
&(Number)
}
\colnumbers
\startdata
081009140&\nodata&41.345$\pm$0.264&(n3)b1&-1 to 52&[-20 to 10, 80 to 100]&23(21)&2(2)\\
081215784&\nodata&5.568$\pm$0.143&n9(na)nbb1&-1 to 10&[-30 to 10, 40 to 60]&27(22)&3(3)\\
090131090&\nodata&35.073$\pm$1.056&(n9)nab1&-1 to 40&[-20 to 10, 60 to 80]&30(16)&3(2)\\
090618353&0.54&112.386$\pm$1.086&(n4)b0&-1 to 150&[-20 to 10, 200 to 220]&28(25)&3(2)\\ 
091127976&0.49&8.701$\pm$0.571&(n6)n9nab1&-1 to 10&[-20 to 10, 50 to 100]&21(17)&3(3)\\ 
100719989&\nodata&21.824$\pm$1.305&(n4)n5b0&-1 to 25&[-30 to 20, 60 to 100]&18(12)&3(3)\\ 
100826957&\nodata&84.993$\pm$0.724&n7n8b1&-1 to 120&[-20 to 10, 200 to 250]&31(26)&3(3)\\ 
101014175&\nodata&449.415$\pm$1.409&n6(n7)b1&-1 to 230&[-30 to 10, 300 to 320]&110(81)&6(5)\\
\enddata 
\vspace{3mm}
\tablecomments{{\it Fermi} burst ID (column 1), redshift (column 2), and $t_{90}$ (column 3), together with the detectors (column 4), source (column 5) and background (column 6) intervals, and total (column 7) and effective ($S\geq$20, parentheses in column 7) time bins using the BBlocks across the source intervals, and the number of total (column 8) and the used (parentheses in column 8) pulses for each burst. Note that the detector in parentheses is the brightest one, used for BBlocks and background fitting.}
\end{deluxetable}\label{table:global}

\clearpage
\startlongtable
\setlength{\tabcolsep}{0.15em}
\begin{deluxetable}{cccccccccccc}
%\rotate
\tablewidth{0pt}
\tabletypesize{\scriptsize}
\tablecaption{Pulsewise Properties of the Sample}
%\tablenum{1}
\tablehead{
\colhead{GRB}
&\colhead{Pulse}
&\colhead{$\Delta T_{\rm src}$}
&\colhead{$N_{\rm tot}$($N_{\rm S\geq20}$)}
&\colhead{Pulse}
&\colhead{Spectral Model}
&\colhead{Physical Model}
&\colhead{$E_{\rm p}$}
&\colhead{$\alpha$}
&\colhead{$F-\alpha$}
&\colhead{$F-E_{\rm p}$}
&\colhead{$\alpha-E_{\rm p}$}\\
&
&(s)
&(Number)
&(Grade)
&(Preferred)
&(Preferred)
&(Evolution)
&(Evolution)
&
Type(r)
&
Type(r)
&
Type(r)
}
\colnumbers
\startdata
081009140&$P_{1}$&0 to 10&18(15)&Gold&Band&Above SCS line&{\it f.}-{\it t.}&{\it f.}-{\it t.}&2p(0.11)&1p(0.41)&3(-0.55)\\
081009140&$P_{2}$&35 to 52&10(6)&Gold&CPL&Above SCS line&{\it f.}-{\it t.}&{\it h.}-t.-{\it s.}&1n(-0.83)&3(0.20)&1p(-0.03)\\
\hline
081215784&$P_{1}$&-1 to 2.9&10(8)&Silver&Band&$Ph$&{\it h.}-t.-{\it s.}&{\it f.}-{\it t.}&1p(0.69)&2p(0.50)&2p(-0.12)\\
081215784&$P_{2}$&2.9 to 4.4&5(5)&Silver&CPL&Above SCS line&{\it f.}-{\it t.}&{\it f.}-{\it t.}&1p(0.70)&1p(0.70)&3(0.20)\\
081215784&$P_{3}$&4.4 to 10&12(9)&Silver&Band&$Ph$&{\it f.}-{\it t.}&{\it f.}-{\it t.}&1p(0.90)&1p(1.00)&1p(0.90)\\
\hline
090131090&$P_{1}$&0 to 5.5&7(2)&Bronze&CPL&Above SCS line&...&...&...&...&...\\ 
090131090&$P_{2}$&5.5 to 10&10(6)&Silver&Band&*&{\it h.}-t.-{\it s.}&{\it f.}-{\it t.}&1p(0.83)&1p(0.77)&1p(0.37)\\ 
090131090&$P_{3}$&22 to 40&13(8)&Silver&Band&*&{\it h.}-t.-{\it s.}&{\it f.}-{\it t.}&1p(0.75)&1n(-0.14)&1n(-0.29)\\
\enddata 
\vspace{3mm}
\tablecomments{{\it Fermi} burst ID (column 1), the pulse group name (column 2), the source interval of each pulse (column 3), the total and effective ($S\geq$20) time bins by using the BBlocks across the source intervals (column 4), the identified grade of pulse (column 5), the preferred spectral model indicated by $\Delta$DIC (column 6), and the preferred physical model (column 7), the pulse-wise $E_{\rm p}$ (column 8) and $\alpha$ (column 9) evolution, and the type of parameter relation with the Spearman's rank coefficient $r$ in the parentheses for the gold and silver samples: $\alpha$-$E_{\rm p}$ relation (column 10), $F$-$E_{\rm p}$ relation (column 11), and $F$-$\alpha$ relation (column 12). In column 7 we make an asterisk for spectra that have $\alpha_{\rm max}$ above the SCS line and $Ph$ for the spectra that are classified as being from the photosphere. Note that $Ph$ implies that the pulse has at least one time bin that is compatible with the photosphere..}
\end{deluxetable}\label{table:Pulse}

\clearpage
\begin{deluxetable}{cccccccc}
%\rotate
\tablewidth{0pt}
\tabletypesize{\scriptsize}
\tablecaption{Results of the Average and Deviation Values of the Parameter Distribution}
\tablehead{
\colhead{Pulse}
&\colhead{Model}
&\colhead{Spectra}
&\colhead{$\alpha$}
&\colhead{$E_{\rm p}$}
&\colhead{$F$}
&\colhead{$\beta$}
&\colhead{$\beta$}\\
(Group)&(Selected)&(Number)&&(keV)&(erg s$^{-1}$ cm$^{-2}$)&(wellconverged)&(unconverged)
}
\colnumbers
\startdata
\hline
$P_{1}$&Best&338&-0.70$\pm$0.29&$\rm log_{10}$(234)$\pm$0.40&$\rm log_{10}$(6.17e-6)$\pm$0.51&\nodata&\nodata\\
$P_{2}$&Best&361&-0.85$\pm$0.37&$\rm log_{10}$(224)$\pm$0.44&$\rm log_{10}$(4.57e-6)$\pm$0.51&\nodata&\nodata\\
$P_{3}$&Best&156&-1.02$\pm$0.33&$\rm log_{10}$(158)$\pm$0.30&$\rm log_{10}$(2.95e-6)$\pm$0.39&\nodata&\nodata\\
$P_{(4+5+6)}$&Best&89&-1.03$\pm$0.19&$\rm log_{10}$(202)$\pm$0.34&$\rm log_{10}$(2.92e-5)$\pm$0.39&\nodata&\nodata\\
Overall&Best&944&-0.84$\pm$0.35&$\rm log_{10}$(214)$\pm$0.42&$\rm log_{10}$(4.57e-6)$\pm$0.47&\nodata&\nodata\\
\enddata 
\vspace{3mm}
\tablecomments{Column 1 lists the pulse name, column 2 lists the used model, column 3 lists the number of the spectra for each pulse, columns 4-8 list the average and its deviation (1$\sigma$) for the spectral parameters and energy flux.} 
\end{deluxetable}\label{table:distribution}

\clearpage
\begin{deluxetable}{ccccc|cccccc}
%\rotate
\tablewidth{0pt}
%\tabletypesize{}
\tablecaption{Probability of the Kolmogorov-Smirnov (K-S) Test for the Spectral Parameter Distributions}
\tablehead{}
%\colnumbers
\startdata
\hline
\\
&$\alpha$($P_{1}$)&$\alpha$($P_{2}$)&$\alpha$($P_{3}$)&$\alpha$[$P_{(4+5+6)}$]&&$E_{\rm p}$($P_{1}$)&$E_{\rm p}$($P_{2}$)&$E_{\rm p}$($P_{3}$)&$E_{\rm p}$[$P_{(4+5+6)}$]\\
$\alpha$($P_{1}$)&1.&$<10^{-2}$&$<10^{-2}$&$<10^{-2}$&$E_{\rm p}$($P_{1}$)&1.&0.11&$<10^{-2}$&0.37\\
$\alpha$($P_{2}$)&$<10^{-2}$&1.&$<10^{-2}$&$<10^{-2}$&$E_{\rm p}$($P_{2}$)&0.11&1.&$<10^{-2}$&0.43\\
$\alpha$($P_{3}$)&$<10^{-2}$&$<10^{-2}$&1.&0.12&$E_{\rm p}$($P_{3}$)&$<10^{-2}$&$<10^{-2}$&1.&0.03\\
$\alpha$[$P_{(4+5+6)}$]&$<10^{-2}$&$<10^{-2}$&0.12&1.&$E_{\rm p}$[$P_{(4+5+6)}$]&0.37&0.43&0.03&1.\\
\hline
\enddata 
%\vspace{3mm}
\tablecomments{The probability of the K-S Statistic on two samples. The left part is for $\alpha$ distributions, and the right part is for $E_{\rm p}$ distributions.}
\label{tab:KStest}
\end{deluxetable}

\clearpage
\begin{deluxetable}{ccc|cc|cc|cc}
%\rotate
\tablewidth{0pt}
\tabletypesize{\scriptsize}
\tablecaption{Statistical Results of the Pulsewise Spectral Evolution}
\tablehead{
\colhead{Pulse}
&\colhead{Full Sample}
&\colhead{Used Sample}
&\multicolumn{2}{c}{Hard-to-soft}
&\multicolumn{2}{c}{Flux-tracking}
&\multicolumn{2}{c}{Others}\\
\cline{4-9}
(Group)&(Number)&(Number)&(Number)&(Percentage)&(Number)&(Percentage)&(Number)&(Percentage)
}
\colnumbers
\startdata
\hline\noalign{\smallskip}
\multicolumn{9}{c}{$E_{\rm p}$ Evolution}\tabularnewline
\hline\noalign{\smallskip}
$P_{1}$&39&35&12&34\%&20&57\%&3&9\%\\
$P_{2}$&39&38&13&34\%&25&66\%&0&0\%\\
$P_{3}$&24&20&7&35\%&12&60\%&1&5\%\\
$P_{4}$&9&6&1&17\%&4&67\%&1&16\%\\
$P_{5}$&4&3&1&33\%&2&67\%&0&0\%\\
$P_{6}$&2&1&0&0\%&1&100\%&0&0\%\\
Overall&117&103&34&26\%&64&63\%&5&3\%\\
\hline\noalign{\smallskip}
\multicolumn{9}{c}{$\alpha$ Evolution}\tabularnewline
\hline\noalign{\smallskip}
$P_{1}$&39&35&10&29\%&21&60\%&4&11\%\\
$P_{2}$&39&38&10&26\%&25&66\%&3&8\%\\
$P_{3}$&24&20&8&40\%&12&60\%&0&0\%\\
$P_{4}$&9&6&0&0\%&5&83\%&1&17\%\\
$P_{5}$&4&3&0&0\%&3&100\%&0&0\%\\
$P_{6}$&2&1&0&0\%&1&100\%&0&0\%\\
Overall&117&103&28&27\%&67&65\%&8&8\%\\
\enddata 
\tablecomments{Column 1 lists the pulse groups; column 2 and 3 list the number of the full (gold+silver+bronze) and the used (gold+silver) samples for each pulse group, and columns 4-9 list the number and percentage of the hard-to-soft, Flux-tracking, and the other patterns, respectively.}\label{tab:SepctralPatterns}
\end{deluxetable}

\clearpage
\begin{deluxetable}{ccccccccc}
%\rotate
\tablewidth{0pt}
\tabletypesize{\scriptsize}
\tablecaption{Comparison between the Multipulse Sample in This Paper and the Single-pulse Sample in \cite{Yu2019}}
\tablehead{
\colhead{Sample}
&\colhead{Sample Source}
&\colhead{Sample Size}
&\colhead{$\rm log_{10}$($t_{90}$)}
&\colhead{$\rm log_{10}$(Fluence)}
&\colhead{$\rm log_{10}$(Peak flux)}
&\colhead{$\rm log_{10}$($E_{\rm p}$)}
&\colhead{$\alpha$}
&\colhead{$\beta$}\\
&&(Number)&(s)&(erg cm$^{-2}$)&&(keV)
}
\colnumbers
\startdata
\hline
Single-pulse sample&\cite{Yu2019}&37 bursts&$\rm log_{10}$(16.6)$\pm$0.33&$\rm log_{10}$(3.24e-5)$\pm$0.40&$\rm log_{10}$(25.1)$\pm$0.28&$\rm log_{10}$(182)$\pm$0.34&-0.54$\pm$0.30&-2.34$\pm$0.55\\
Multi-pulse sample&this paper&39 bursts&$\rm log_{10}$(36.3)$\pm$0.39&$\rm log_{10}$(7.94e-5)$\pm$0.33&$\rm log_{10}$(51.3)$\pm$0.23&$\rm log_{10}$(219)$\pm$0.36&-0.69$\pm$0.22&-2.28$\pm$0.36\\
\enddata 
\tablecomments{Column 1 lists the samples, column 2 lists the sample source, column 3 list the sample size, columns 4-9 list the average and its deviation (1$\sigma$) value of the distributions, including the duration, fluence, peak flux, $E_{\rm p}$, $\alpha$, and $\beta$.}
\label{tab:Comparison}
\end{deluxetable}

\clearpage
\vspace{35mm}
\appendix
\setcounter{figure}{0}    
\setcounter{section}{0}
\setcounter{table}{0}
\renewcommand{\thesection}{A\arabic{section}}
\renewcommand{\thefigure}{A\arabic{figure}}
\renewcommand{\thetable}{A\arabic{table}}
\renewcommand{\theequation}{A\arabic{equation}}

In this appendix, we provide additional figures and tables and present the definition of the fitting models, as well as the discarded sample.

\subsection{Best Model-based Parameter Evolution in Different pulses}

\begin{figure*}
\includegraphics[angle=0,scale=0.3]{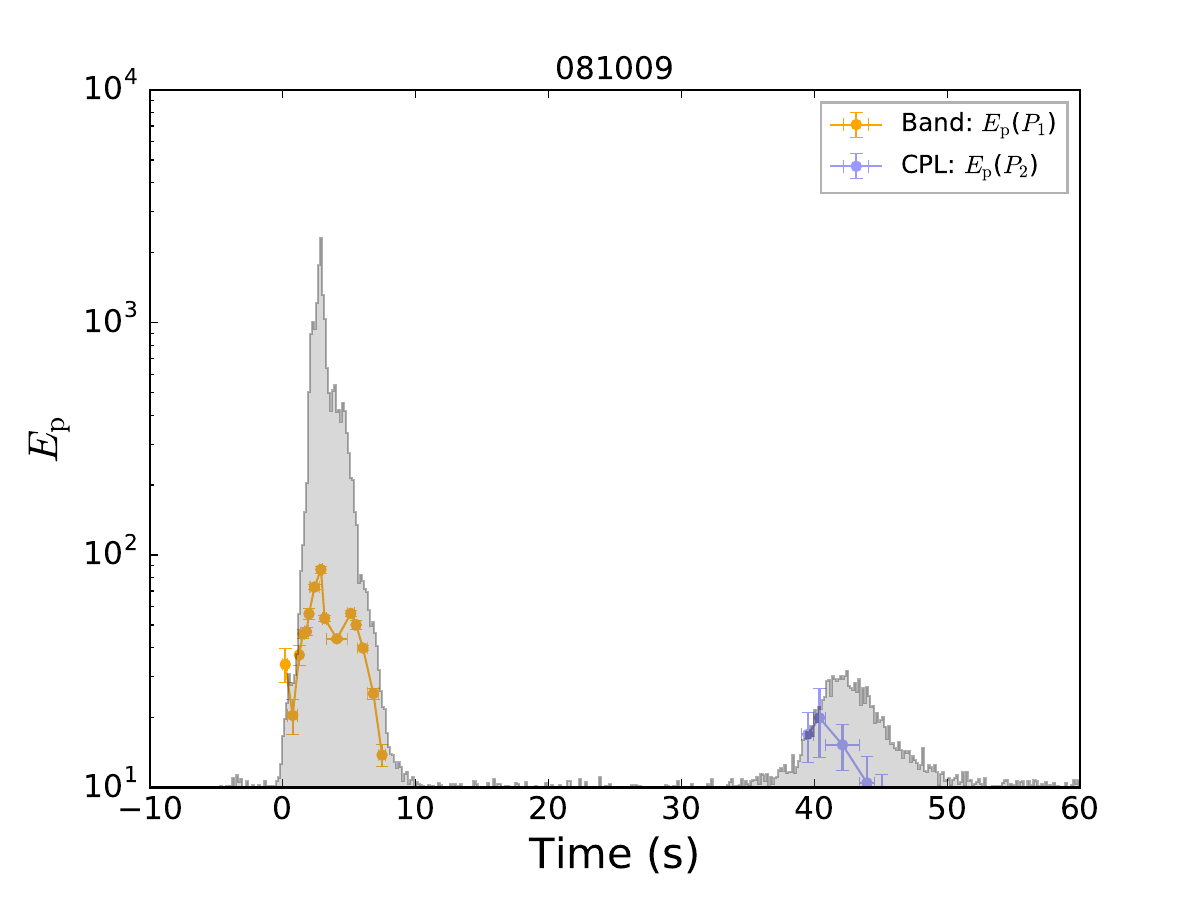}
\includegraphics[angle=0,scale=0.3]{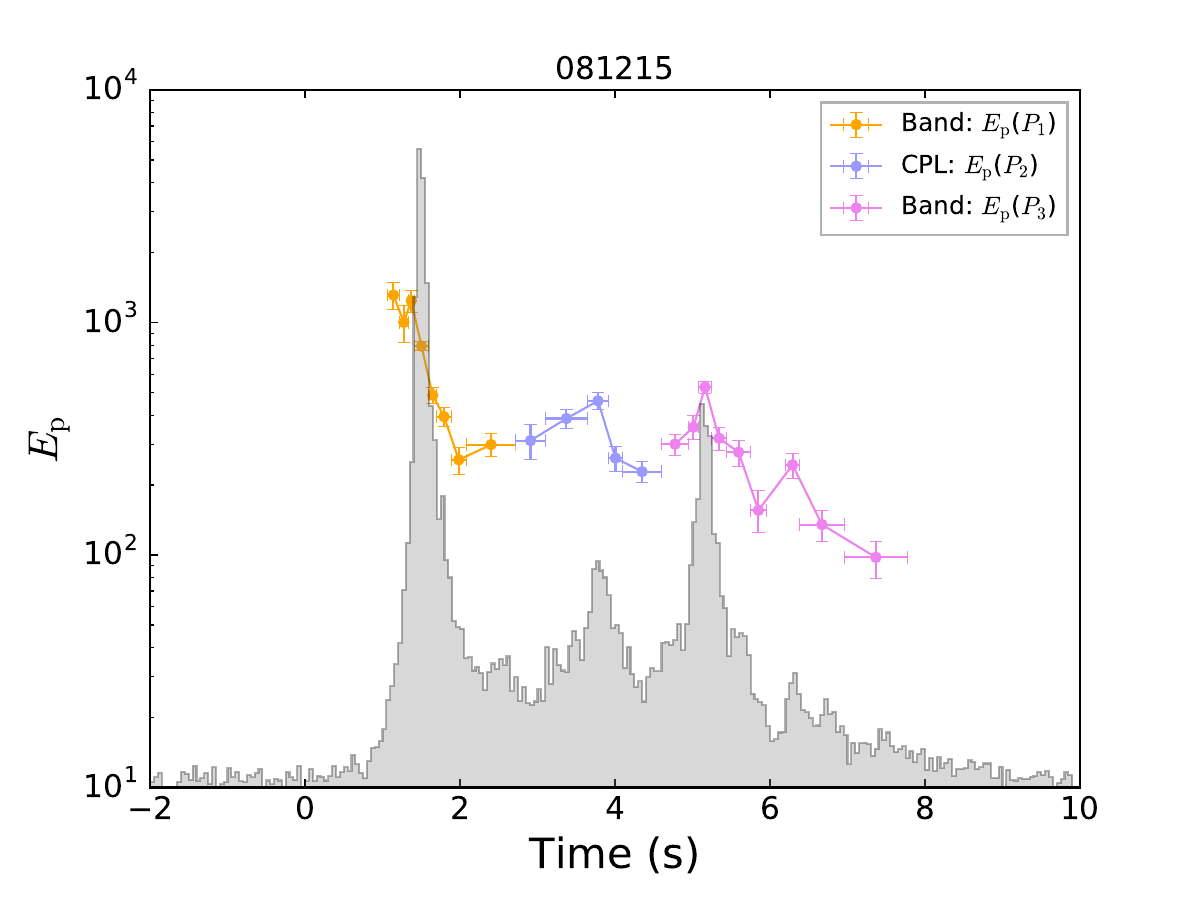}
\includegraphics[angle=0,scale=0.3]{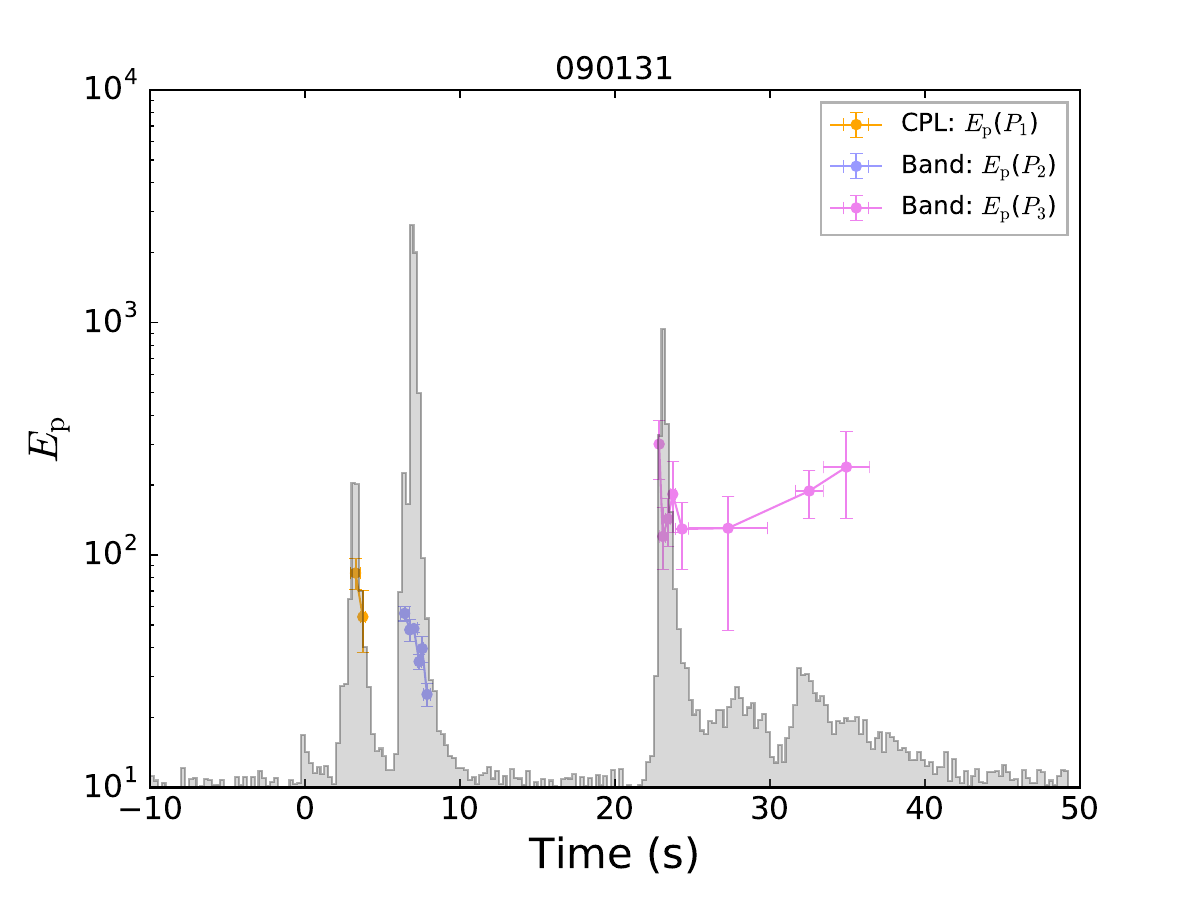}
\includegraphics[angle=0,scale=0.3]{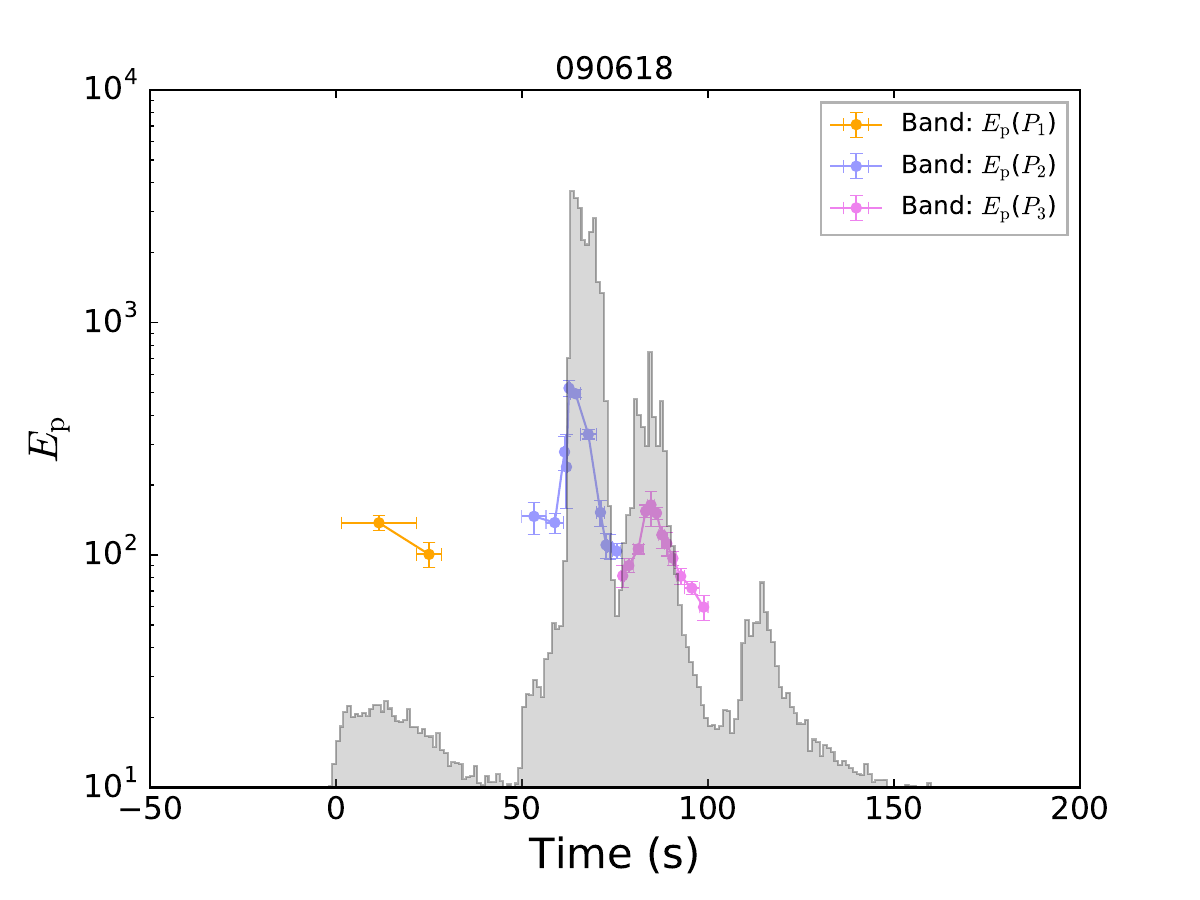}
\includegraphics[angle=0,scale=0.3]{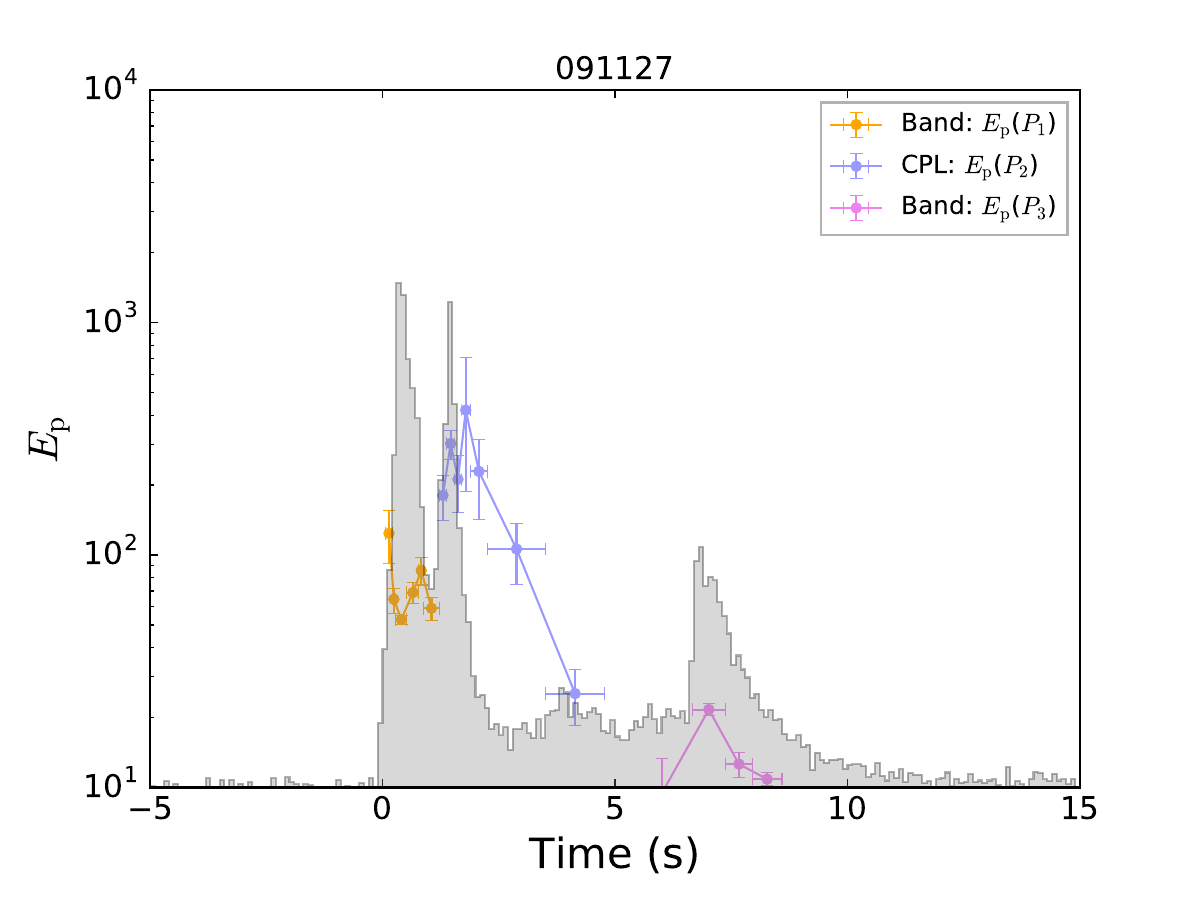}
\includegraphics[angle=0,scale=0.3]{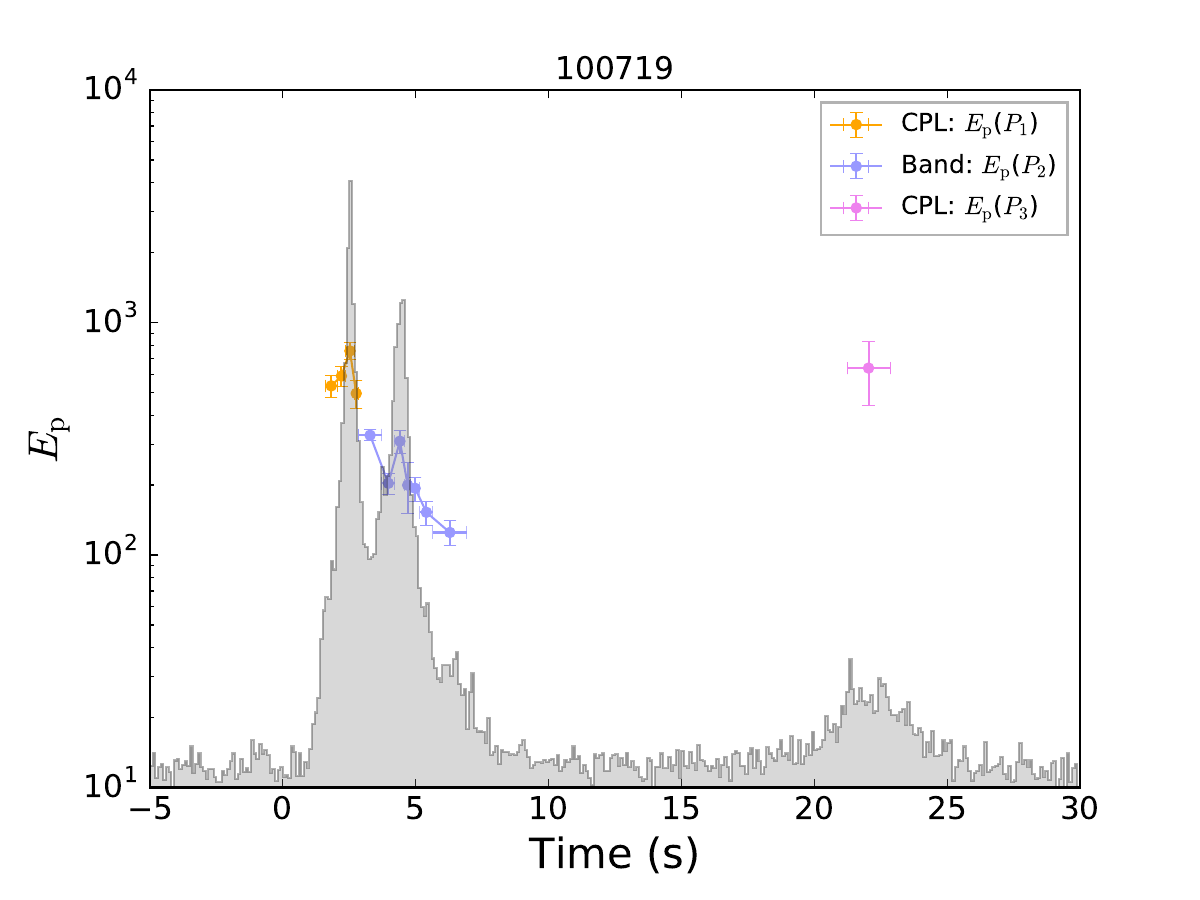}
\includegraphics[angle=0,scale=0.3]{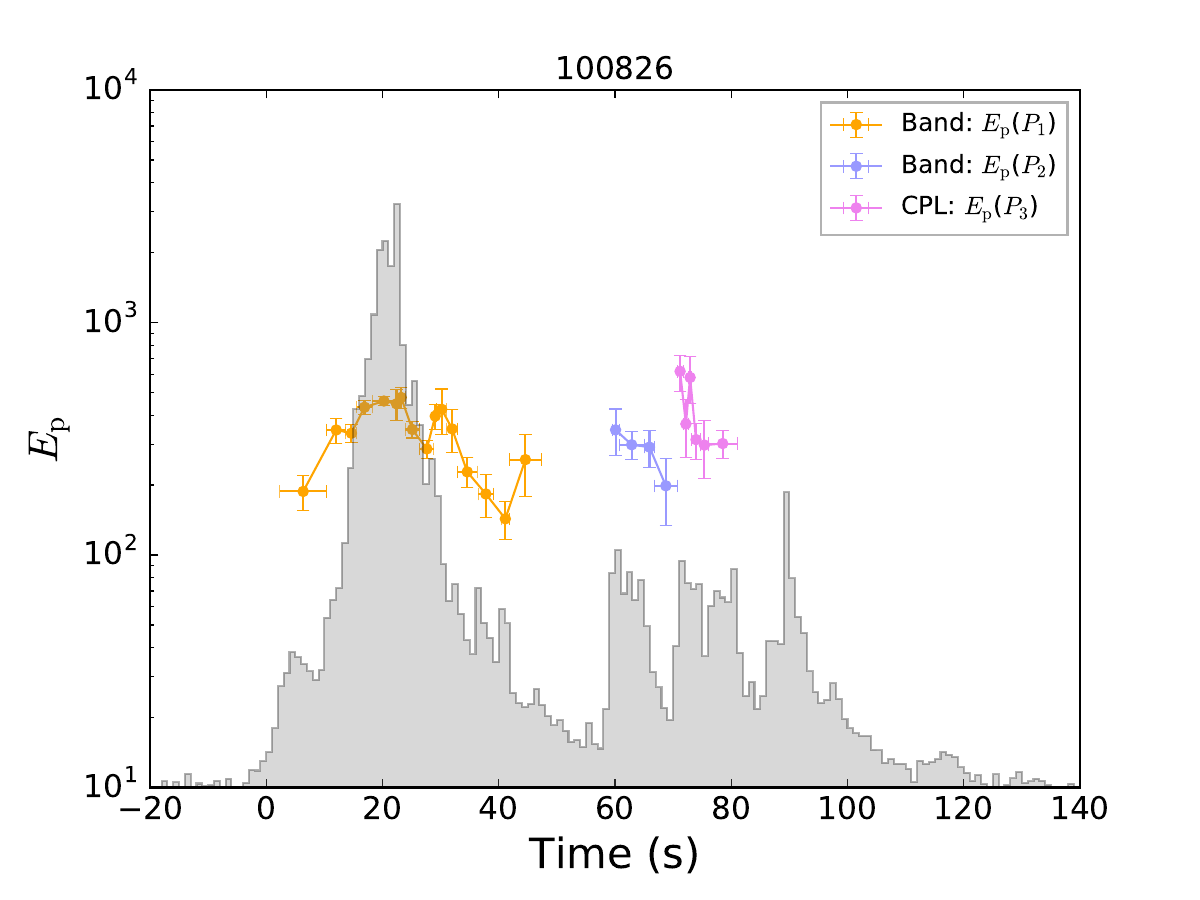}
\includegraphics[angle=0,scale=0.3]{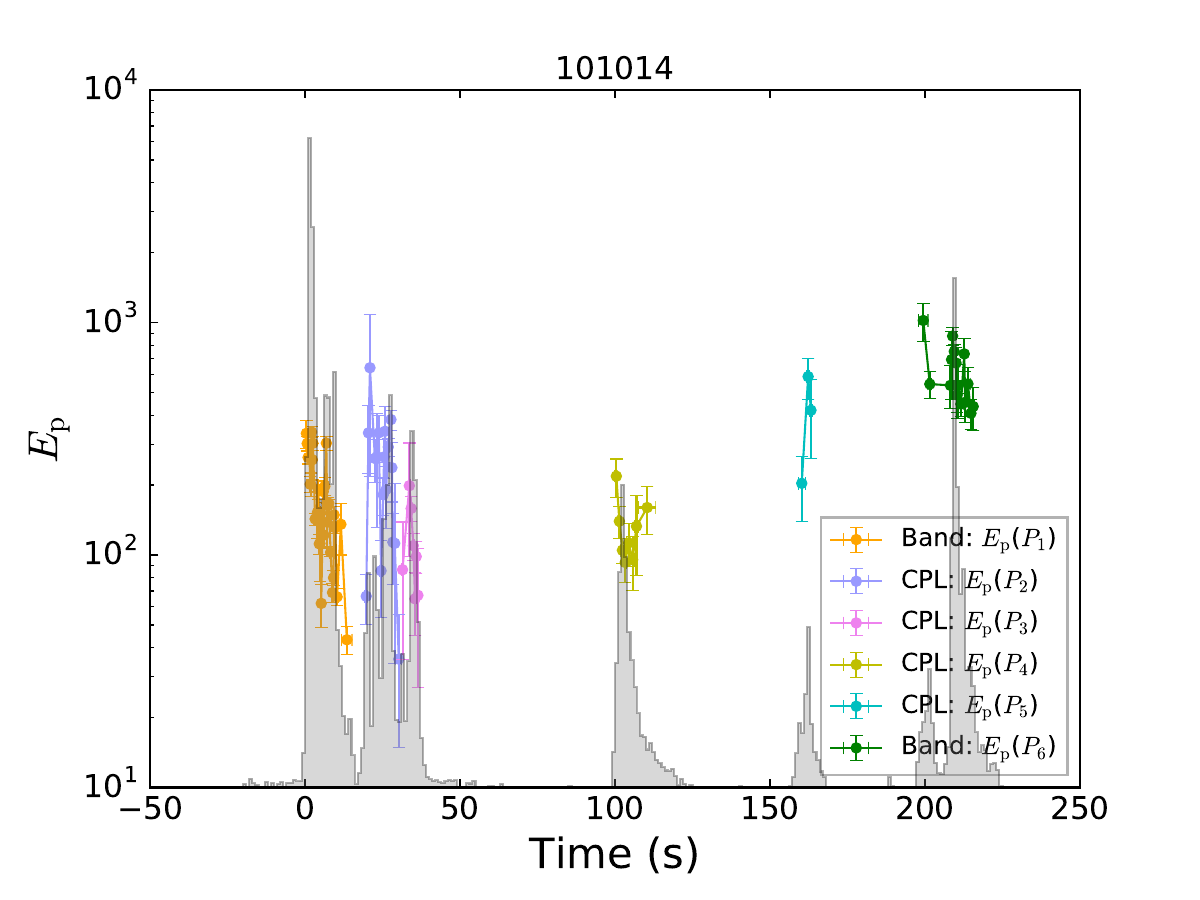}
\includegraphics[angle=0,scale=0.3]{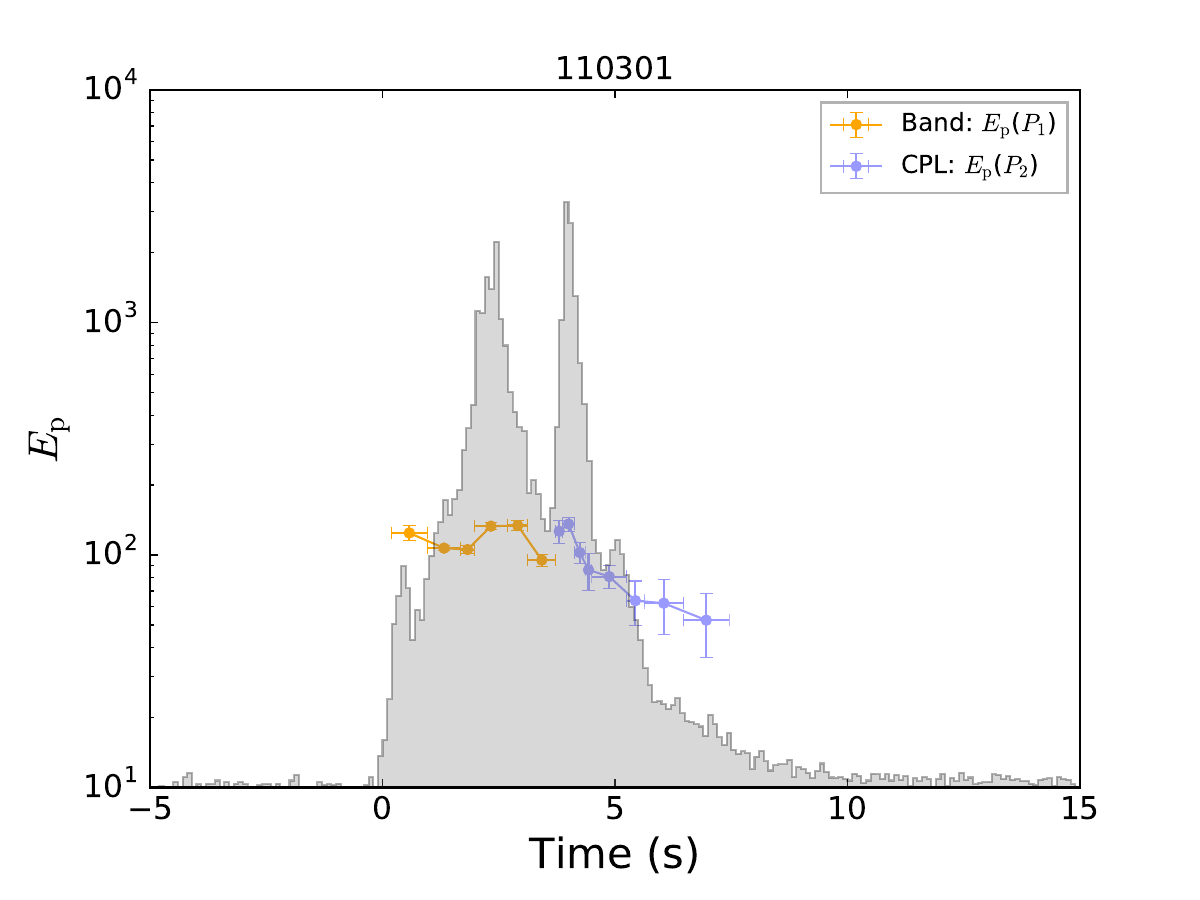}
\includegraphics[angle=0,scale=0.3]{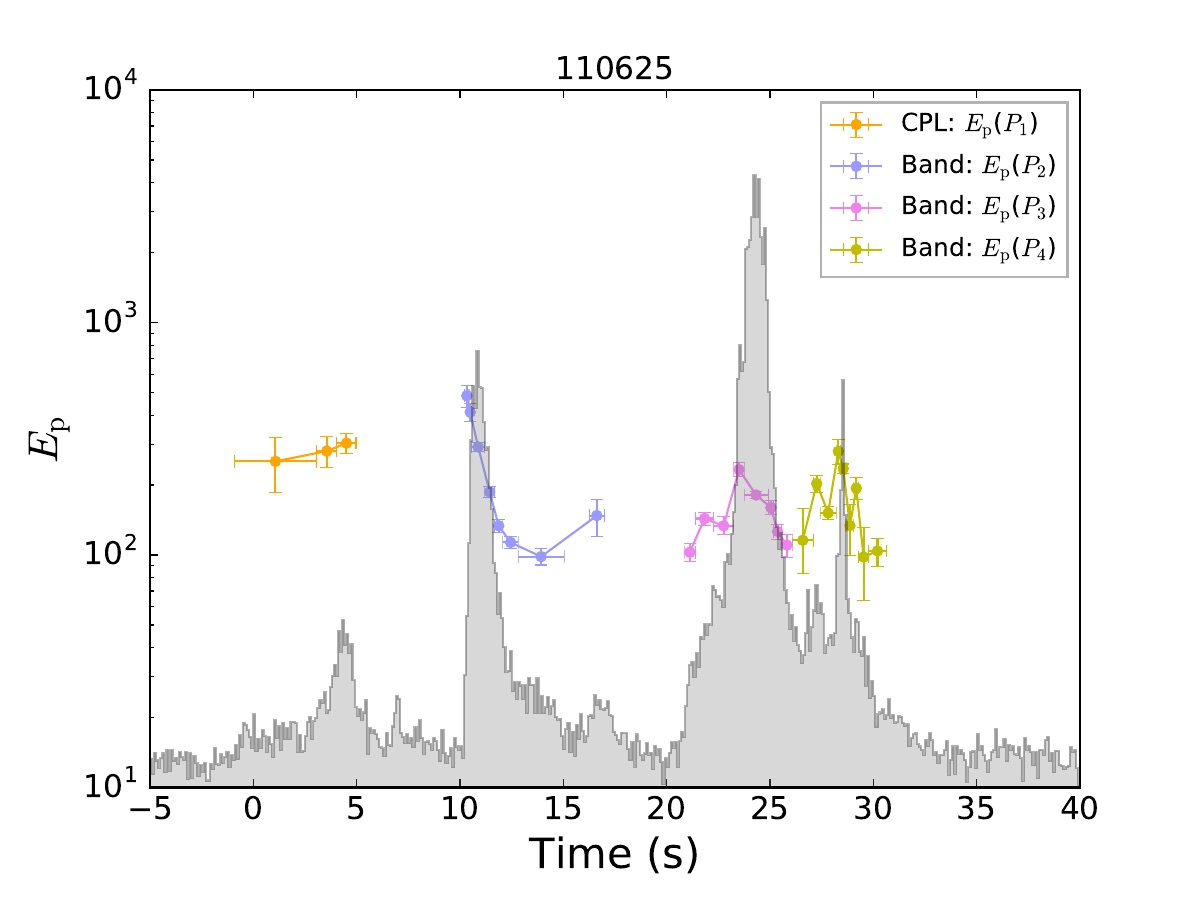}
\includegraphics[angle=0,scale=0.3]{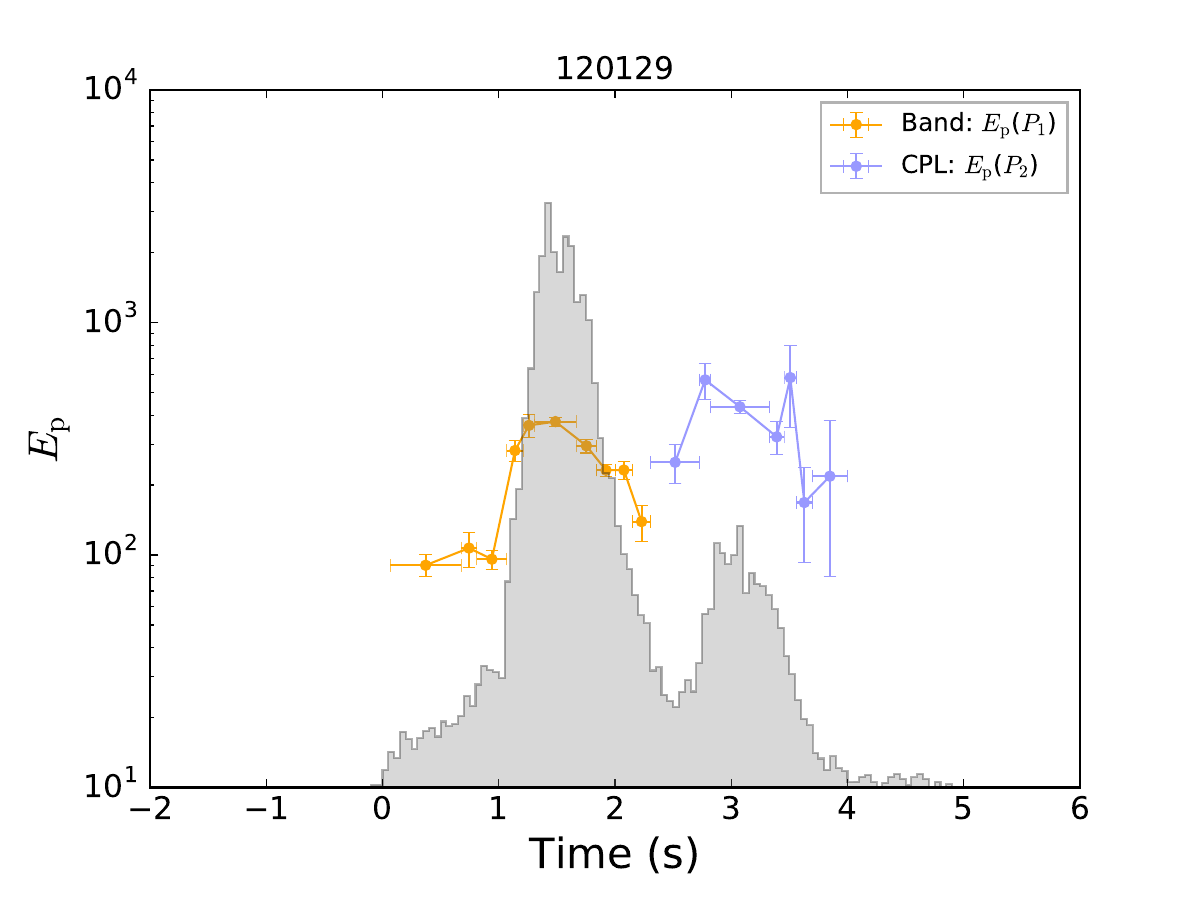}
\includegraphics[angle=0,scale=0.3]{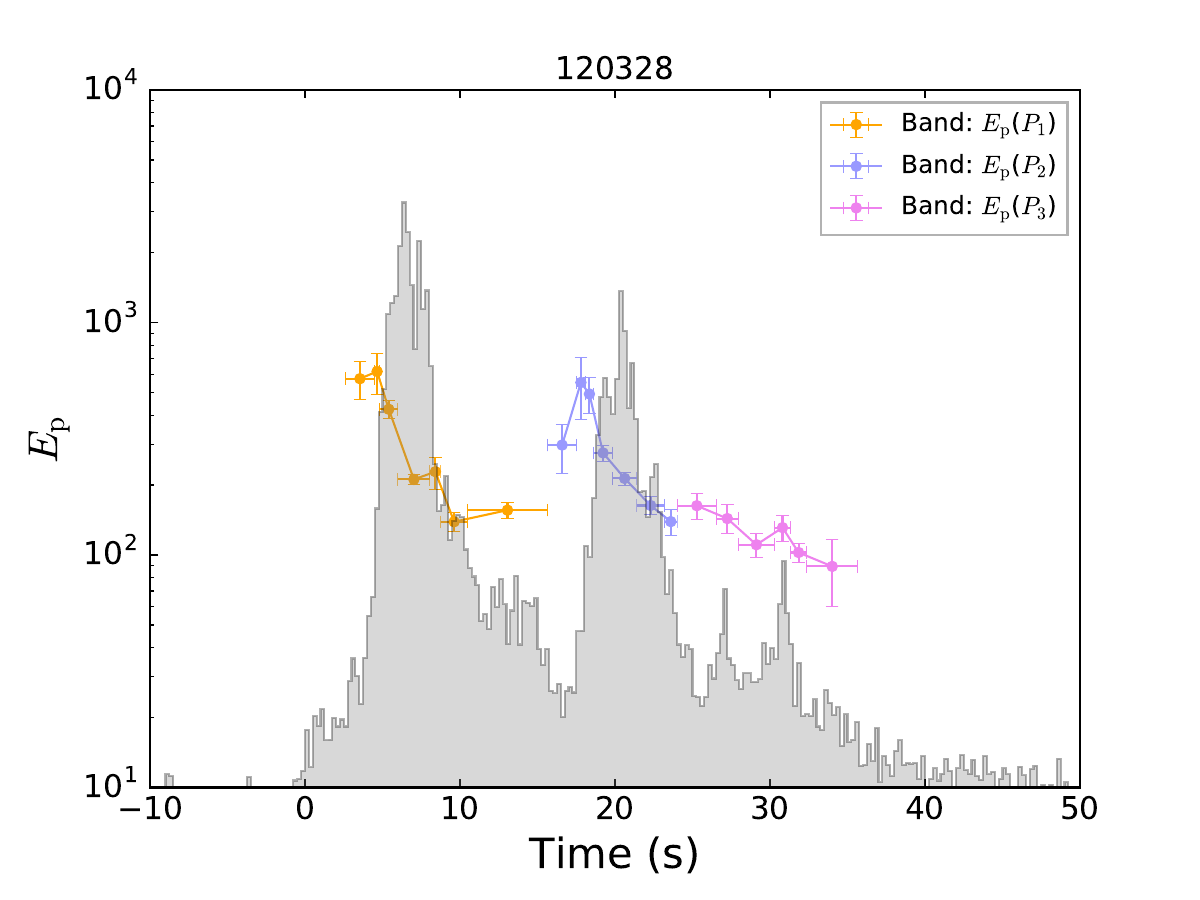}
\caption{Temporal evolution of the $E_{\rm p}$. Data points with solid orange, blue-magenta, violet, yellow, cyan, and green colors indicate time-series pulses of the $P_{1}$, $P_{2}$, $P_{3}$, $P_{4}$, $P_{5}$ ,and $P_{6}$, respectively. Count rate lightcurves are overlaid in gray. All data points correspond to a statistical significance $S \geq 20$.}\label{fig:Ep_Best}
\end{figure*}

In Figures \ref{fig:Ep_Best}-\ref{fig:Beta}, we provide the evolution of the spectral parameters $E_{\rm p}$, $\alpha$, $\nu$$F_{\nu}$ flux, and $\beta$ for all the bursts based on the best models defined in Section \ref{sec:bestmodel}.

\clearpage
\begin{figure*}
\includegraphics[angle=0,scale=0.3]{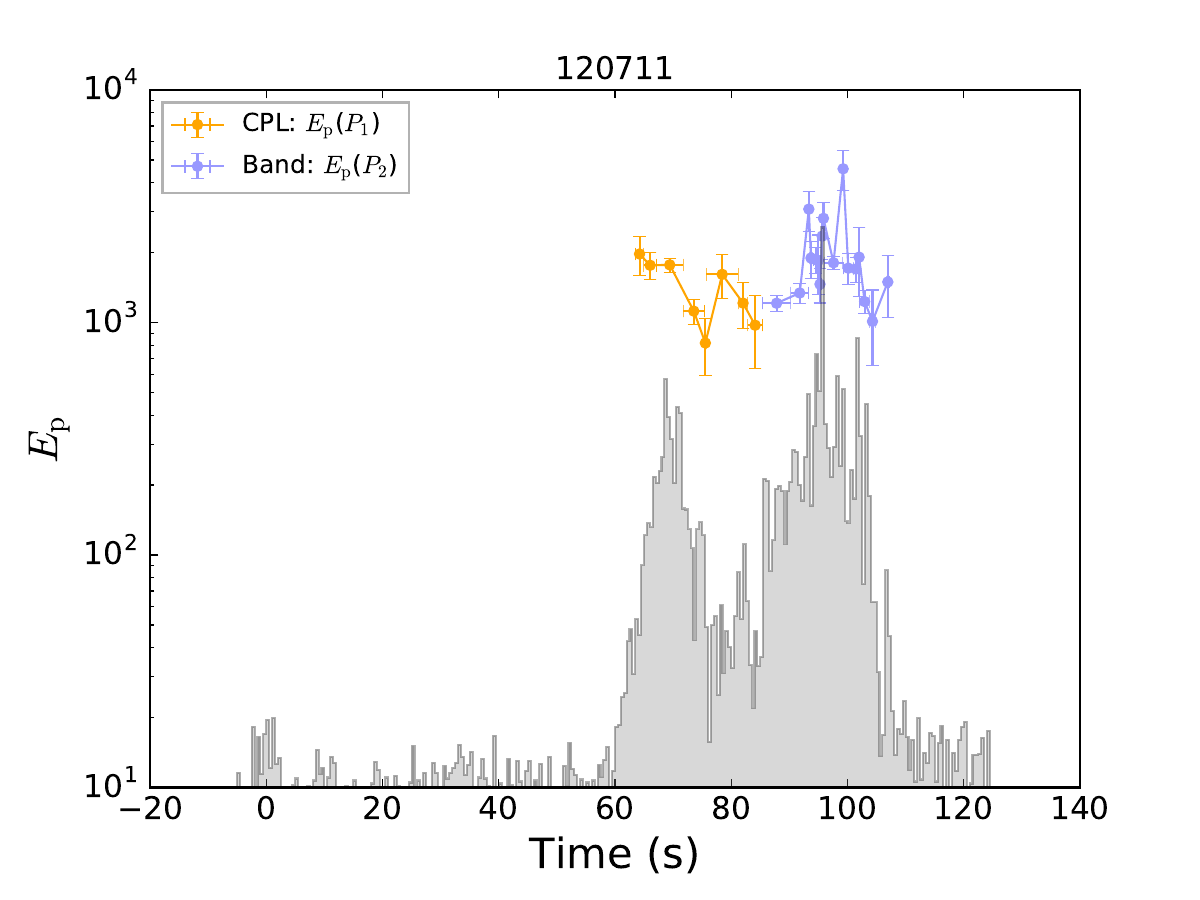}
\includegraphics[angle=0,scale=0.3]{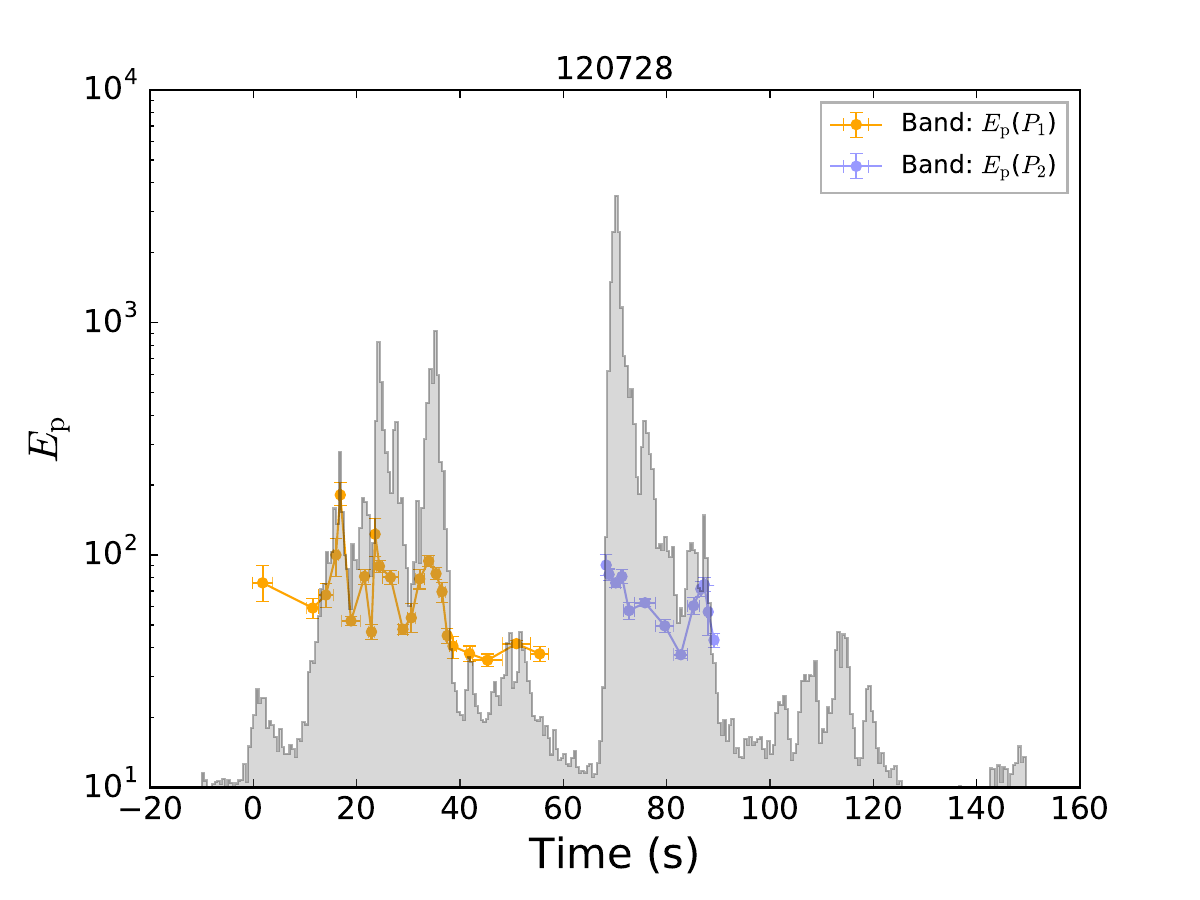}
\includegraphics[angle=0,scale=0.3]{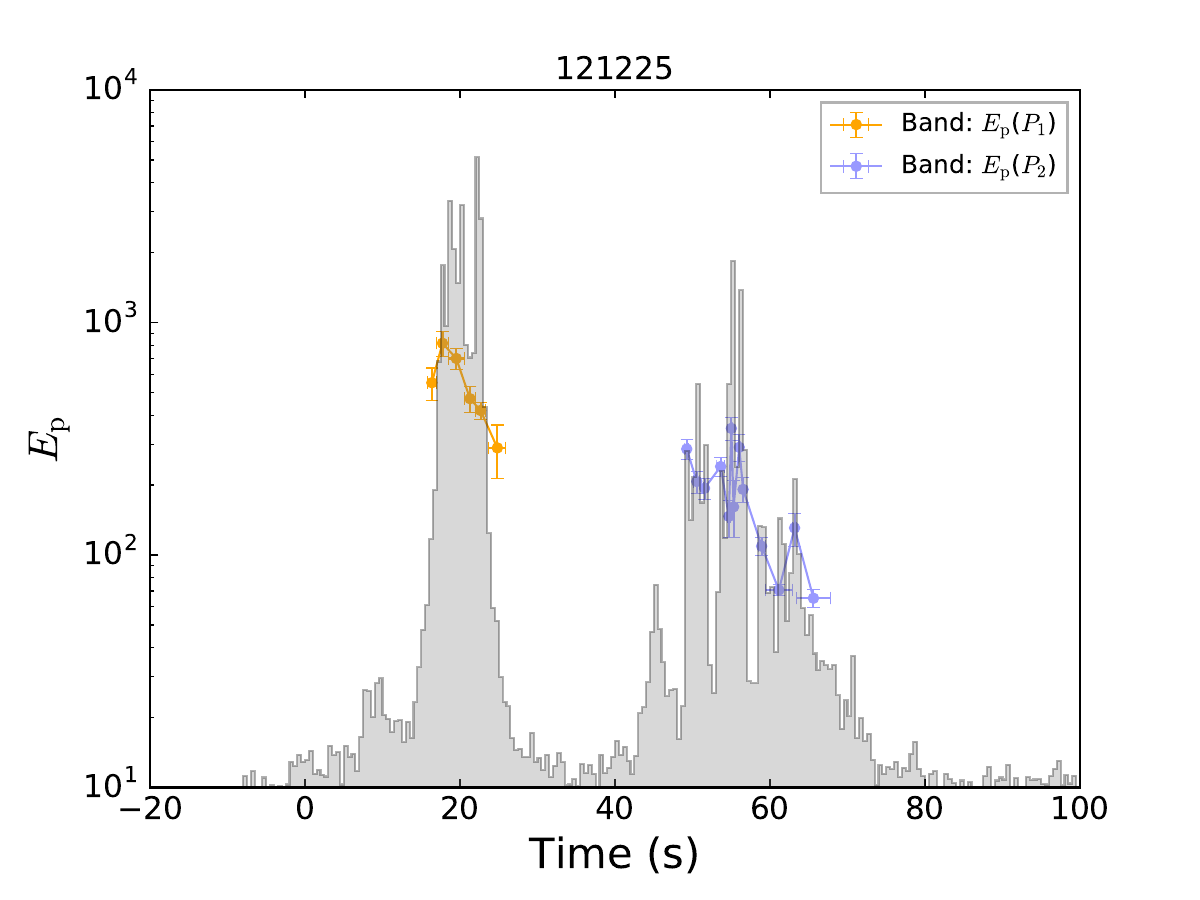}
\includegraphics[angle=0,scale=0.3]{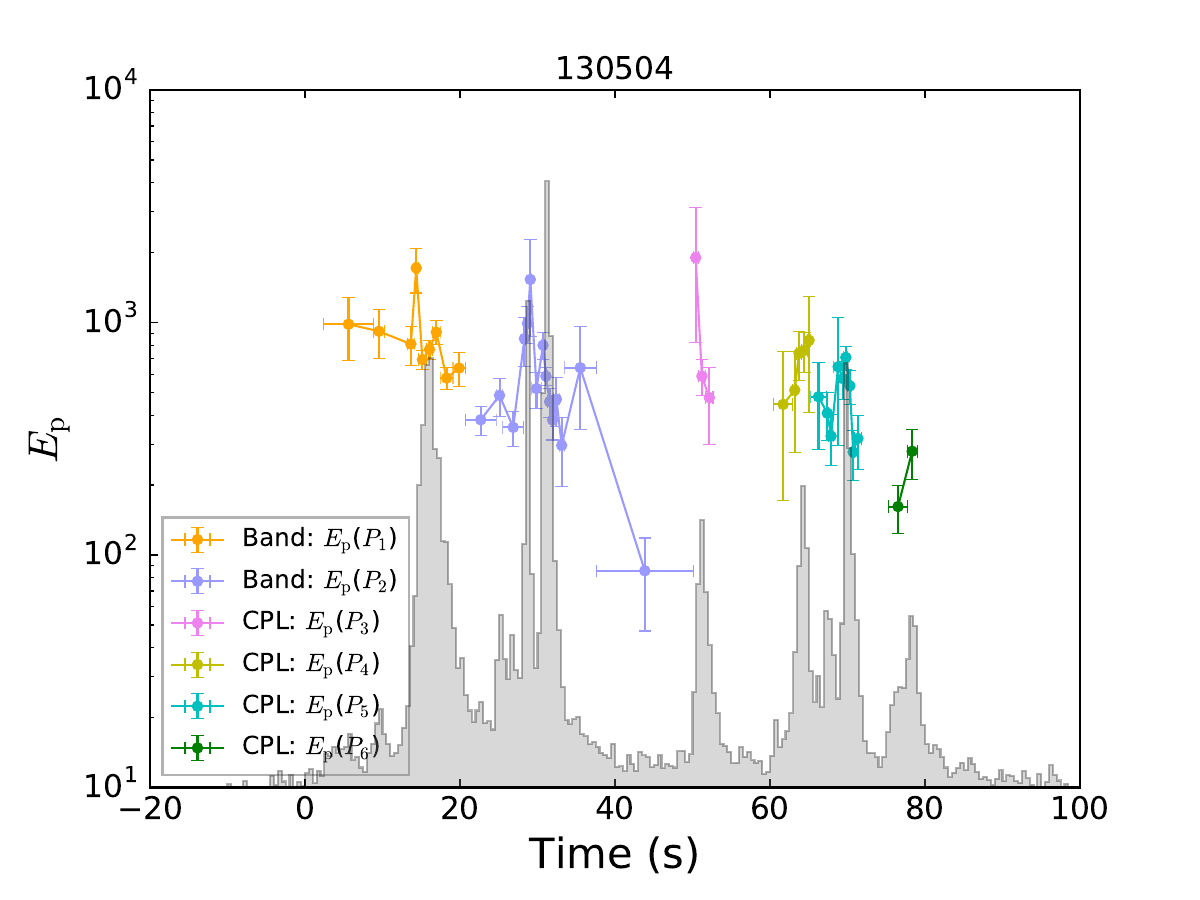}
\includegraphics[angle=0,scale=0.3]{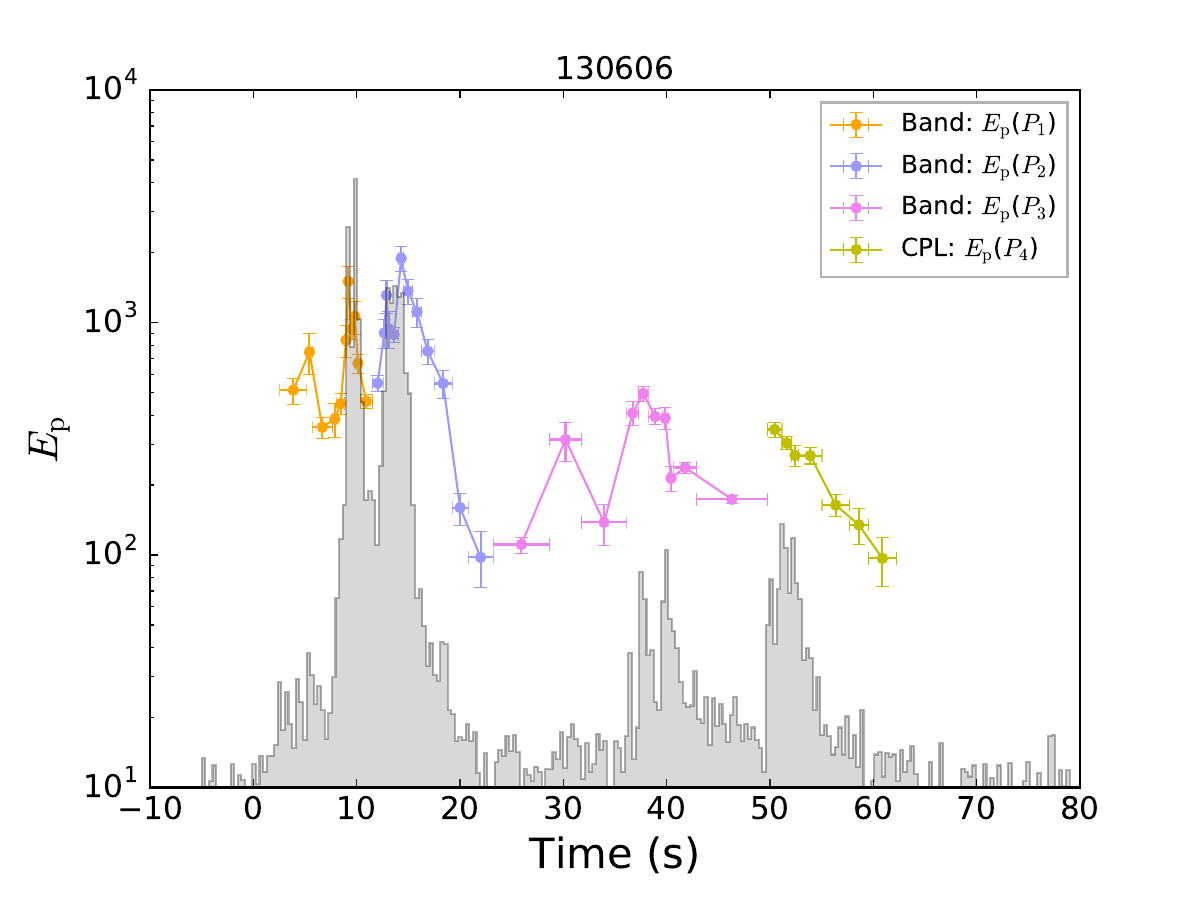}
\includegraphics[angle=0,scale=0.3]{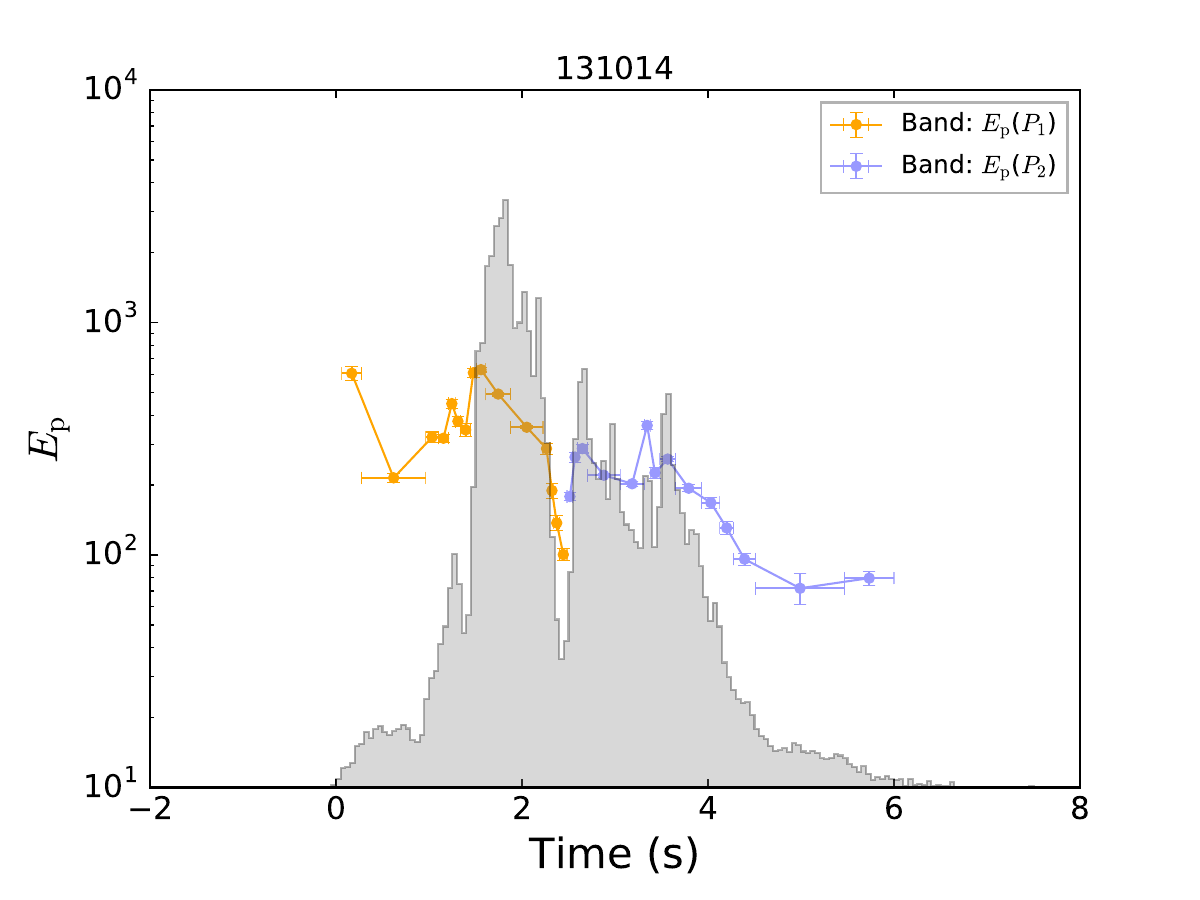}
\includegraphics[angle=0,scale=0.3]{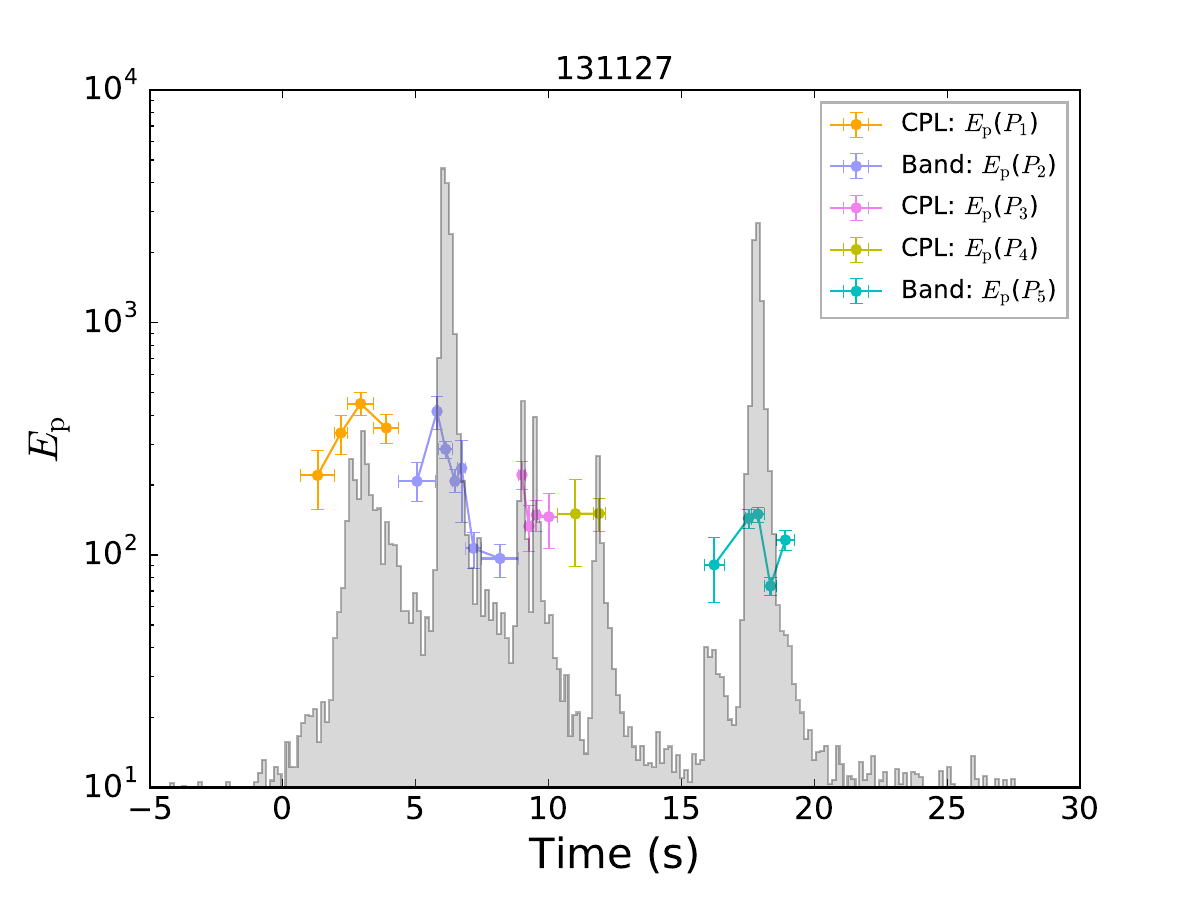}
\includegraphics[angle=0,scale=0.3]{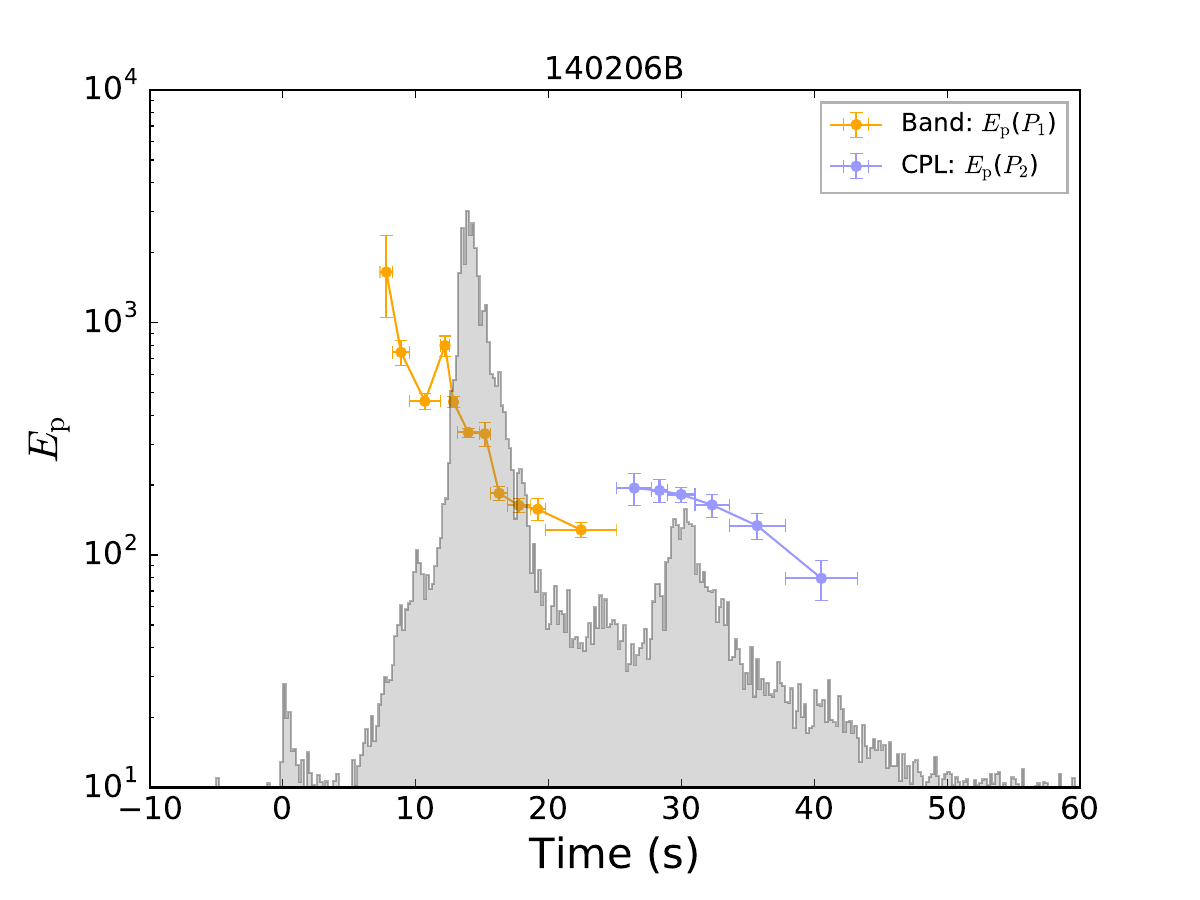}
\includegraphics[angle=0,scale=0.3]{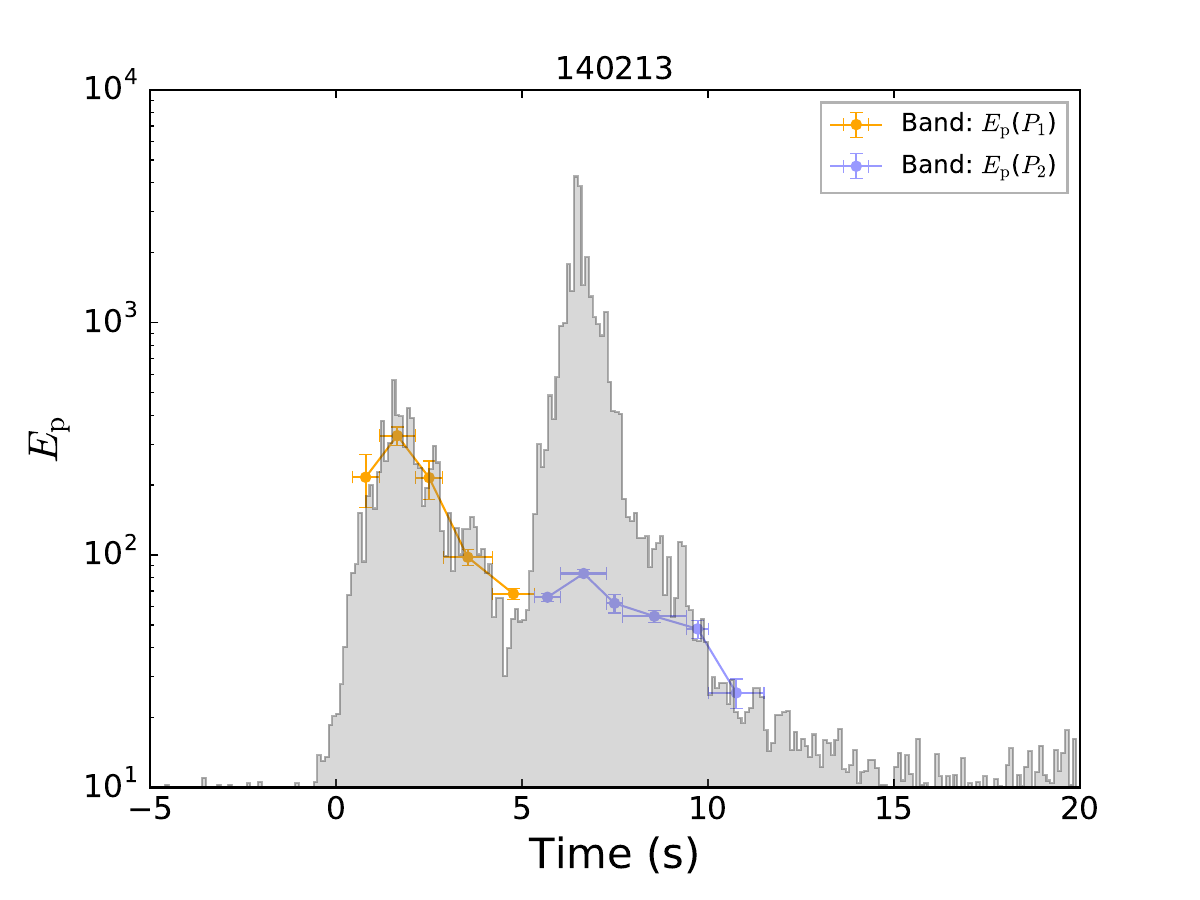}
\includegraphics[angle=0,scale=0.3]{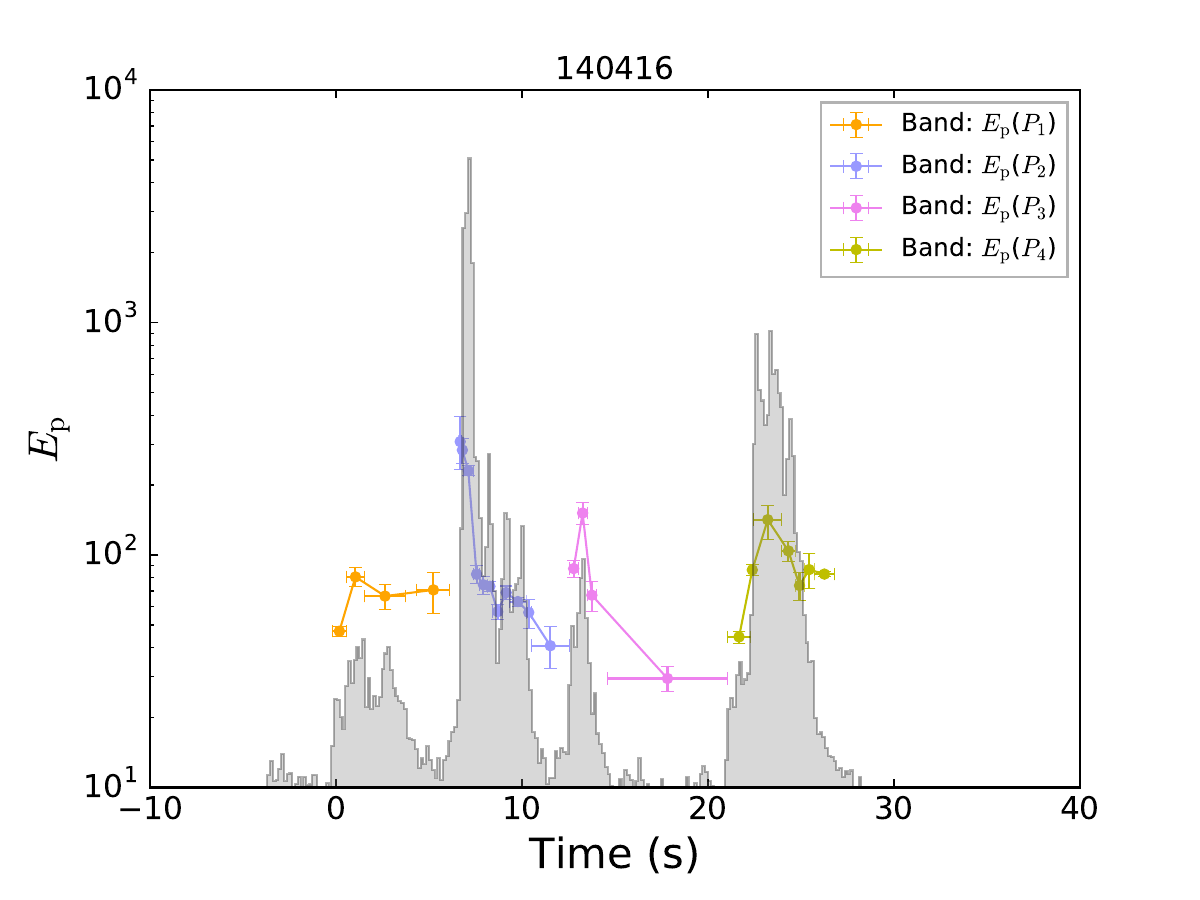}
\includegraphics[angle=0,scale=0.3]{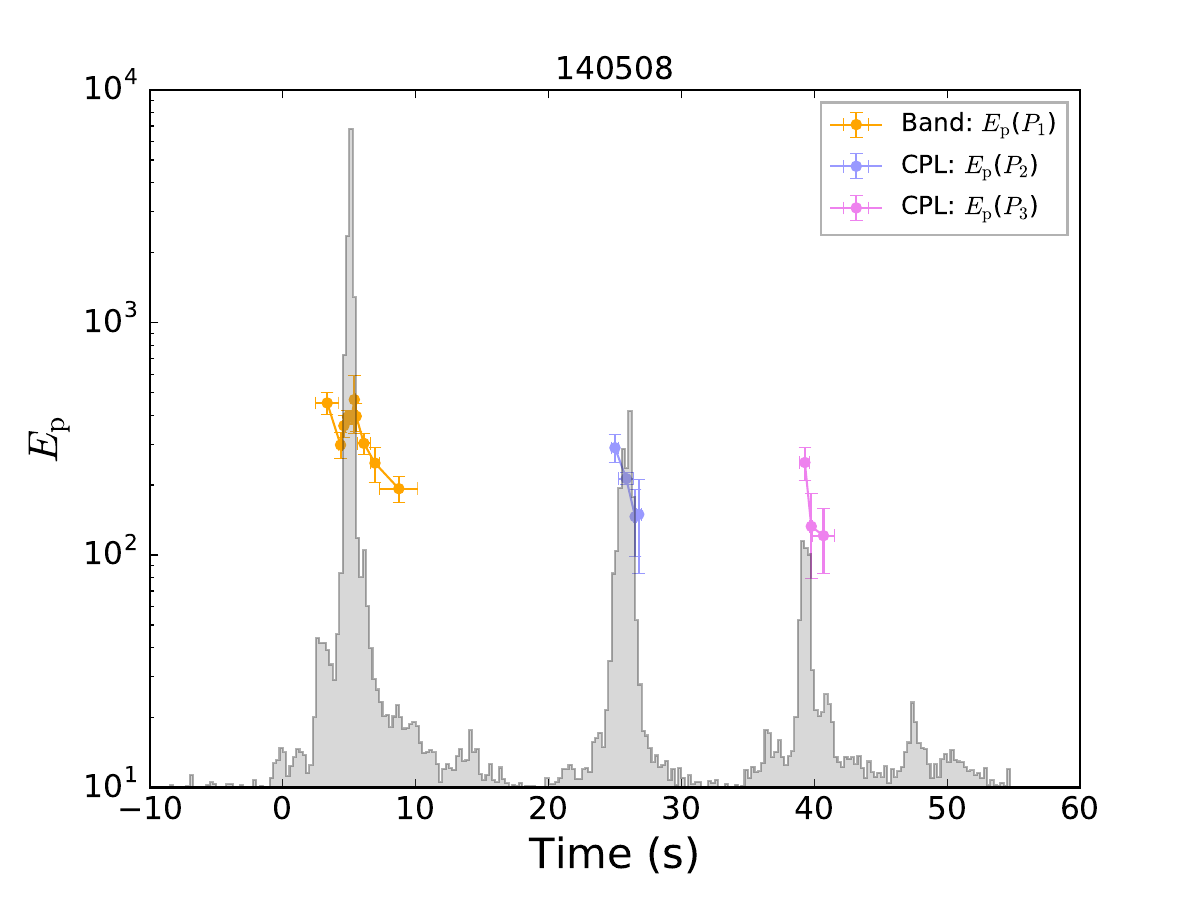}
\includegraphics[angle=0,scale=0.3]{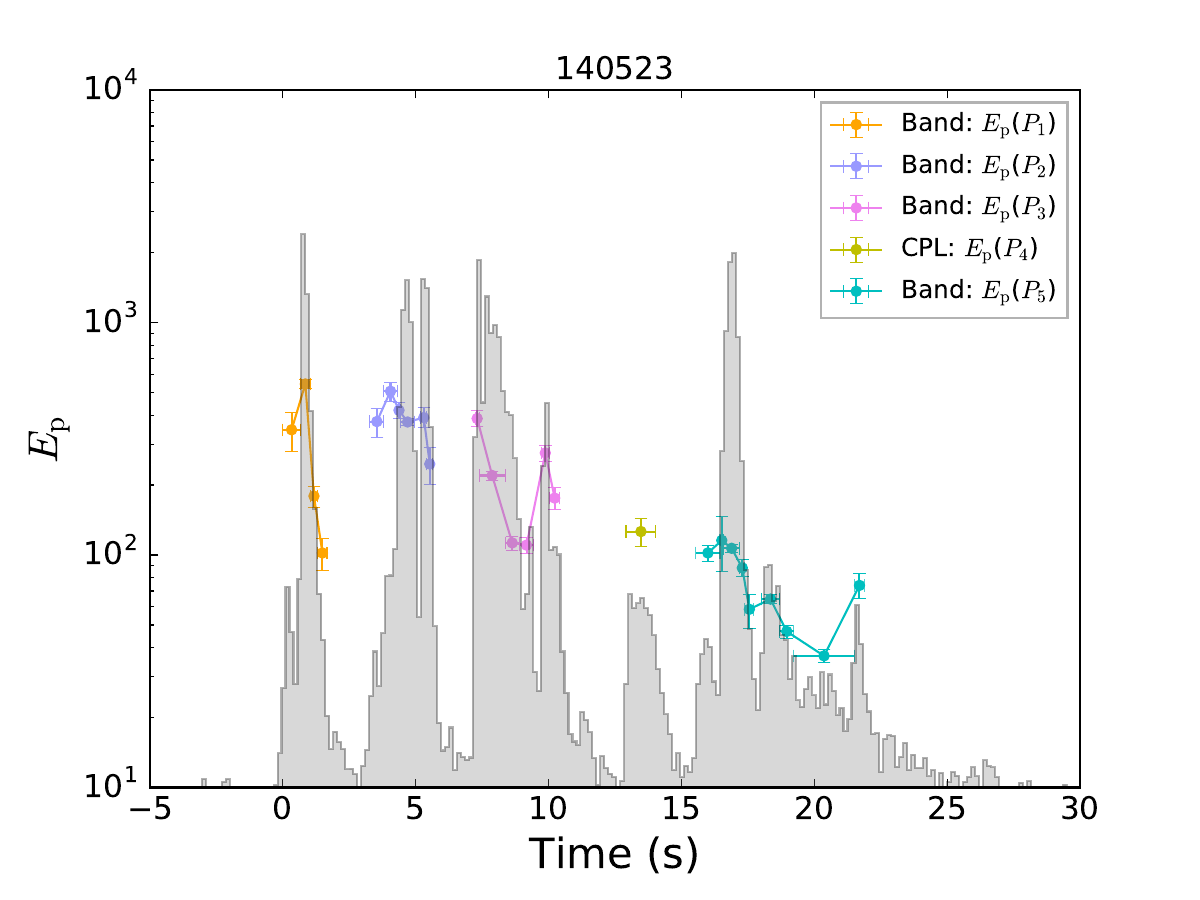}
\includegraphics[angle=0,scale=0.3]{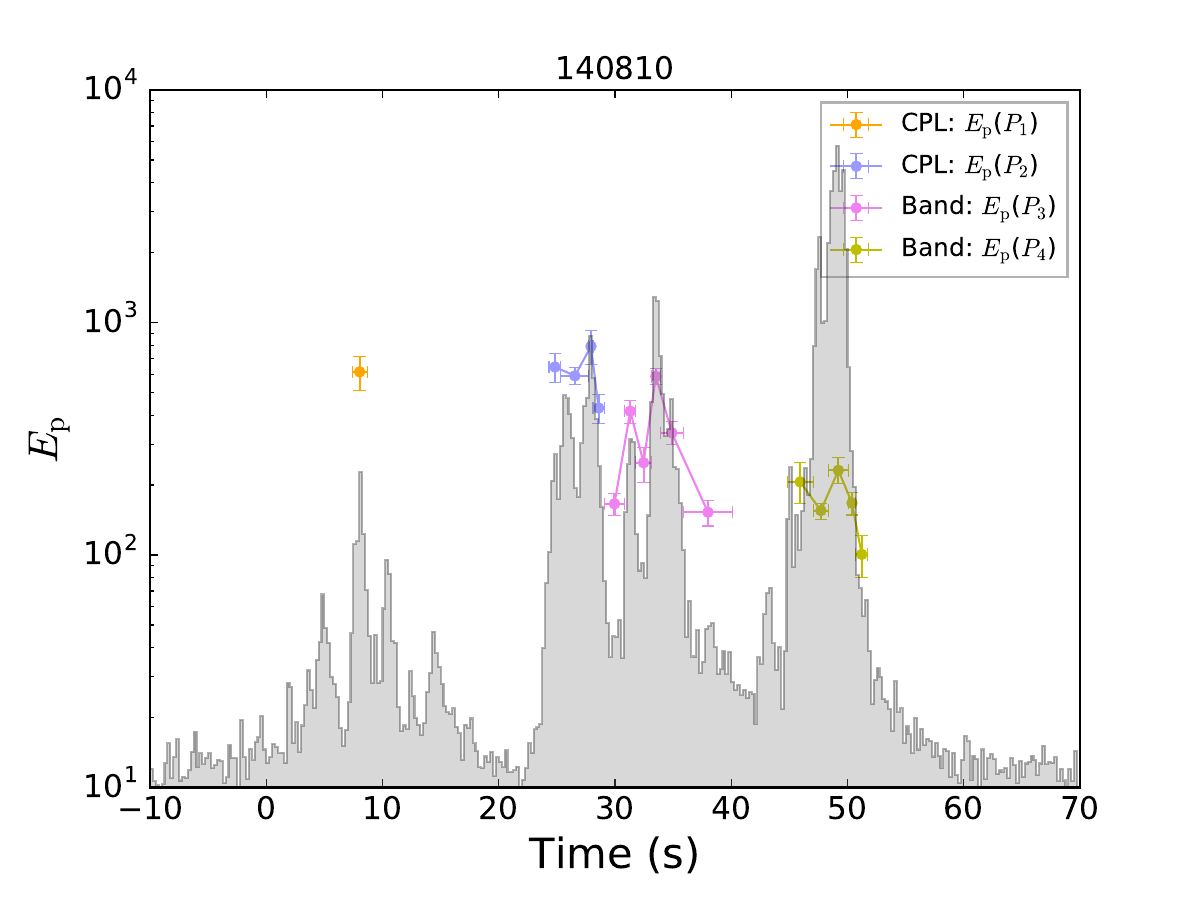}
\includegraphics[angle=0,scale=0.3]{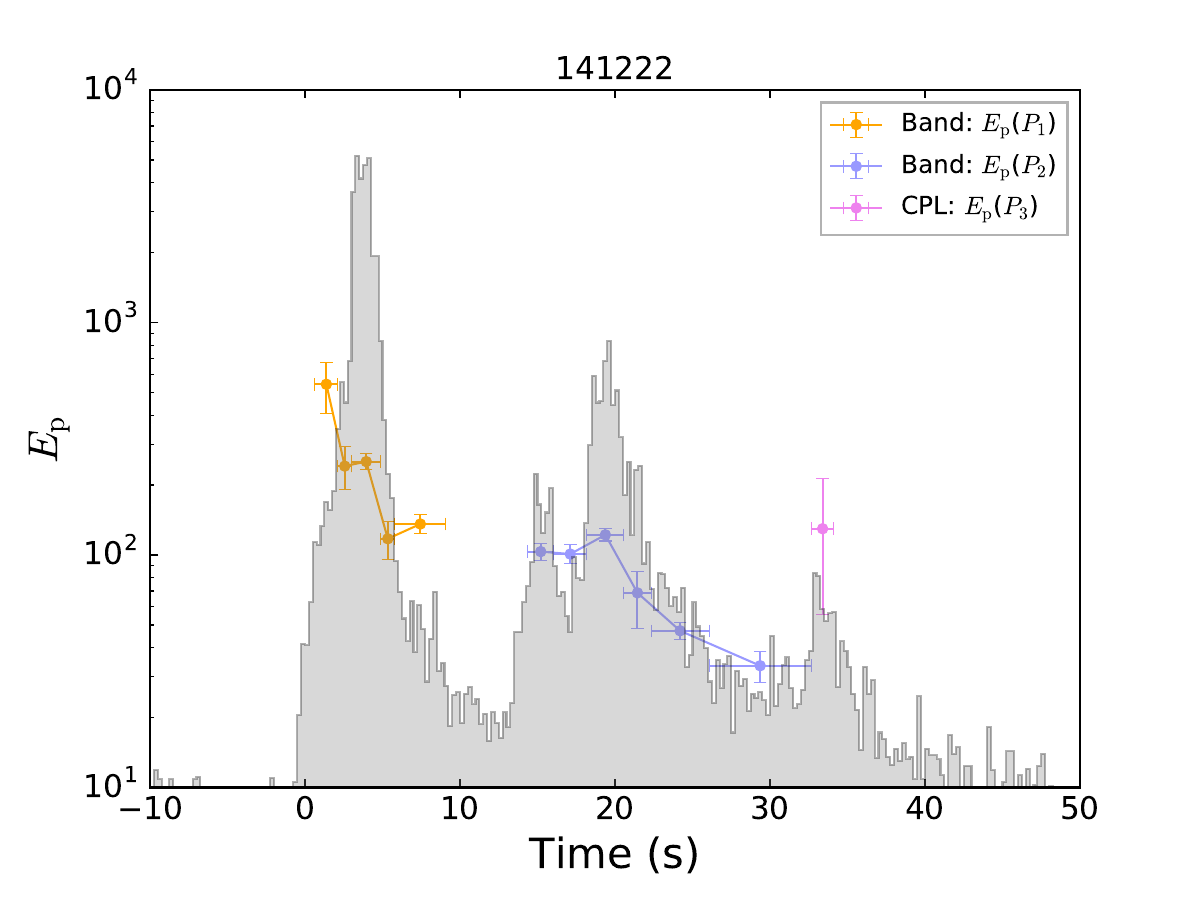}
\includegraphics[angle=0,scale=0.3]{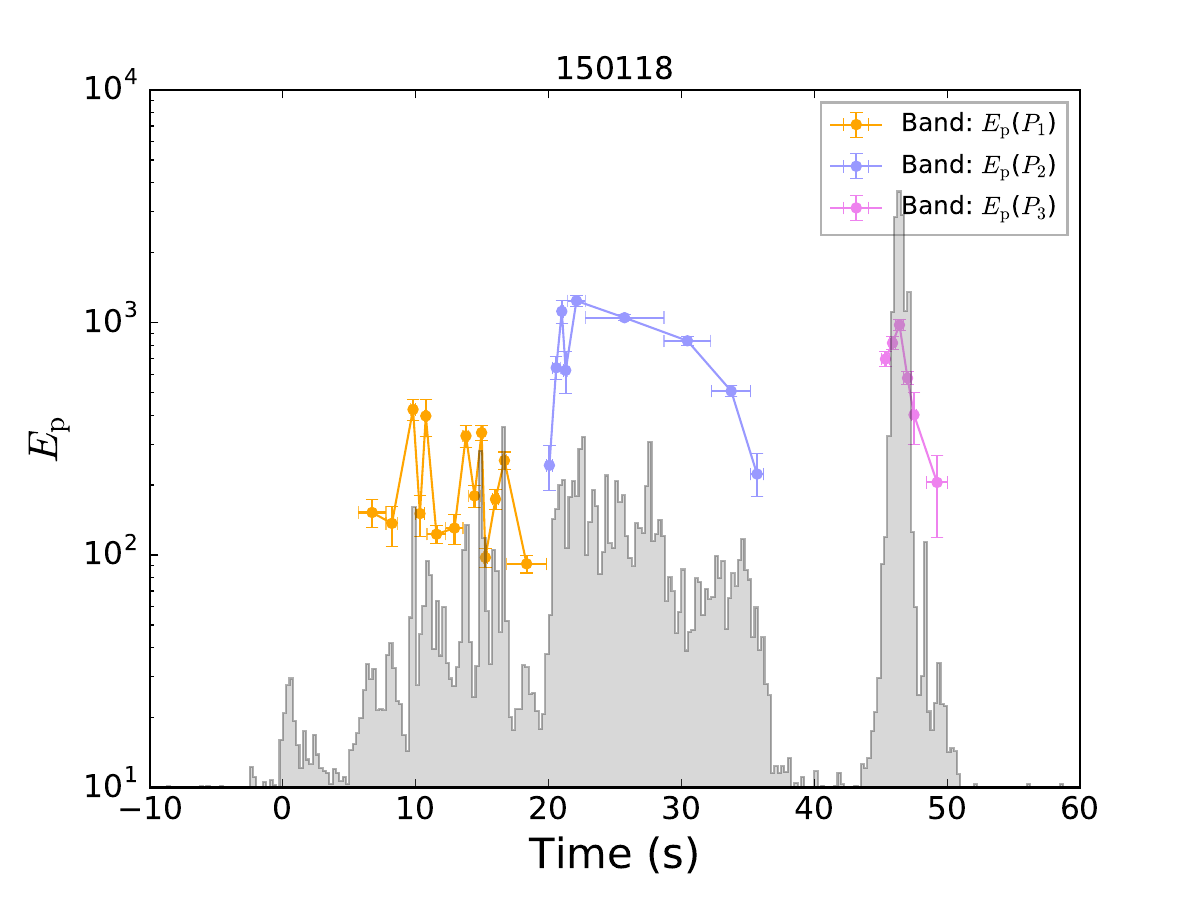}
\center{Fig. \ref{fig:Ep_Best}--- Continued}
\end{figure*}
\begin{figure*}
\includegraphics[angle=0,scale=0.3]{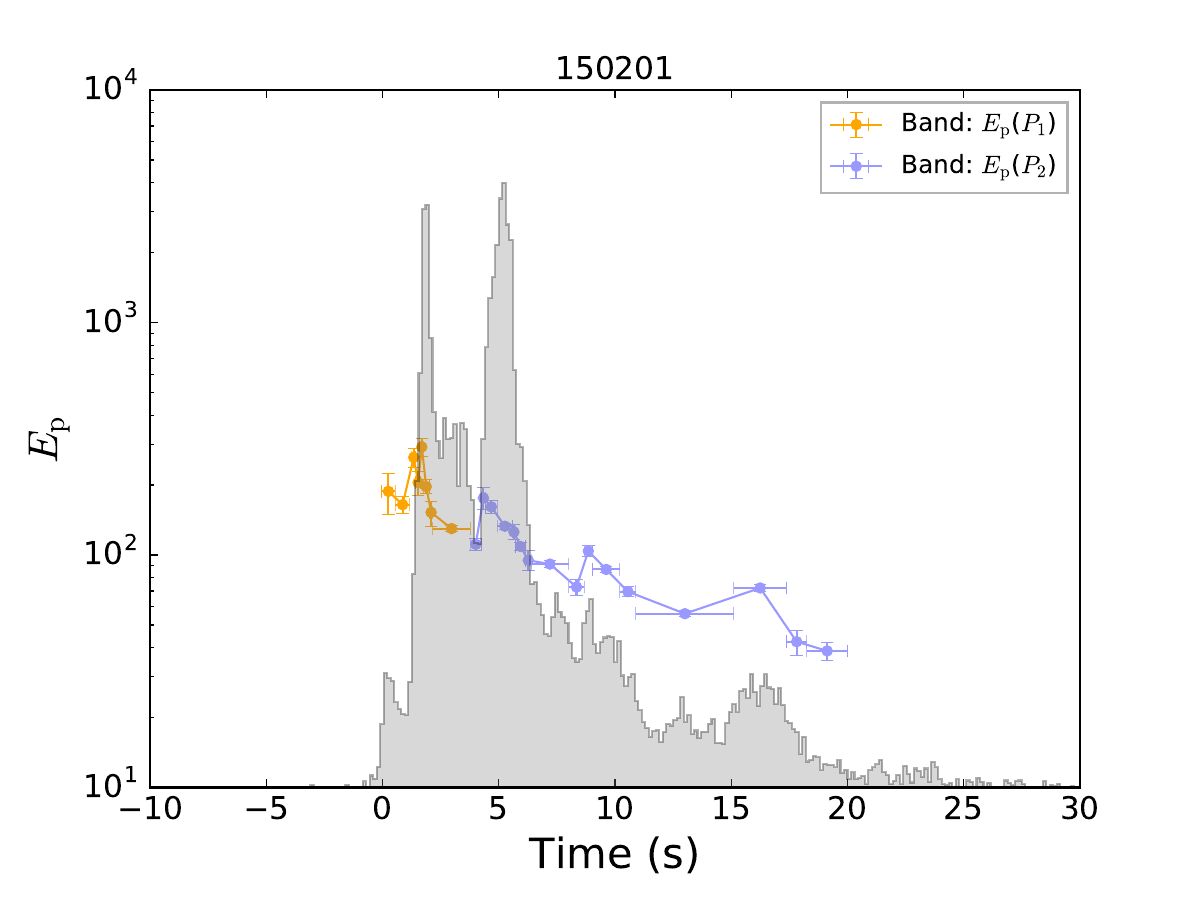}
\includegraphics[angle=0,scale=0.3]{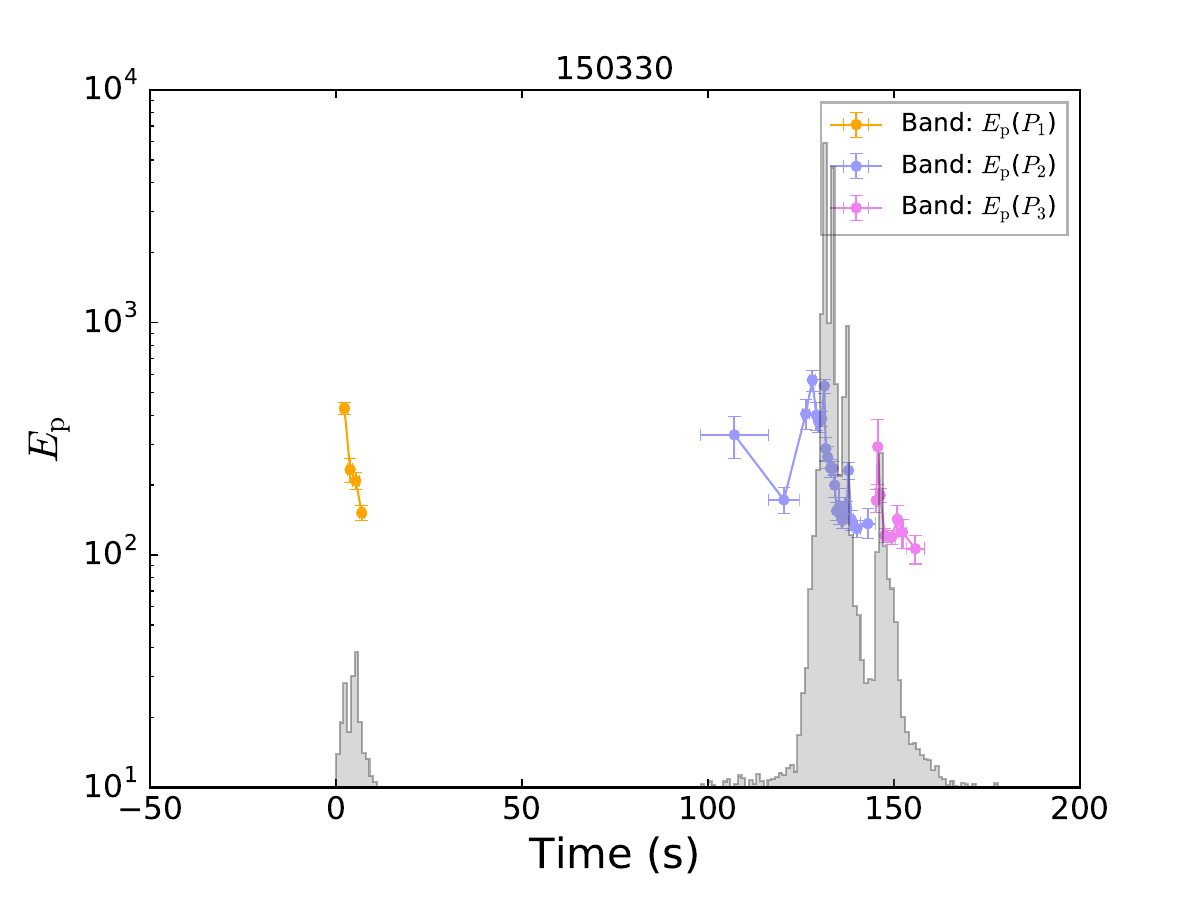}
\includegraphics[angle=0,scale=0.3]{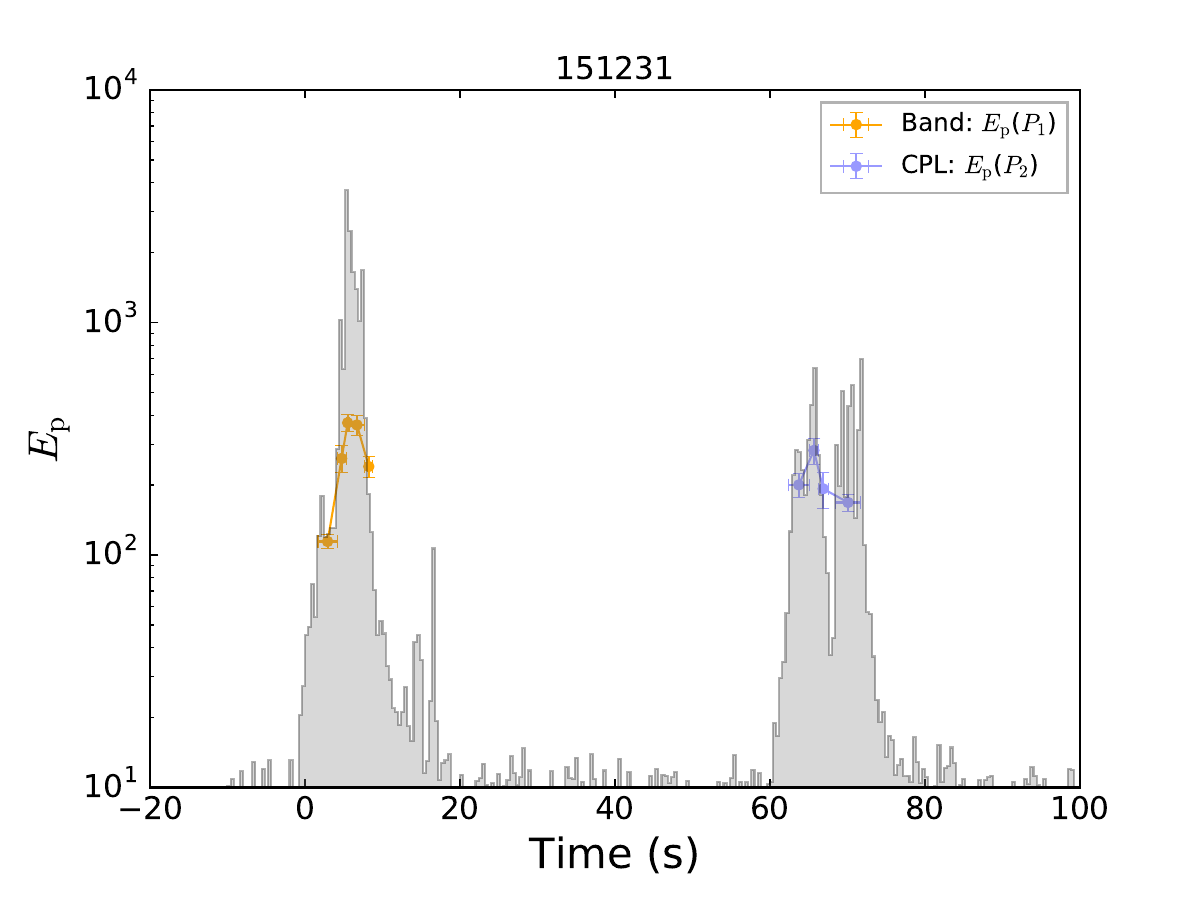}
\includegraphics[angle=0,scale=0.3]{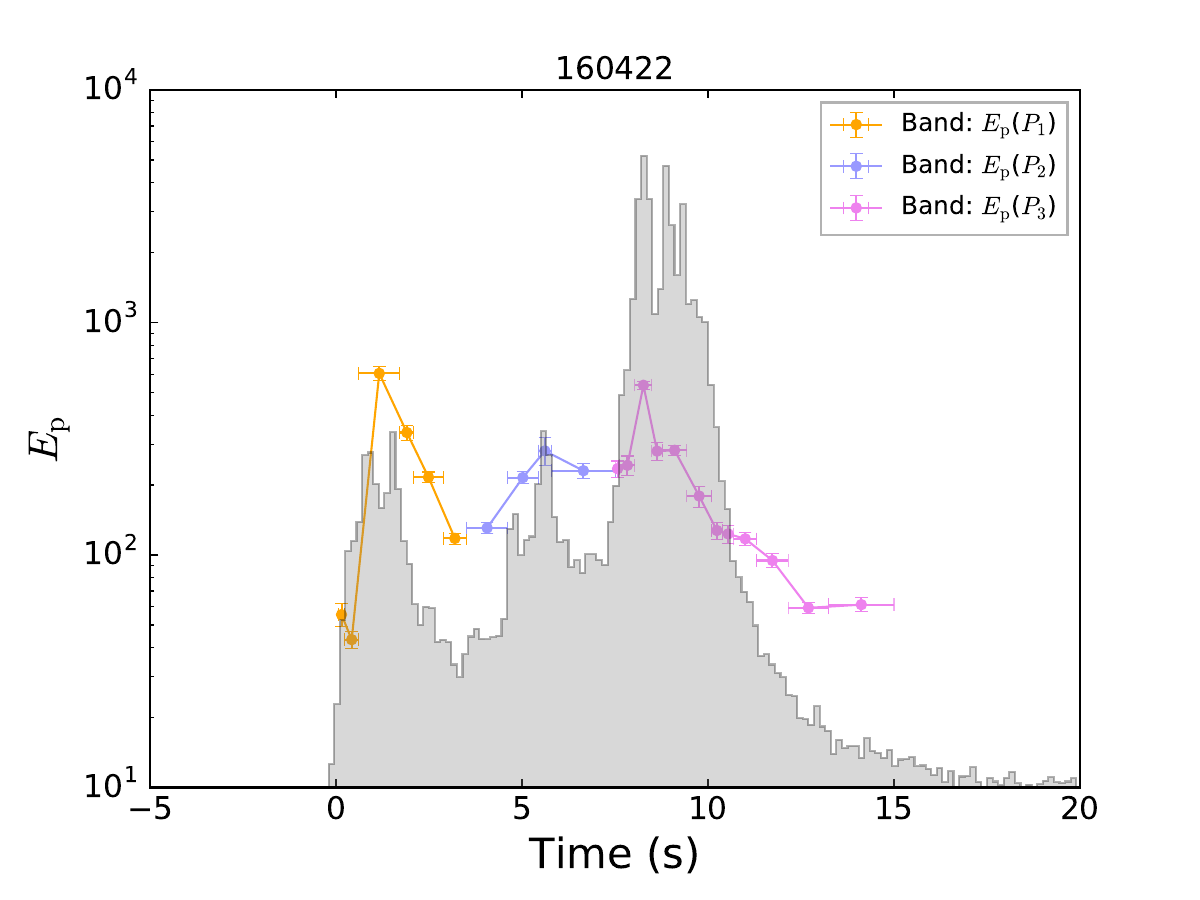}
\includegraphics[angle=0,scale=0.3]{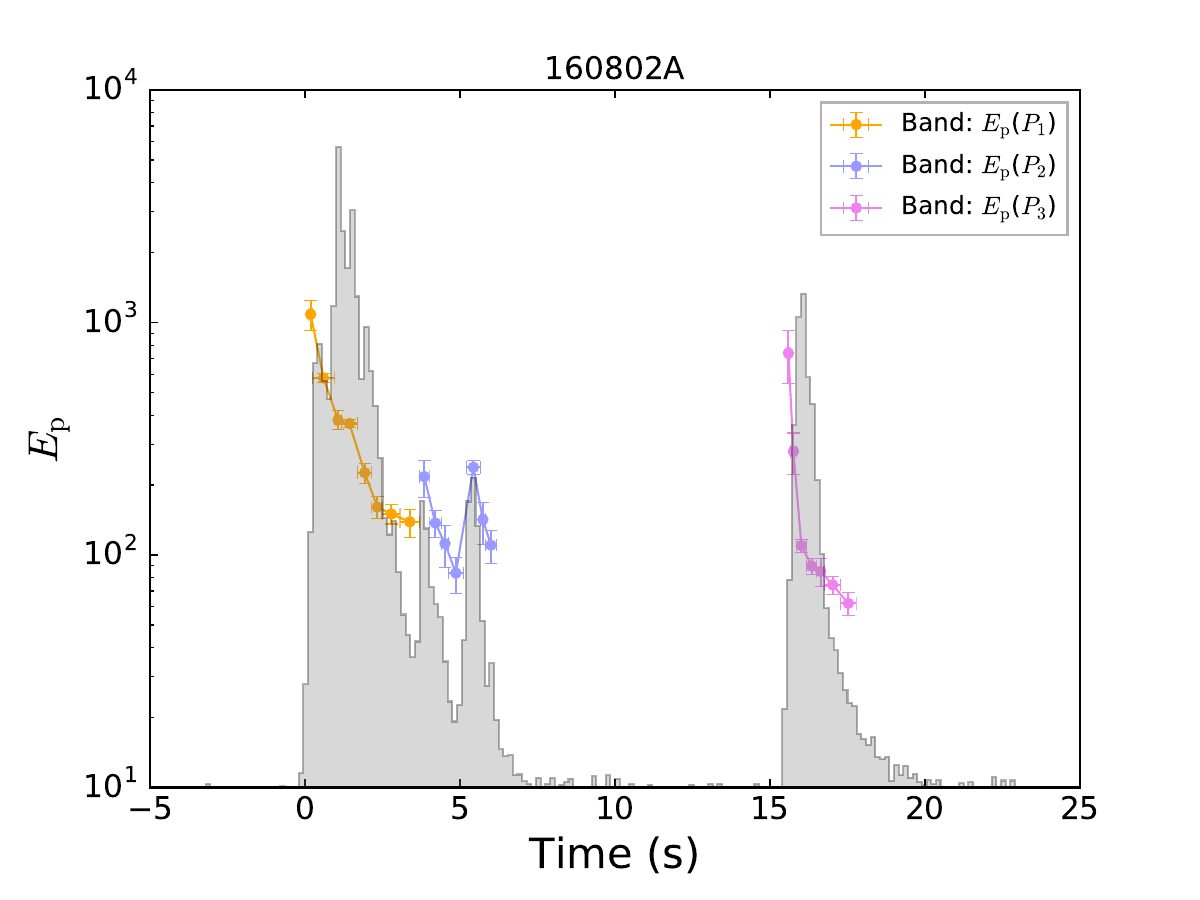}
\includegraphics[angle=0,scale=0.3]{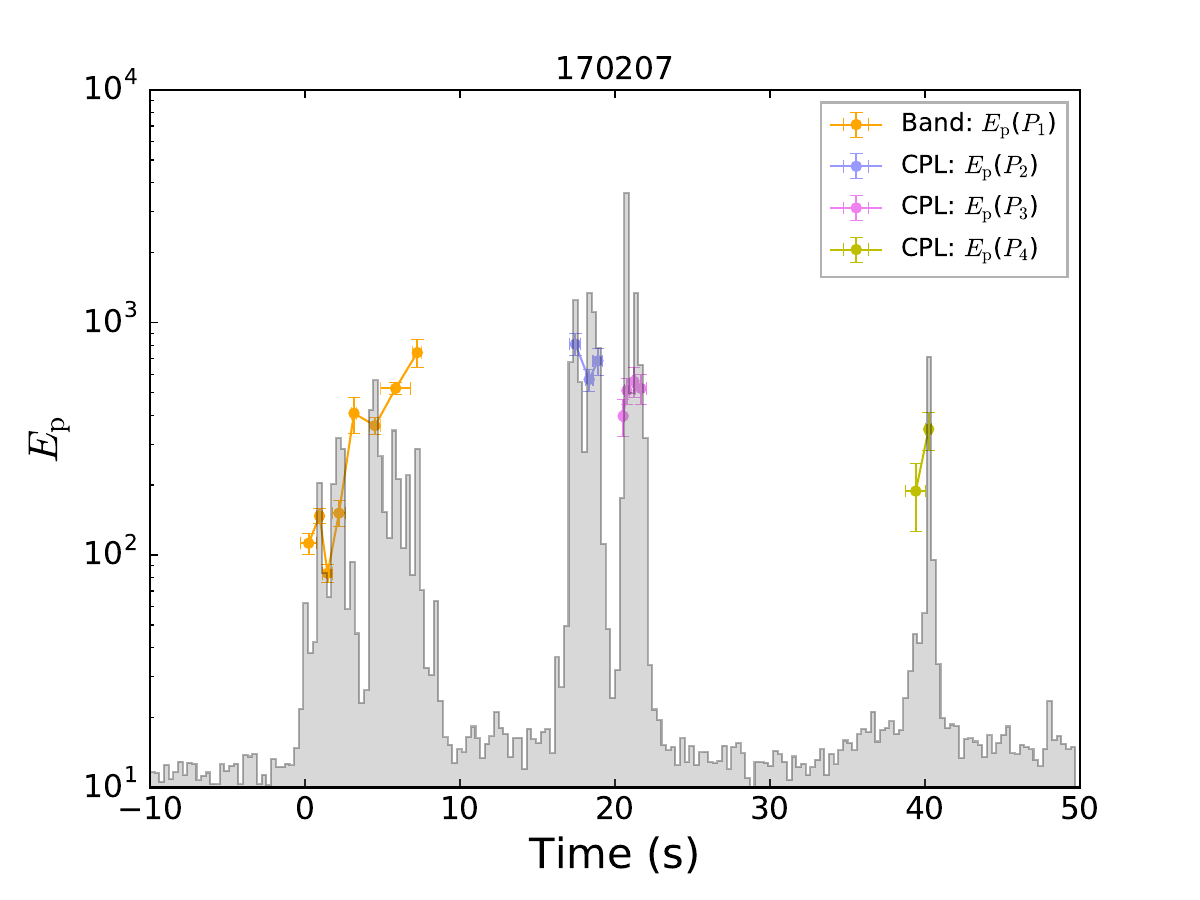}
\includegraphics[angle=0,scale=0.3]{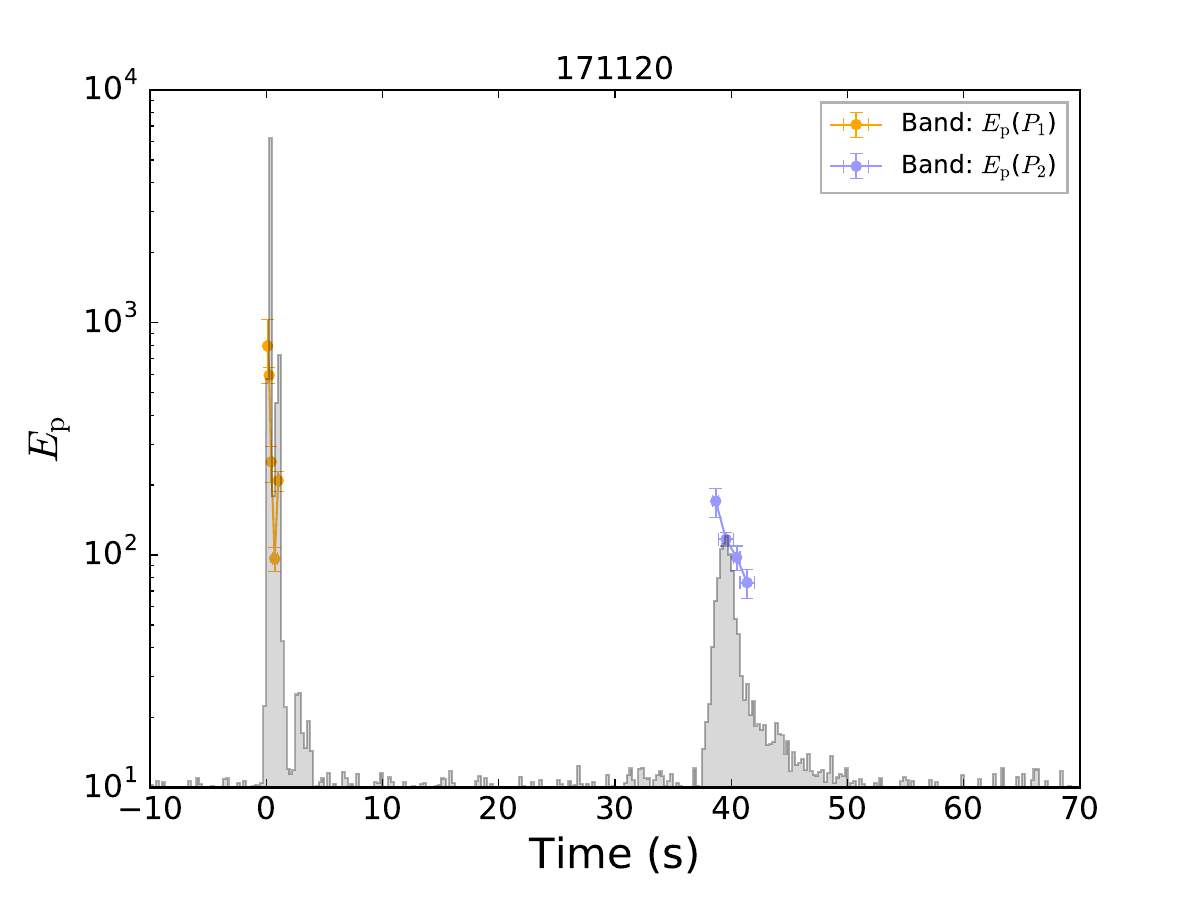}
\includegraphics[angle=0,scale=0.3]{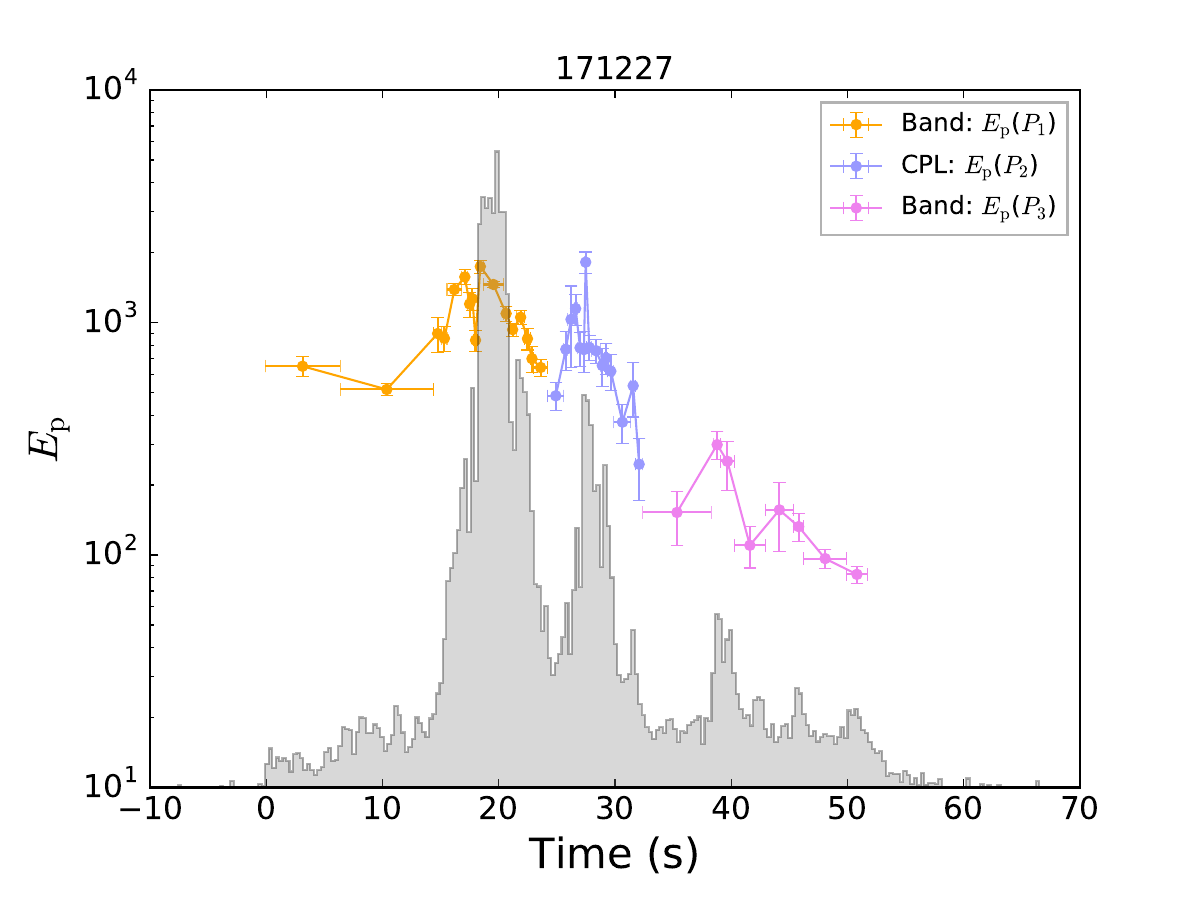}
\includegraphics[angle=0,scale=0.3]{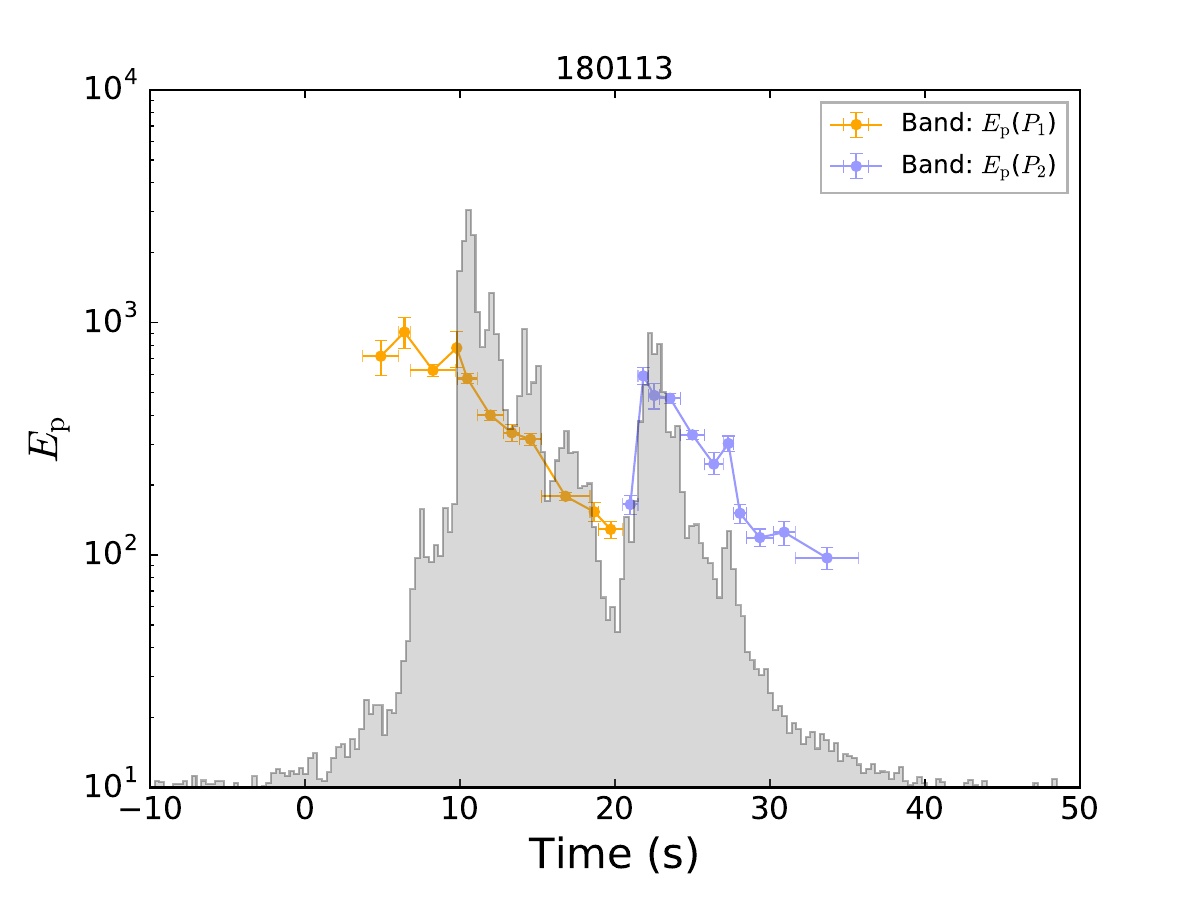}
\includegraphics[angle=0,scale=0.3]{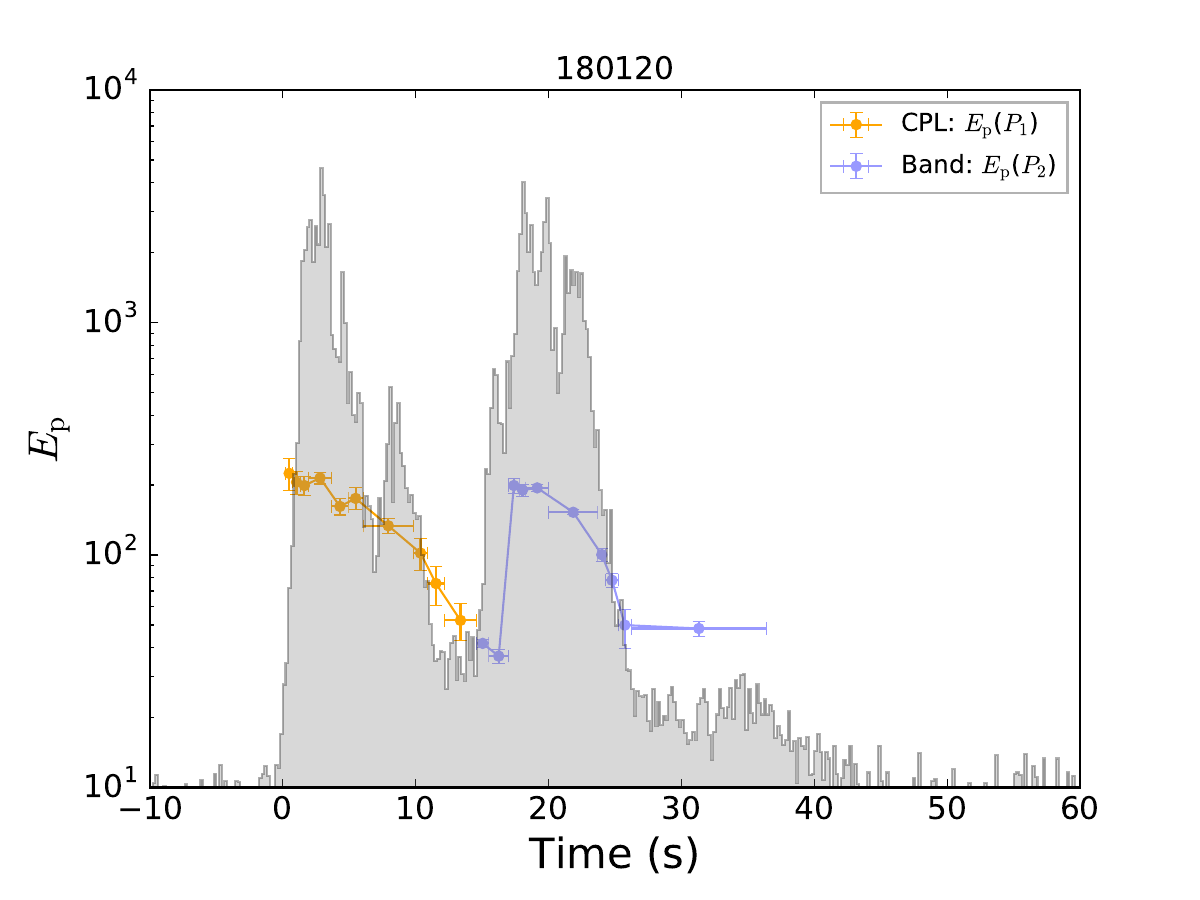}
\includegraphics[angle=0,scale=0.3]{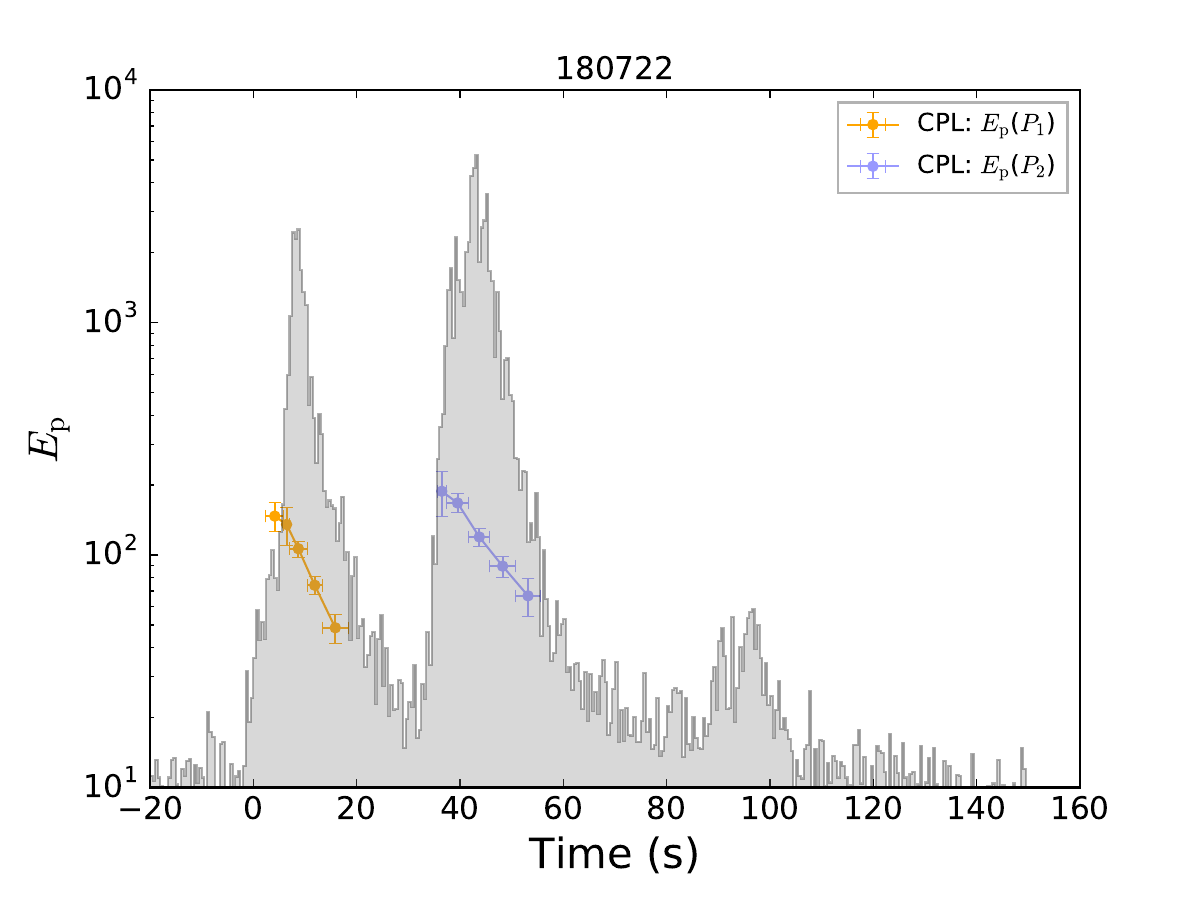}
\includegraphics[angle=0,scale=0.3]{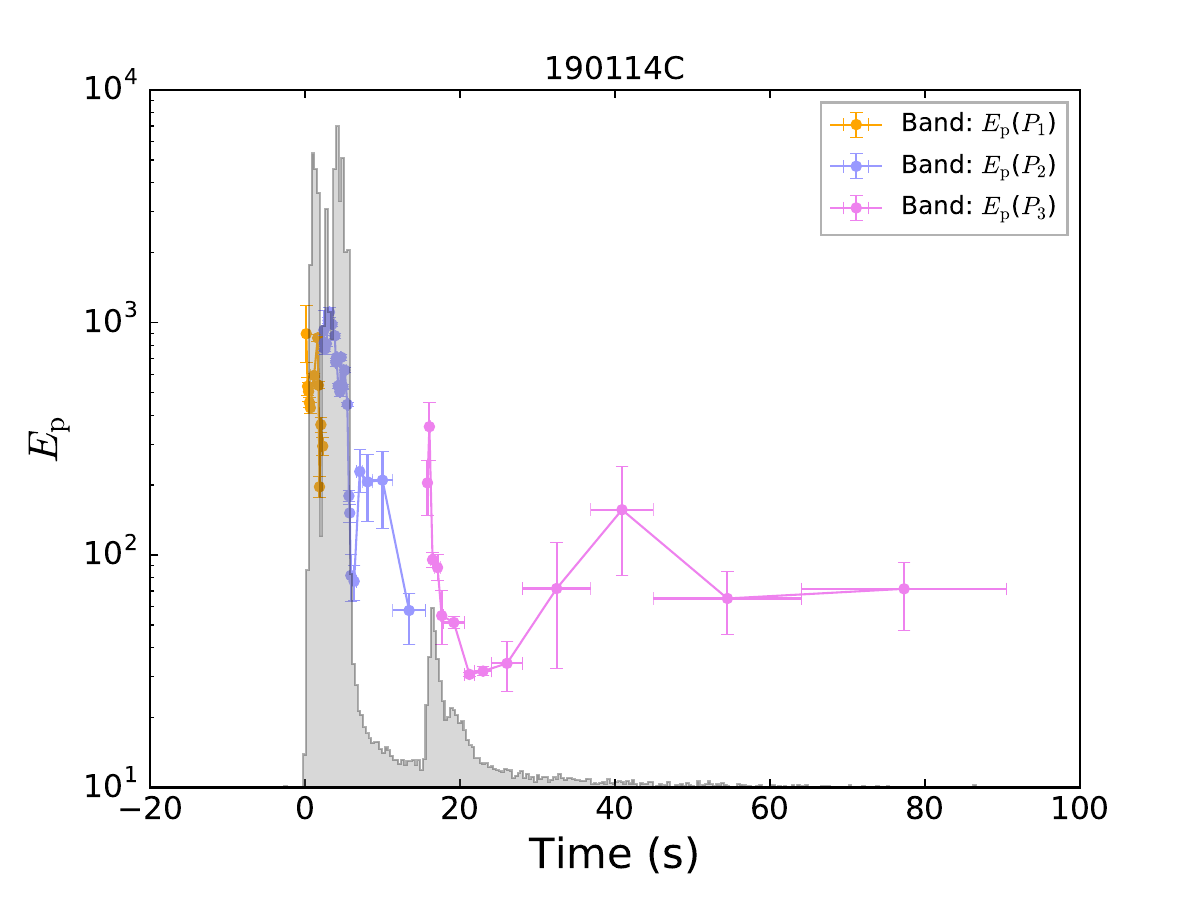}
\center{Fig. \ref{fig:Ep_Best}--- Continued}
\end{figure*}

\clearpage
\vspace{5mm}
\begin{figure*}
\includegraphics[angle=0,scale=0.3]{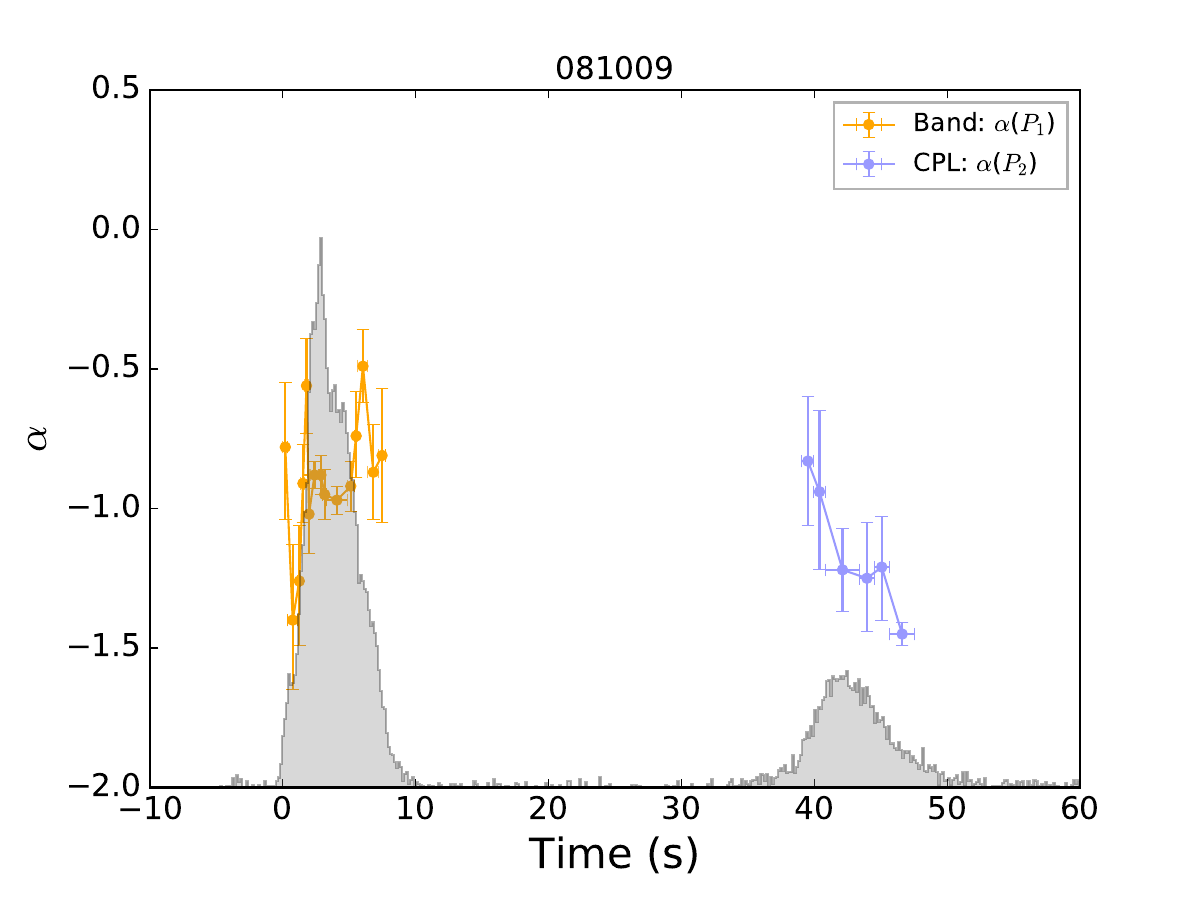}
\includegraphics[angle=0,scale=0.3]{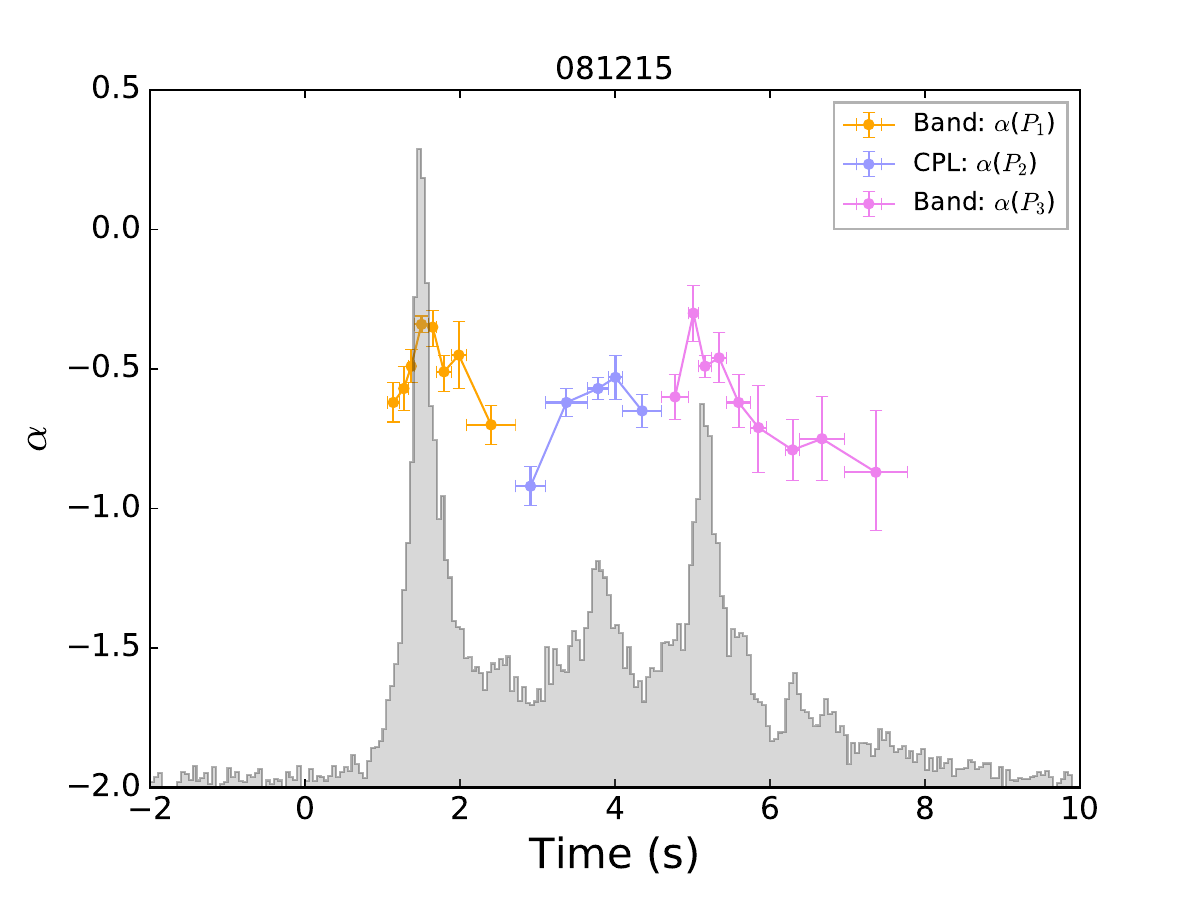}
\includegraphics[angle=0,scale=0.3]{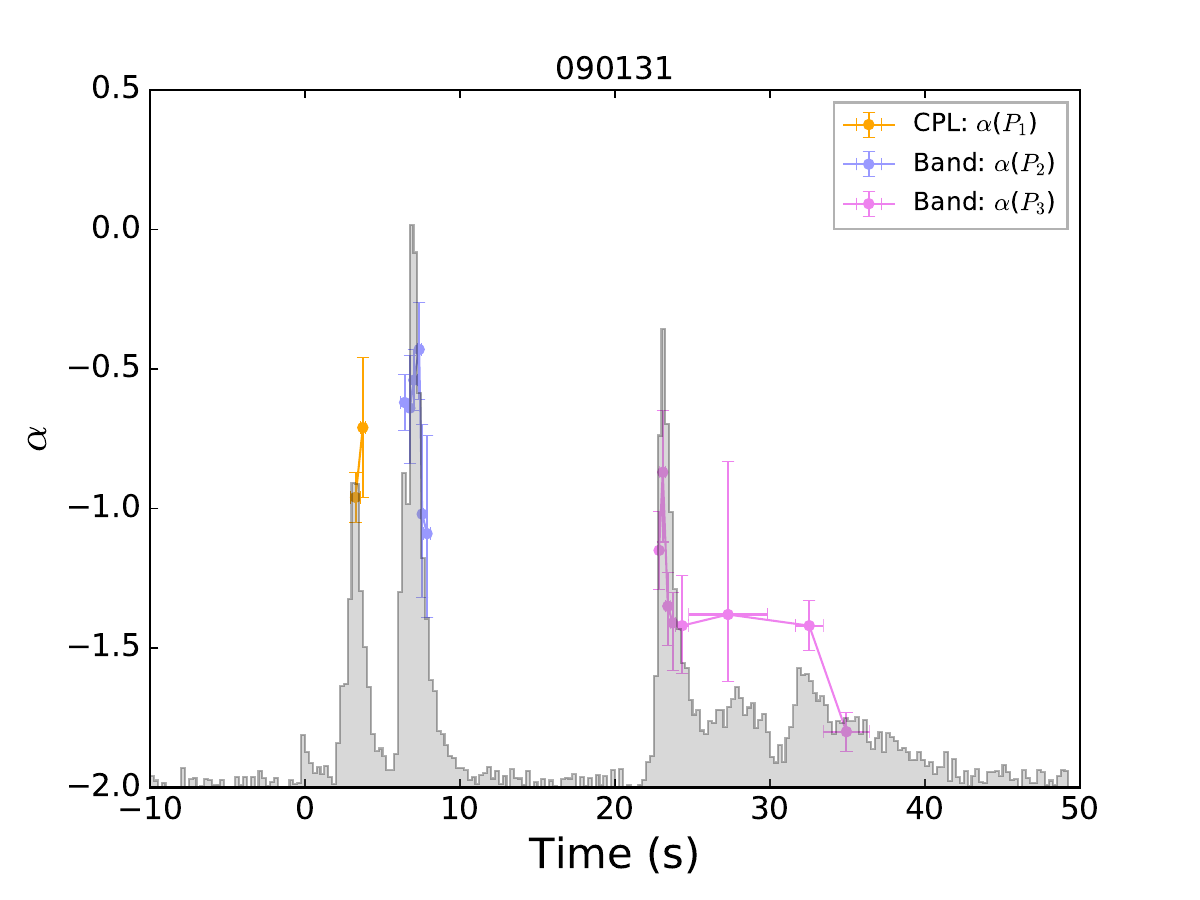}
\includegraphics[angle=0,scale=0.3]{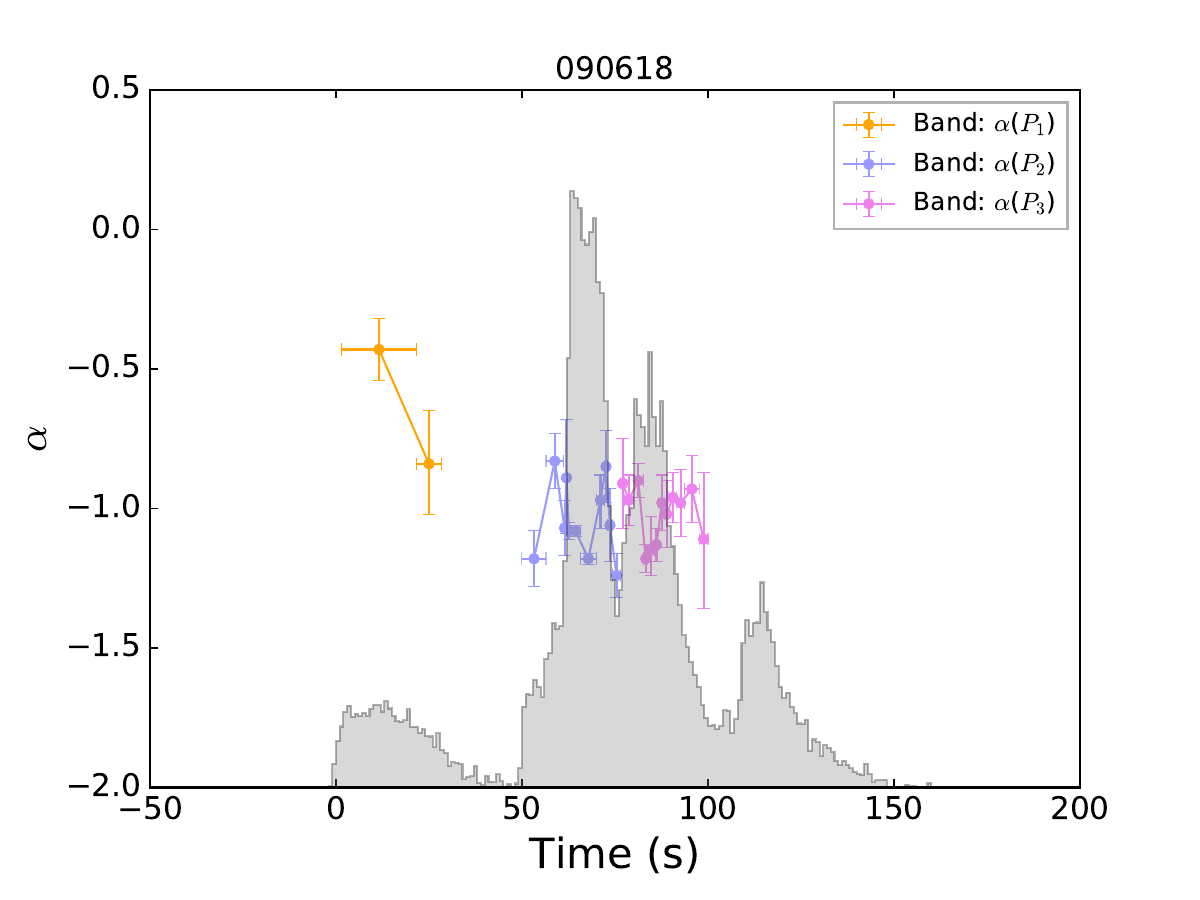}
\includegraphics[angle=0,scale=0.3]{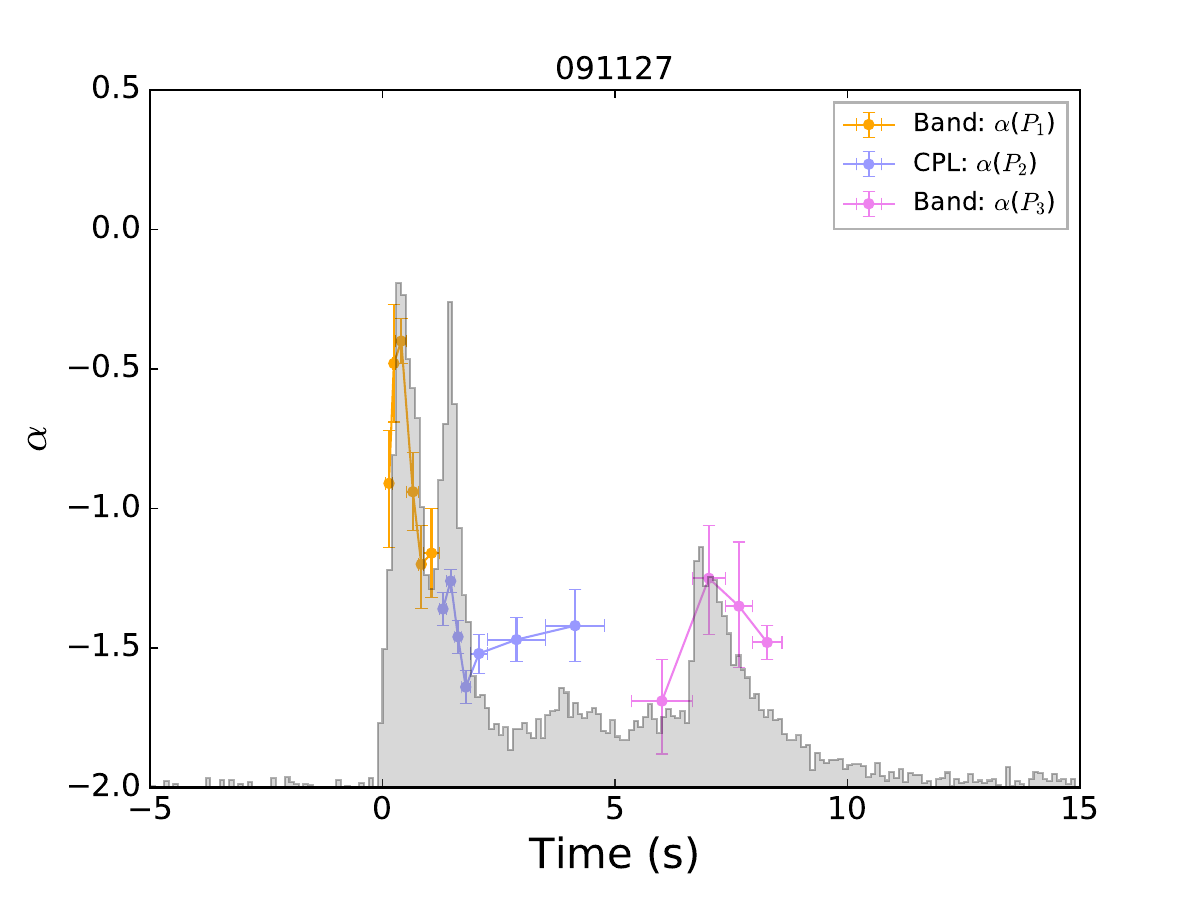}
\includegraphics[angle=0,scale=0.3]{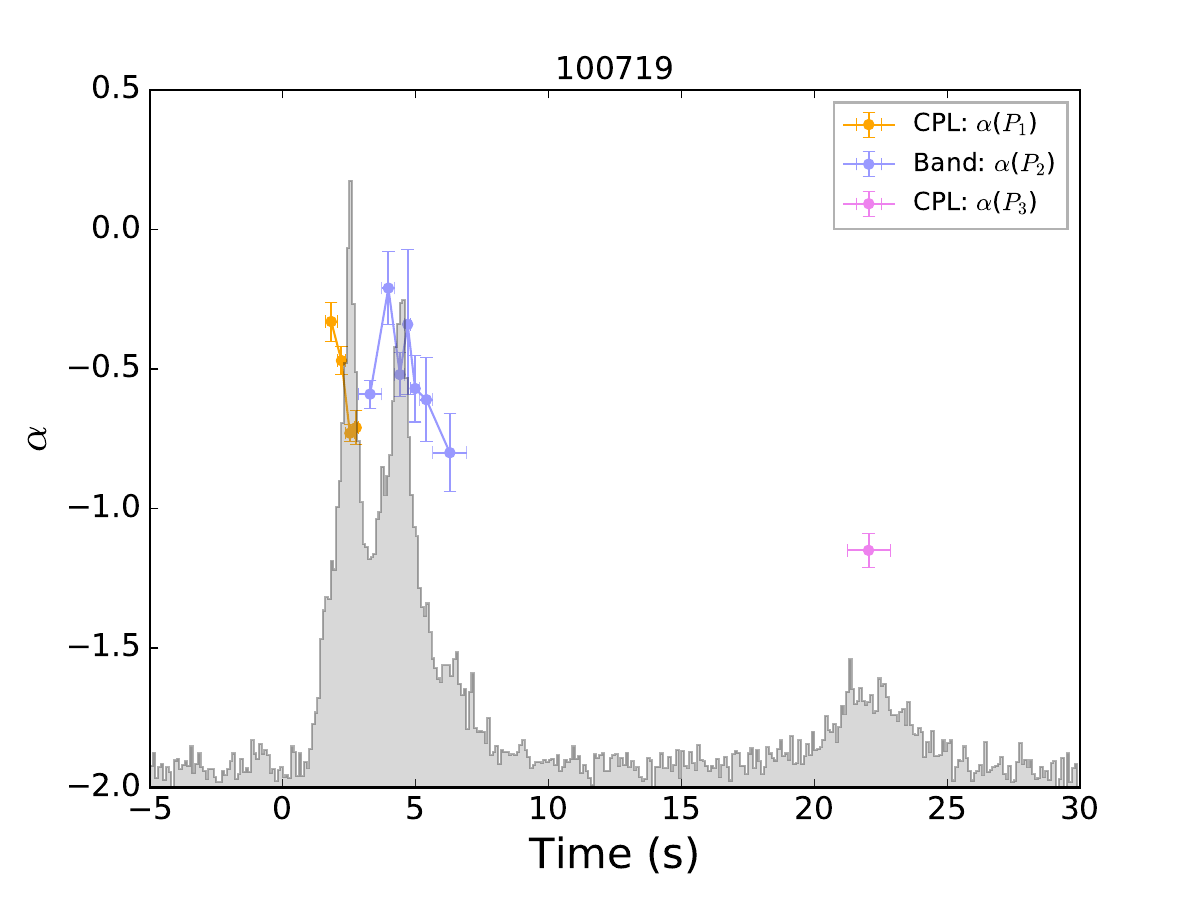}
\includegraphics[angle=0,scale=0.3]{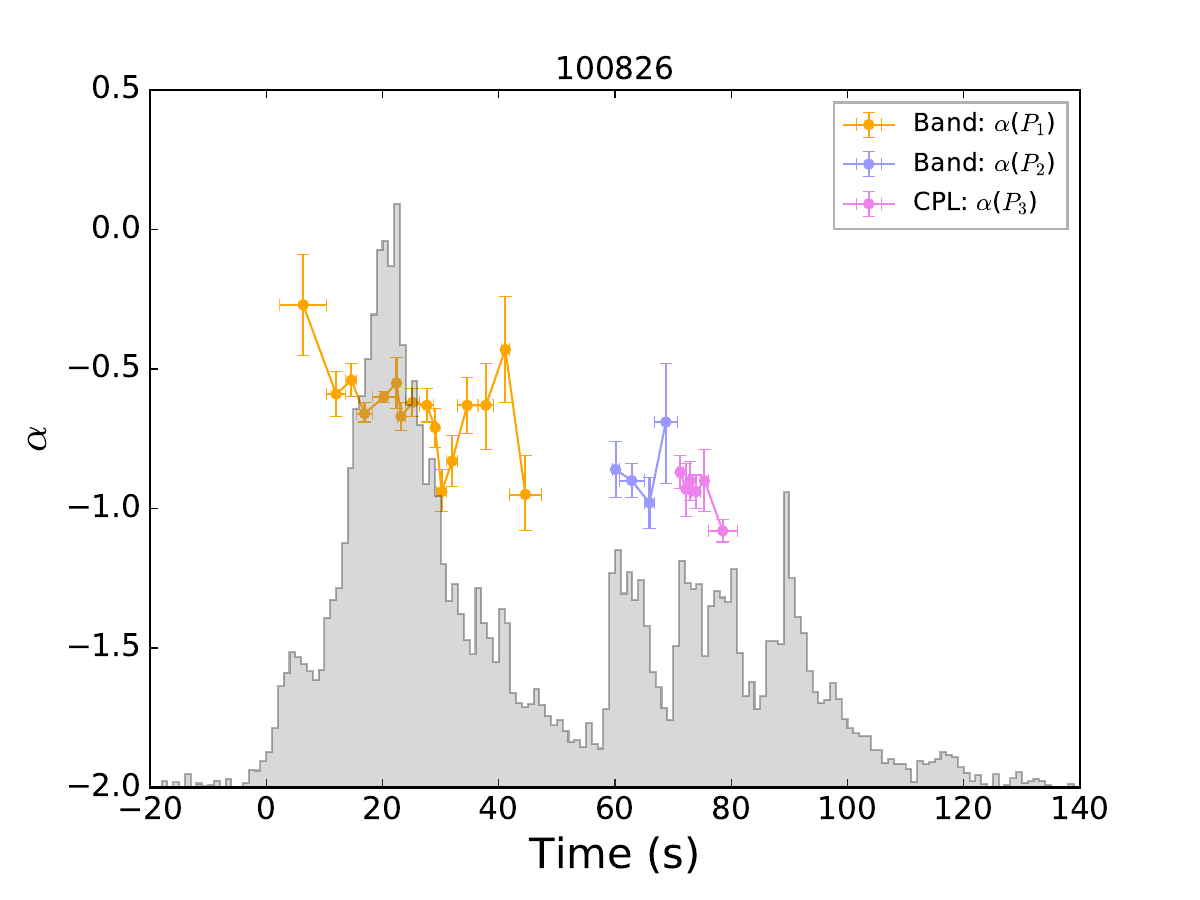}
\includegraphics[angle=0,scale=0.3]{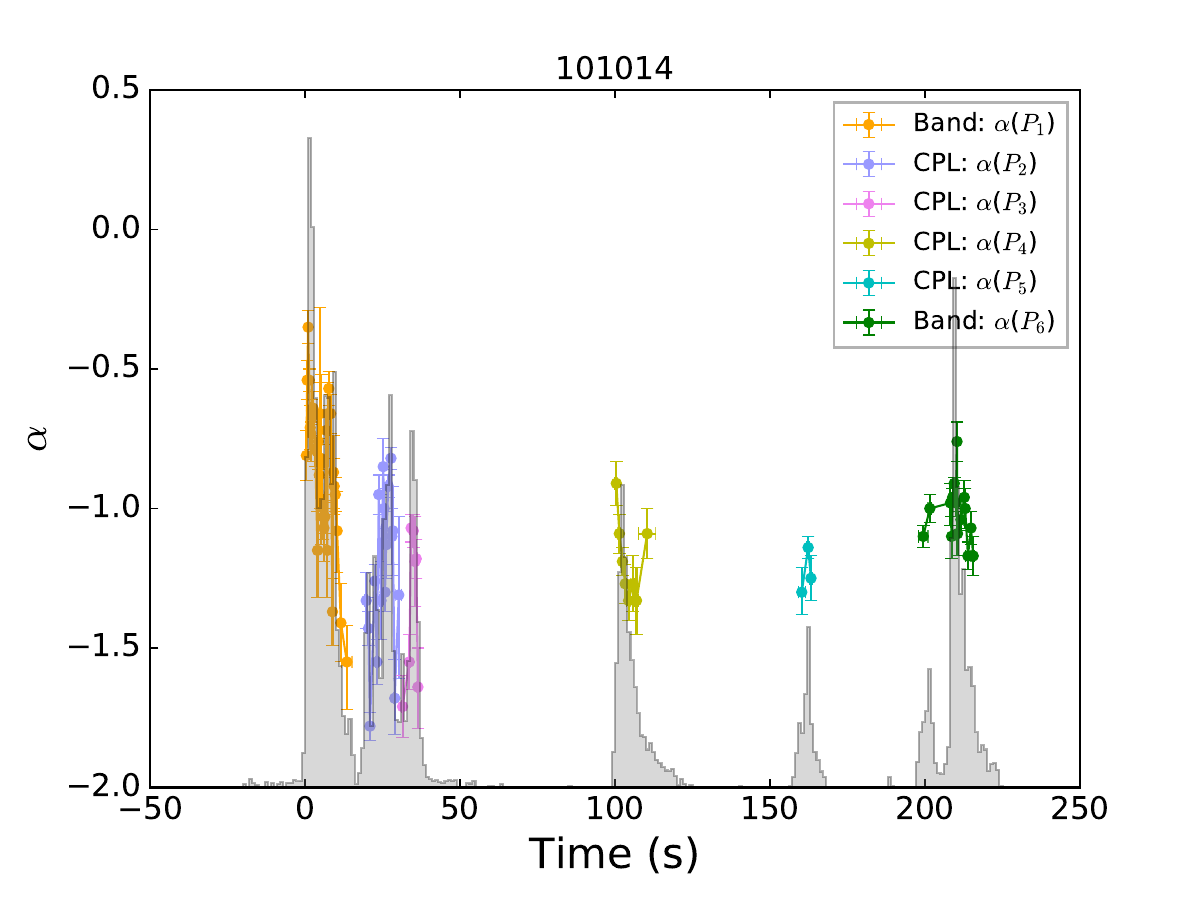}
\includegraphics[angle=0,scale=0.3]{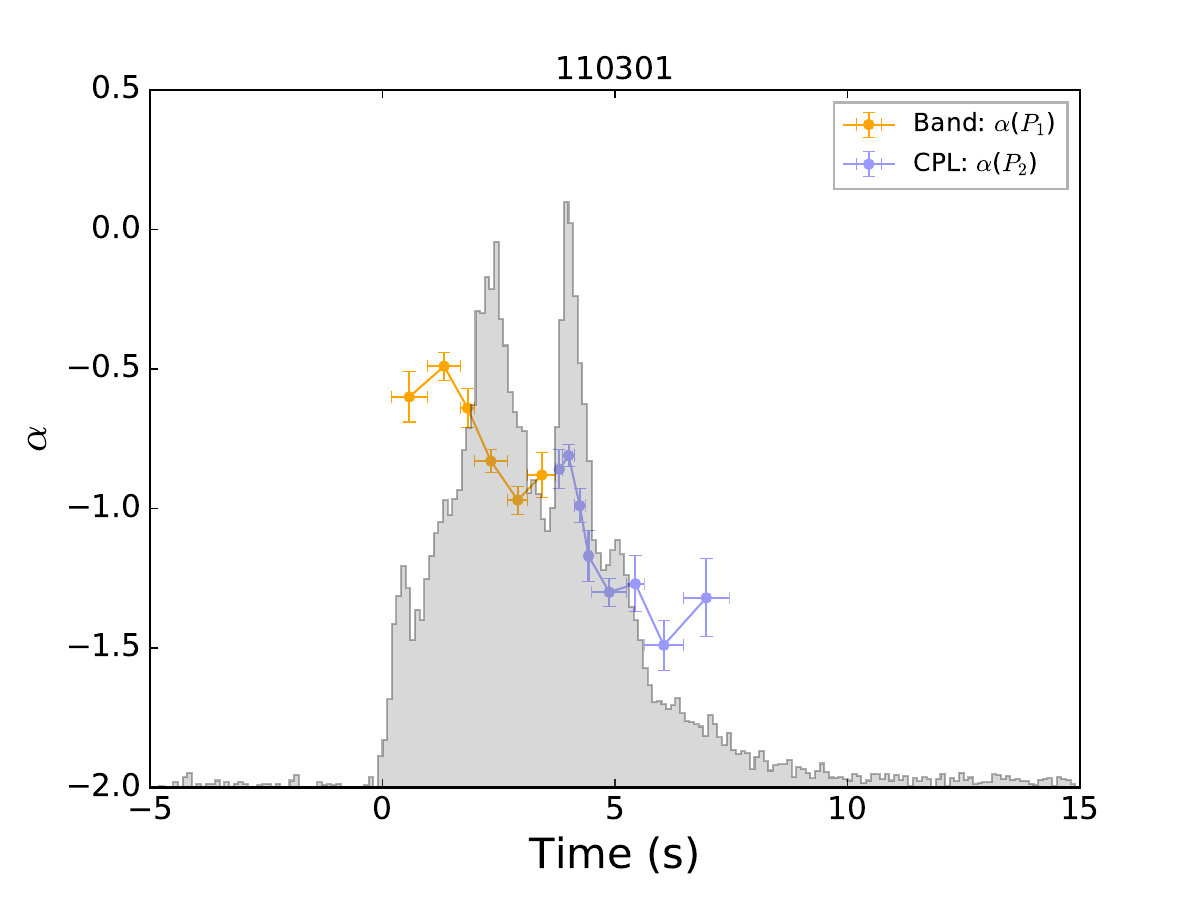}
\includegraphics[angle=0,scale=0.3]{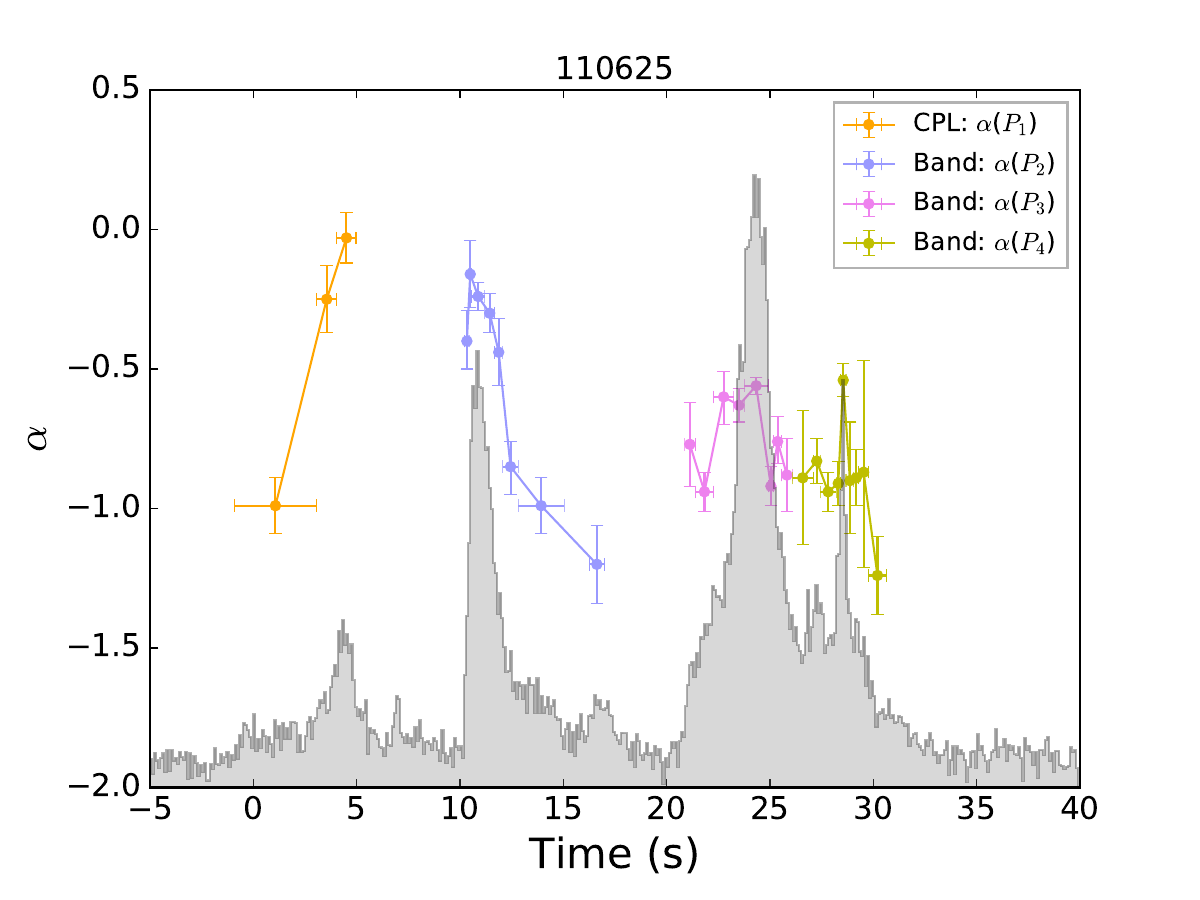}
\includegraphics[angle=0,scale=0.3]{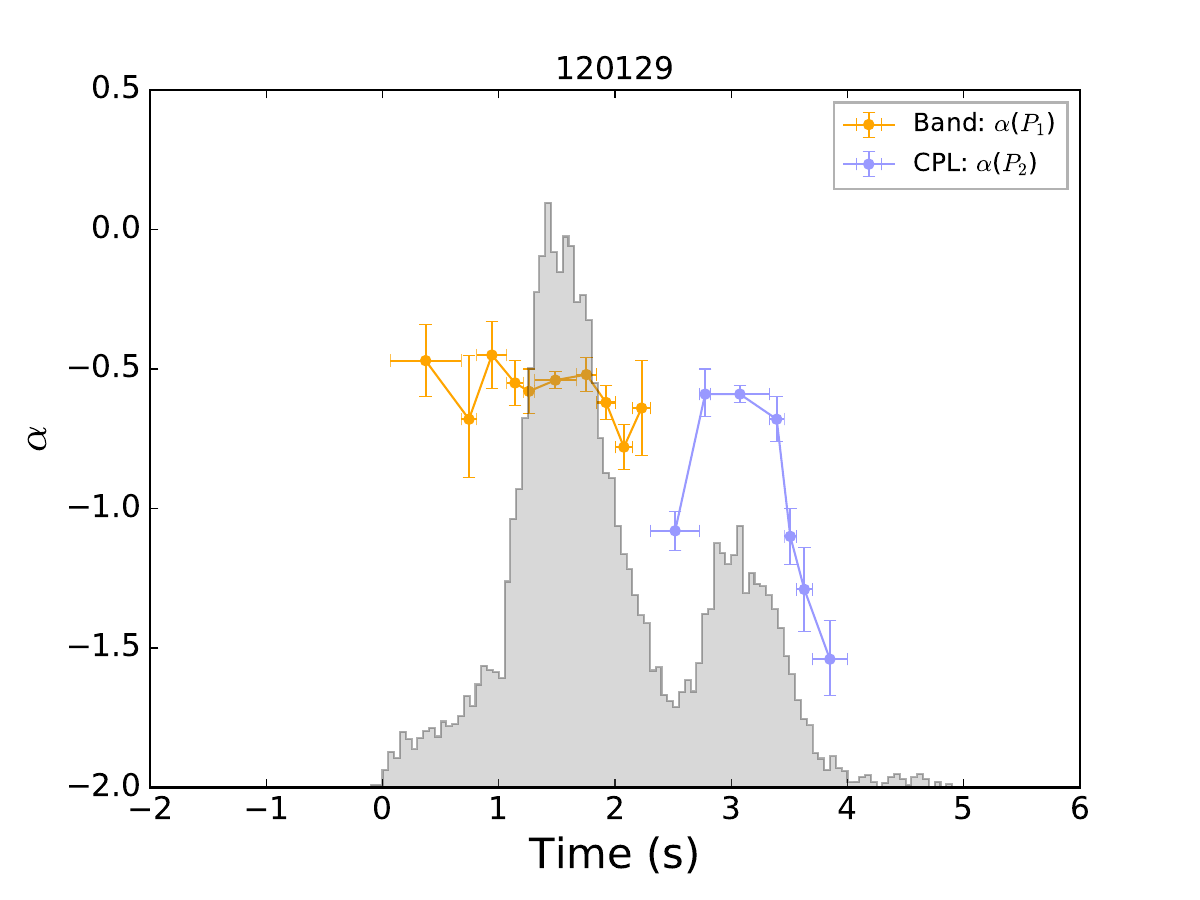}
\includegraphics[angle=0,scale=0.3]{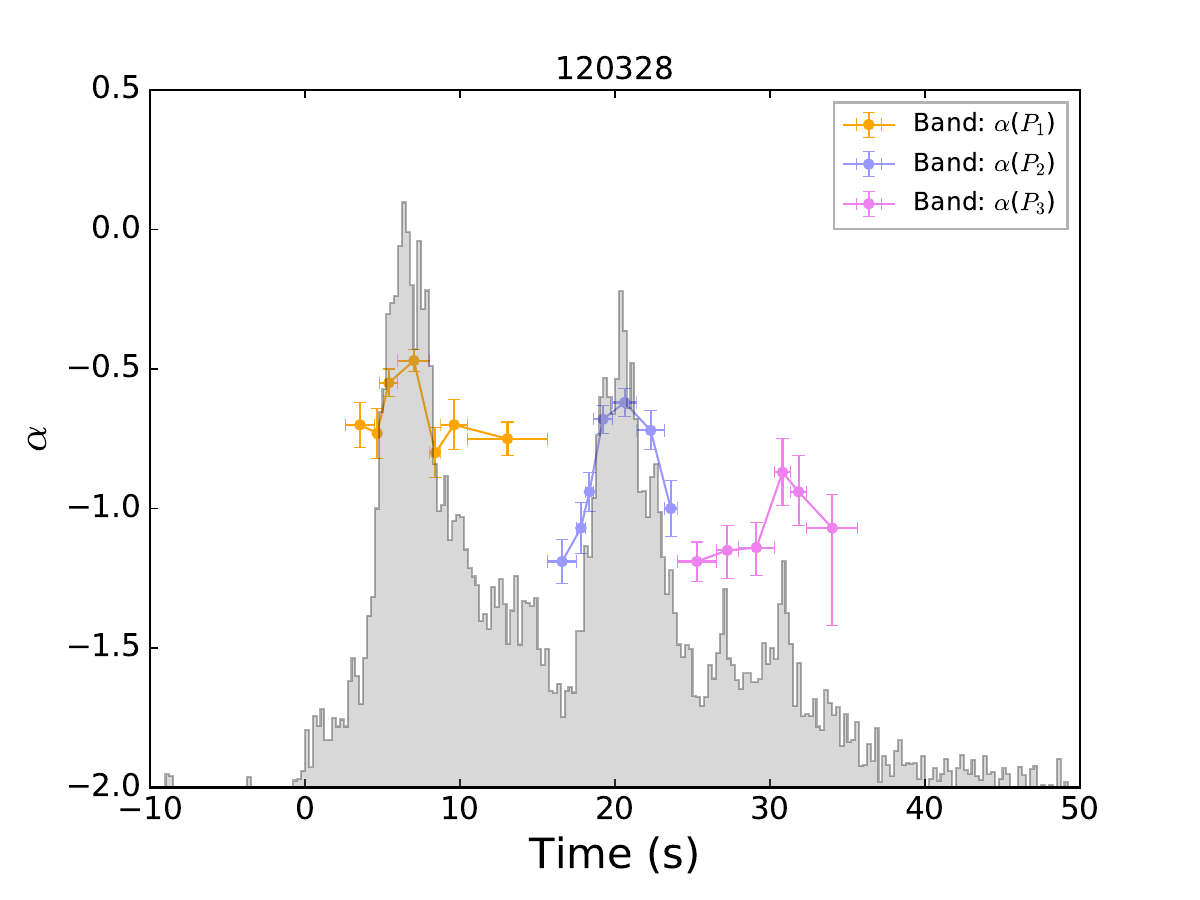}
\caption{Temporal evolution of the $\alpha$ index. The symbols and colors are the same as in Figure \ref{fig:Ep_Best}.}\label{fig:Alpha_Best}
\end{figure*}
\begin{figure*}
\includegraphics[angle=0,scale=0.3]{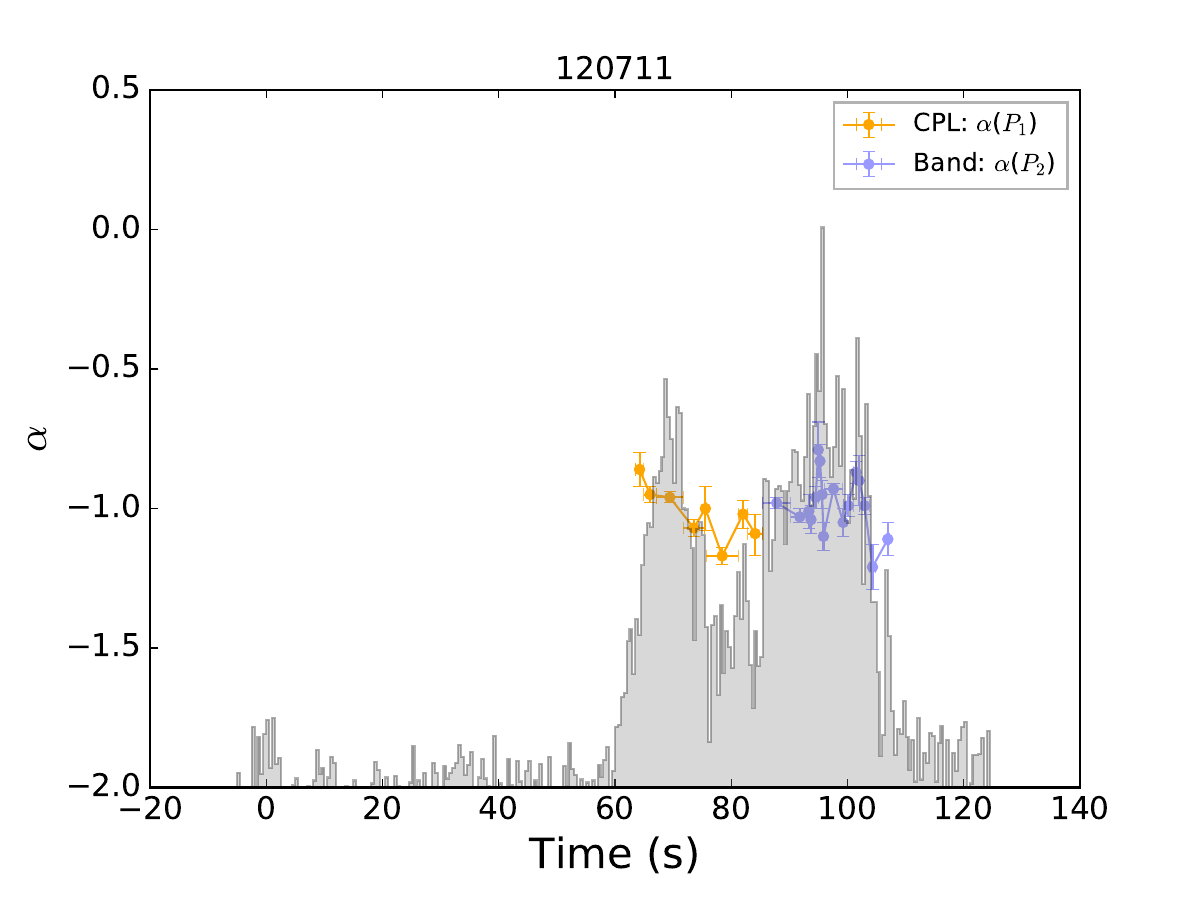}
\includegraphics[angle=0,scale=0.3]{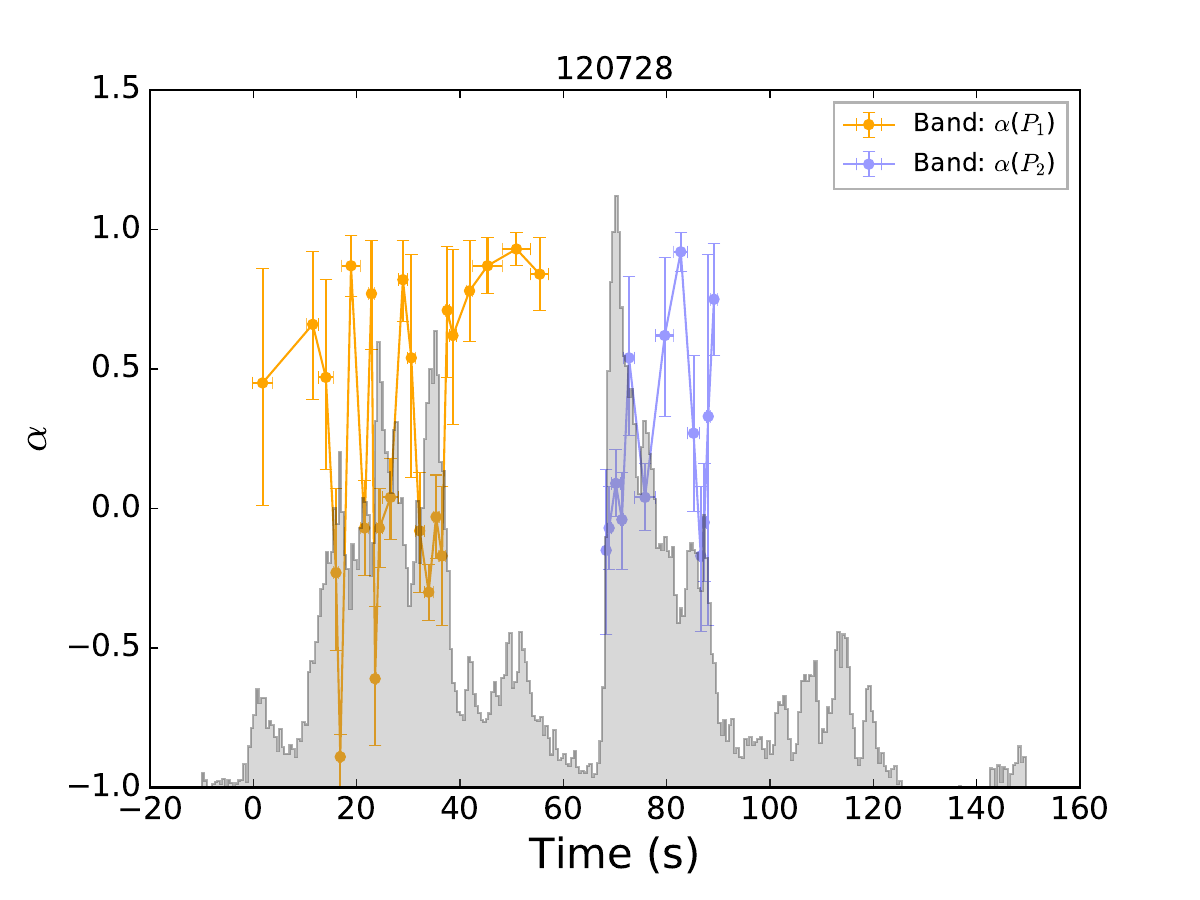}
\includegraphics[angle=0,scale=0.3]{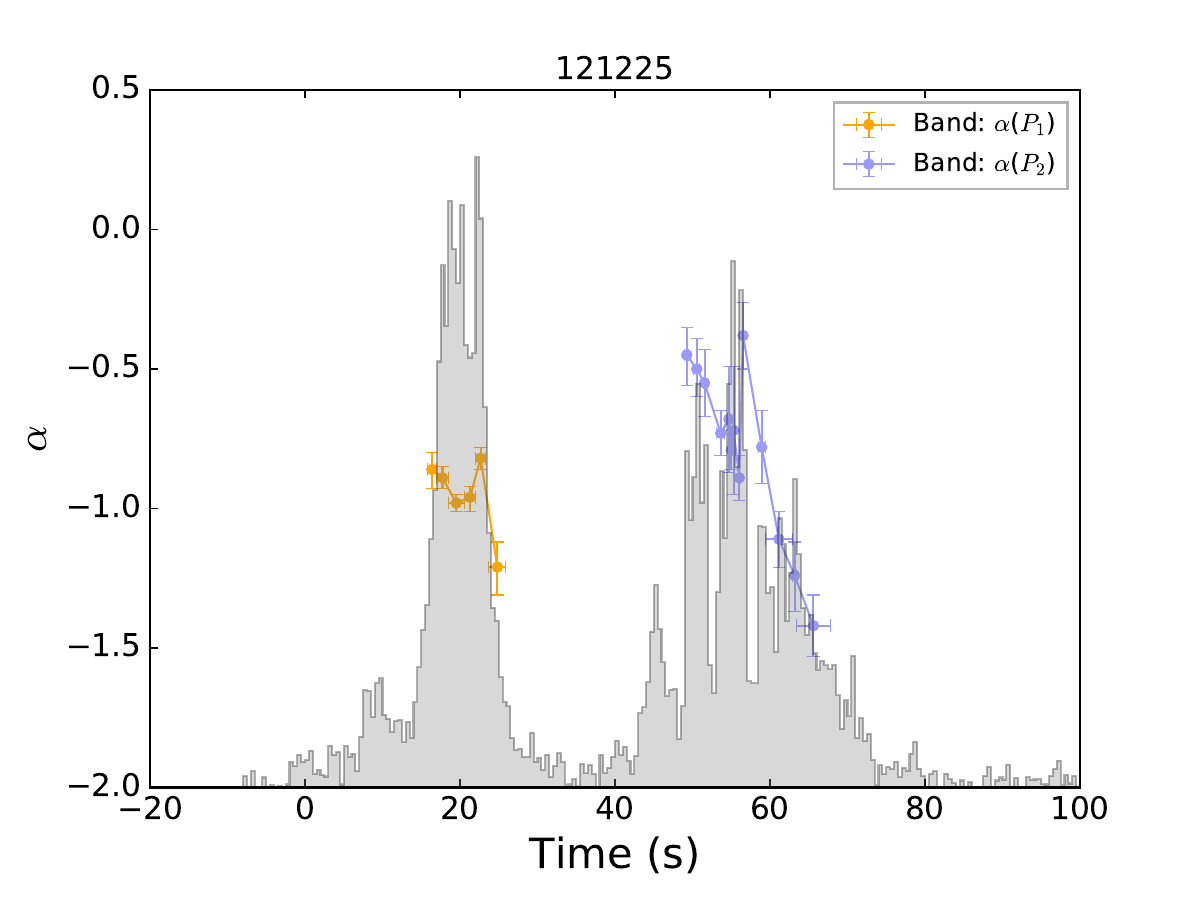}
\includegraphics[angle=0,scale=0.3]{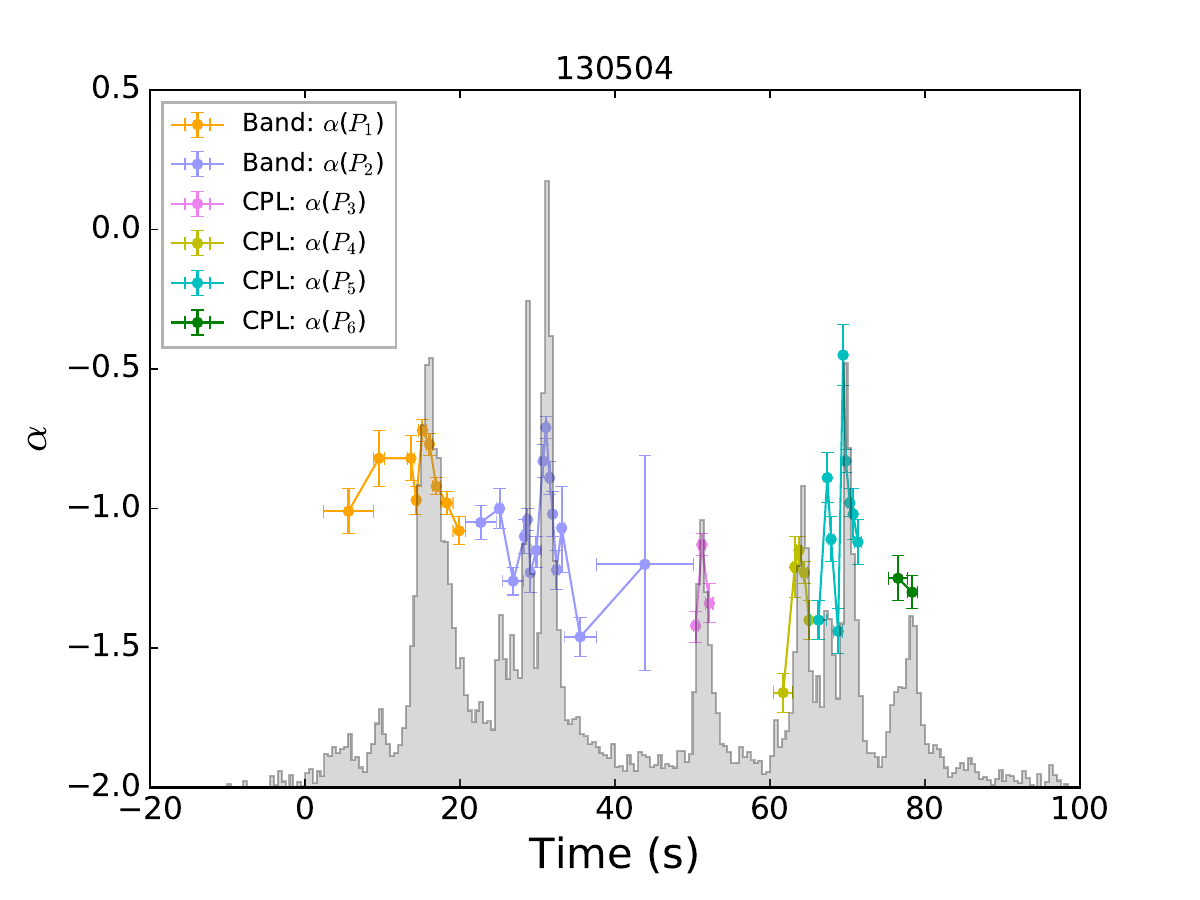}
\includegraphics[angle=0,scale=0.3]{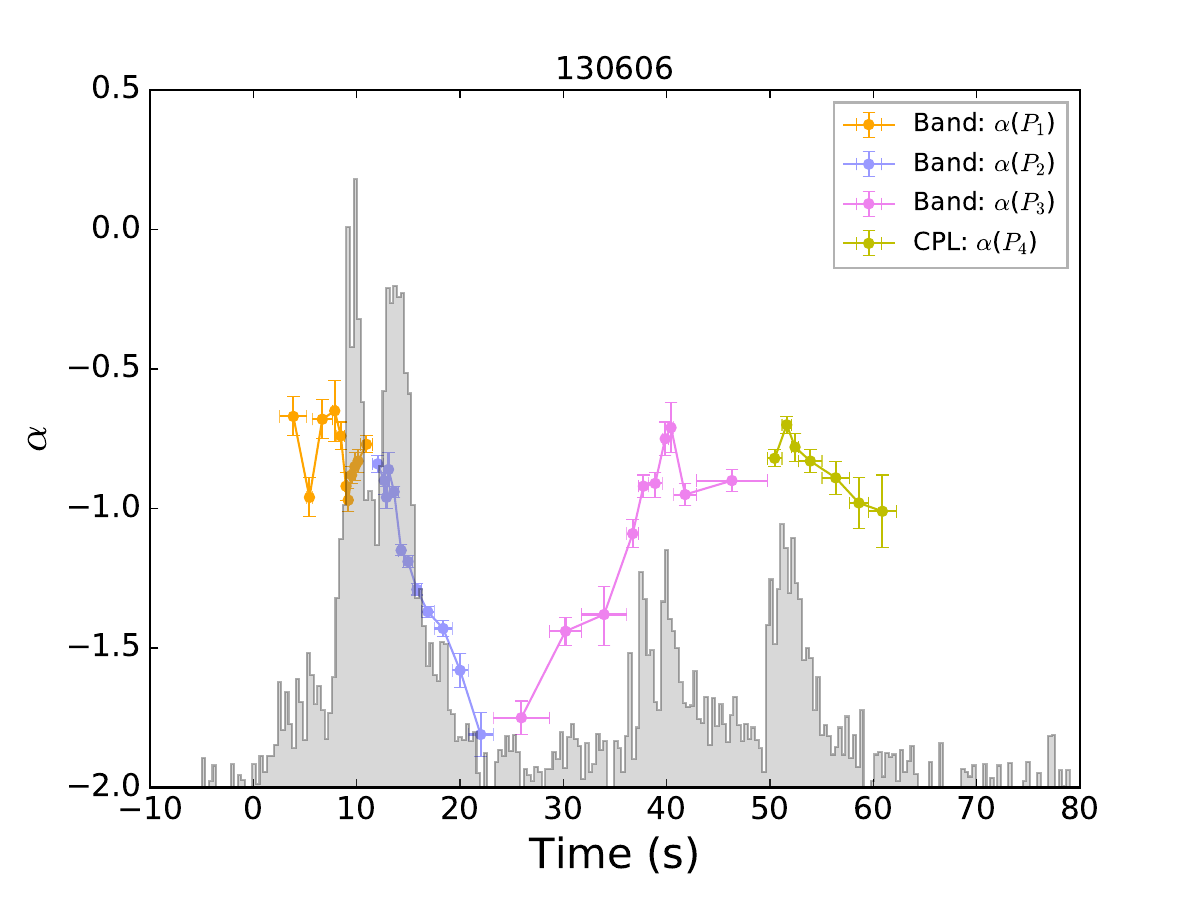}
\includegraphics[angle=0,scale=0.3]{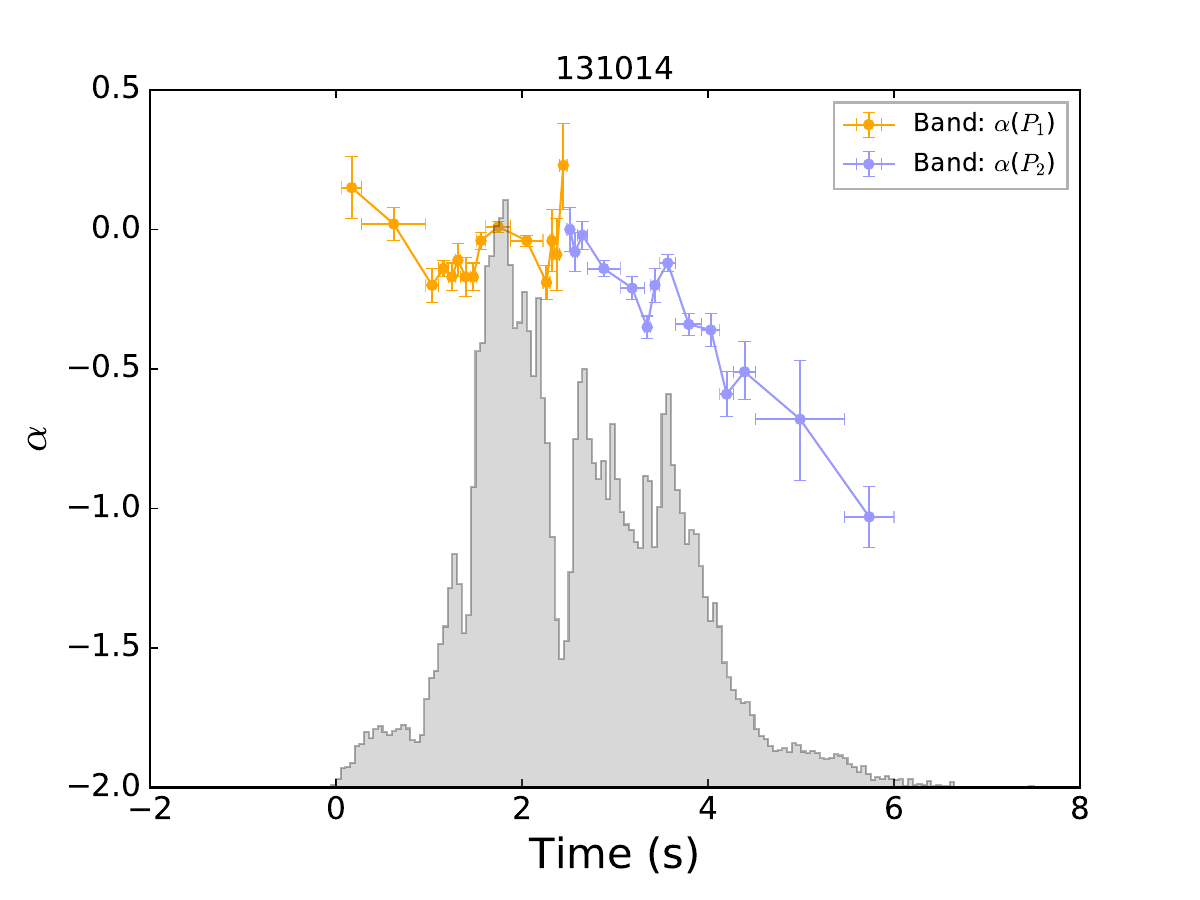}
\includegraphics[angle=0,scale=0.3]{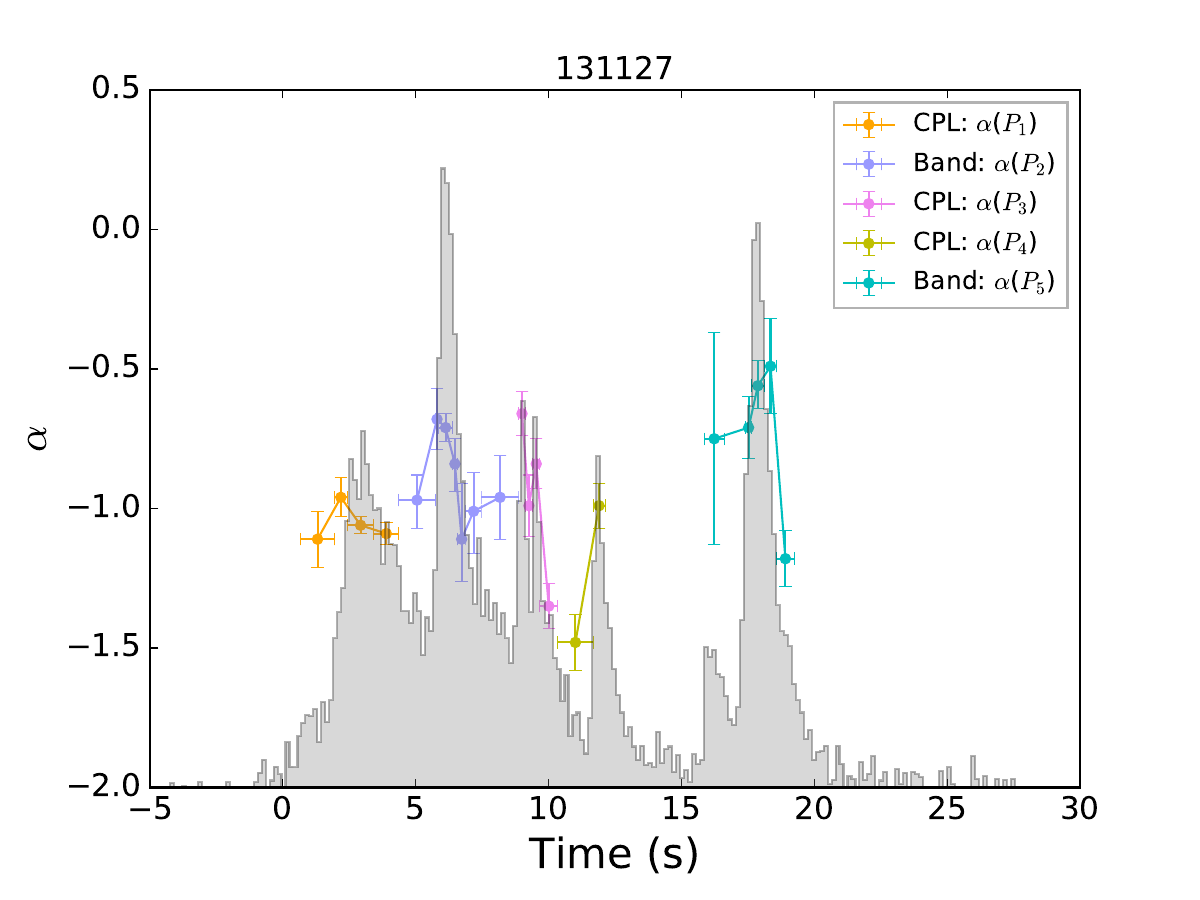}
\includegraphics[angle=0,scale=0.3]{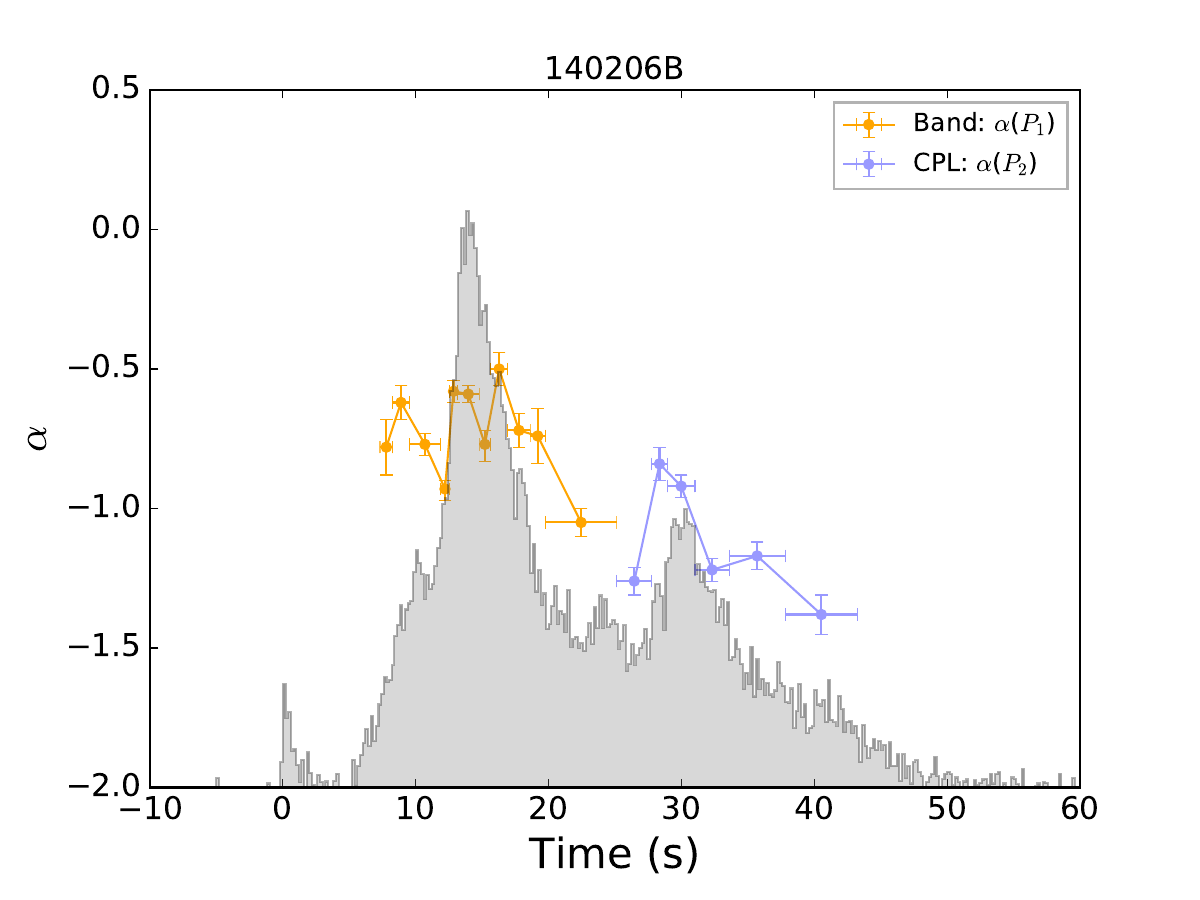}
\includegraphics[angle=0,scale=0.3]{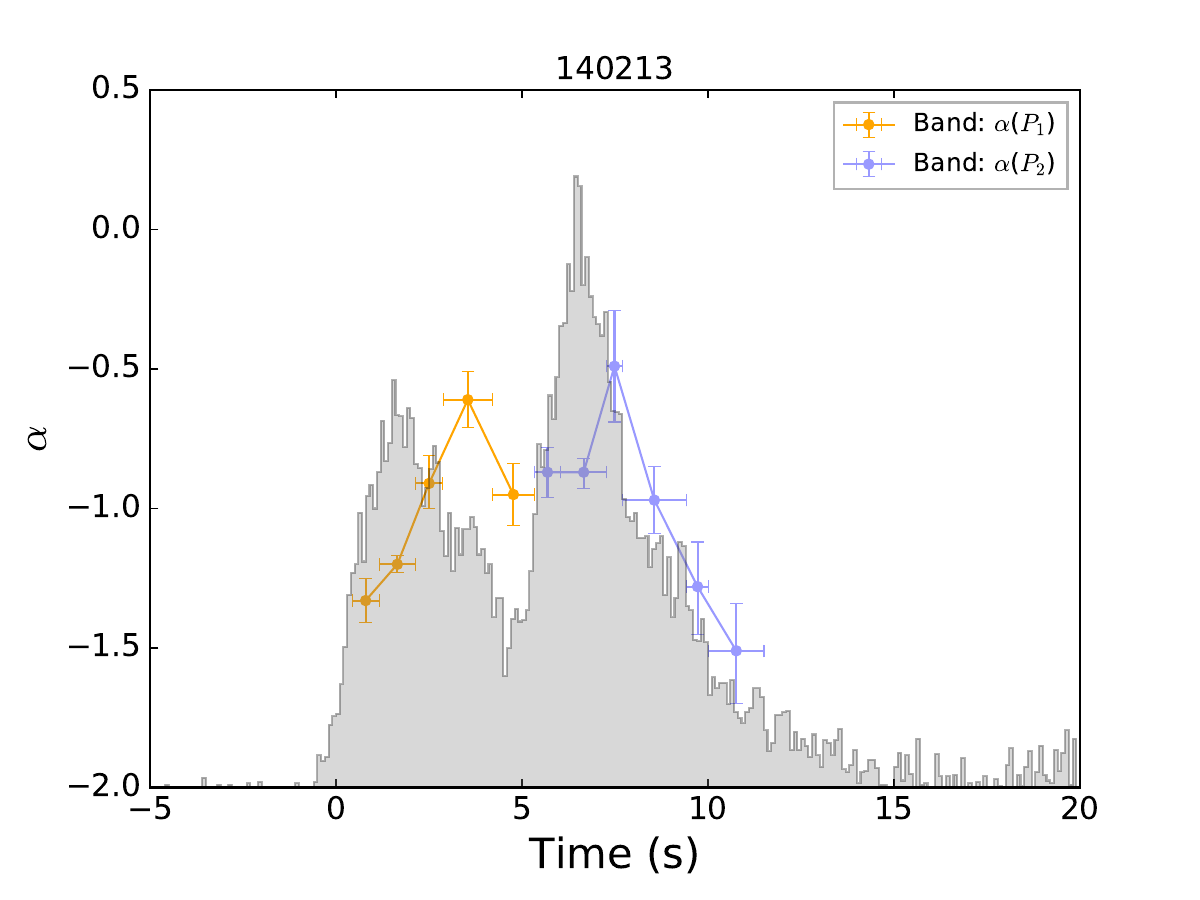}
\includegraphics[angle=0,scale=0.3]{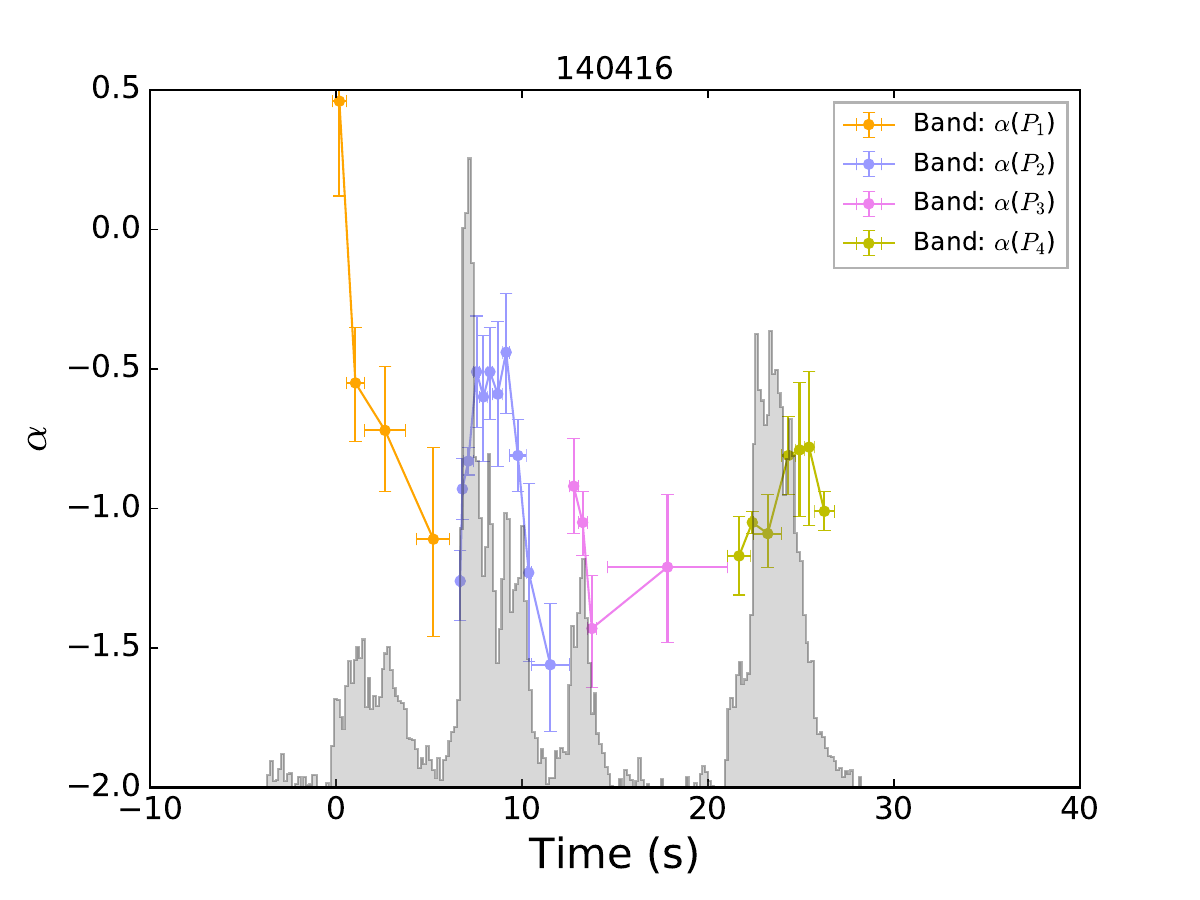}
\includegraphics[angle=0,scale=0.3]{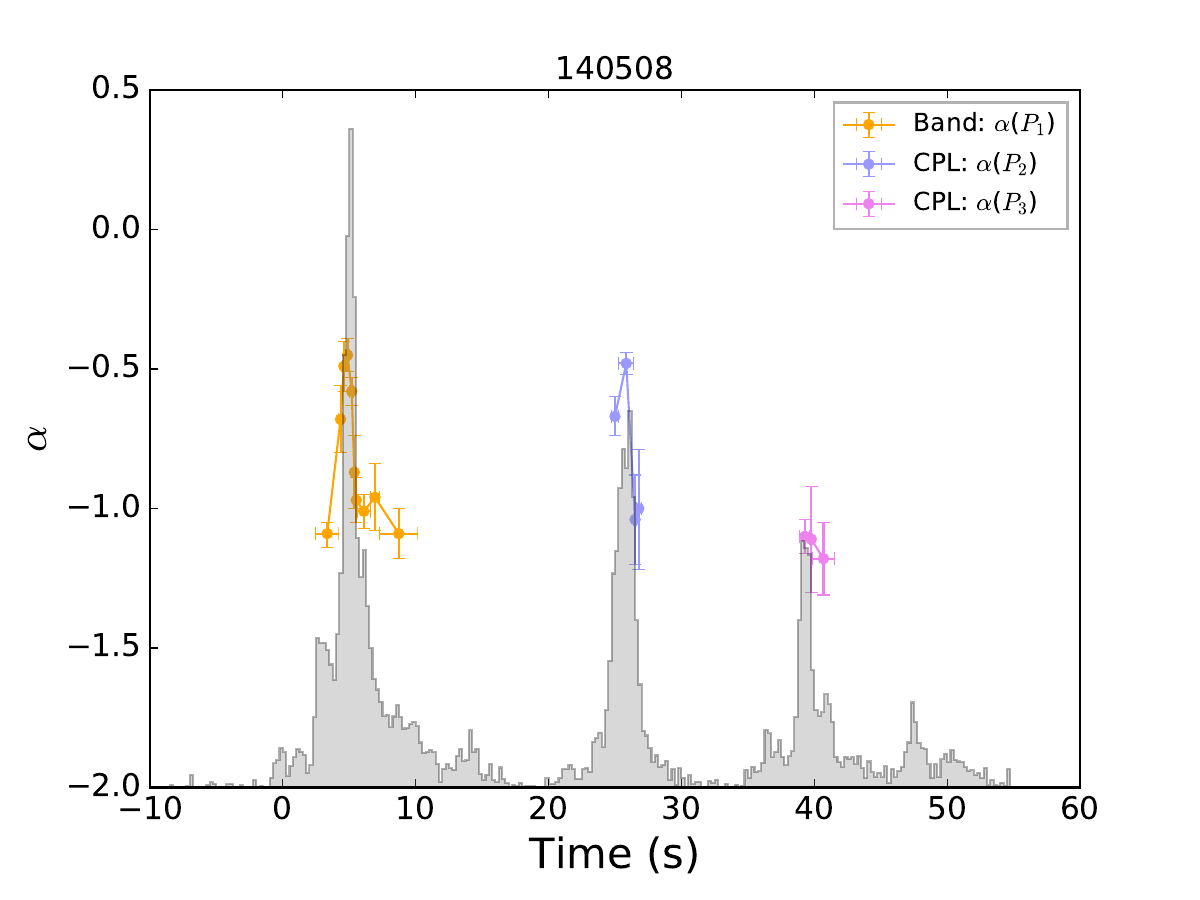}
\includegraphics[angle=0,scale=0.3]{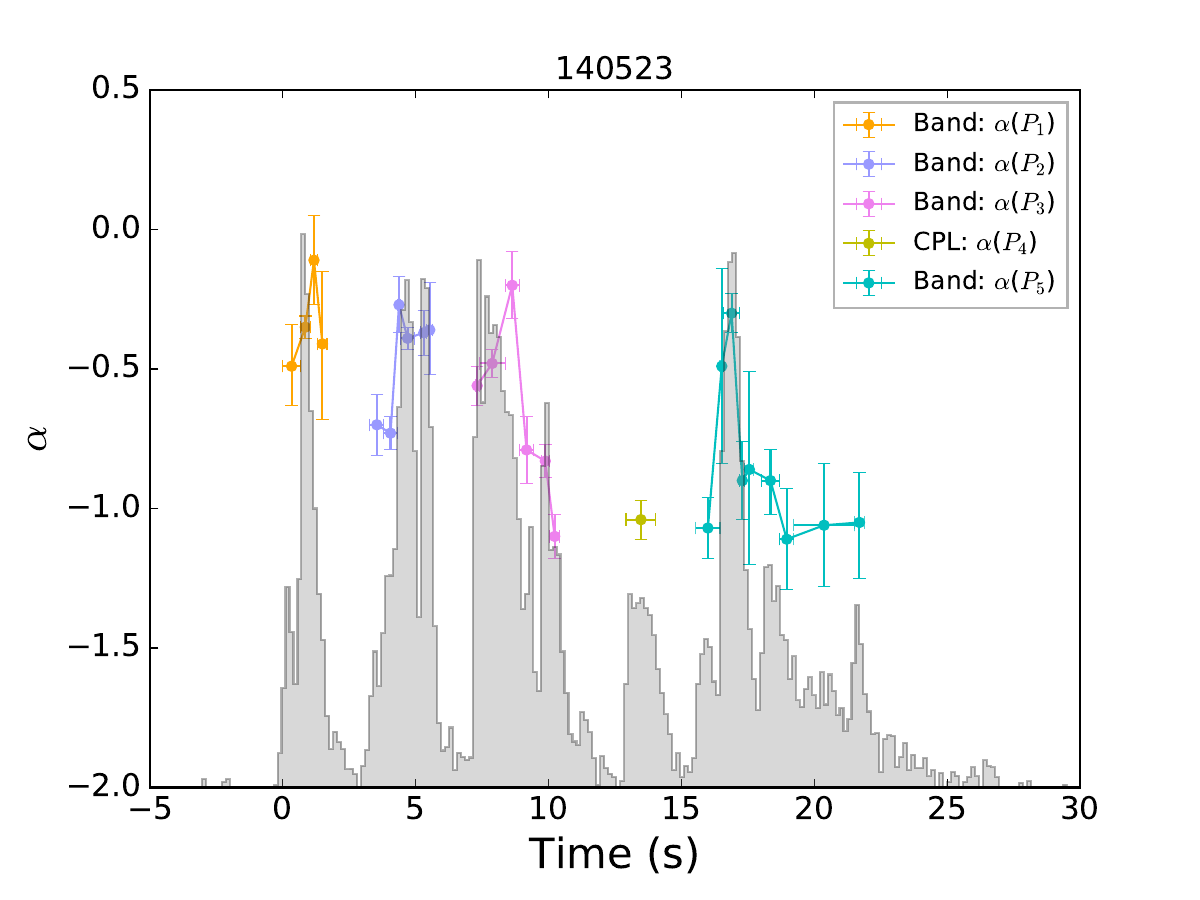}
\includegraphics[angle=0,scale=0.3]{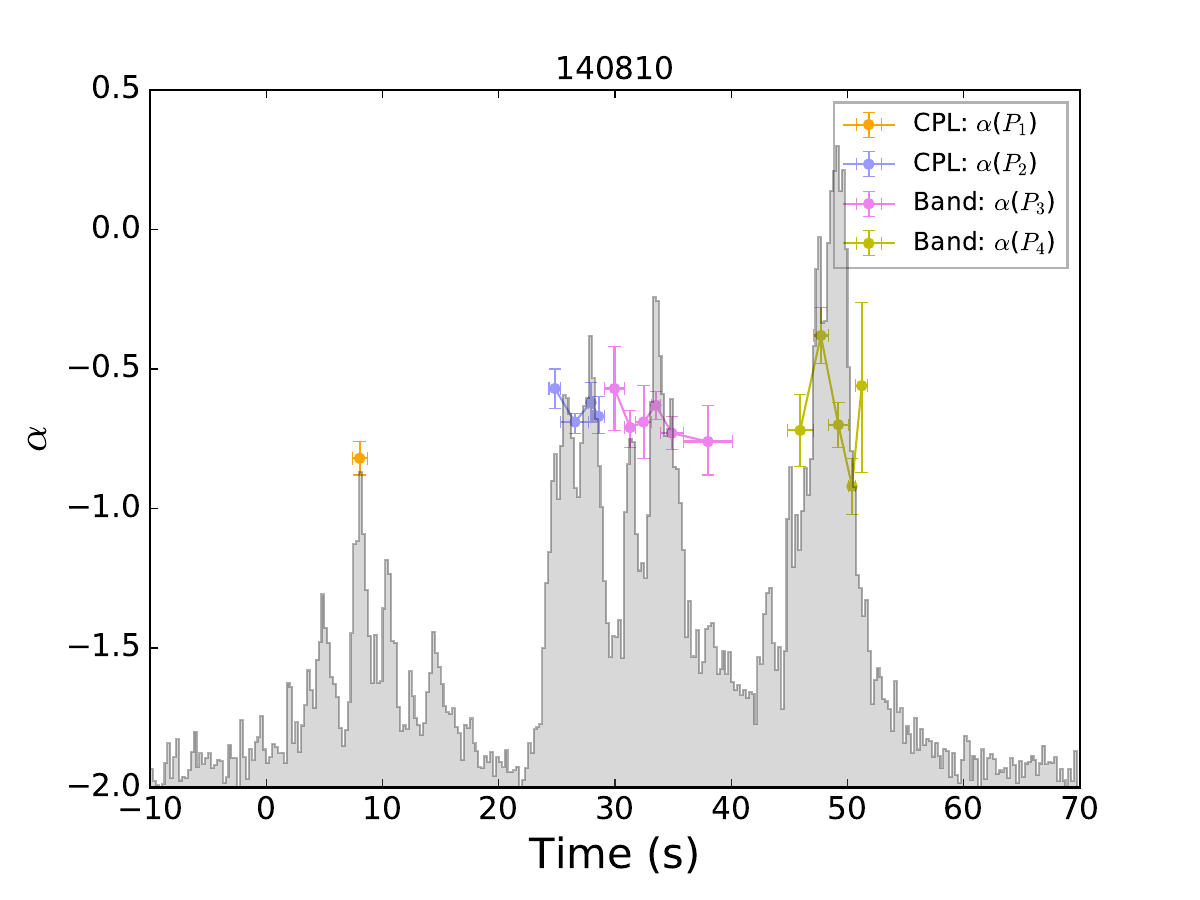}
\includegraphics[angle=0,scale=0.3]{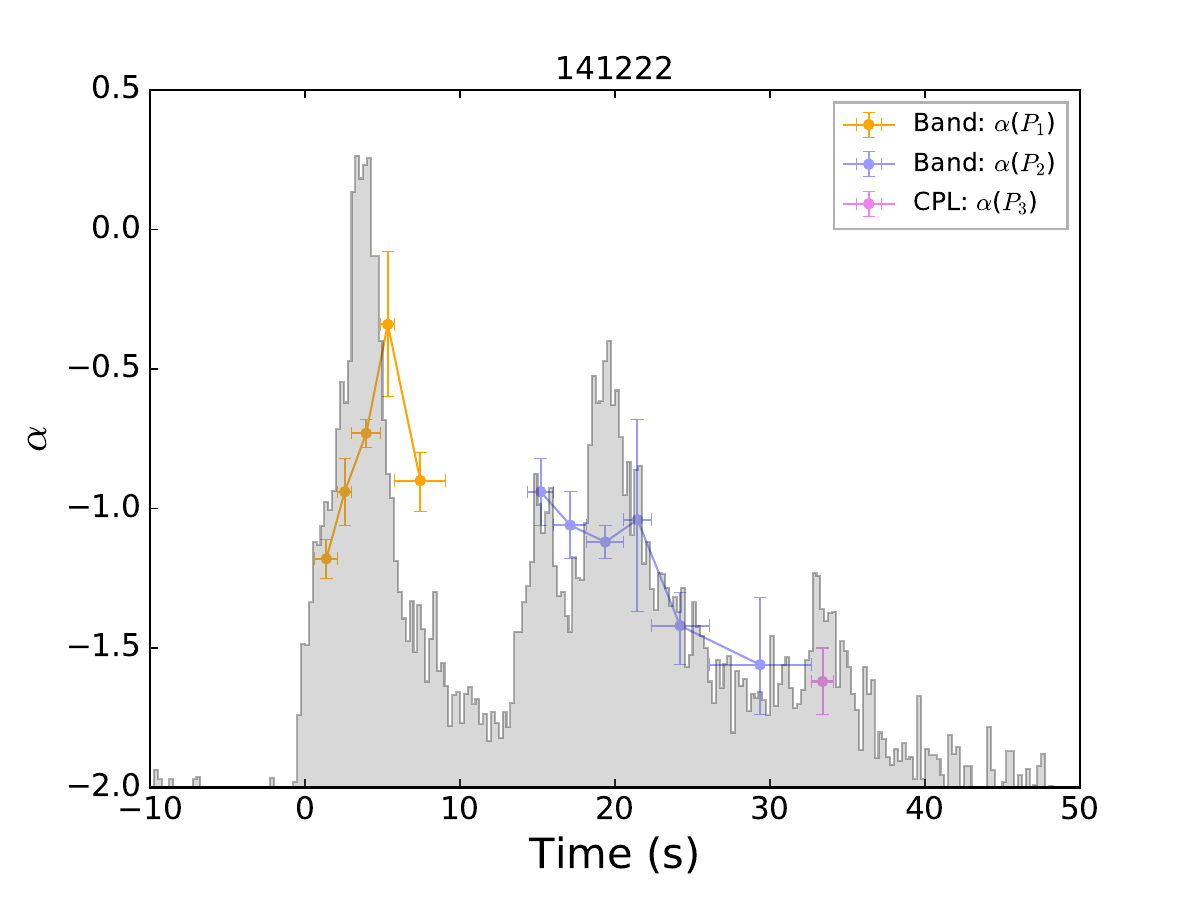}
\includegraphics[angle=0,scale=0.3]{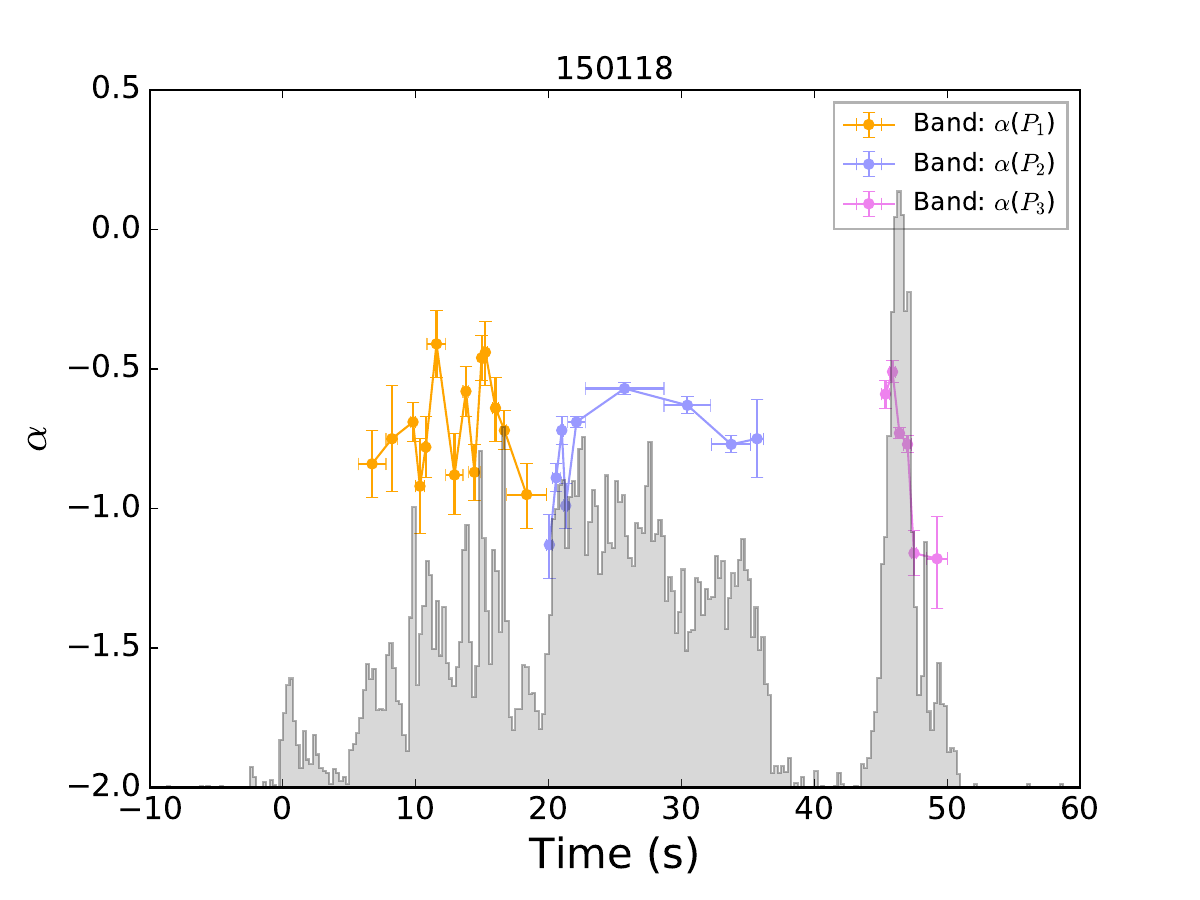}
\center{Fig. \ref{fig:Alpha_Best}--- Continued}
\end{figure*}
\begin{figure*}
\includegraphics[angle=0,scale=0.3]{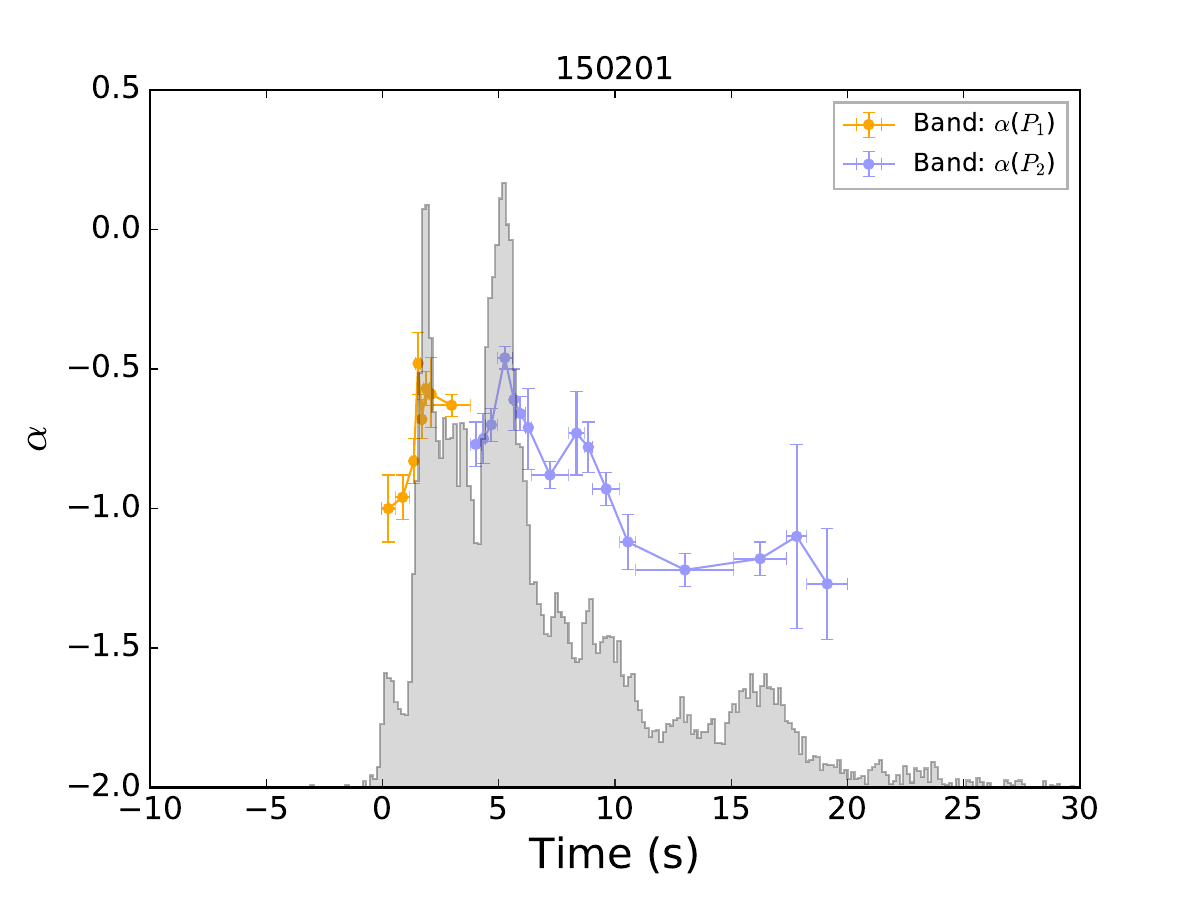}
\includegraphics[angle=0,scale=0.3]{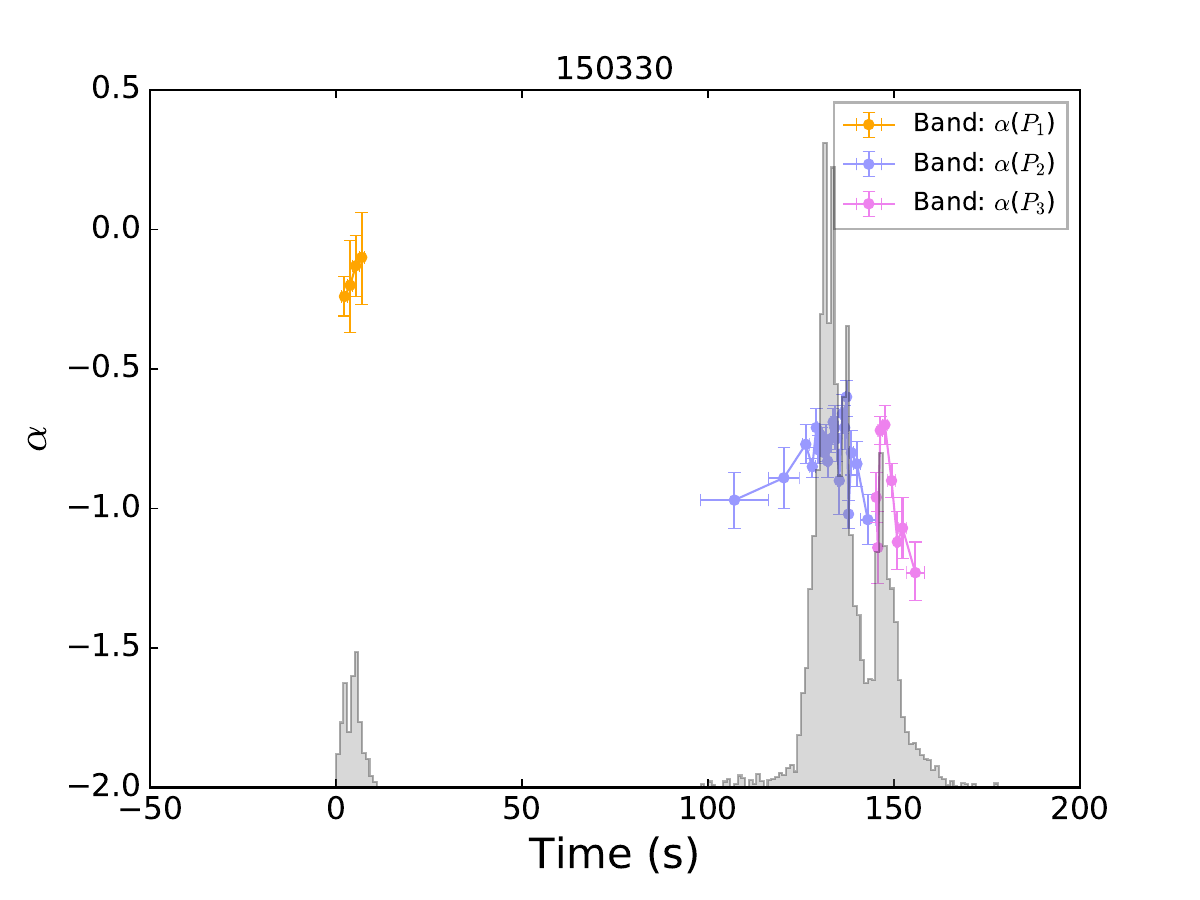}
\includegraphics[angle=0,scale=0.3]{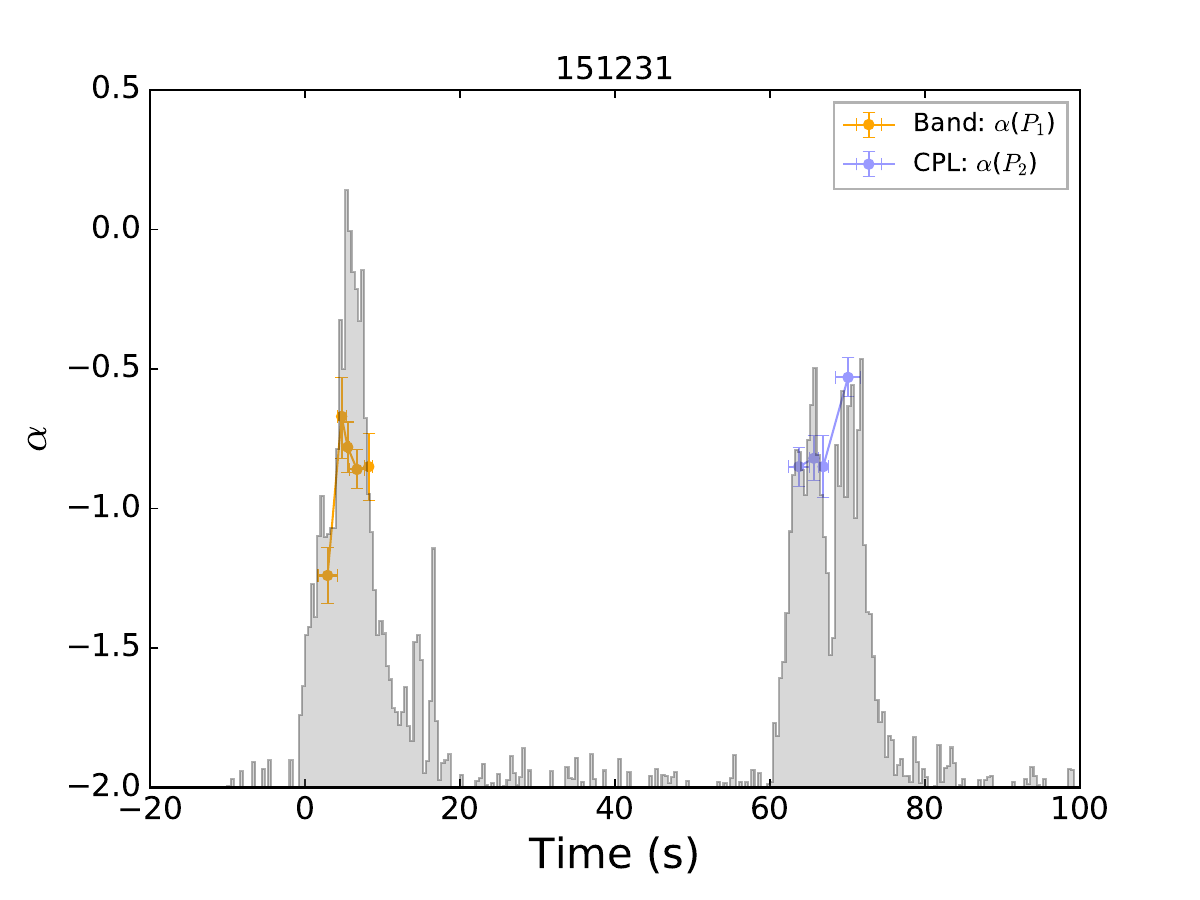}
\includegraphics[angle=0,scale=0.3]{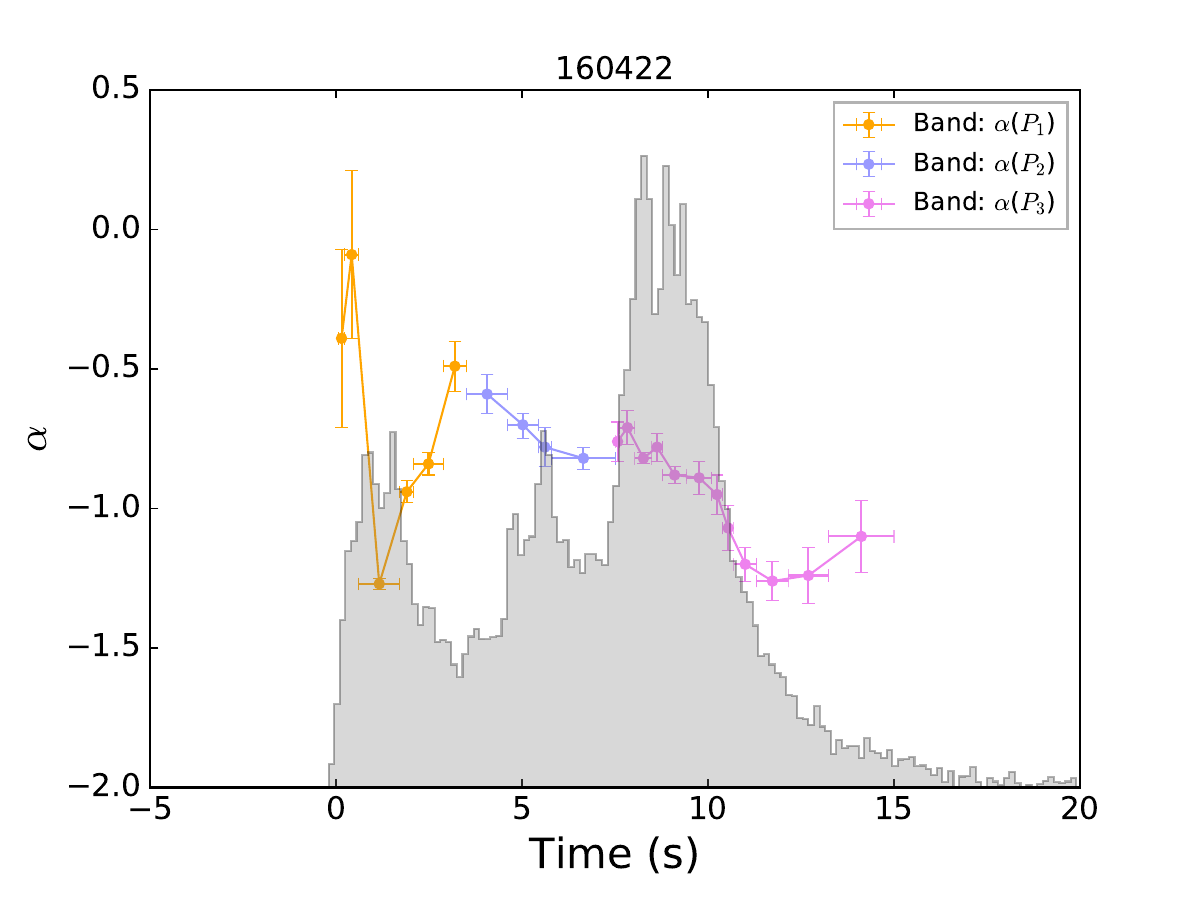}
\includegraphics[angle=0,scale=0.3]{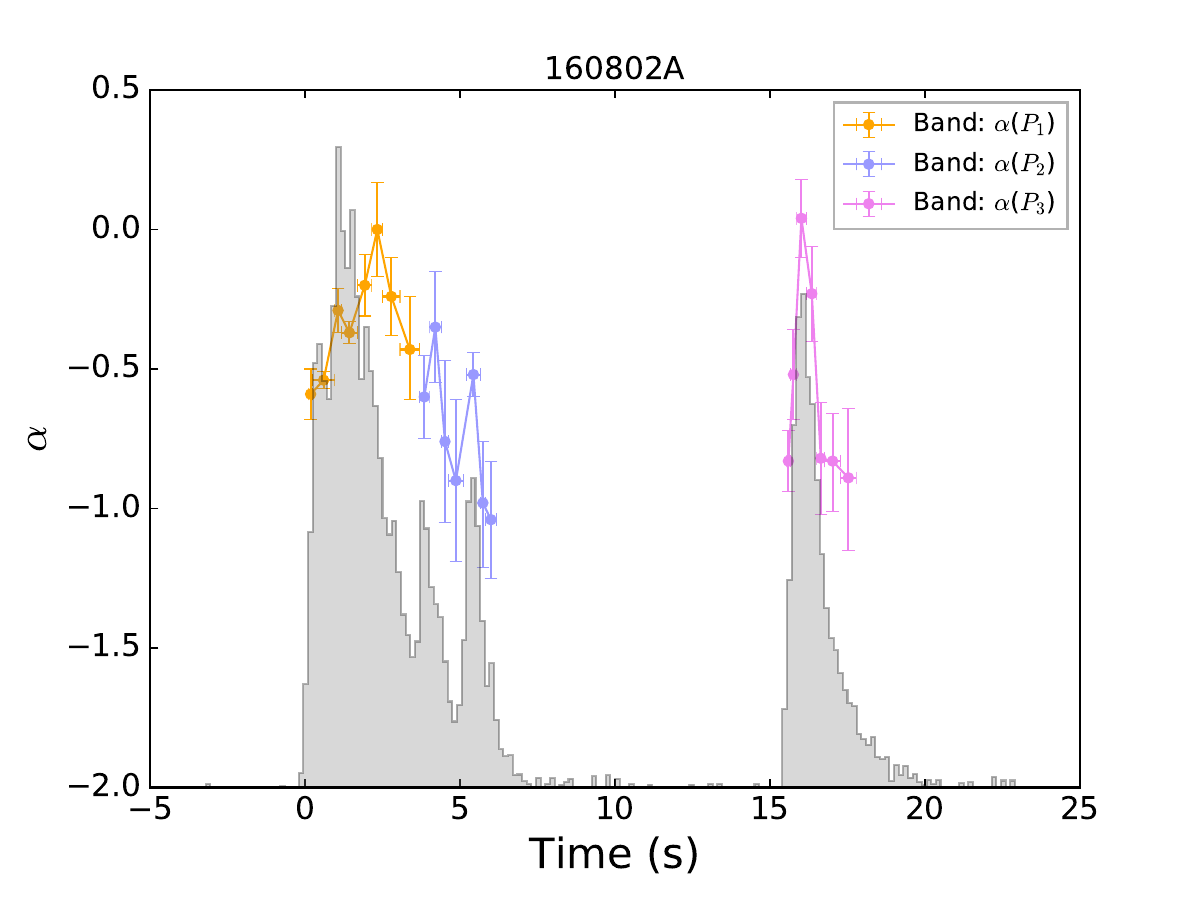}
\includegraphics[angle=0,scale=0.3]{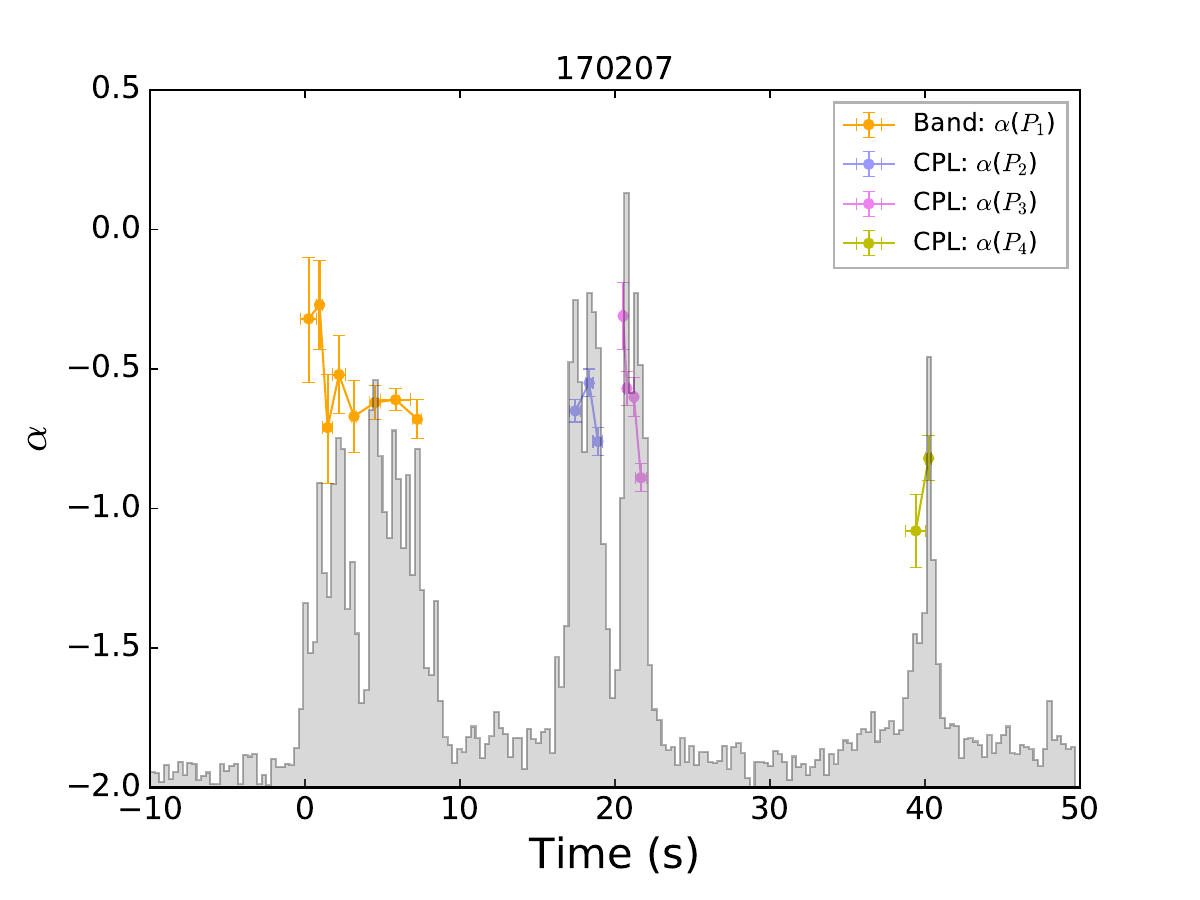}
\includegraphics[angle=0,scale=0.3]{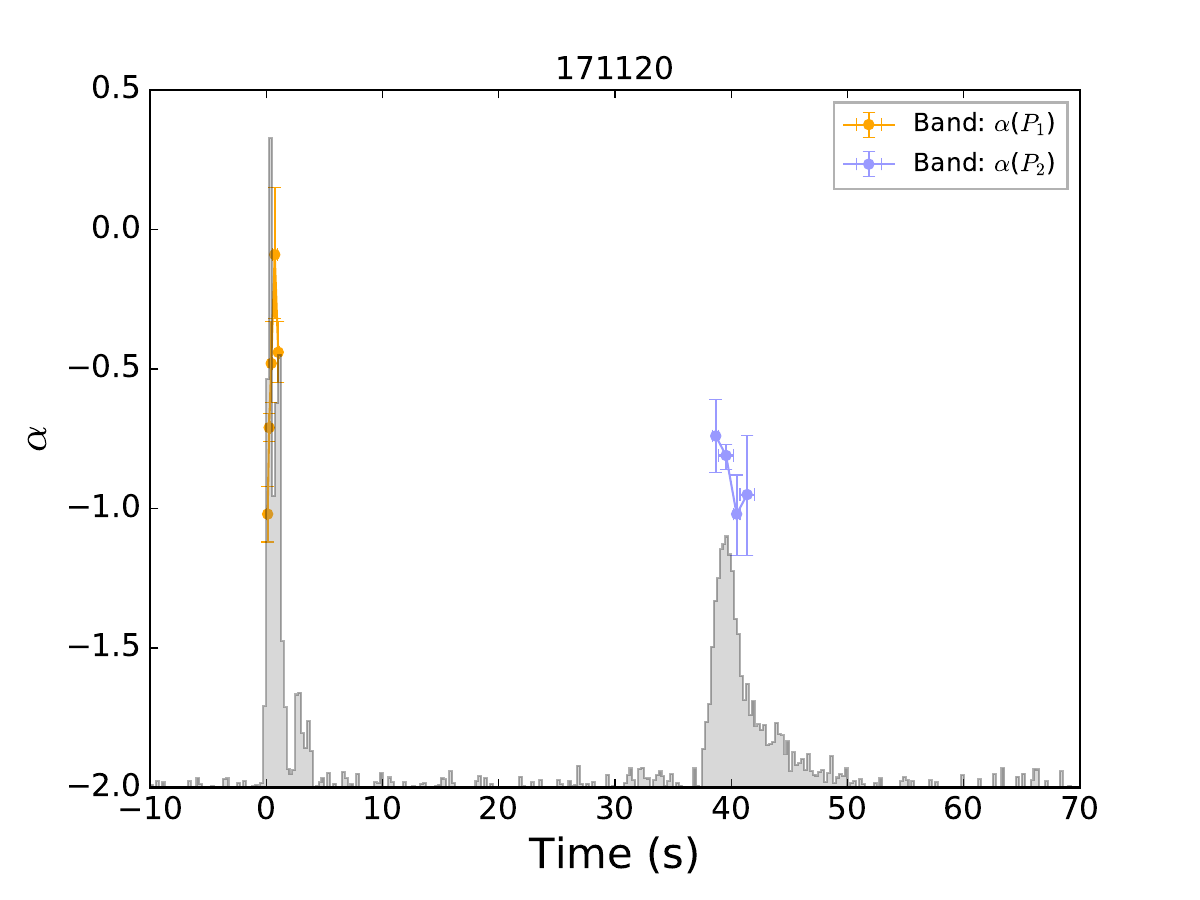}
\includegraphics[angle=0,scale=0.3]{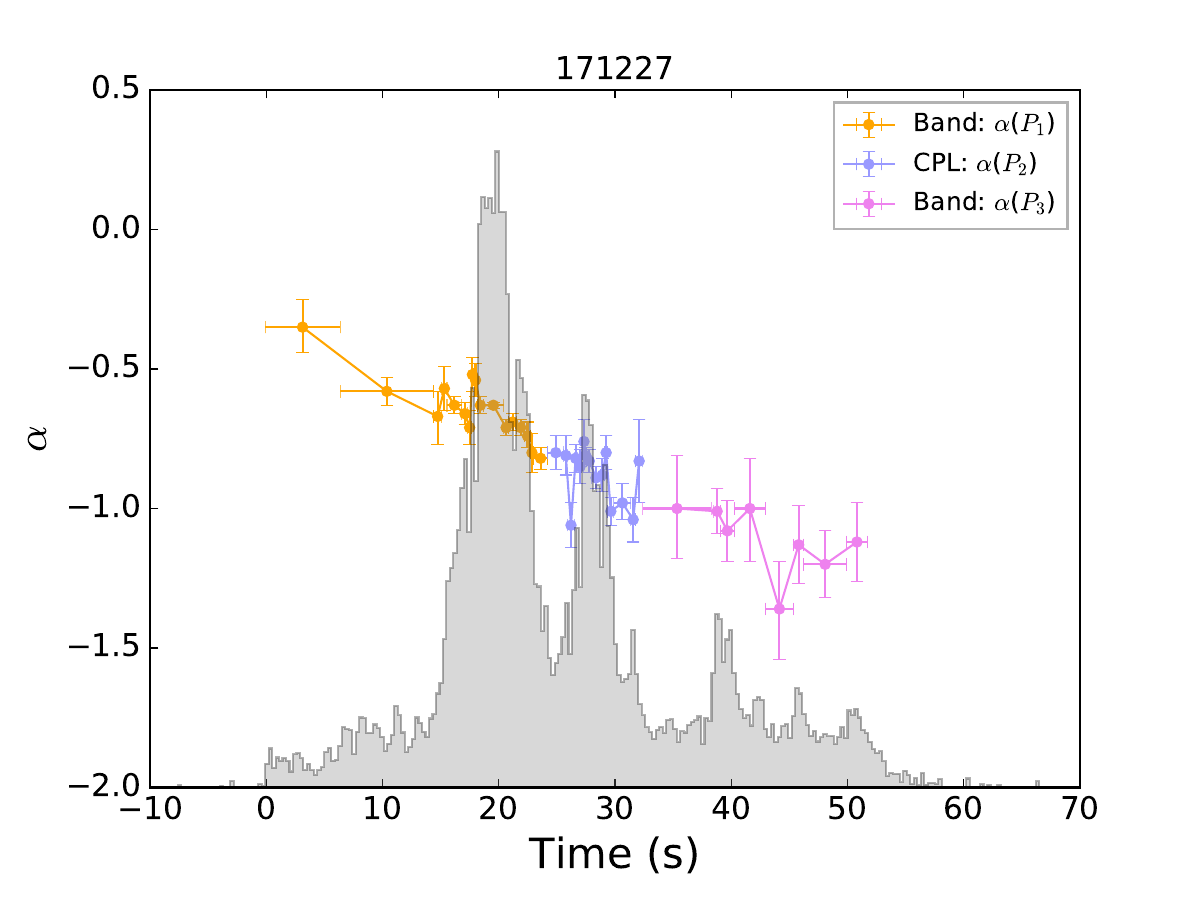}
\includegraphics[angle=0,scale=0.3]{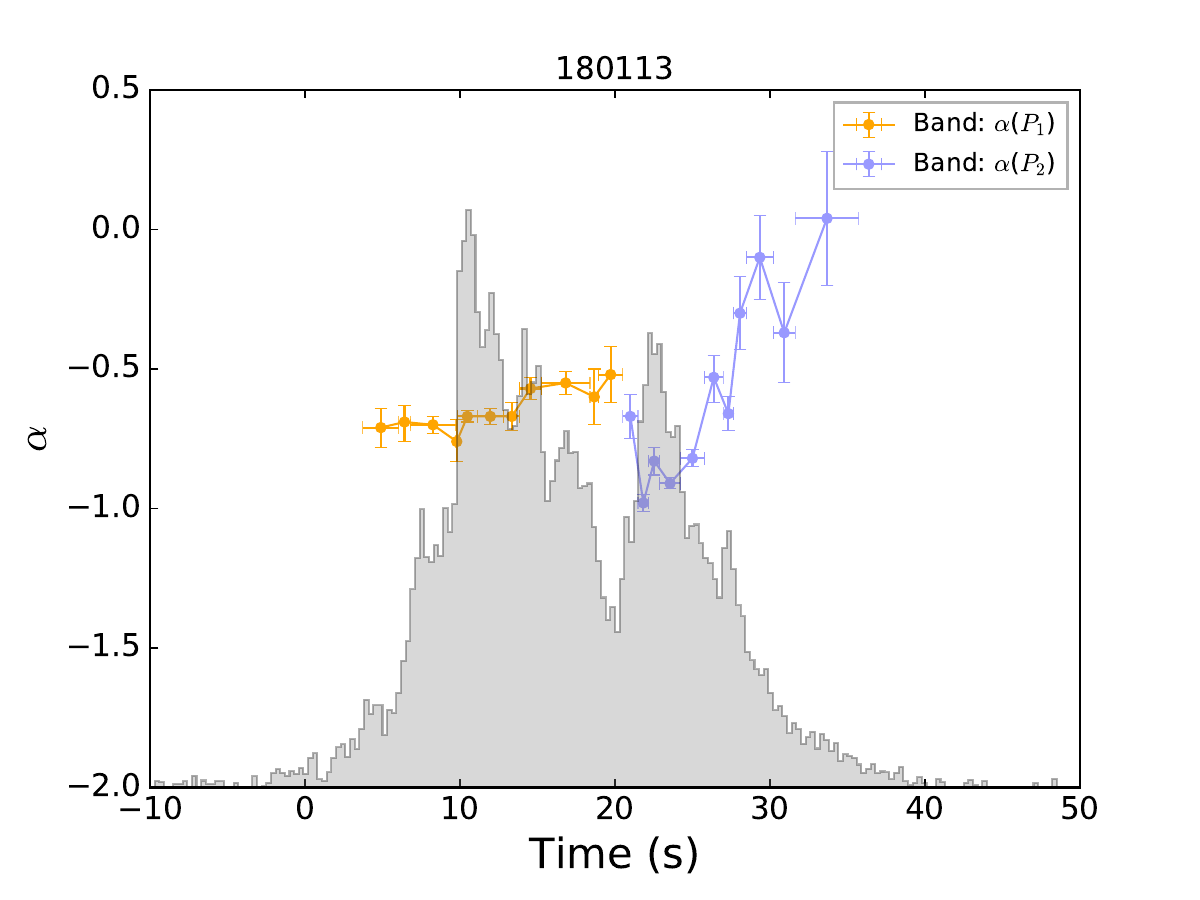}
\includegraphics[angle=0,scale=0.3]{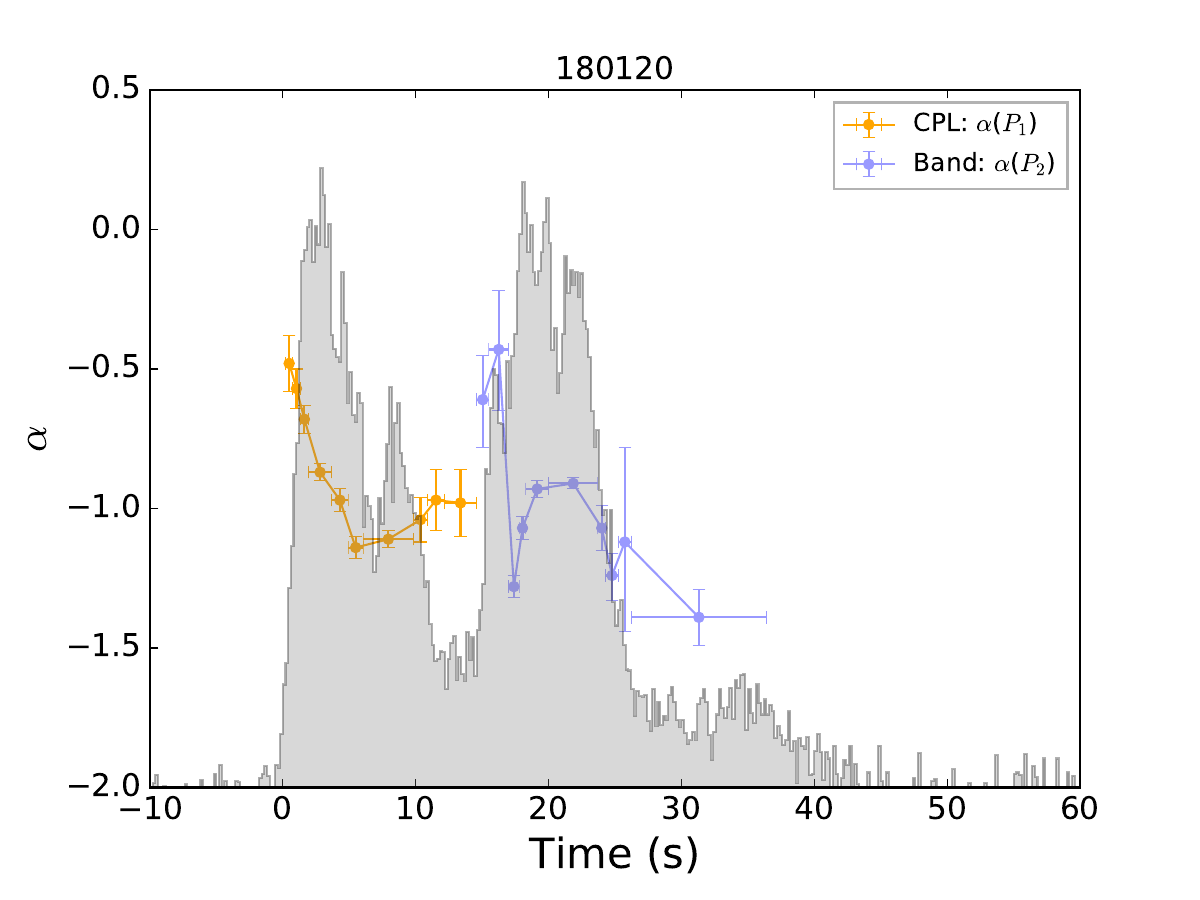}
\includegraphics[angle=0,scale=0.3]{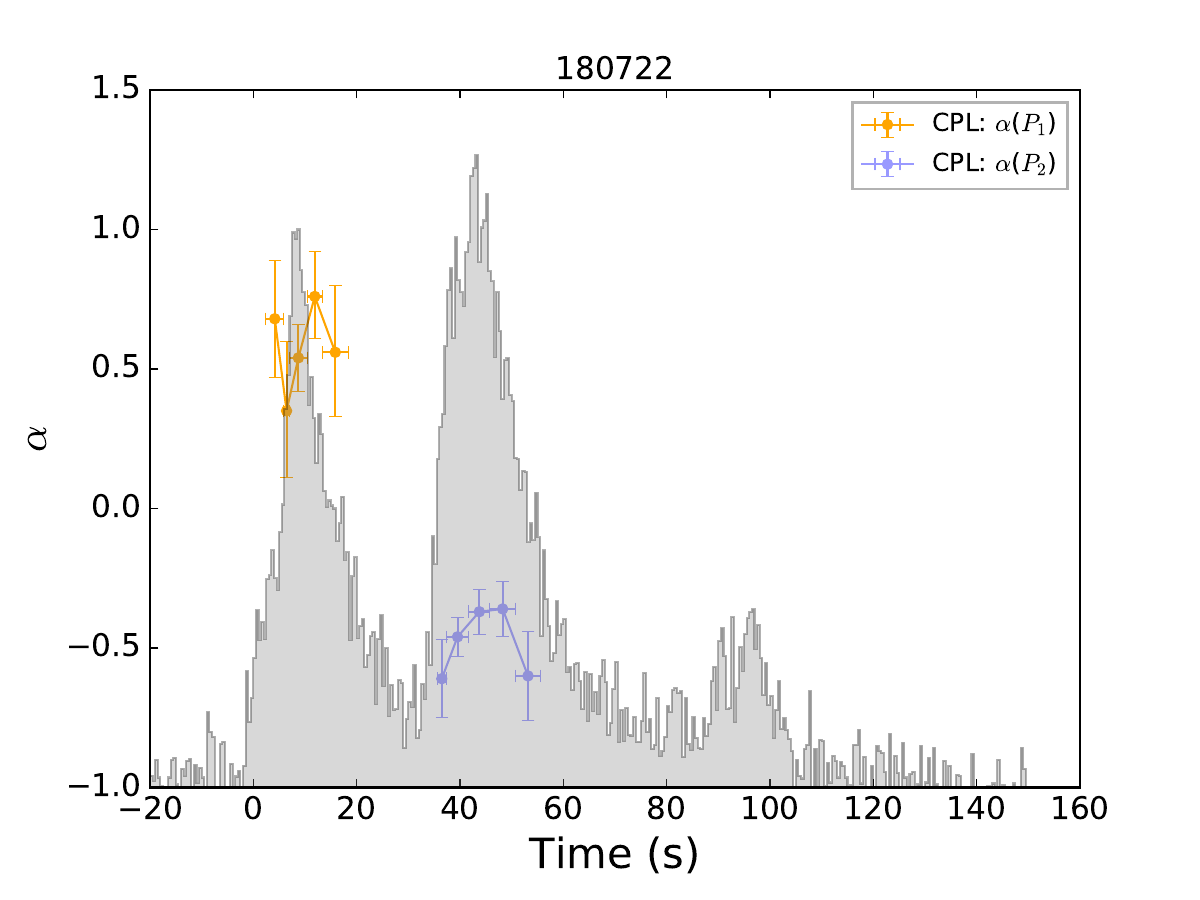}
\includegraphics[angle=0,scale=0.3]{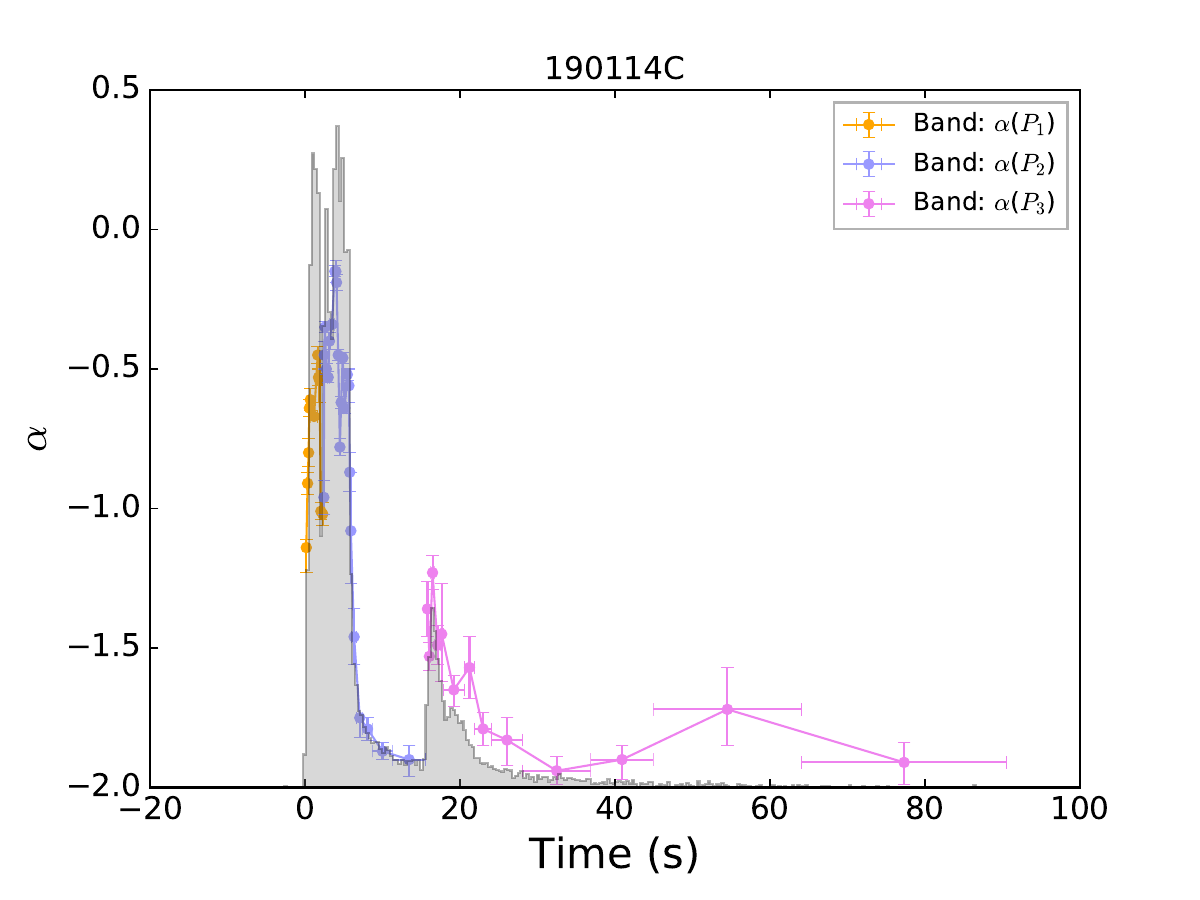}
\center{Fig. \ref{fig:Alpha_Best}--- Continued}
\end{figure*}

\clearpage
\begin{figure*}
\includegraphics[angle=0,scale=0.3]{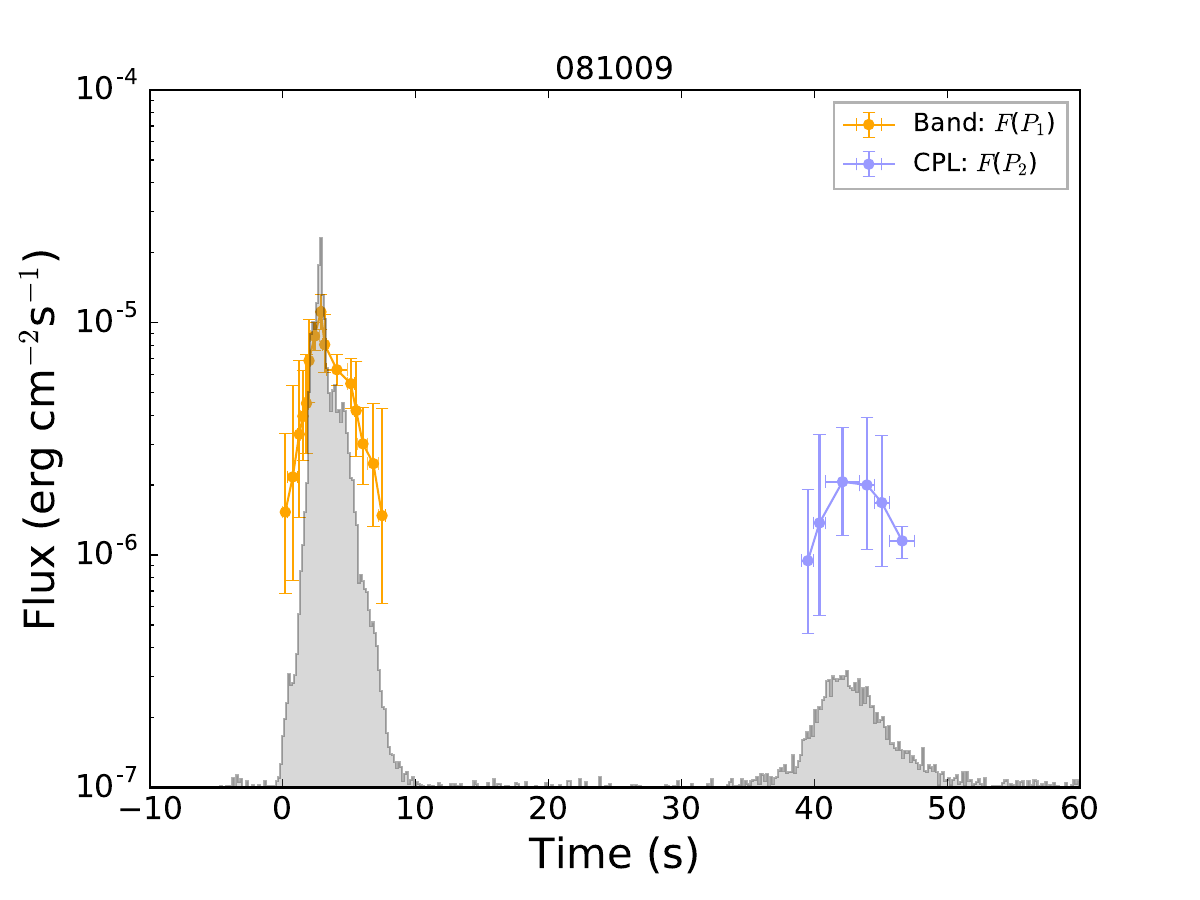}
\includegraphics[angle=0,scale=0.3]{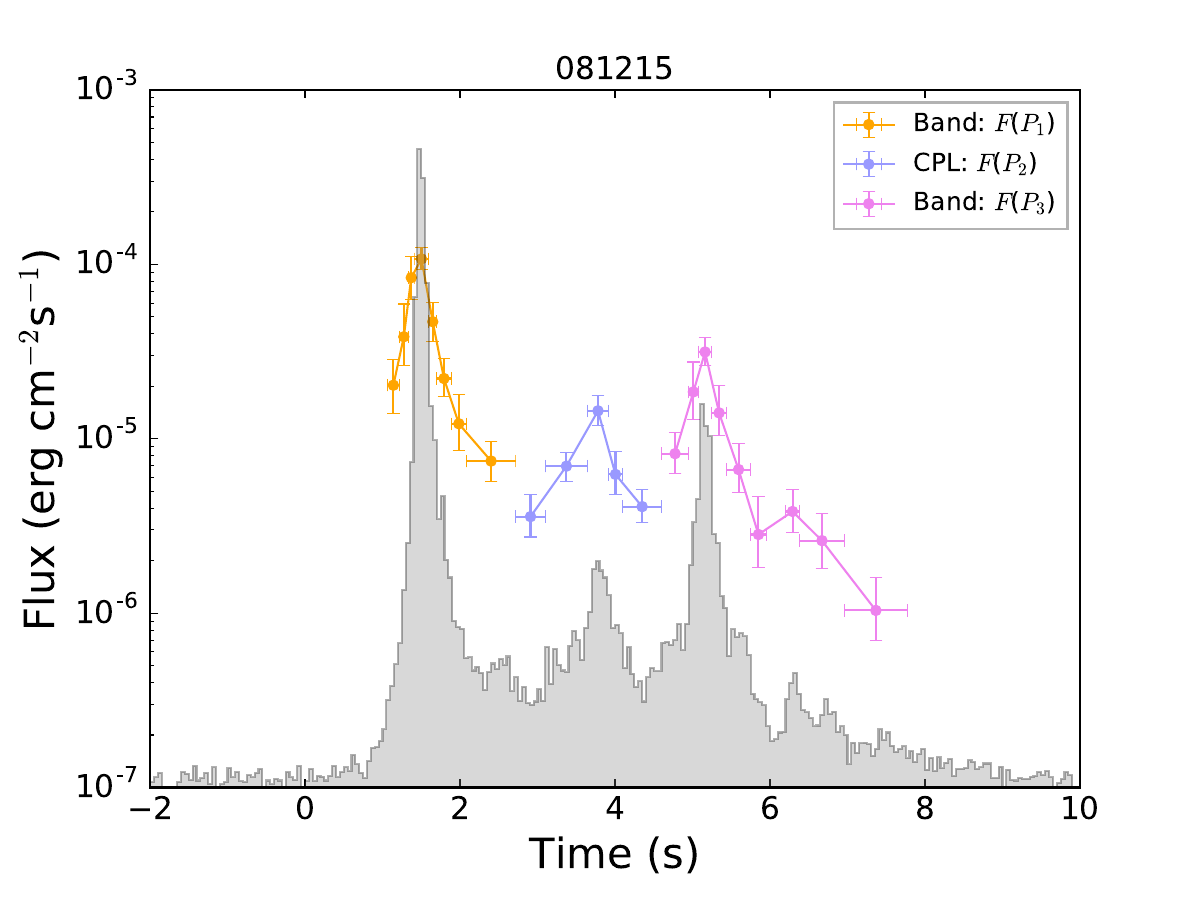}
\includegraphics[angle=0,scale=0.3]{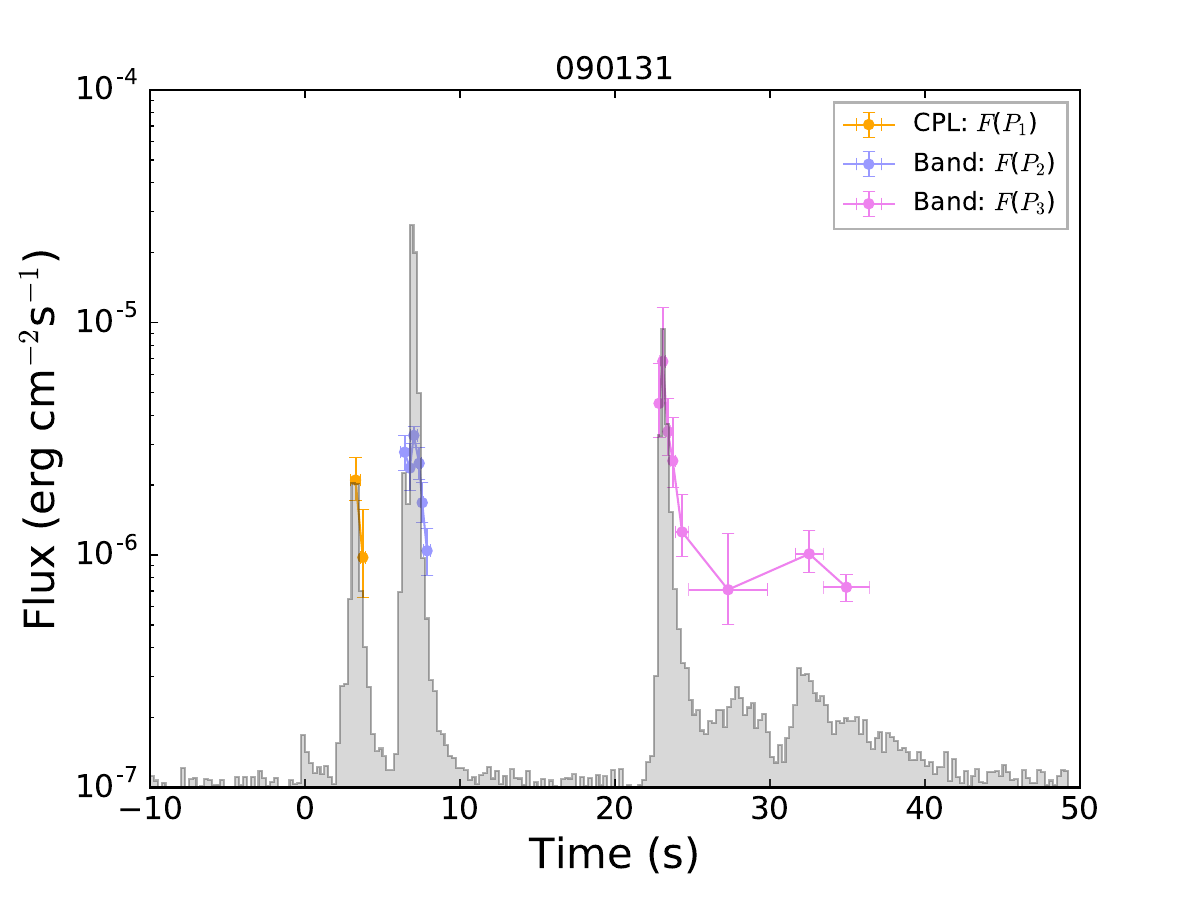}
\includegraphics[angle=0,scale=0.3]{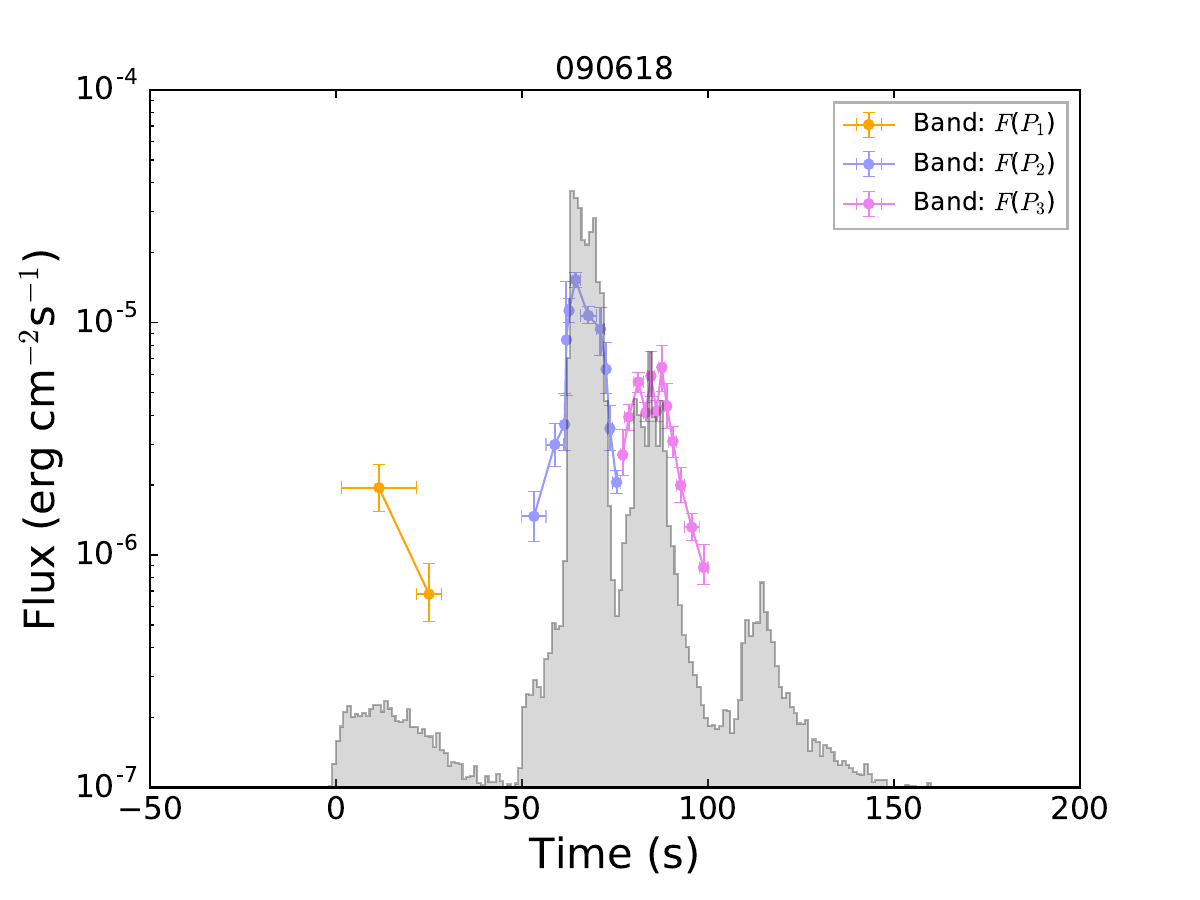}
\includegraphics[angle=0,scale=0.3]{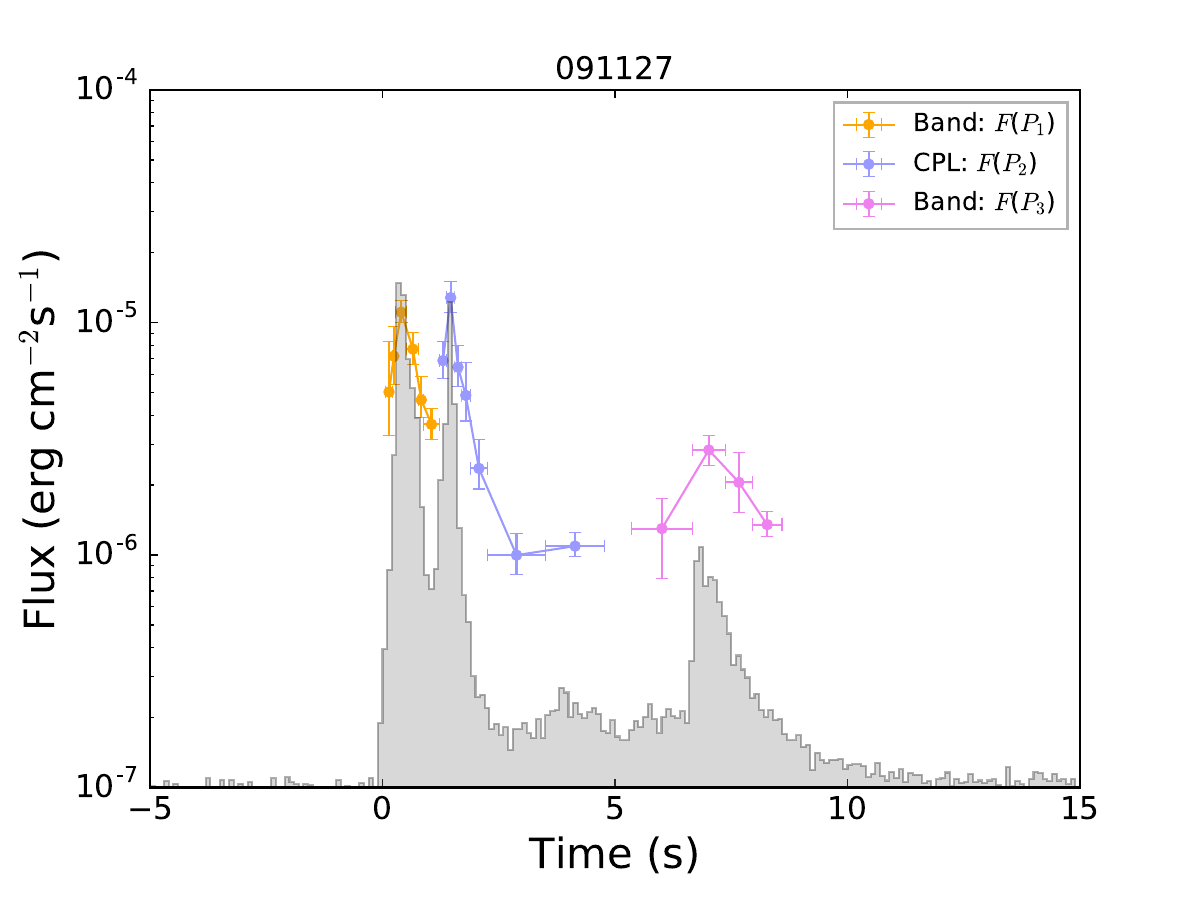}
\includegraphics[angle=0,scale=0.3]{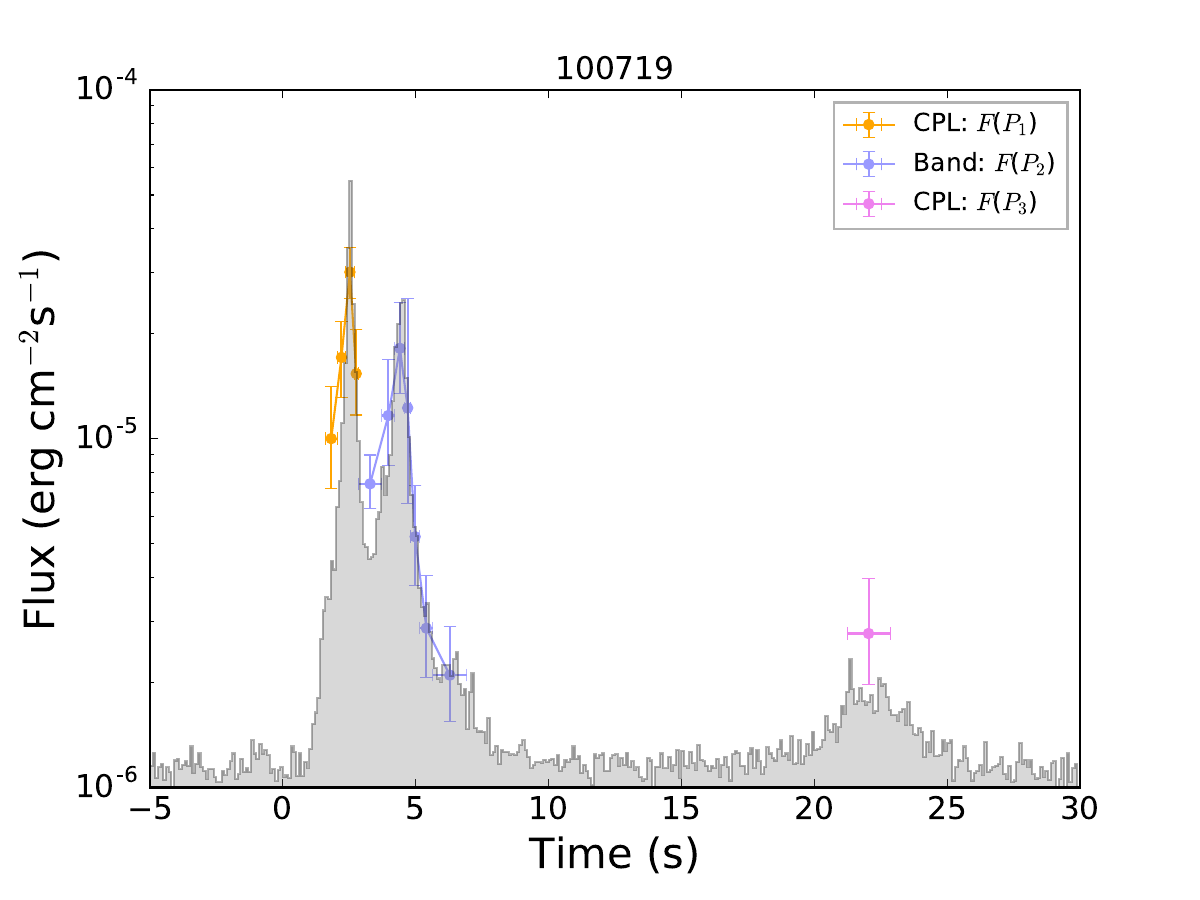}
\includegraphics[angle=0,scale=0.3]{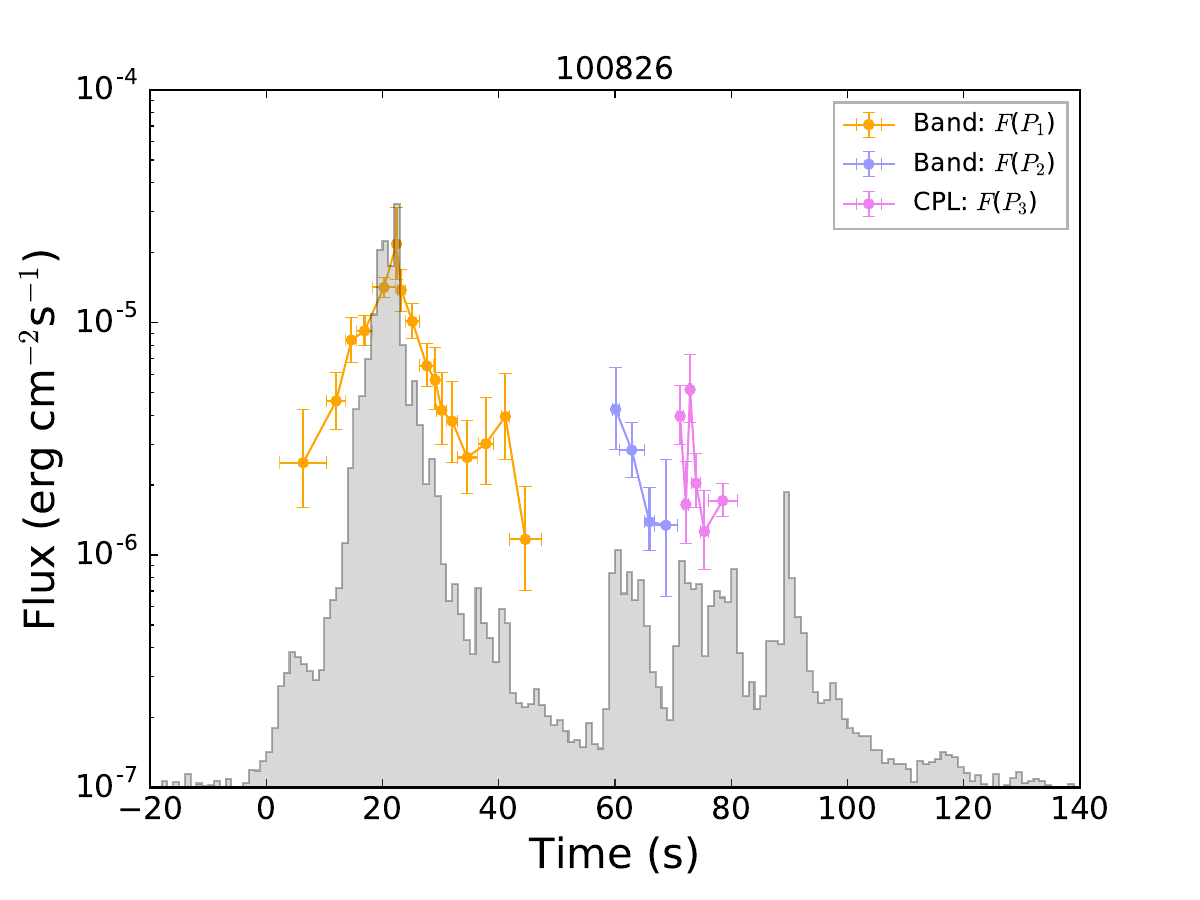}
\includegraphics[angle=0,scale=0.3]{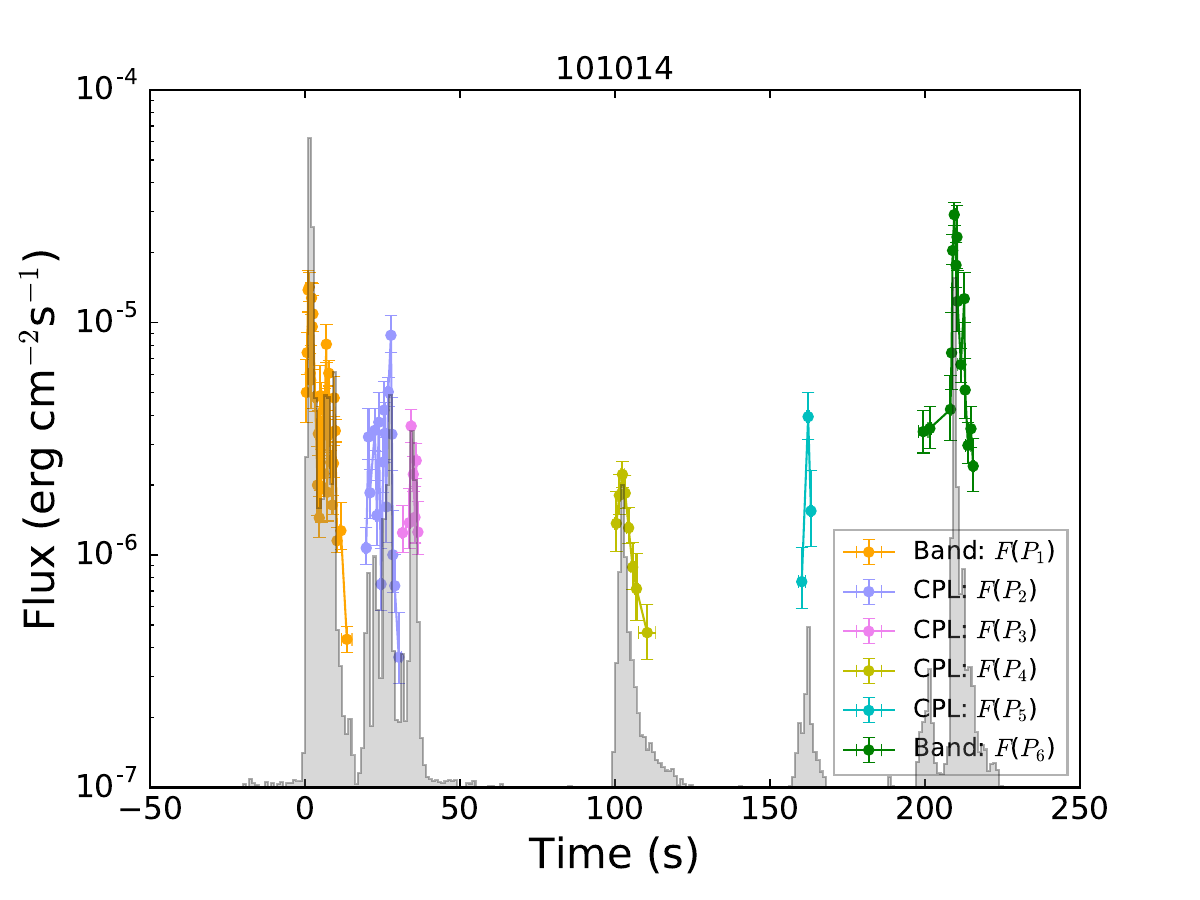}
\includegraphics[angle=0,scale=0.3]{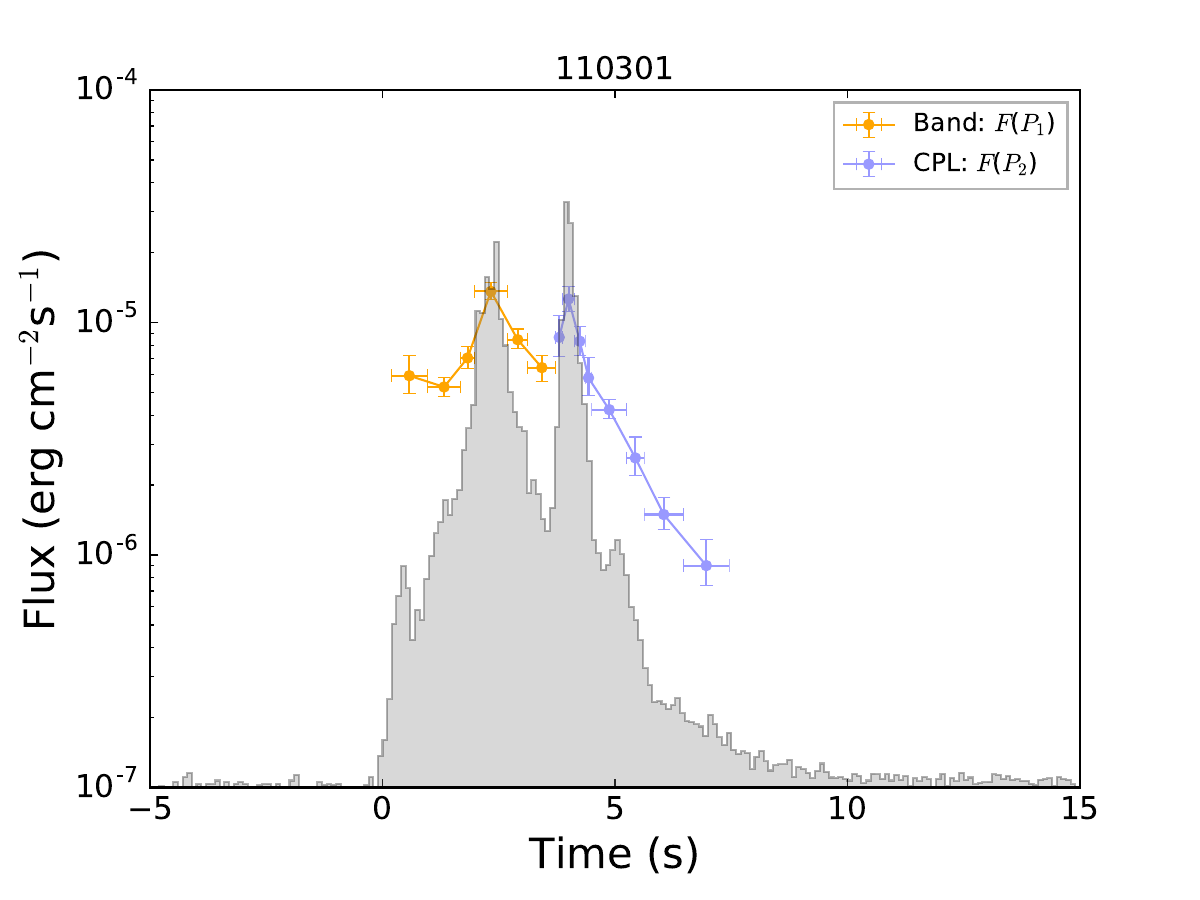}
\includegraphics[angle=0,scale=0.3]{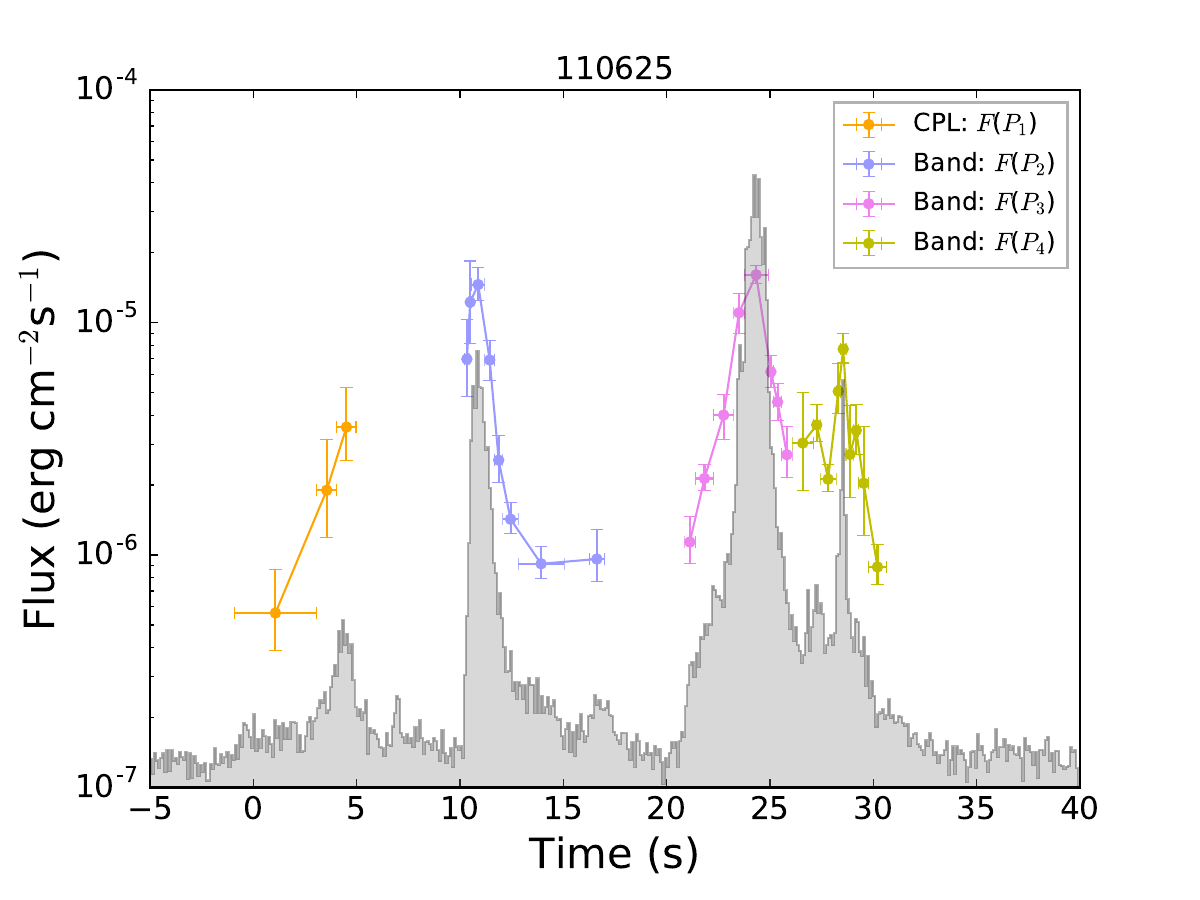}
\includegraphics[angle=0,scale=0.3]{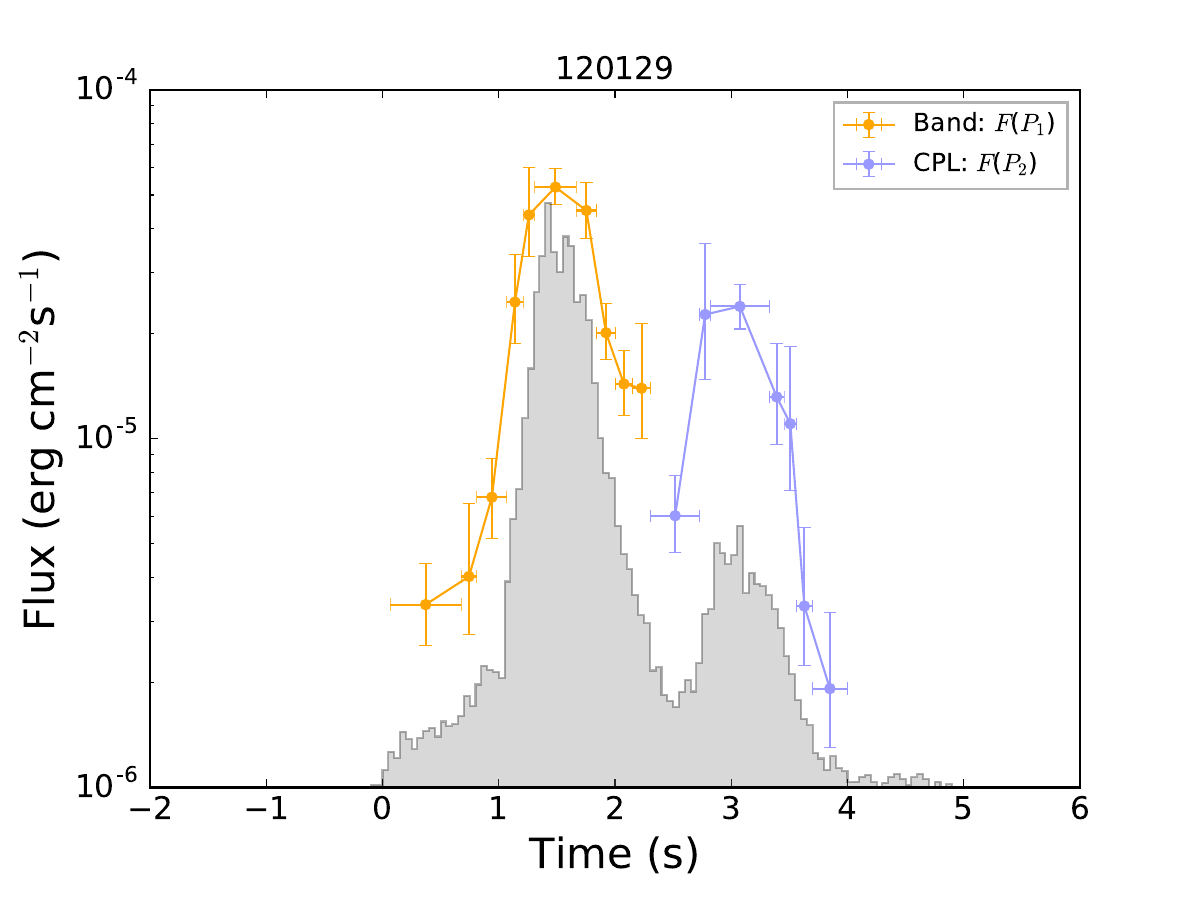}
\includegraphics[angle=0,scale=0.3]{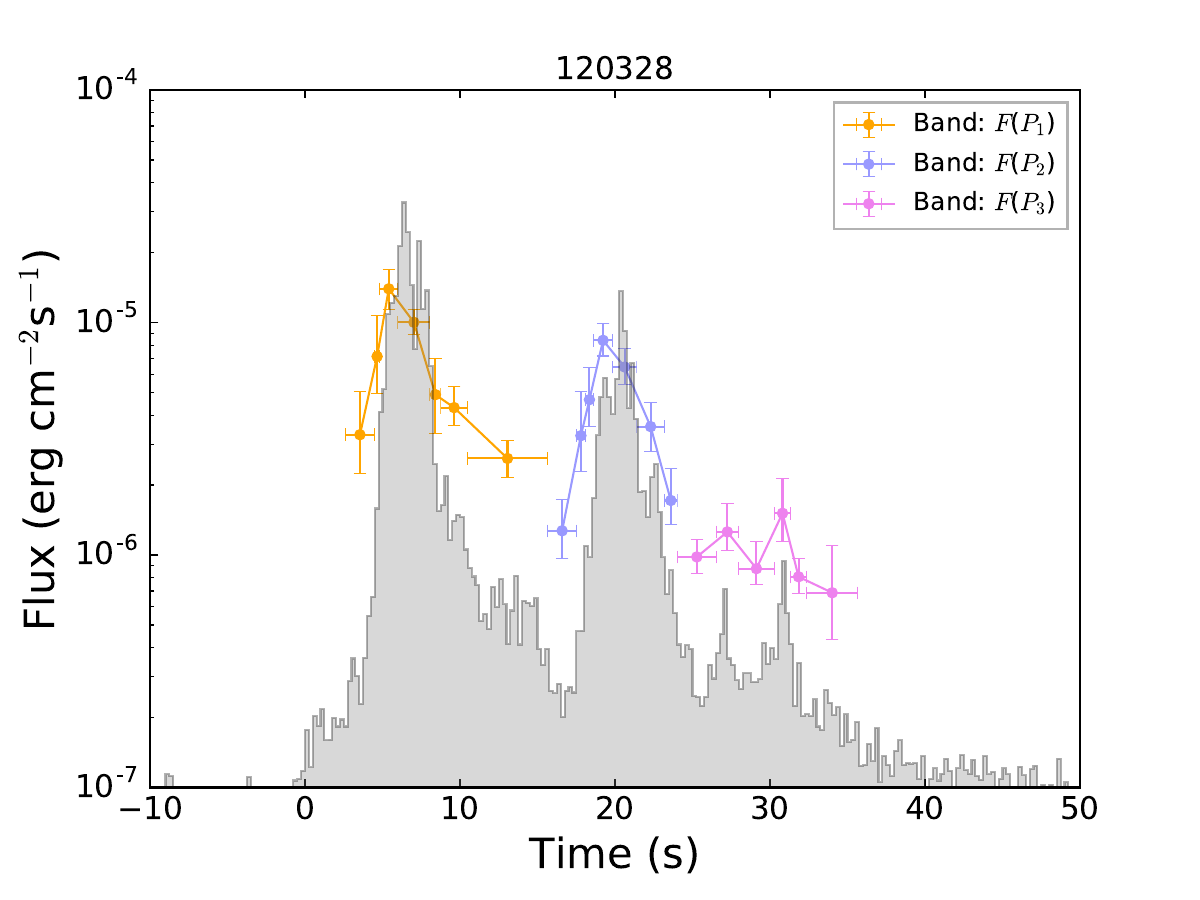}
\caption{Temporal evolution of energy flux $F$. The symbols and colors are the same as in Figure \ref{fig:Ep_Best}.}\label{fig:Flux_Best}
\end{figure*}
\begin{figure*}
\includegraphics[angle=0,scale=0.3]{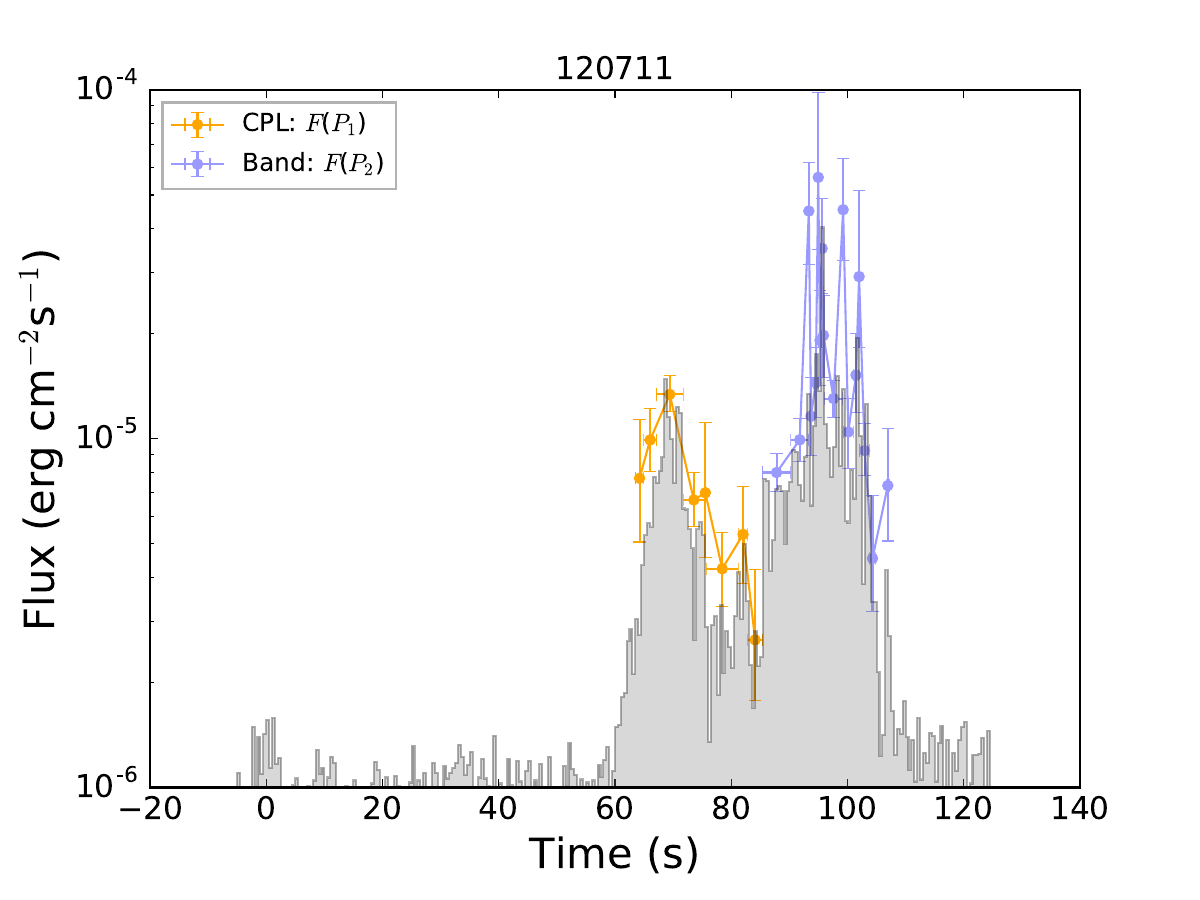}
\includegraphics[angle=0,scale=0.3]{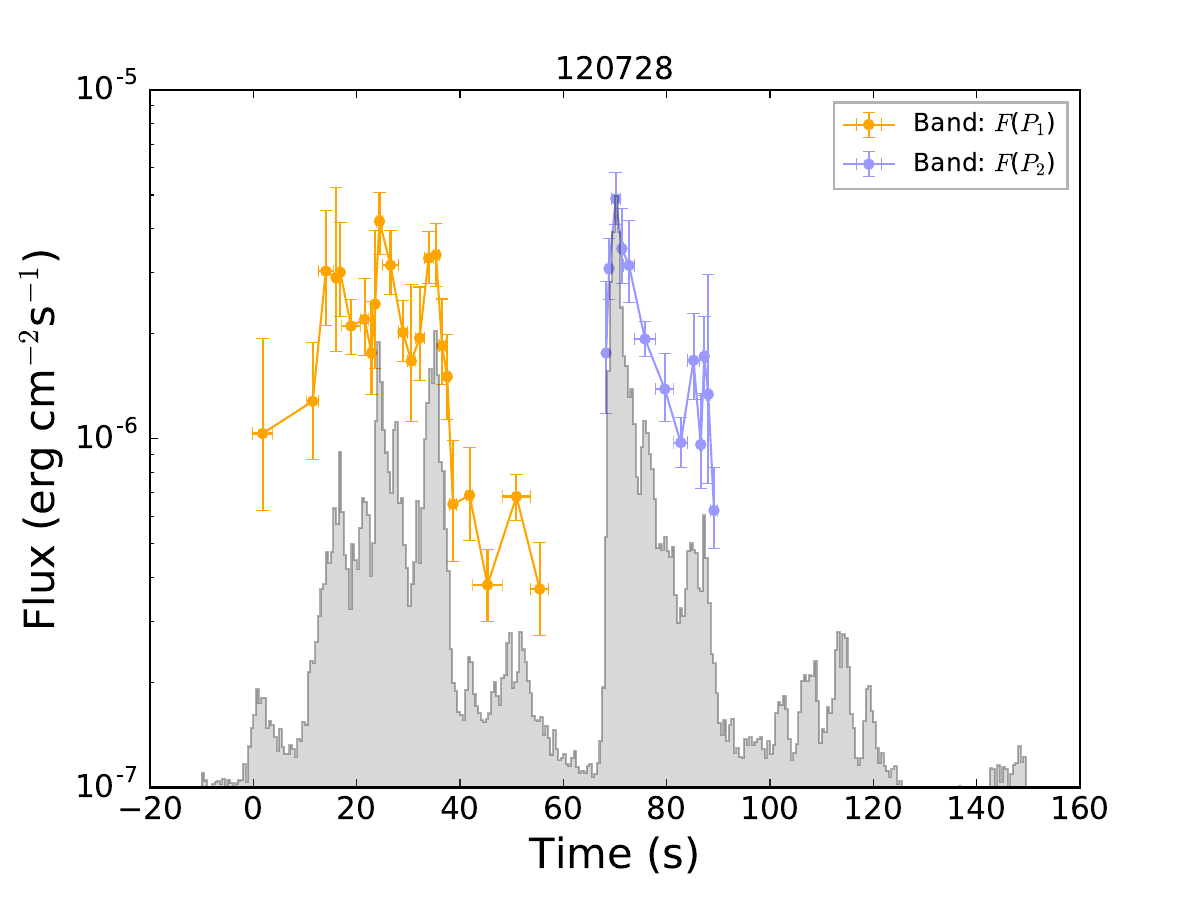}
\includegraphics[angle=0,scale=0.3]{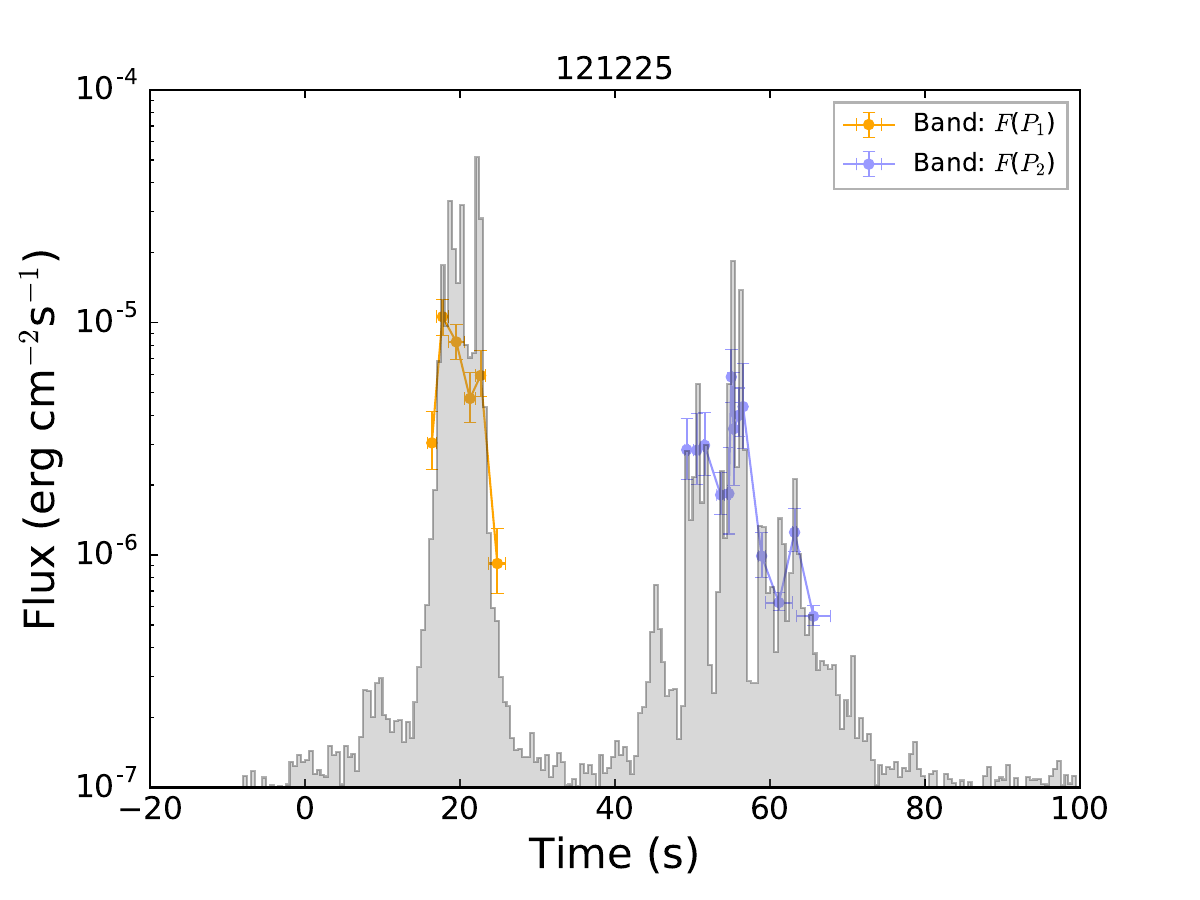}
\includegraphics[angle=0,scale=0.3]{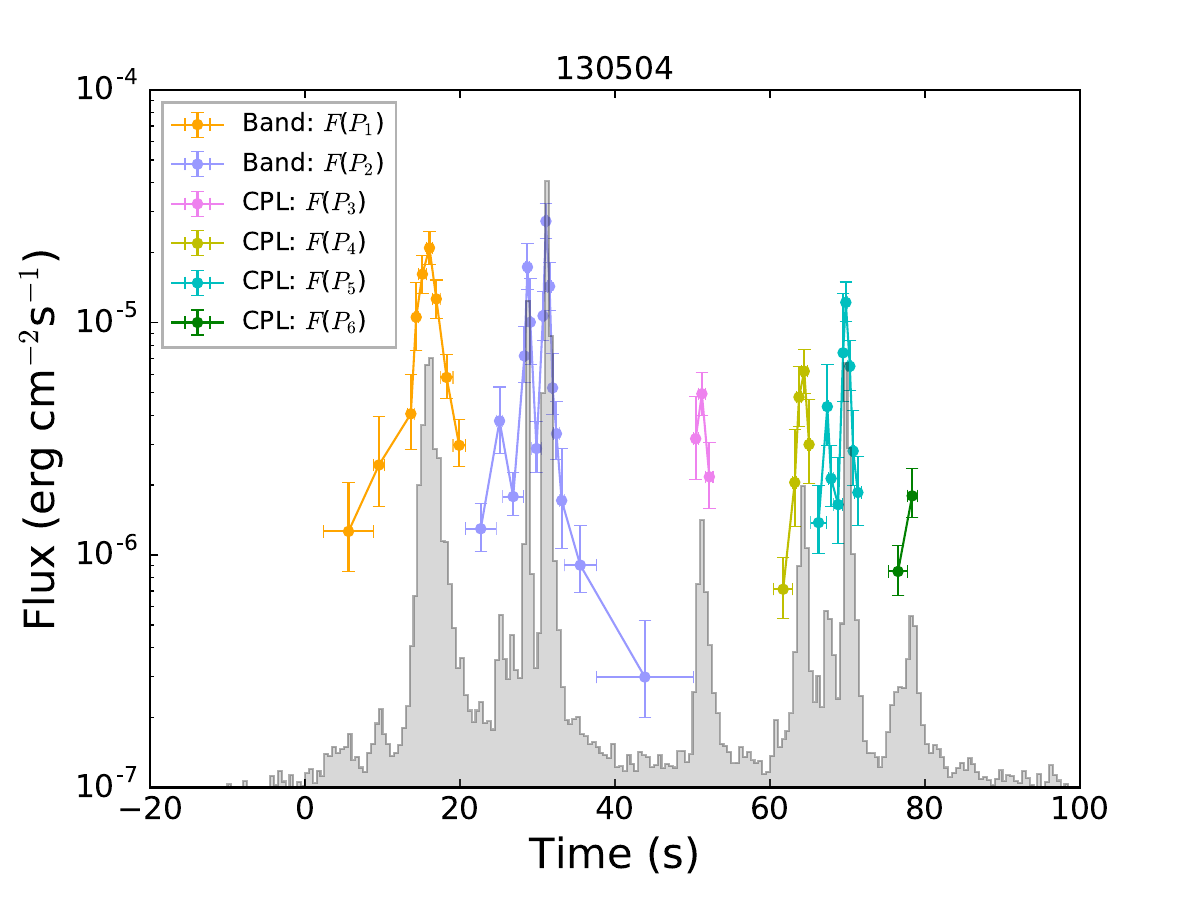}
\includegraphics[angle=0,scale=0.3]{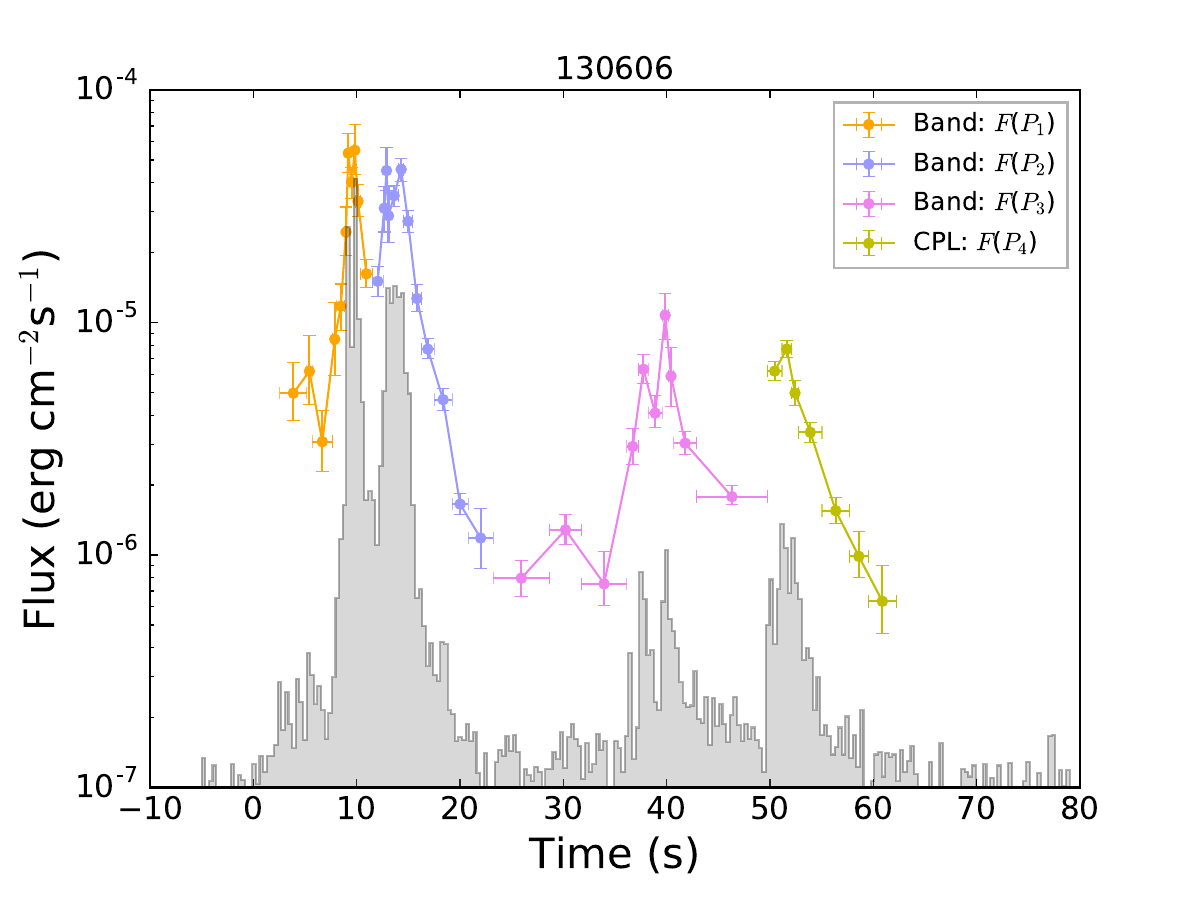}
\includegraphics[angle=0,scale=0.3]{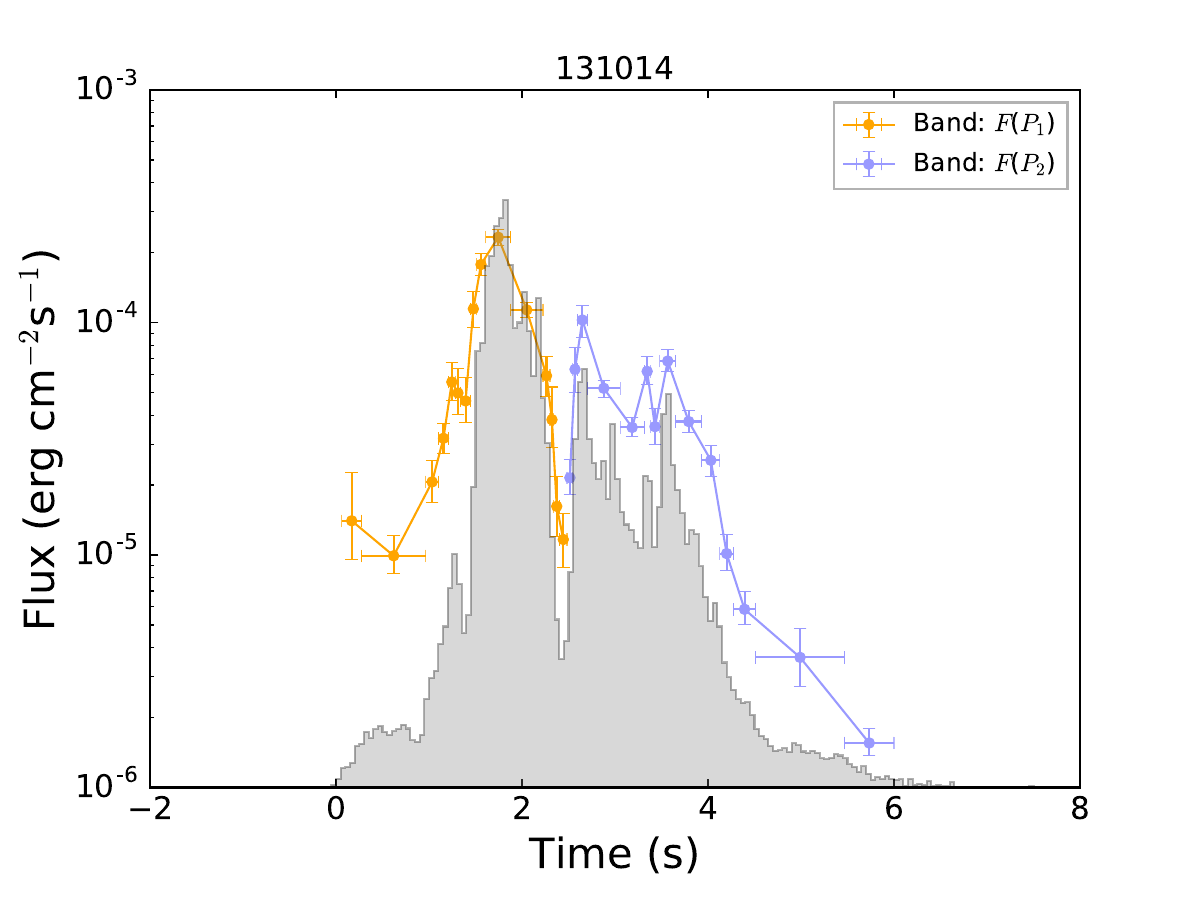}
\includegraphics[angle=0,scale=0.3]{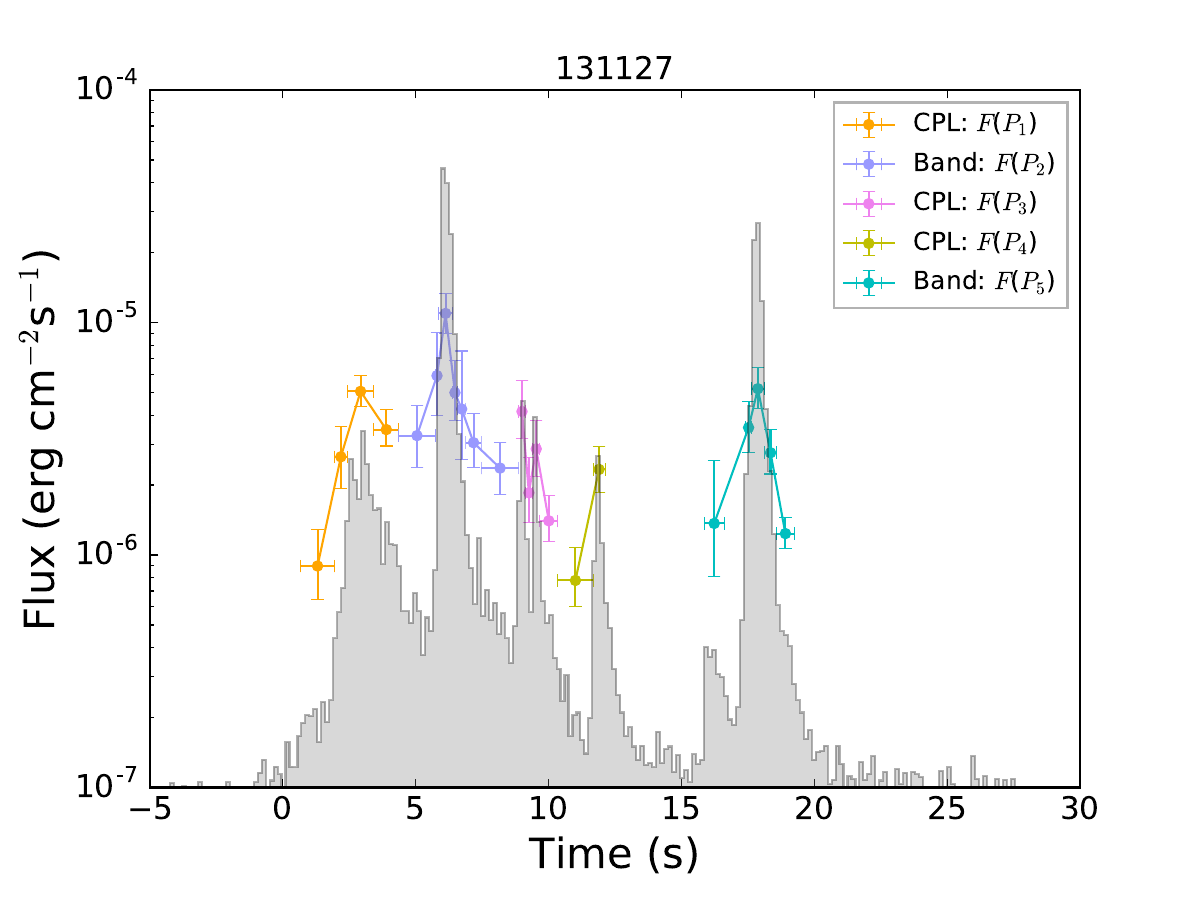}
\includegraphics[angle=0,scale=0.3]{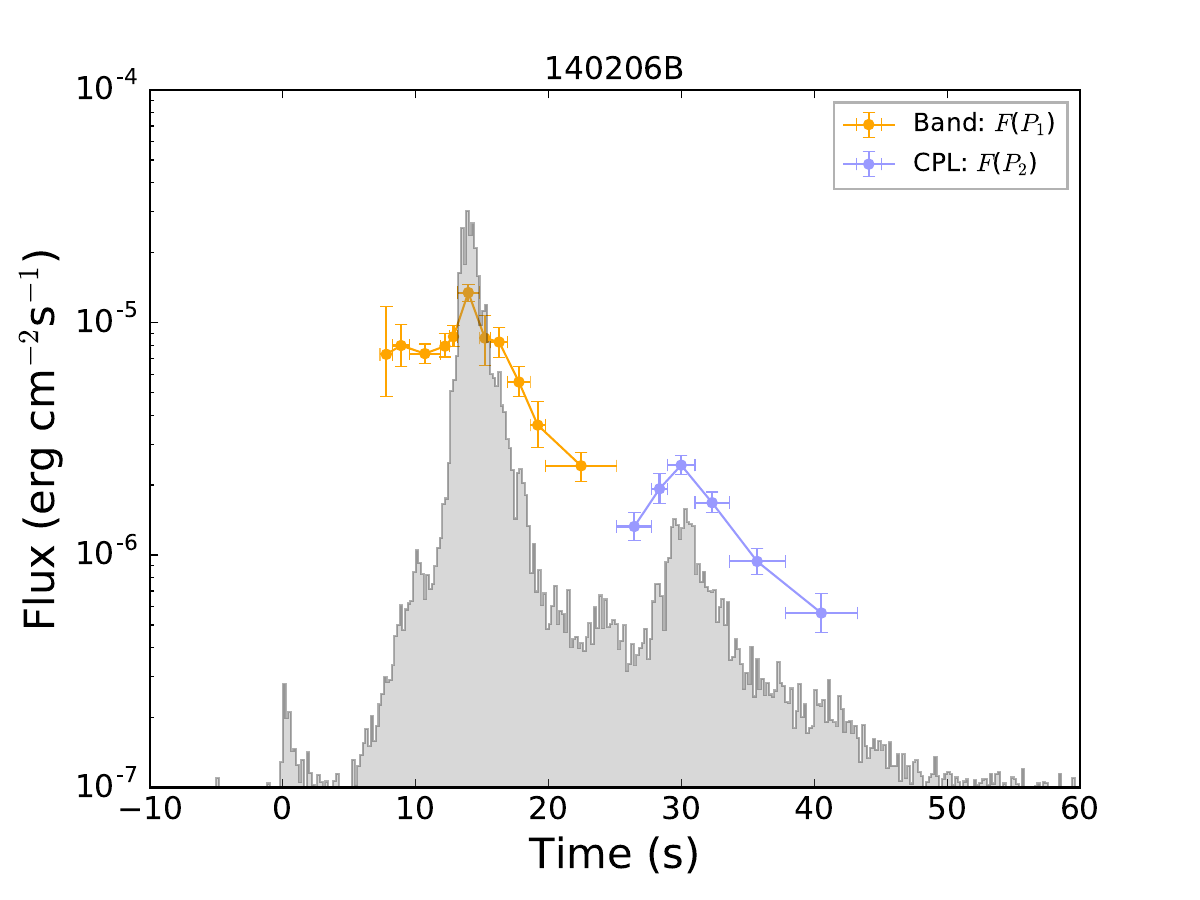}
\includegraphics[angle=0,scale=0.3]{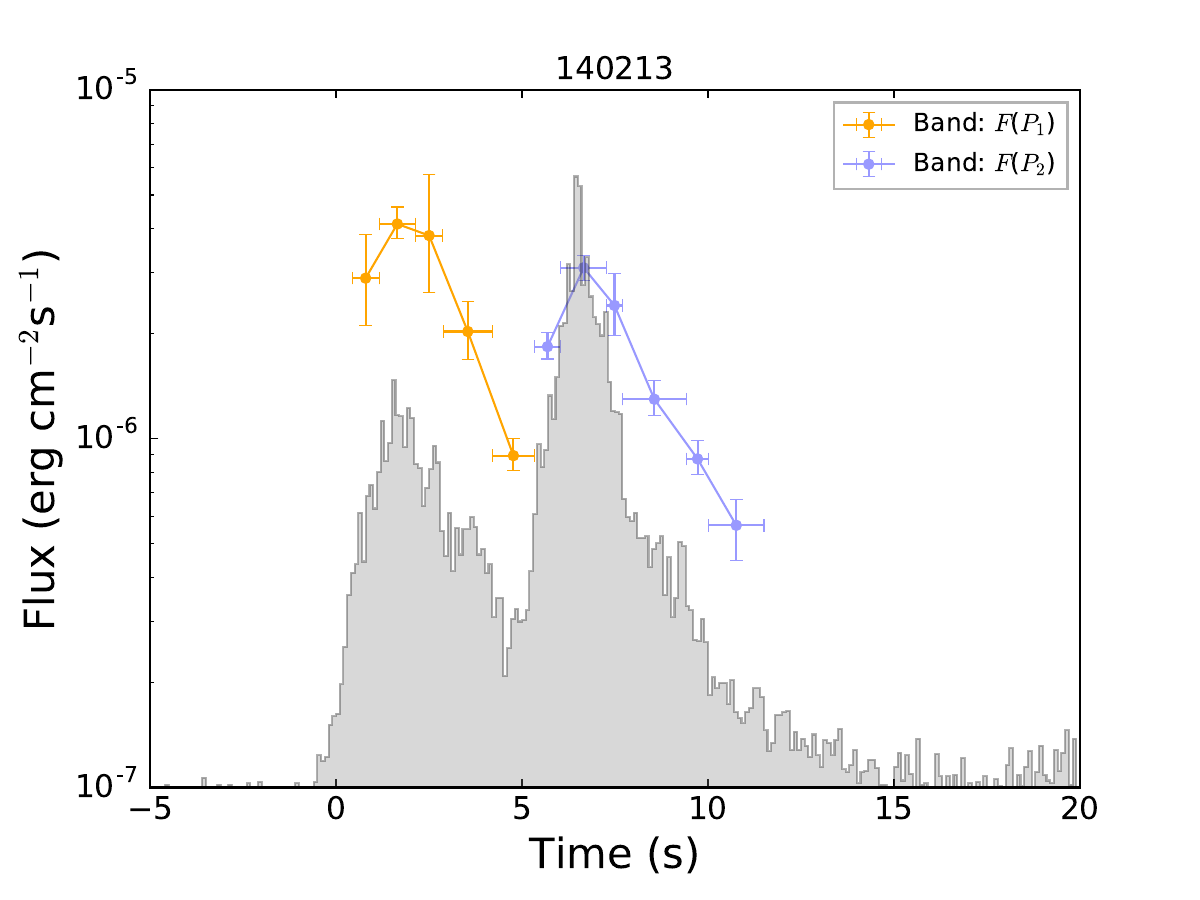}
\includegraphics[angle=0,scale=0.3]{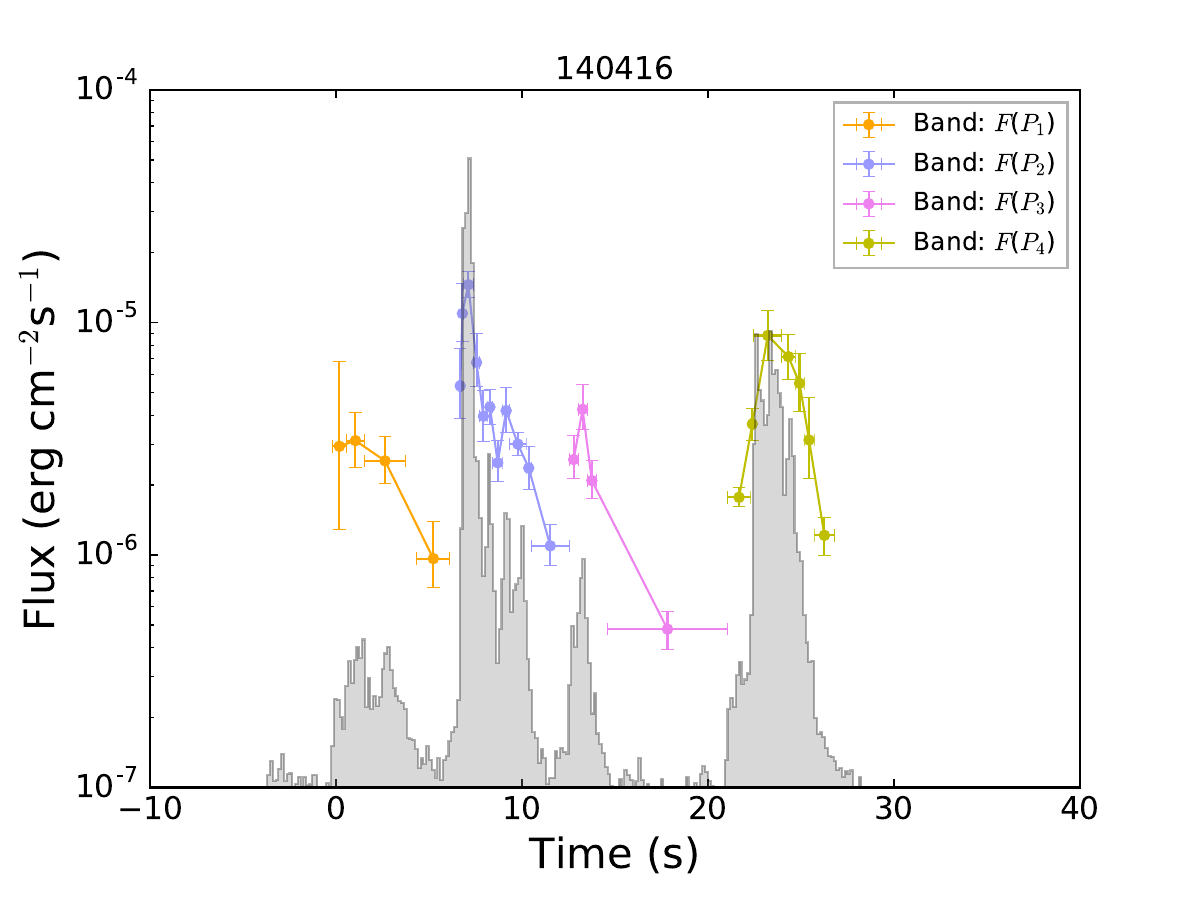}
\includegraphics[angle=0,scale=0.3]{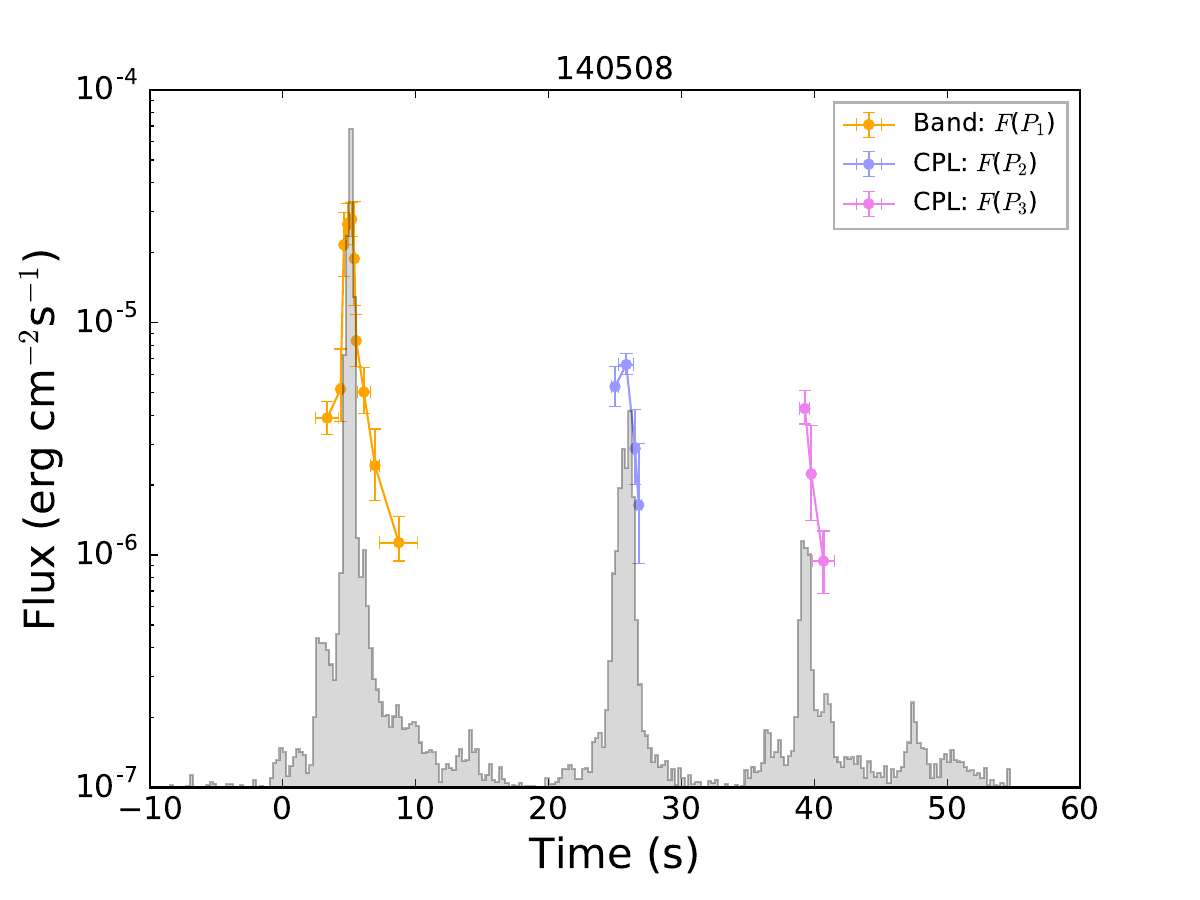}
\includegraphics[angle=0,scale=0.3]{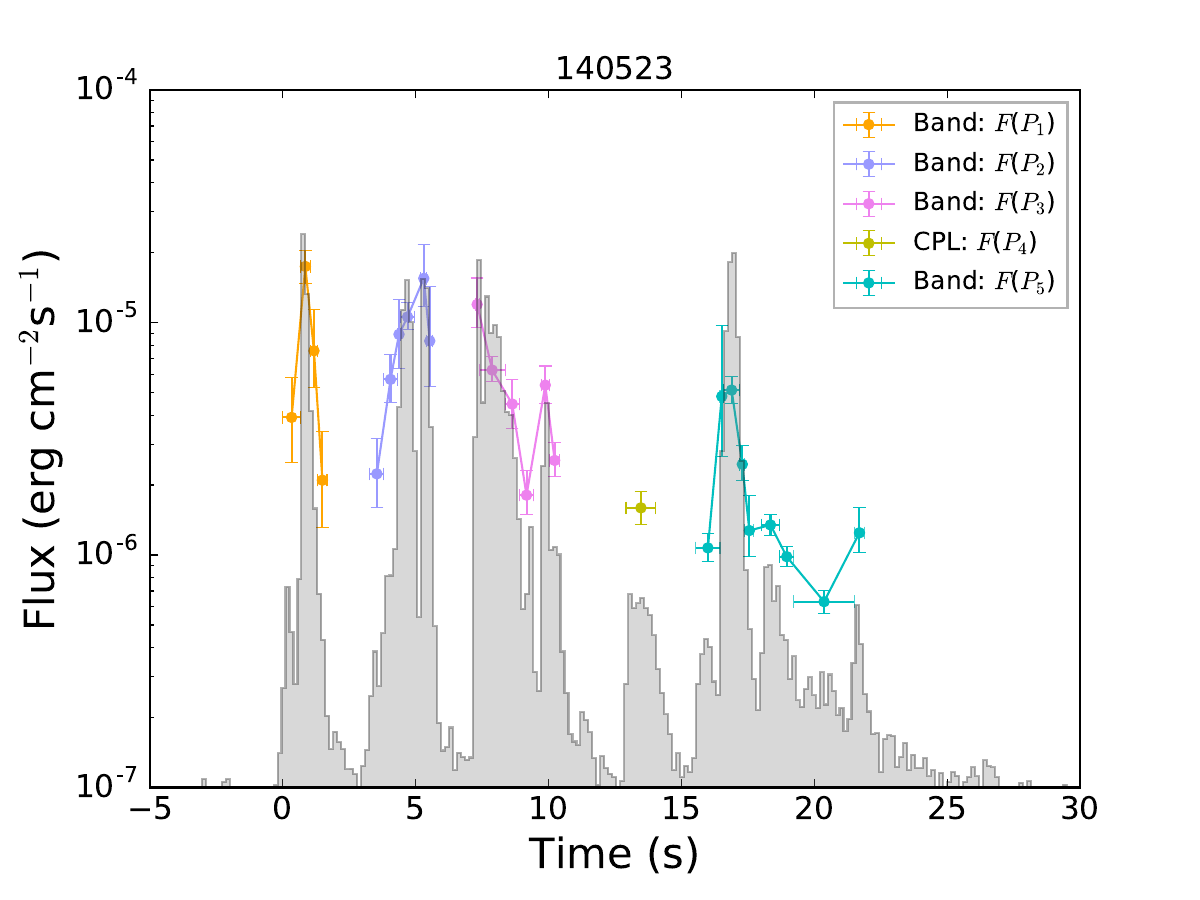}
\includegraphics[angle=0,scale=0.3]{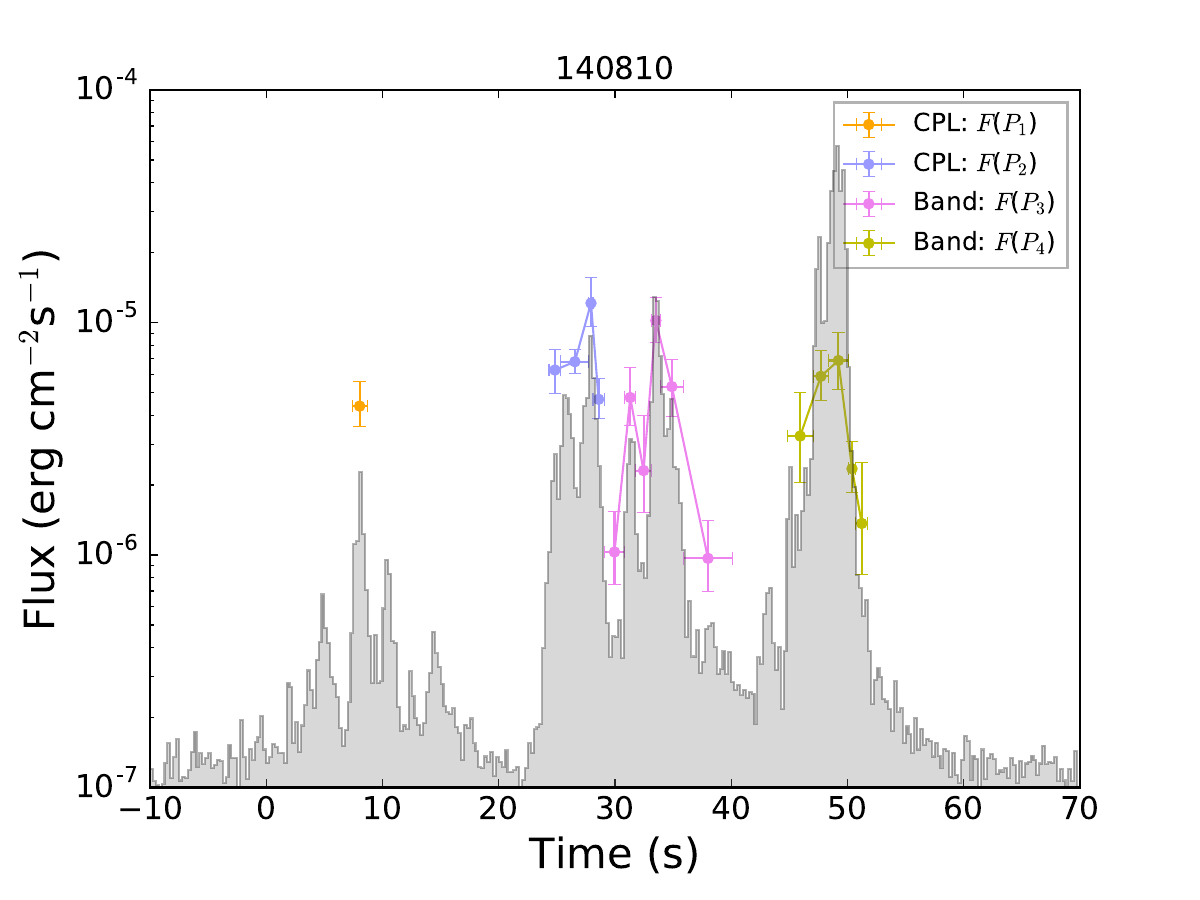}
\includegraphics[angle=0,scale=0.3]{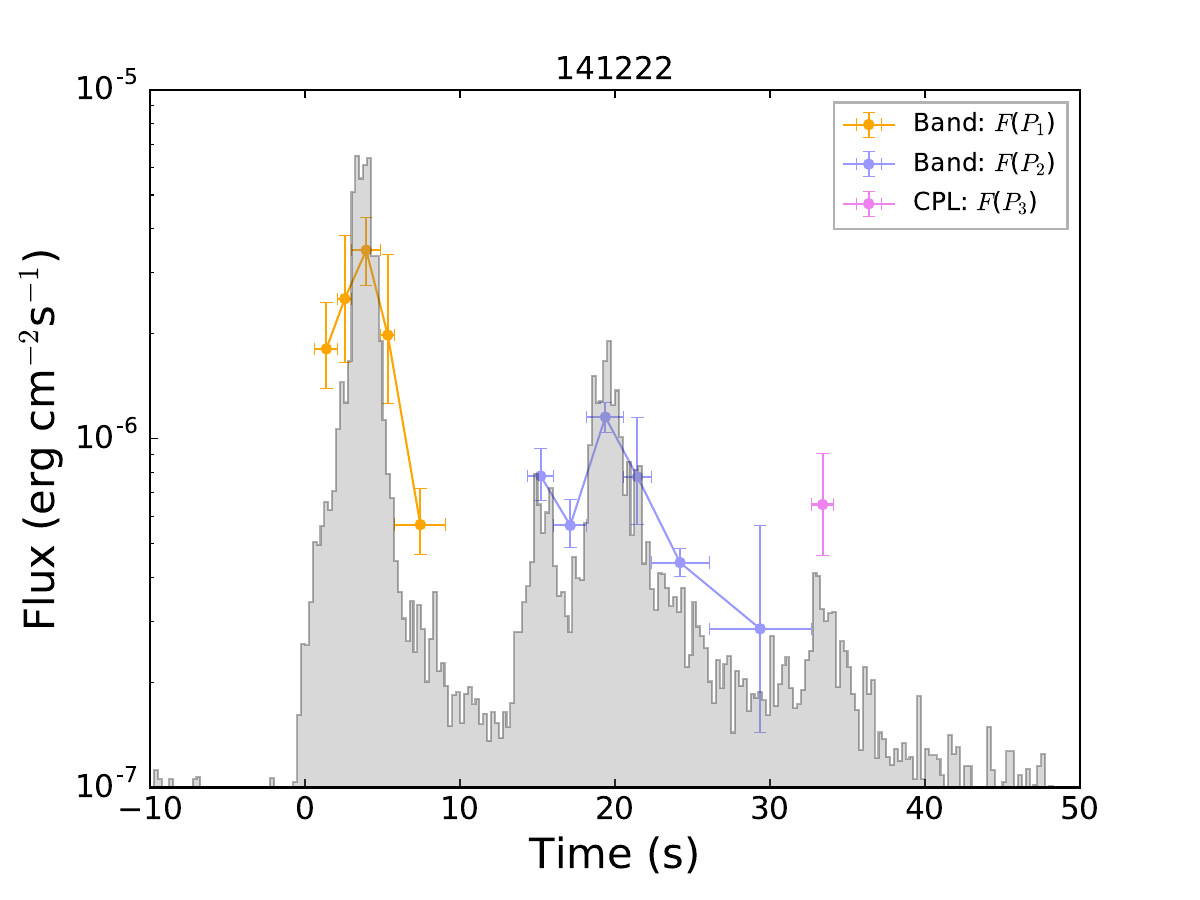}
\includegraphics[angle=0,scale=0.3]{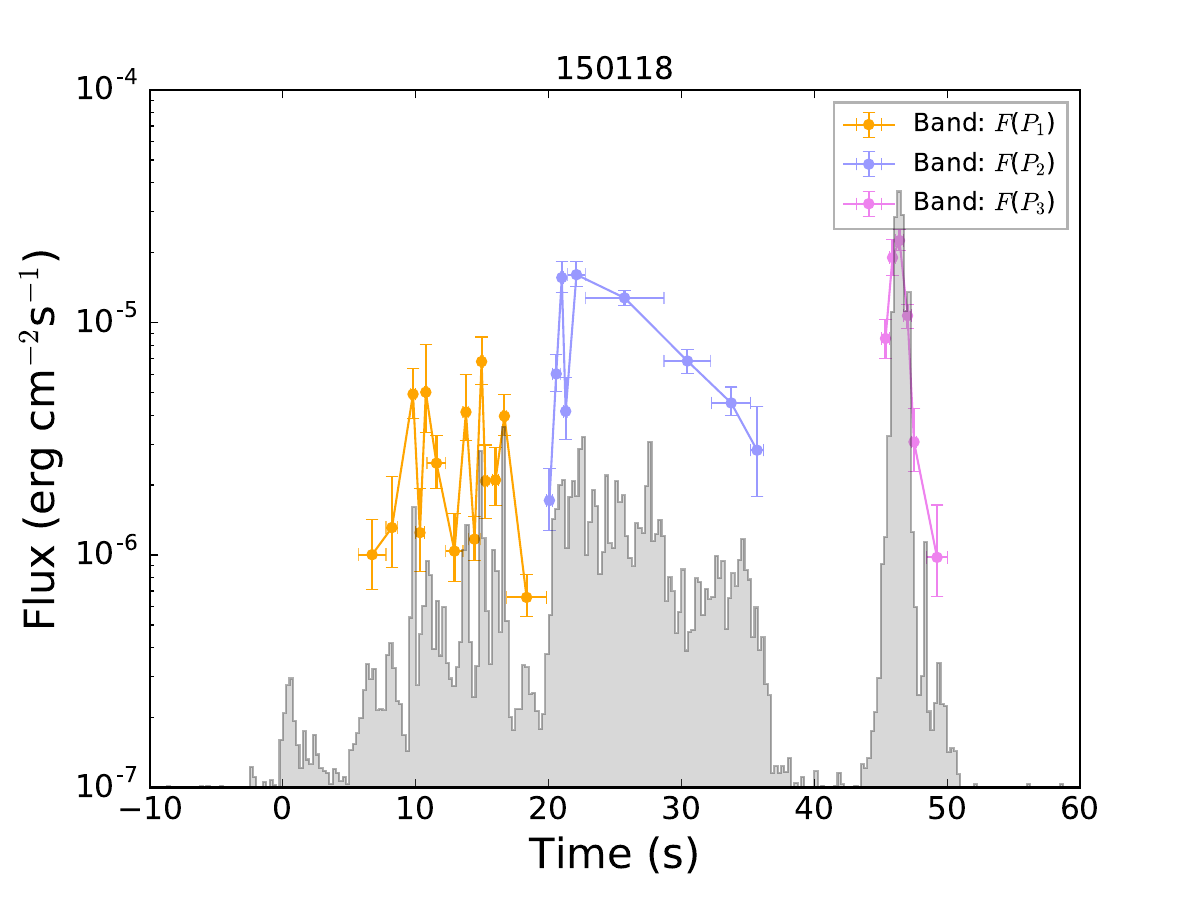}
\center{Fig. \ref{fig:Flux_Best}--- Continued}
\end{figure*}
\begin{figure*}
\includegraphics[angle=0,scale=0.3]{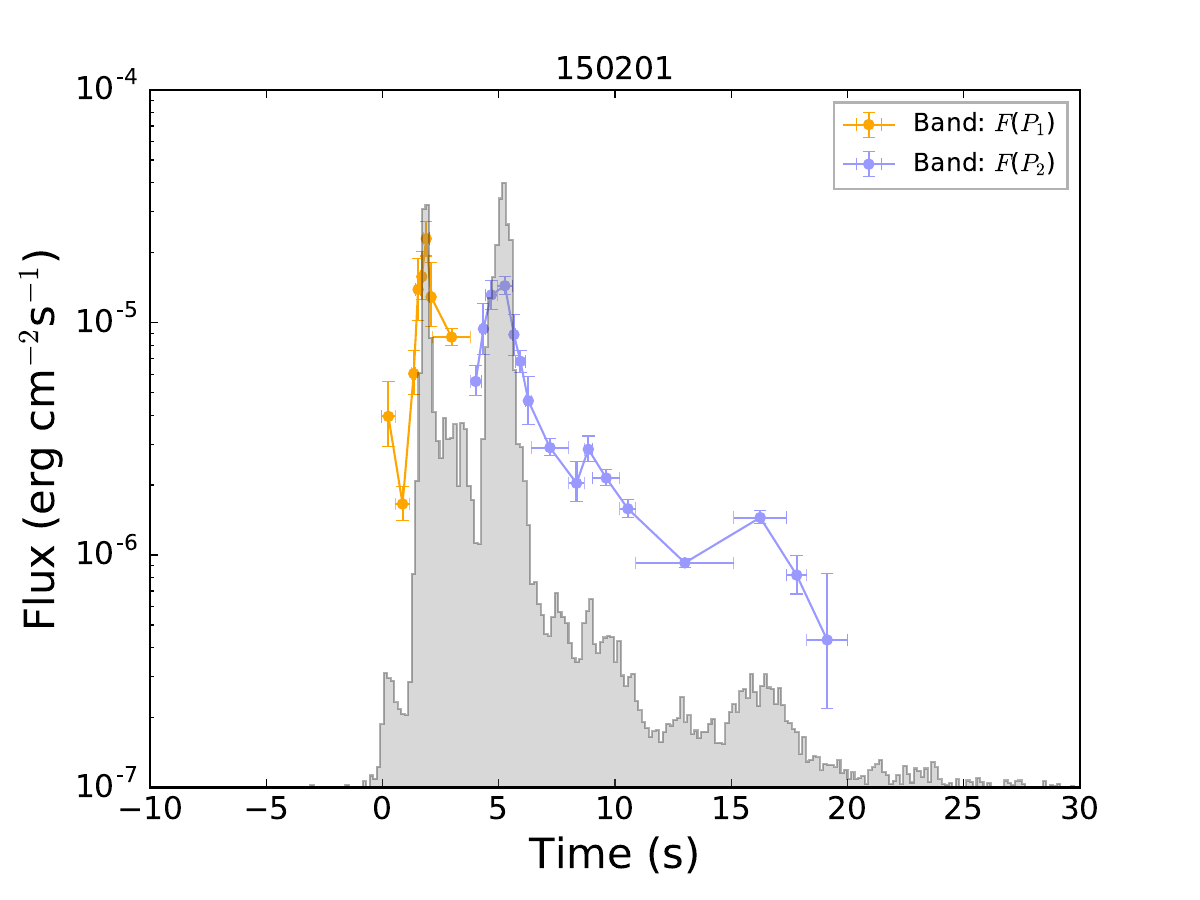}
\includegraphics[angle=0,scale=0.3]{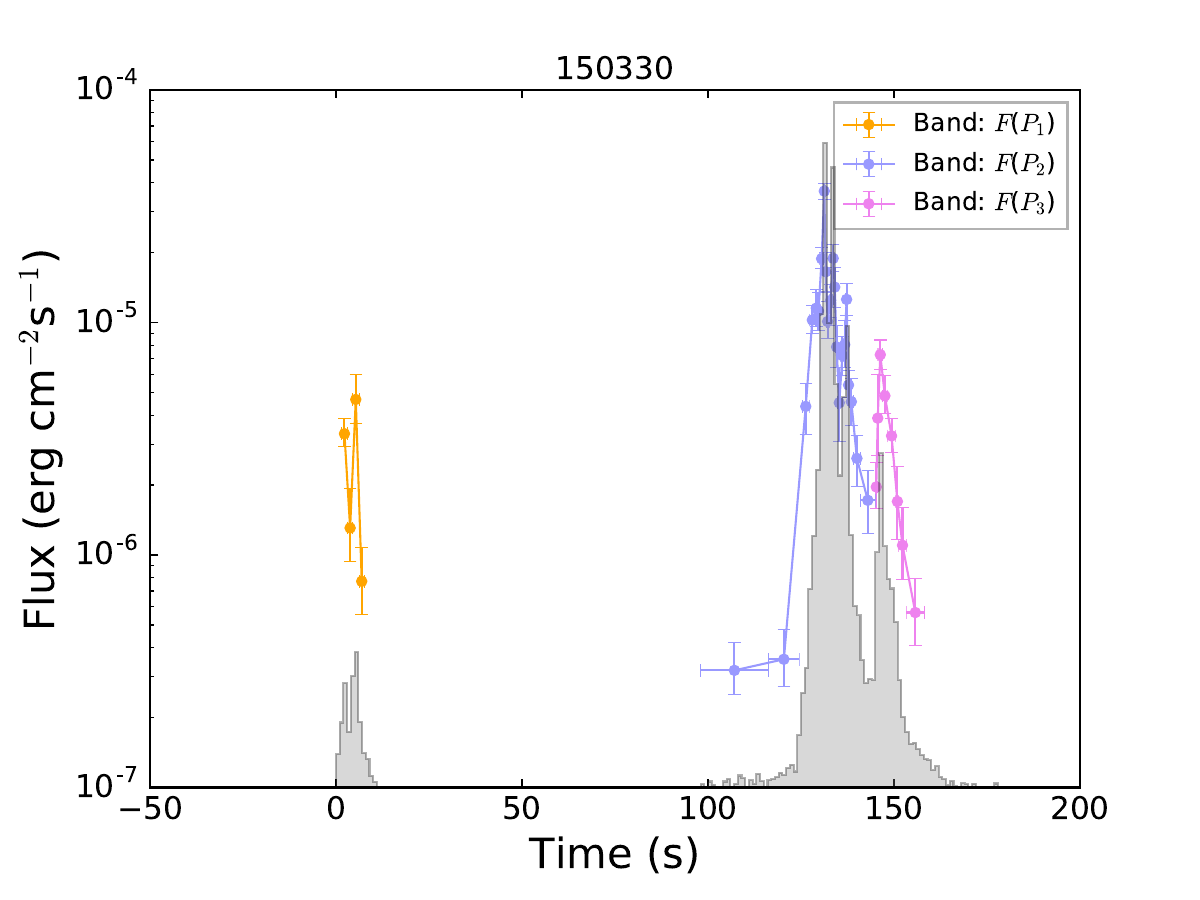}
\includegraphics[angle=0,scale=0.3]{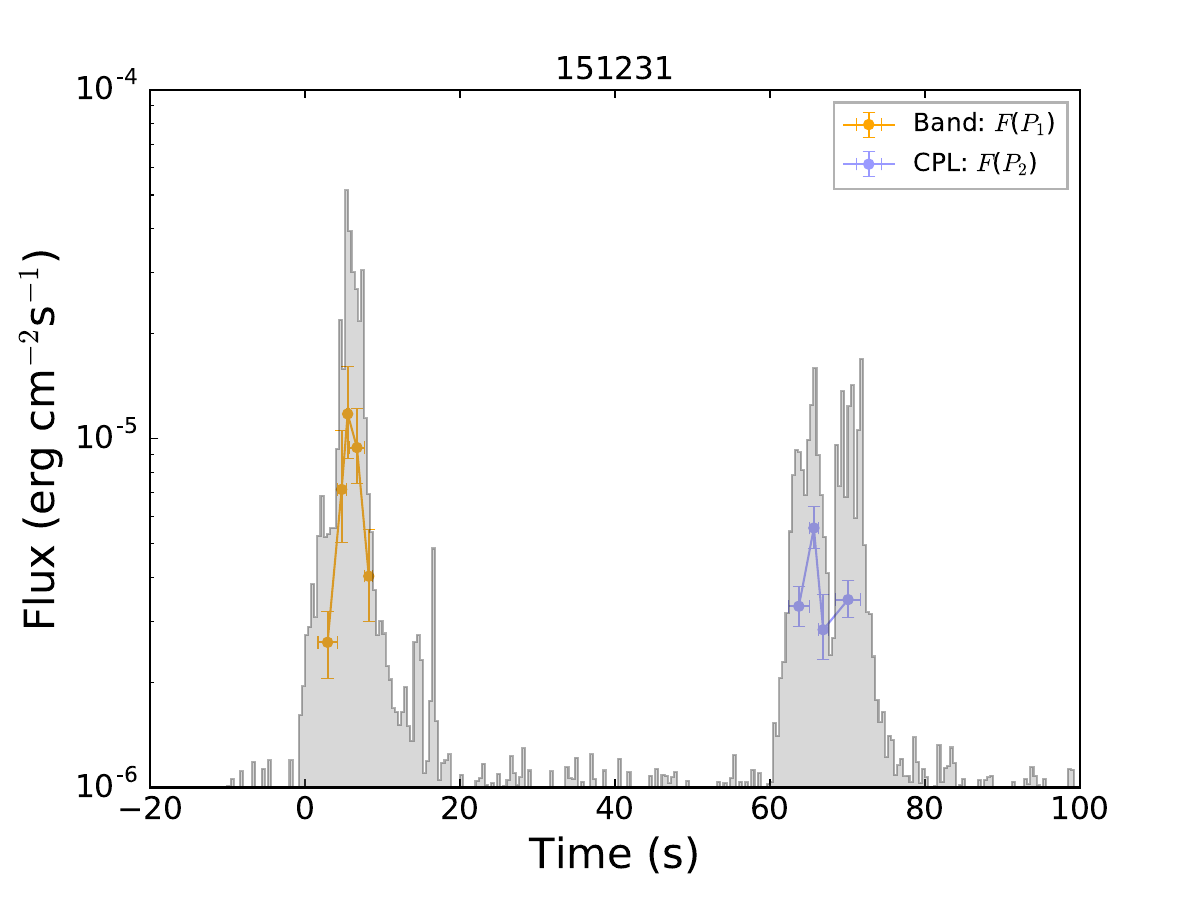}
\includegraphics[angle=0,scale=0.3]{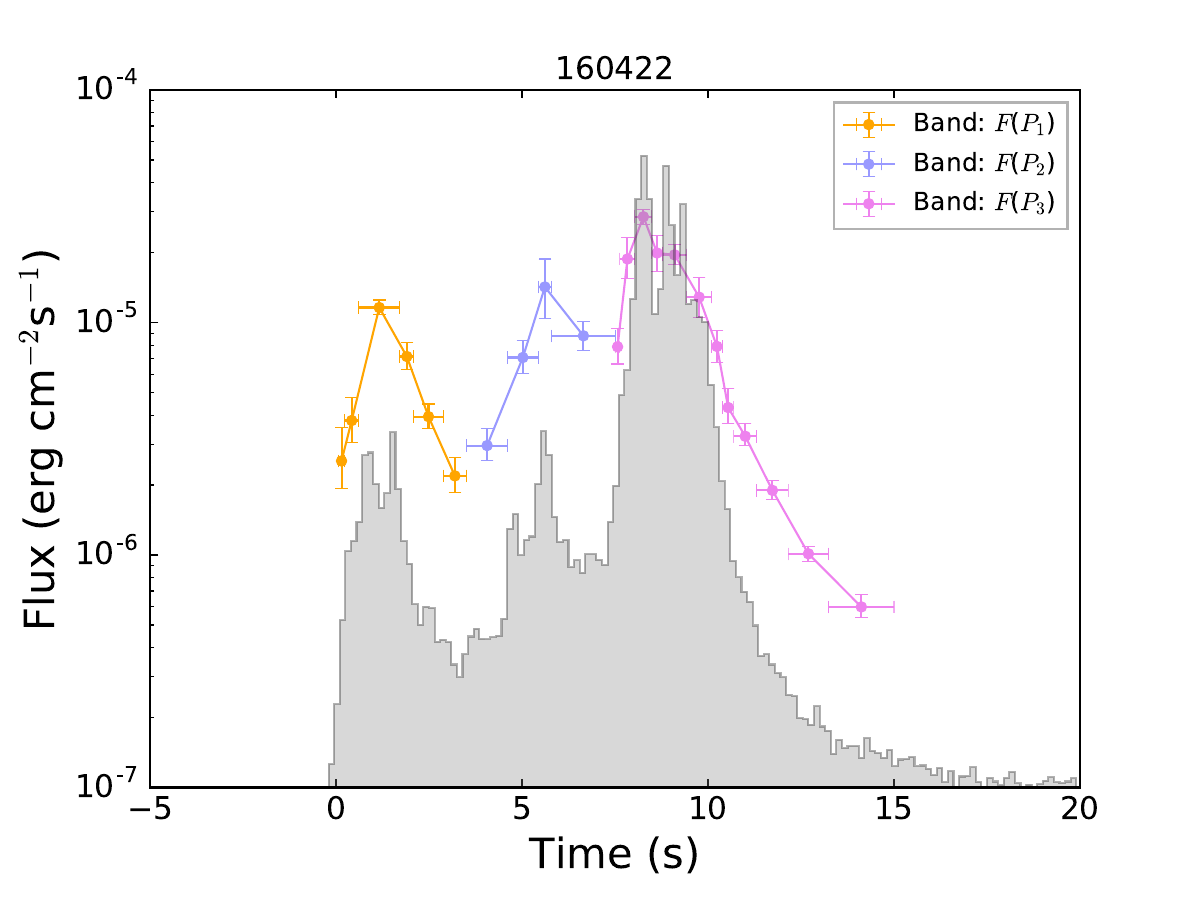}
\includegraphics[angle=0,scale=0.3]{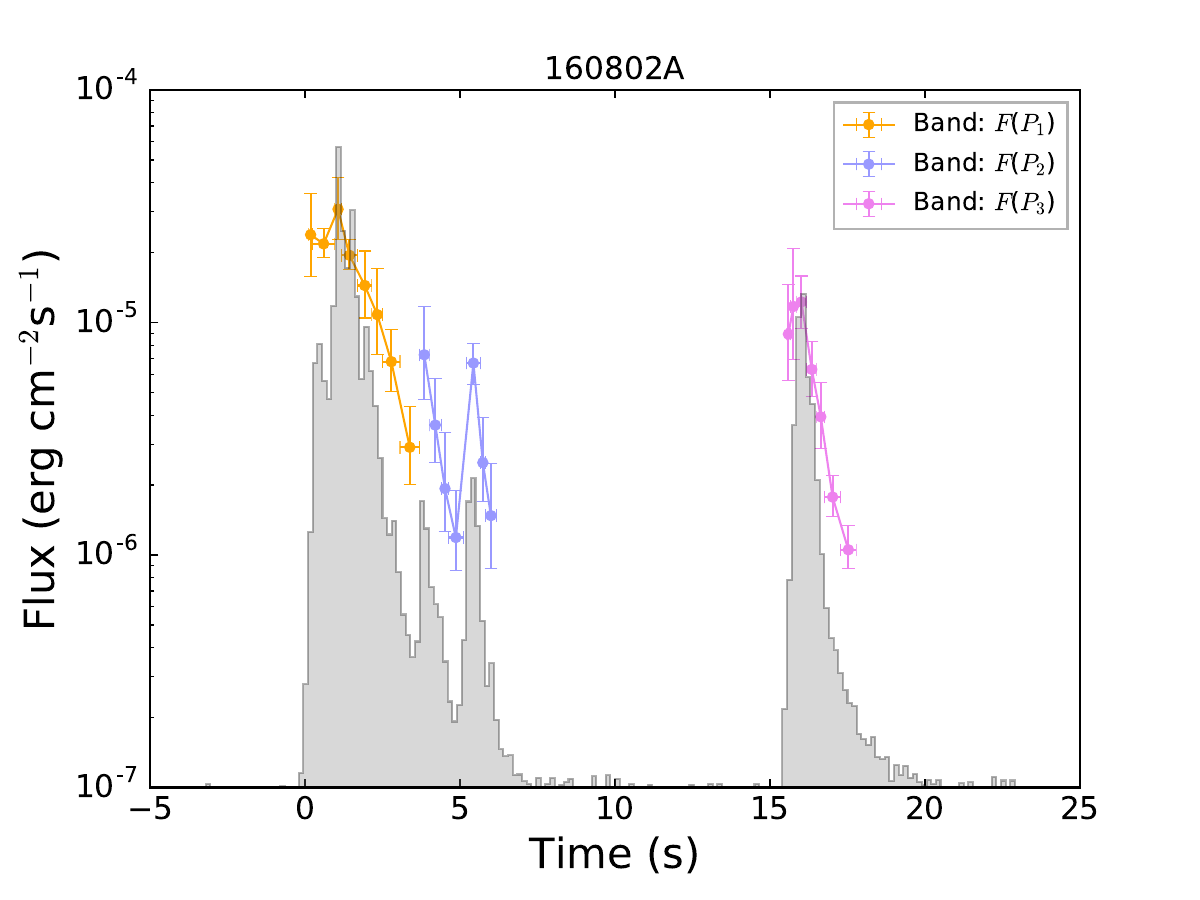}
\includegraphics[angle=0,scale=0.3]{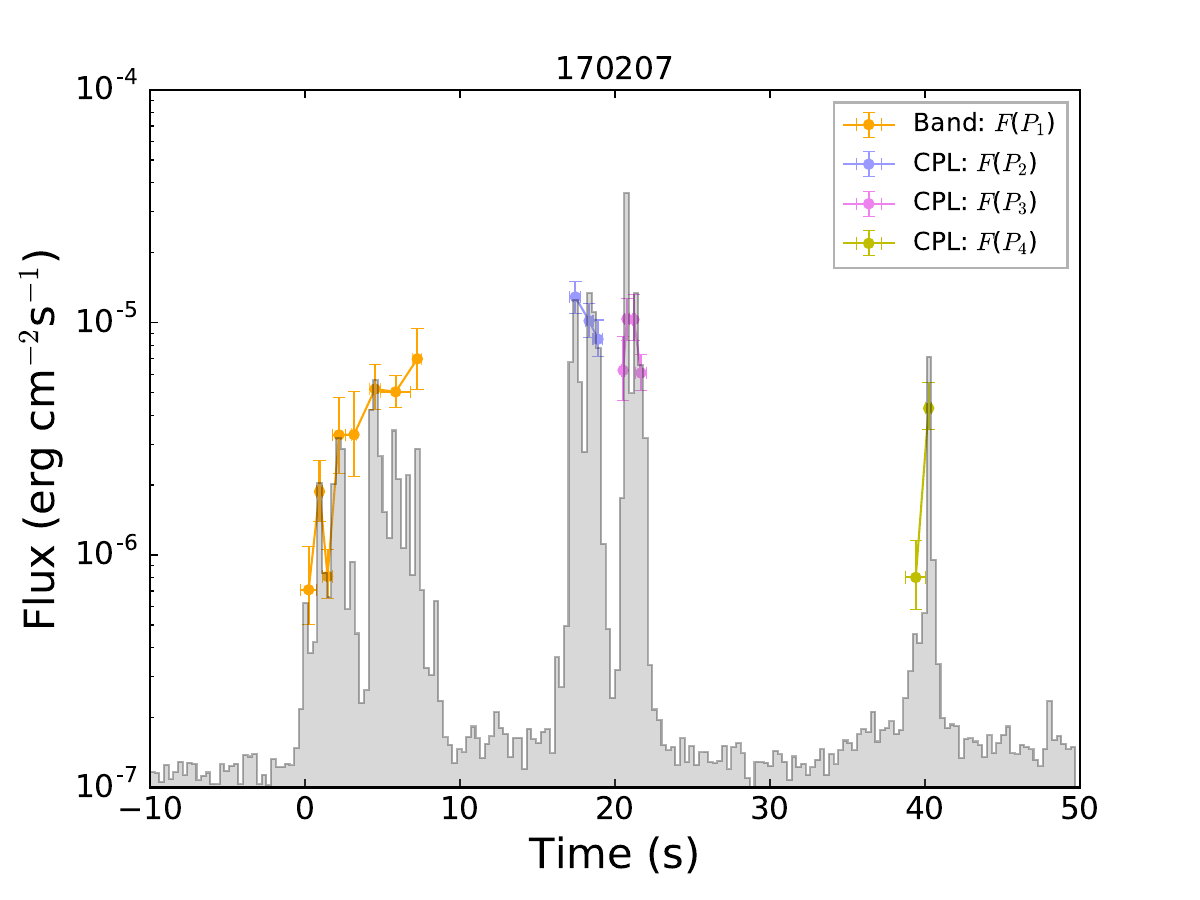}
\includegraphics[angle=0,scale=0.3]{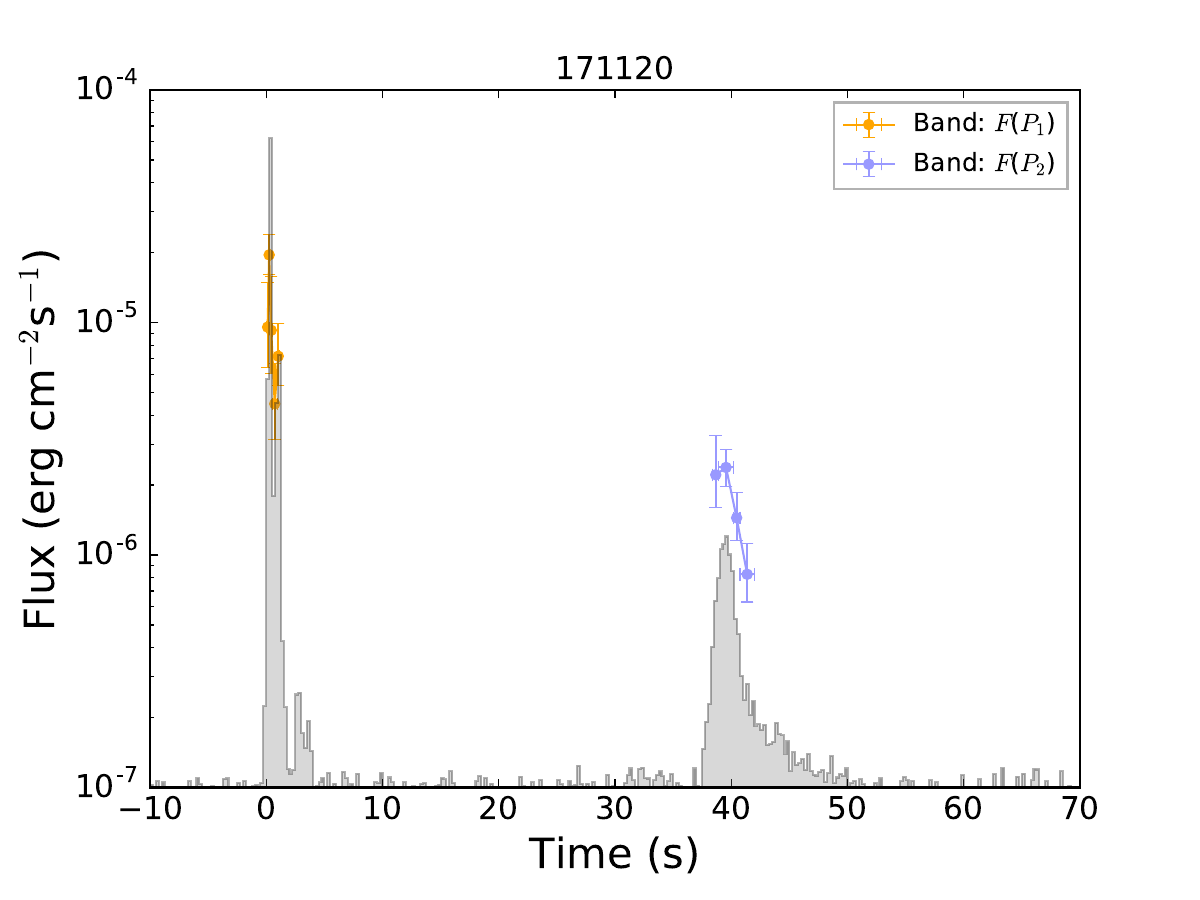}
\includegraphics[angle=0,scale=0.3]{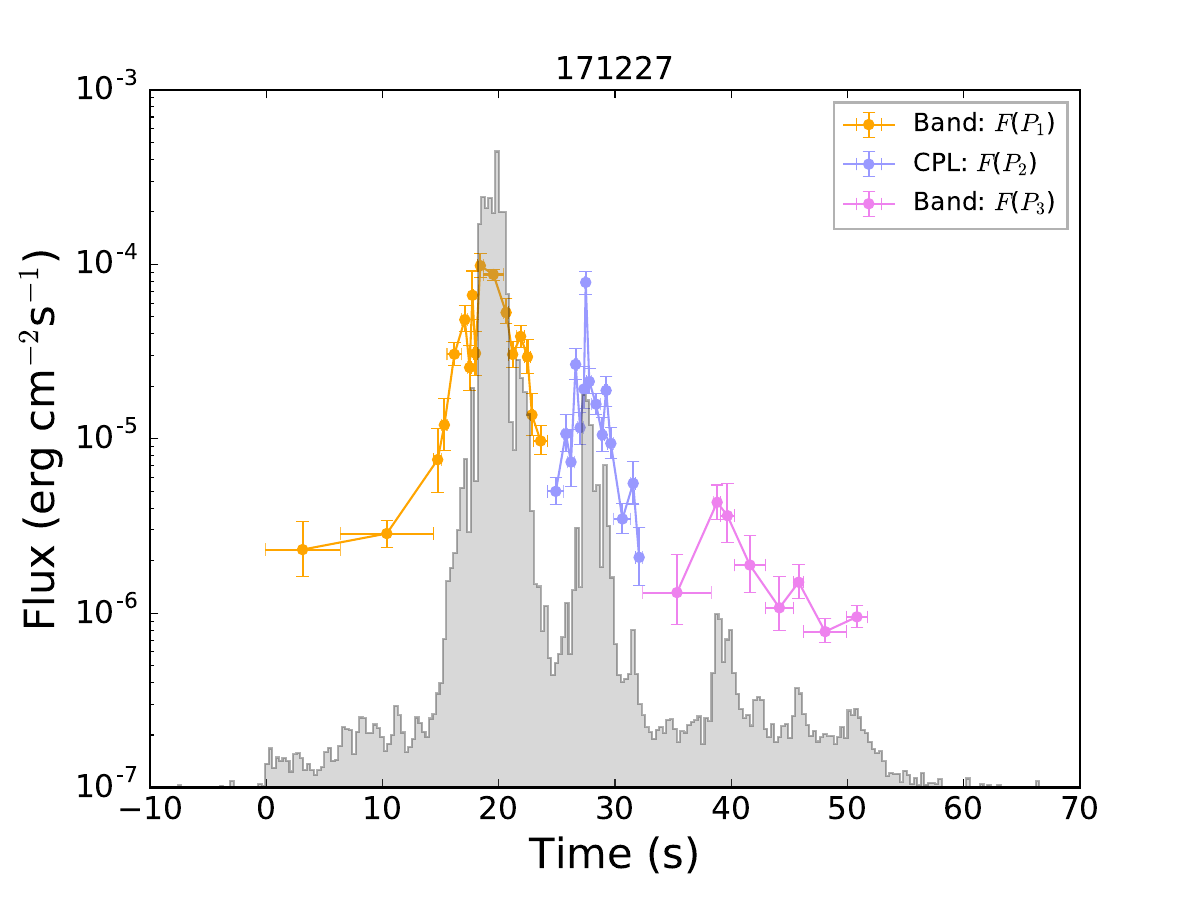}
\includegraphics[angle=0,scale=0.3]{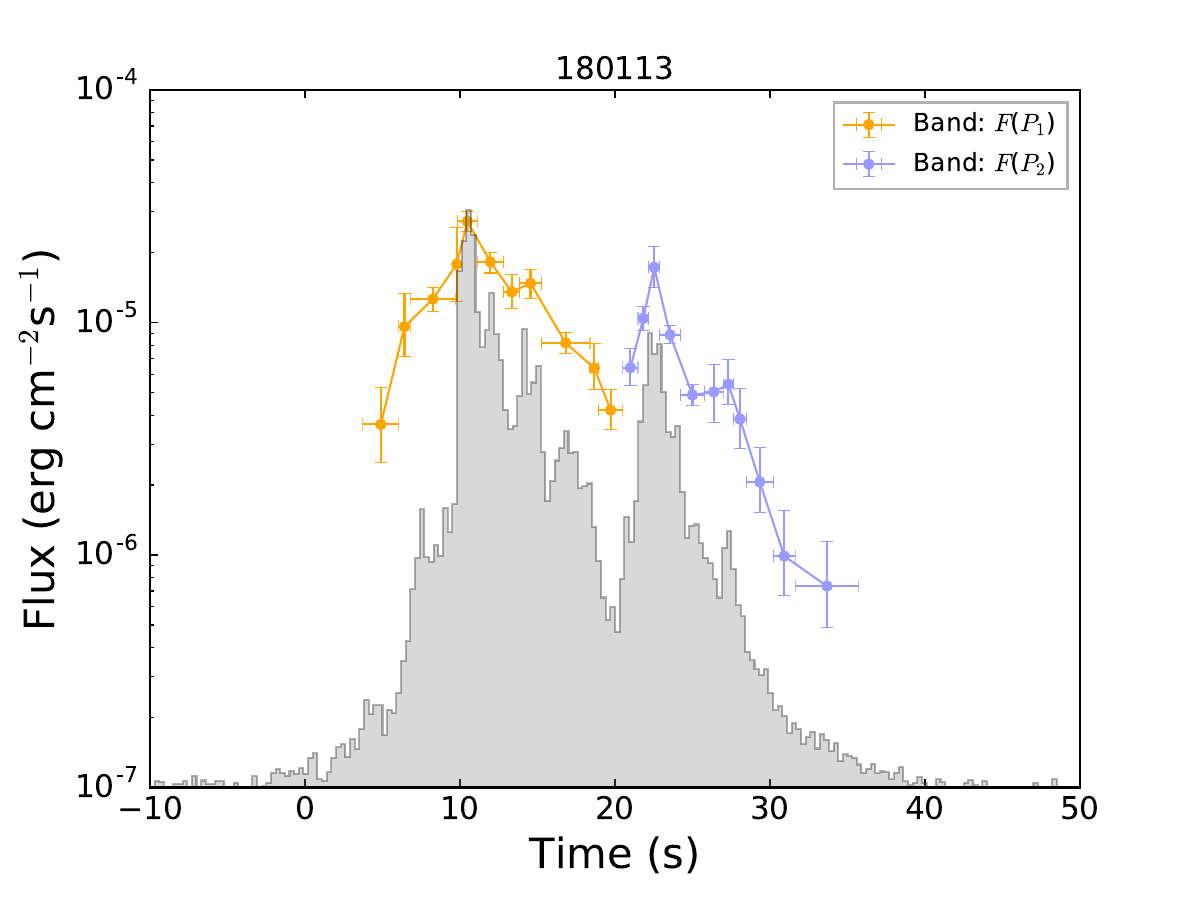}
\includegraphics[angle=0,scale=0.3]{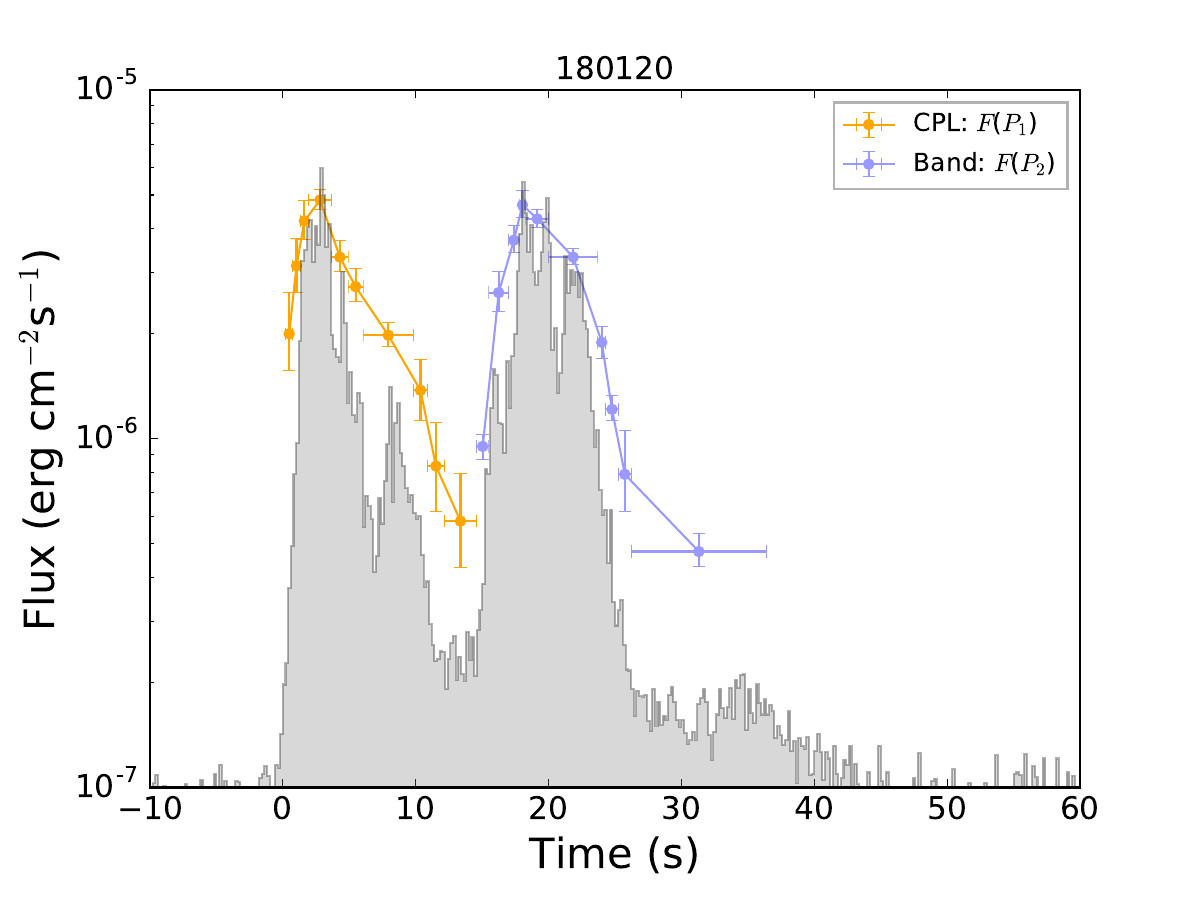}
\includegraphics[angle=0,scale=0.3]{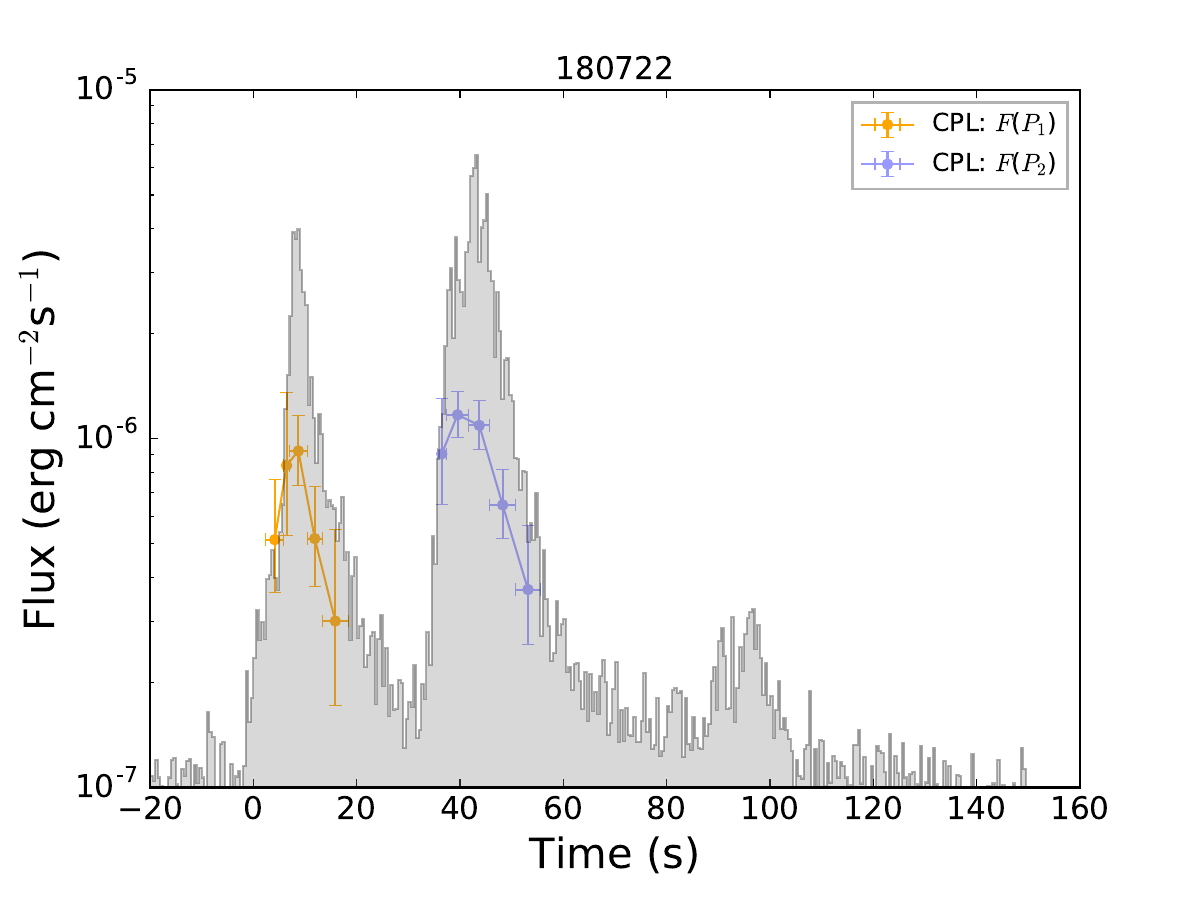}
\includegraphics[angle=0,scale=0.3]{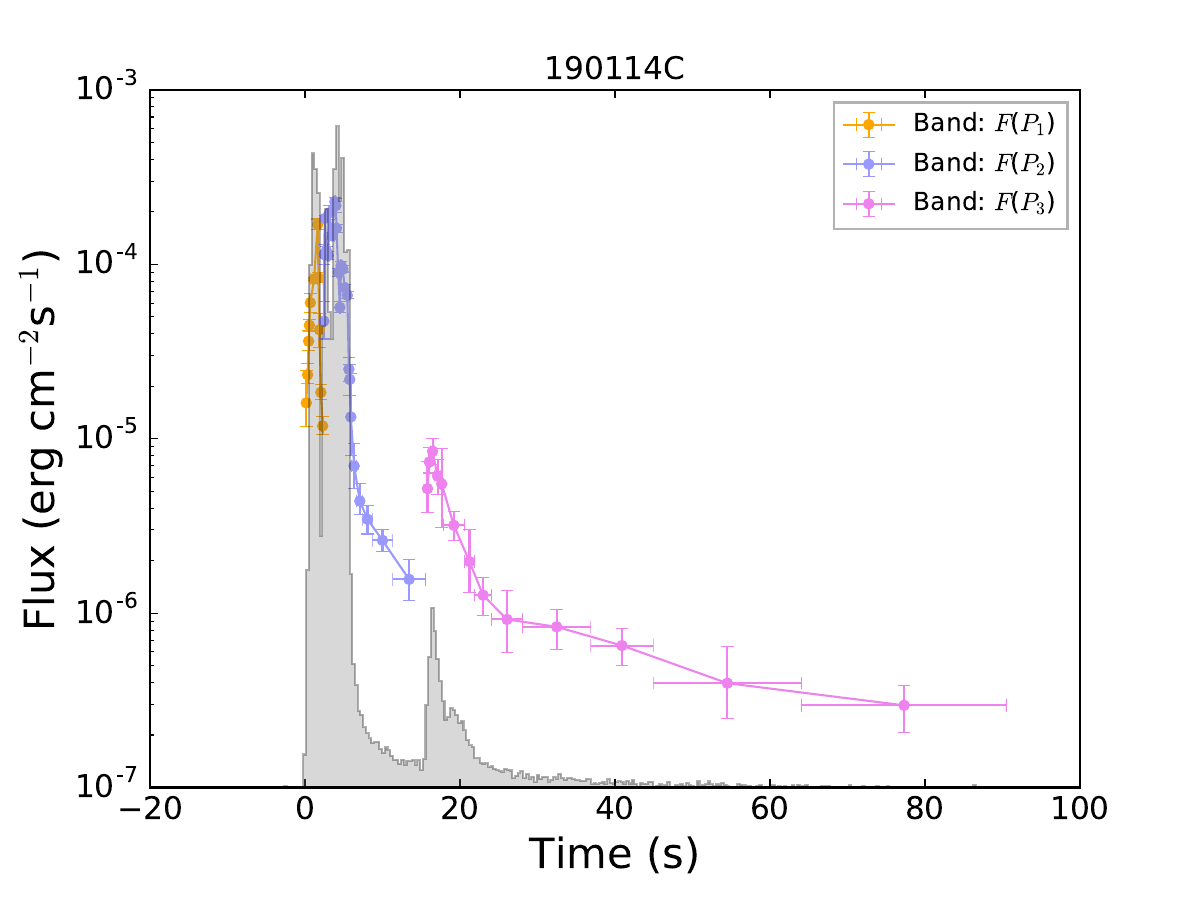}
\center{Fig. \ref{fig:Flux_Best}--- Continued}
\end{figure*}

\clearpage
\begin{figure*}
\includegraphics[angle=0,scale=0.3]{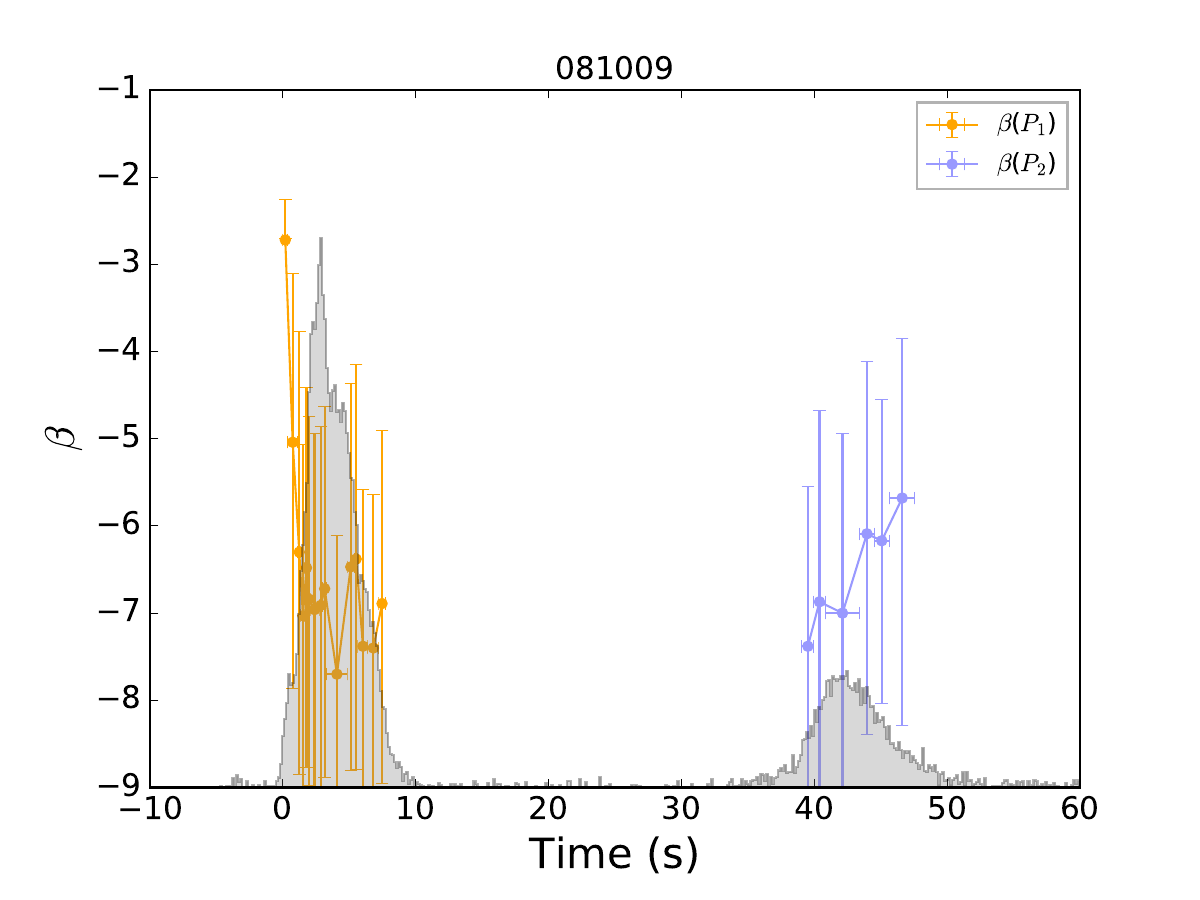}
\includegraphics[angle=0,scale=0.3]{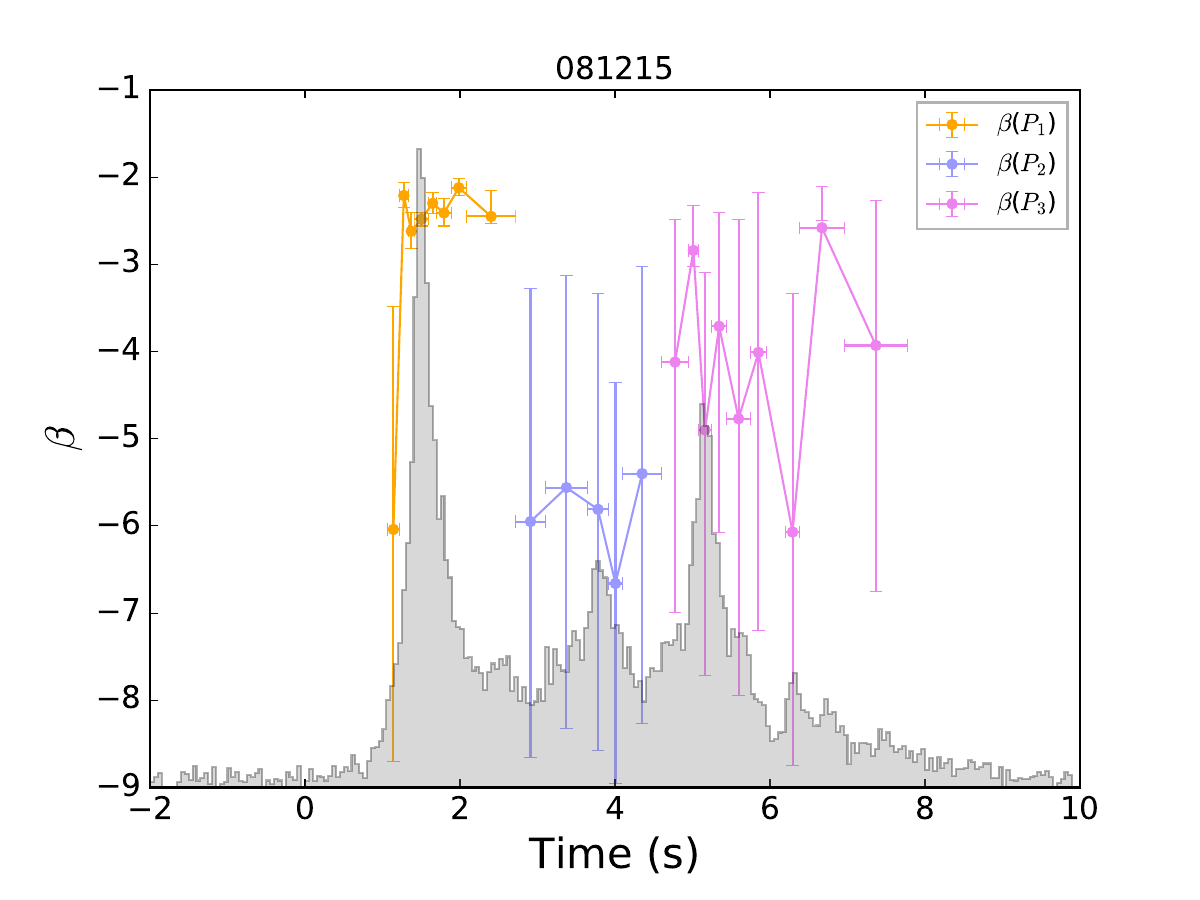}
\includegraphics[angle=0,scale=0.3]{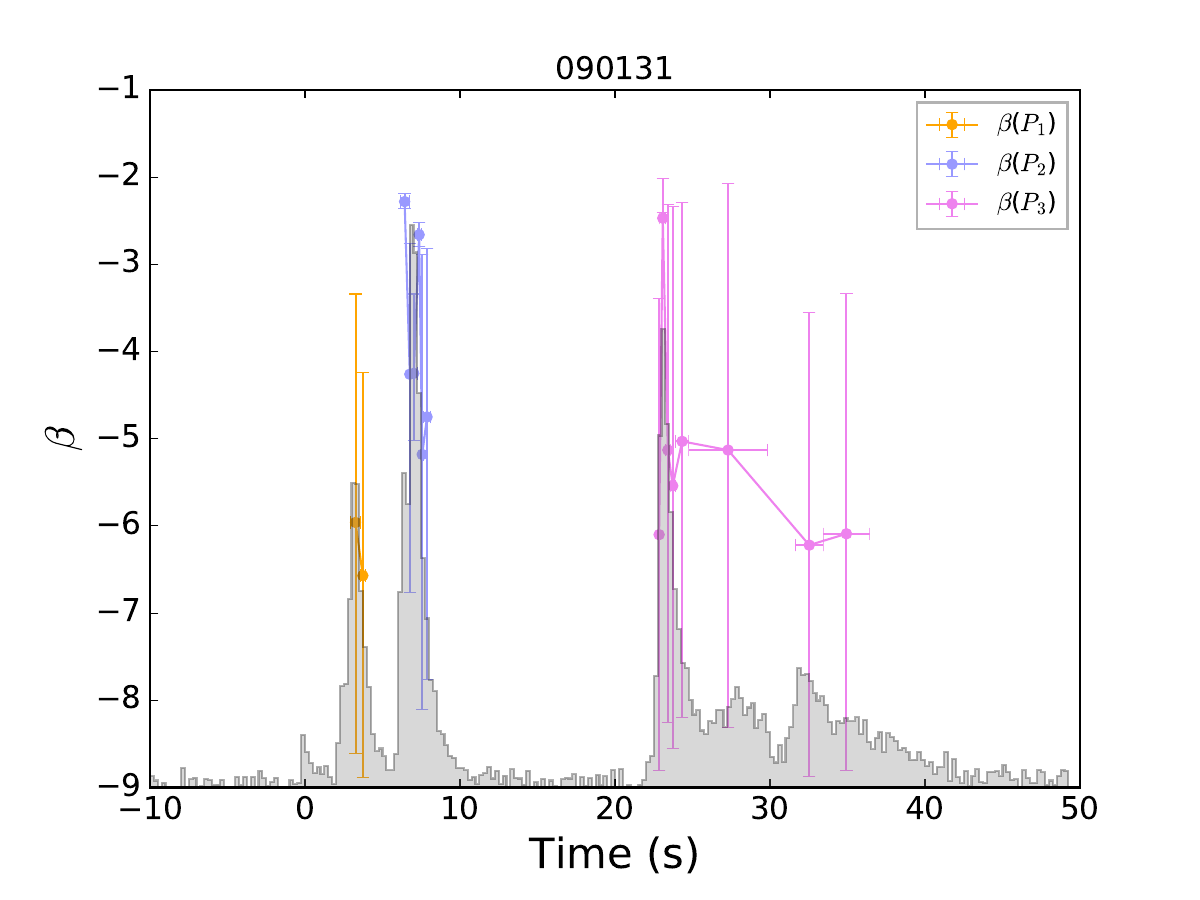}
\includegraphics[angle=0,scale=0.3]{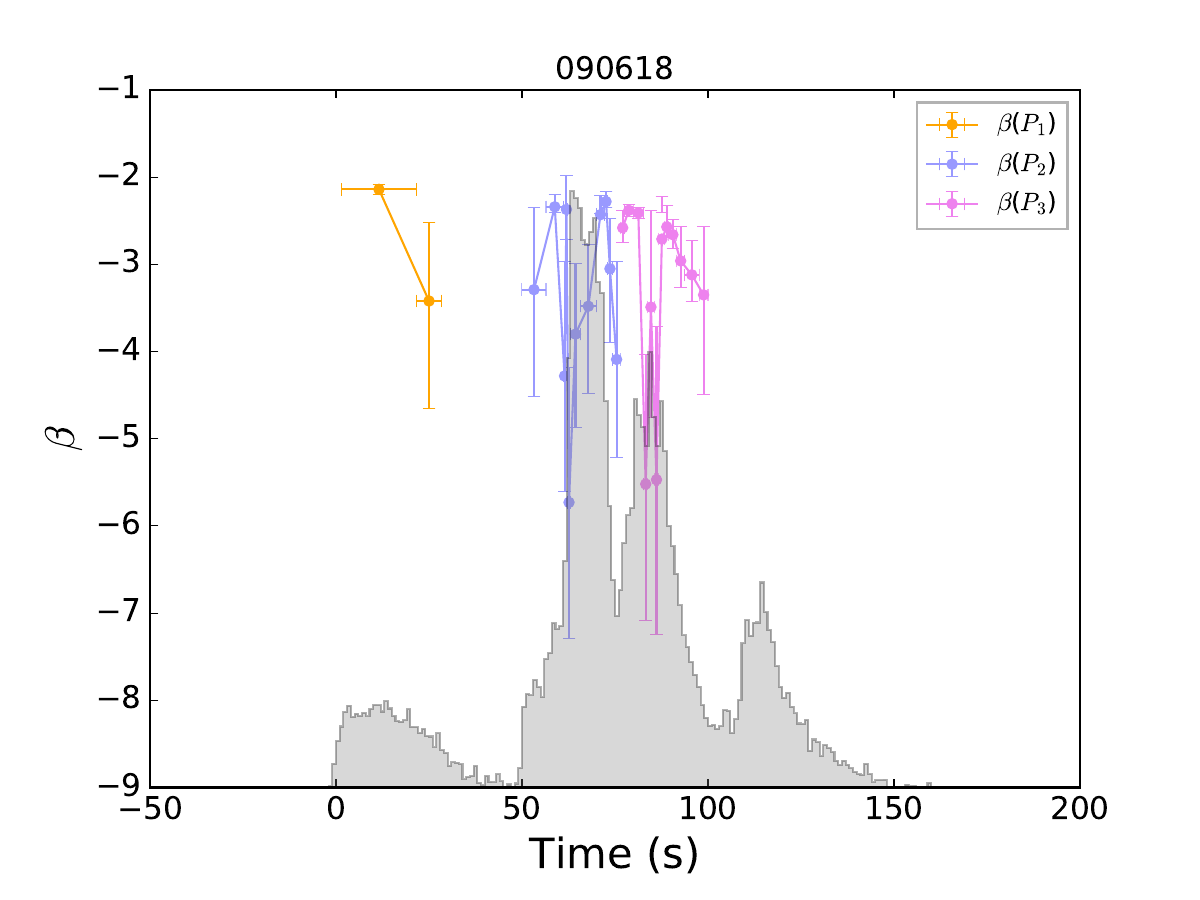}
\includegraphics[angle=0,scale=0.3]{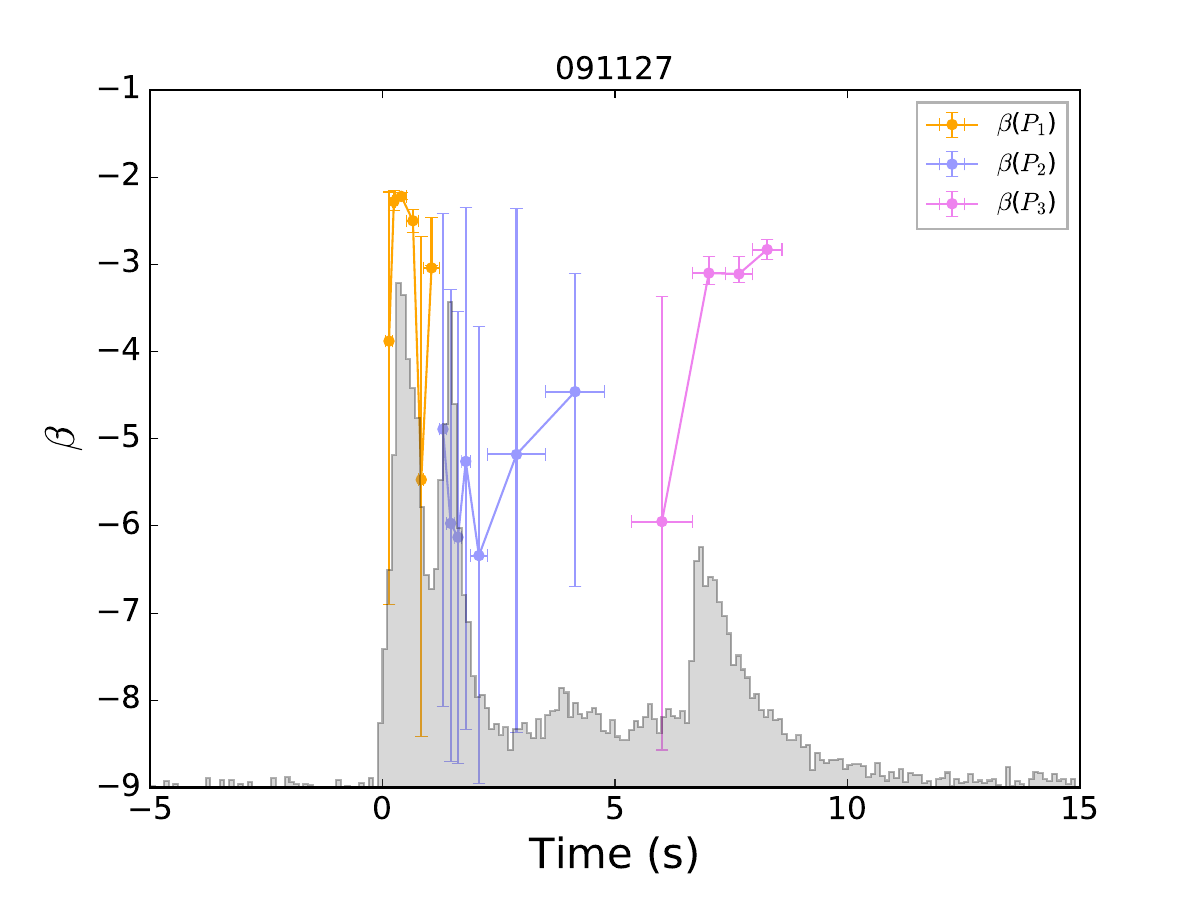}
\includegraphics[angle=0,scale=0.3]{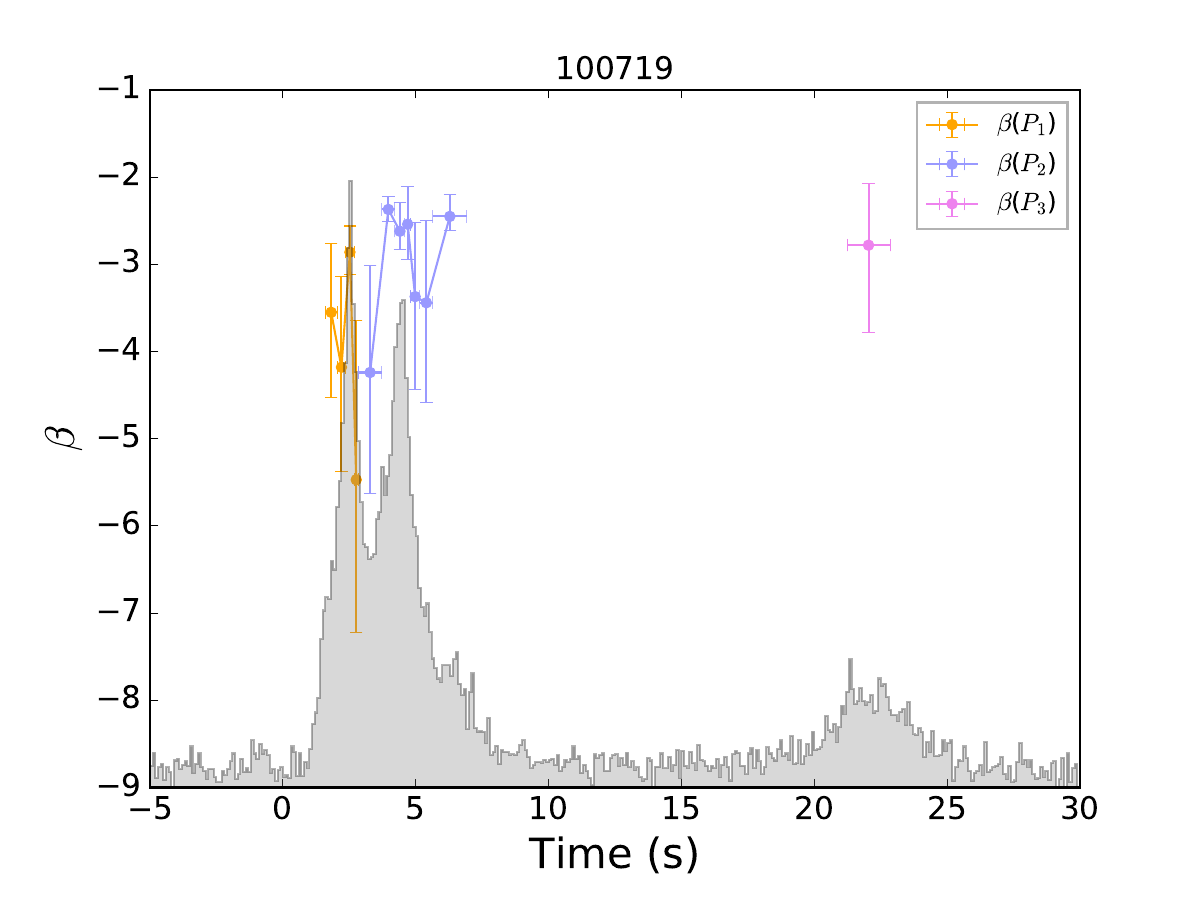}
\includegraphics[angle=0,scale=0.3]{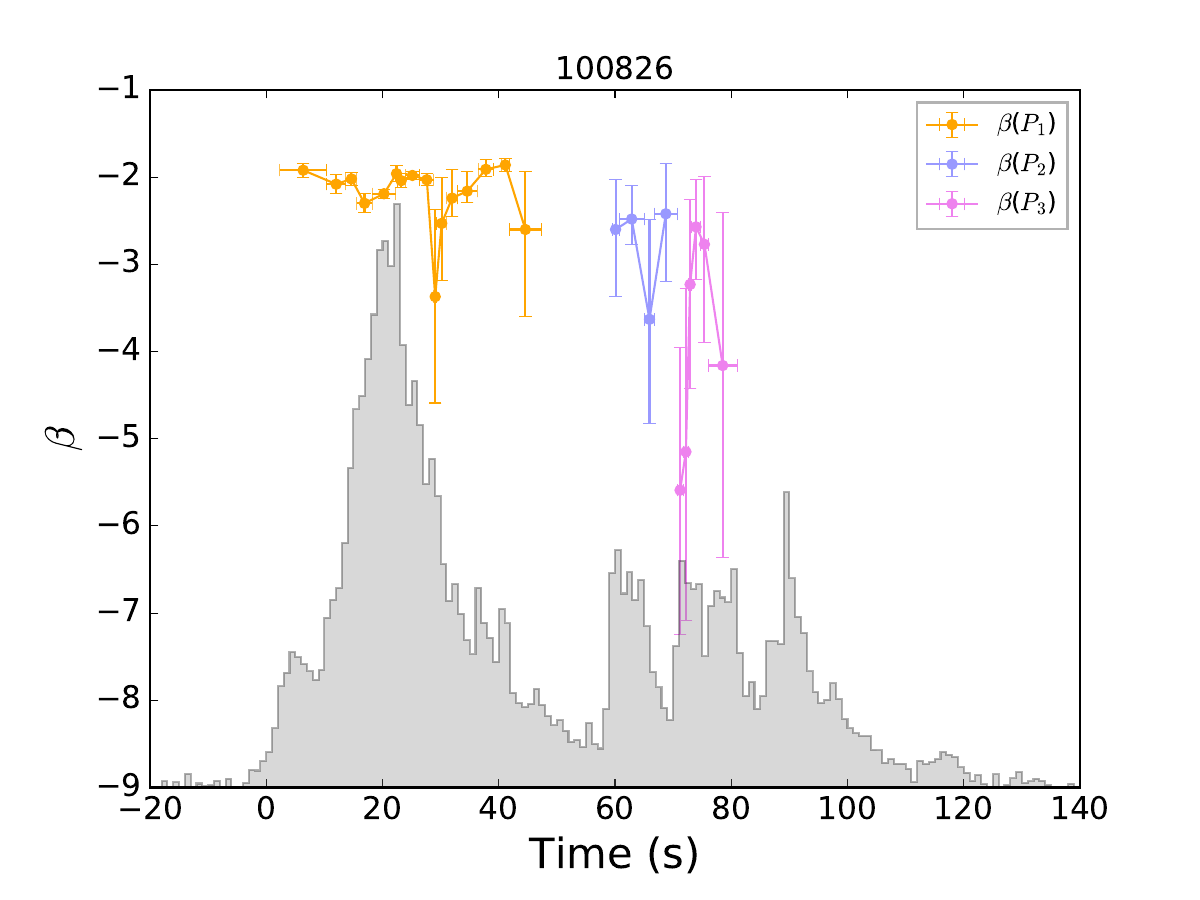}
\includegraphics[angle=0,scale=0.3]{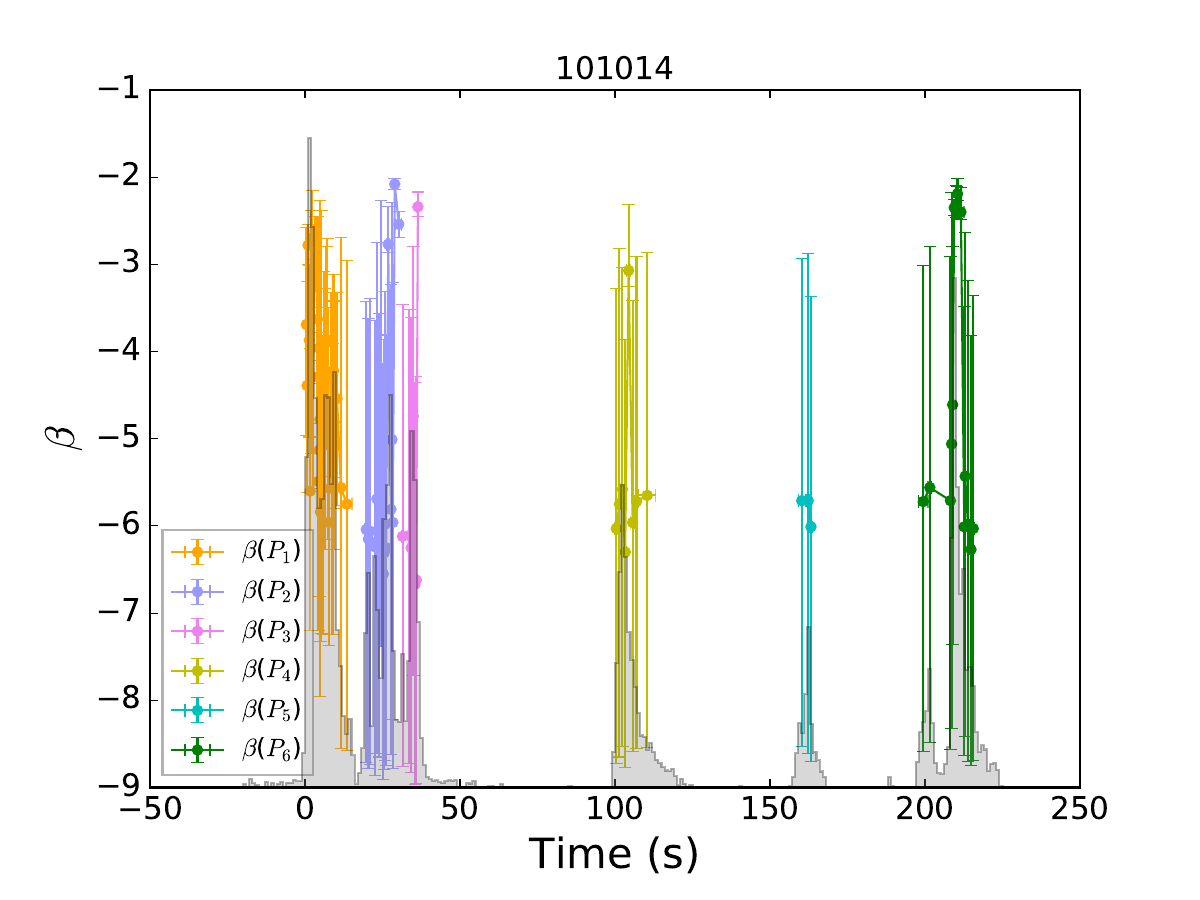}
\includegraphics[angle=0,scale=0.3]{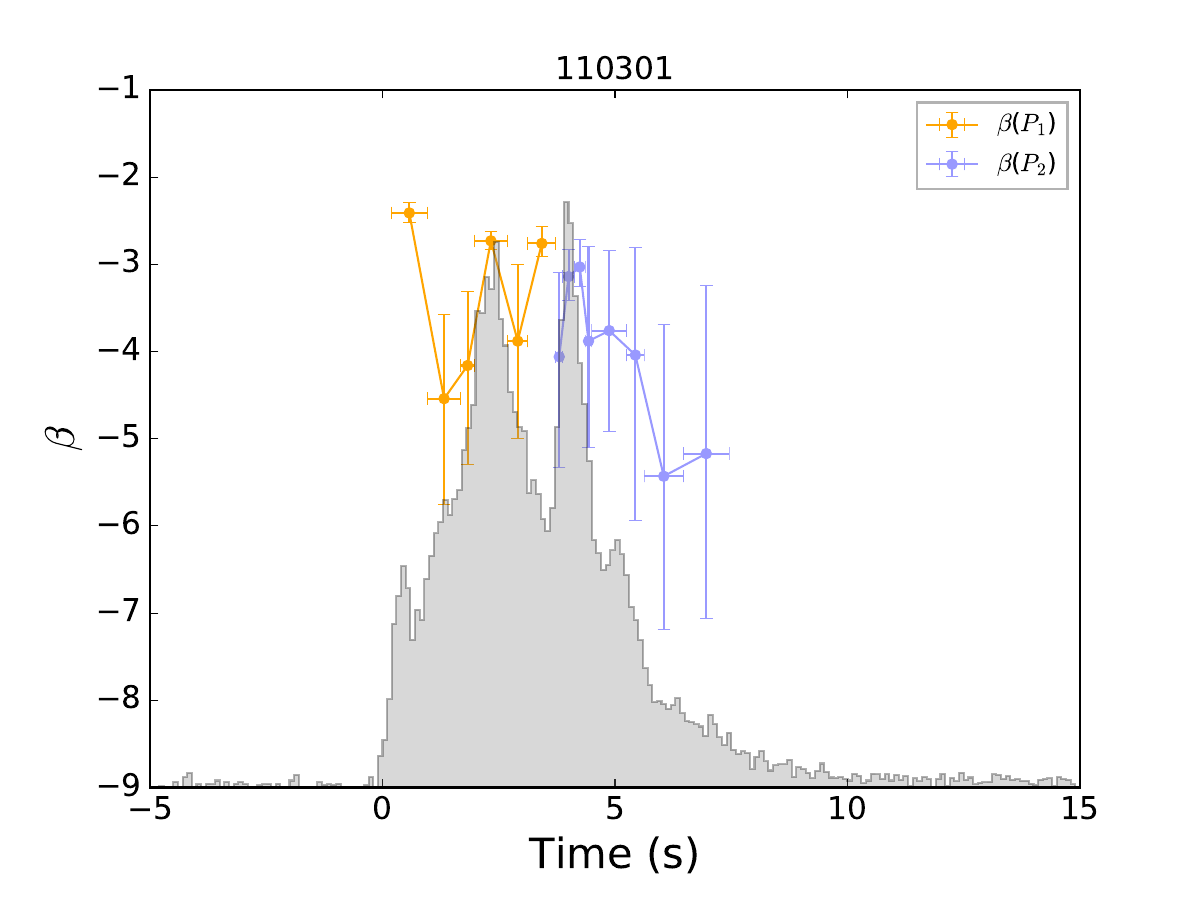}
\includegraphics[angle=0,scale=0.3]{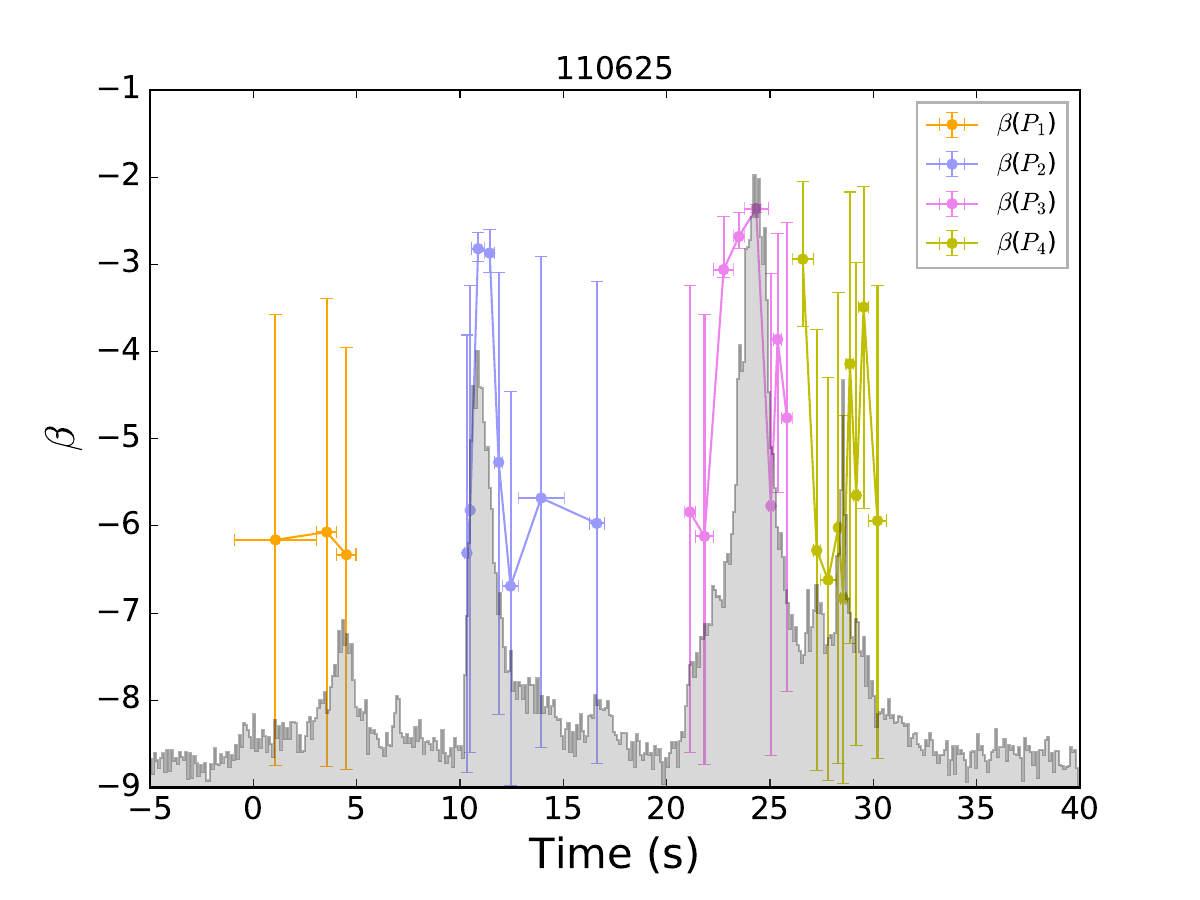}
\includegraphics[angle=0,scale=0.3]{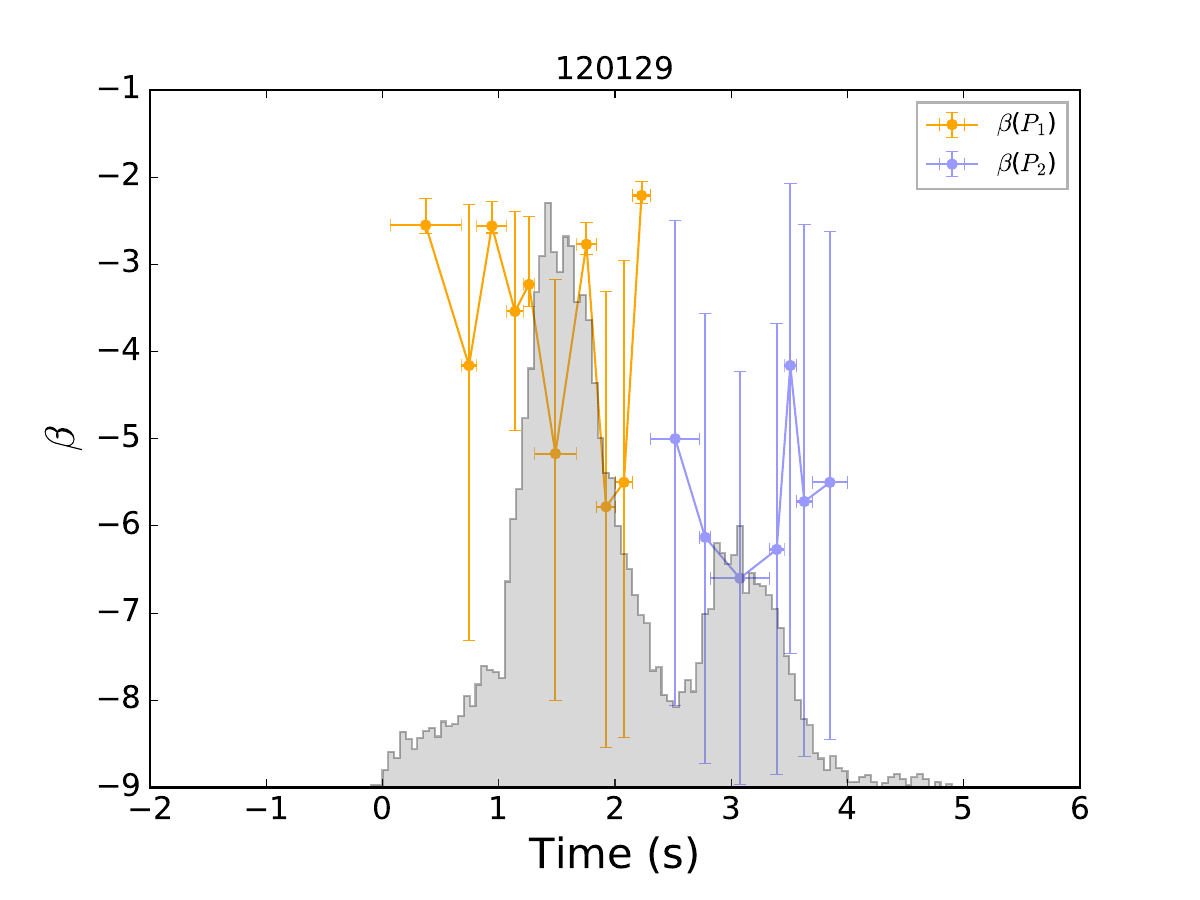}
\includegraphics[angle=0,scale=0.3]{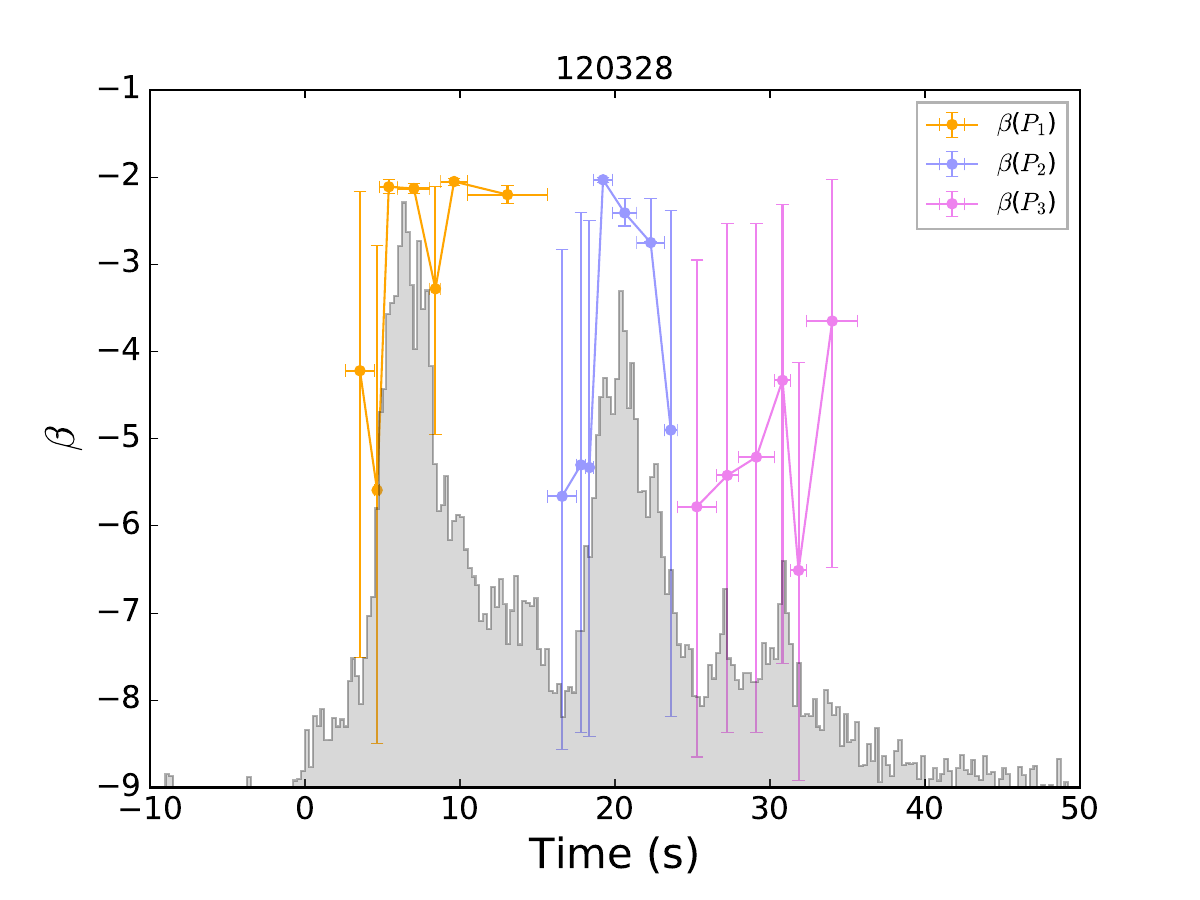}
\caption{Temporal evolution of the $\beta$ index of the Band model. The symbols and colors are the same as in Figure \ref{fig:Alpha_Best}.}\label{fig:Beta}
\end{figure*}
\begin{figure*}
\includegraphics[angle=0,scale=0.3]{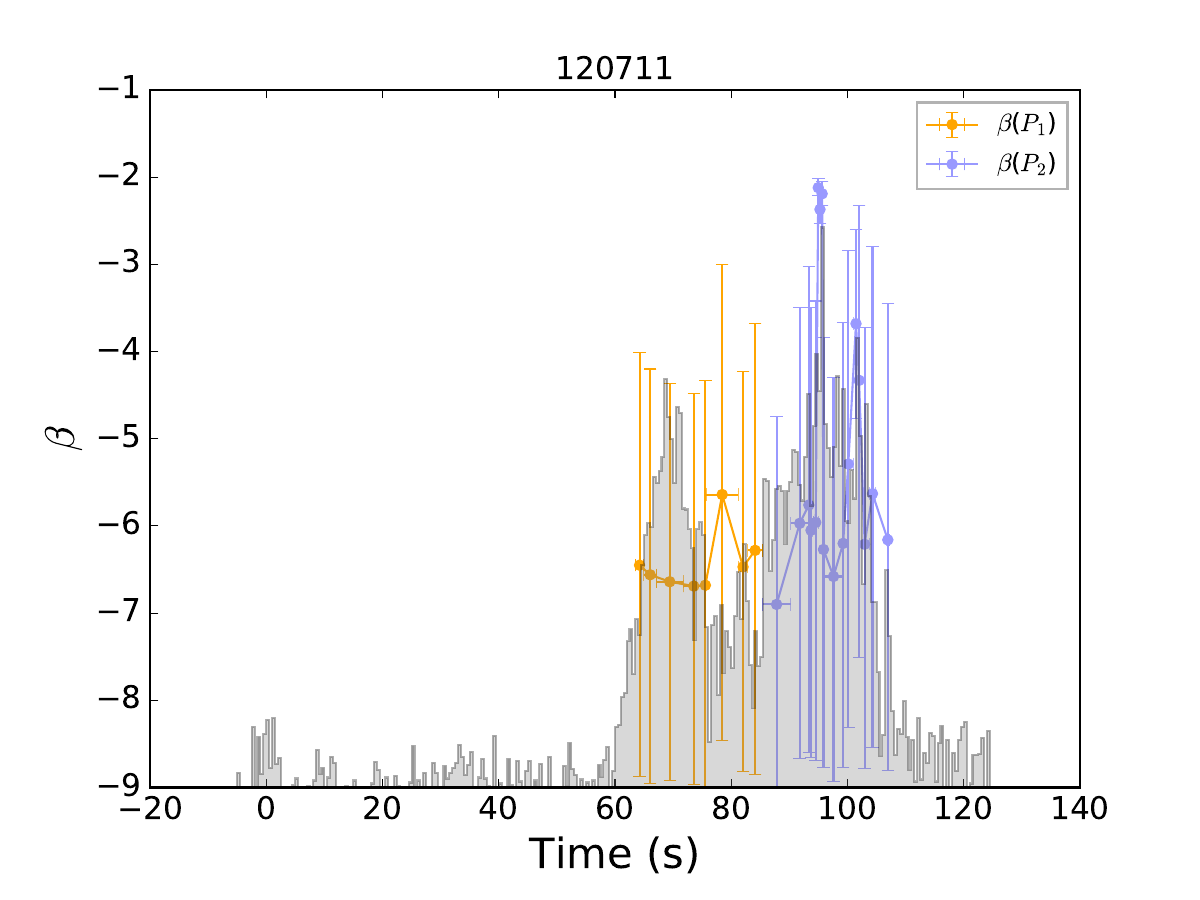}
\includegraphics[angle=0,scale=0.3]{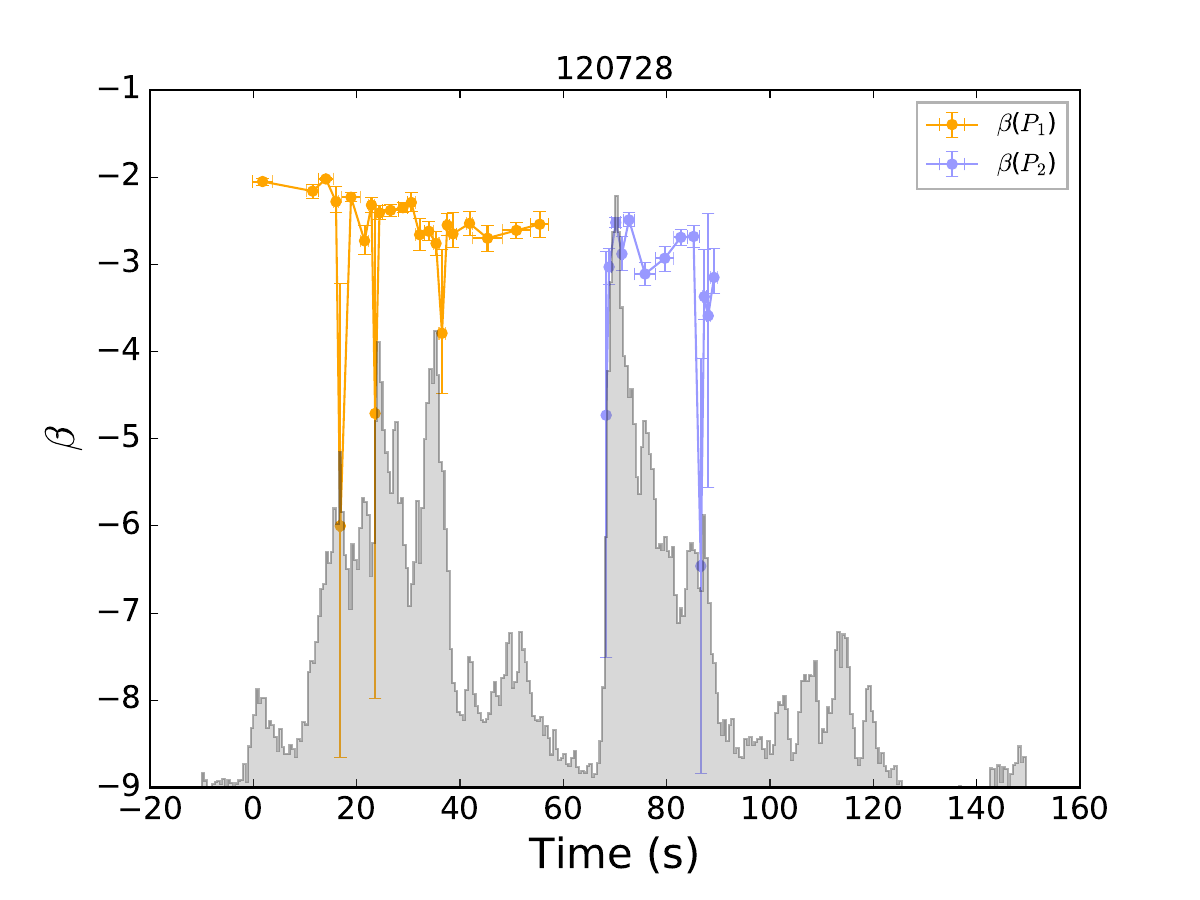}
\includegraphics[angle=0,scale=0.3]{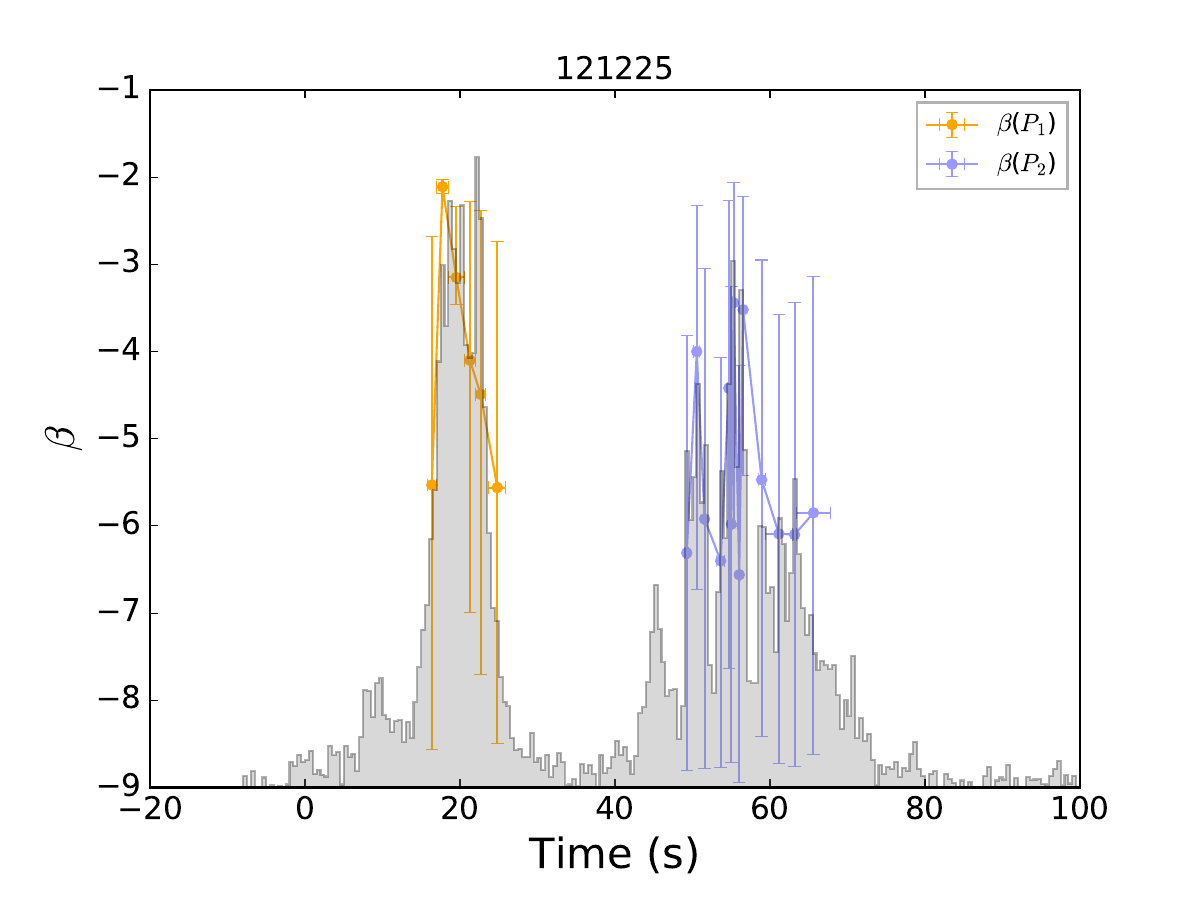}
\includegraphics[angle=0,scale=0.3]{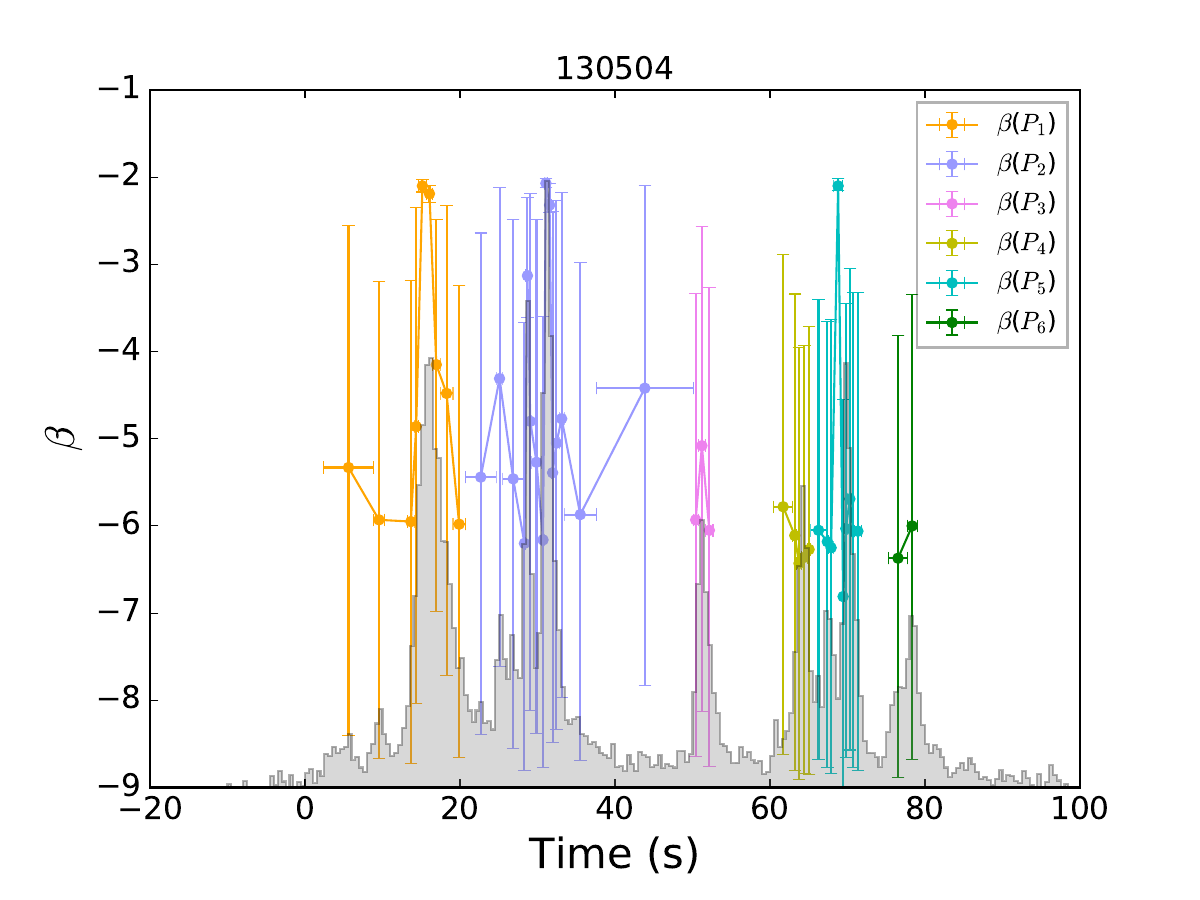}
\includegraphics[angle=0,scale=0.3]{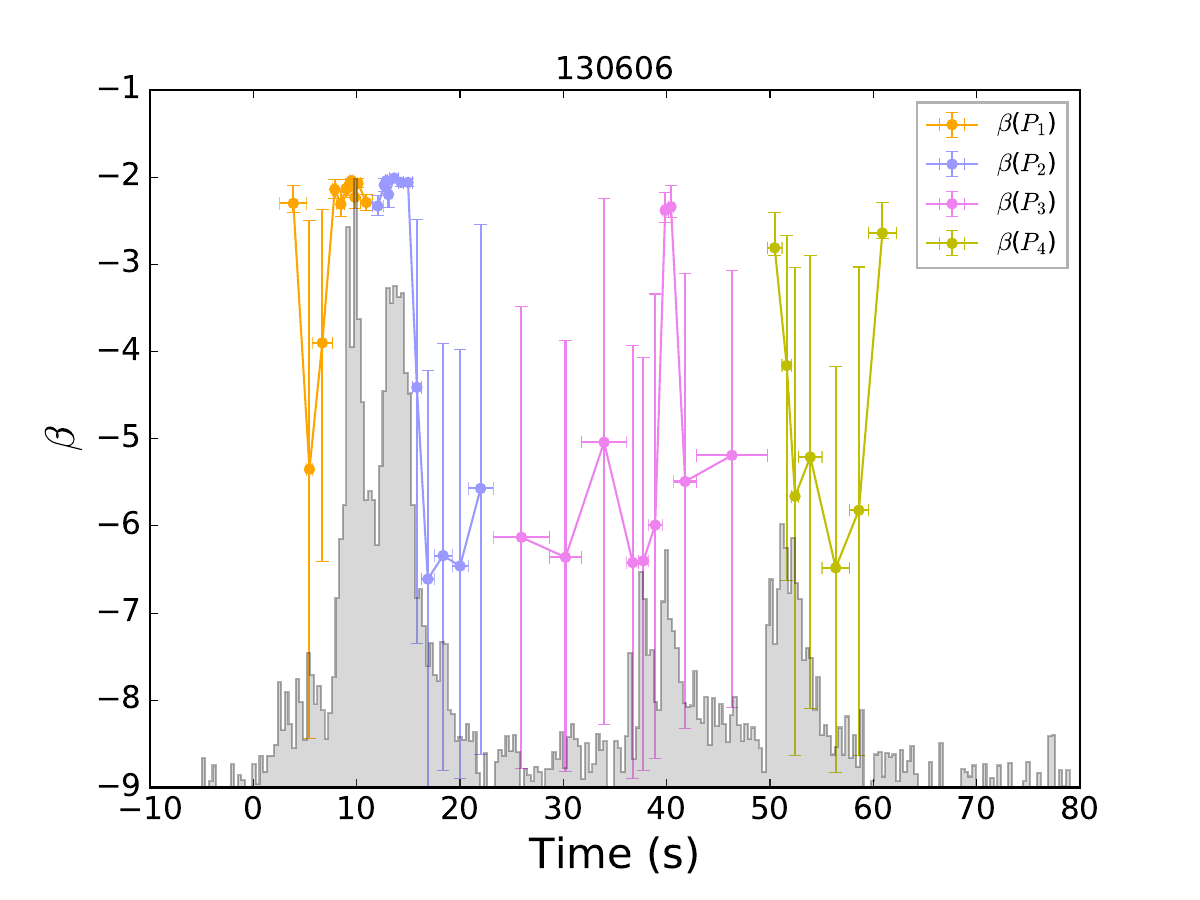}
\includegraphics[angle=0,scale=0.3]{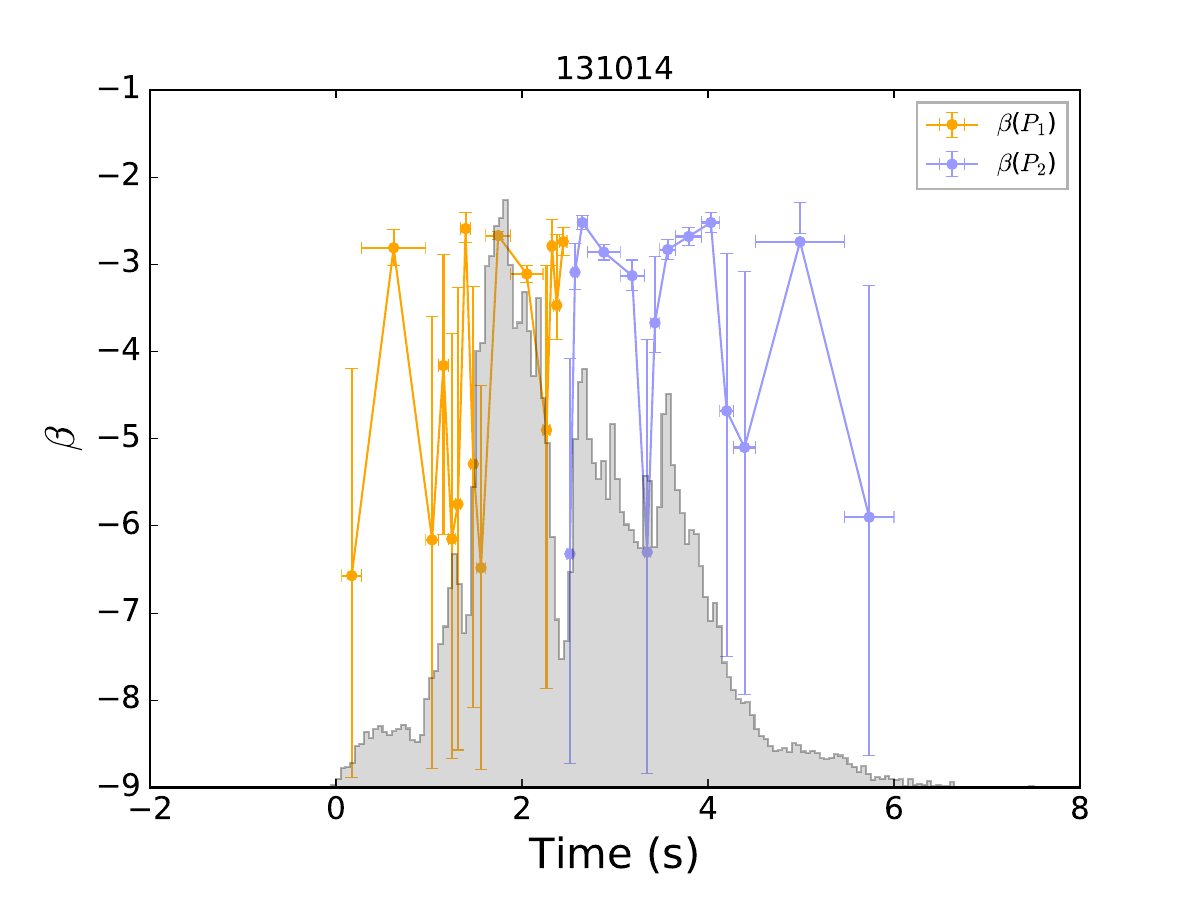}
\includegraphics[angle=0,scale=0.3]{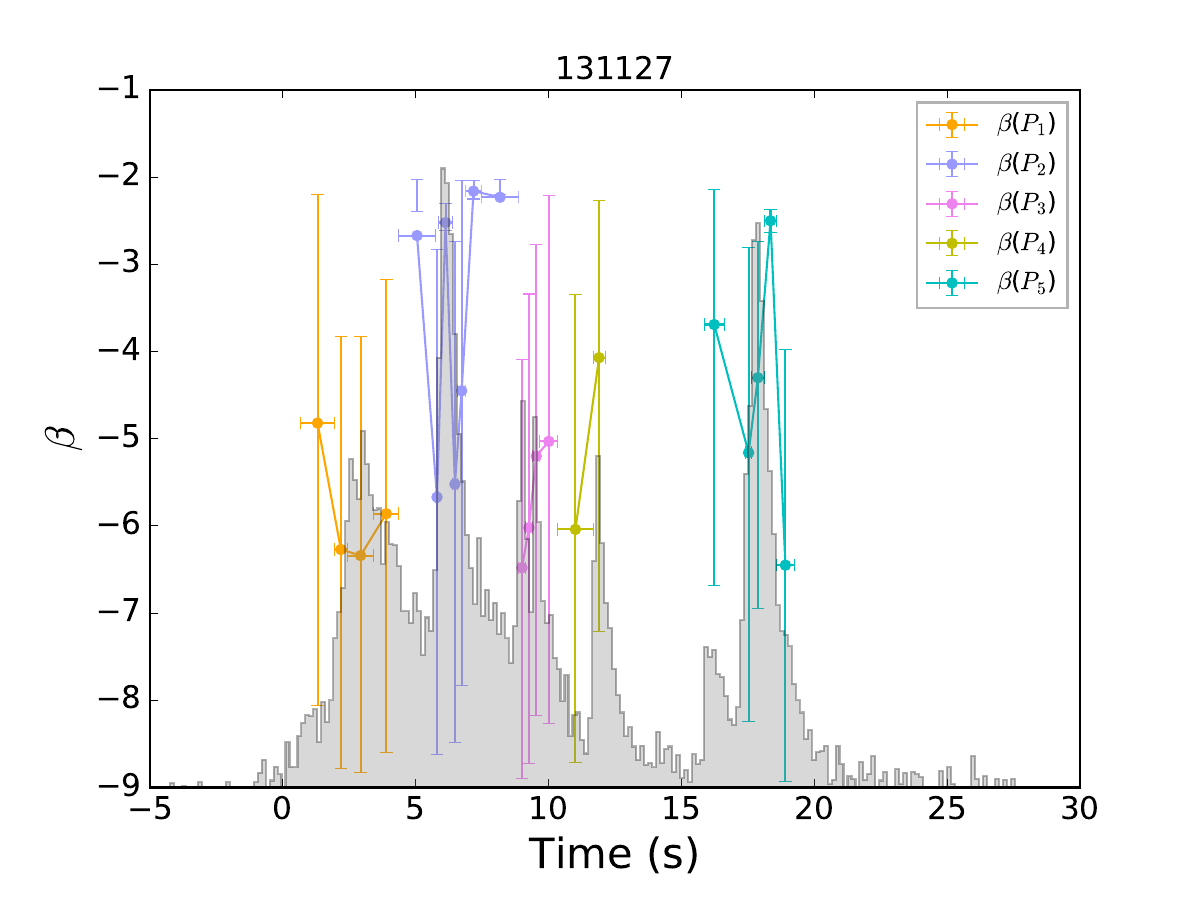}
\includegraphics[angle=0,scale=0.3]{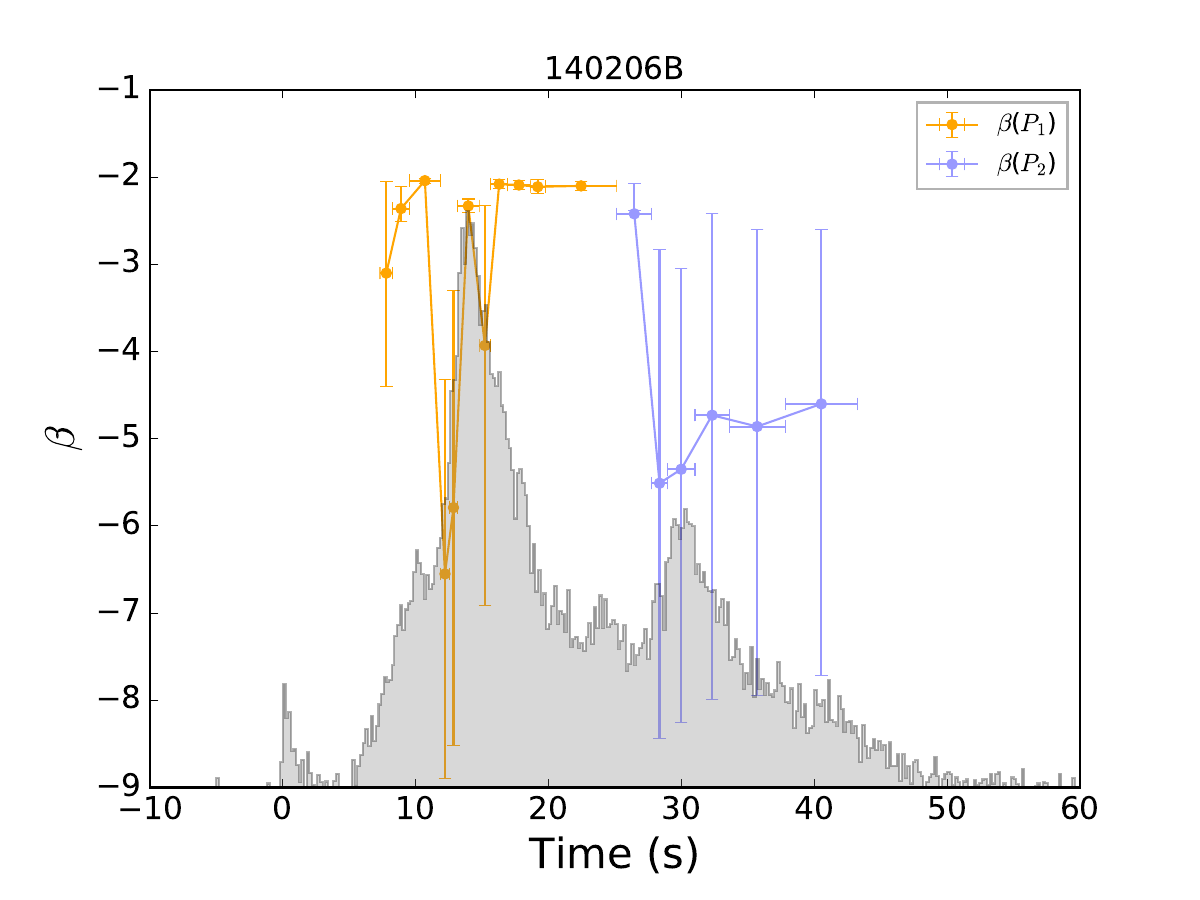}
\includegraphics[angle=0,scale=0.3]{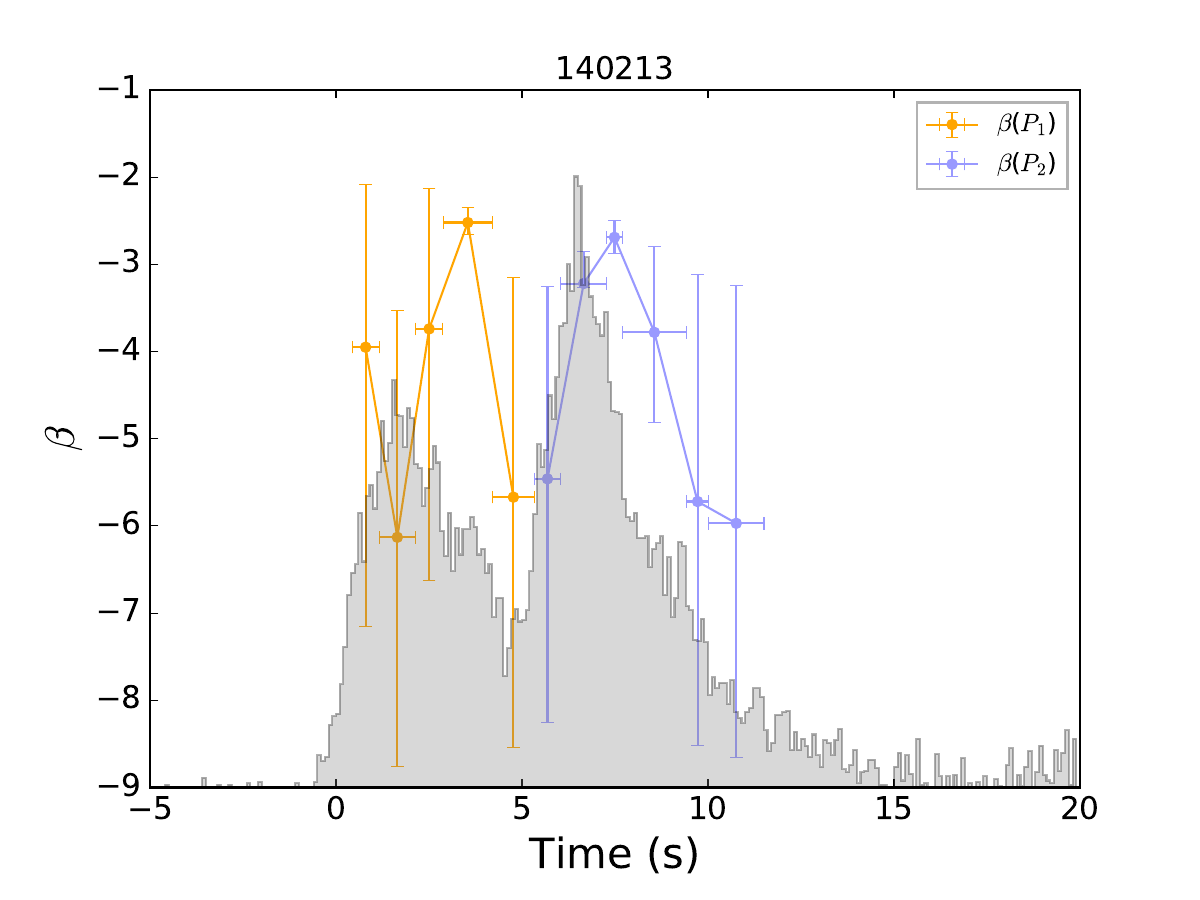}
\includegraphics[angle=0,scale=0.3]{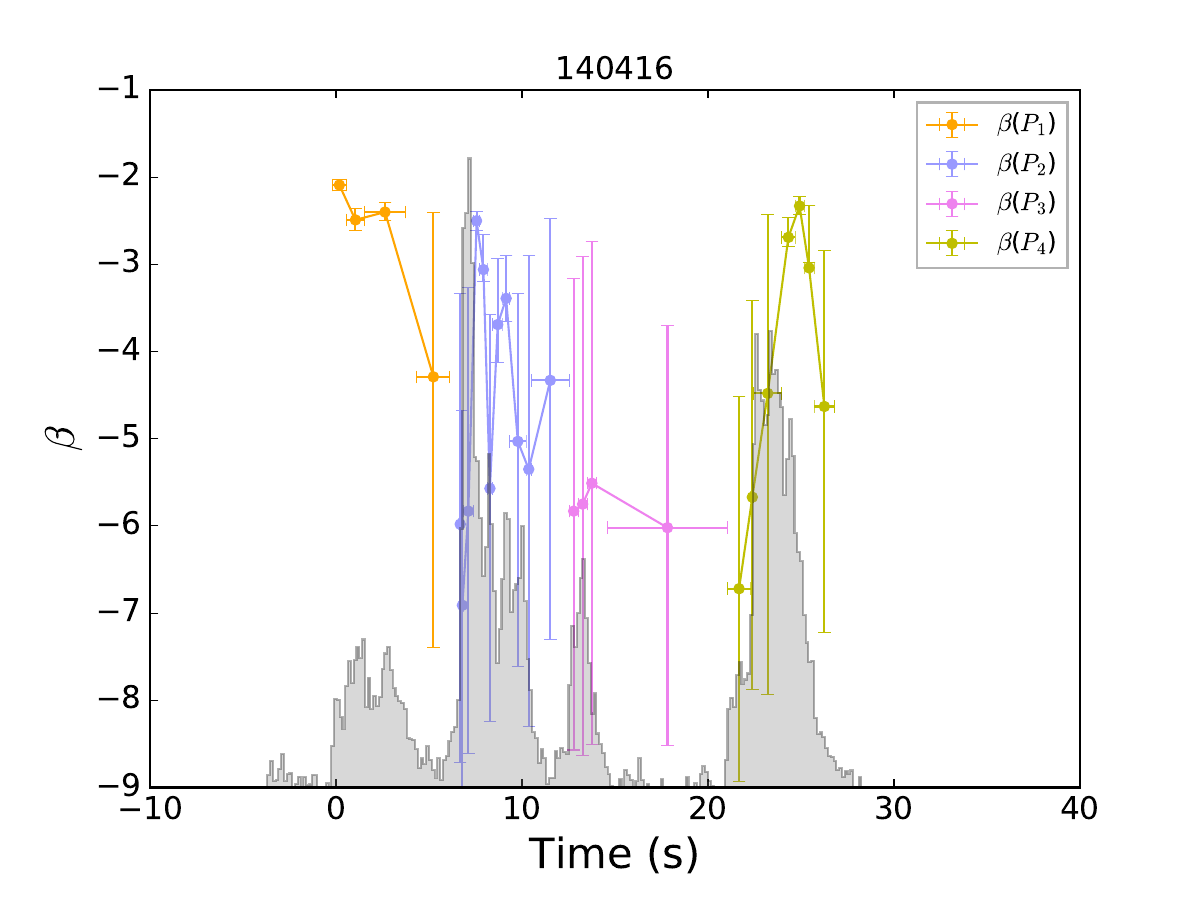}
\includegraphics[angle=0,scale=0.3]{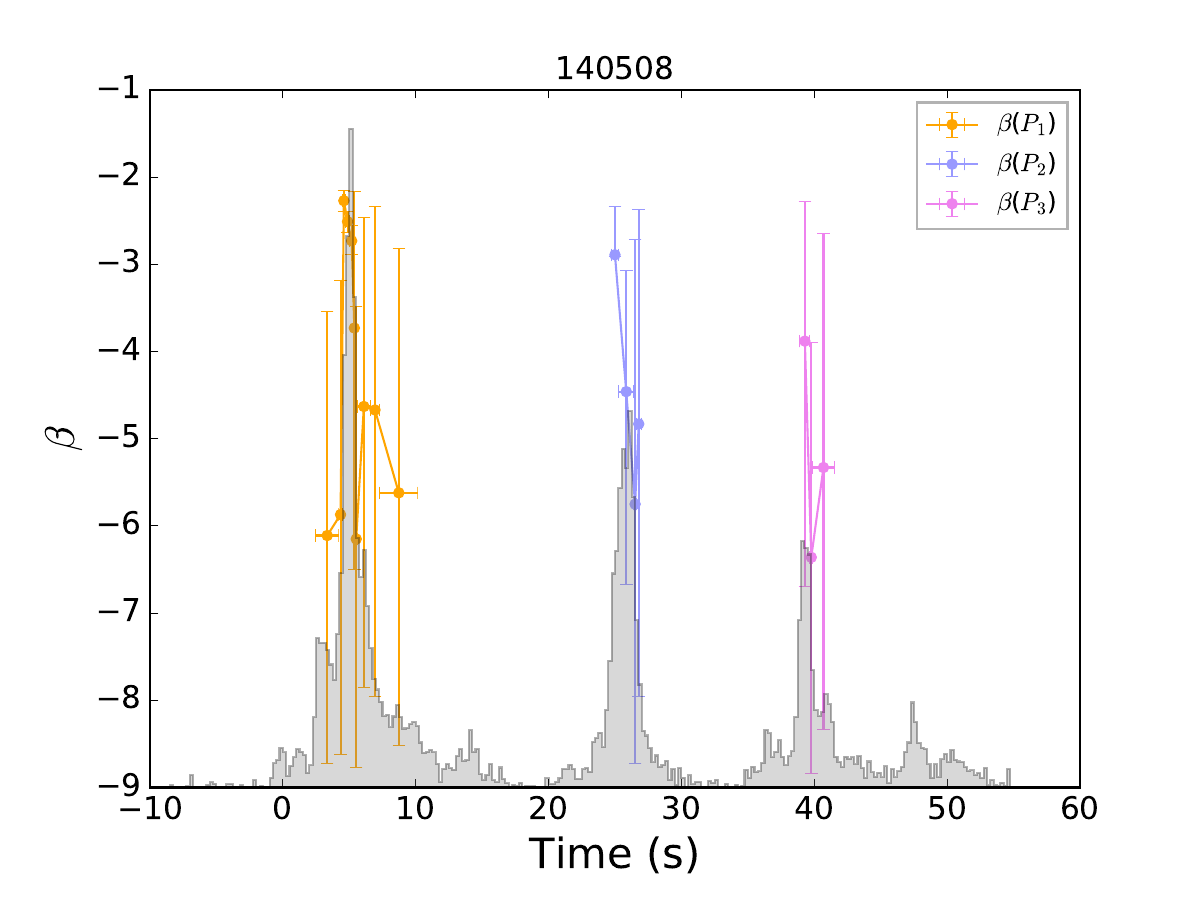}
\includegraphics[angle=0,scale=0.3]{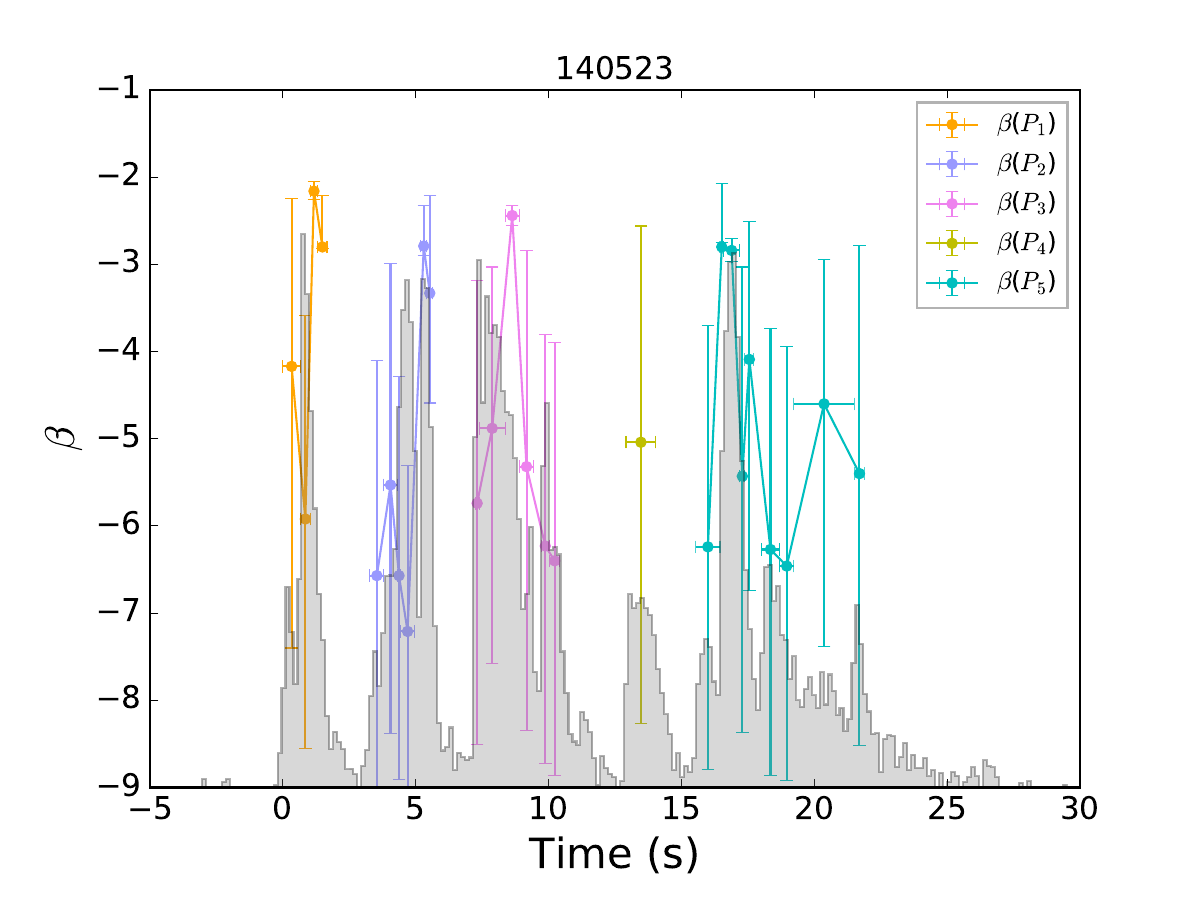}
\includegraphics[angle=0,scale=0.3]{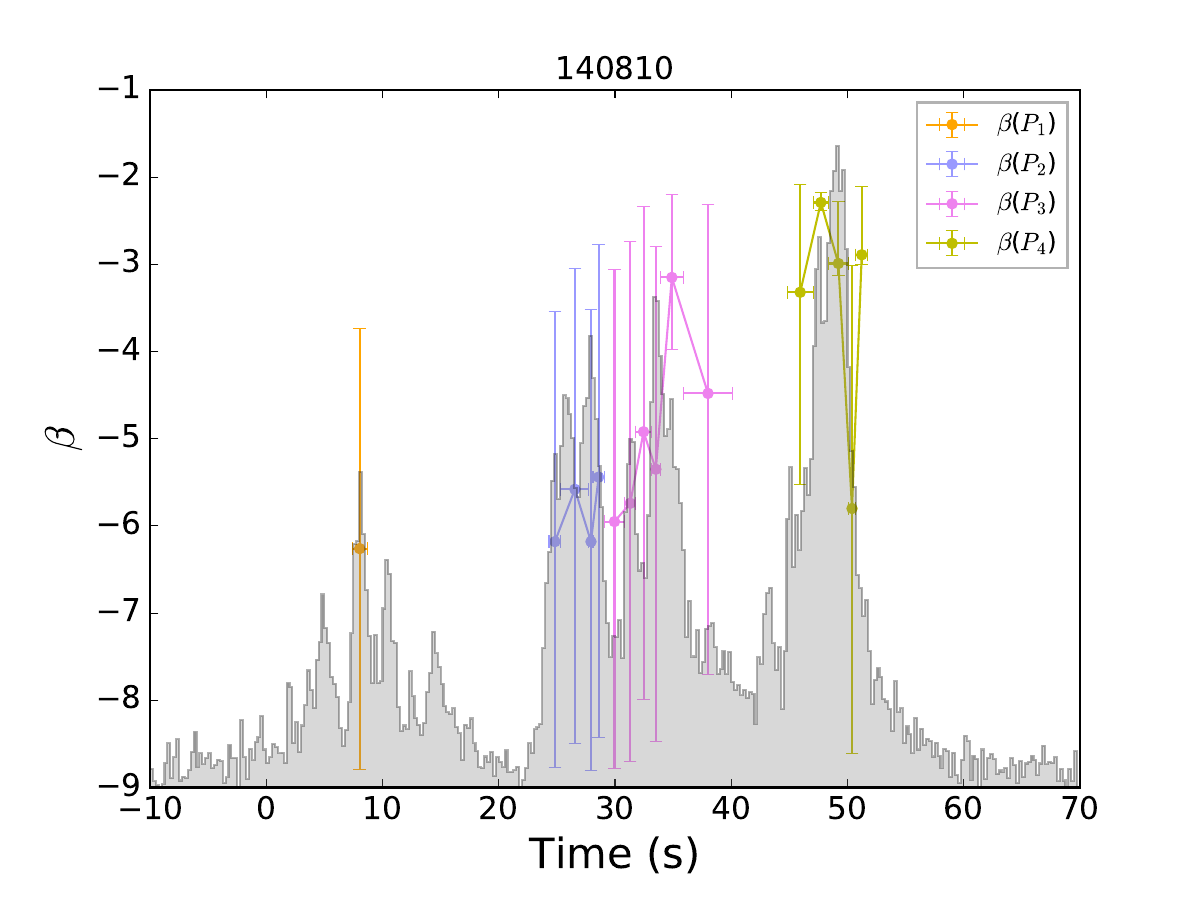}
\includegraphics[angle=0,scale=0.3]{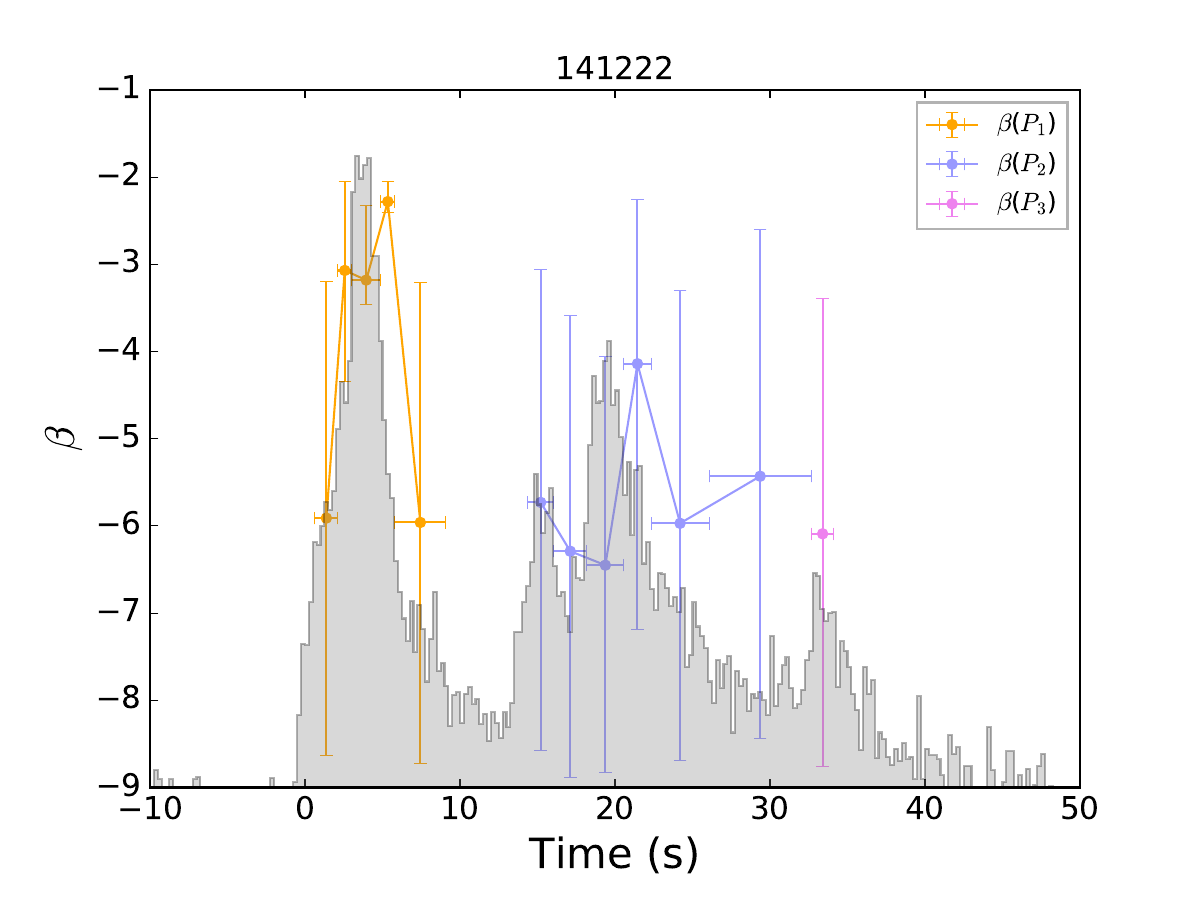}
\includegraphics[angle=0,scale=0.3]{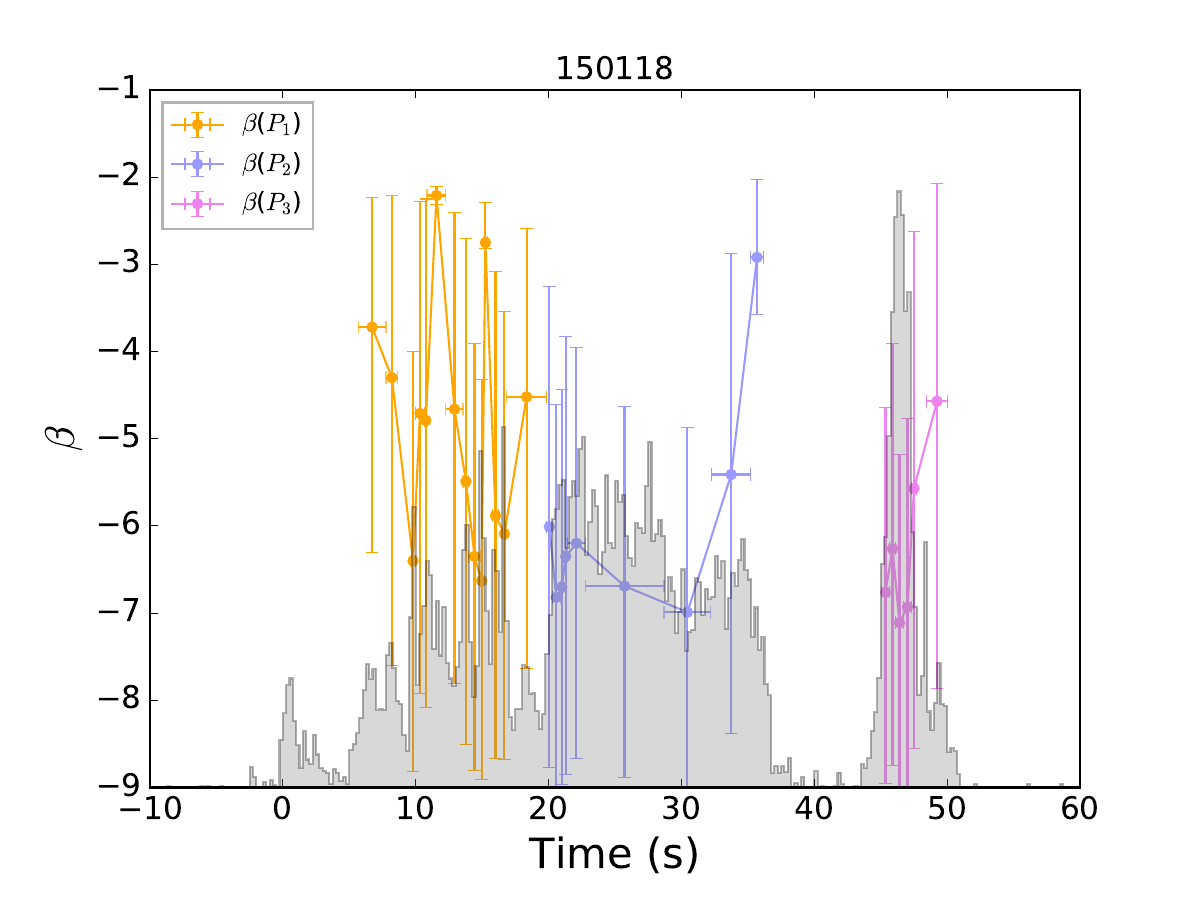}
\center{Fig. \ref{fig:Beta}--- Continued}
\end{figure*}
\begin{figure*}
\includegraphics[angle=0,scale=0.3]{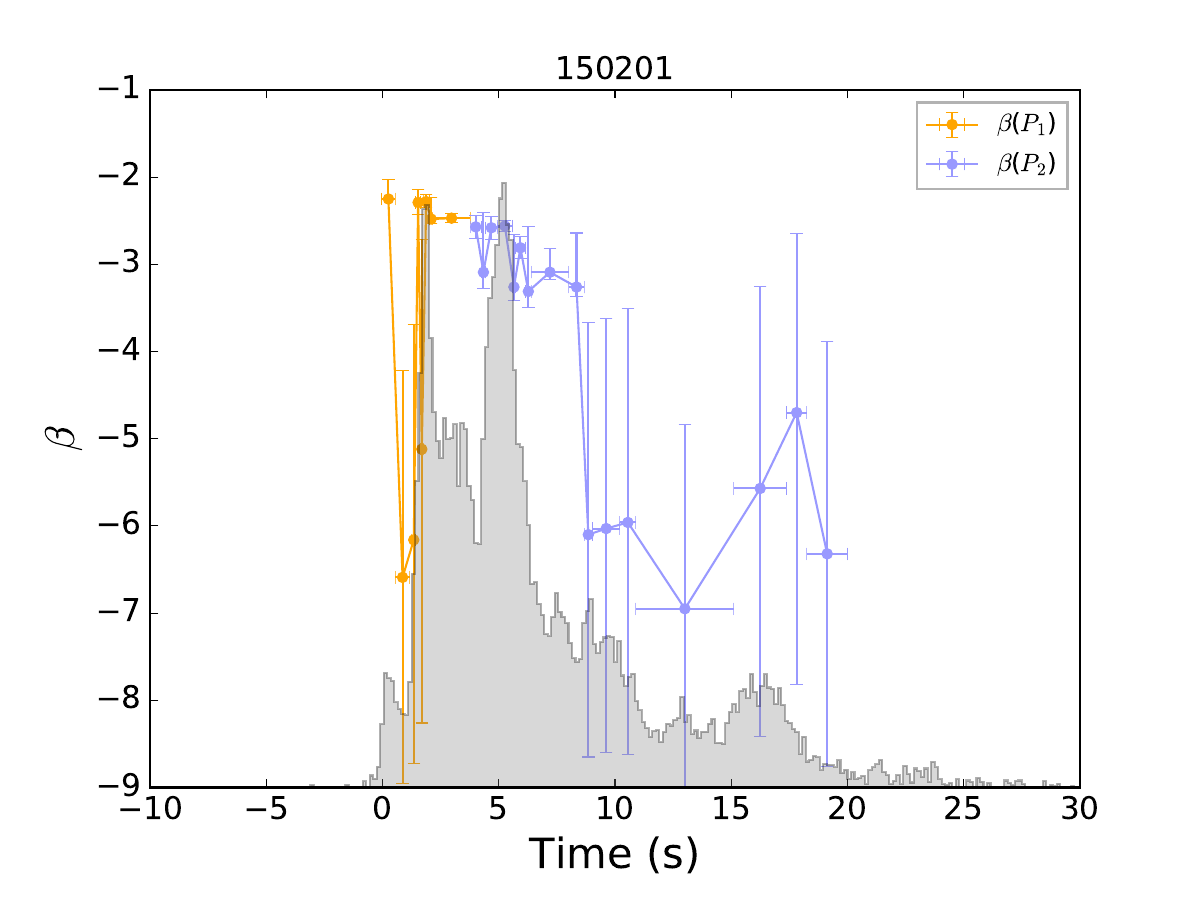}
\includegraphics[angle=0,scale=0.3]{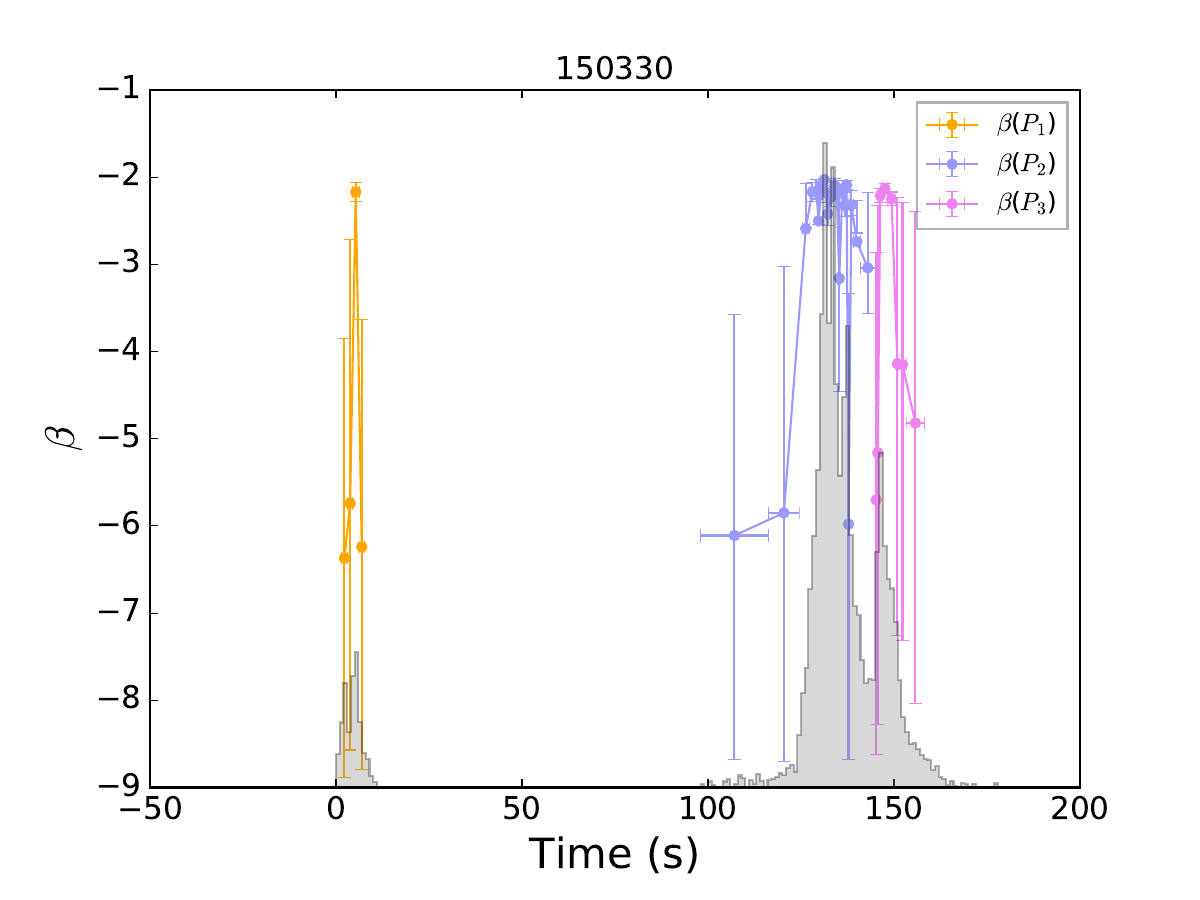}
\includegraphics[angle=0,scale=0.3]{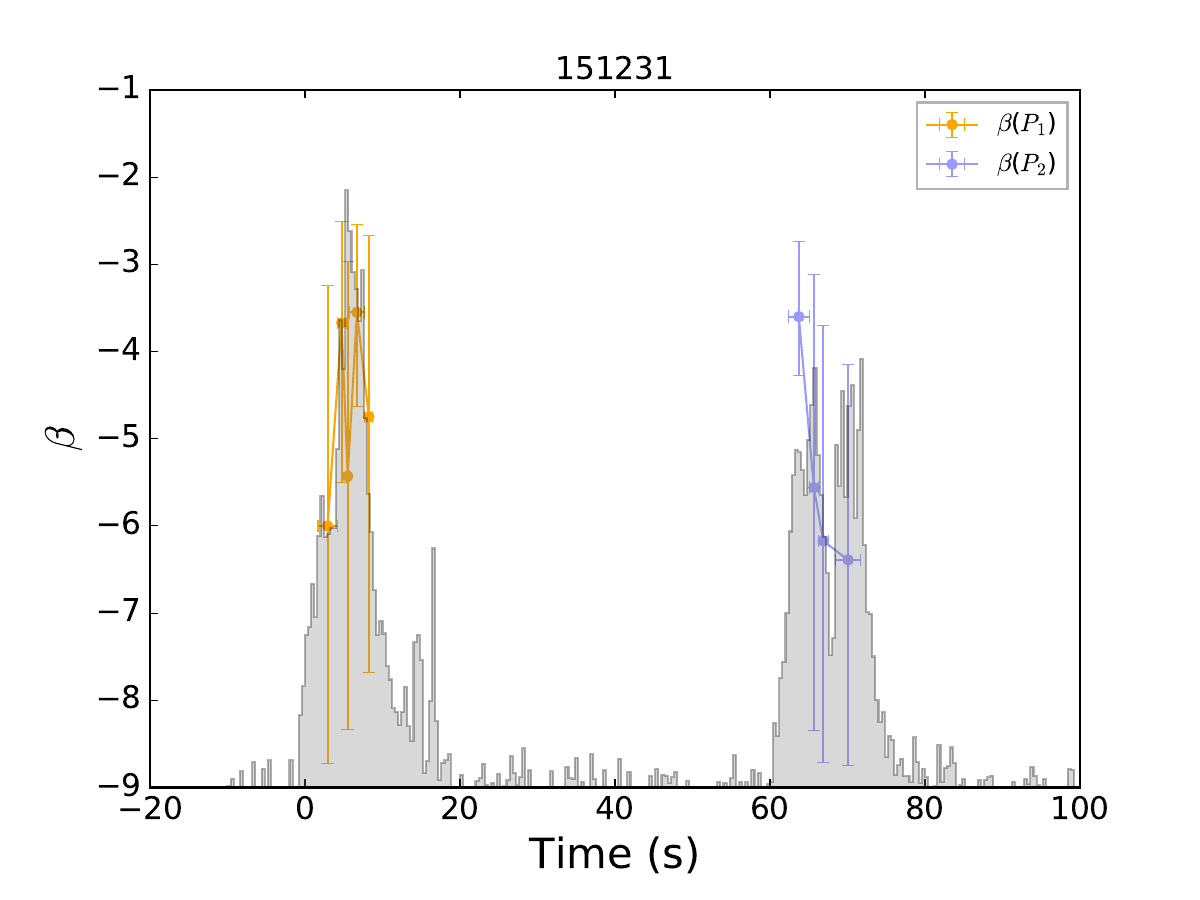}
\includegraphics[angle=0,scale=0.3]{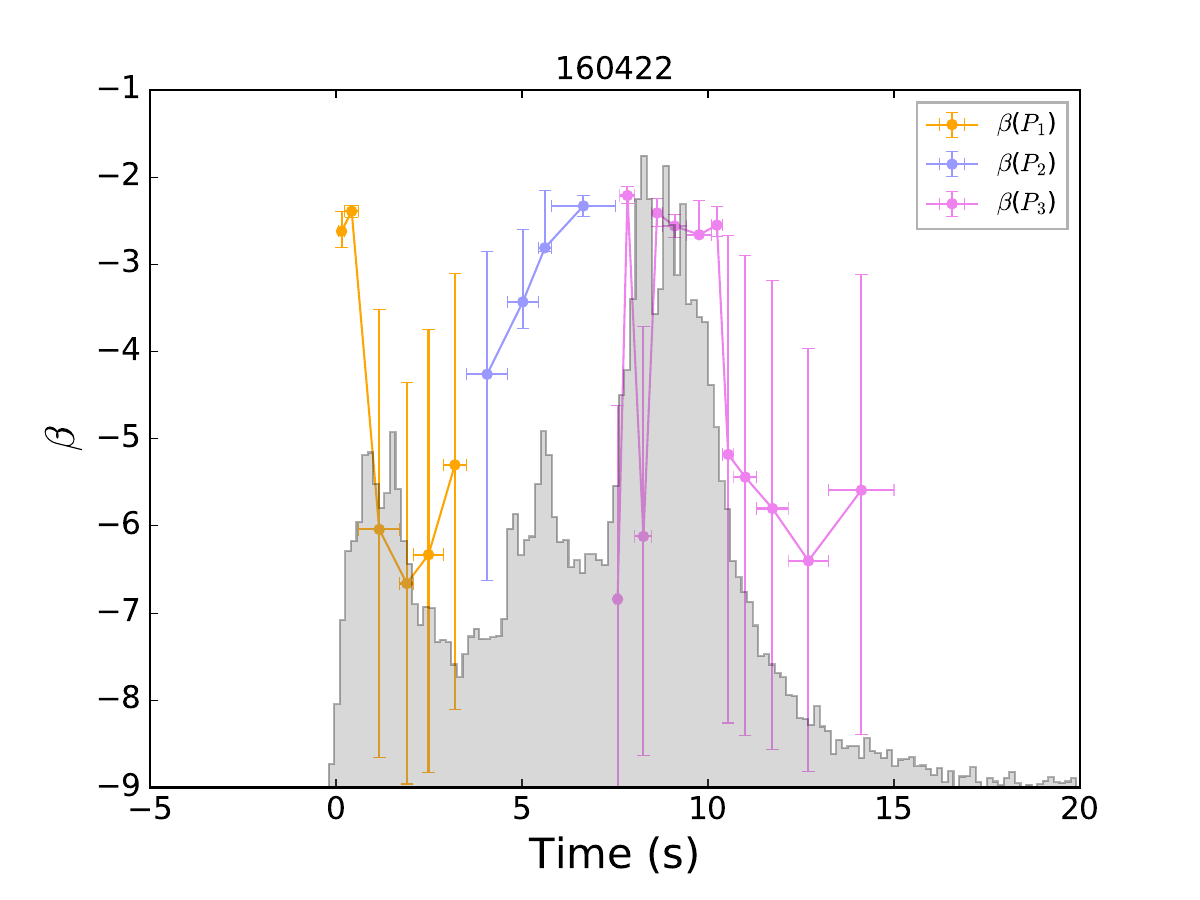}
\includegraphics[angle=0,scale=0.3]{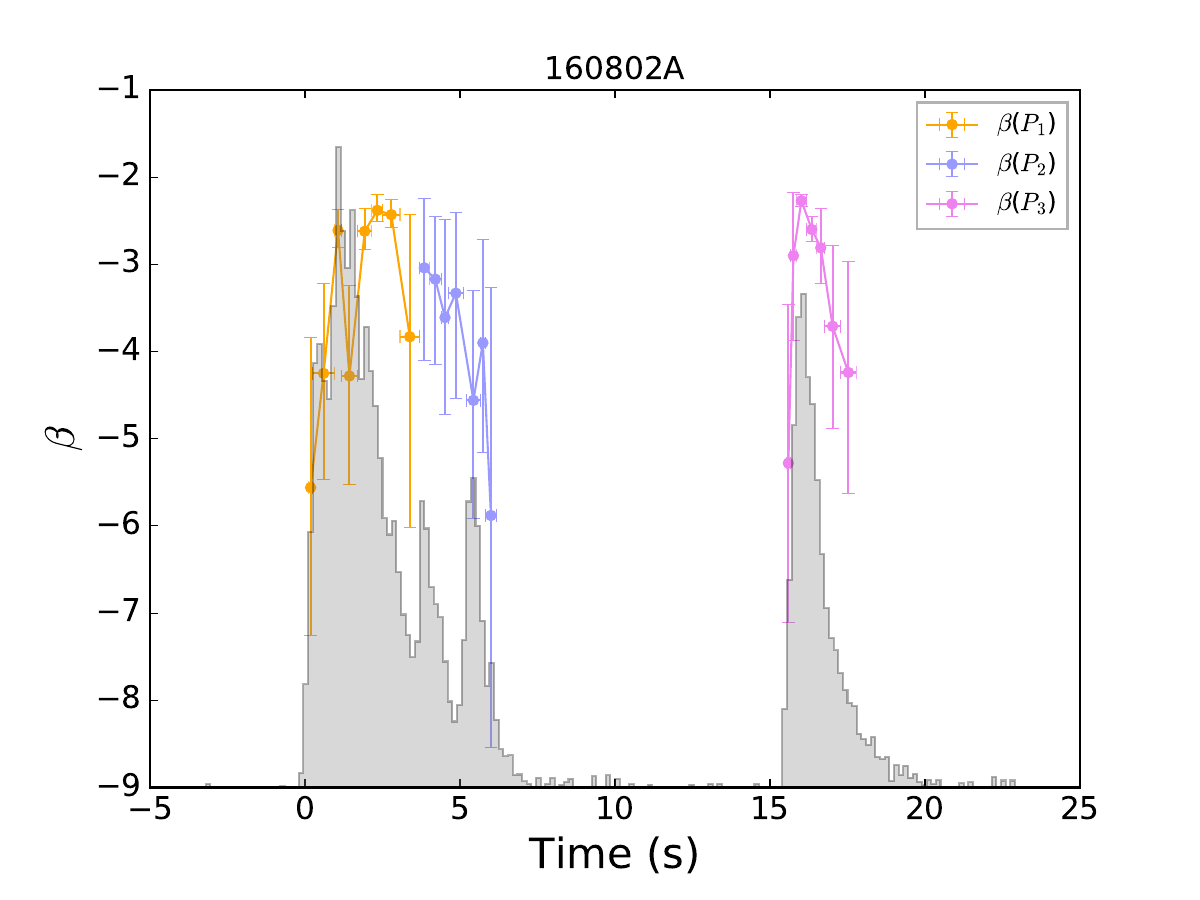}
\includegraphics[angle=0,scale=0.3]{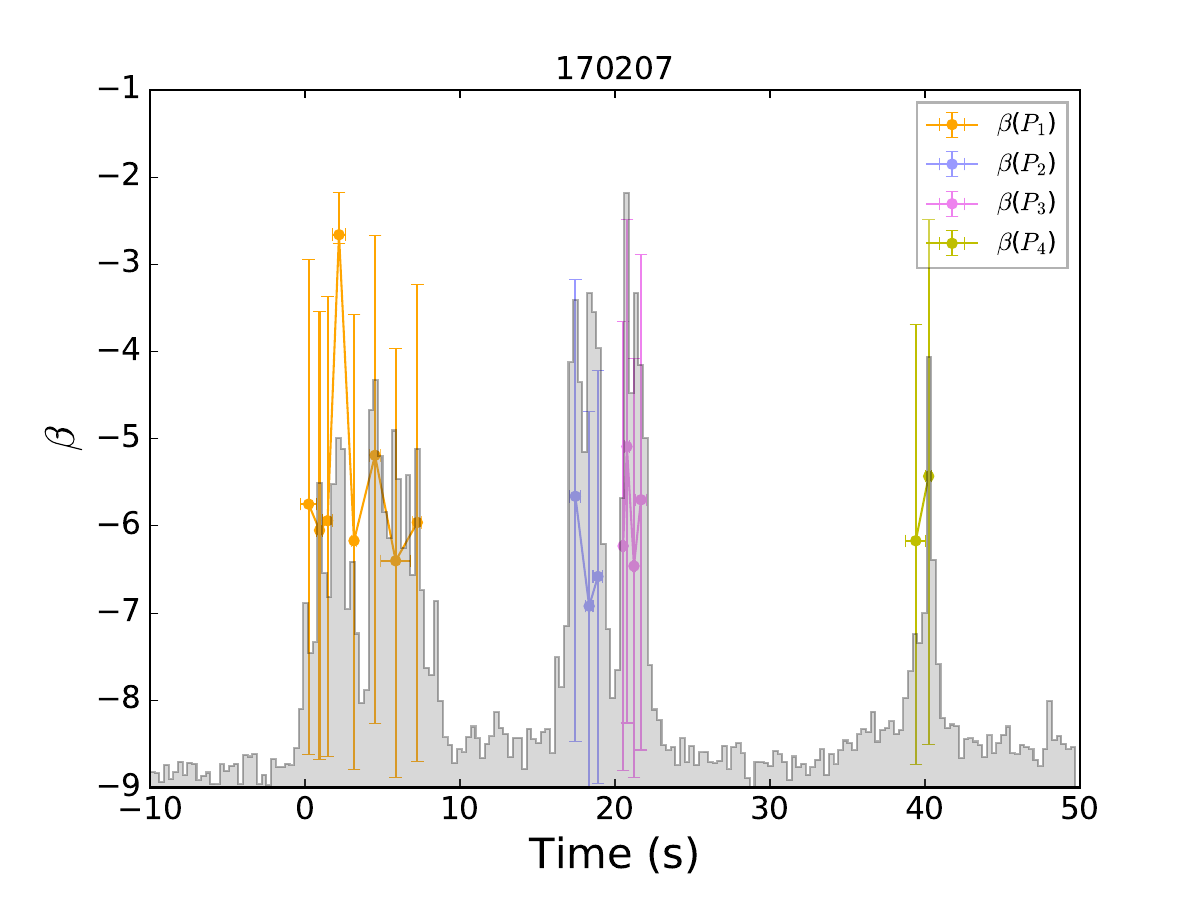}
\includegraphics[angle=0,scale=0.3]{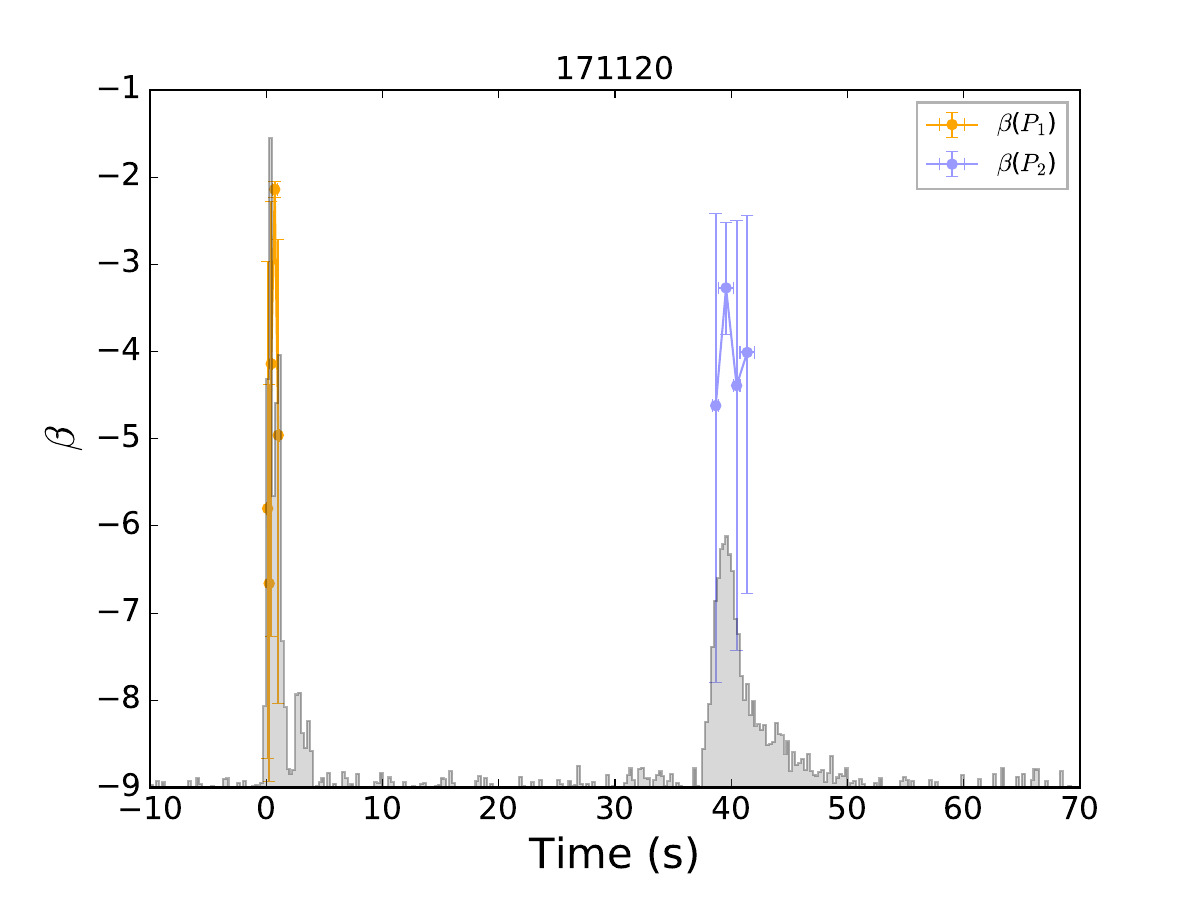}
\includegraphics[angle=0,scale=0.3]{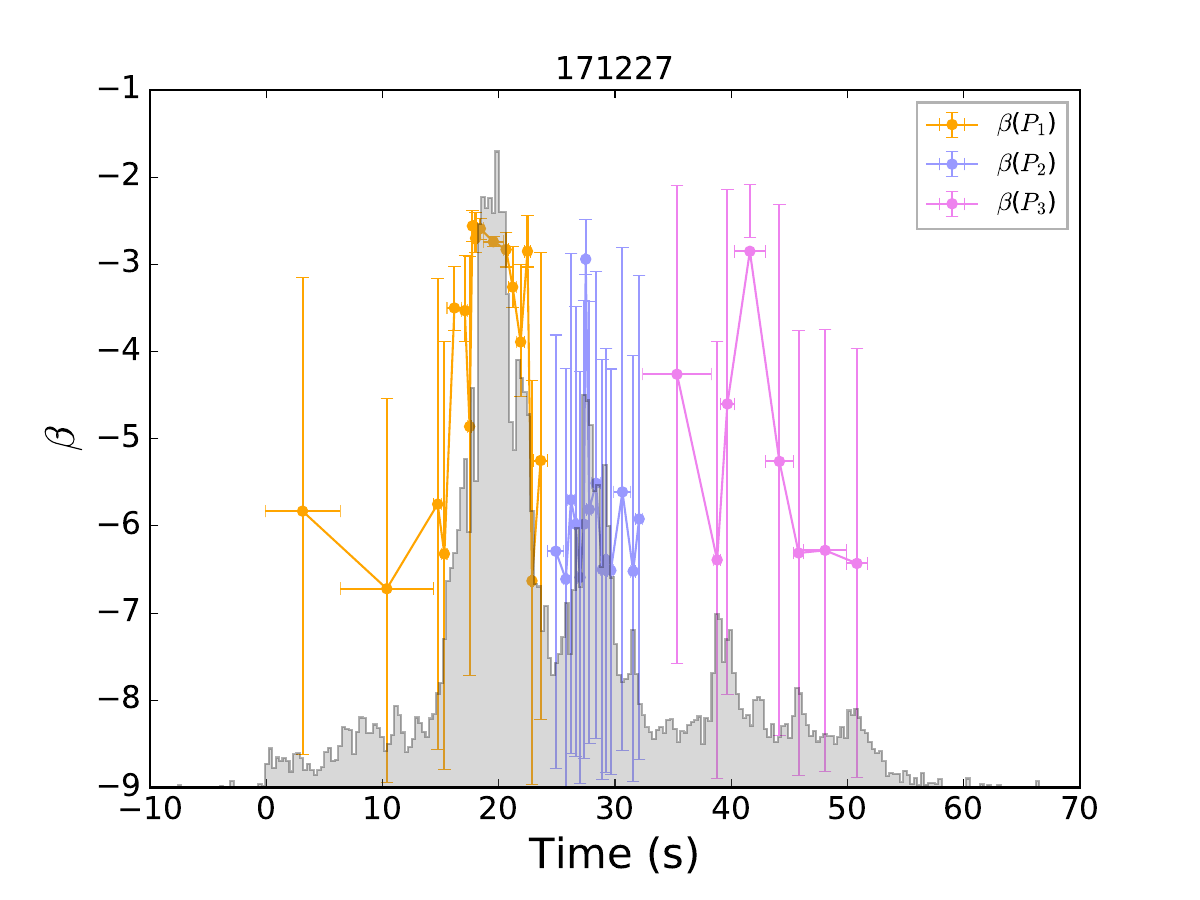}
\includegraphics[angle=0,scale=0.3]{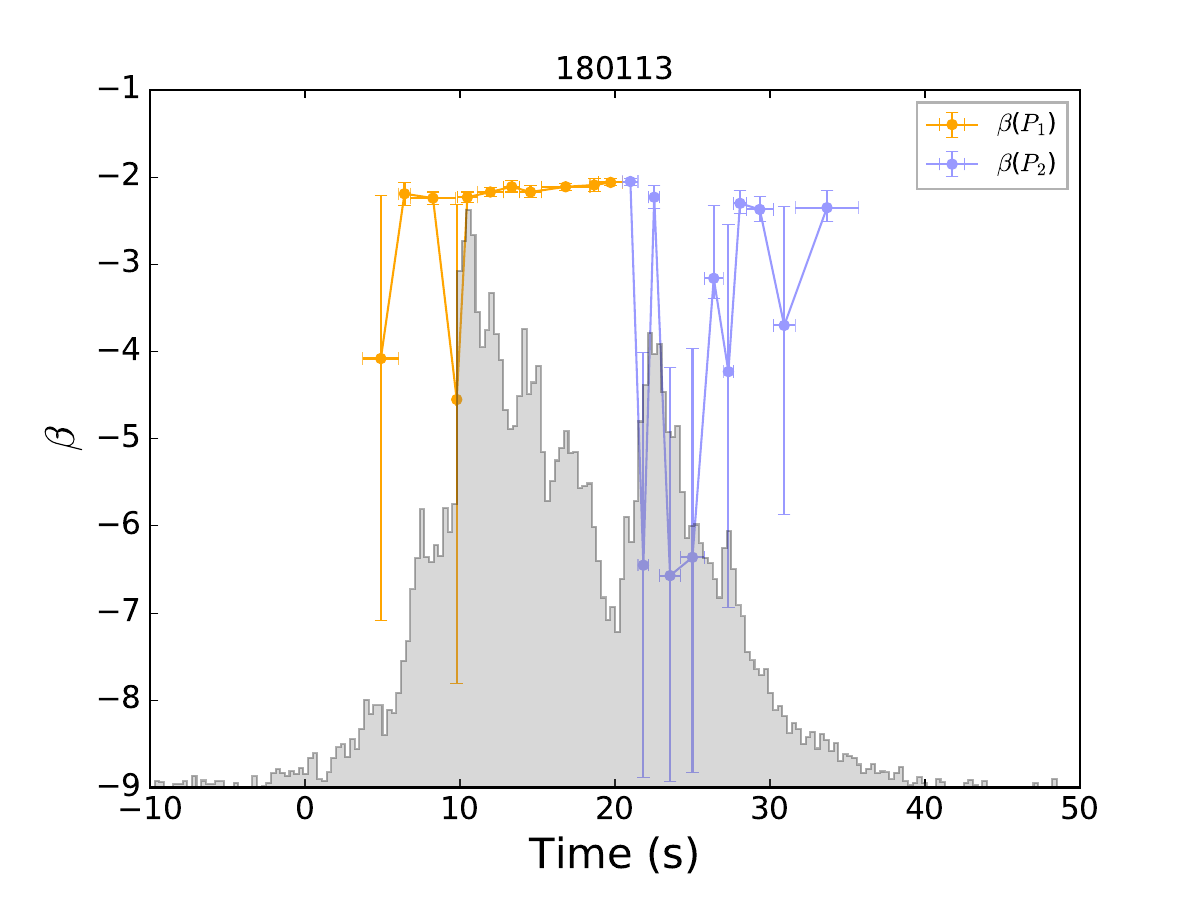}
\includegraphics[angle=0,scale=0.3]{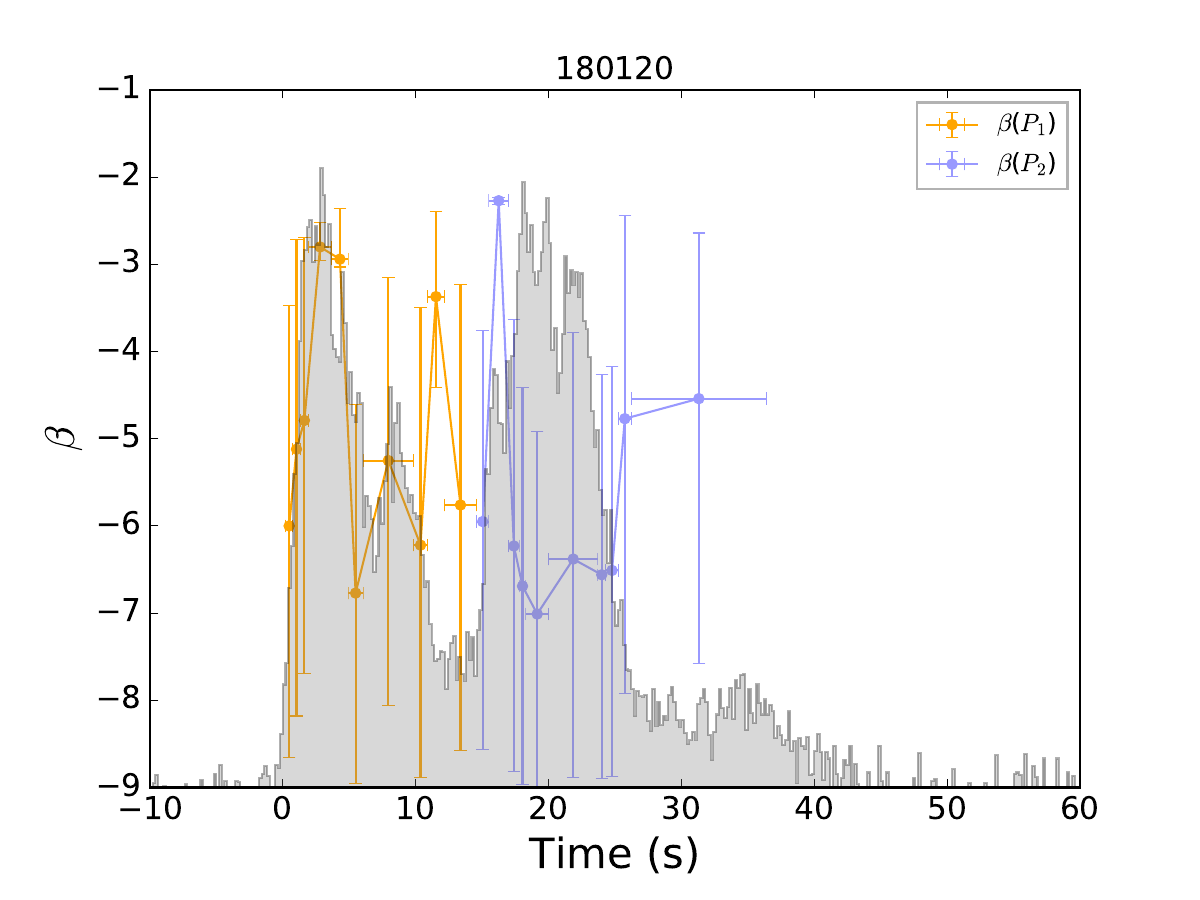}
\includegraphics[angle=0,scale=0.3]{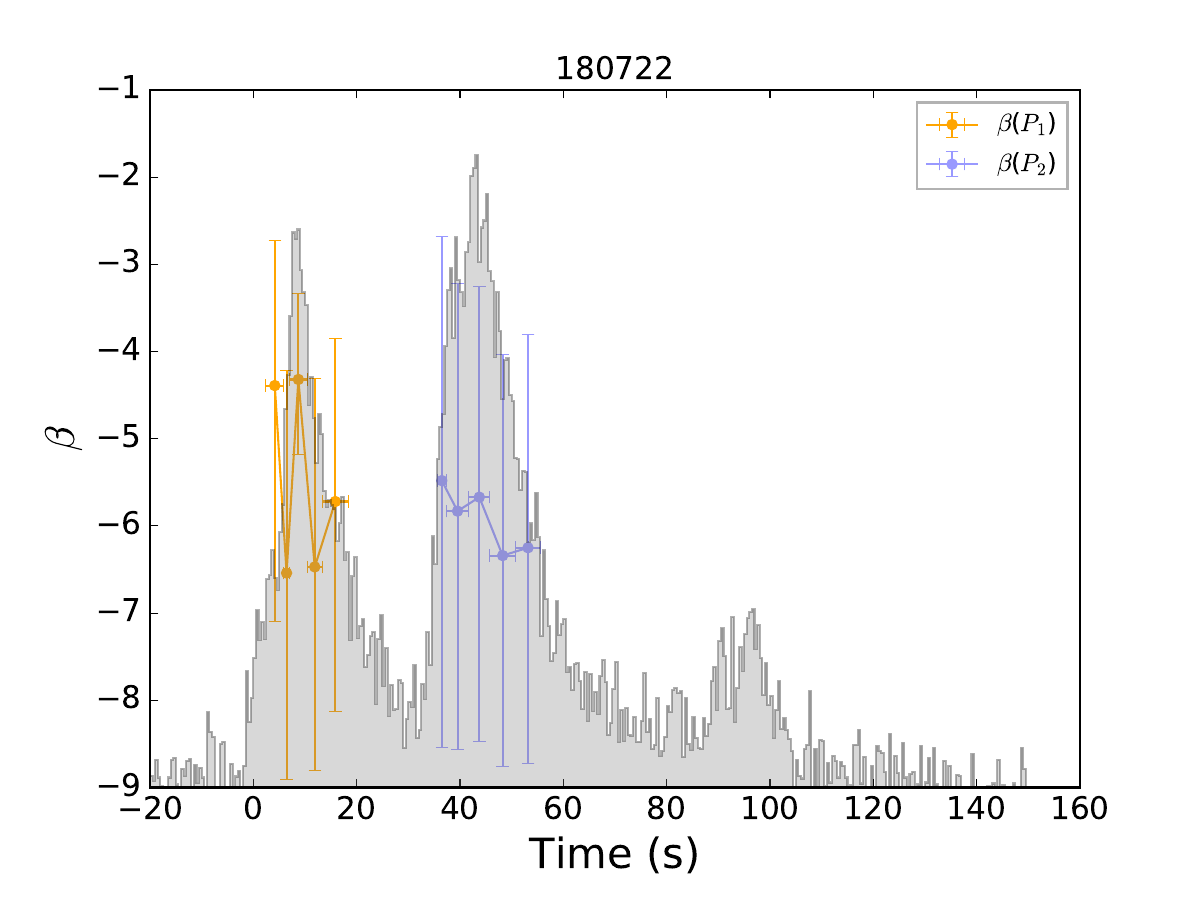}
\includegraphics[angle=0,scale=0.3]{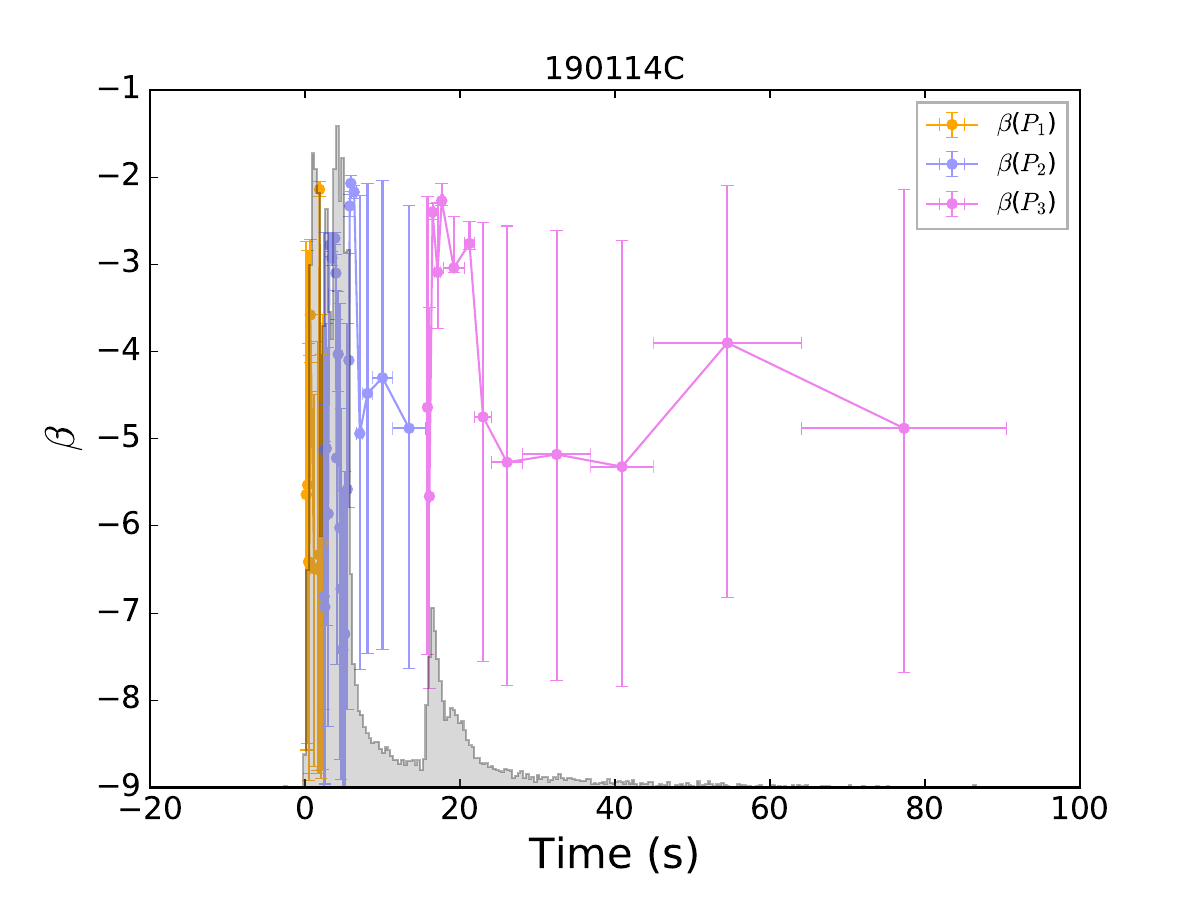}
\center{Fig. \ref{fig:Beta}--- Continued}
\end{figure*}

\clearpage
\subsection{Best Model-based Parameter Relations in Different pulses}

\begin{figure*}
\includegraphics[angle=0,scale=0.3]{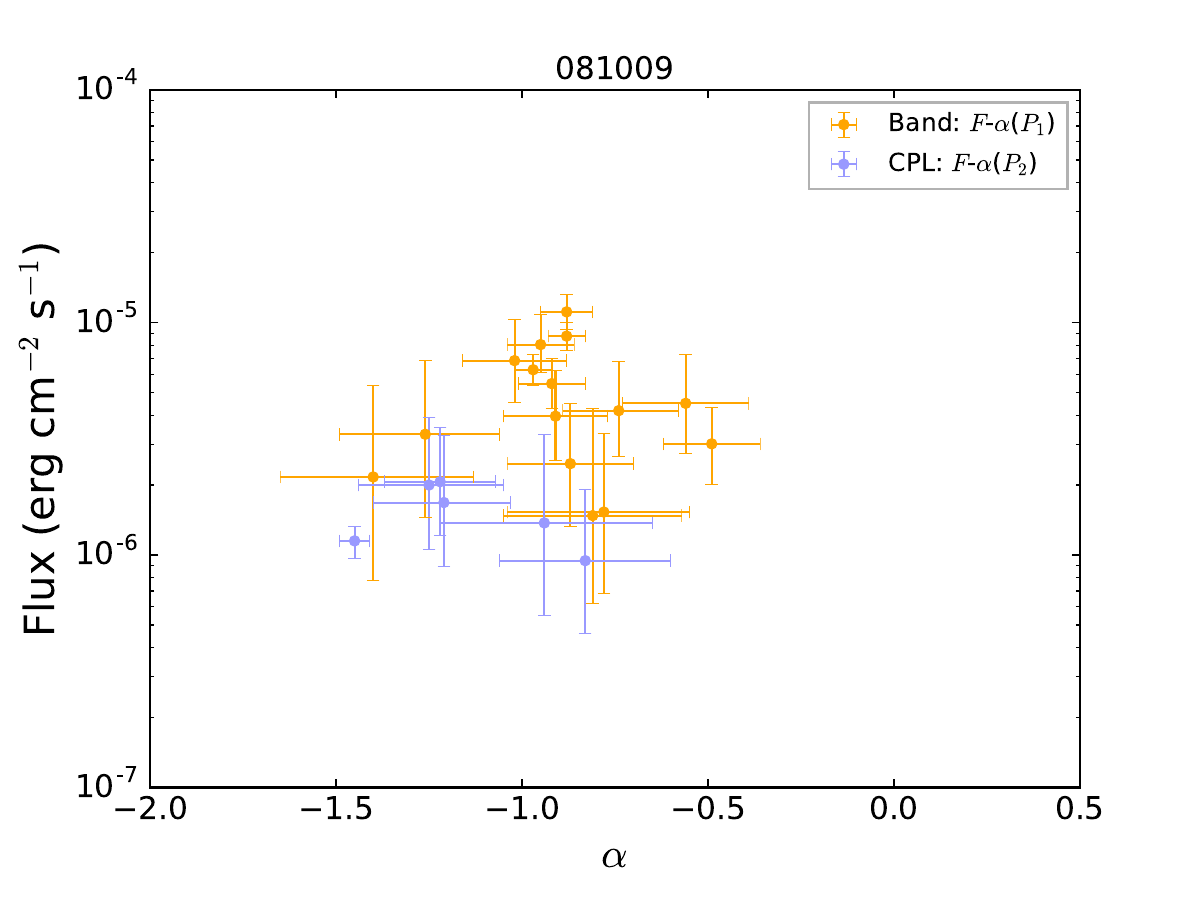}
\includegraphics[angle=0,scale=0.3]{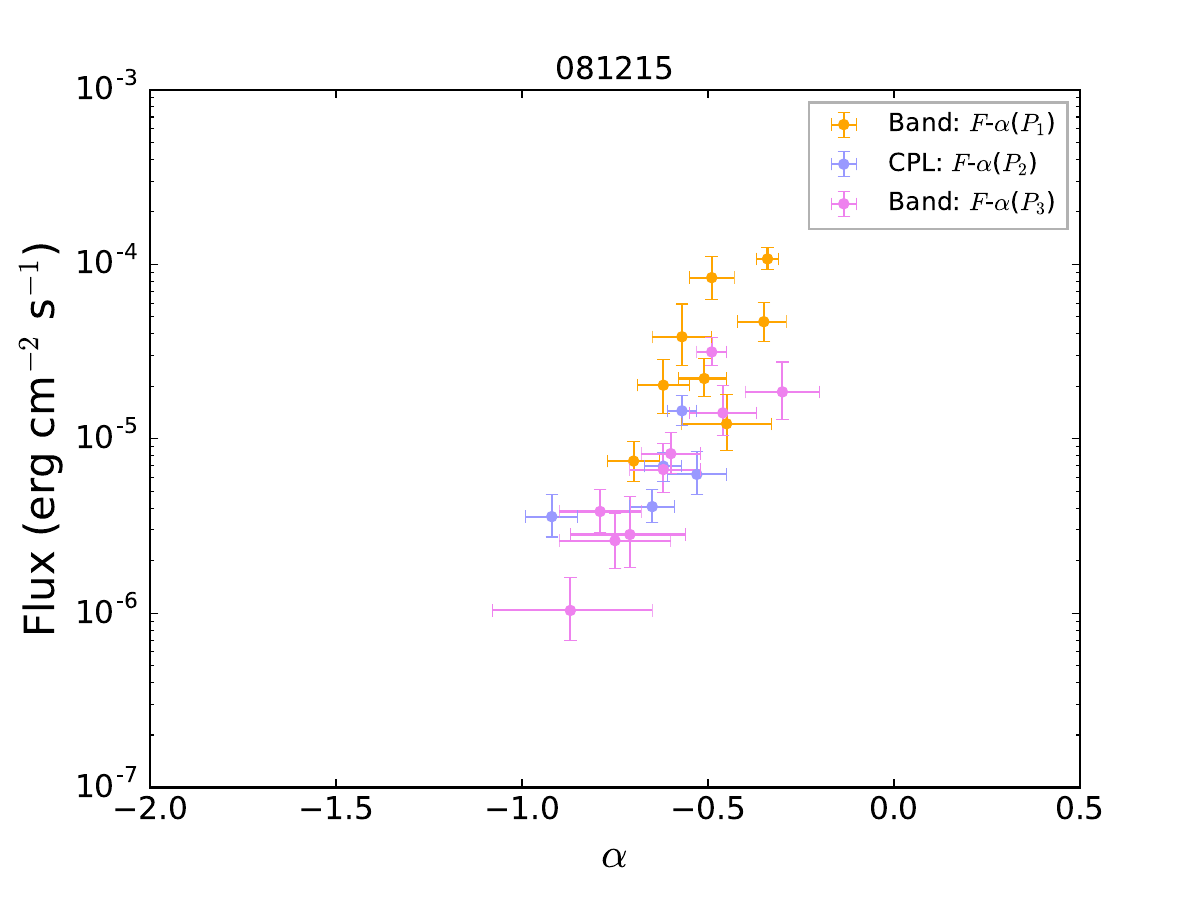}
\includegraphics[angle=0,scale=0.3]{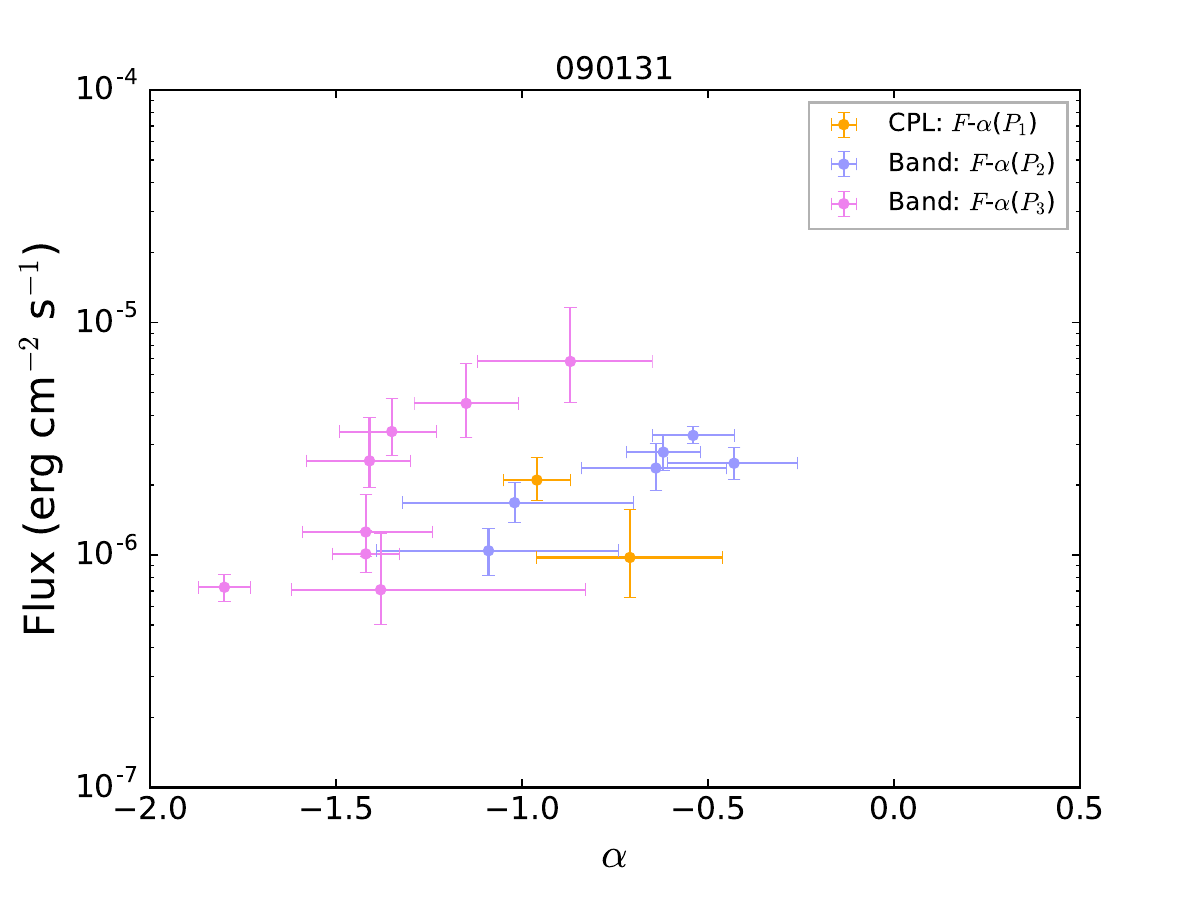}
\includegraphics[angle=0,scale=0.3]{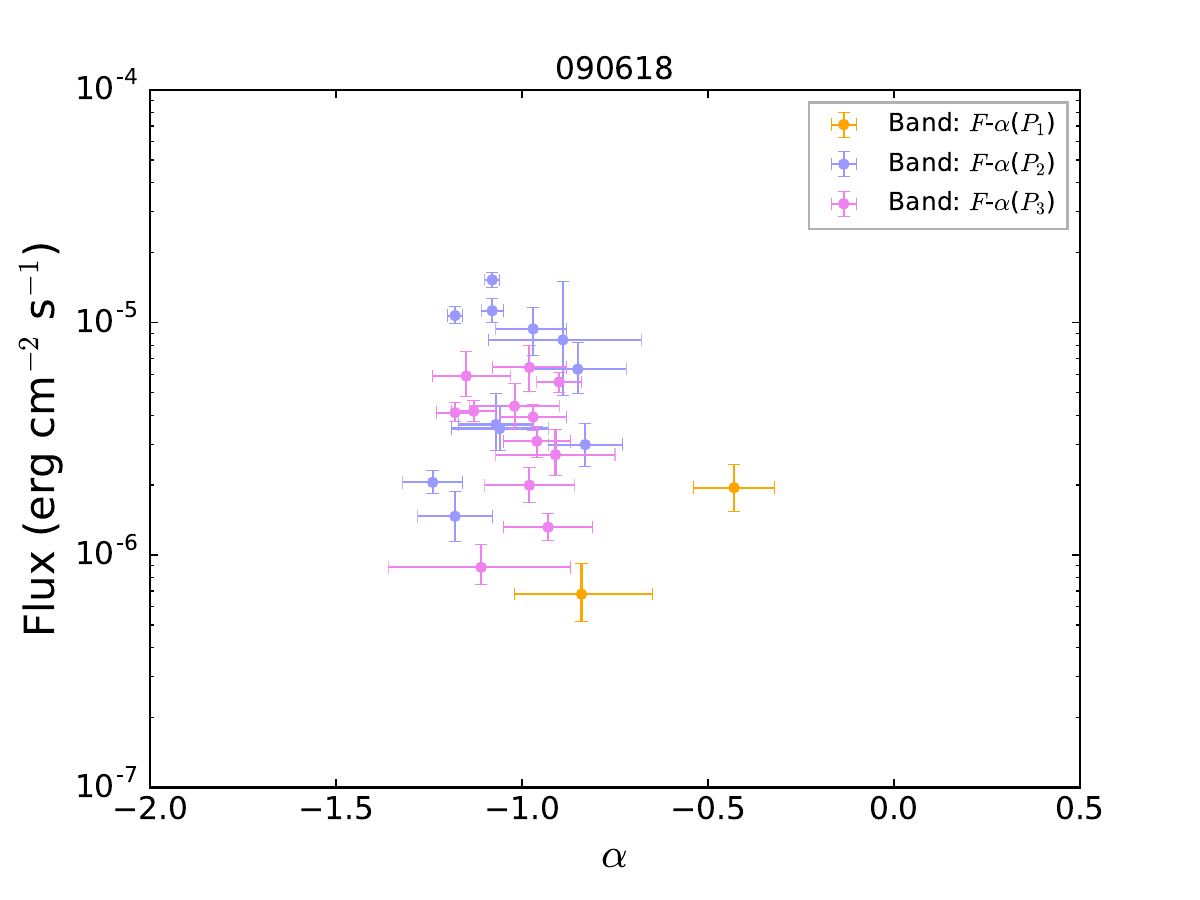}
\includegraphics[angle=0,scale=0.3]{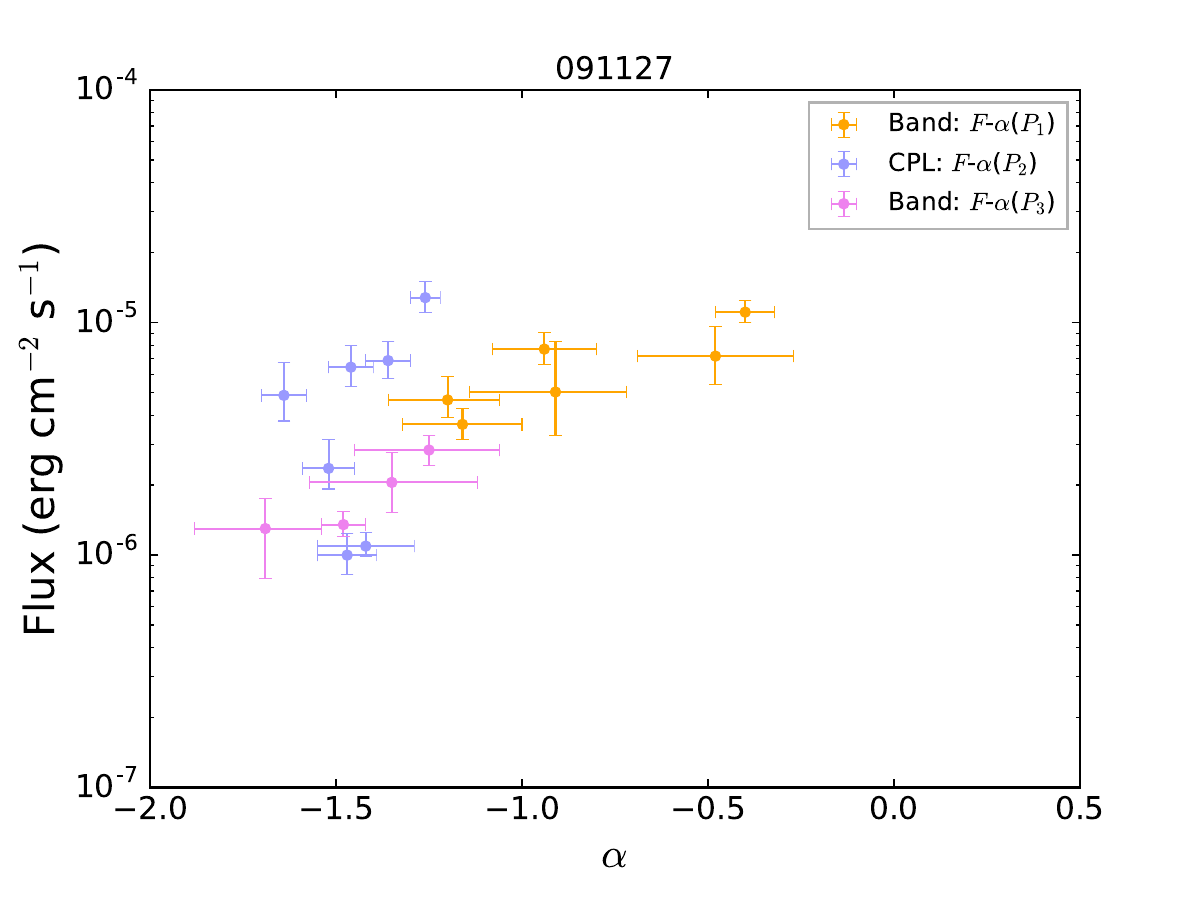}
\includegraphics[angle=0,scale=0.3]{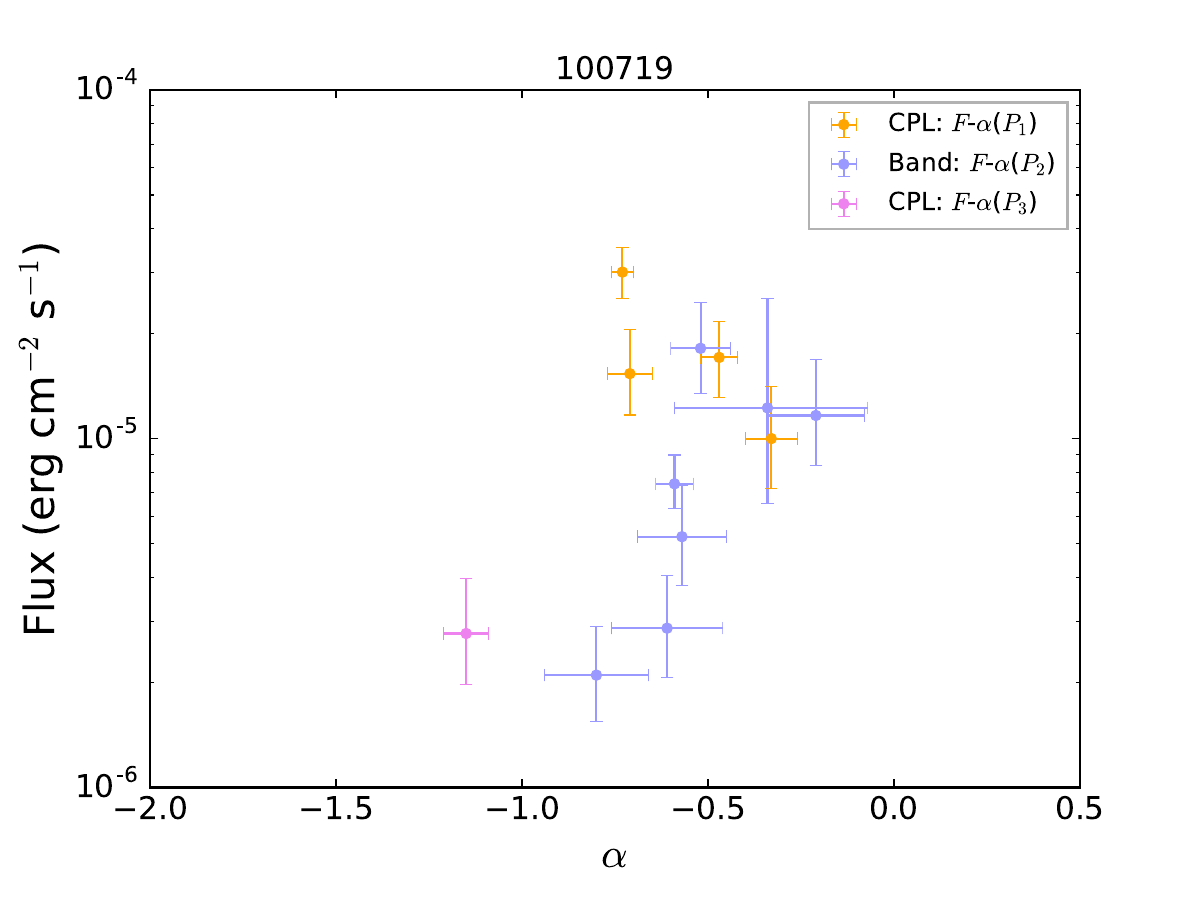}
\includegraphics[angle=0,scale=0.3]{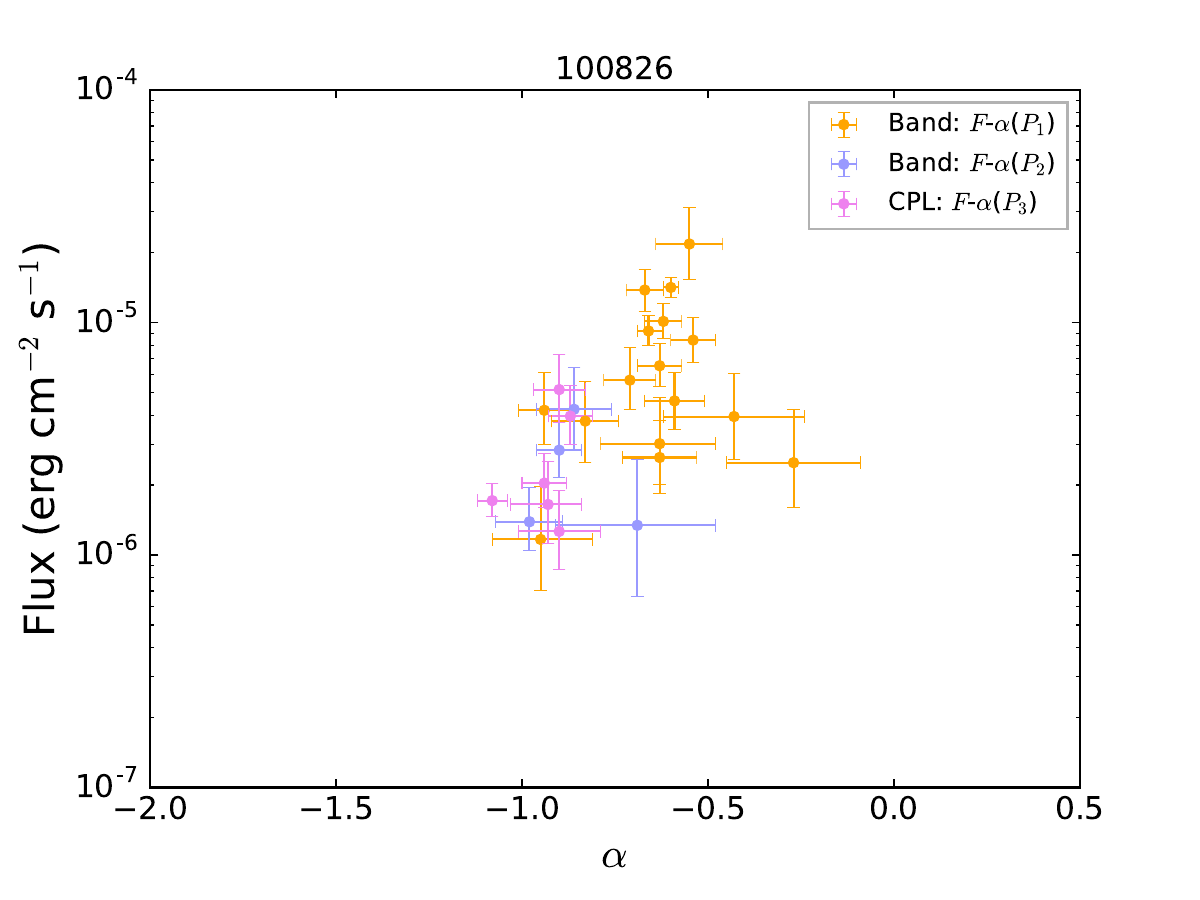}
\includegraphics[angle=0,scale=0.3]{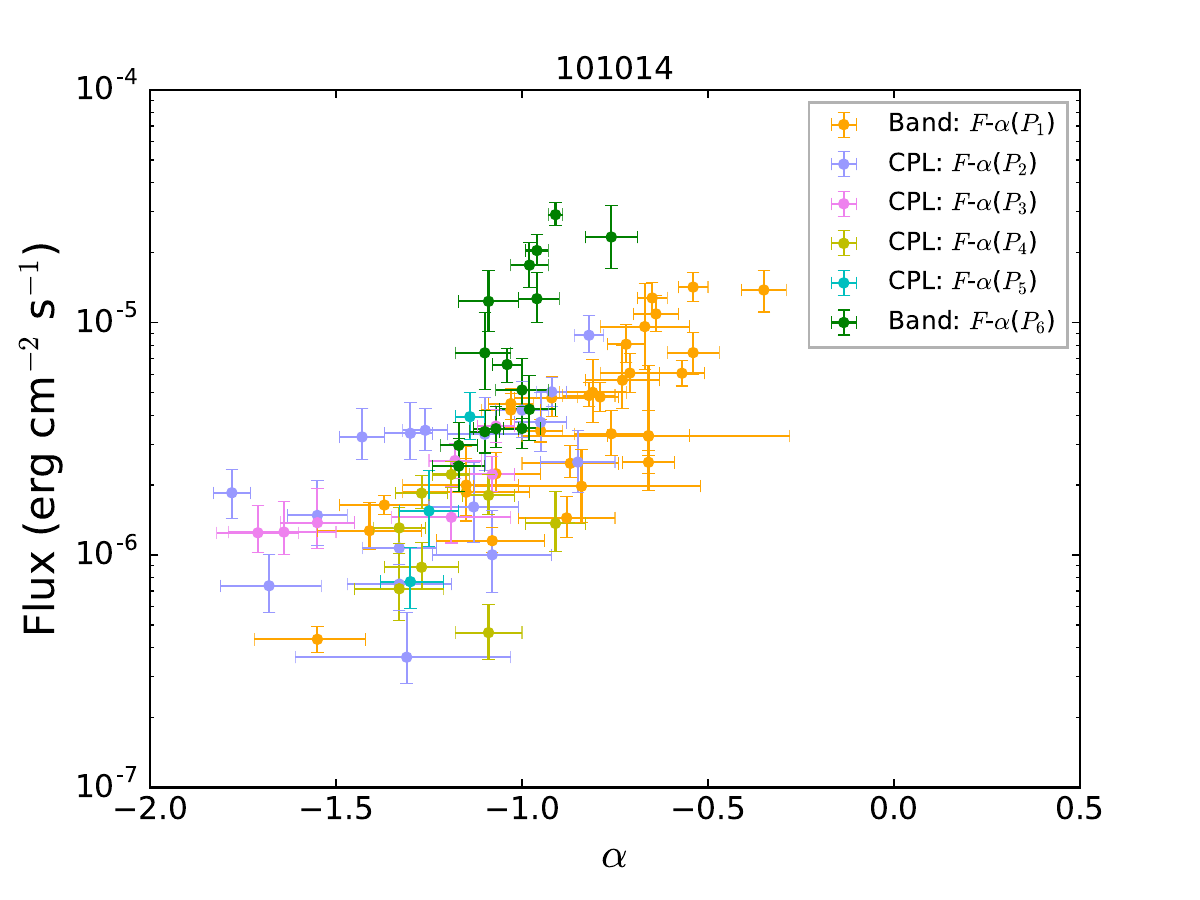}
\includegraphics[angle=0,scale=0.3]{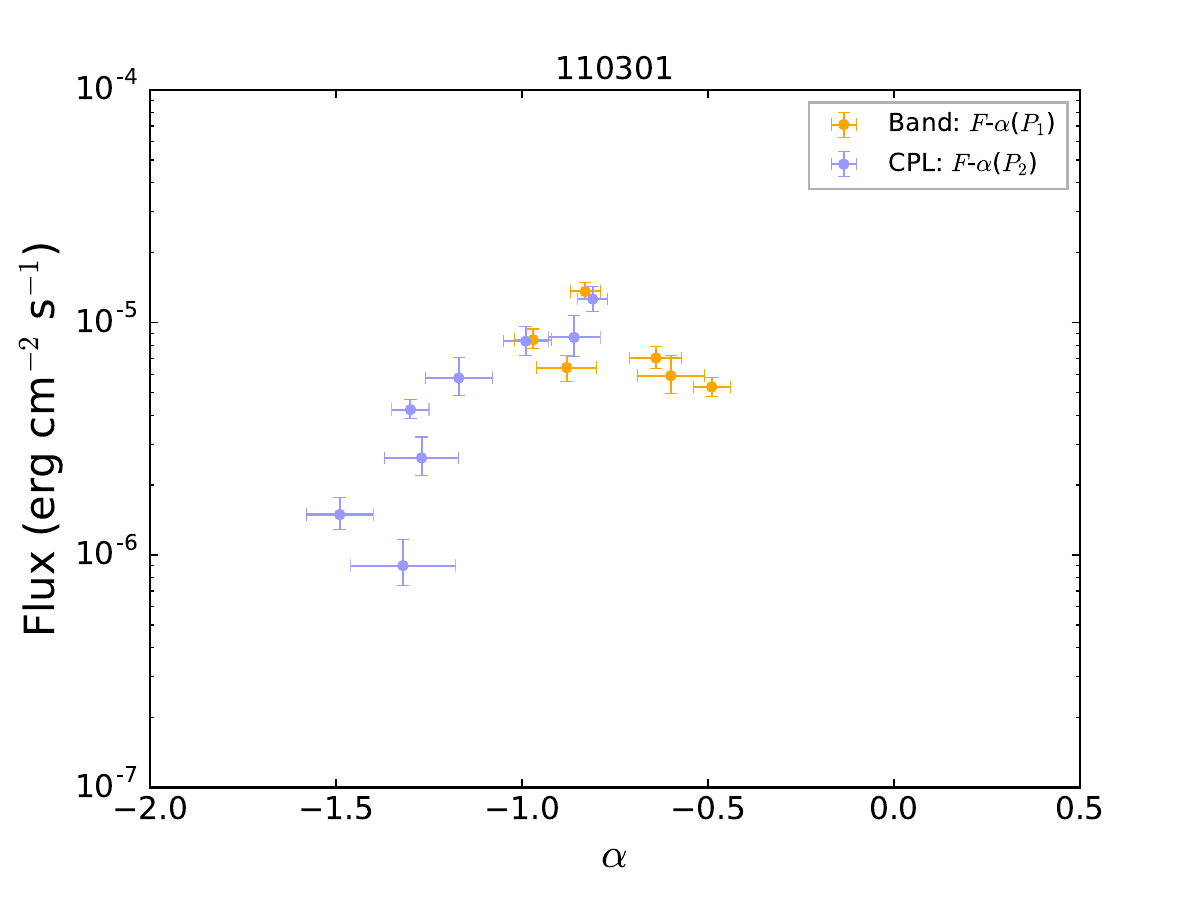}
\includegraphics[angle=0,scale=0.3]{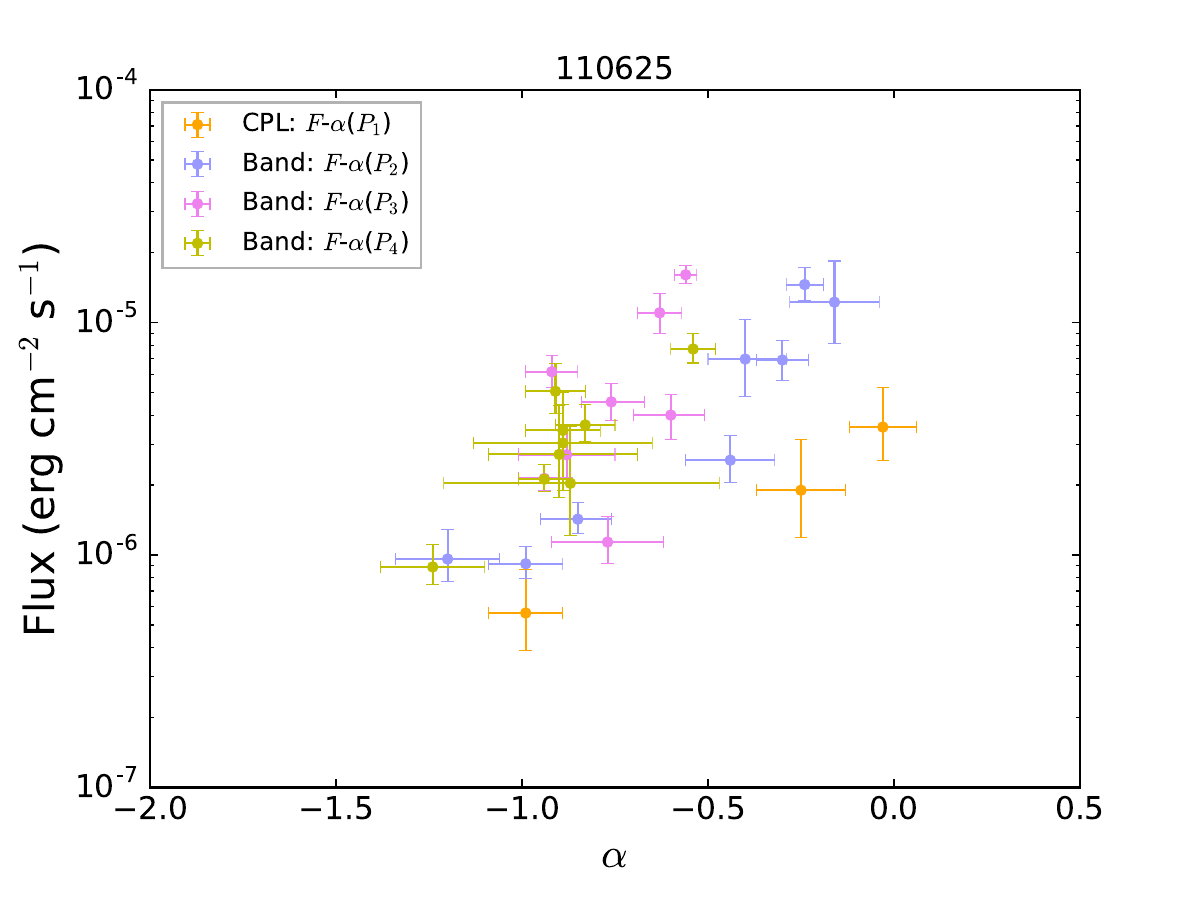}
\includegraphics[angle=0,scale=0.3]{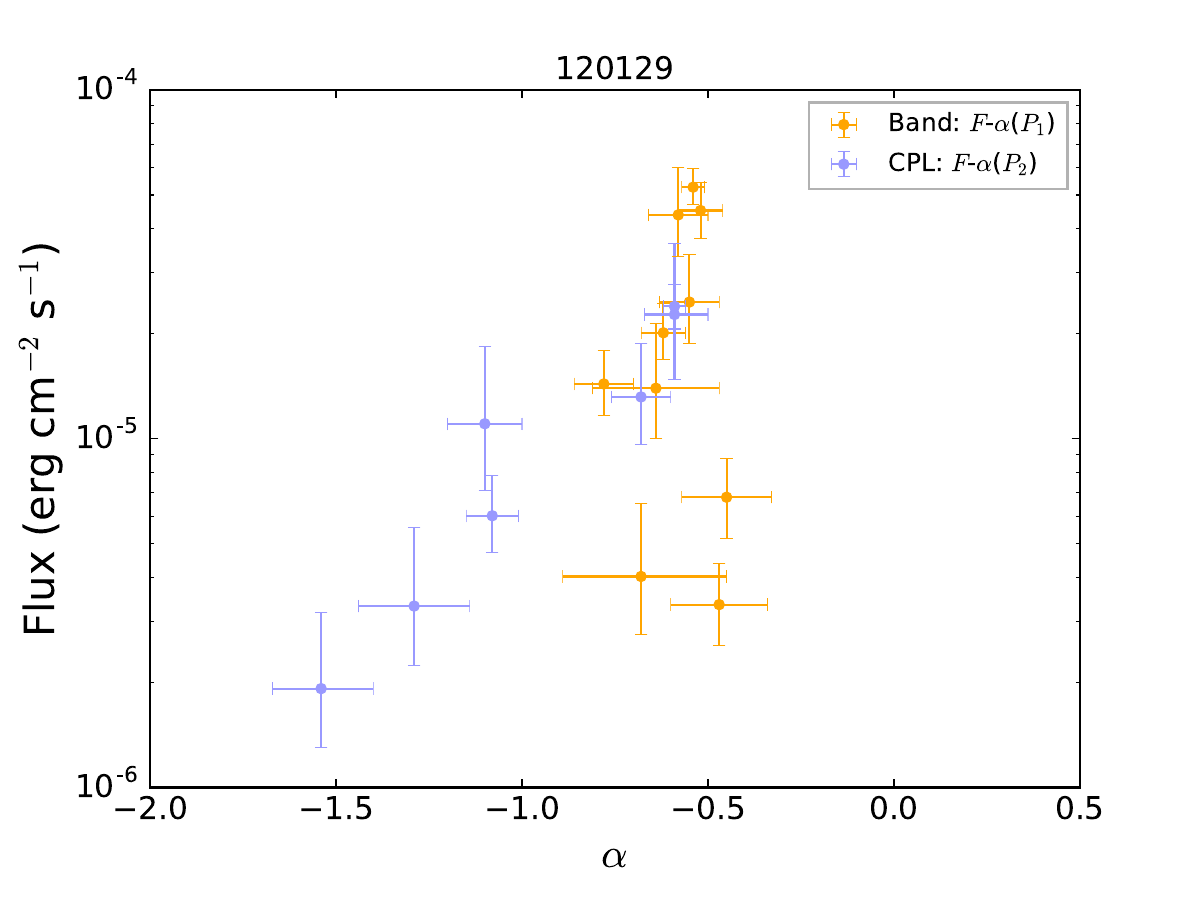}
\includegraphics[angle=0,scale=0.3]{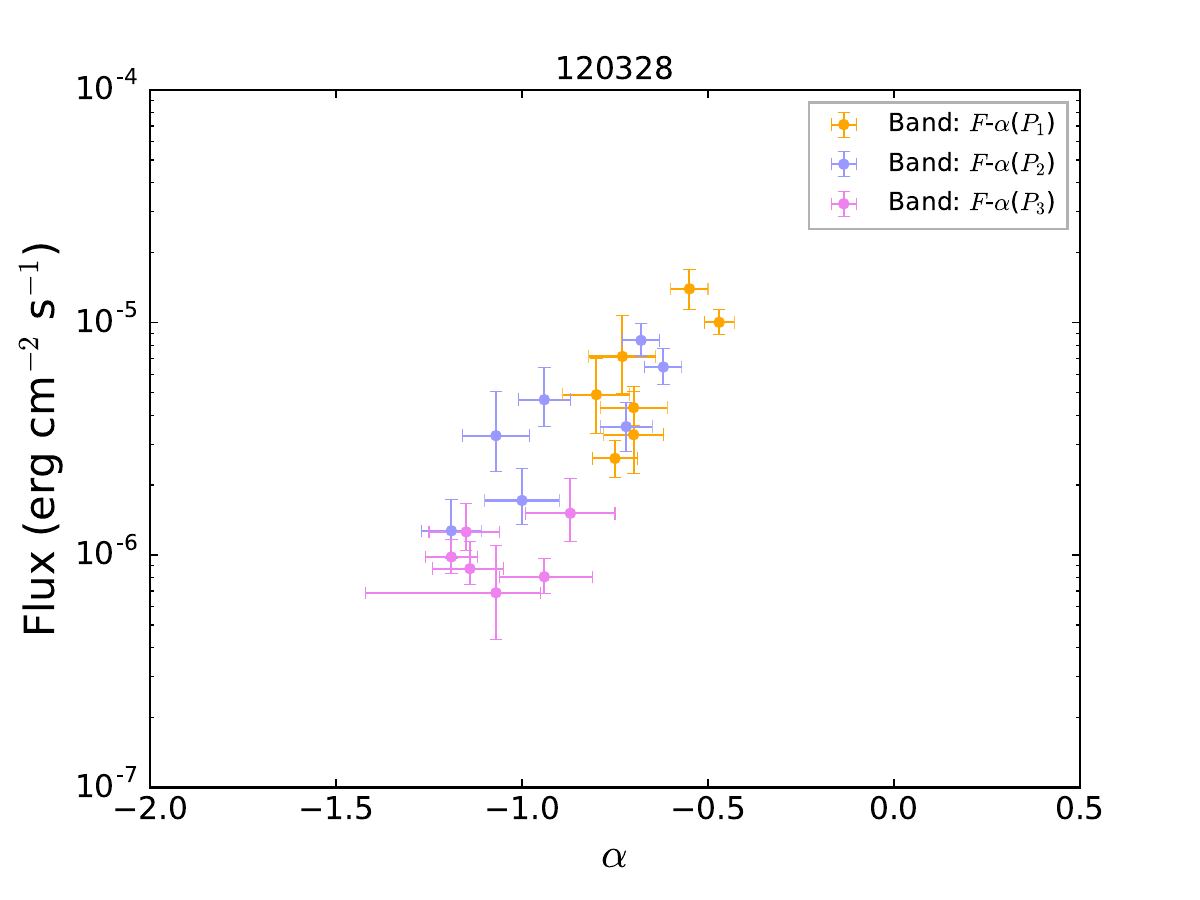}
\caption{The $F$-$\alpha$ relation. Data points with solid orange, blue-magenta, violet, yellow, cyan, and green colors indicate time-series pulses of the $P_{1}$, $P_{2}$, $P_{3}$, $P_{4}$, $P_{5}$, and $P_{6}$, respectively. All data points correspond to a statistical significance $S \geq 20$.}\label{fig:FluxAlpha_Best}
\end{figure*}

In Figures \ref{fig:FluxAlpha_Best}-\ref{fig:EpAlpha_Best}, we provide the $F$-$\alpha$, $F$-$E_{\rm p}$, and $\alpha$-$E_{\rm p}$ relations for all the bursts based on the best models defined in Section \ref{sec:bestmodel}.

\clearpage
\begin{figure*}
\includegraphics[angle=0,scale=0.3]{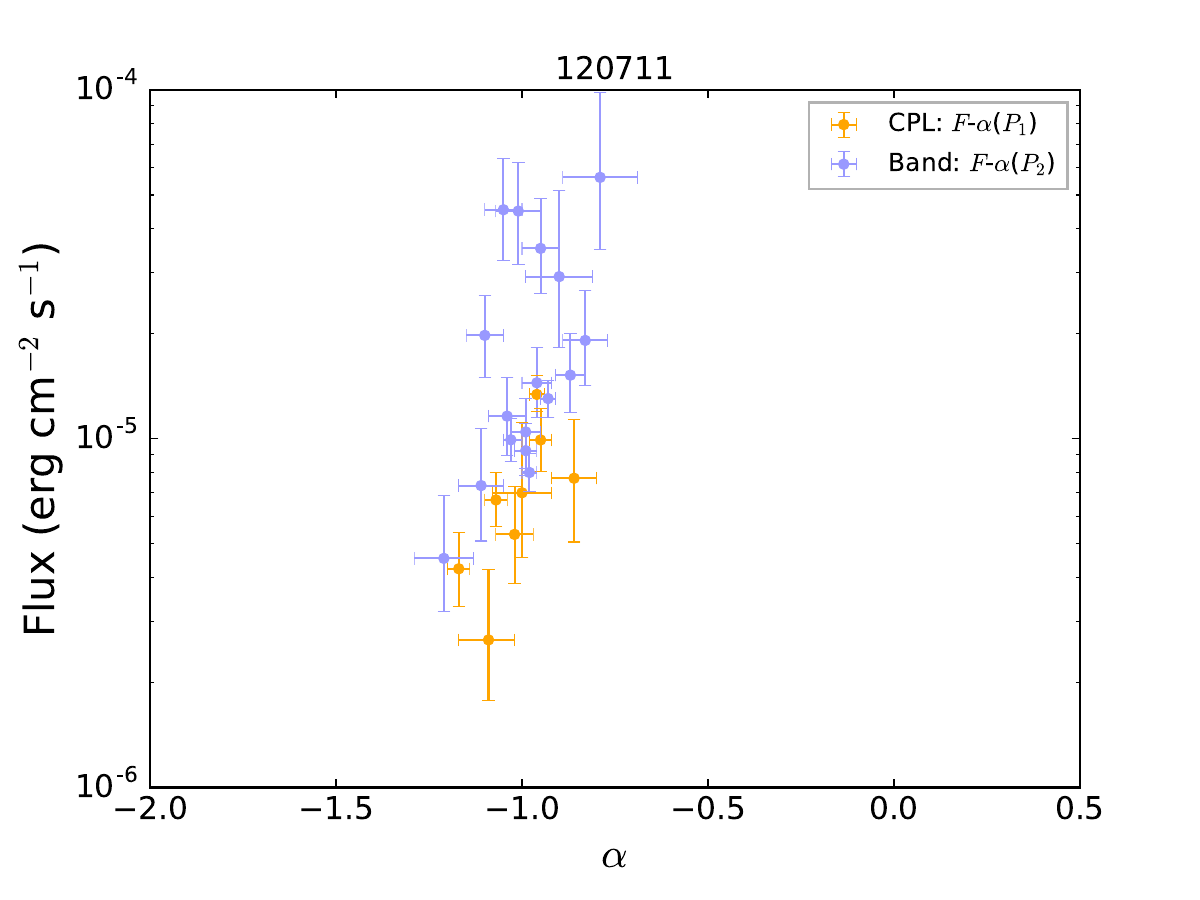}
\includegraphics[angle=0,scale=0.3]{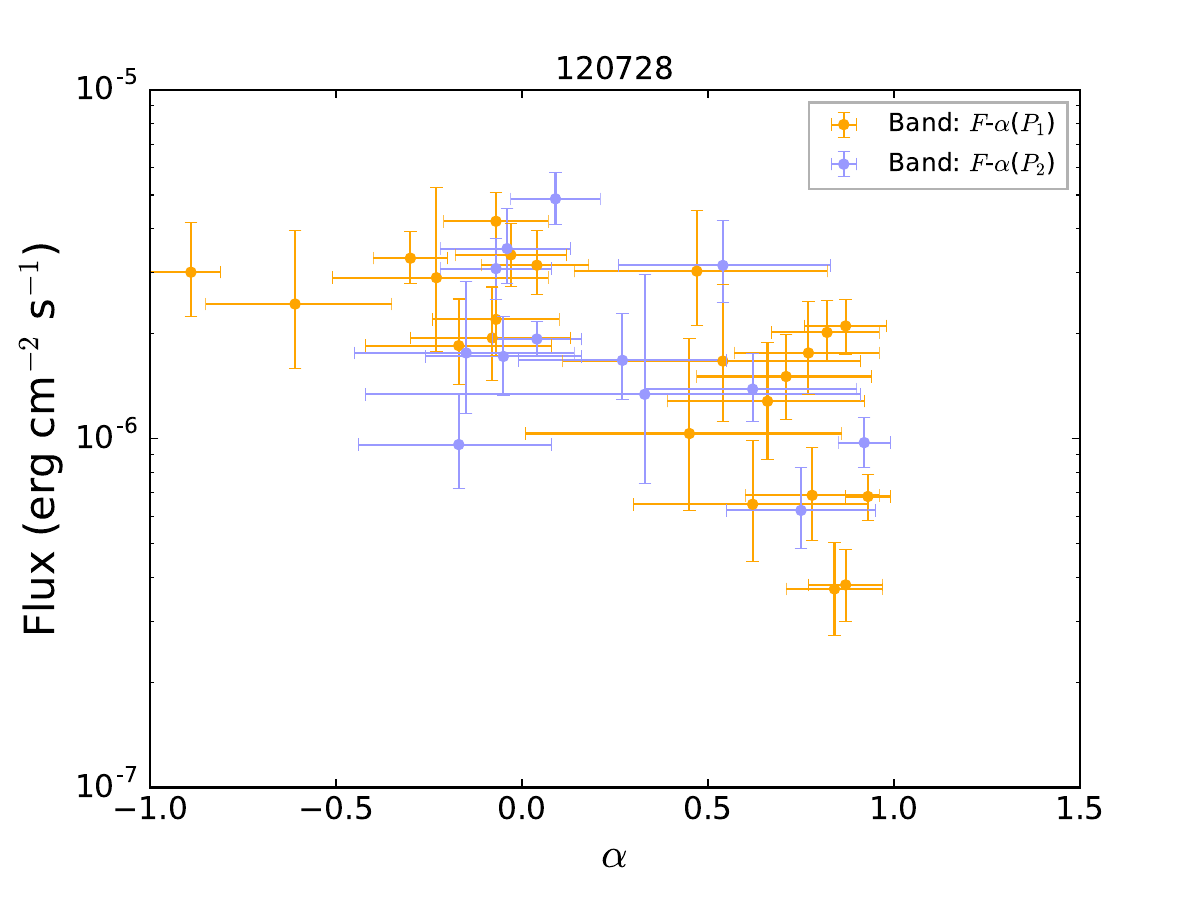}
\includegraphics[angle=0,scale=0.3]{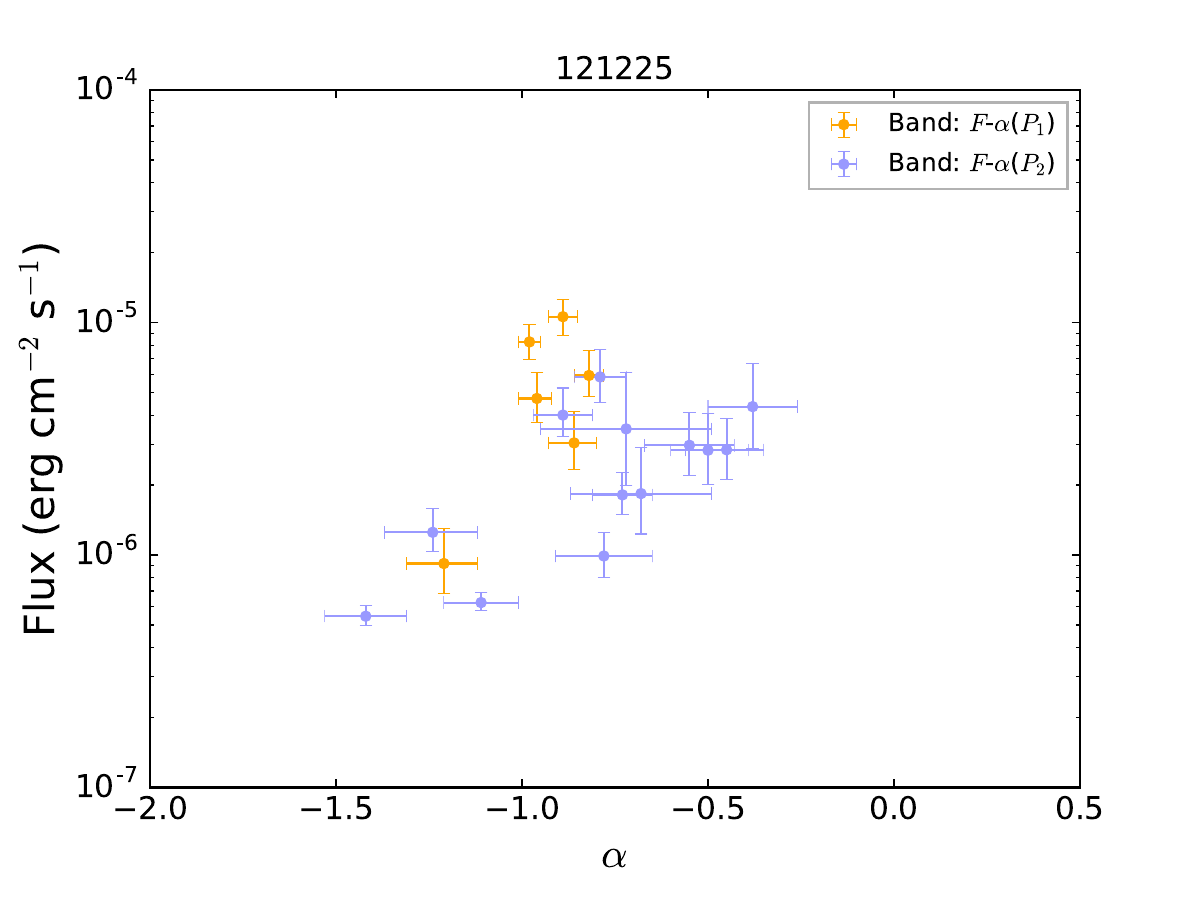}
\includegraphics[angle=0,scale=0.3]{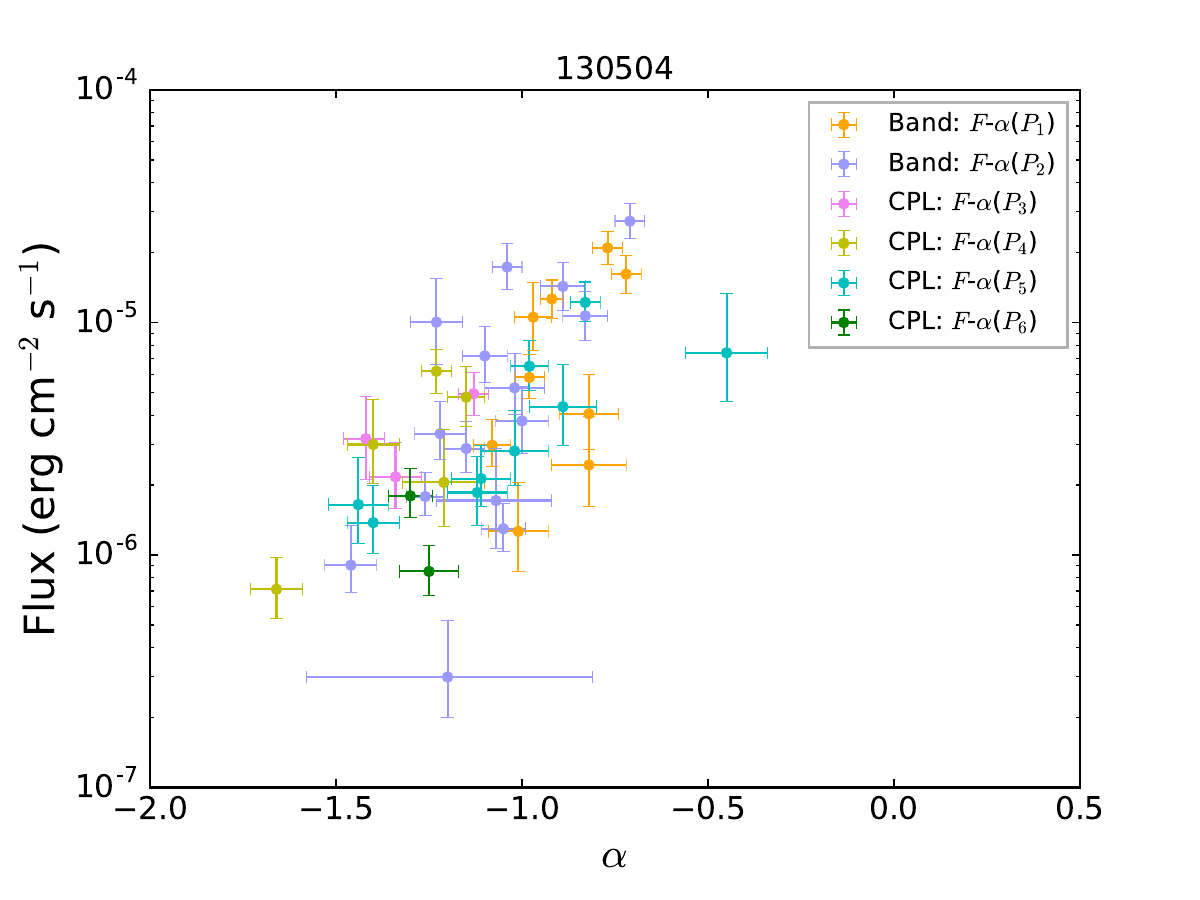}
\includegraphics[angle=0,scale=0.3]{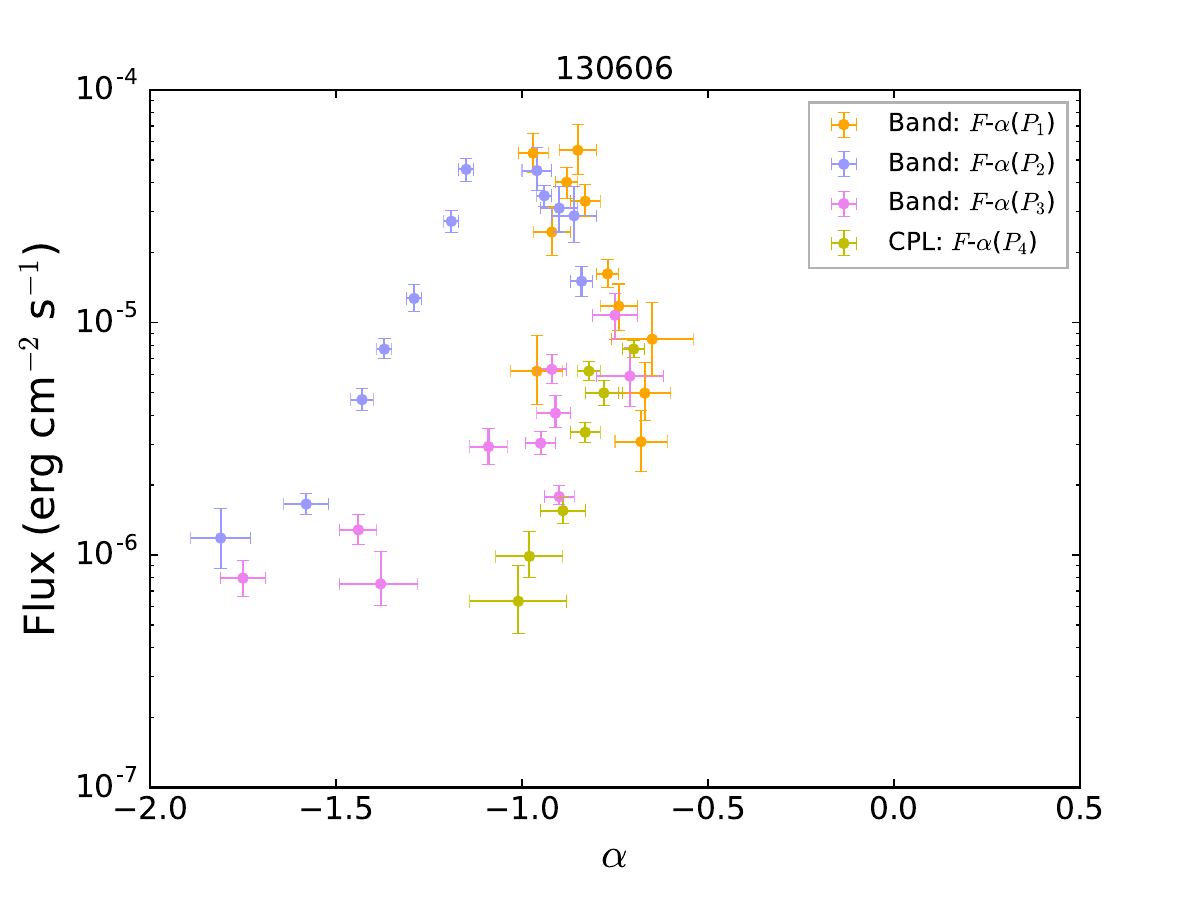}
\includegraphics[angle=0,scale=0.3]{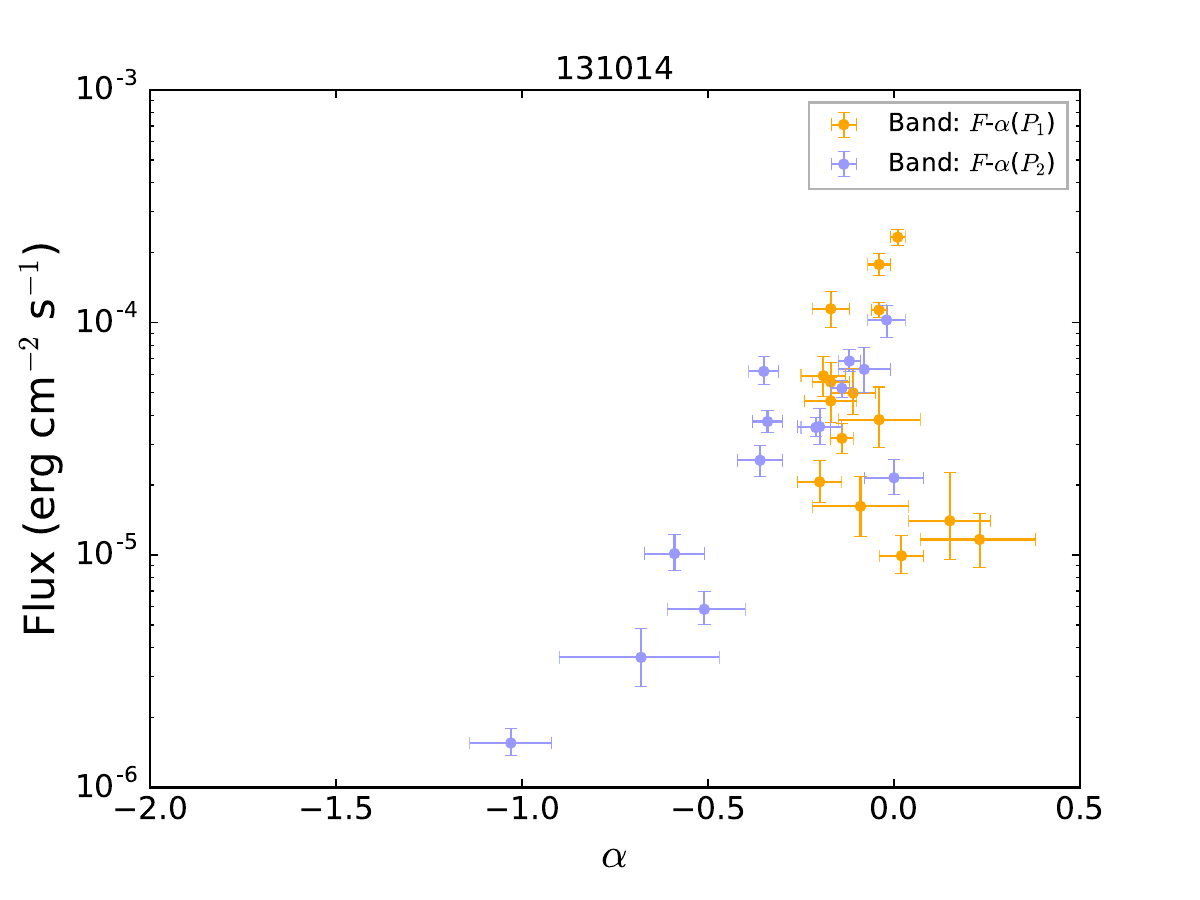}
\includegraphics[angle=0,scale=0.3]{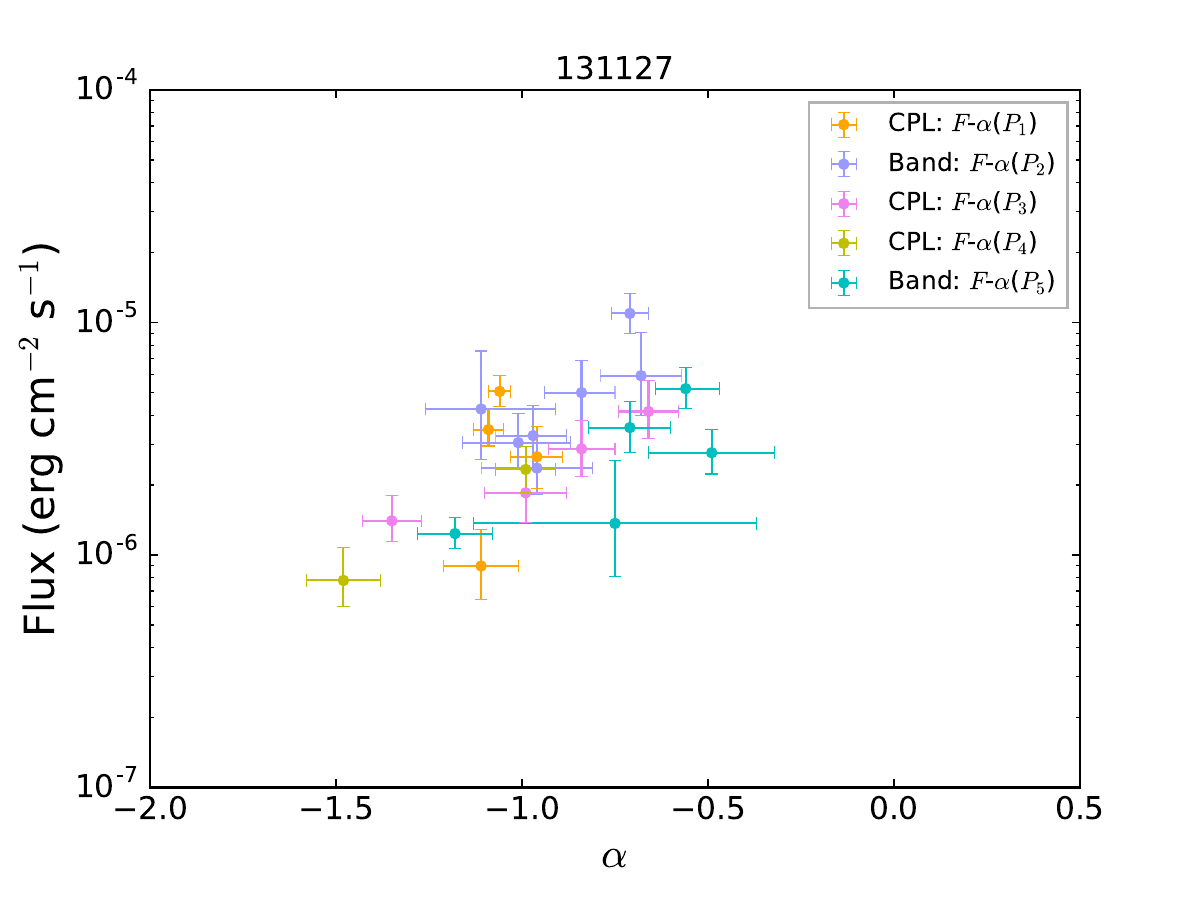}
\includegraphics[angle=0,scale=0.3]{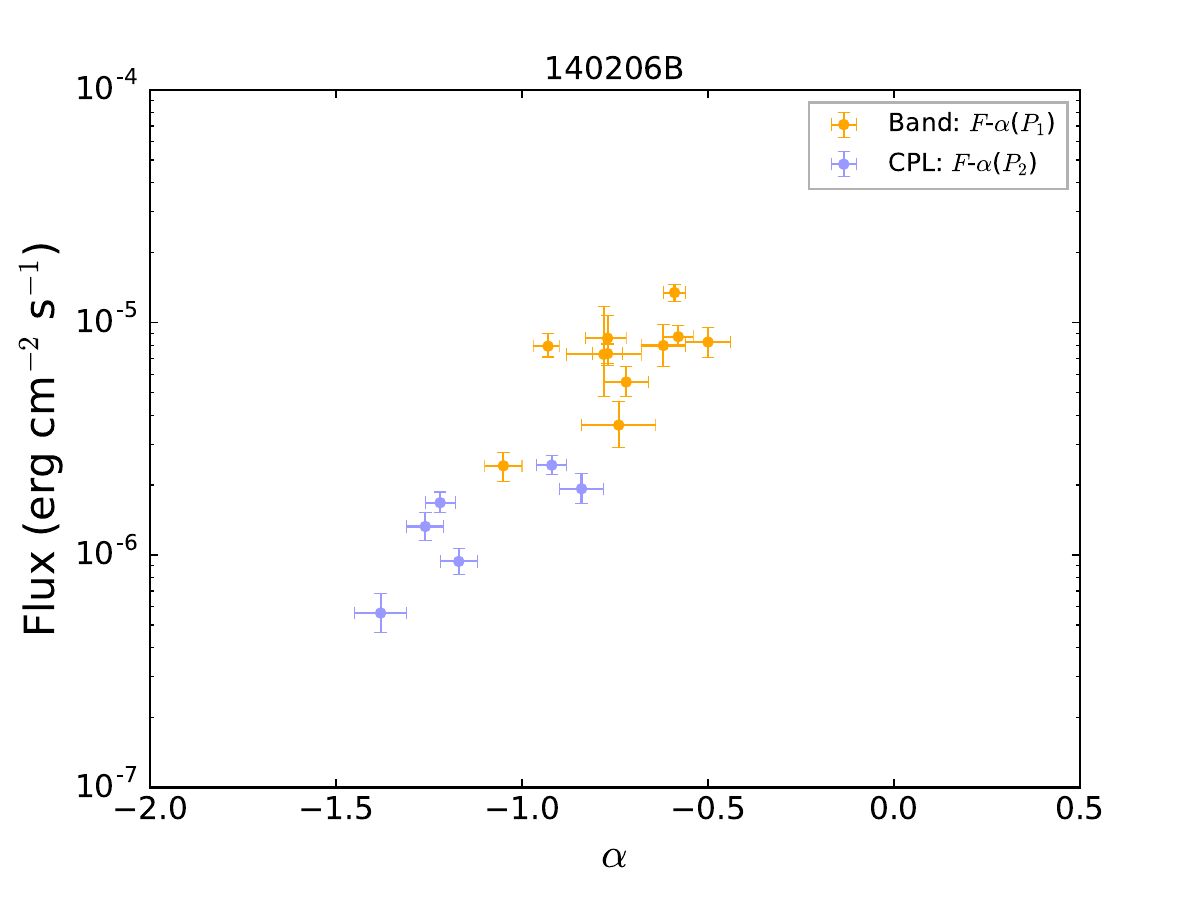}
\includegraphics[angle=0,scale=0.3]{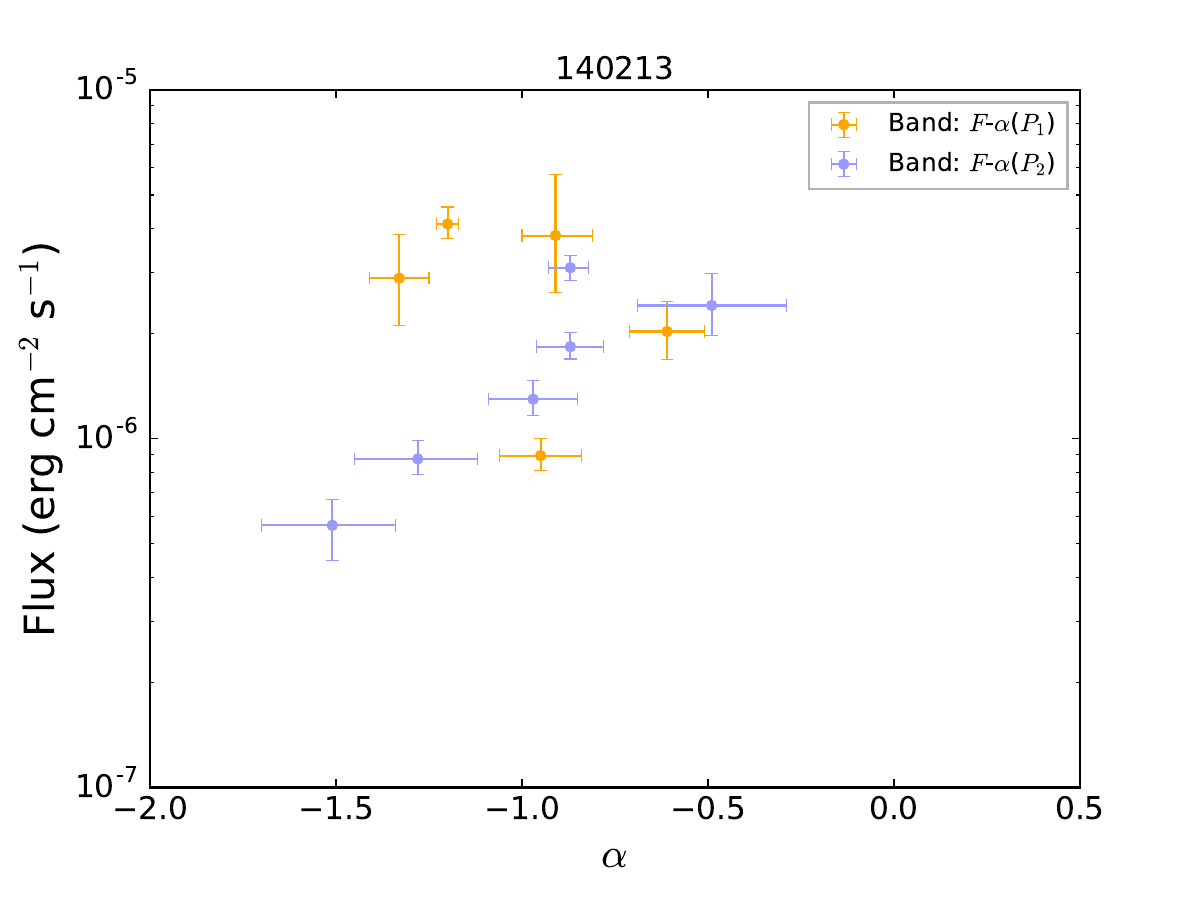}
\includegraphics[angle=0,scale=0.3]{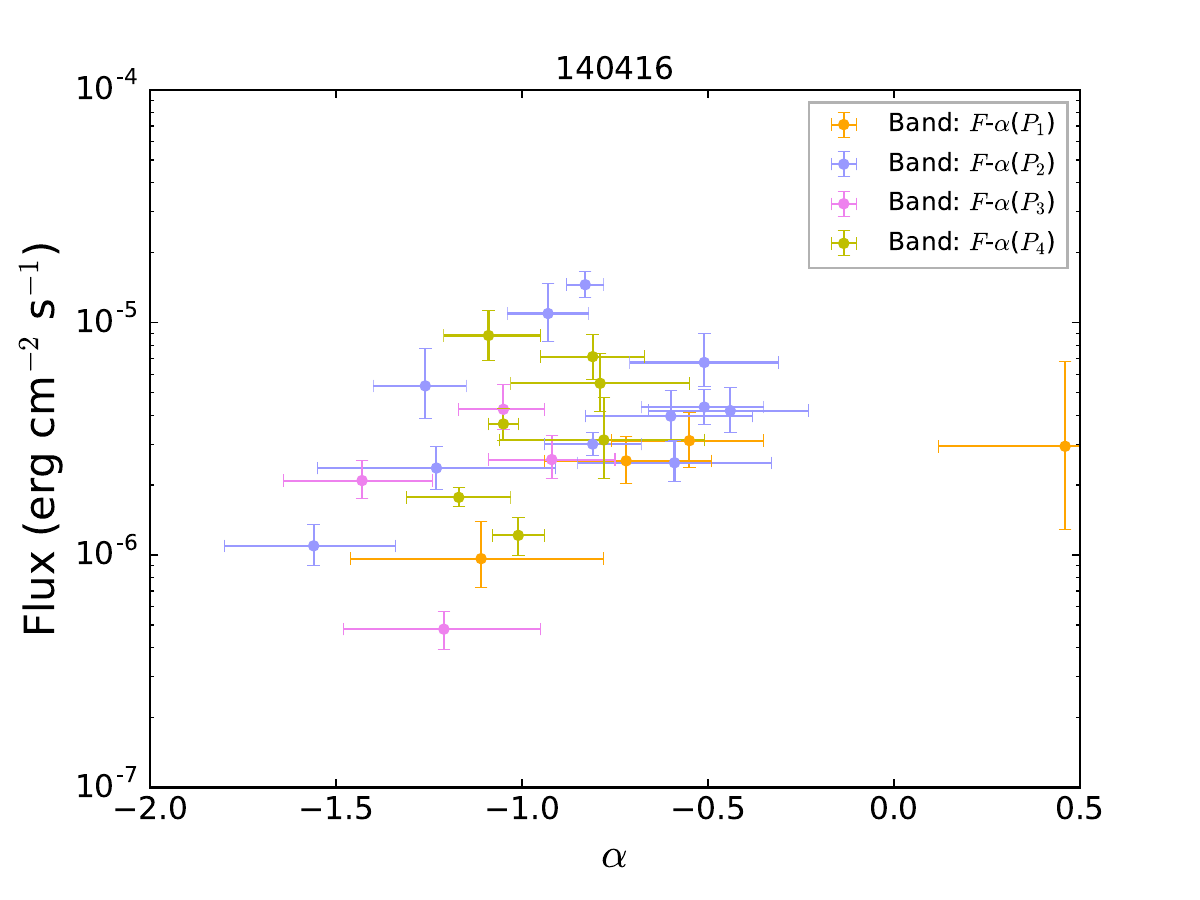}
\includegraphics[angle=0,scale=0.3]{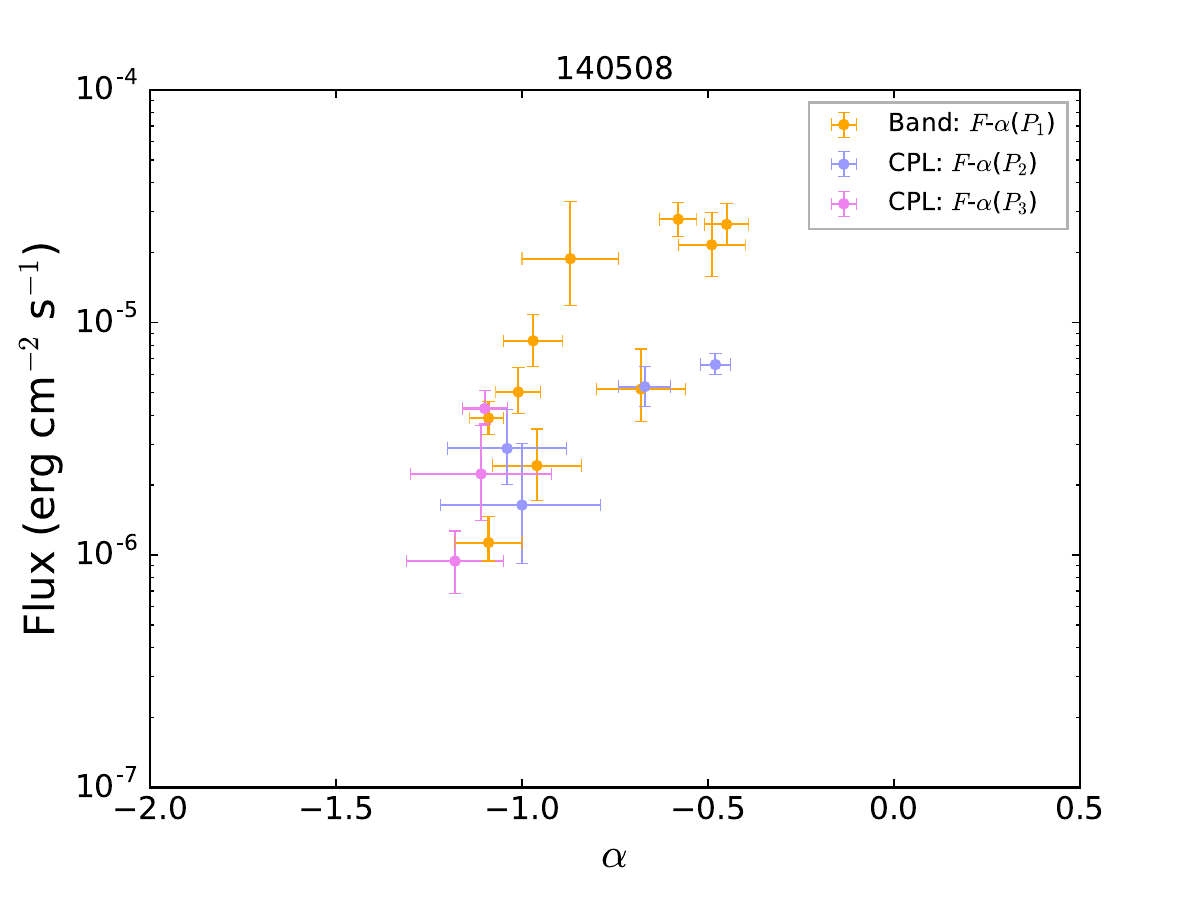}
\includegraphics[angle=0,scale=0.3]{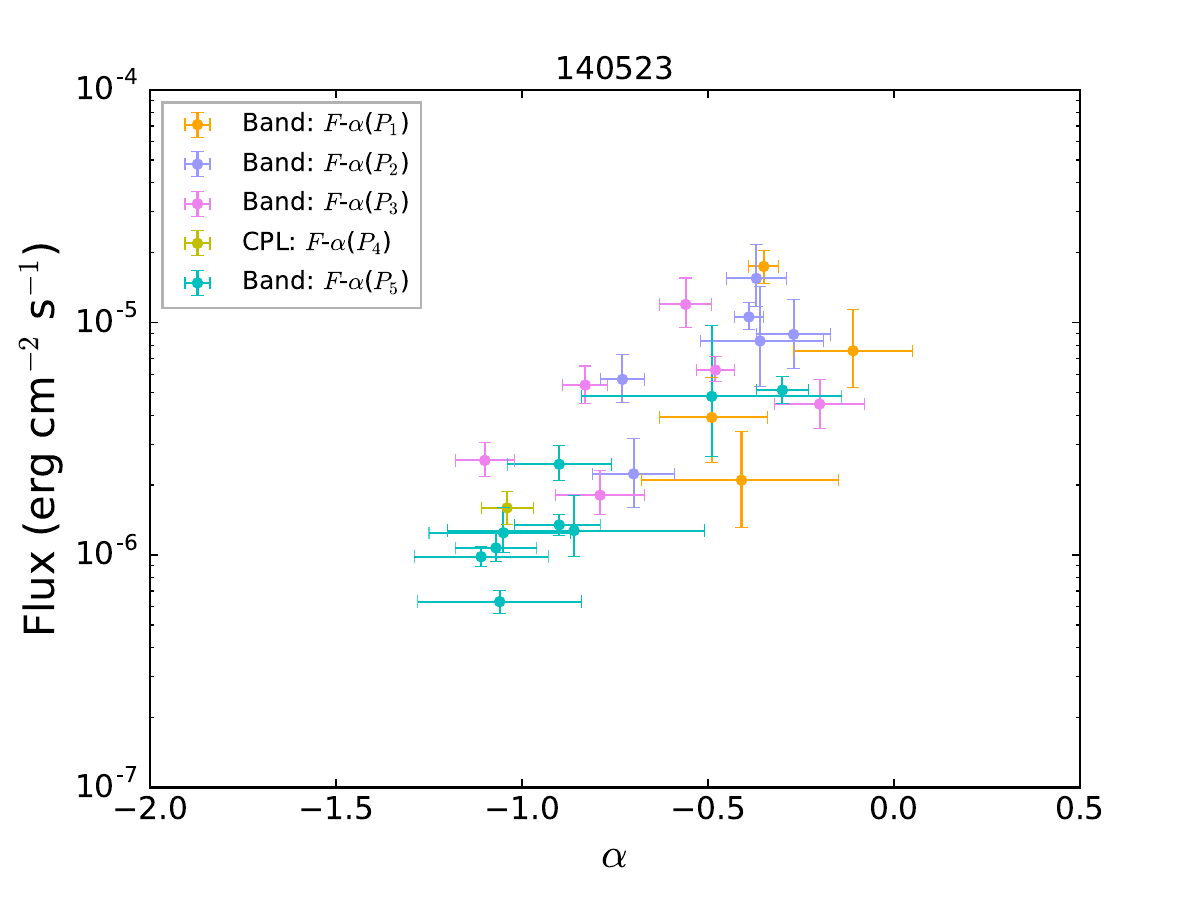}
\includegraphics[angle=0,scale=0.3]{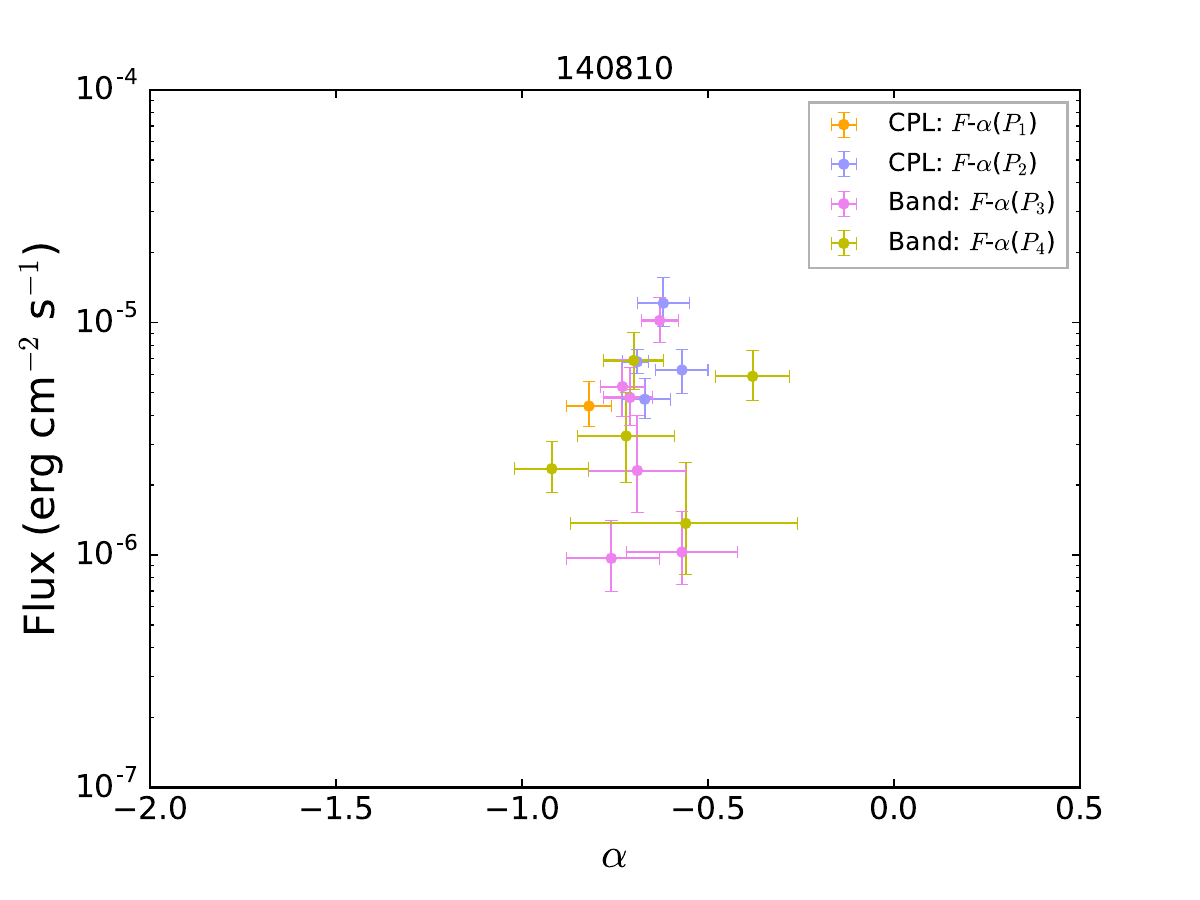}
\includegraphics[angle=0,scale=0.3]{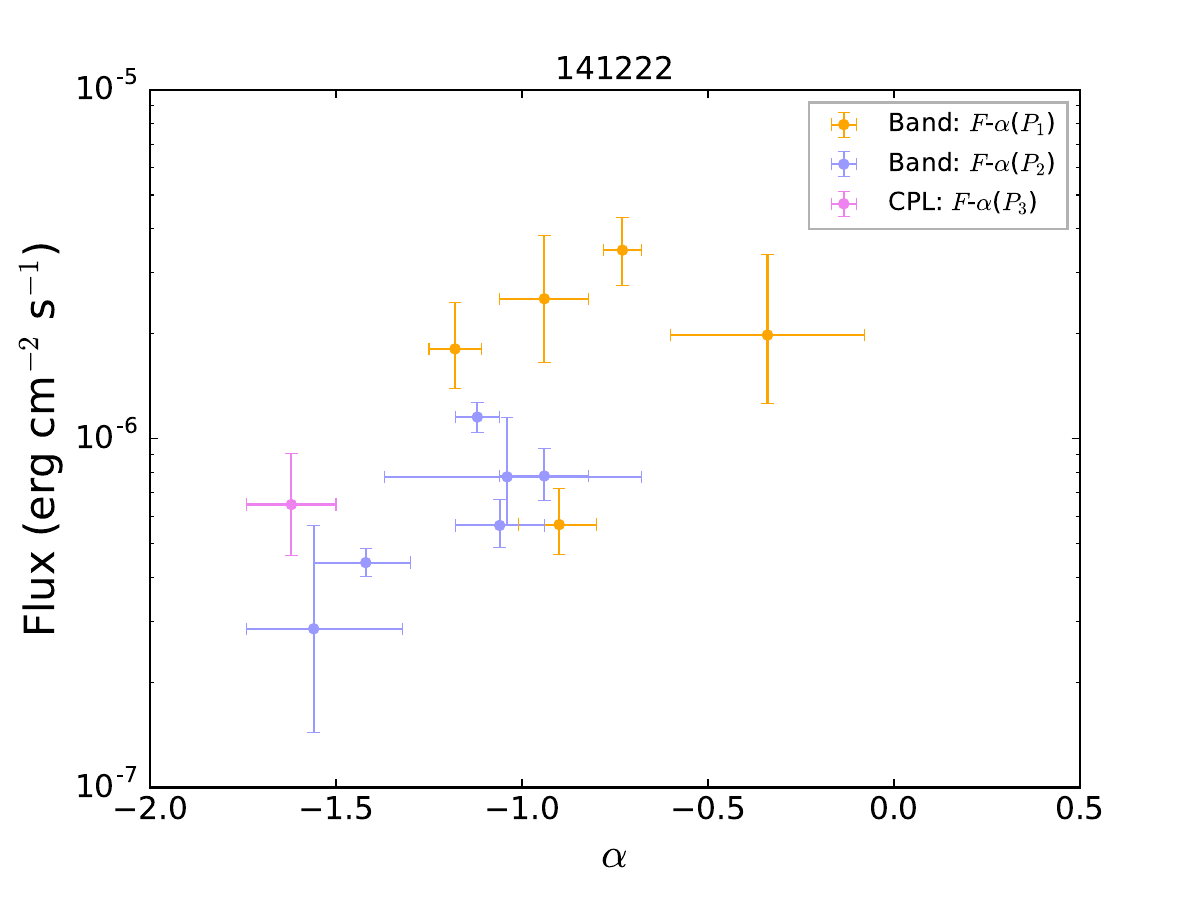}
\includegraphics[angle=0,scale=0.3]{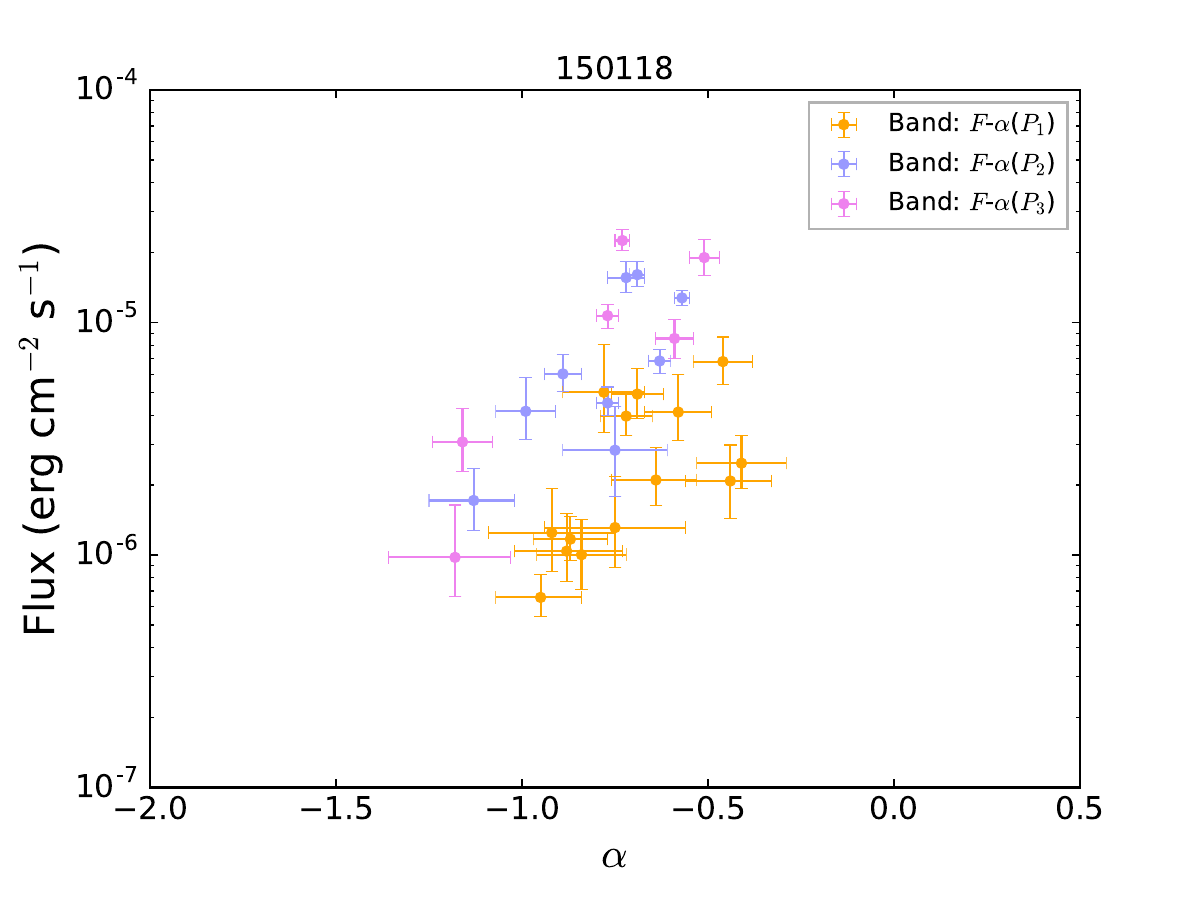}
\center{Fig. \ref{fig:FluxAlpha_Best}--- Continued}
\end{figure*}
\begin{figure*}
\includegraphics[angle=0,scale=0.3]{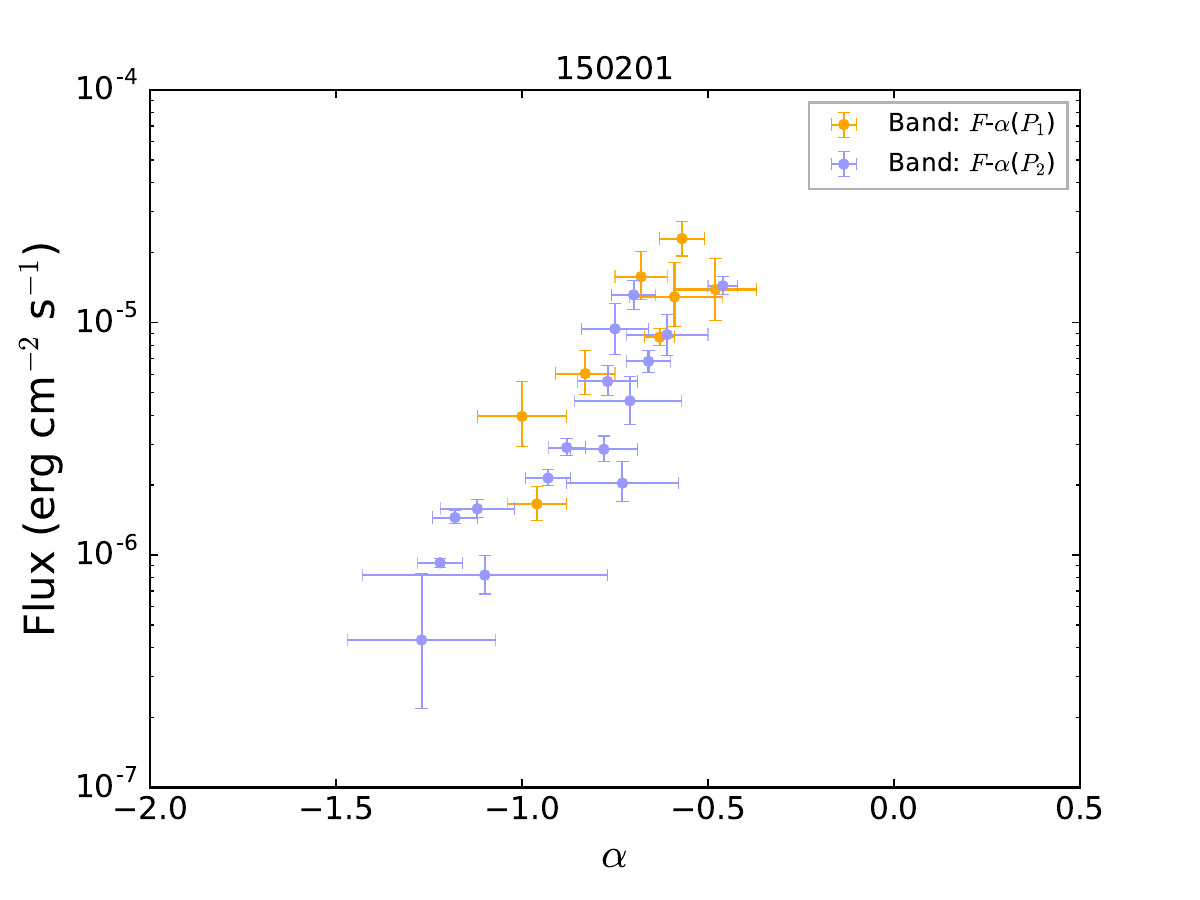}
\includegraphics[angle=0,scale=0.3]{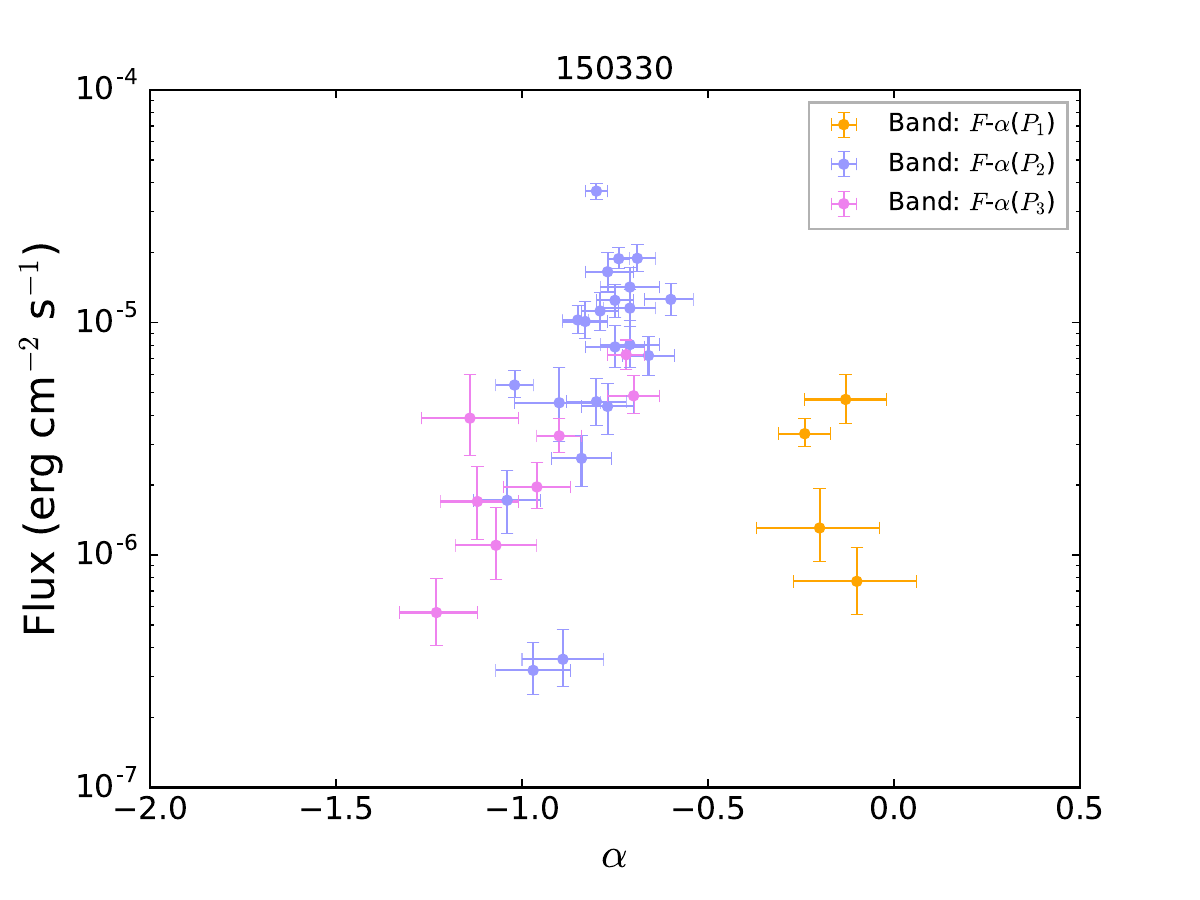}
\includegraphics[angle=0,scale=0.3]{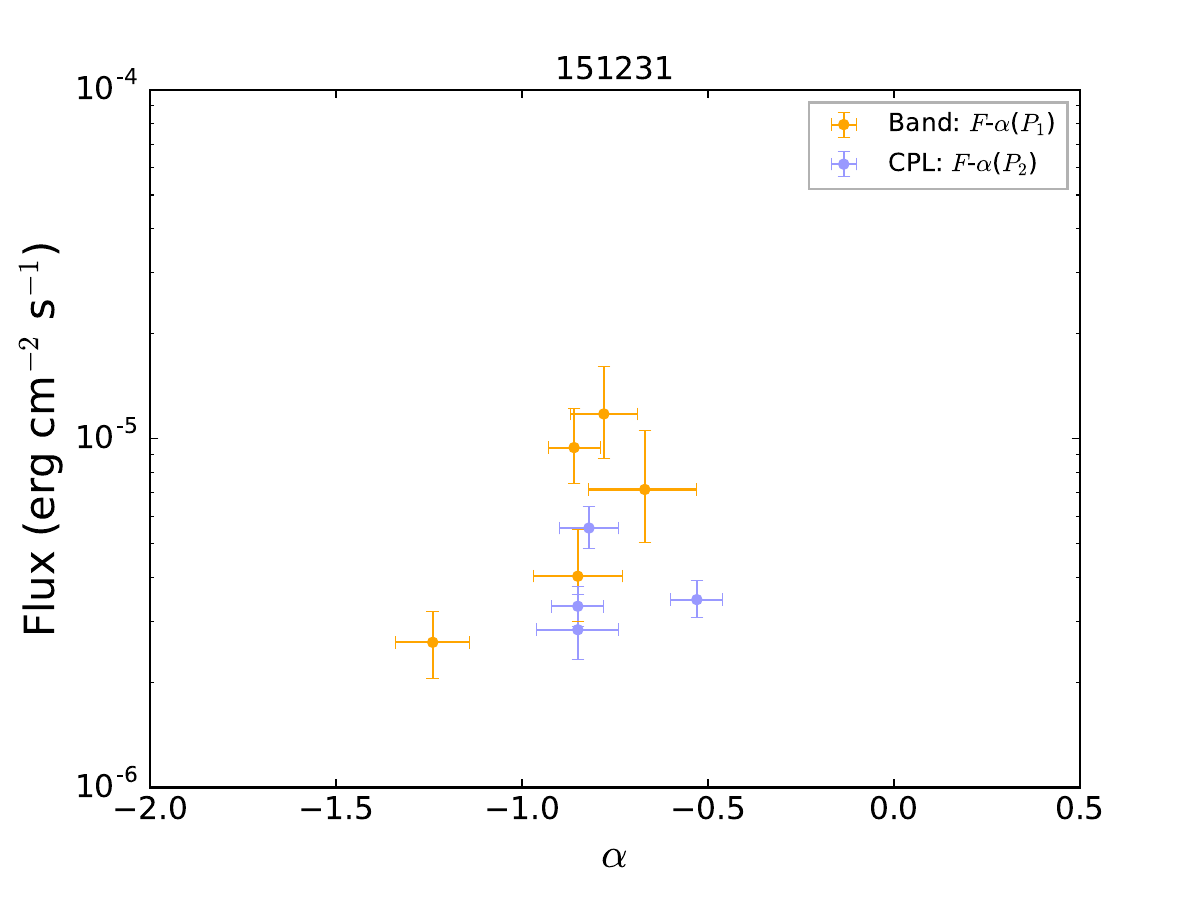}
\includegraphics[angle=0,scale=0.3]{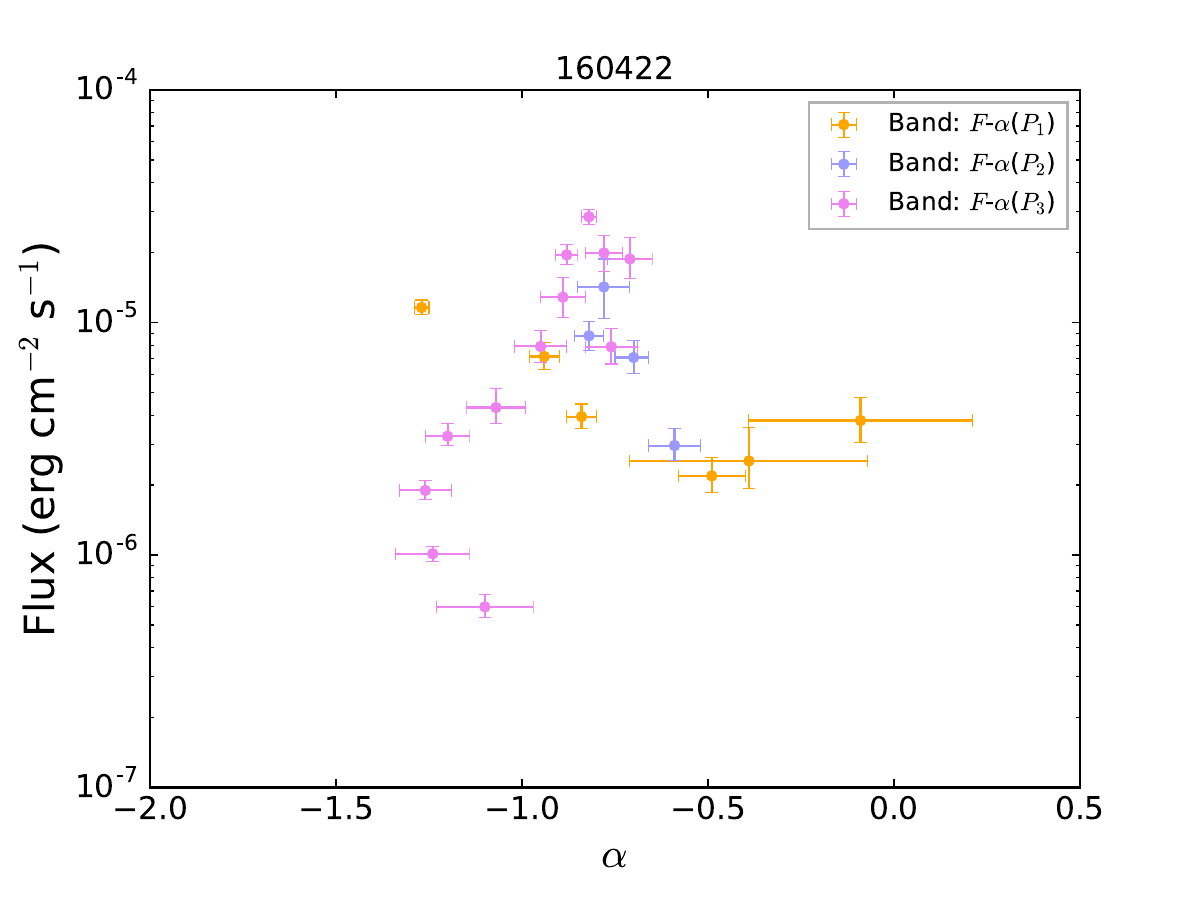}
\includegraphics[angle=0,scale=0.3]{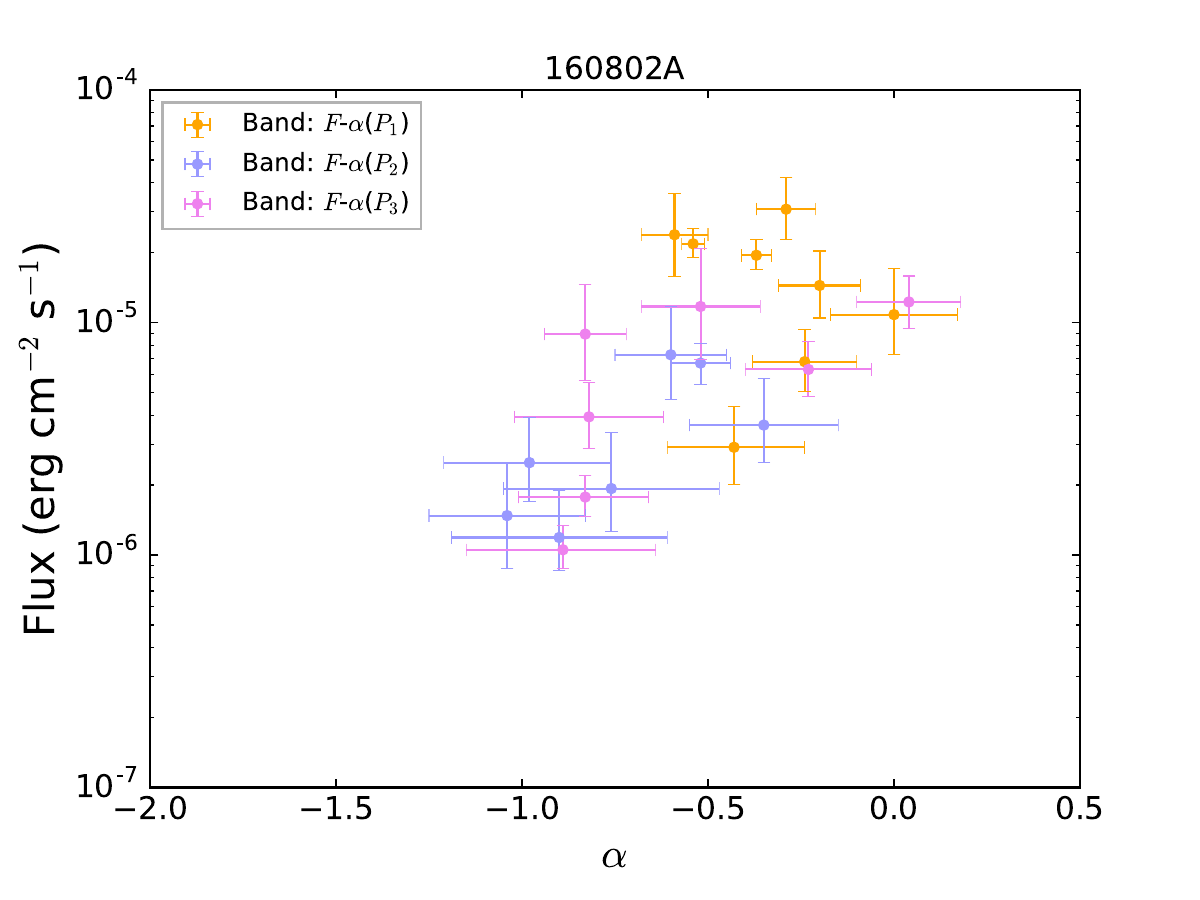}
\includegraphics[angle=0,scale=0.3]{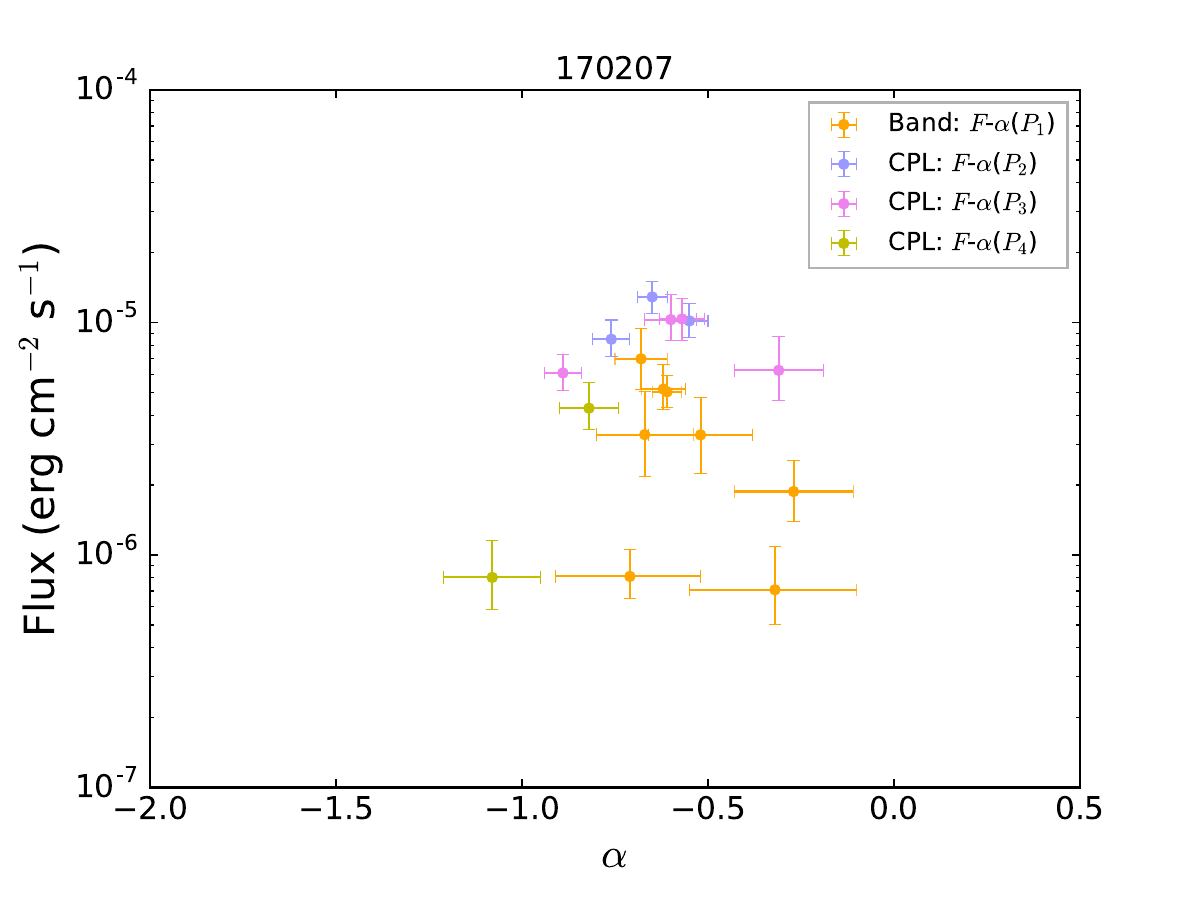}
\includegraphics[angle=0,scale=0.3]{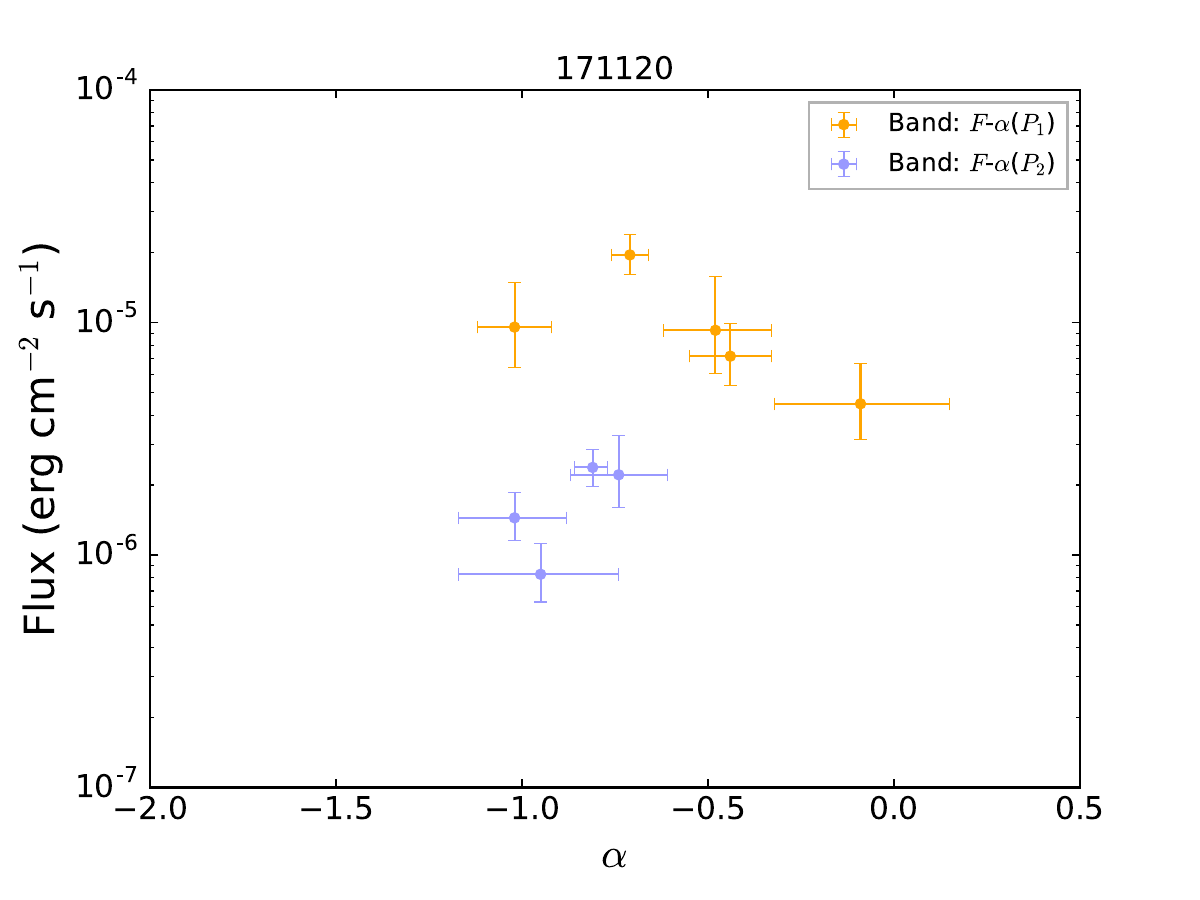}
\includegraphics[angle=0,scale=0.3]{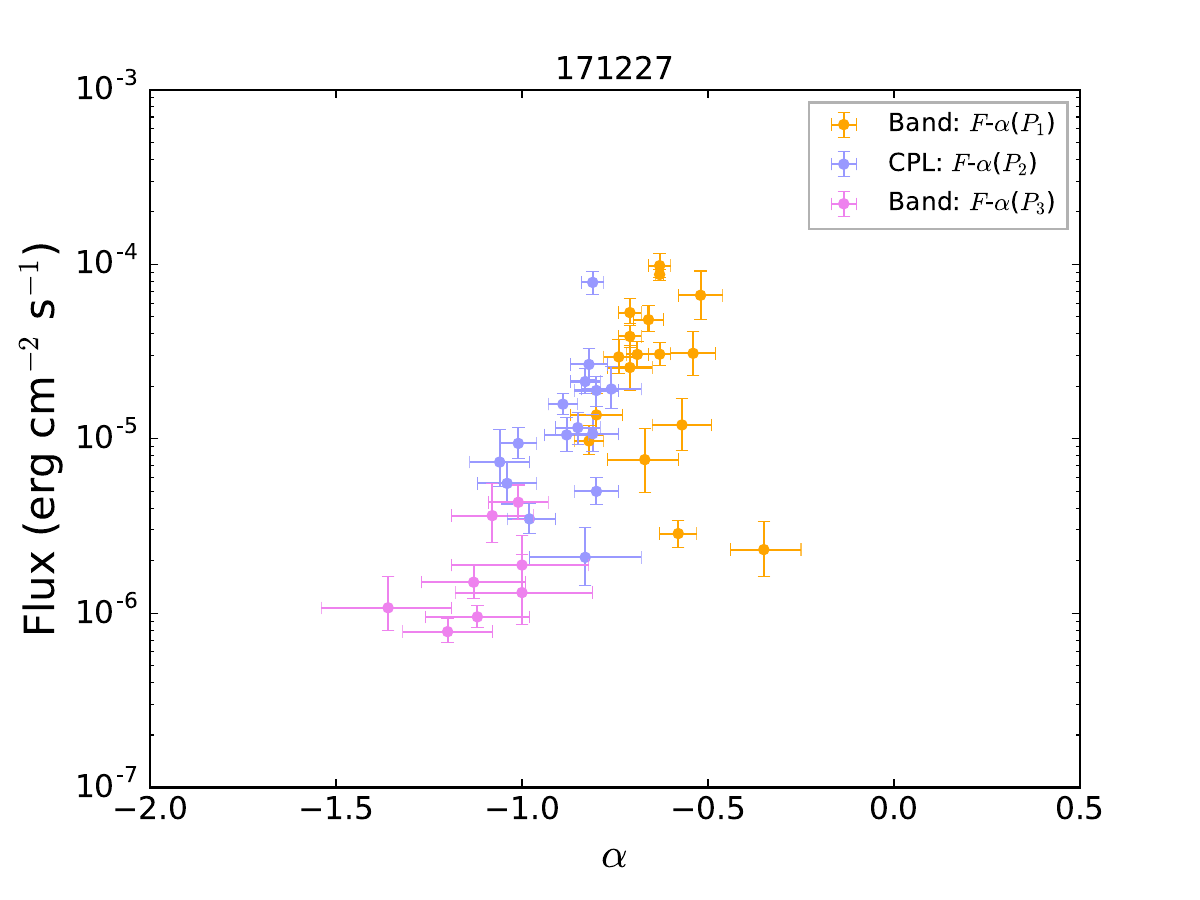}
\includegraphics[angle=0,scale=0.3]{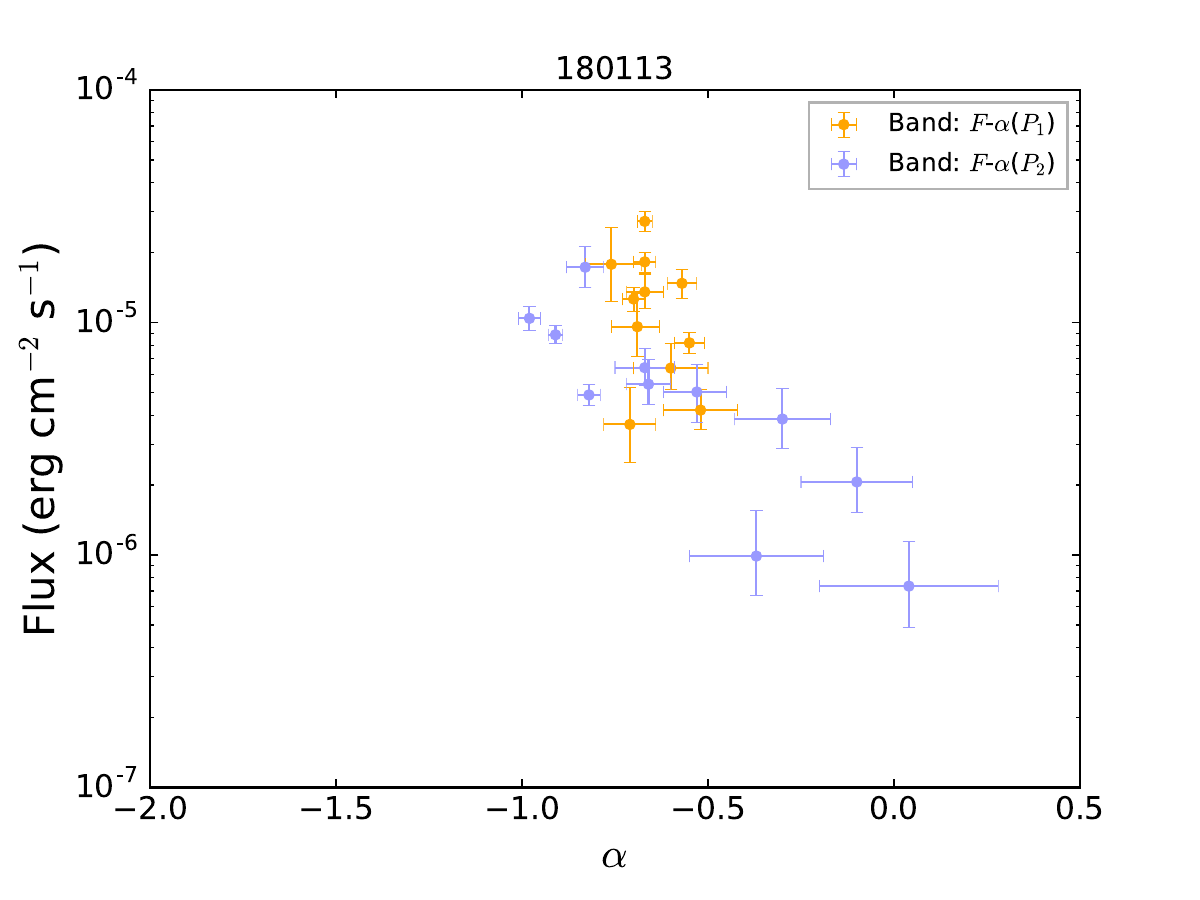}
\includegraphics[angle=0,scale=0.3]{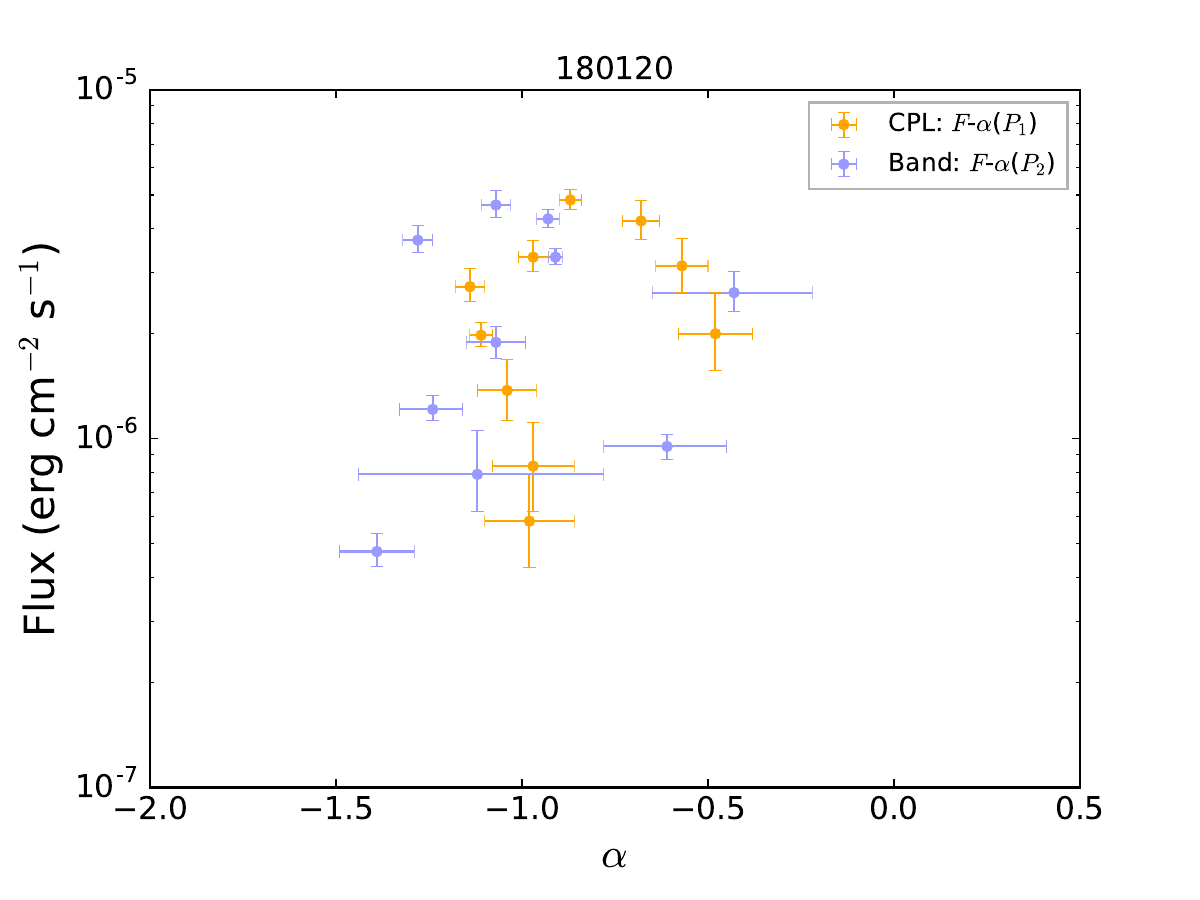}
\includegraphics[angle=0,scale=0.3]{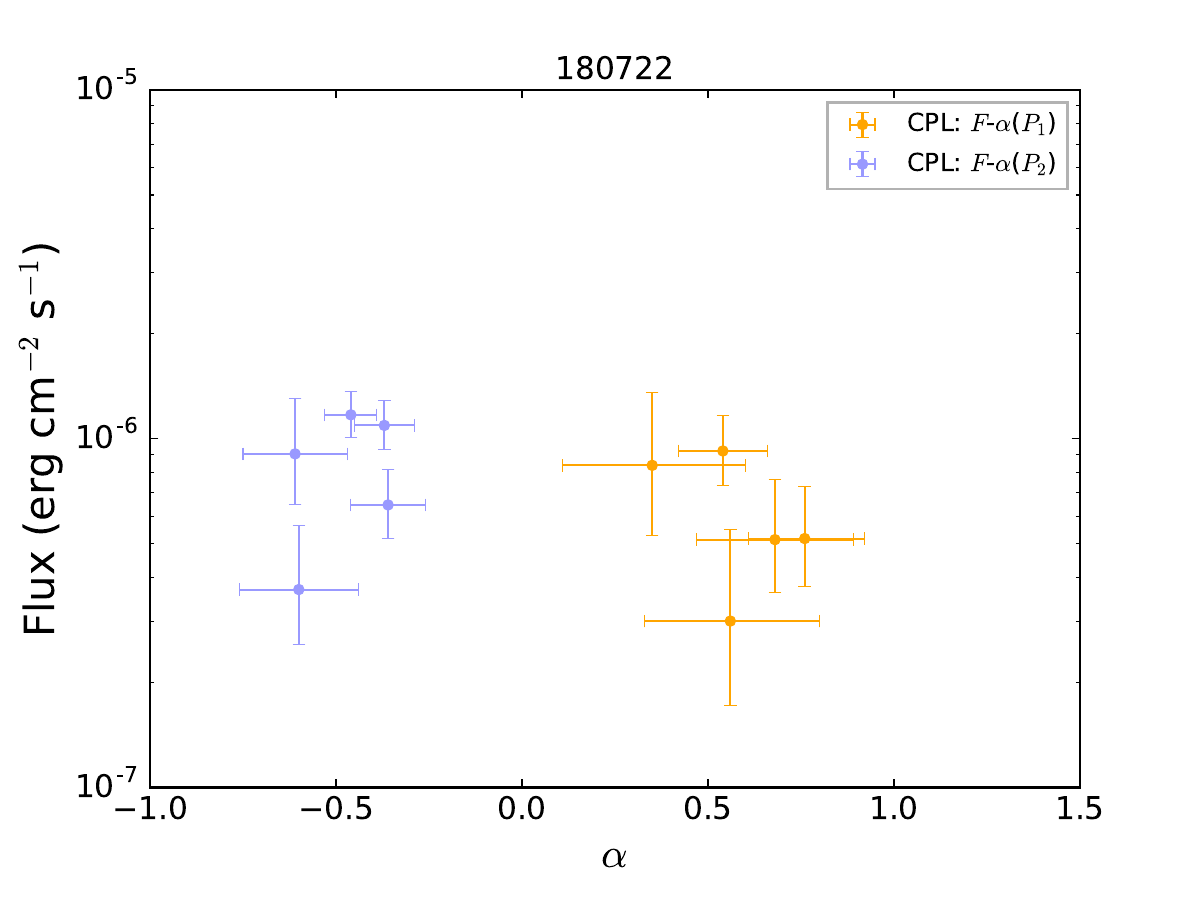}
\includegraphics[angle=0,scale=0.3]{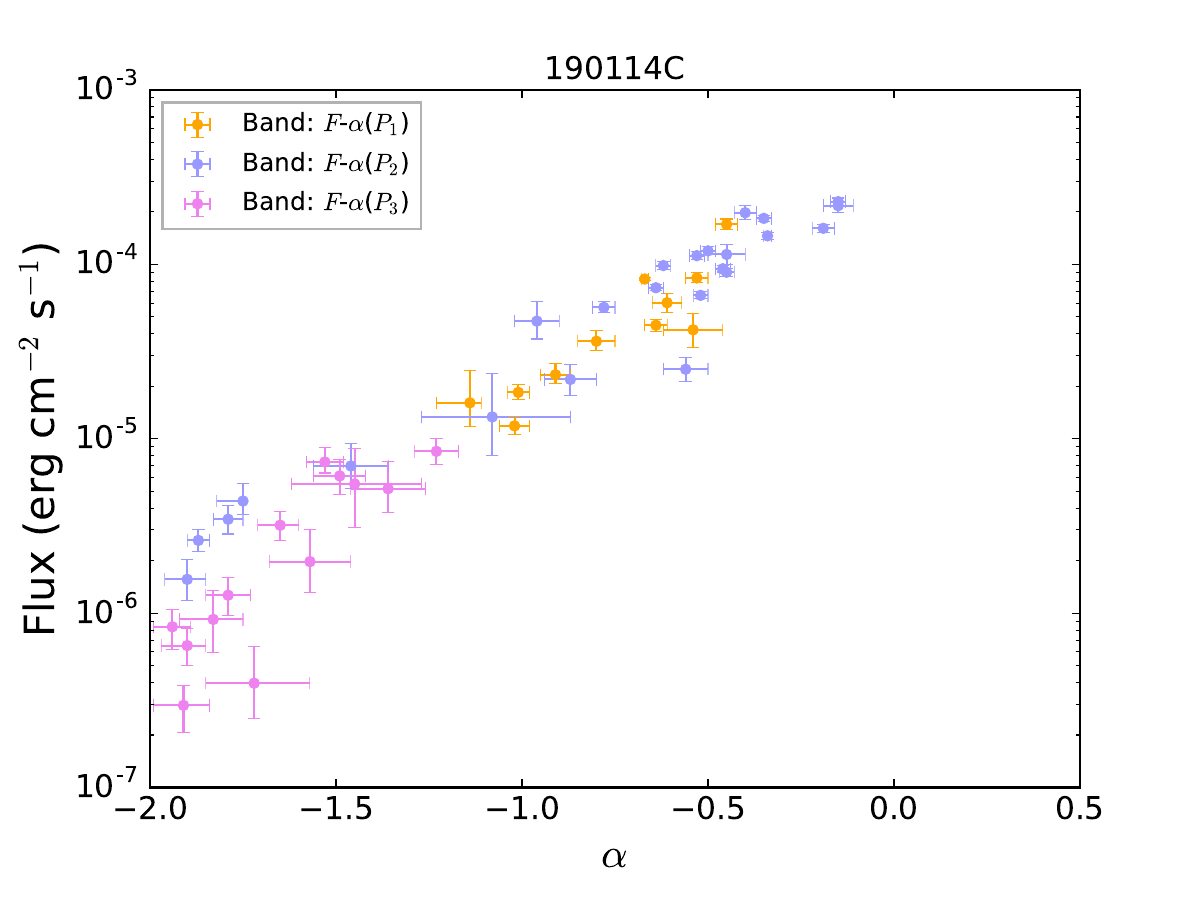}
\center{Fig. \ref{fig:FluxAlpha_Best}--- Continued}
\end{figure*}

\clearpage
\begin{figure*}
\includegraphics[angle=0,scale=0.3]{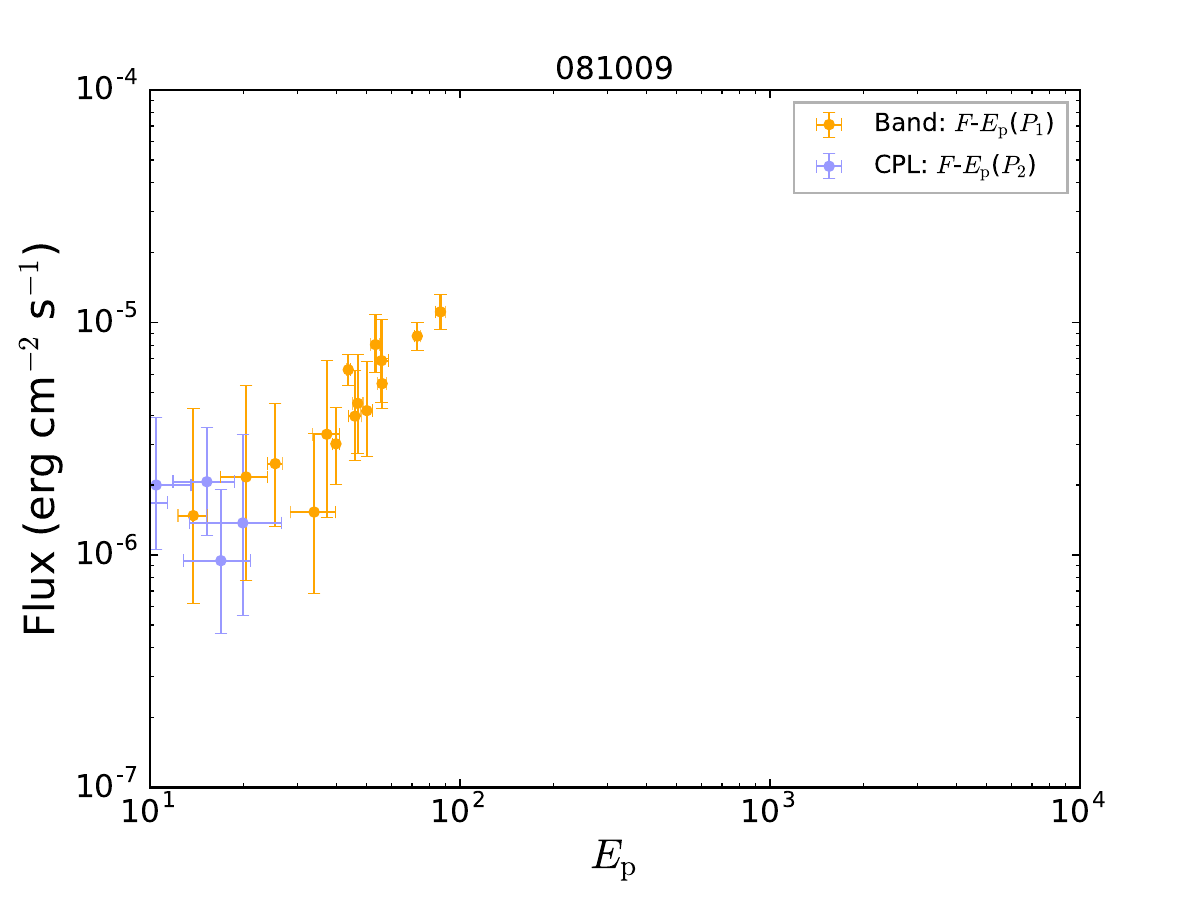}
\includegraphics[angle=0,scale=0.3]{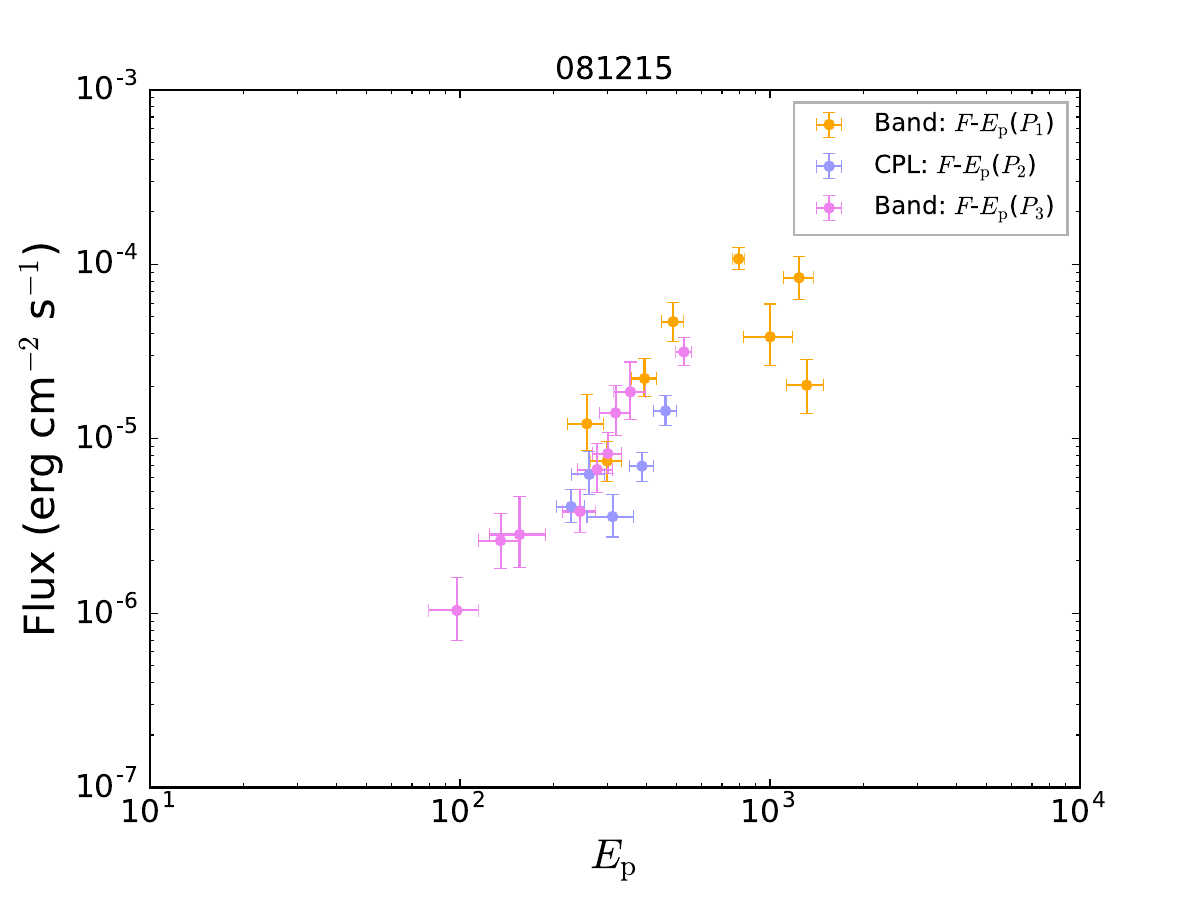}
\includegraphics[angle=0,scale=0.3]{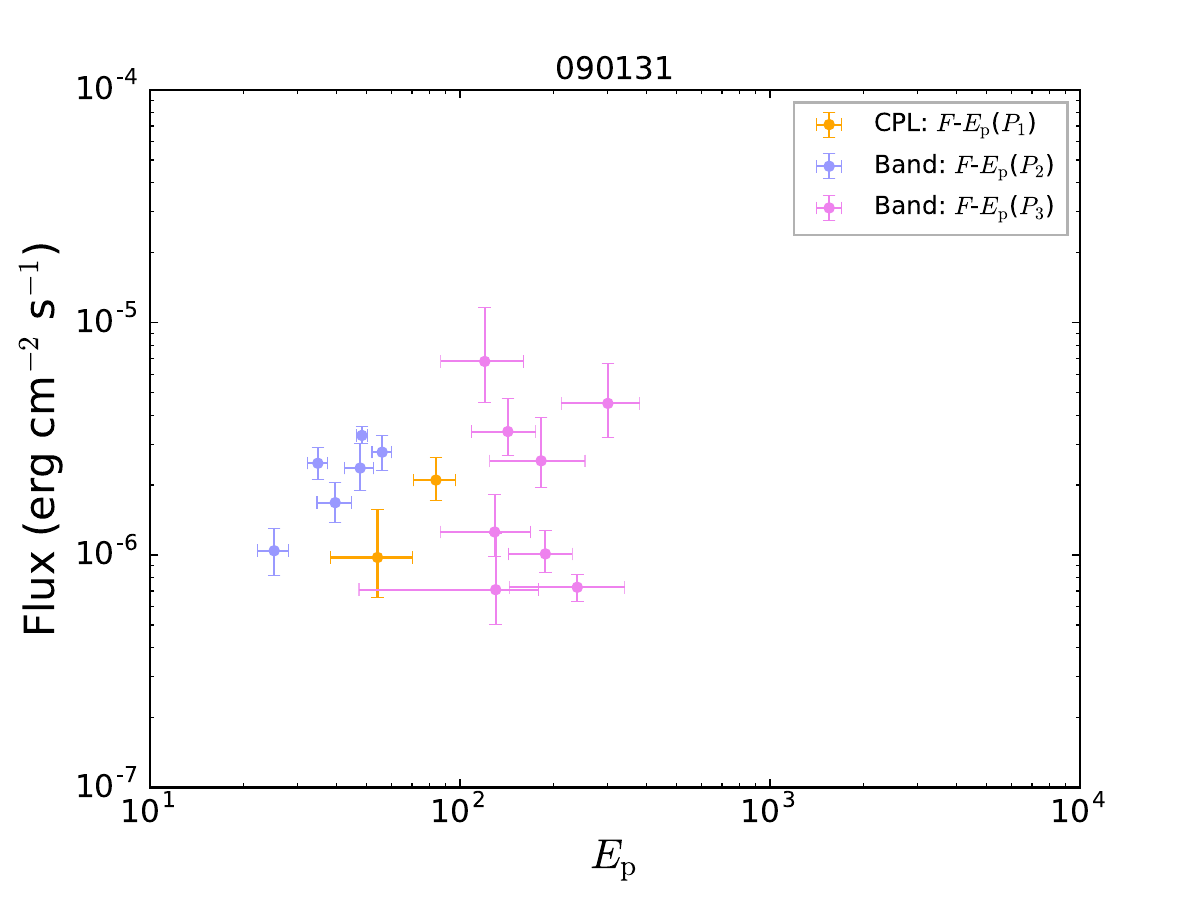}
\includegraphics[angle=0,scale=0.3]{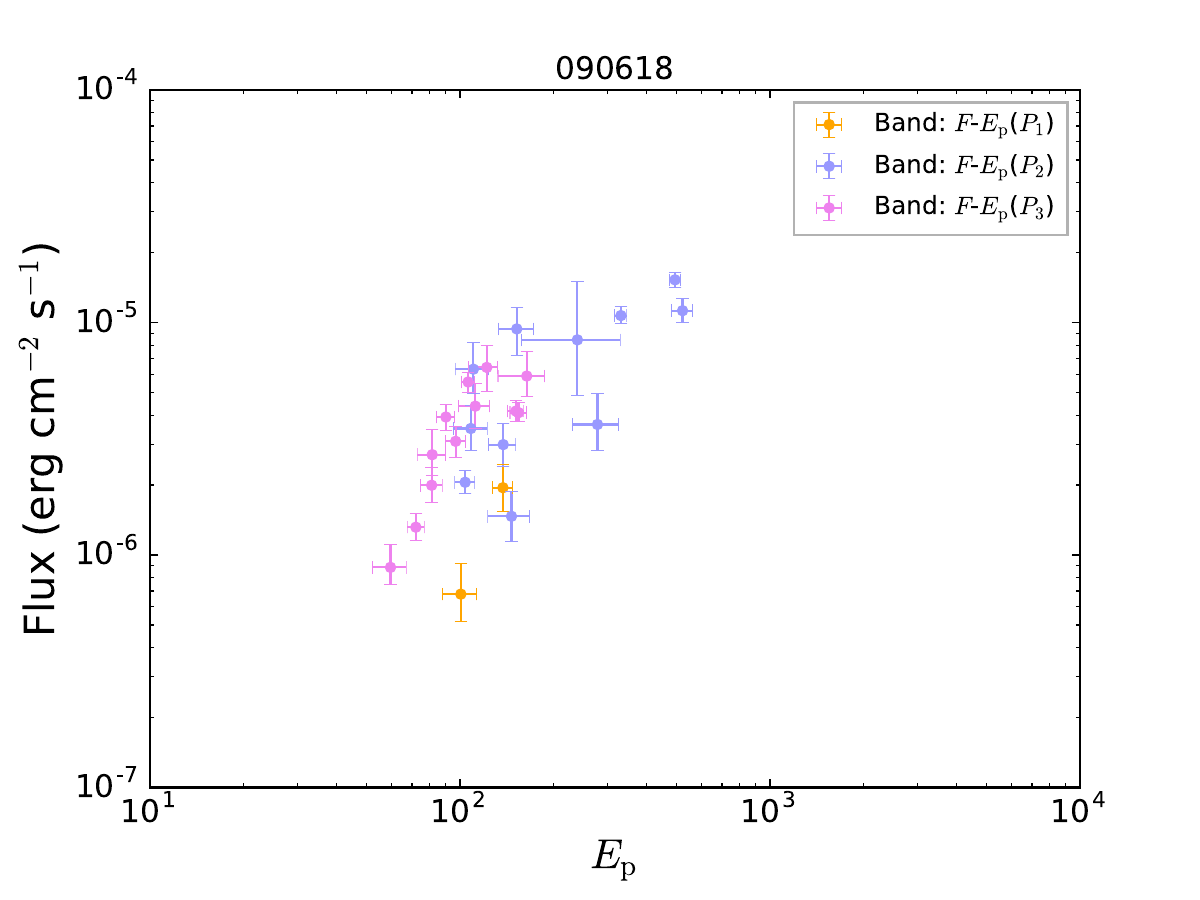}
\includegraphics[angle=0,scale=0.3]{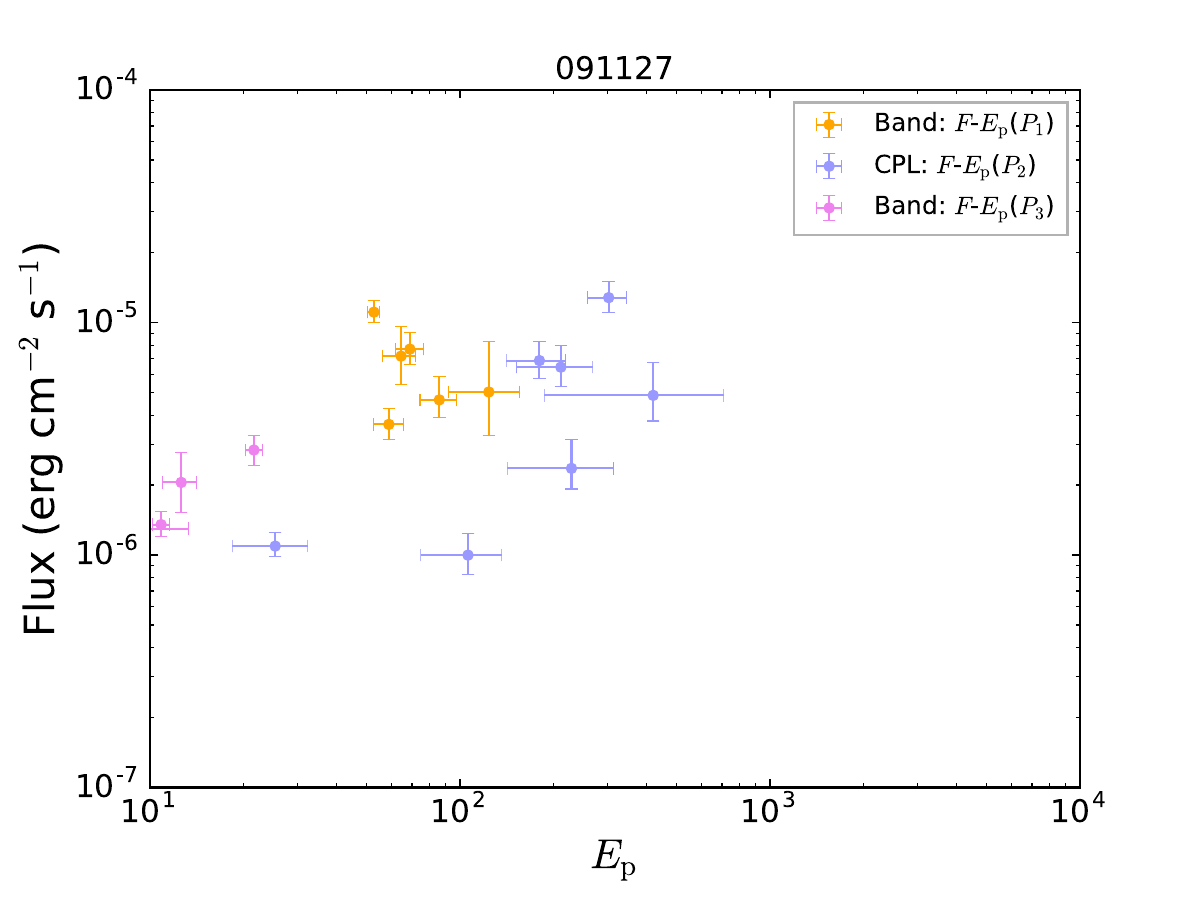}
\includegraphics[angle=0,scale=0.3]{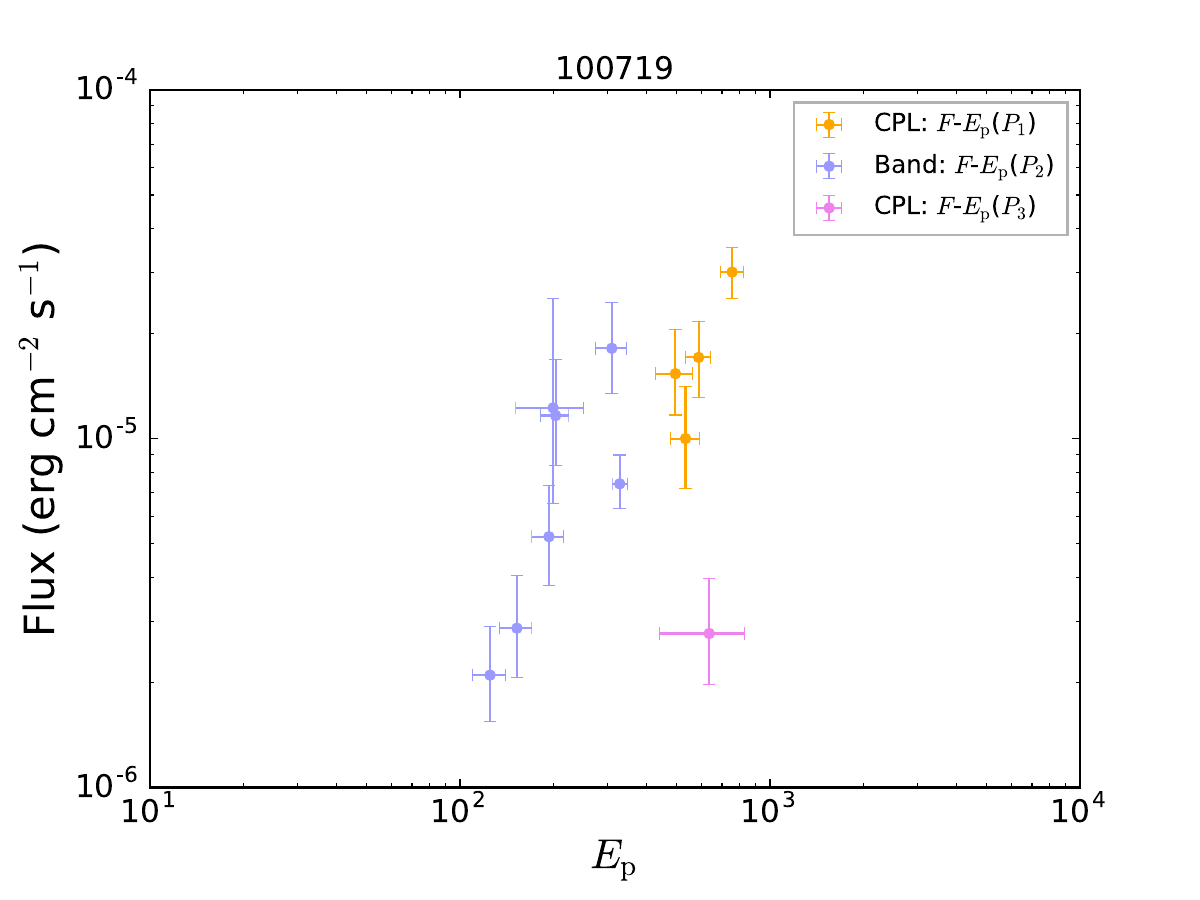}
\includegraphics[angle=0,scale=0.3]{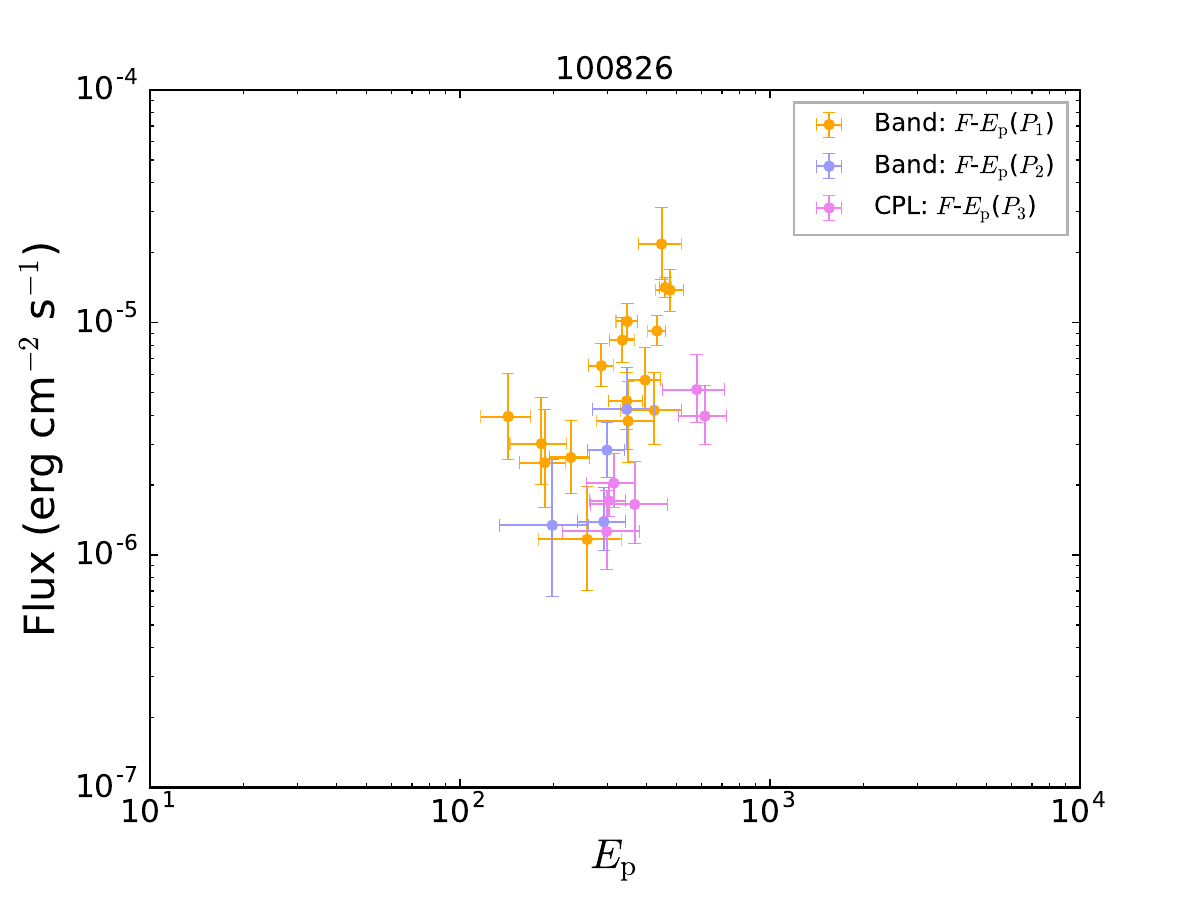}
\includegraphics[angle=0,scale=0.3]{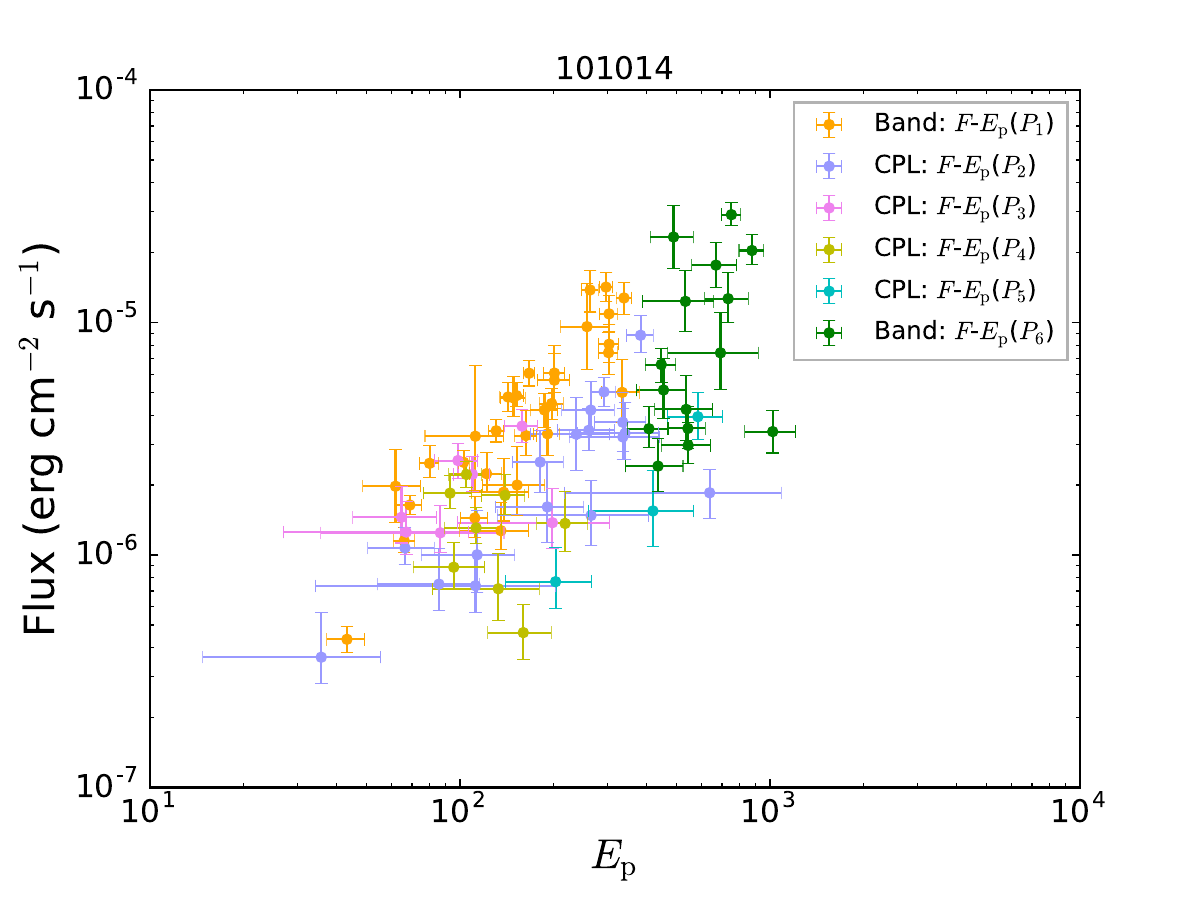}
\includegraphics[angle=0,scale=0.3]{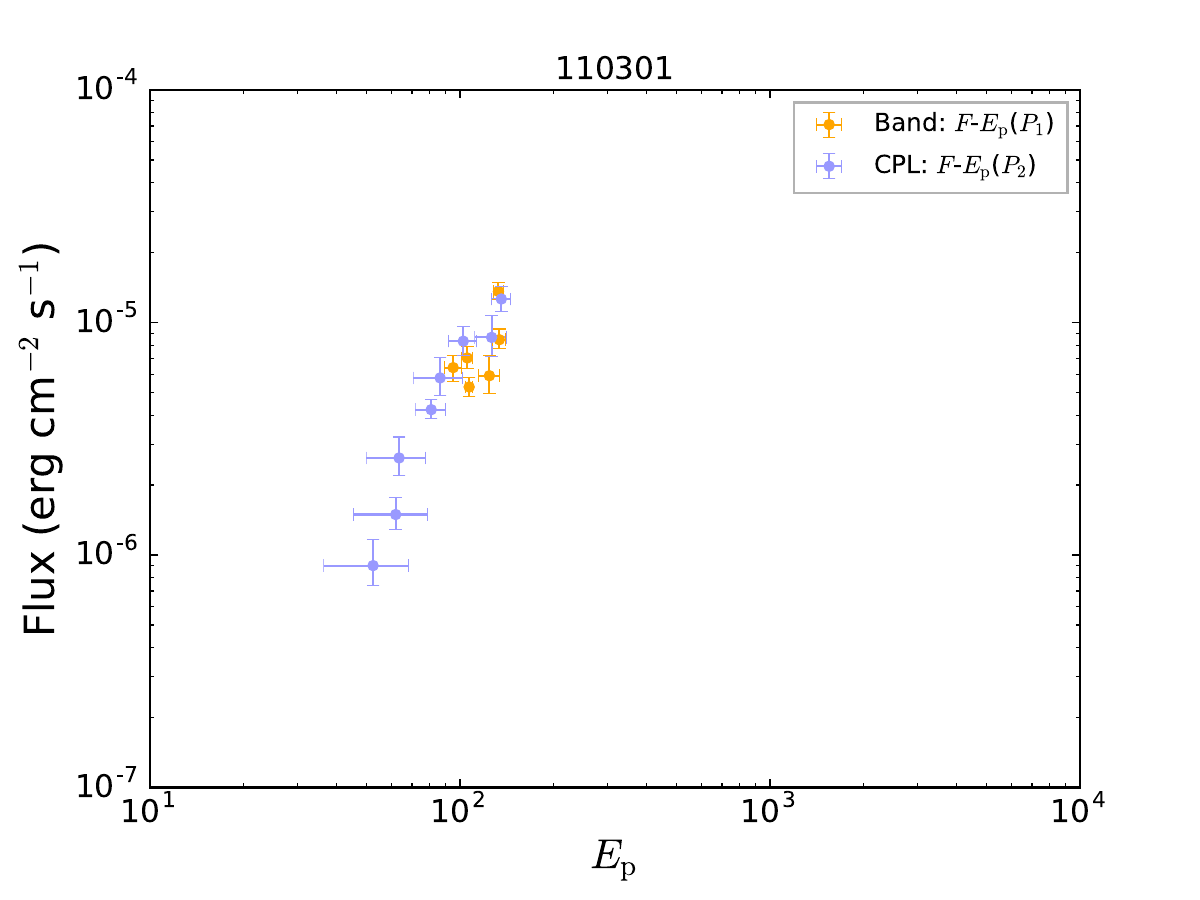}
\includegraphics[angle=0,scale=0.3]{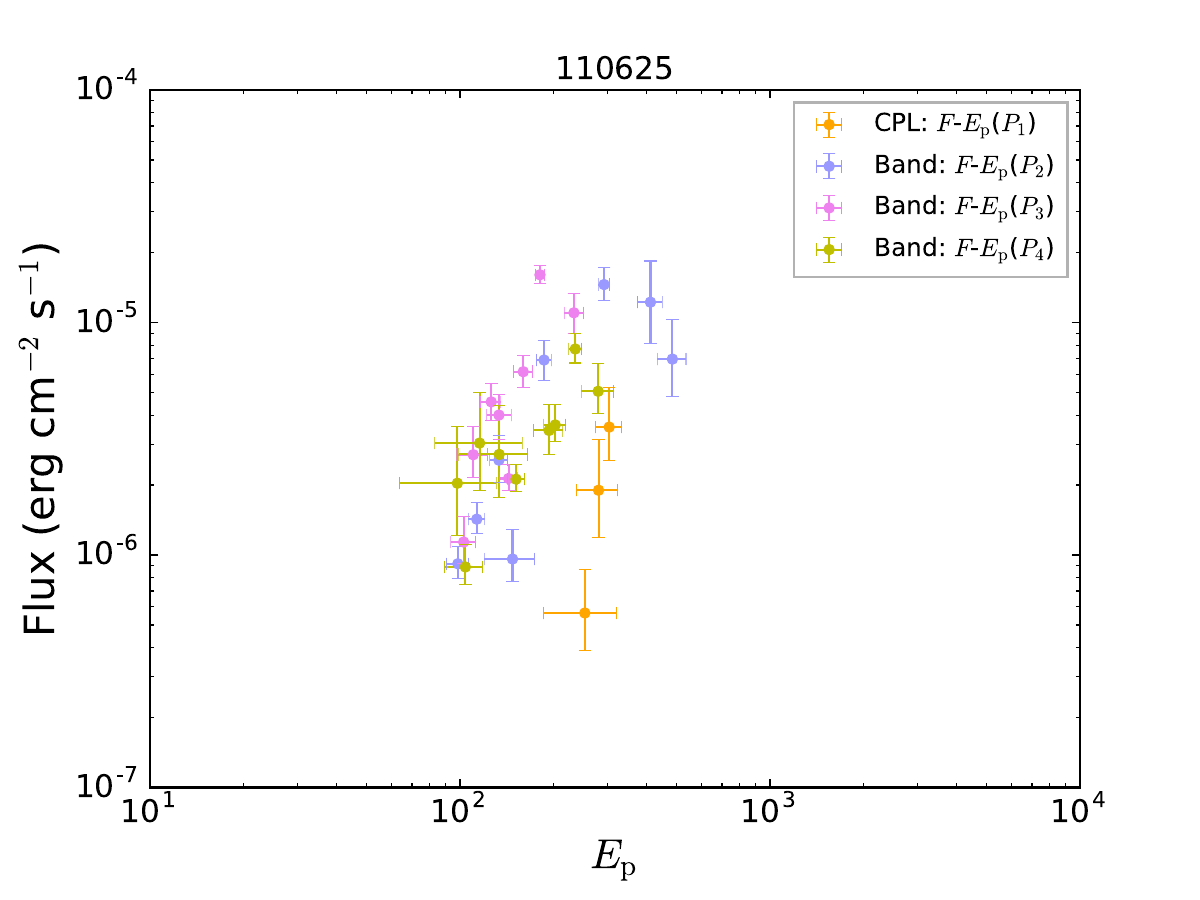}
\includegraphics[angle=0,scale=0.3]{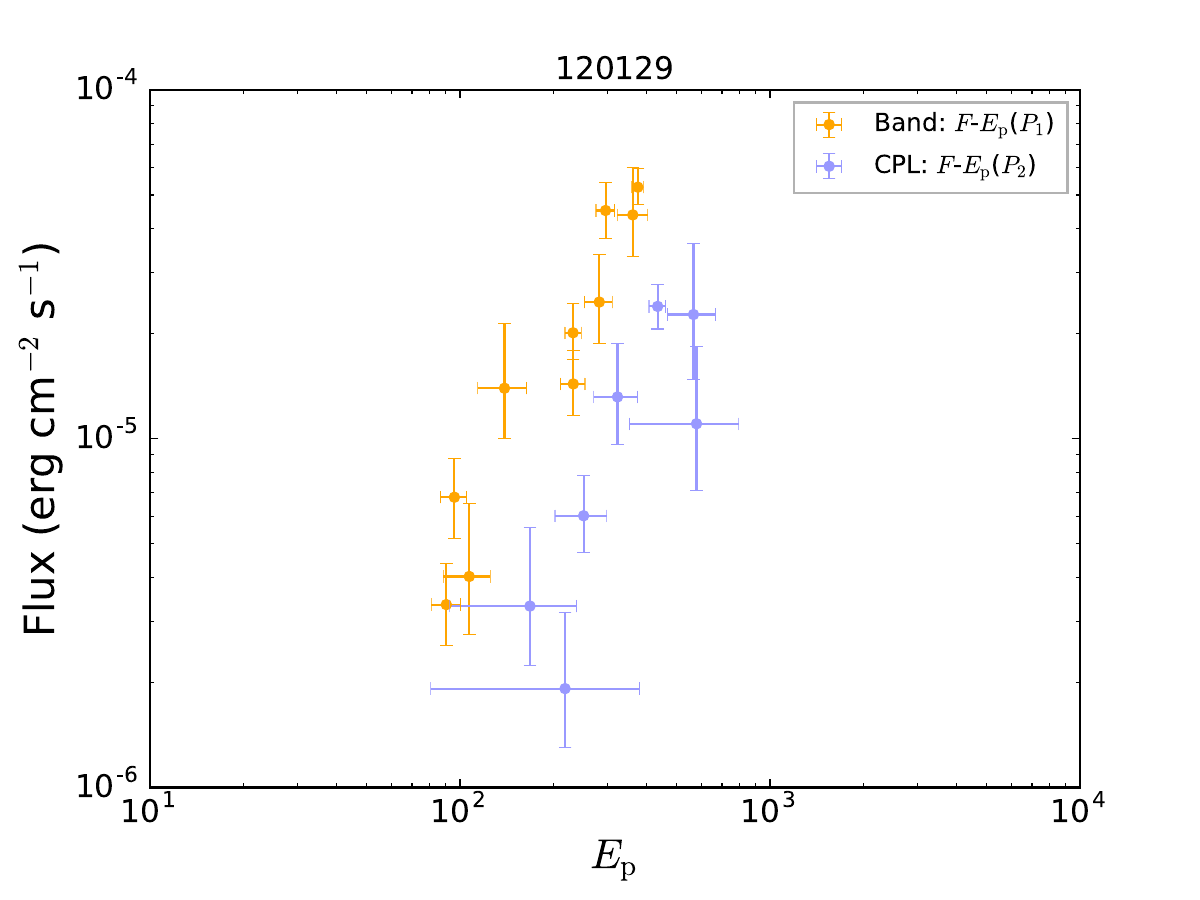}
\includegraphics[angle=0,scale=0.3]{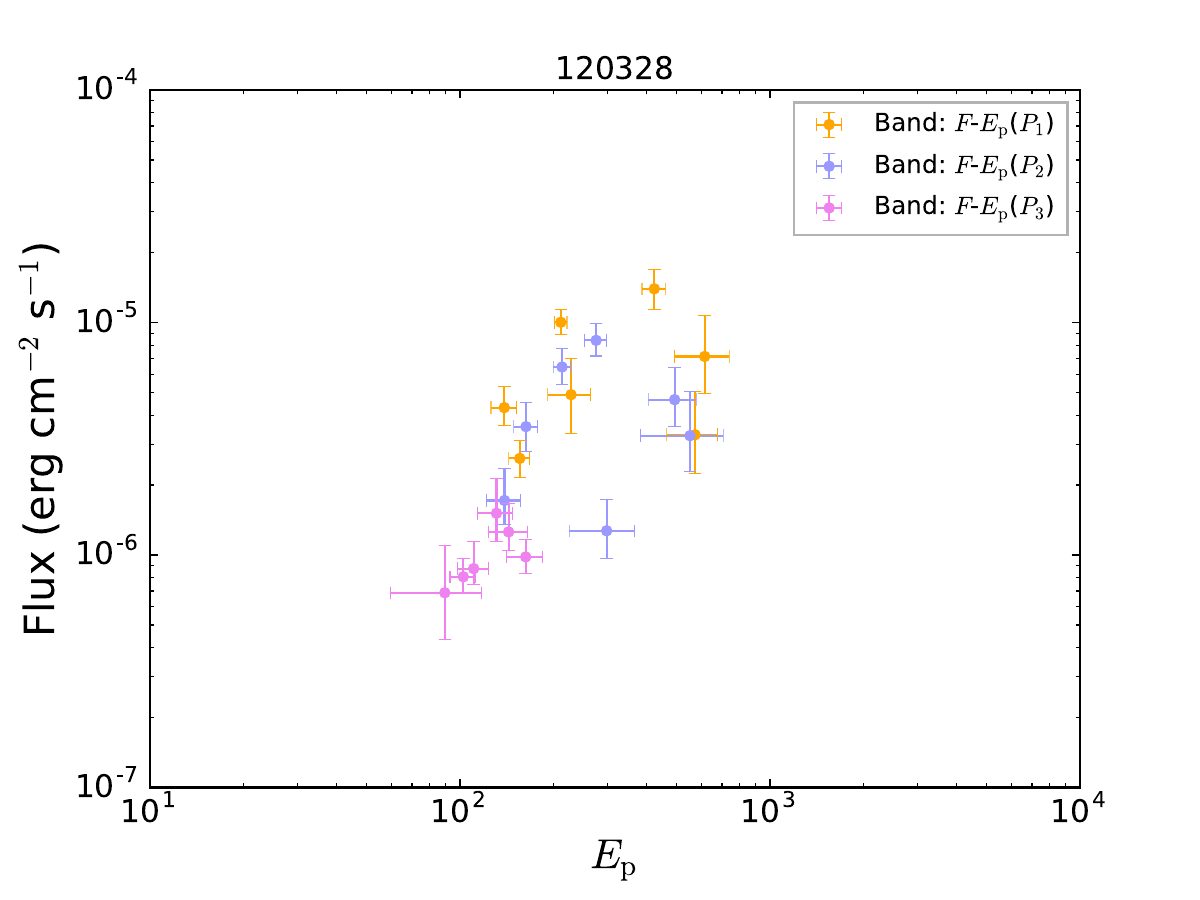}
\caption{The $F$-$E_{\rm p}$ relation. The symbols and colors are the same as in Figure \ref{fig:FluxAlpha_Best}.}\label{fig:FluxEp_Best}
\end{figure*}
\begin{figure*}
\includegraphics[angle=0,scale=0.3]{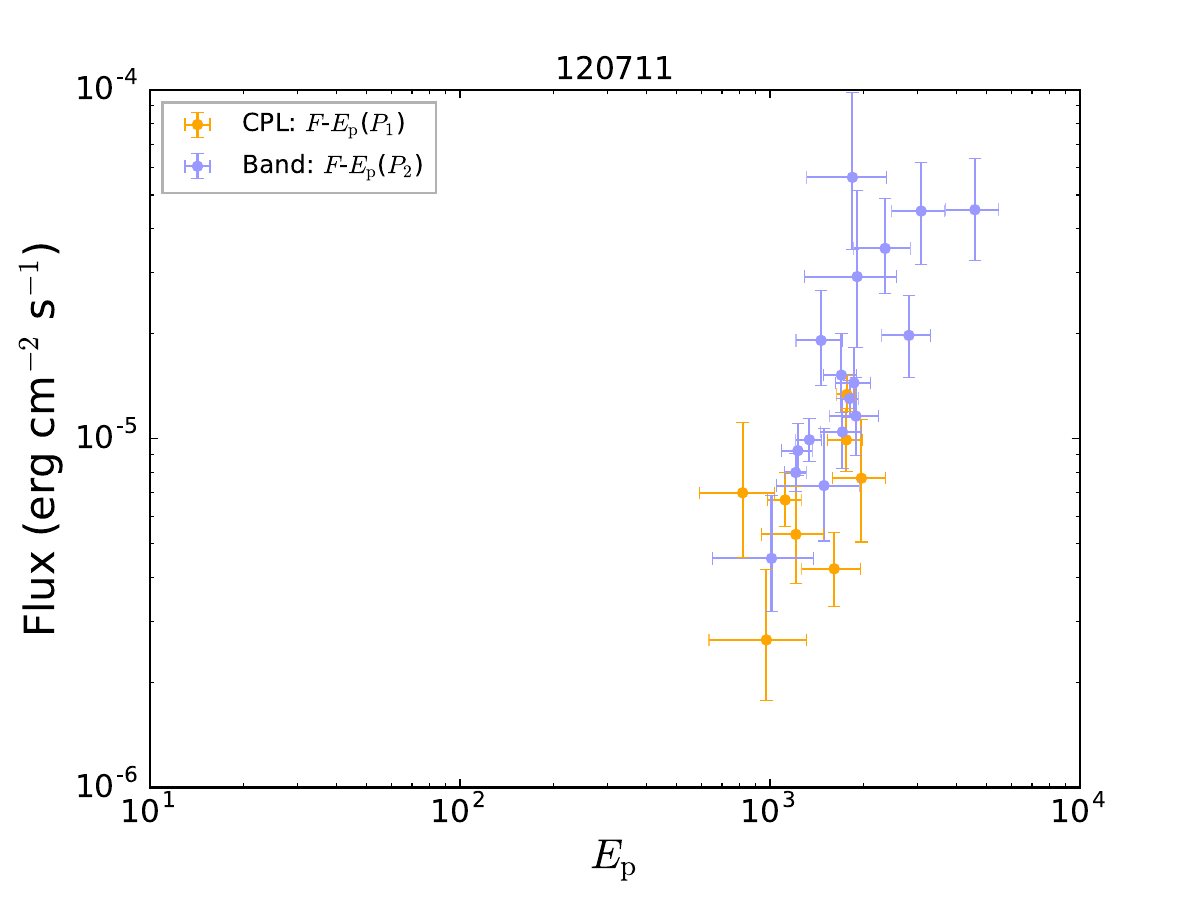}
\includegraphics[angle=0,scale=0.3]{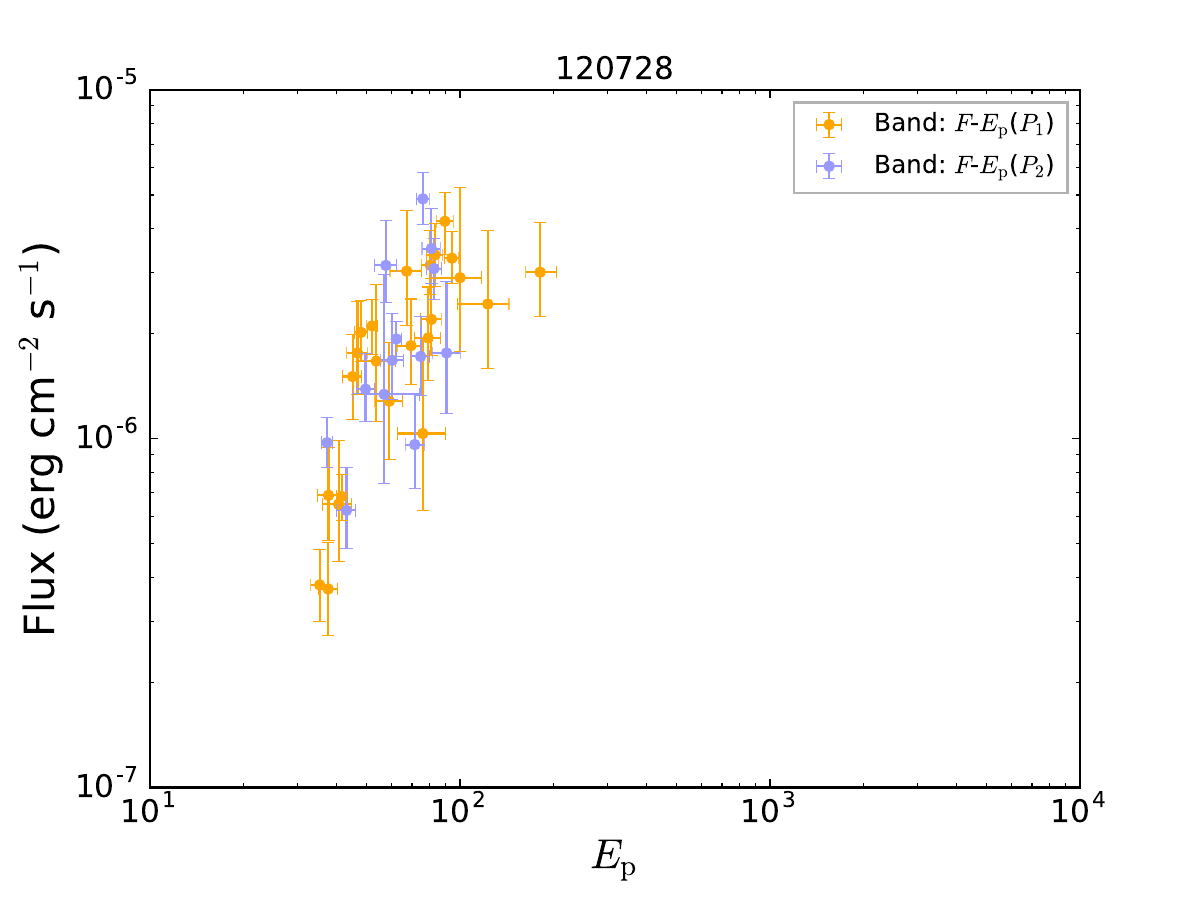}
\includegraphics[angle=0,scale=0.3]{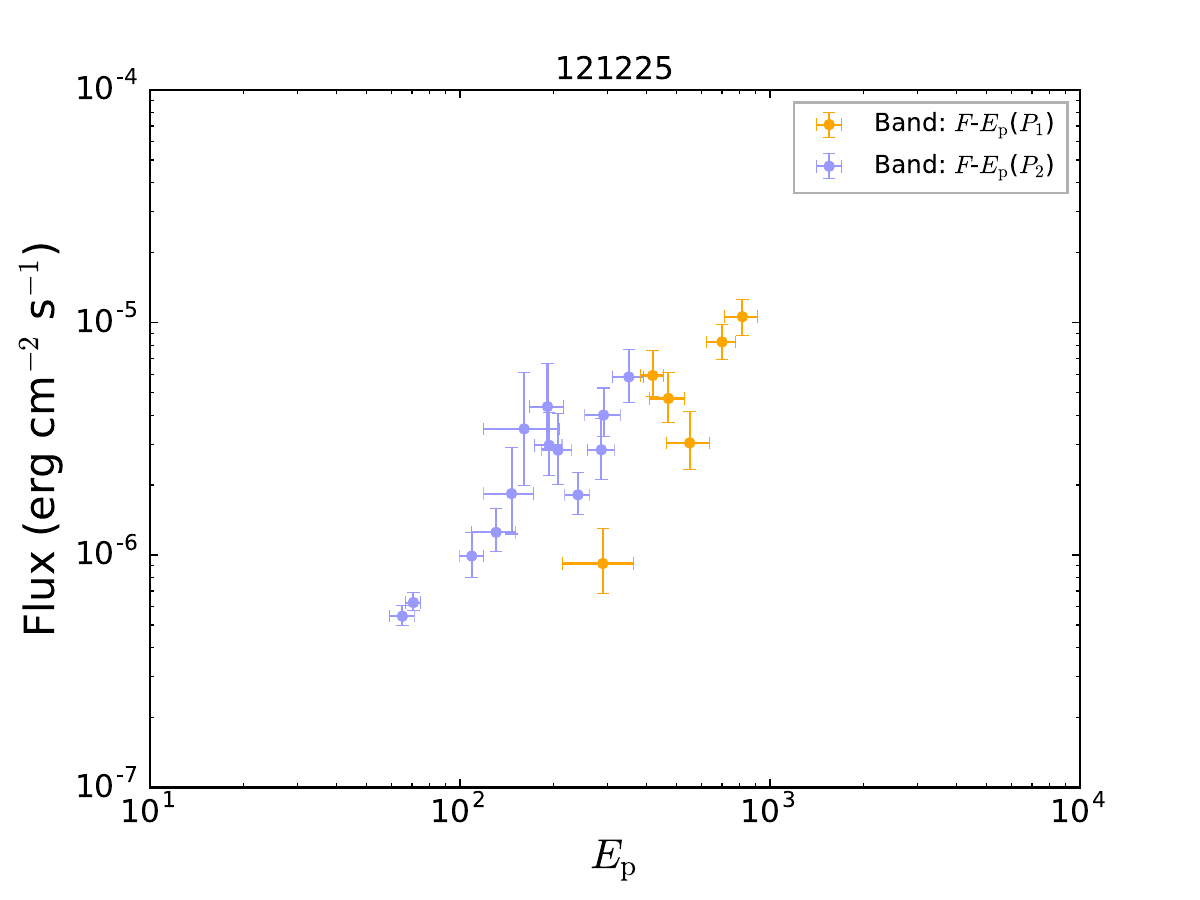}
\includegraphics[angle=0,scale=0.3]{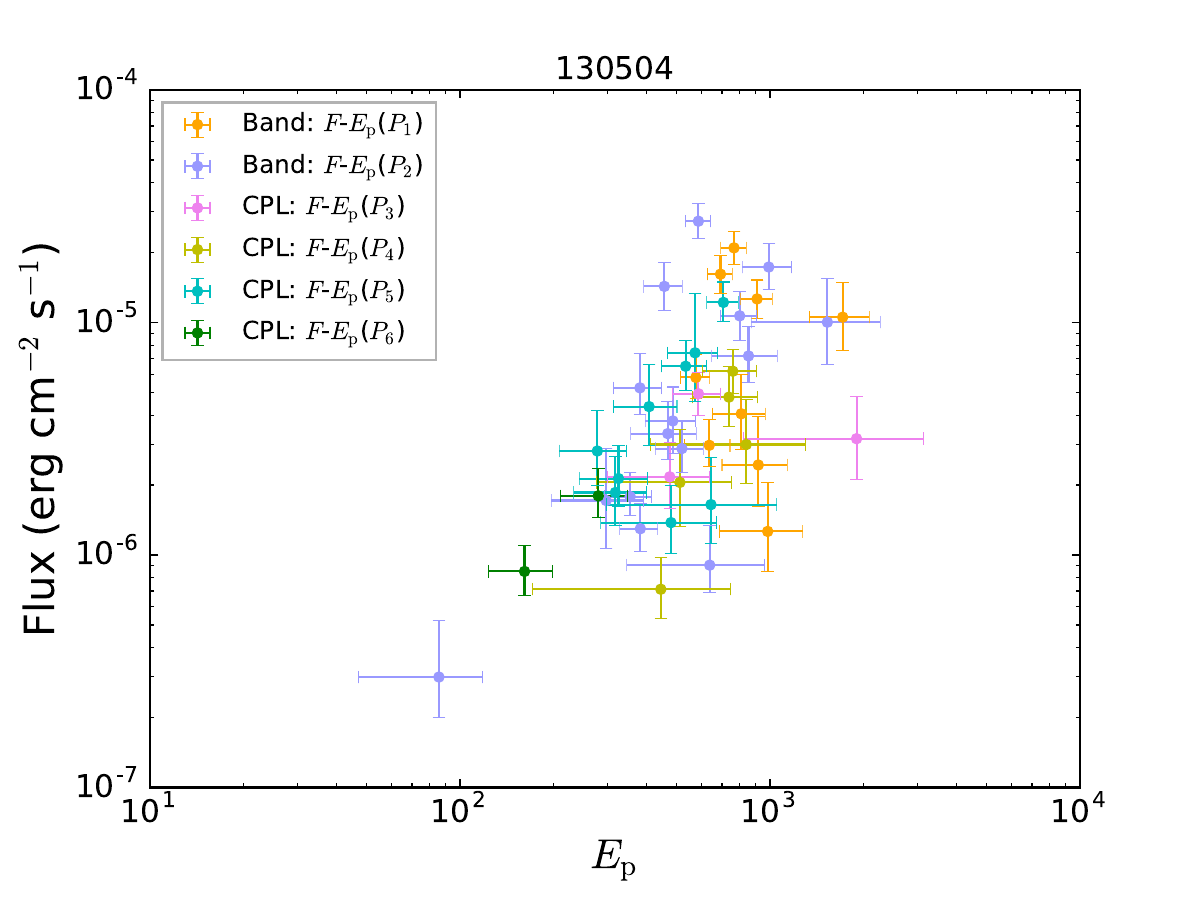}
\includegraphics[angle=0,scale=0.3]{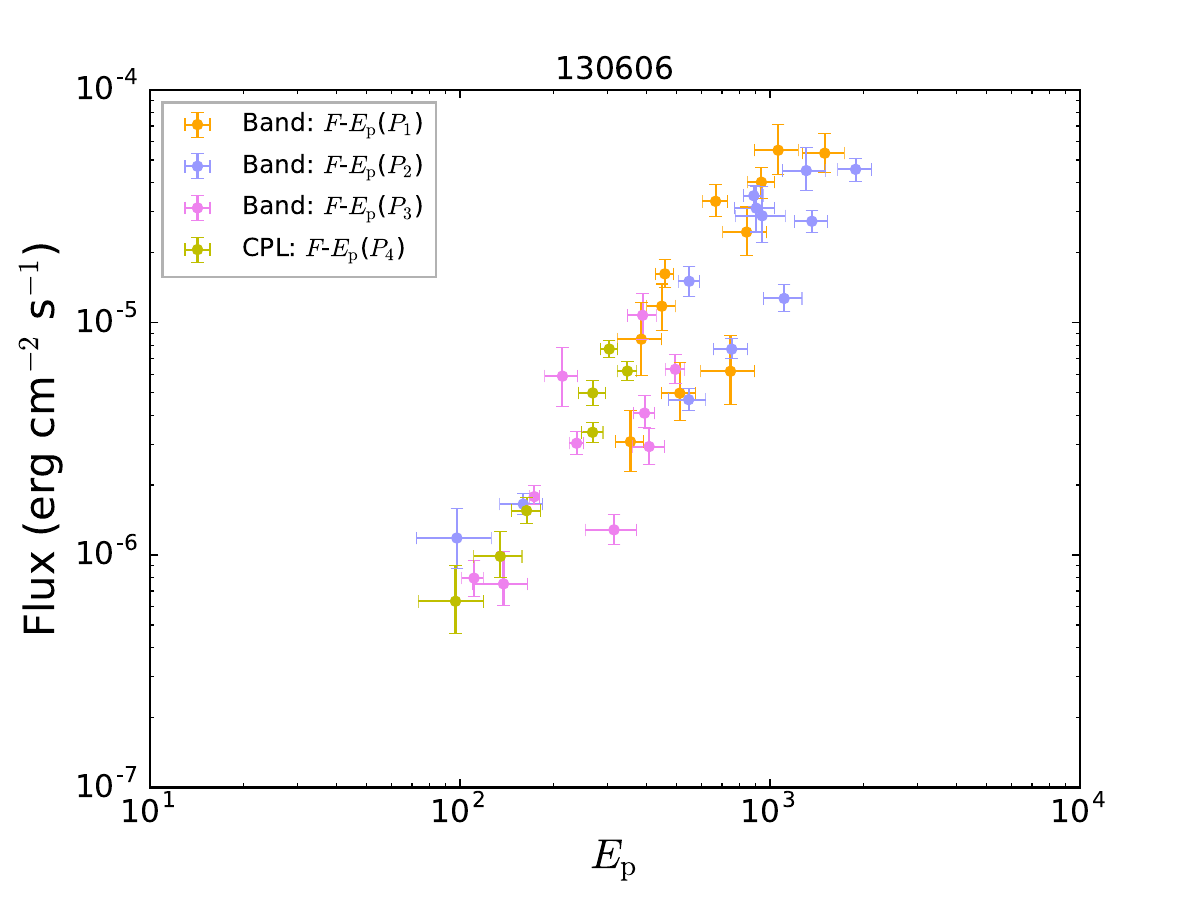}
\includegraphics[angle=0,scale=0.3]{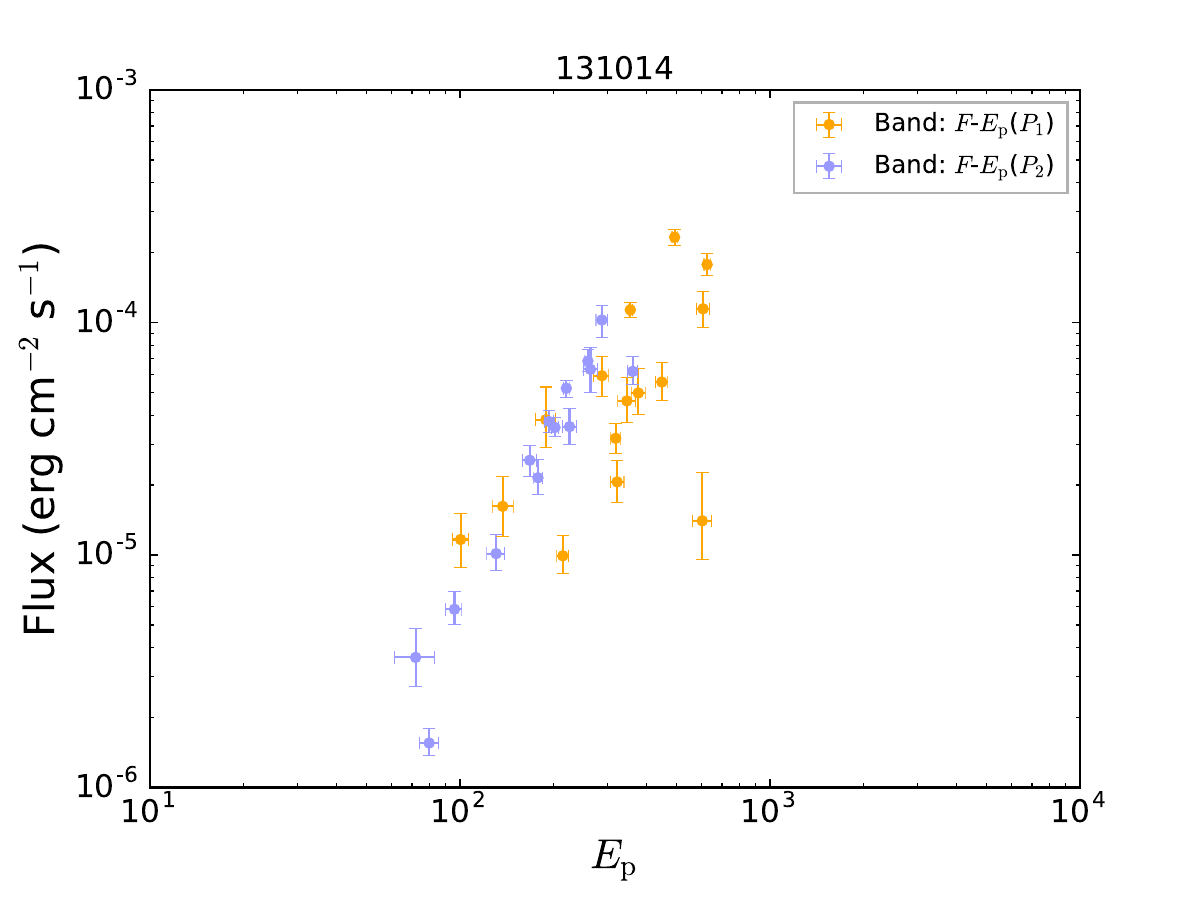}
\includegraphics[angle=0,scale=0.3]{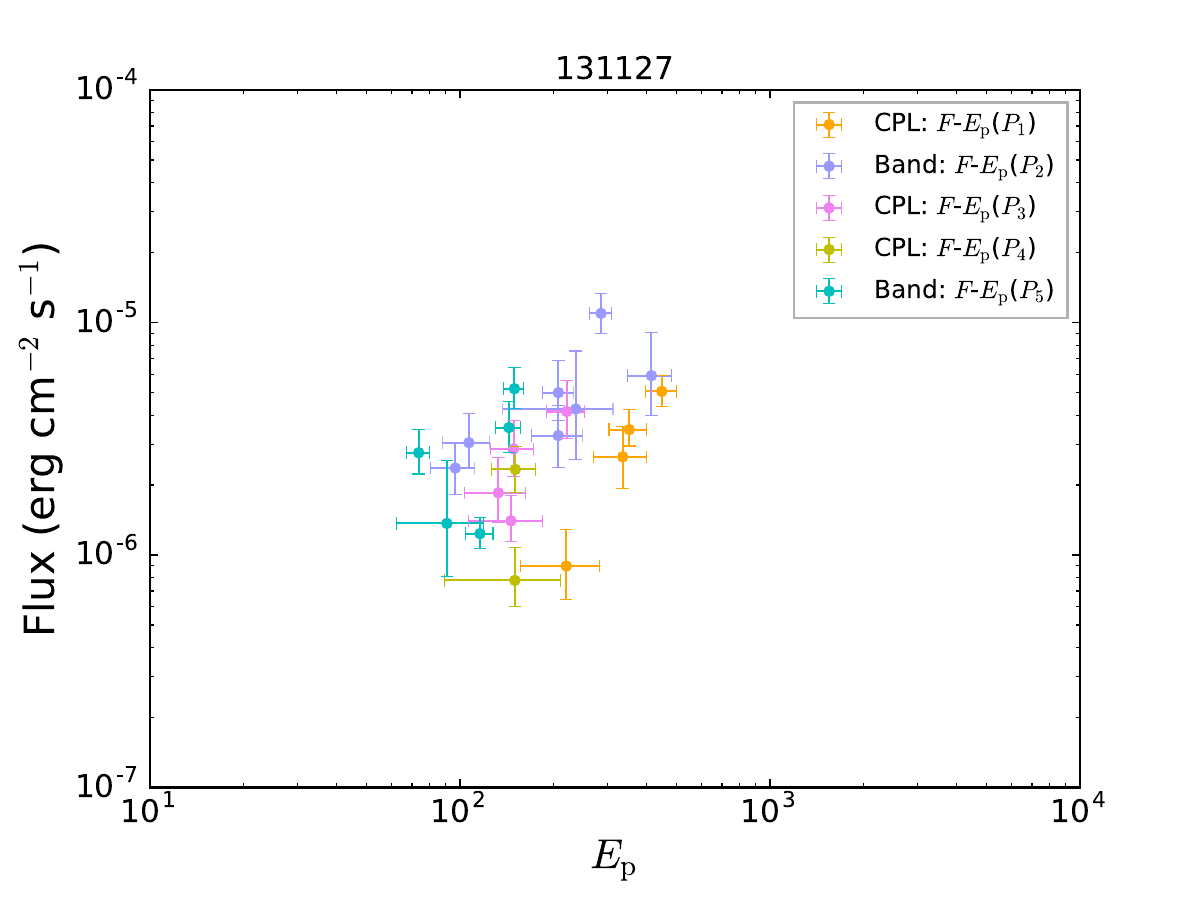}
\includegraphics[angle=0,scale=0.3]{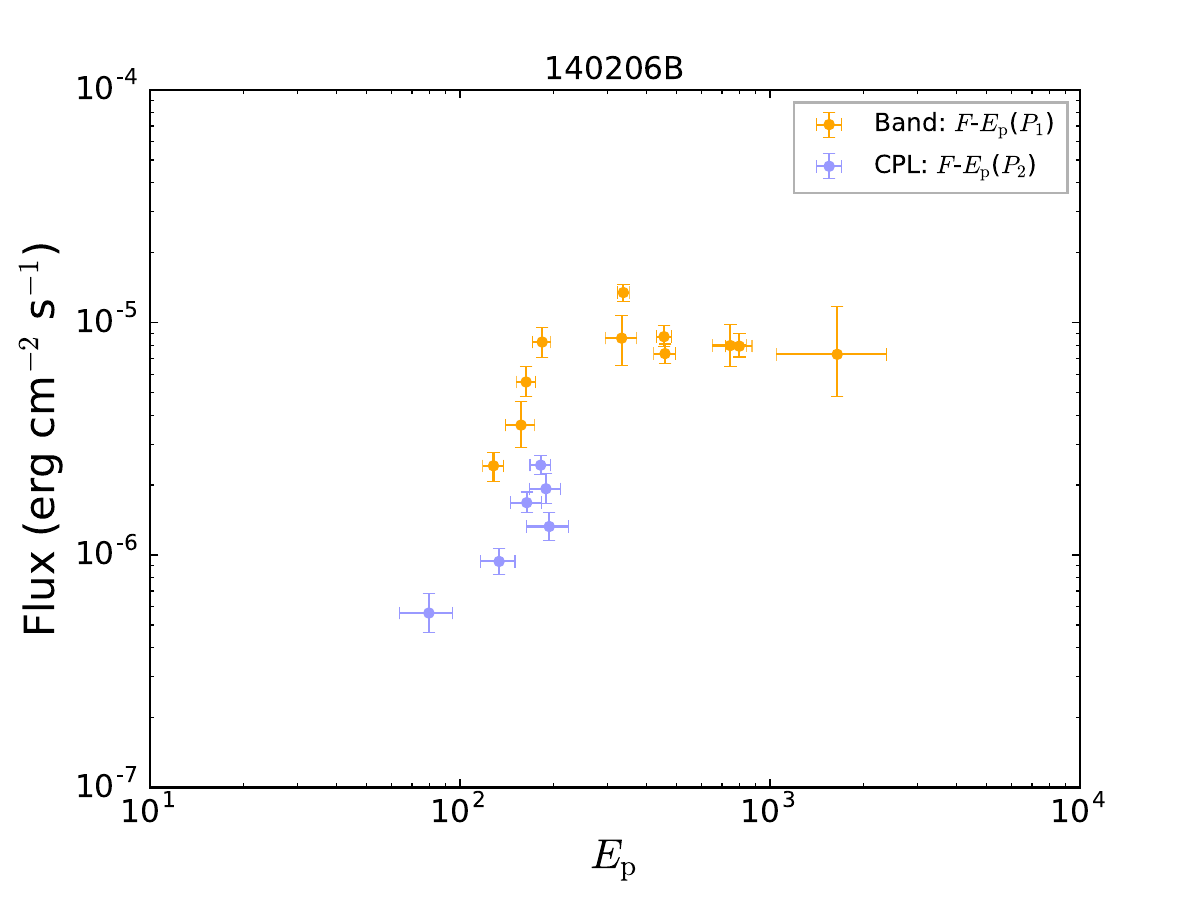}
\includegraphics[angle=0,scale=0.3]{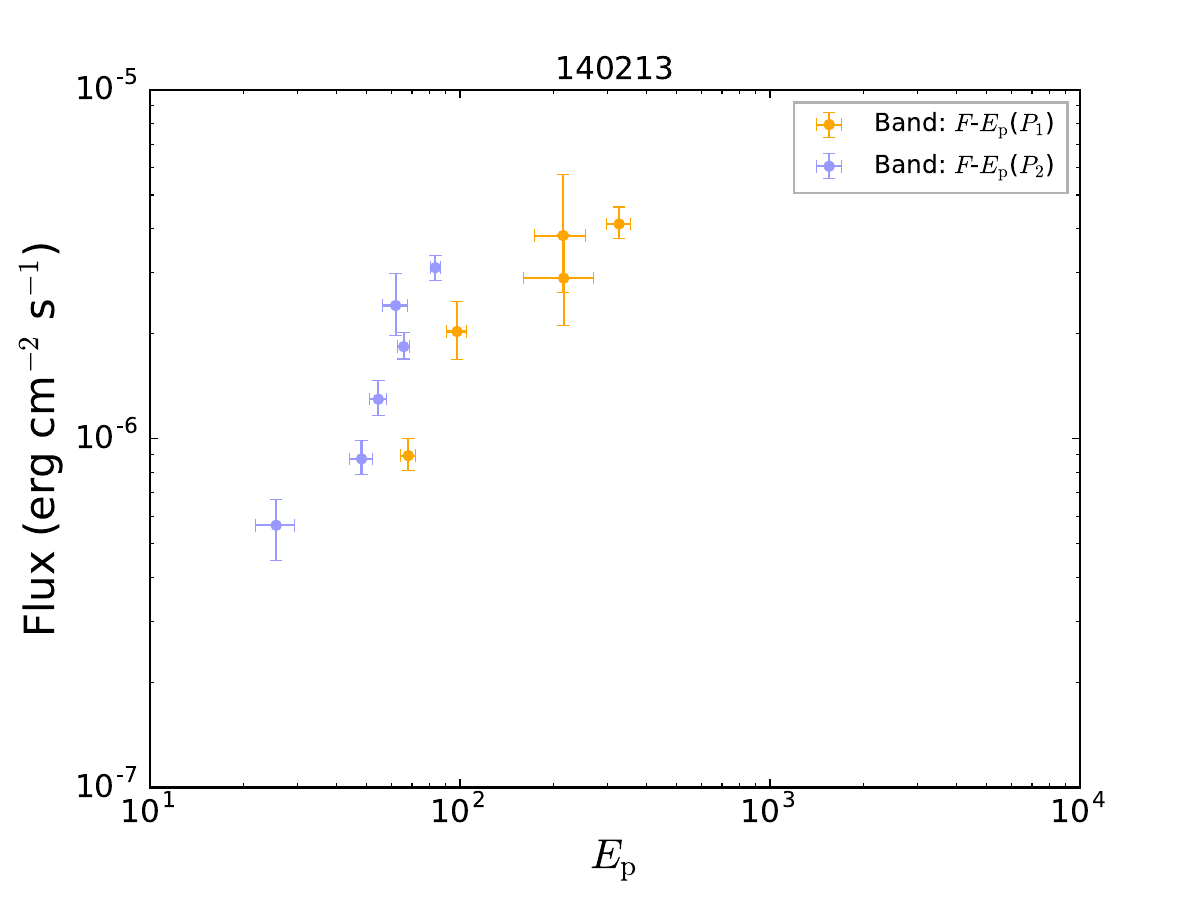}
\includegraphics[angle=0,scale=0.3]{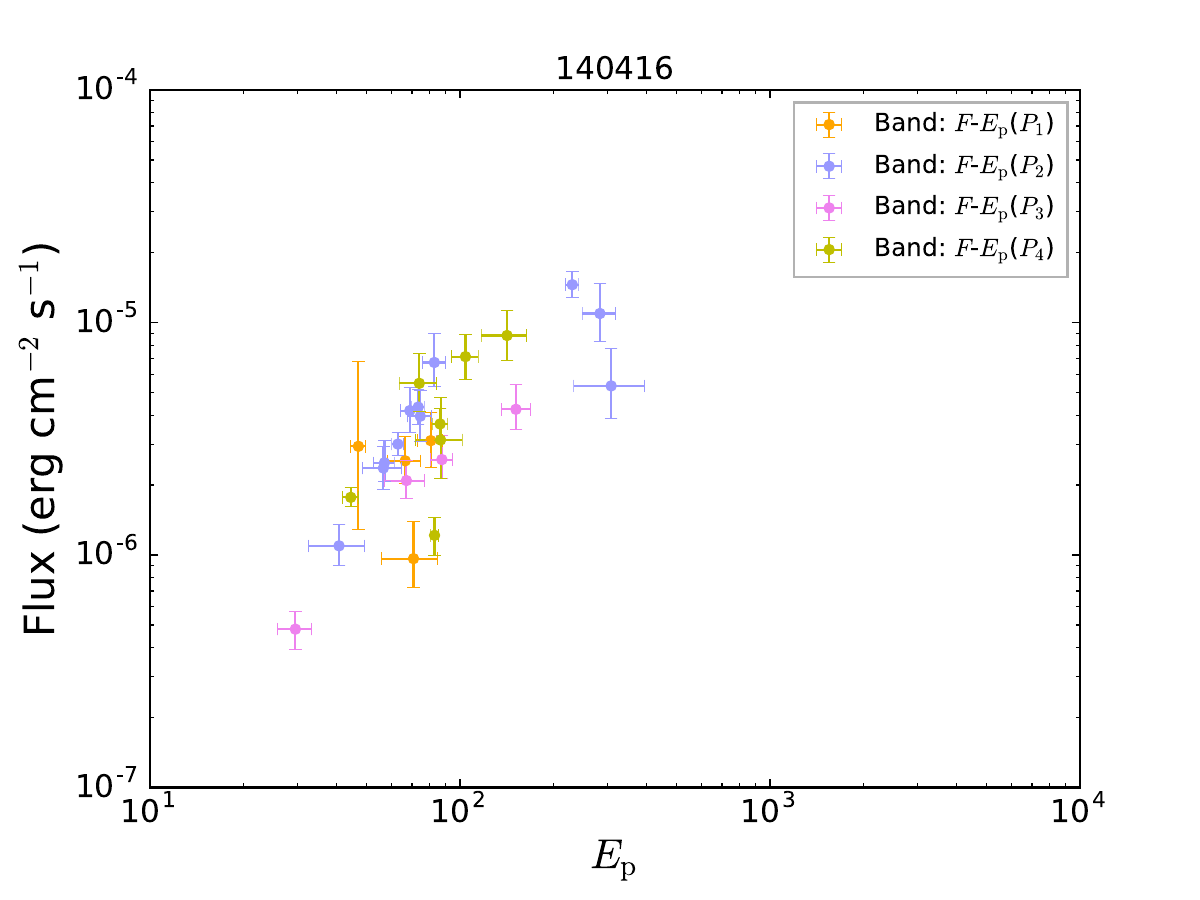}
\includegraphics[angle=0,scale=0.3]{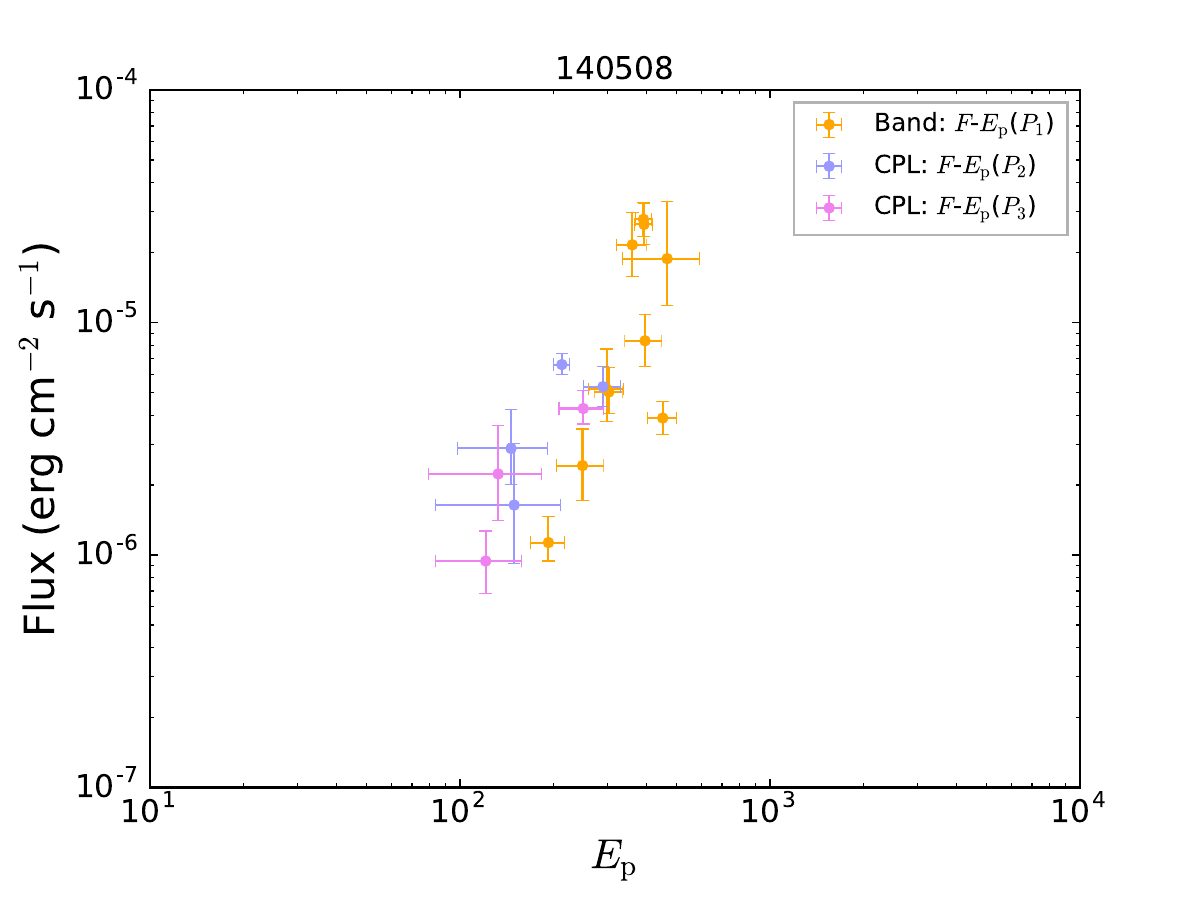}
\includegraphics[angle=0,scale=0.3]{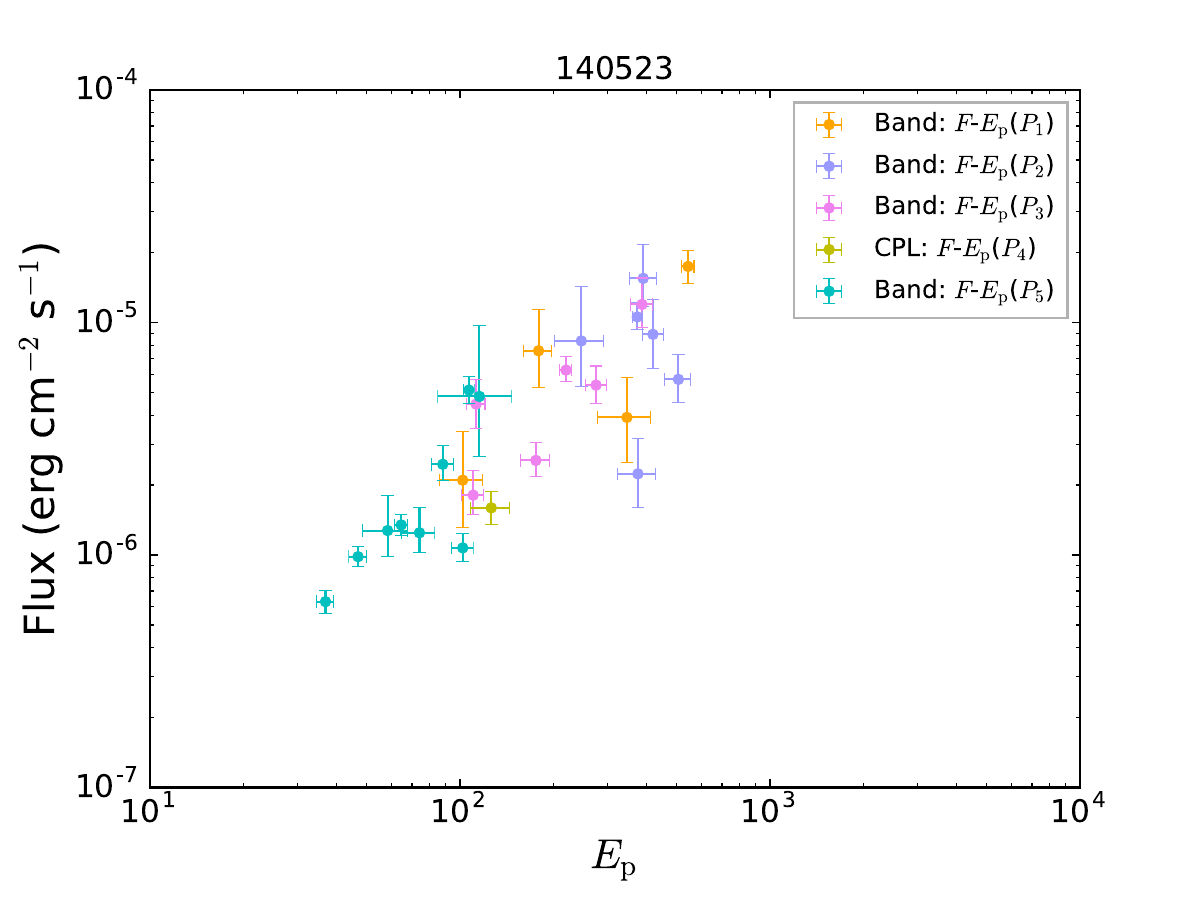}
\includegraphics[angle=0,scale=0.3]{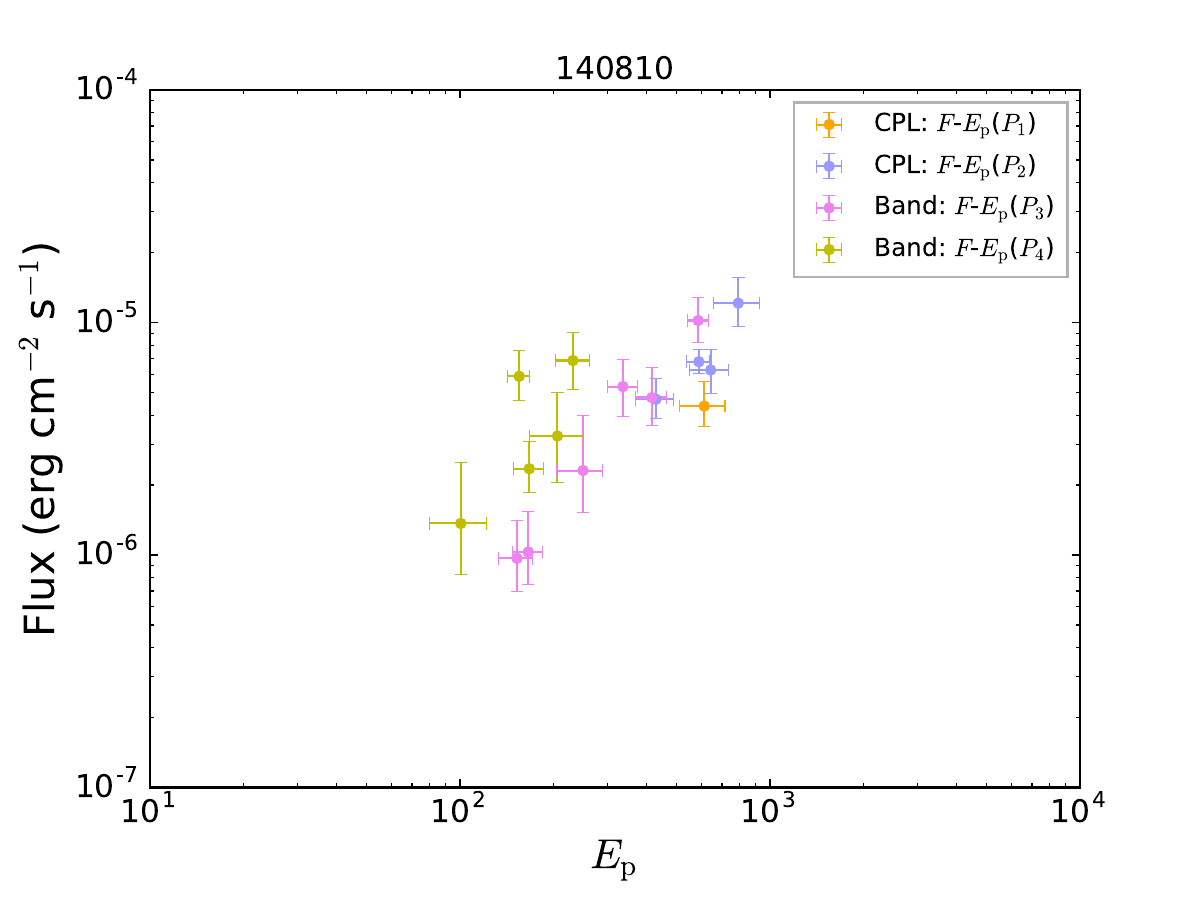}
\includegraphics[angle=0,scale=0.3]{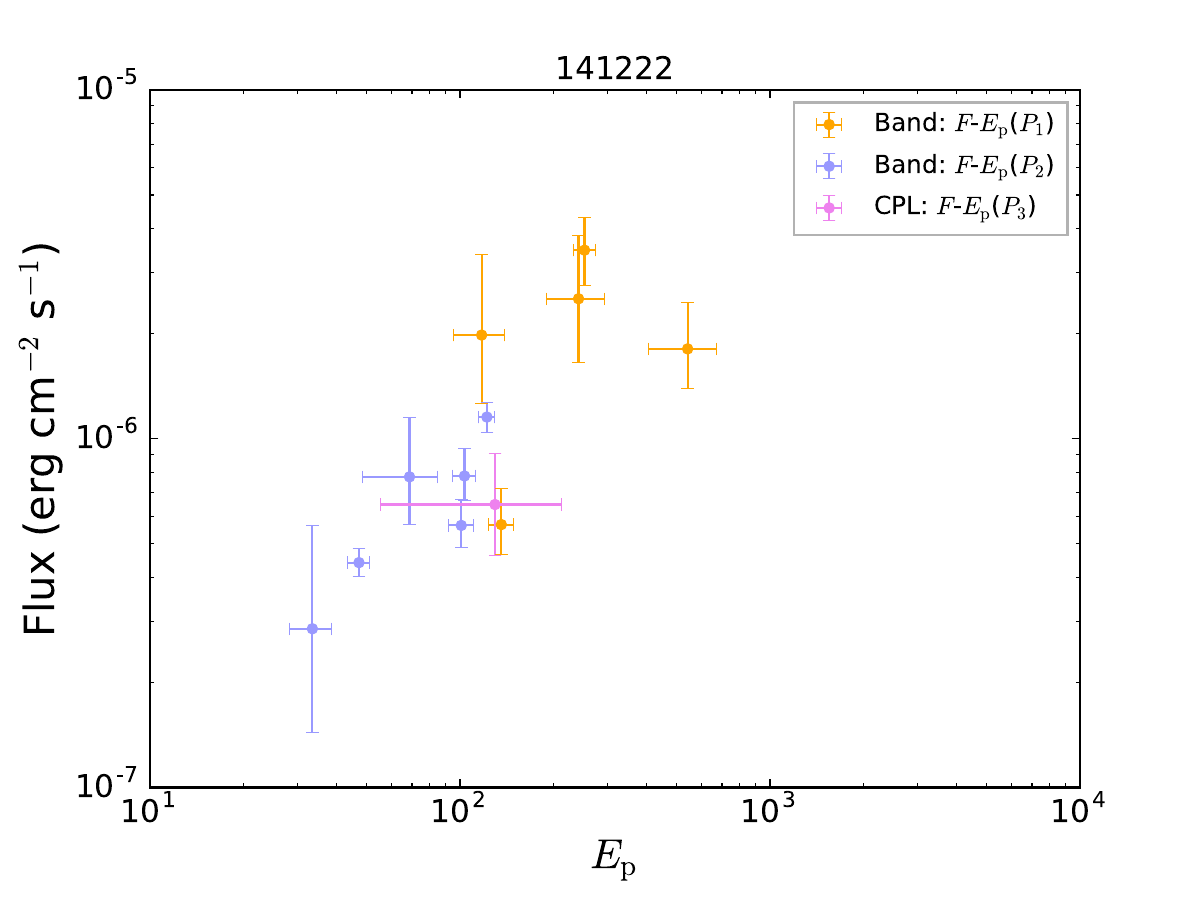}
\includegraphics[angle=0,scale=0.3]{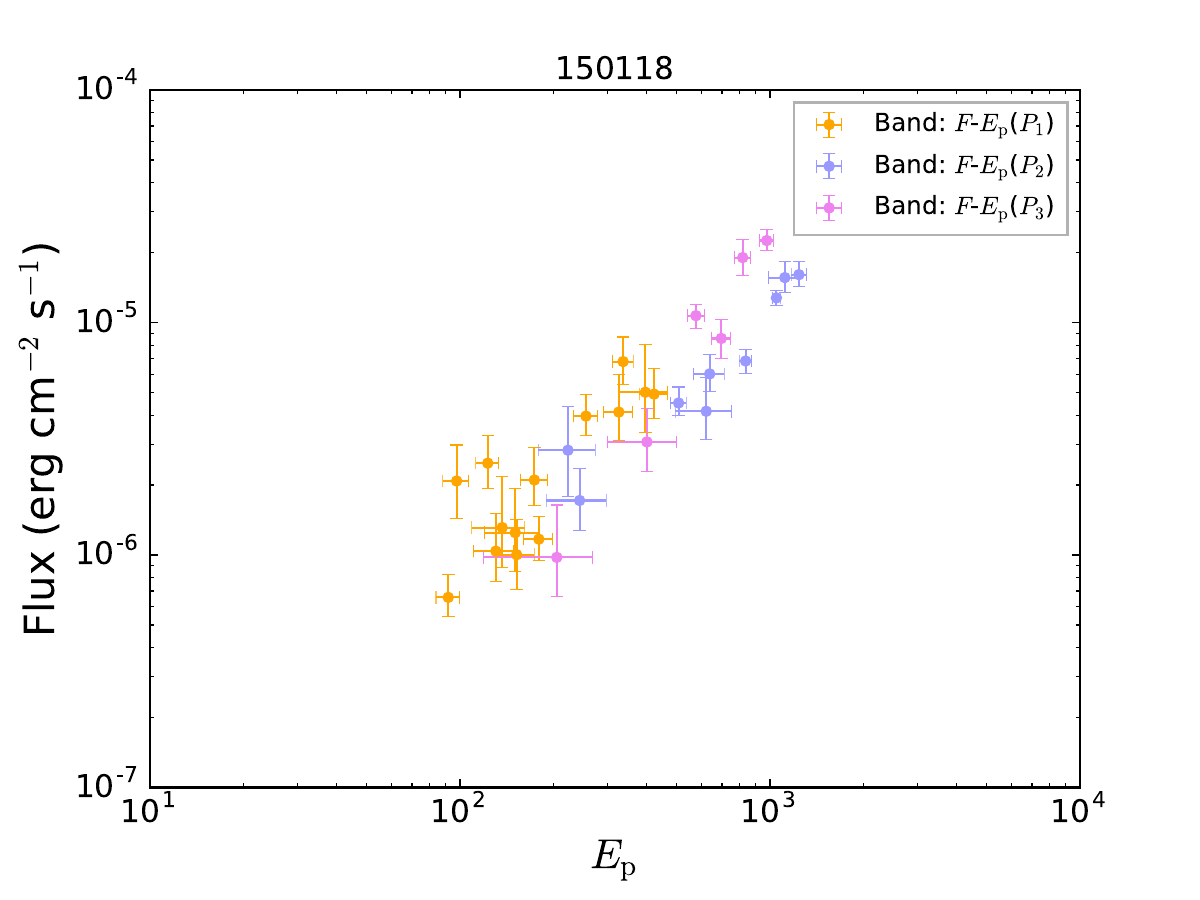}
\center{Fig. \ref{fig:FluxEp_Best}--- Continued}
\end{figure*}
\begin{figure*}
\includegraphics[angle=0,scale=0.3]{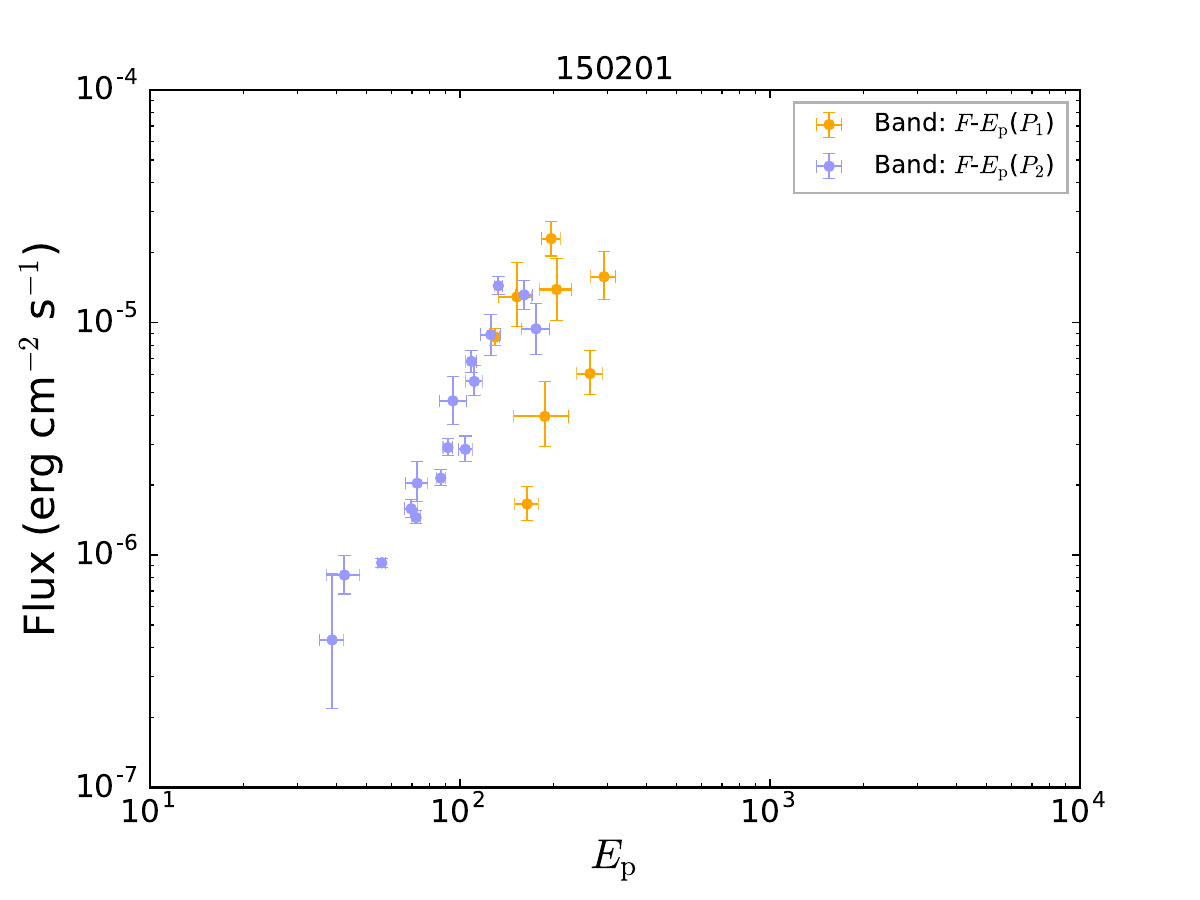}
\includegraphics[angle=0,scale=0.3]{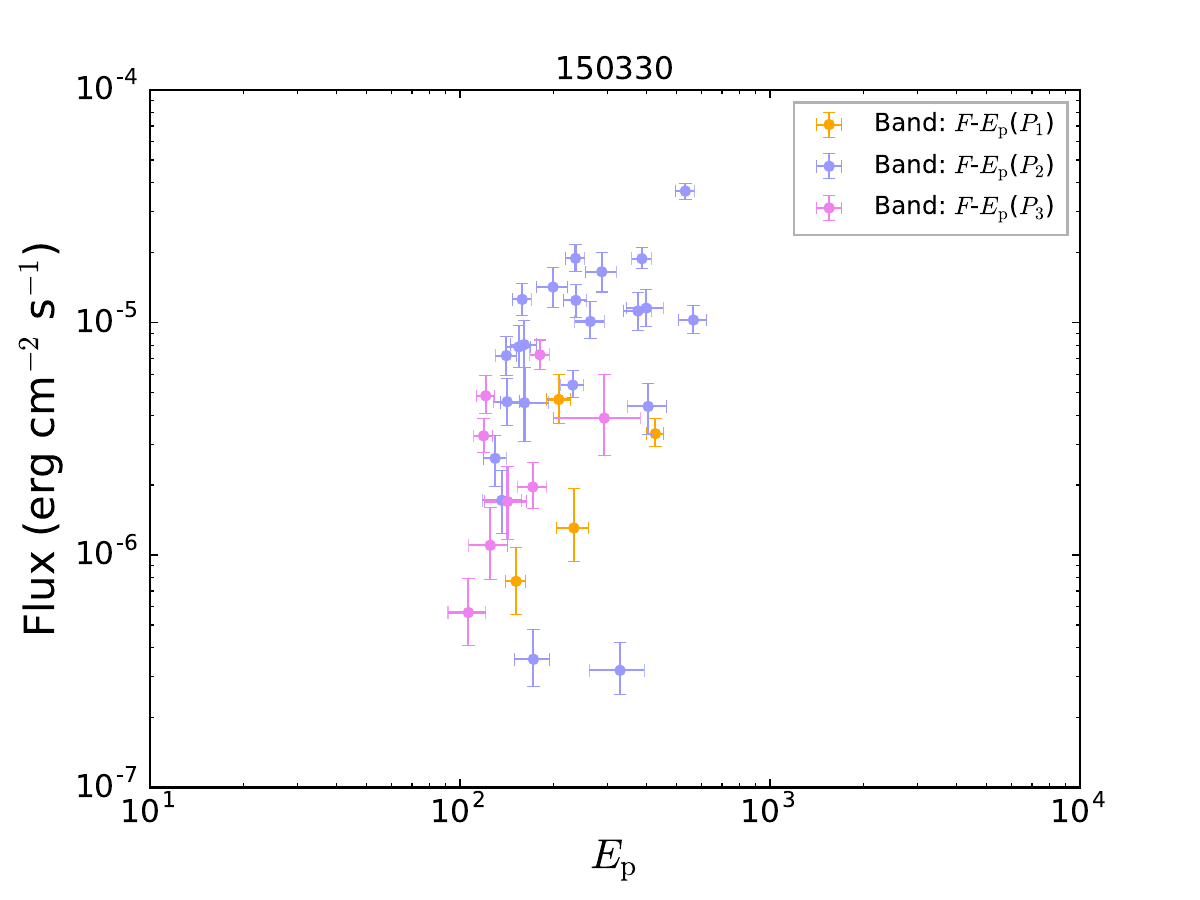}
\includegraphics[angle=0,scale=0.3]{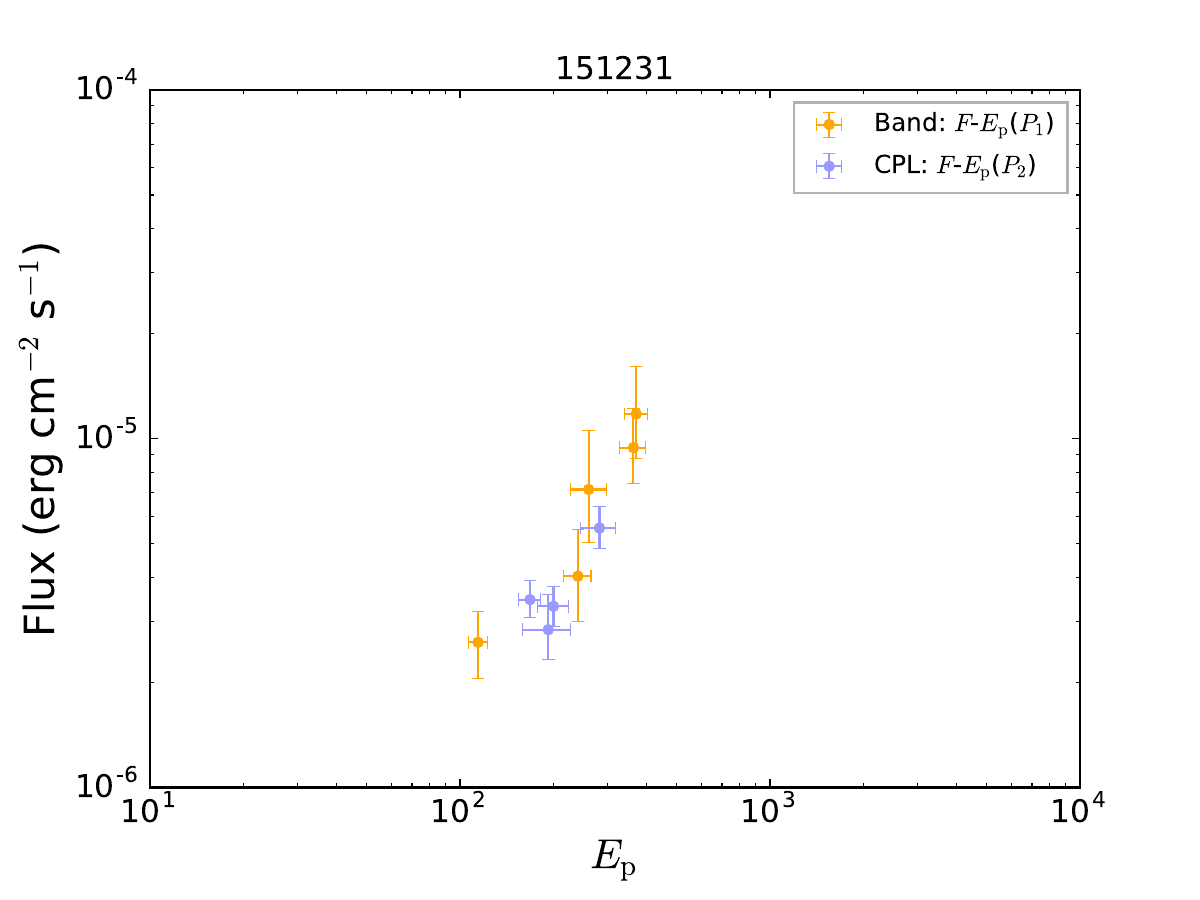}
\includegraphics[angle=0,scale=0.3]{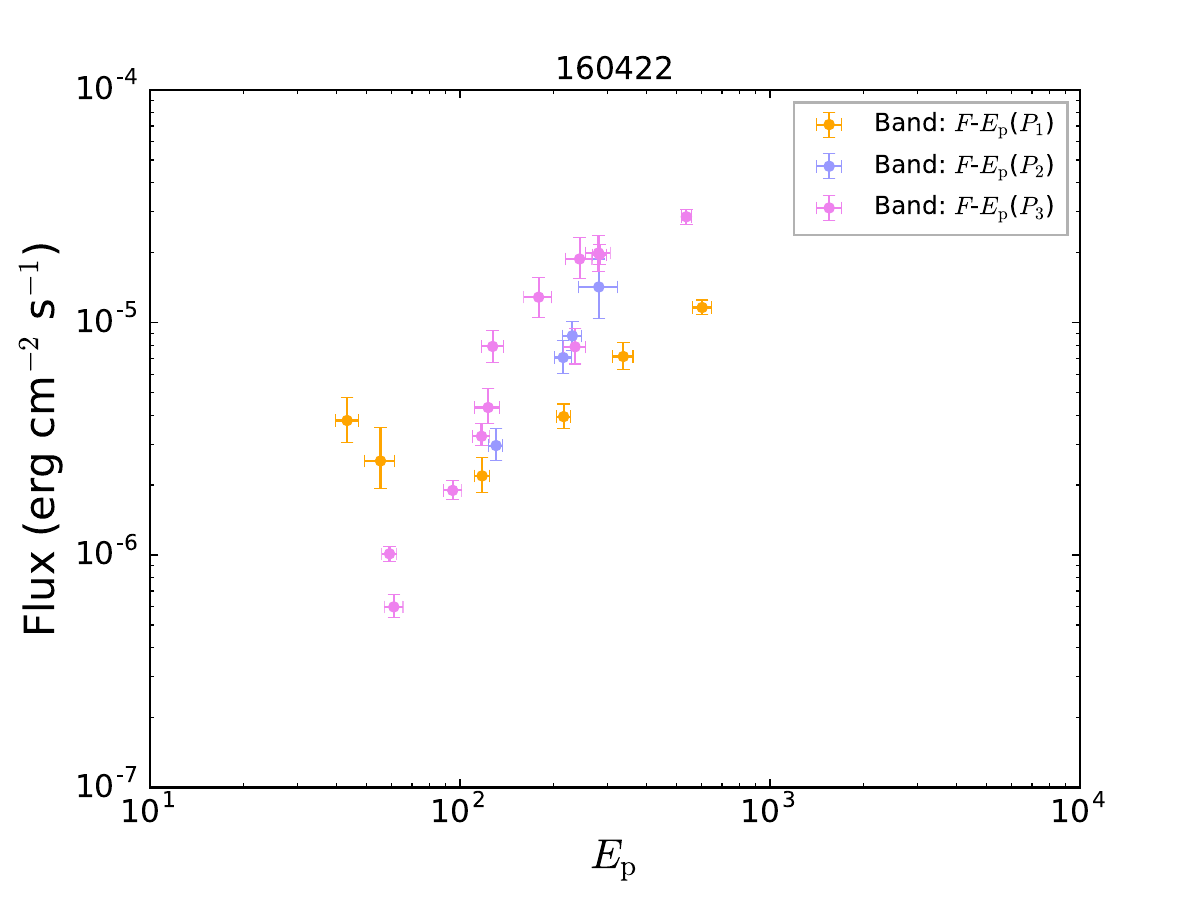}
\includegraphics[angle=0,scale=0.3]{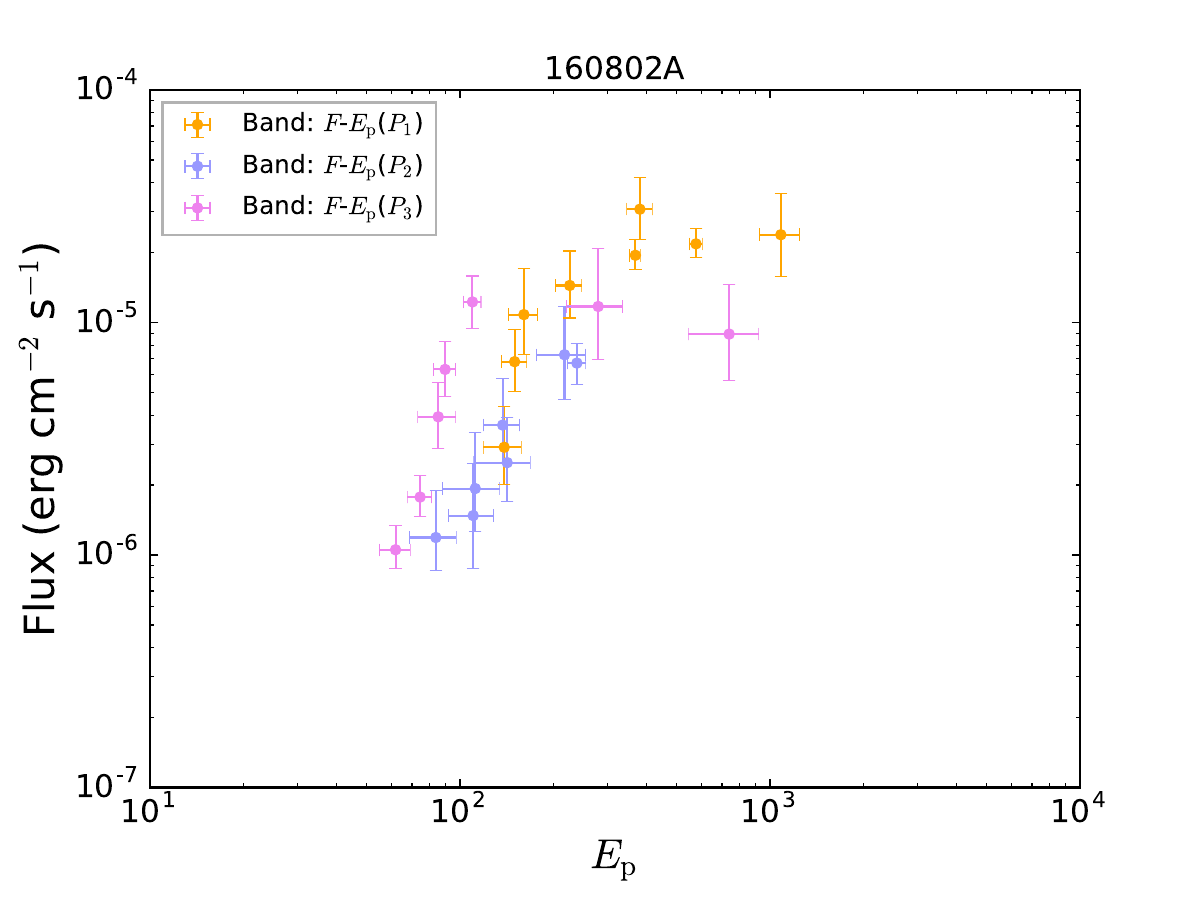}
\includegraphics[angle=0,scale=0.3]{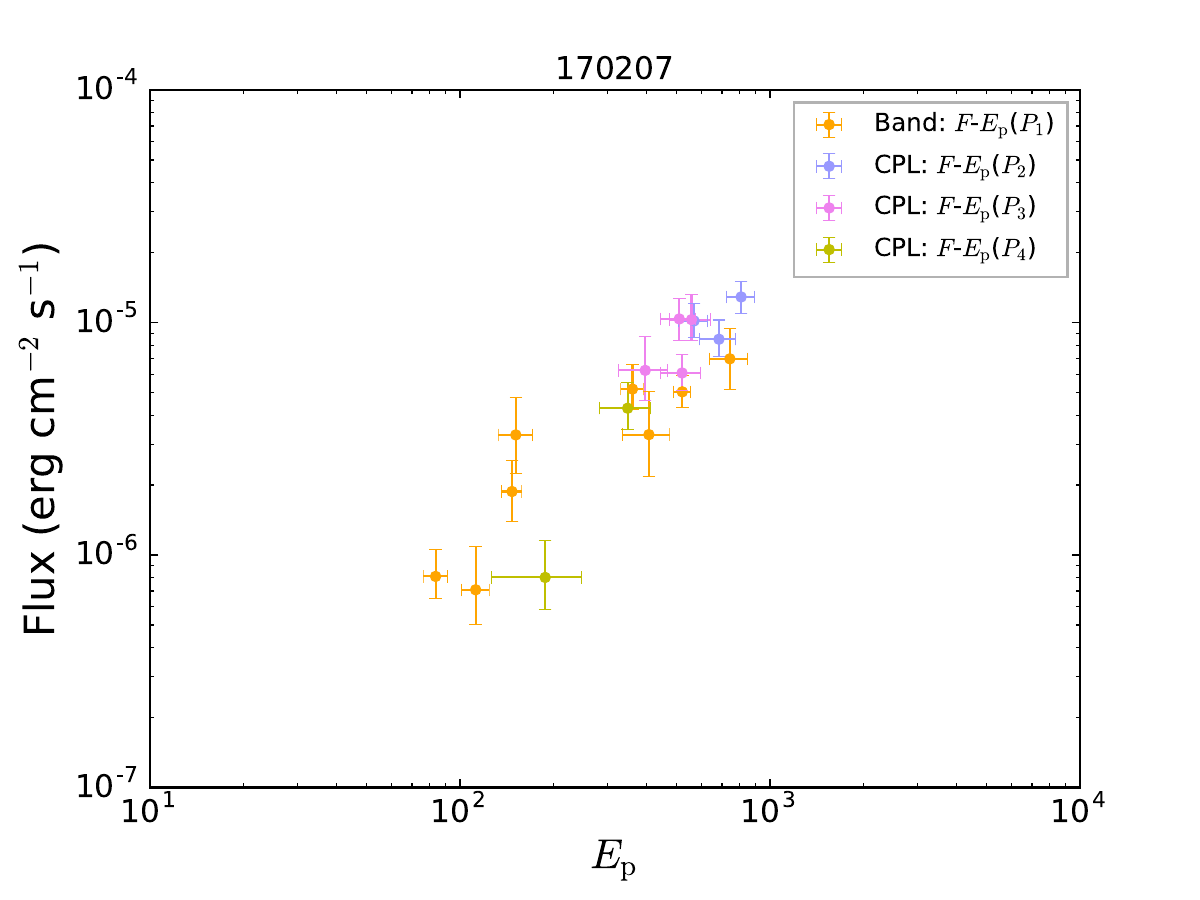}
\includegraphics[angle=0,scale=0.3]{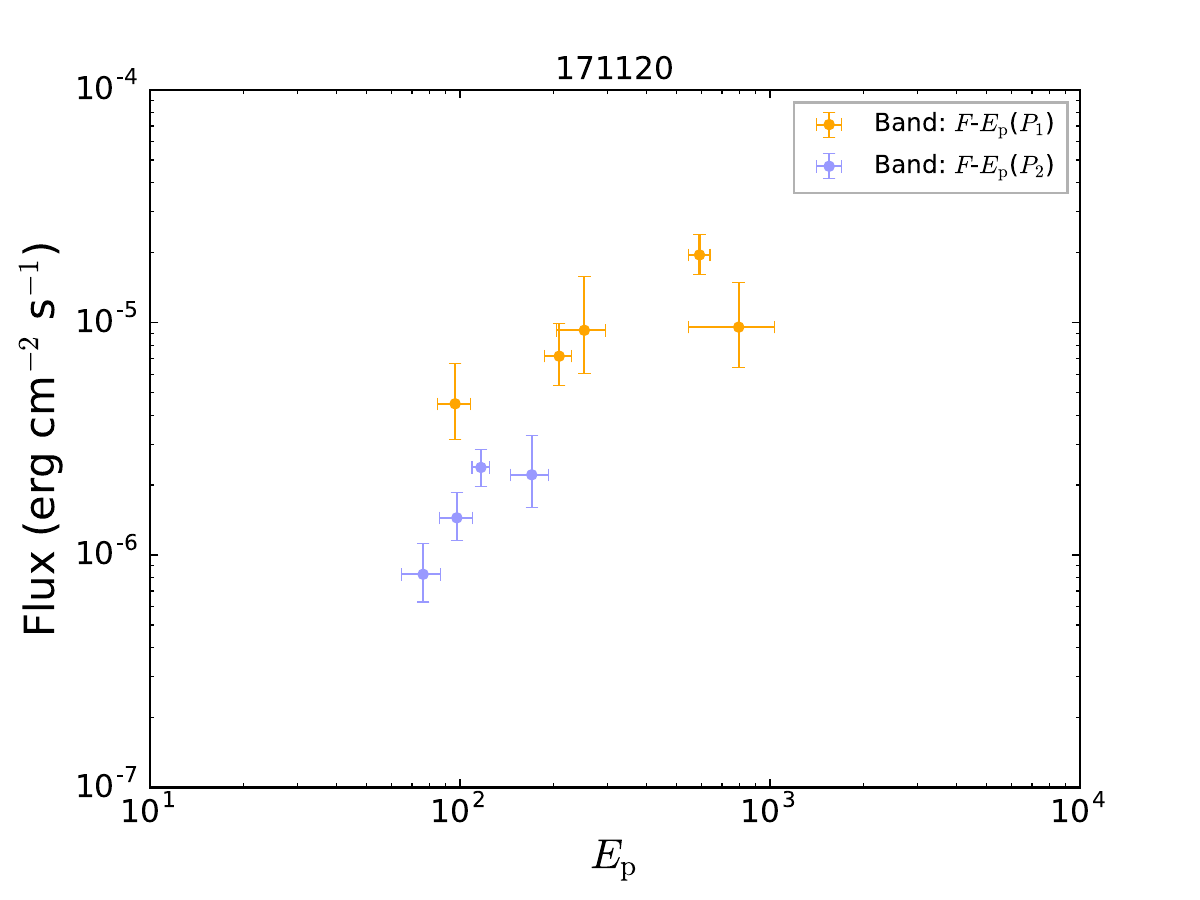}
\includegraphics[angle=0,scale=0.3]{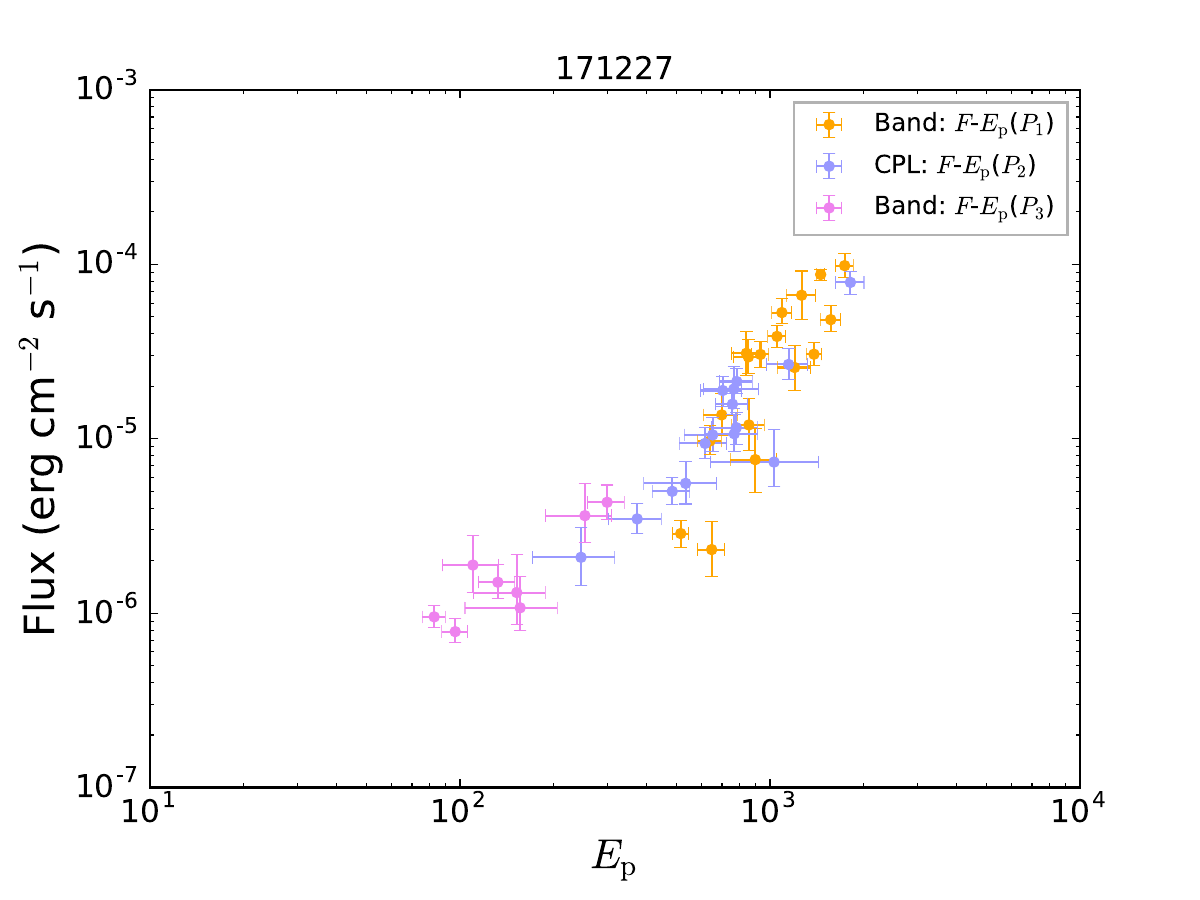}
\includegraphics[angle=0,scale=0.3]{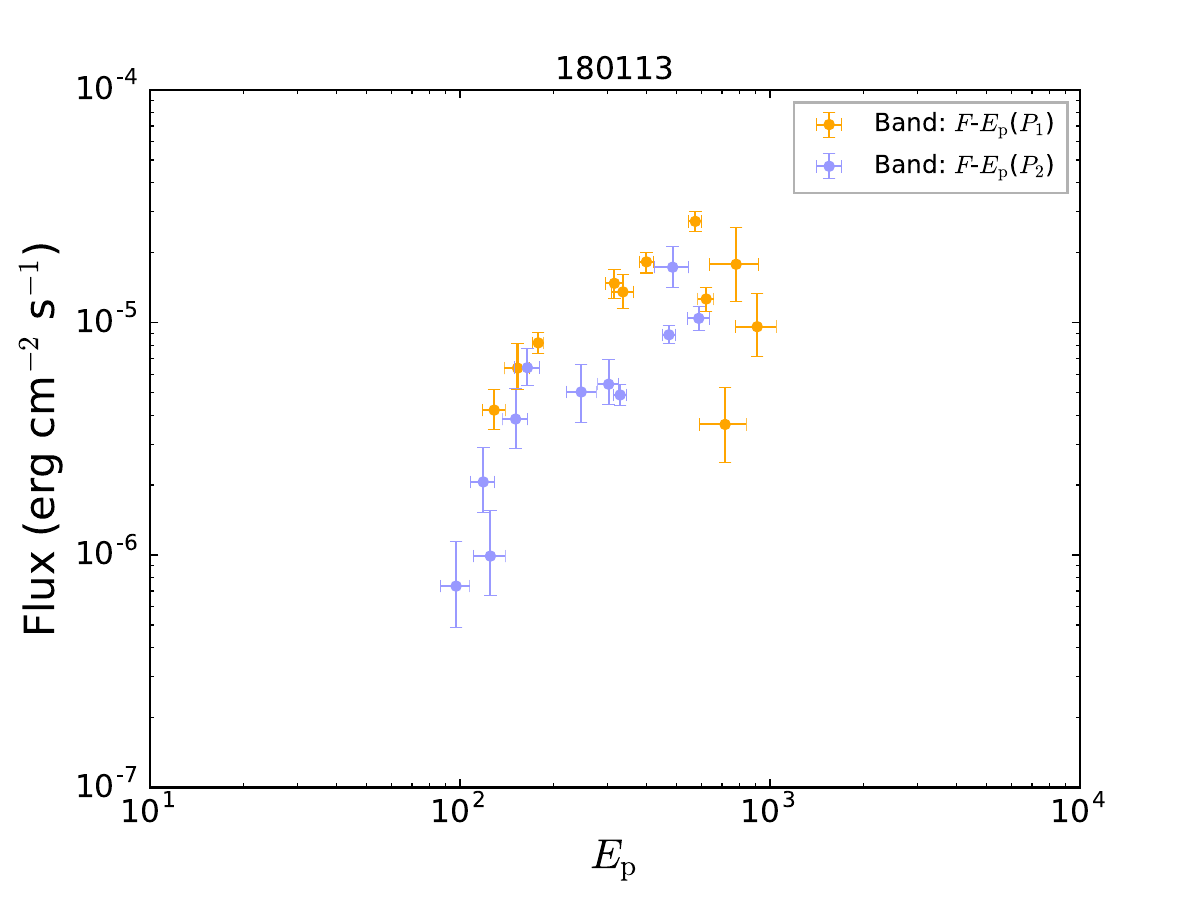}
\includegraphics[angle=0,scale=0.3]{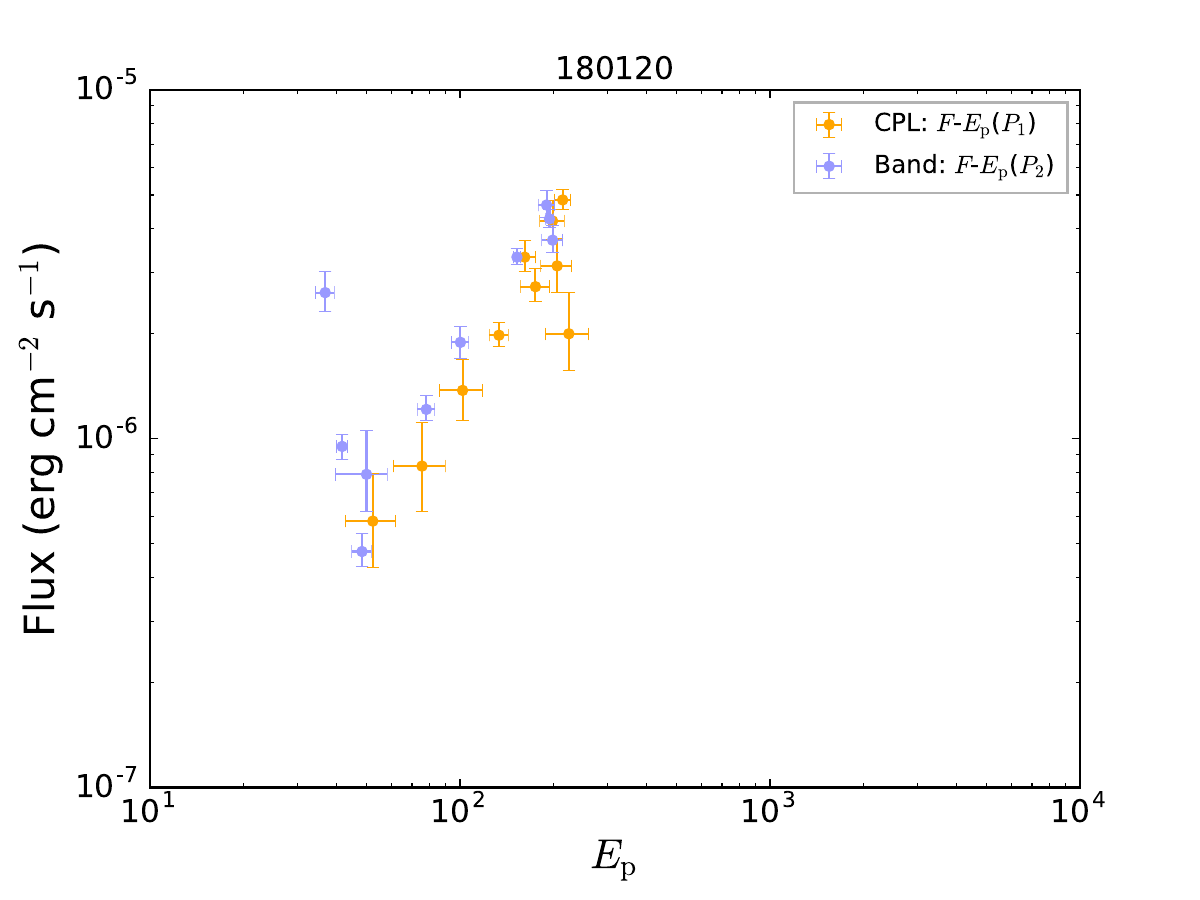}
\includegraphics[angle=0,scale=0.3]{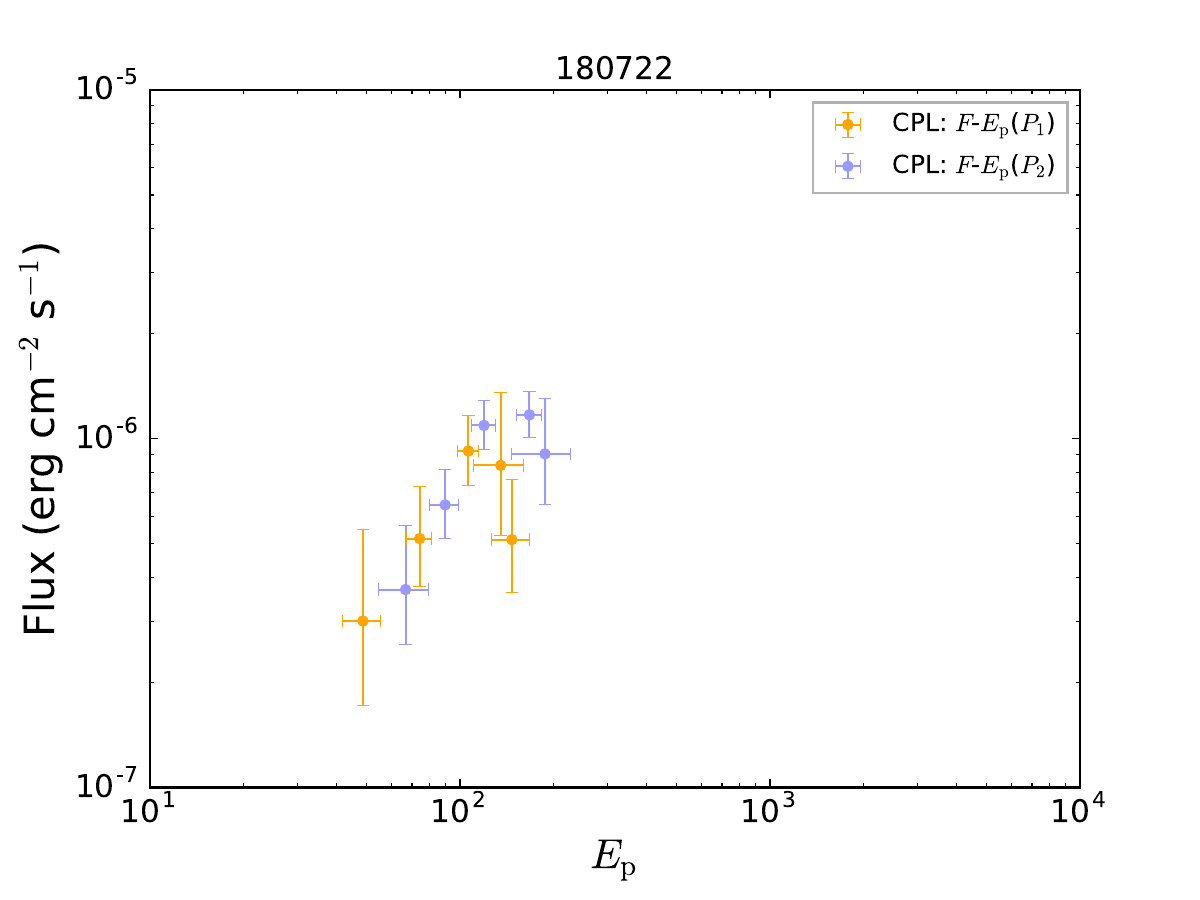}
\includegraphics[angle=0,scale=0.3]{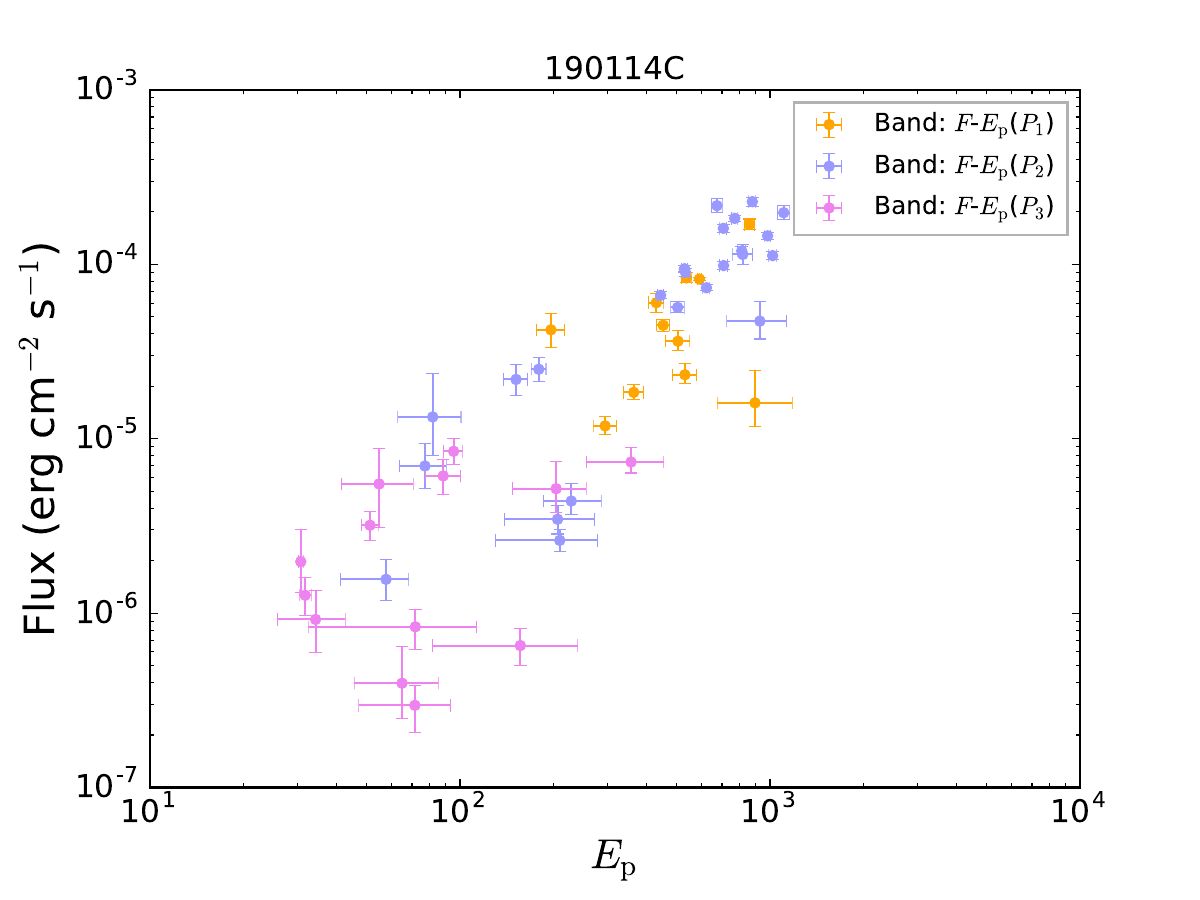}
\center{Fig. \ref{fig:FluxEp_Best}--- Continued}
\end{figure*}

\clearpage
\begin{figure*}
\includegraphics[angle=0,scale=0.3]{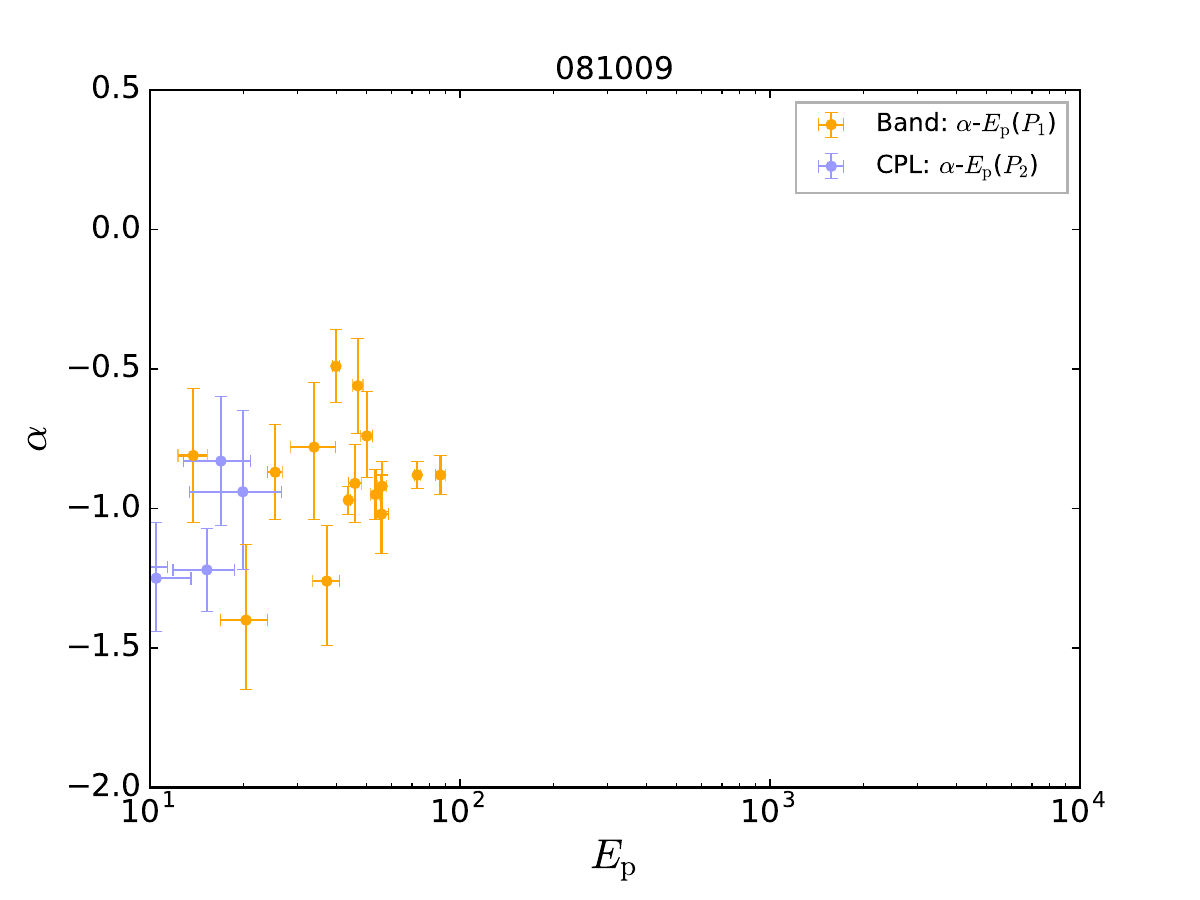}
\includegraphics[angle=0,scale=0.3]{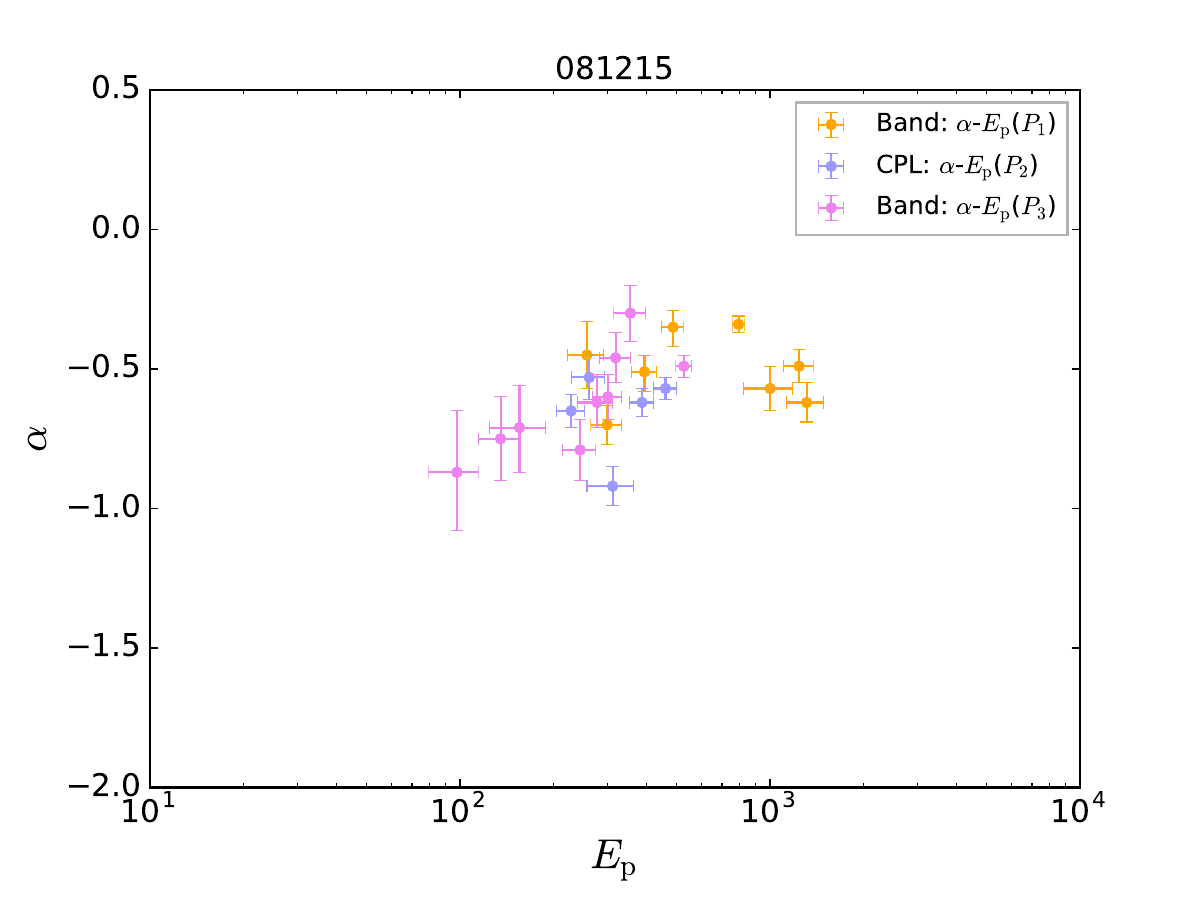}
\includegraphics[angle=0,scale=0.3]{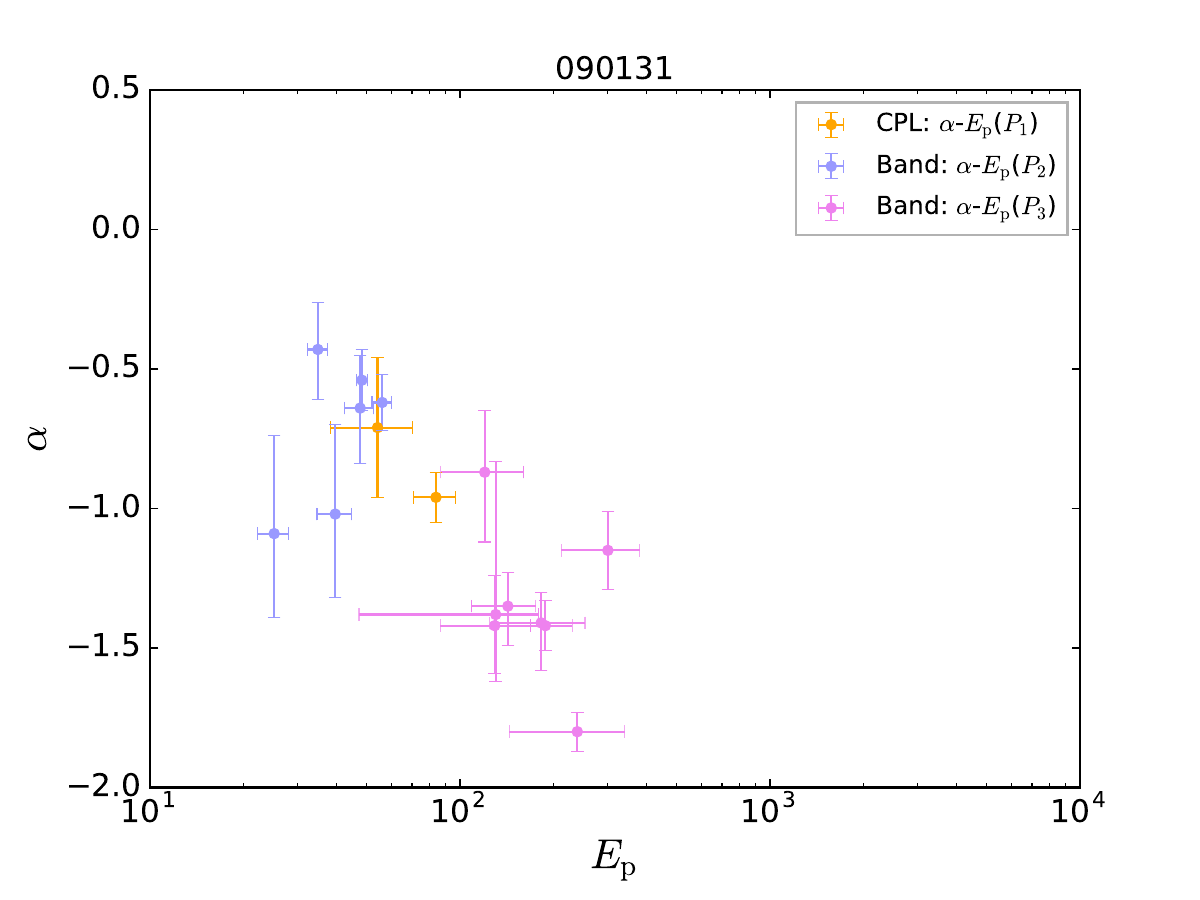}
\includegraphics[angle=0,scale=0.3]{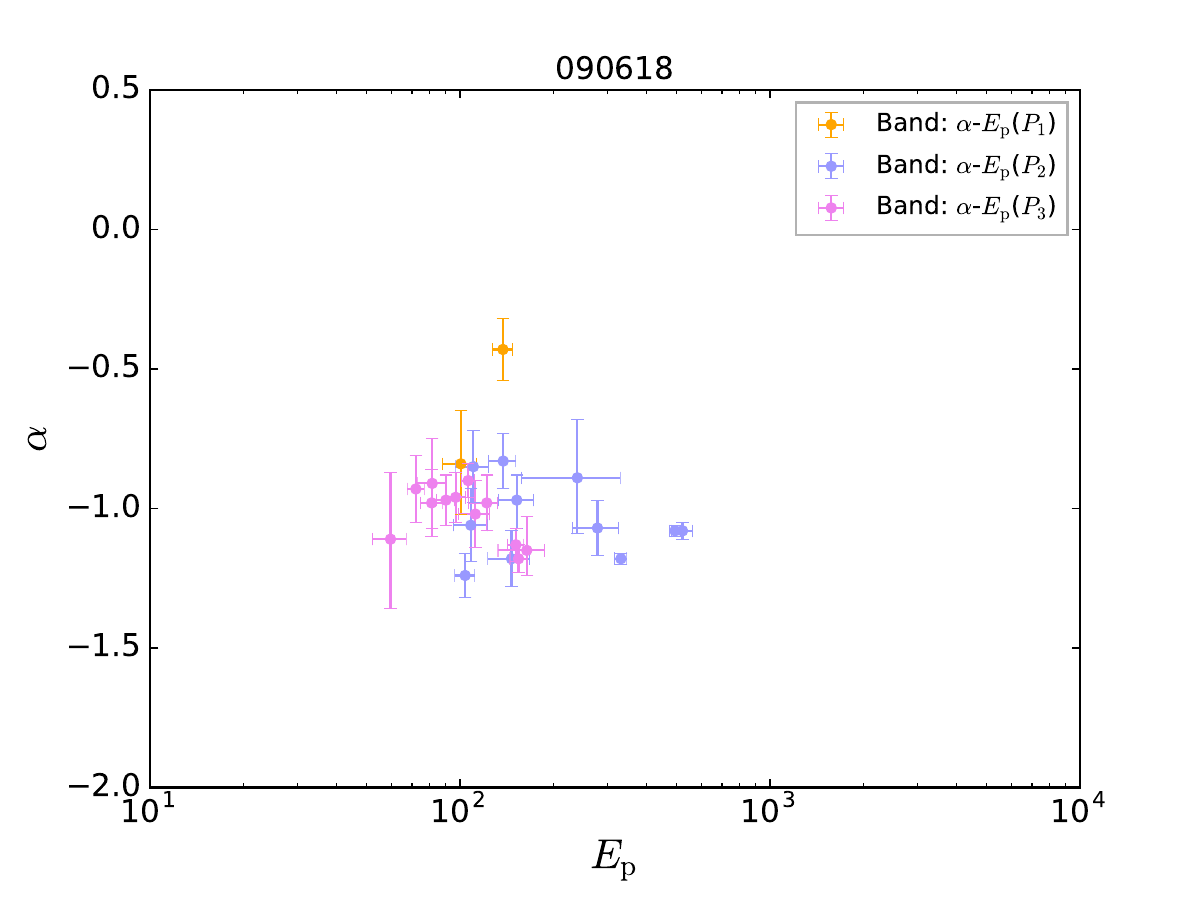}
\includegraphics[angle=0,scale=0.3]{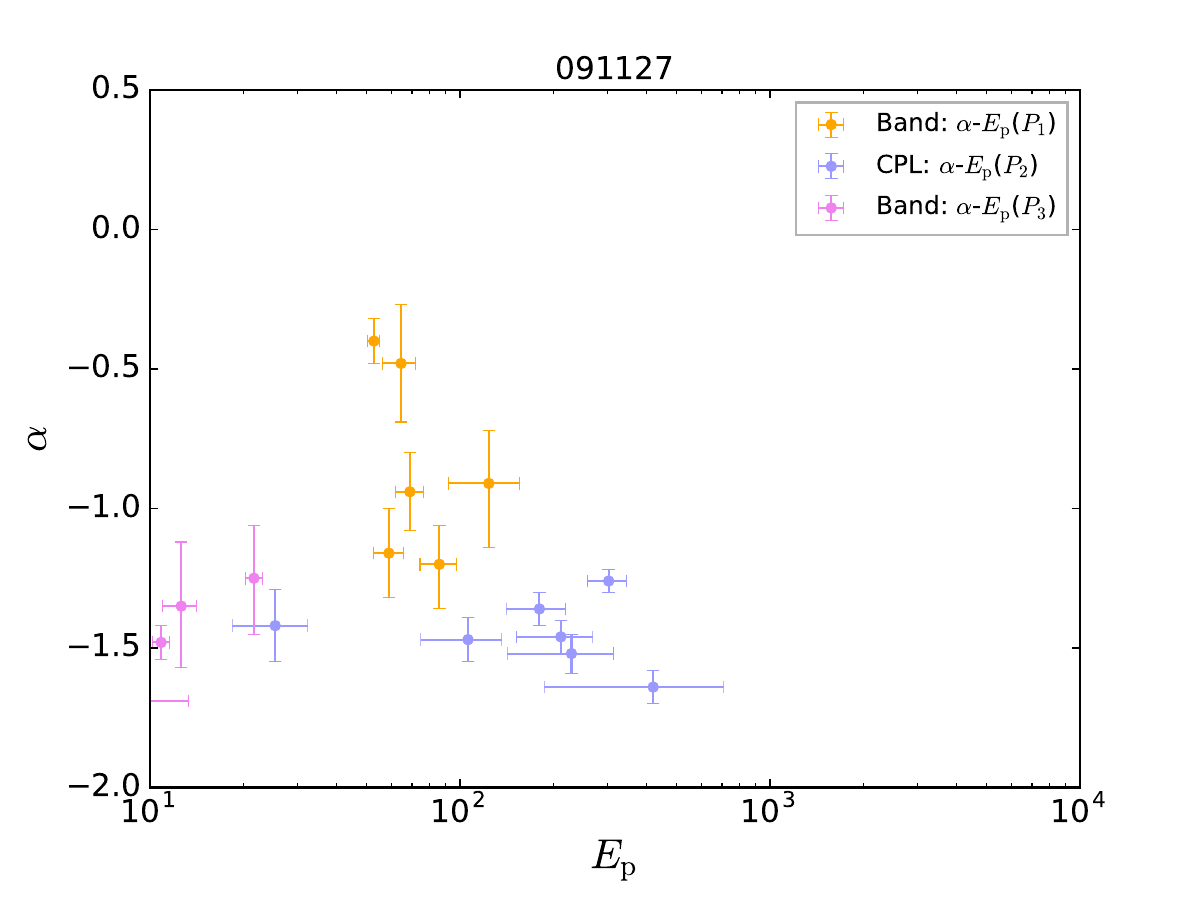}
\includegraphics[angle=0,scale=0.3]{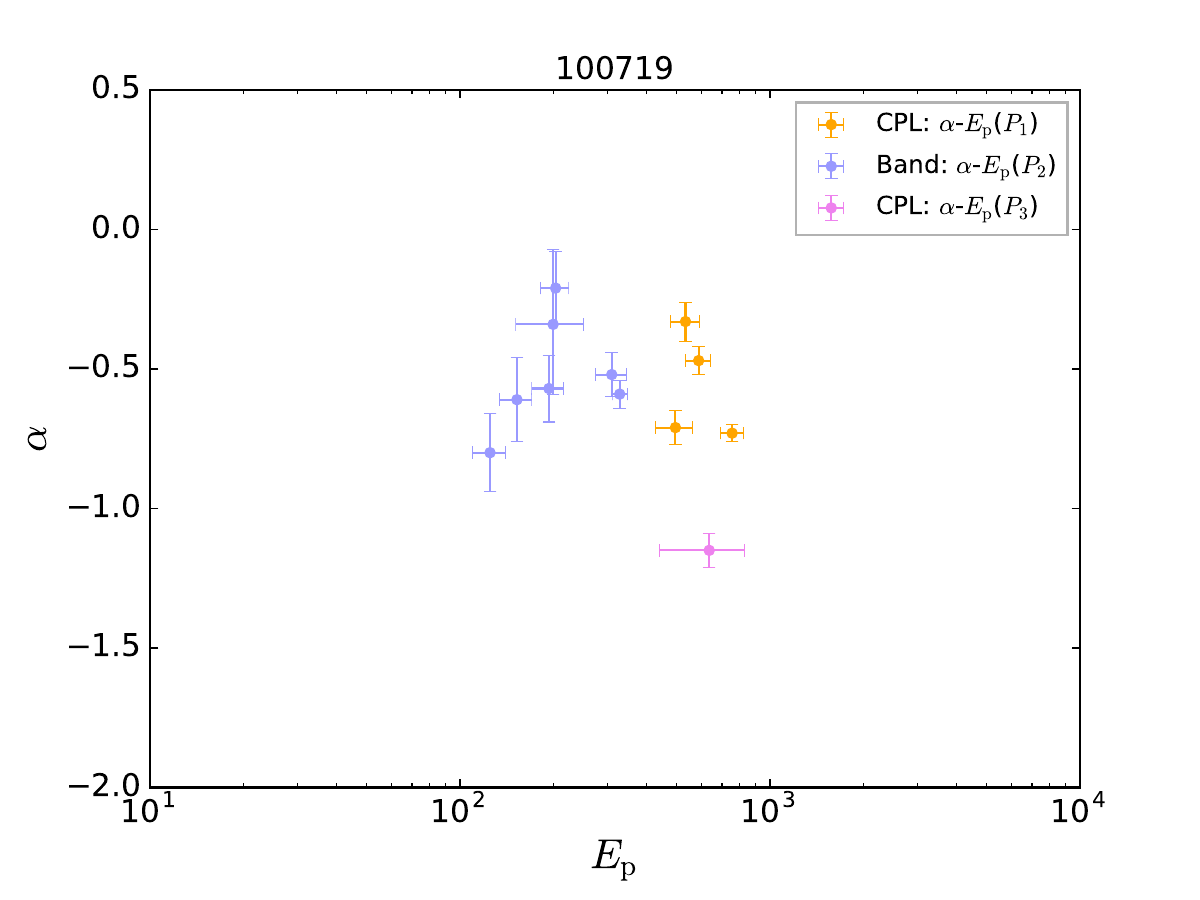}
\includegraphics[angle=0,scale=0.3]{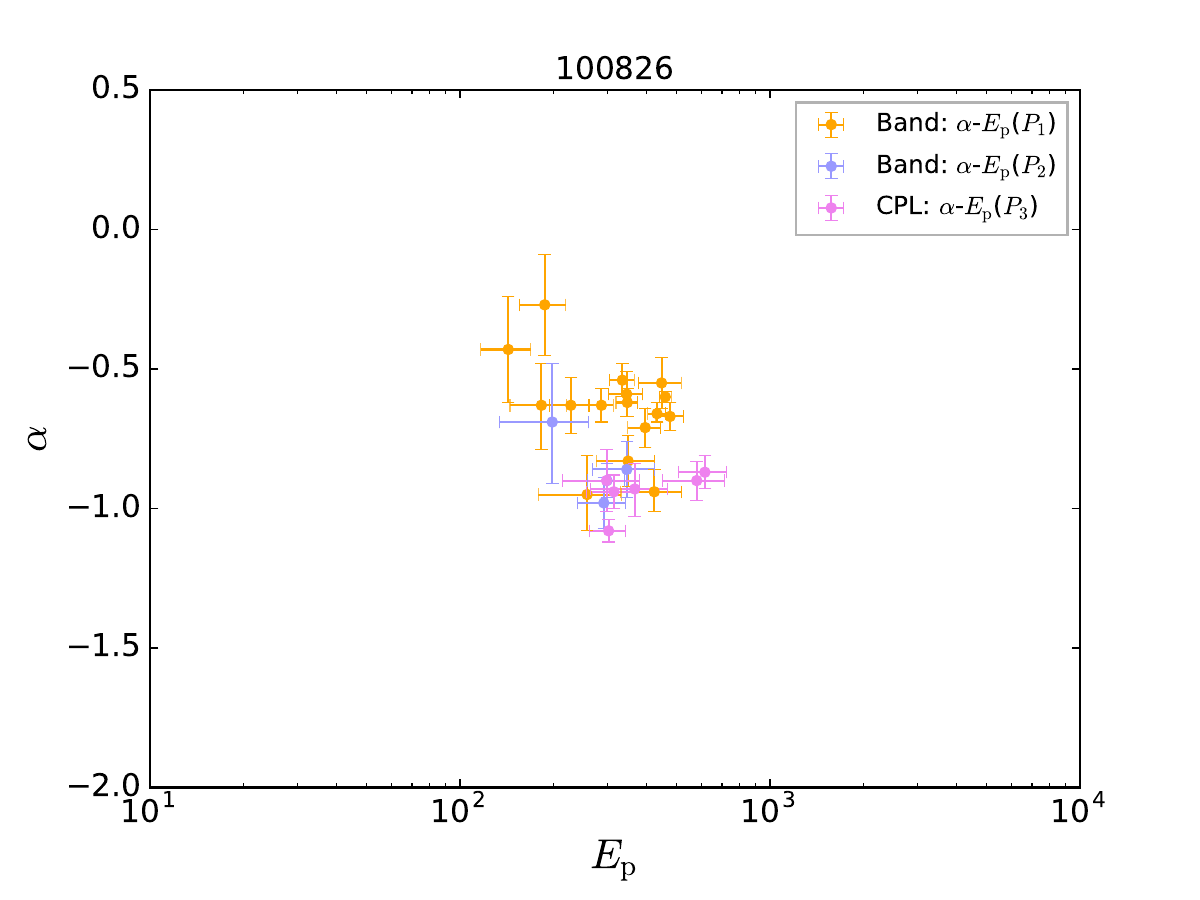}
\includegraphics[angle=0,scale=0.3]{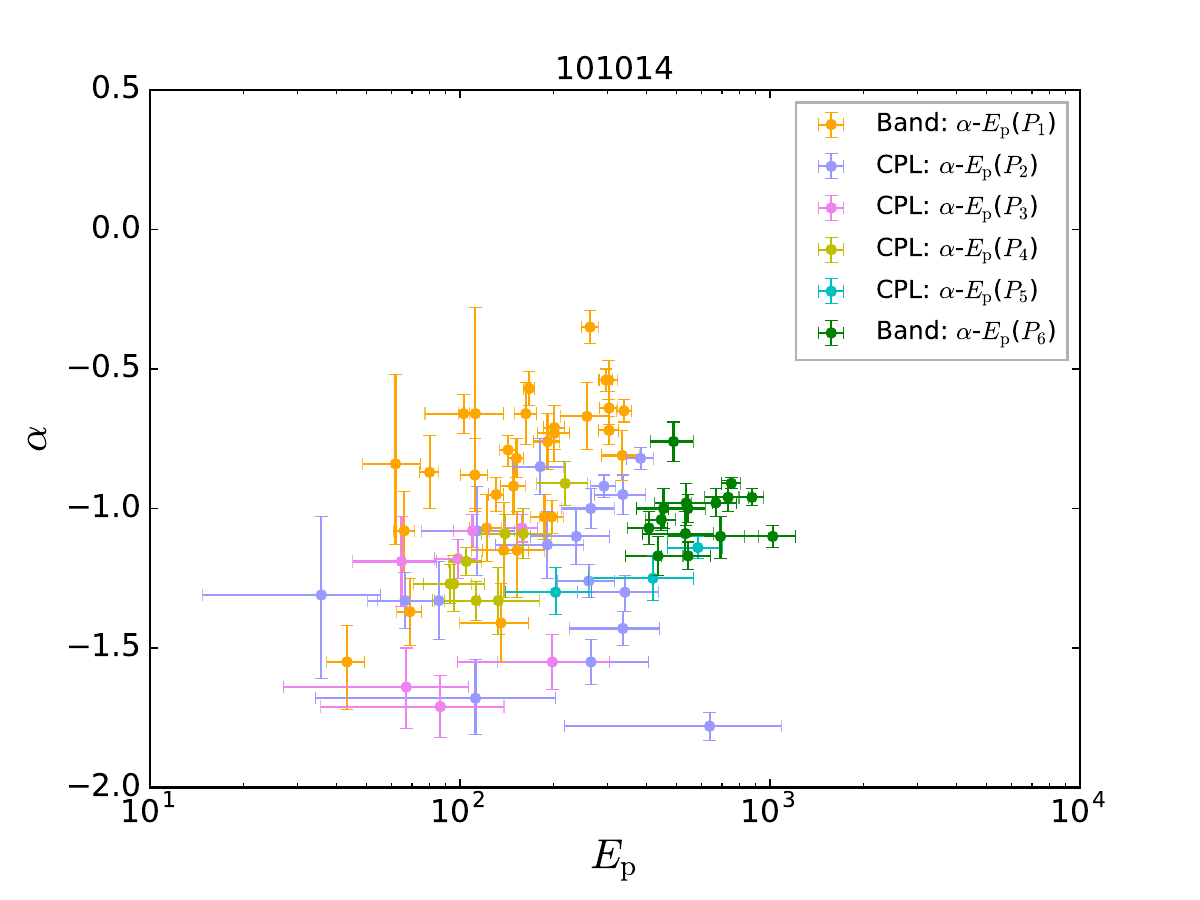}
\includegraphics[angle=0,scale=0.3]{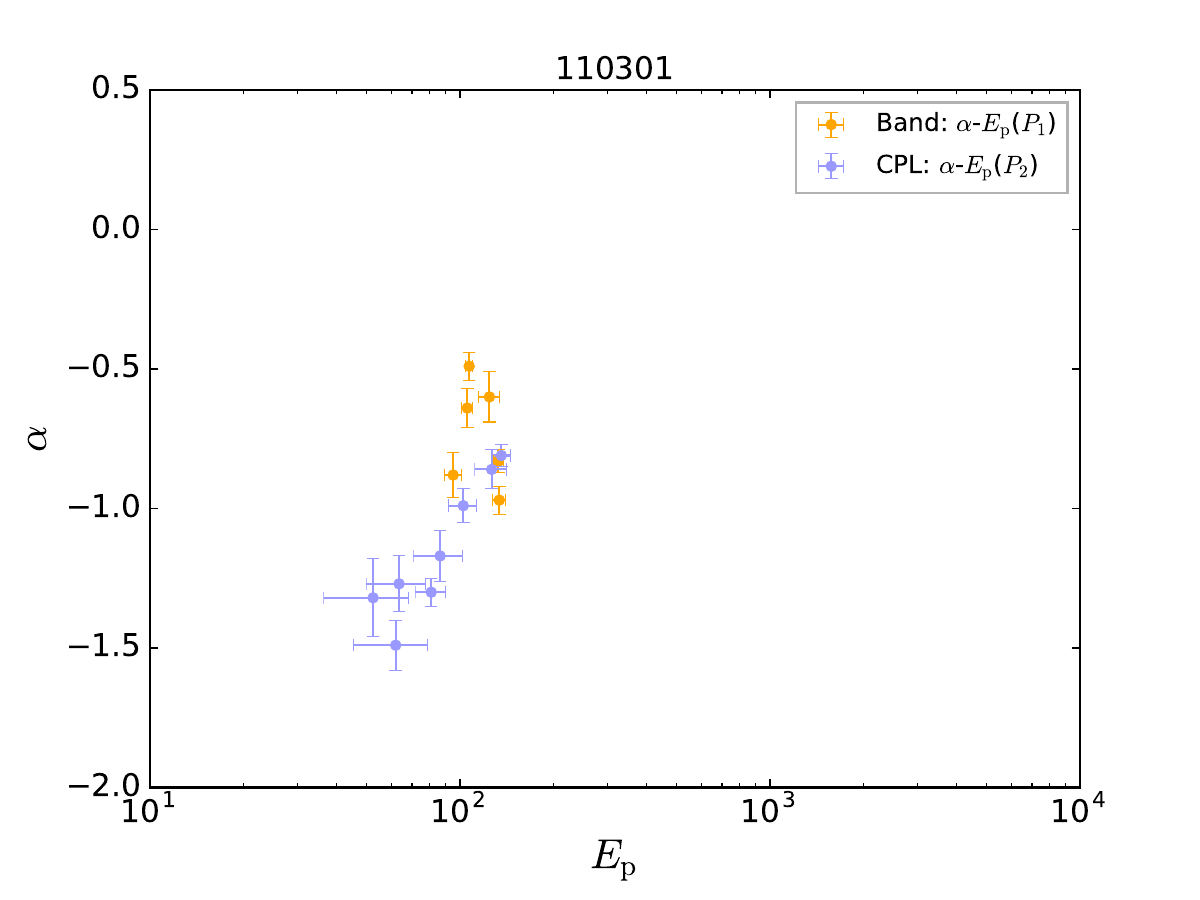}
\includegraphics[angle=0,scale=0.3]{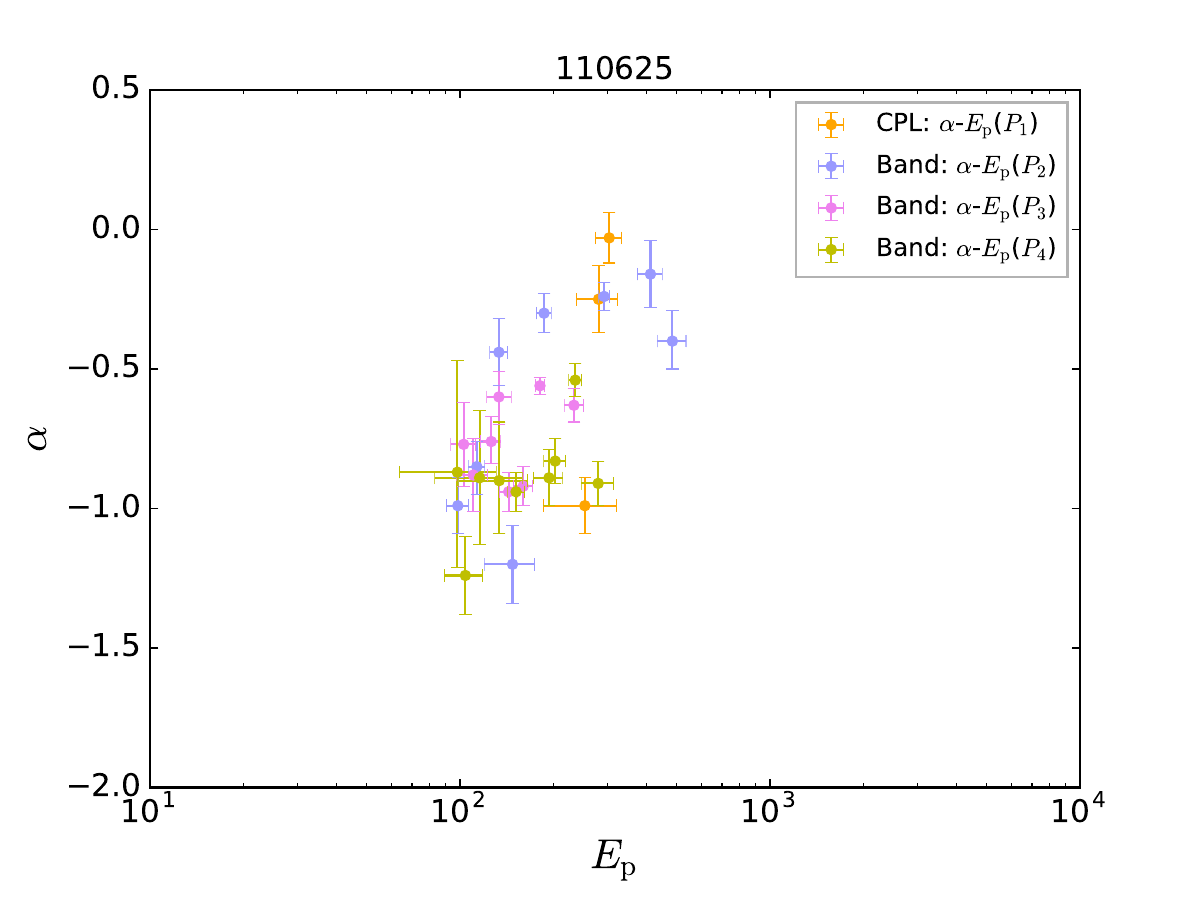}
\includegraphics[angle=0,scale=0.3]{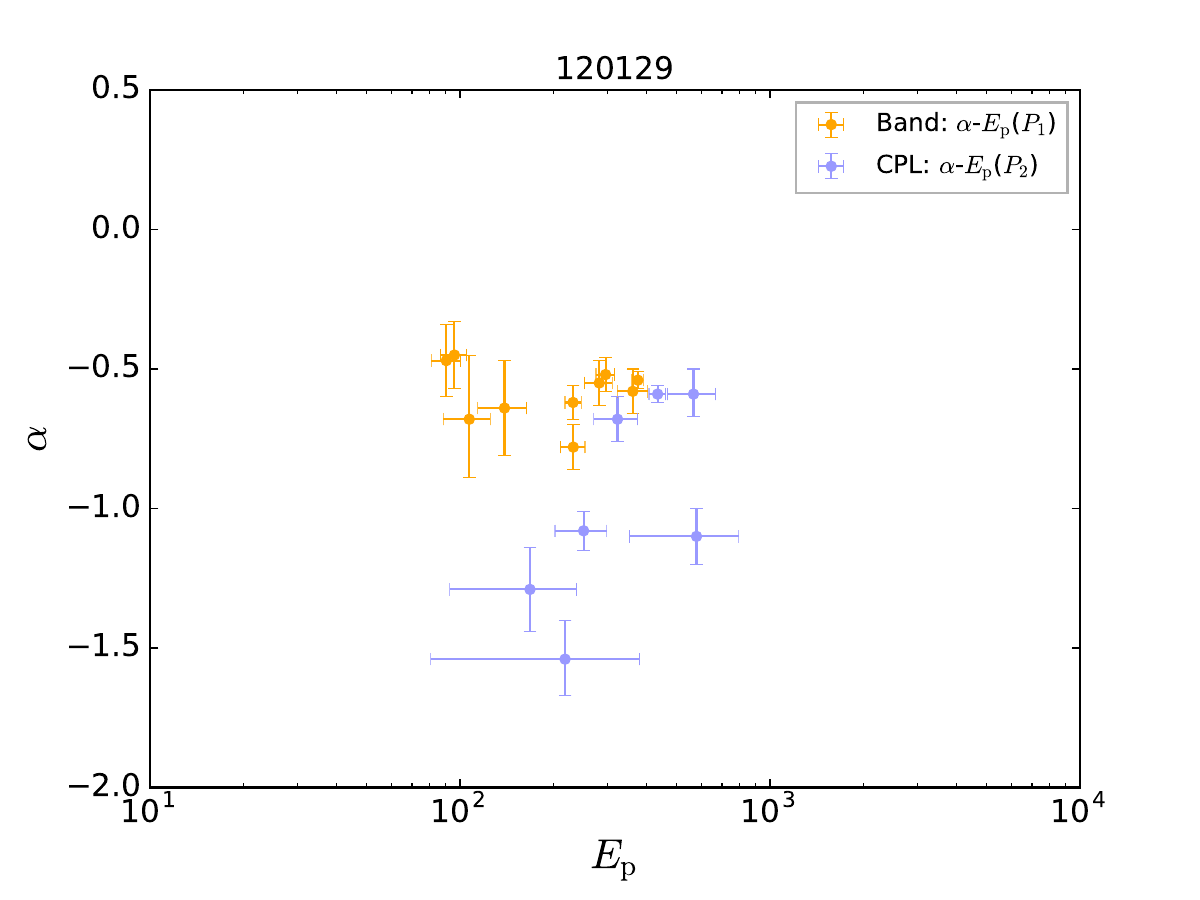}
\includegraphics[angle=0,scale=0.3]{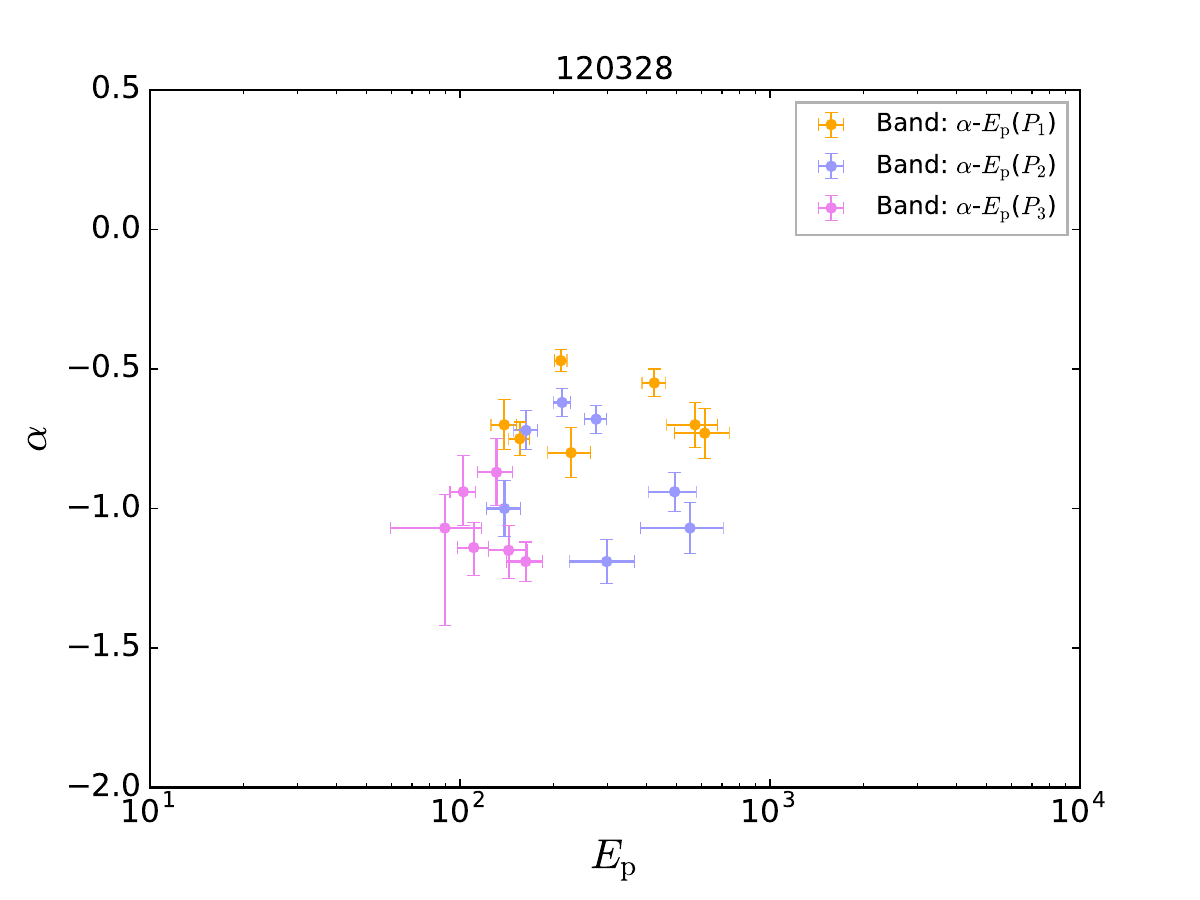}
\caption{The $\alpha$-$E_{\rm p}$ relation. The symbols and colors are the same as in Figure \ref{fig:FluxAlpha_Best}.}\label{fig:EpAlpha_Best}
\end{figure*}
\begin{figure*}
\includegraphics[angle=0,scale=0.3]{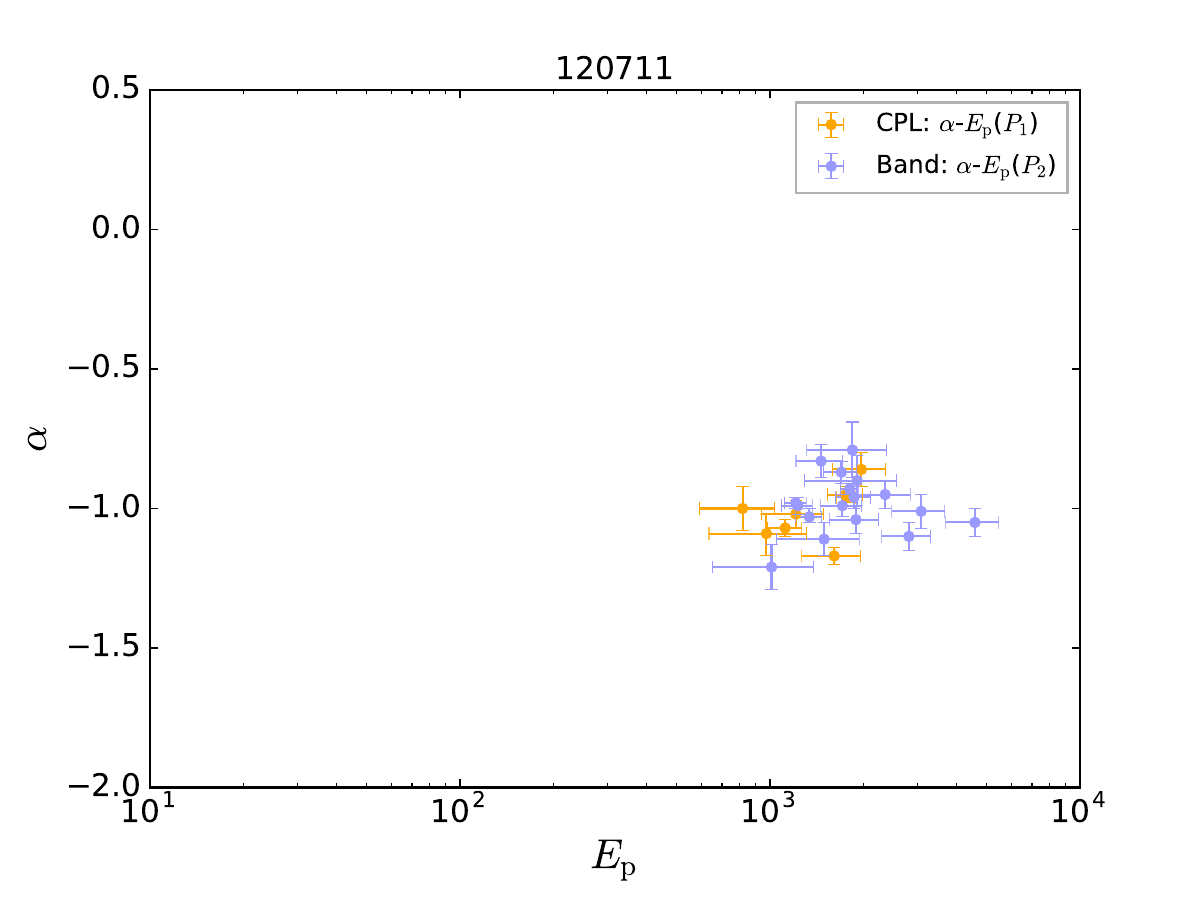}
\includegraphics[angle=0,scale=0.3]{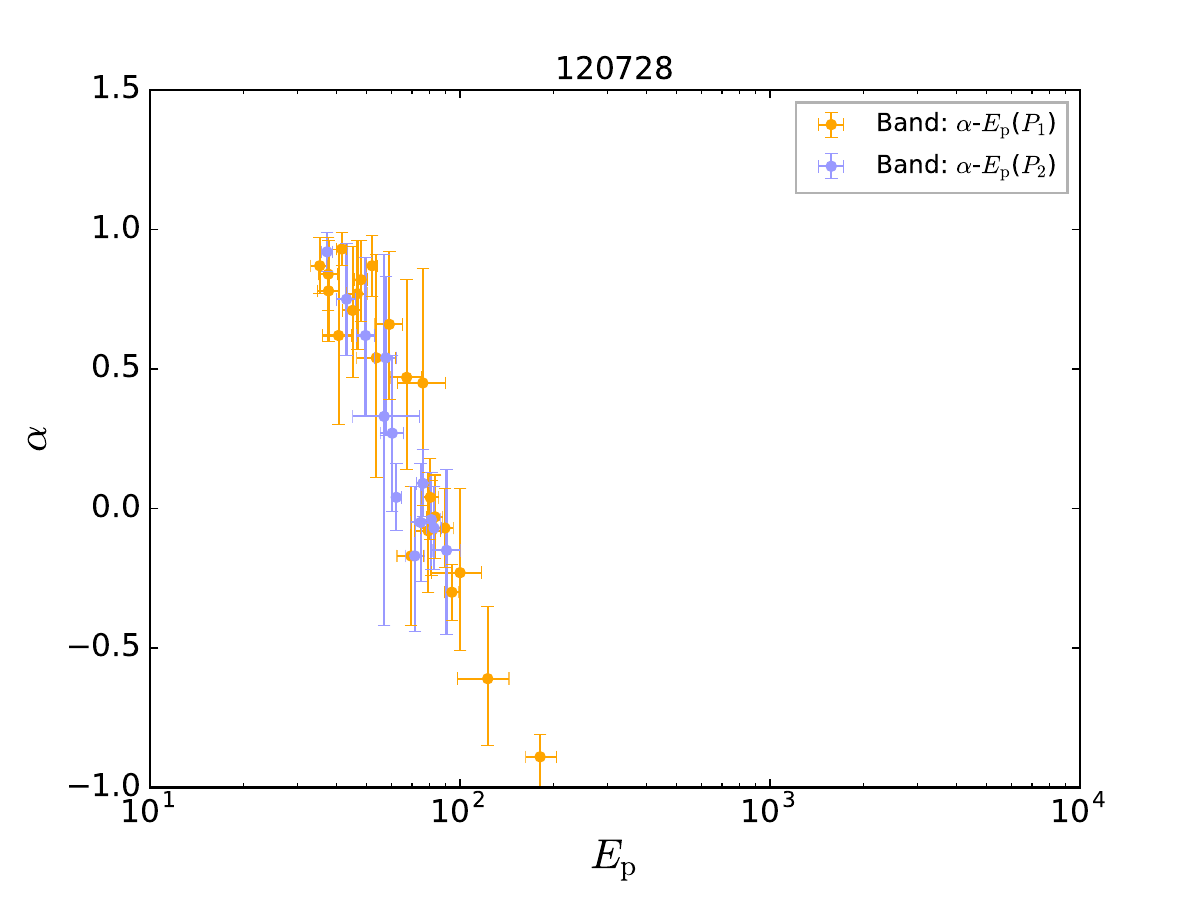}
\includegraphics[angle=0,scale=0.3]{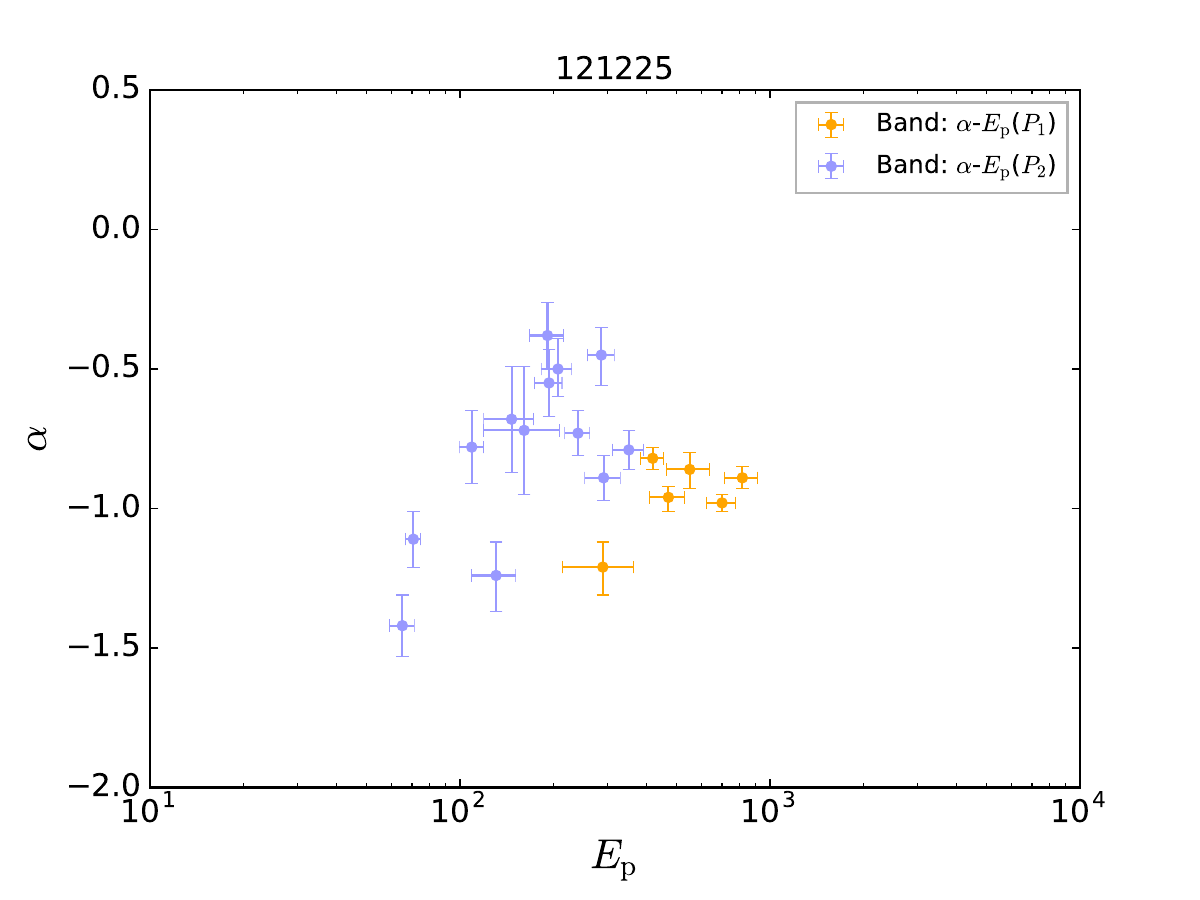}
\includegraphics[angle=0,scale=0.3]{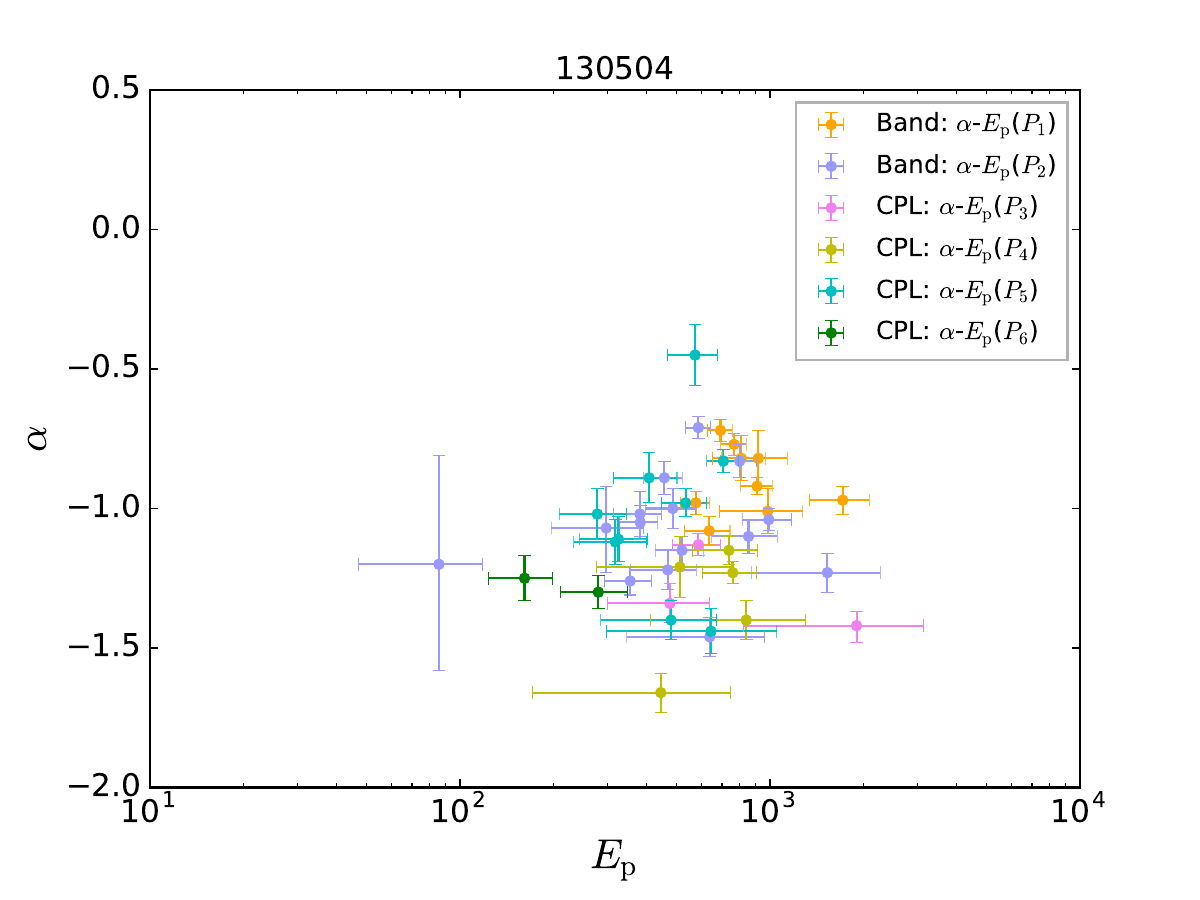}
\includegraphics[angle=0,scale=0.3]{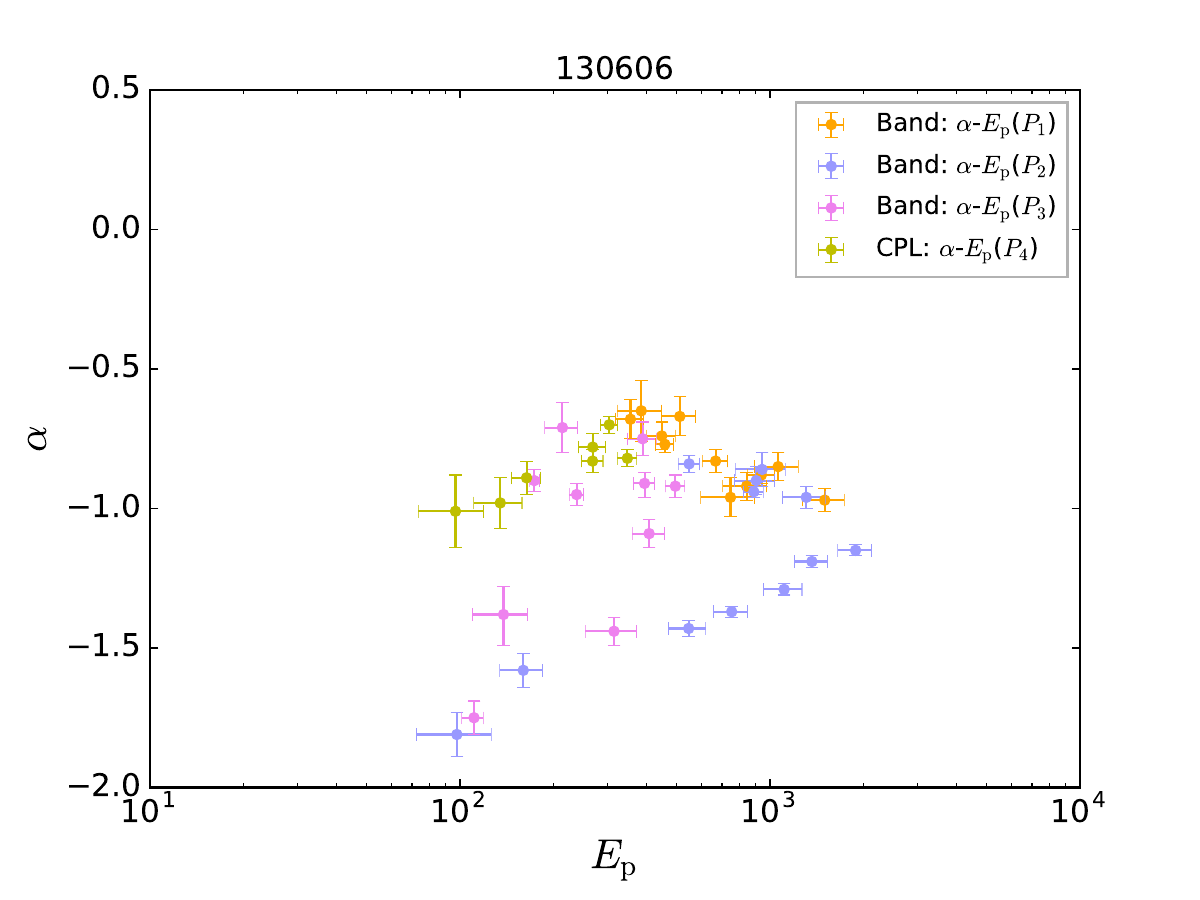}
\includegraphics[angle=0,scale=0.3]{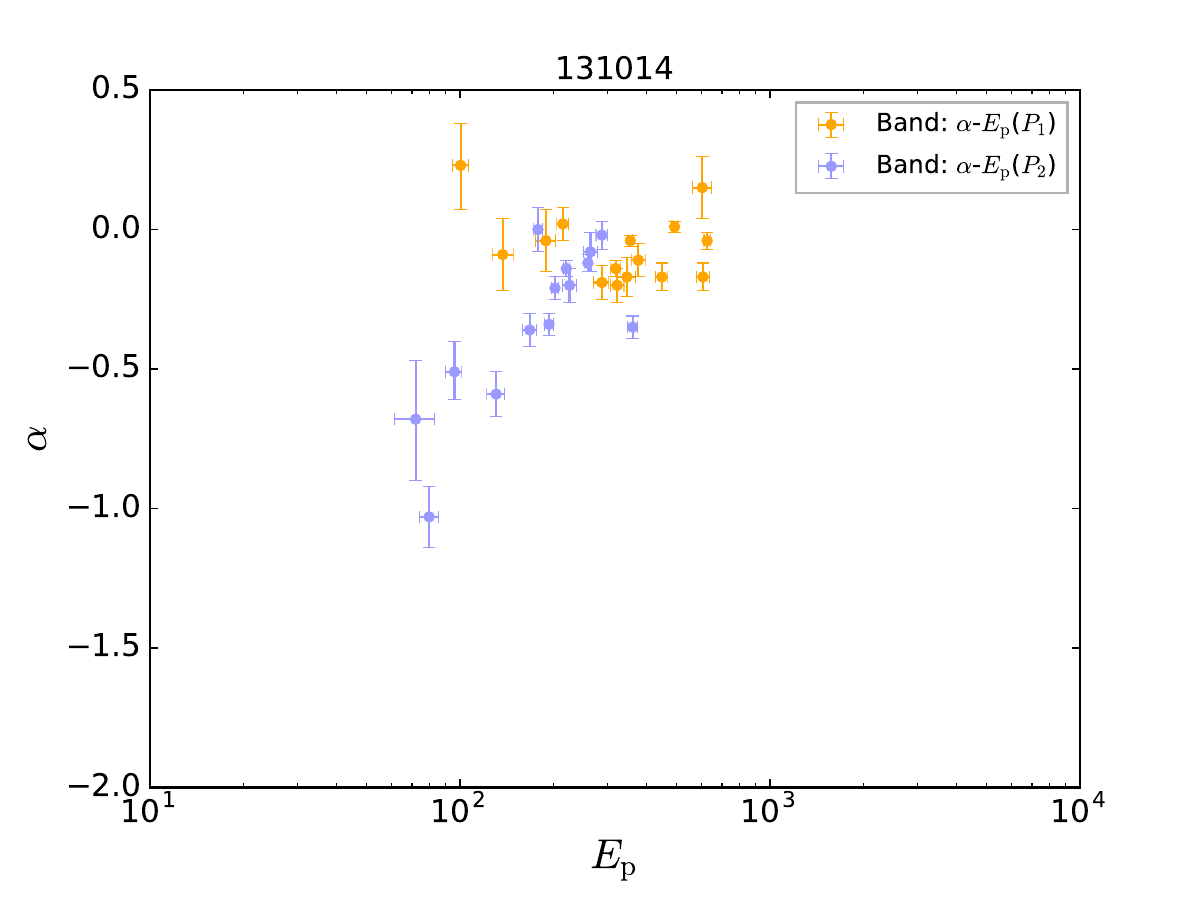}
\includegraphics[angle=0,scale=0.3]{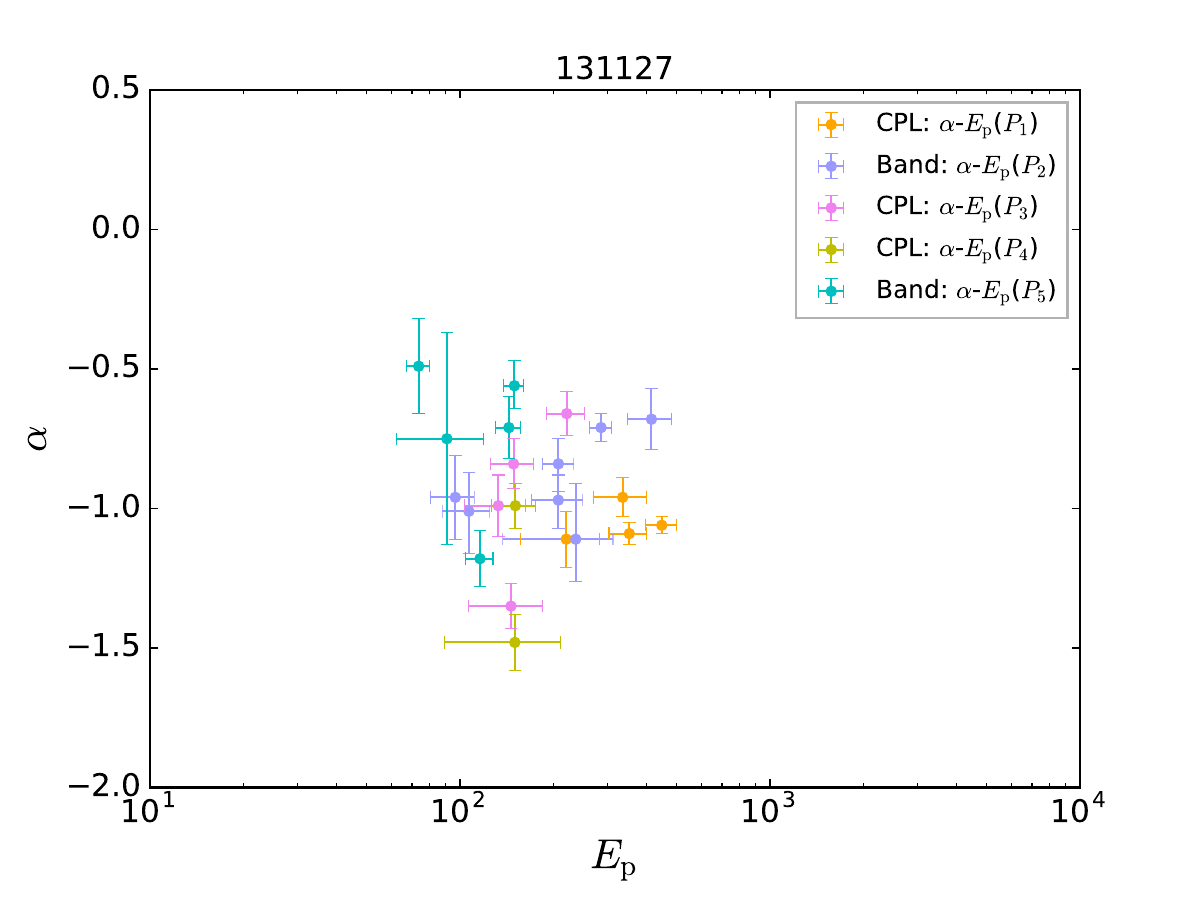}
\includegraphics[angle=0,scale=0.3]{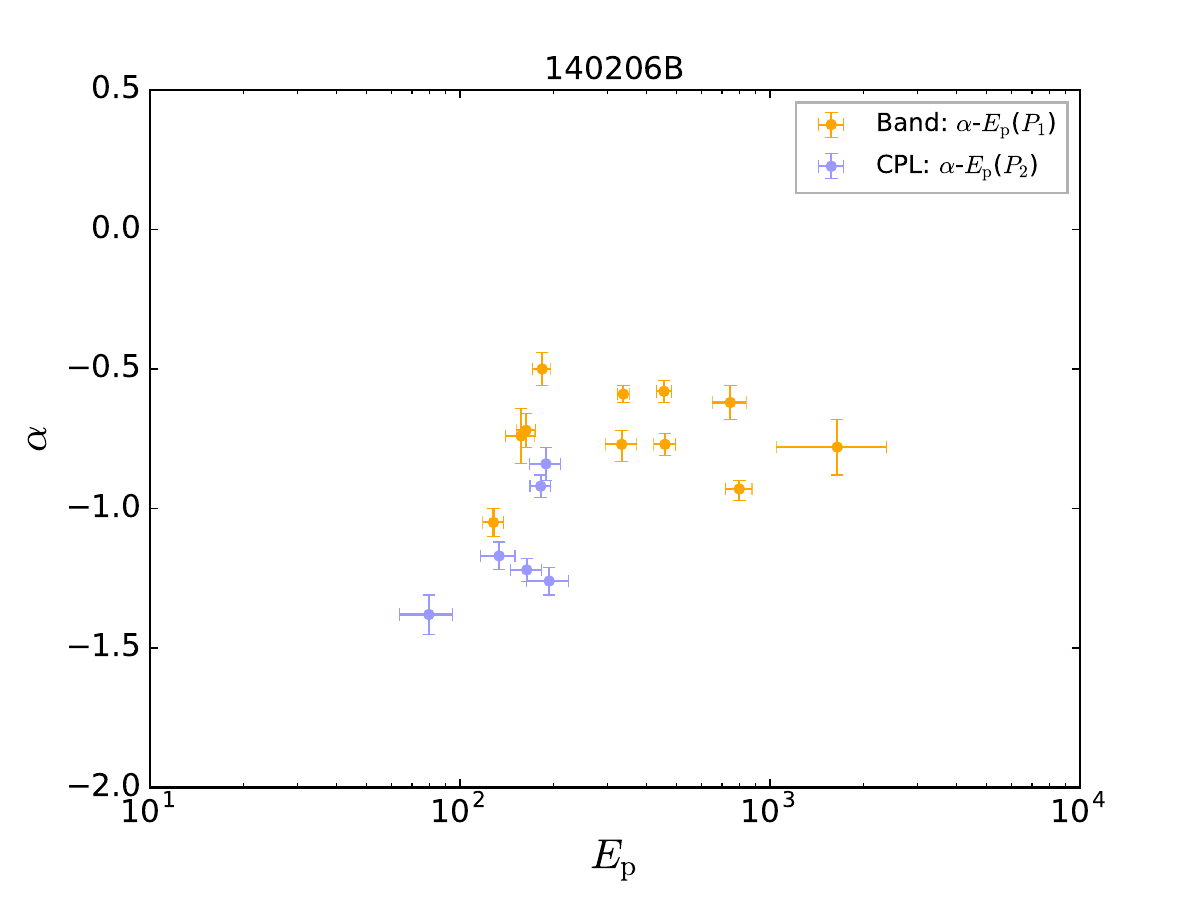}
\includegraphics[angle=0,scale=0.3]{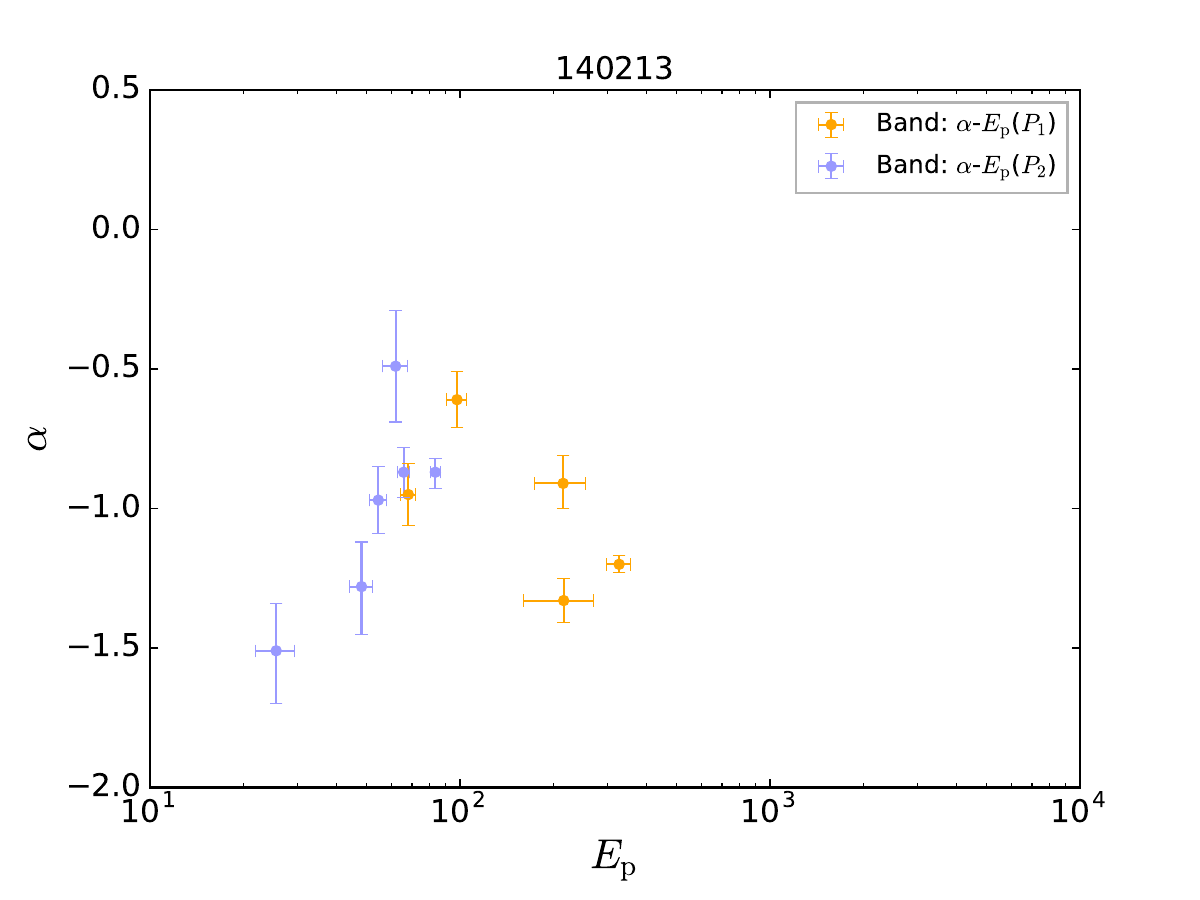}
\includegraphics[angle=0,scale=0.3]{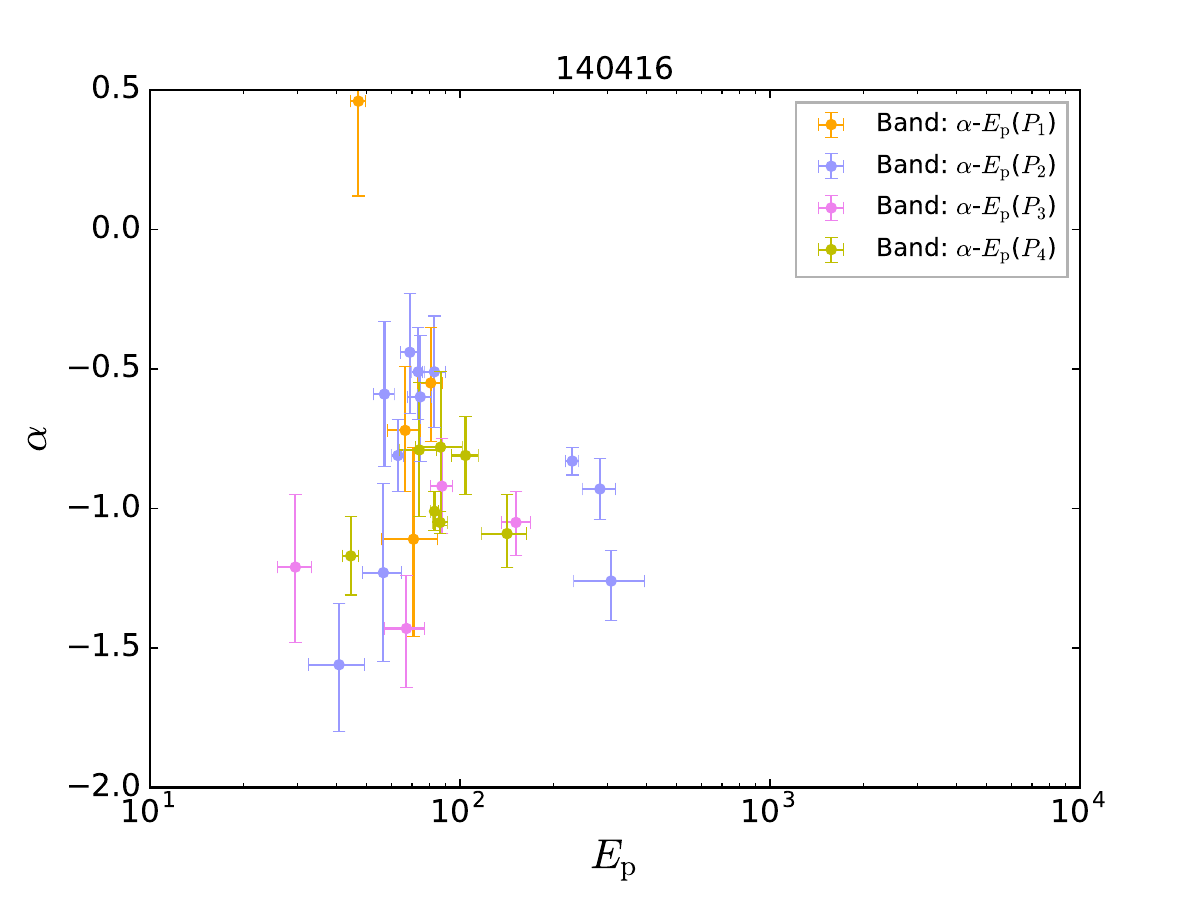}
\includegraphics[angle=0,scale=0.3]{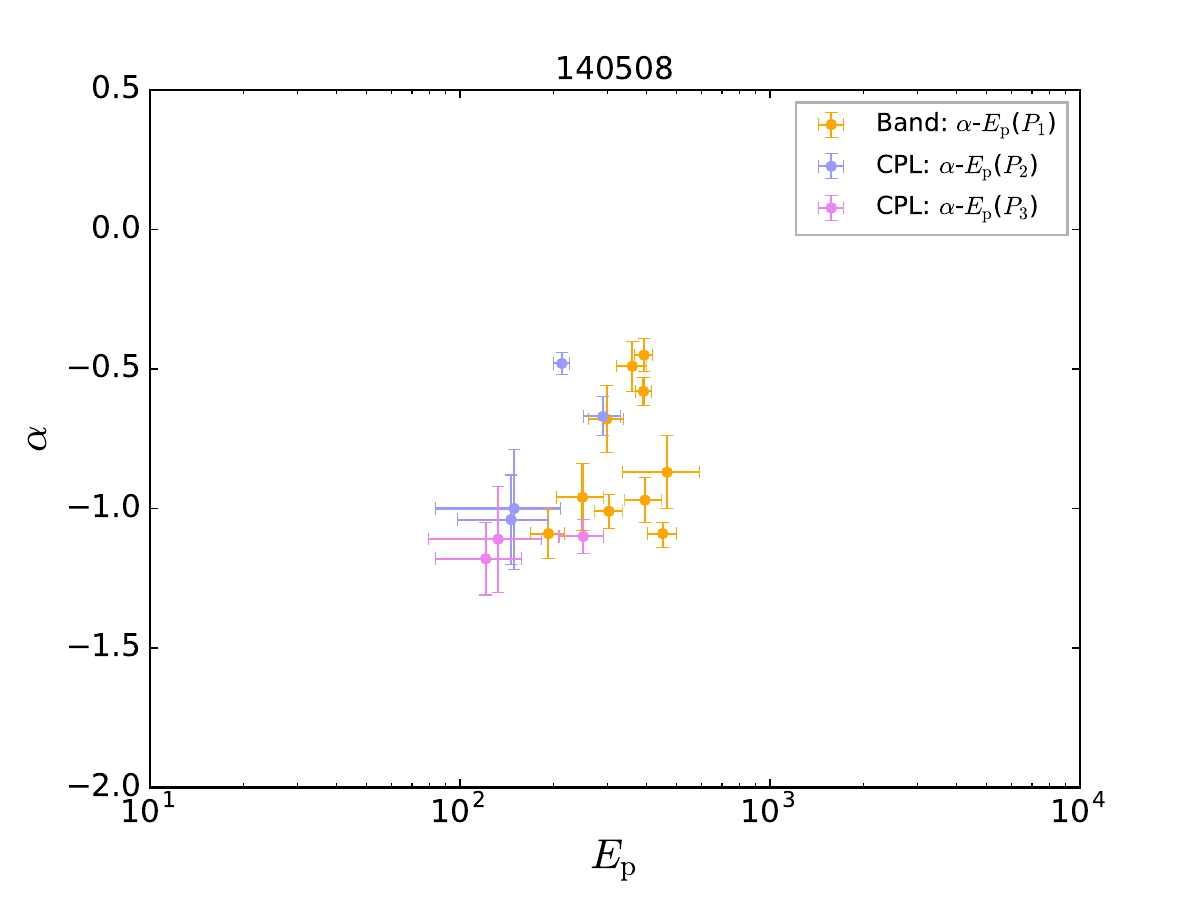}
\includegraphics[angle=0,scale=0.3]{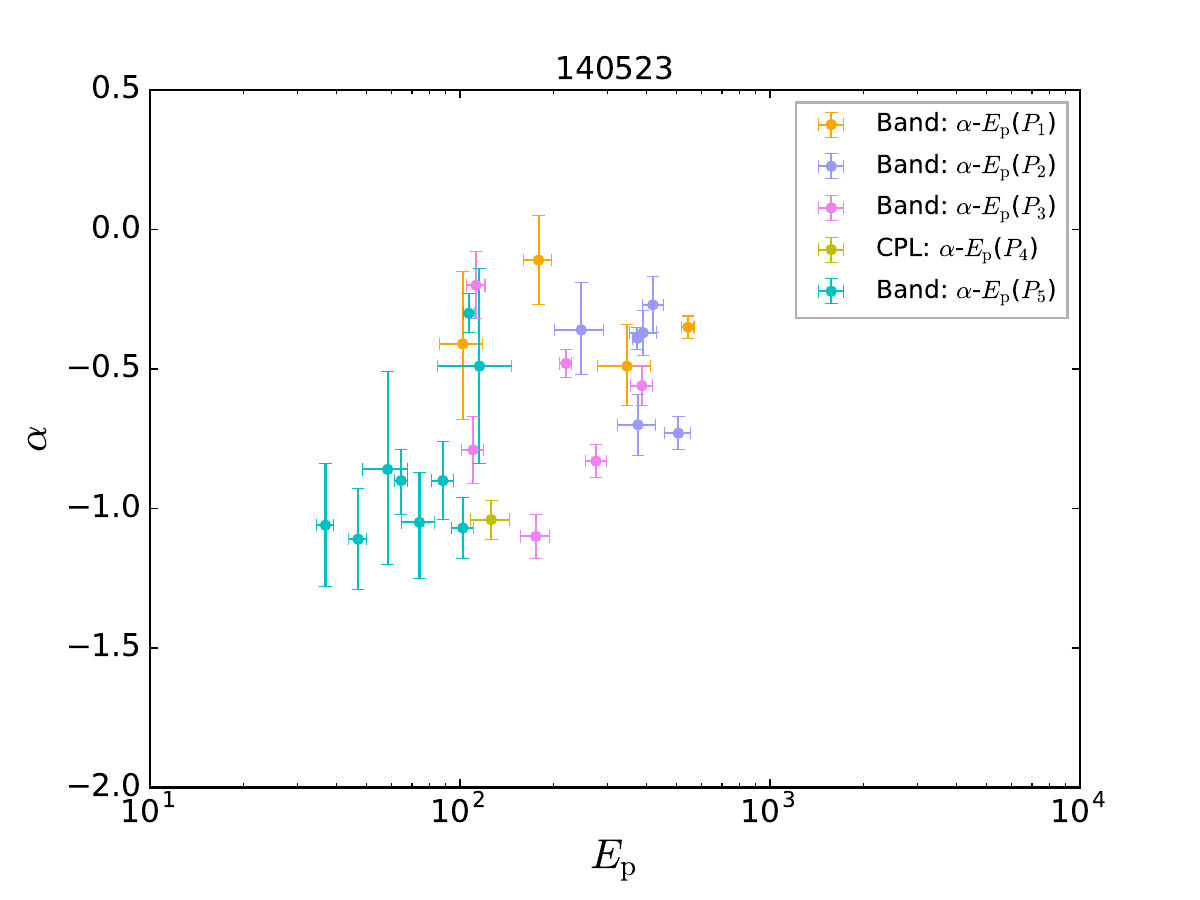}
\includegraphics[angle=0,scale=0.3]{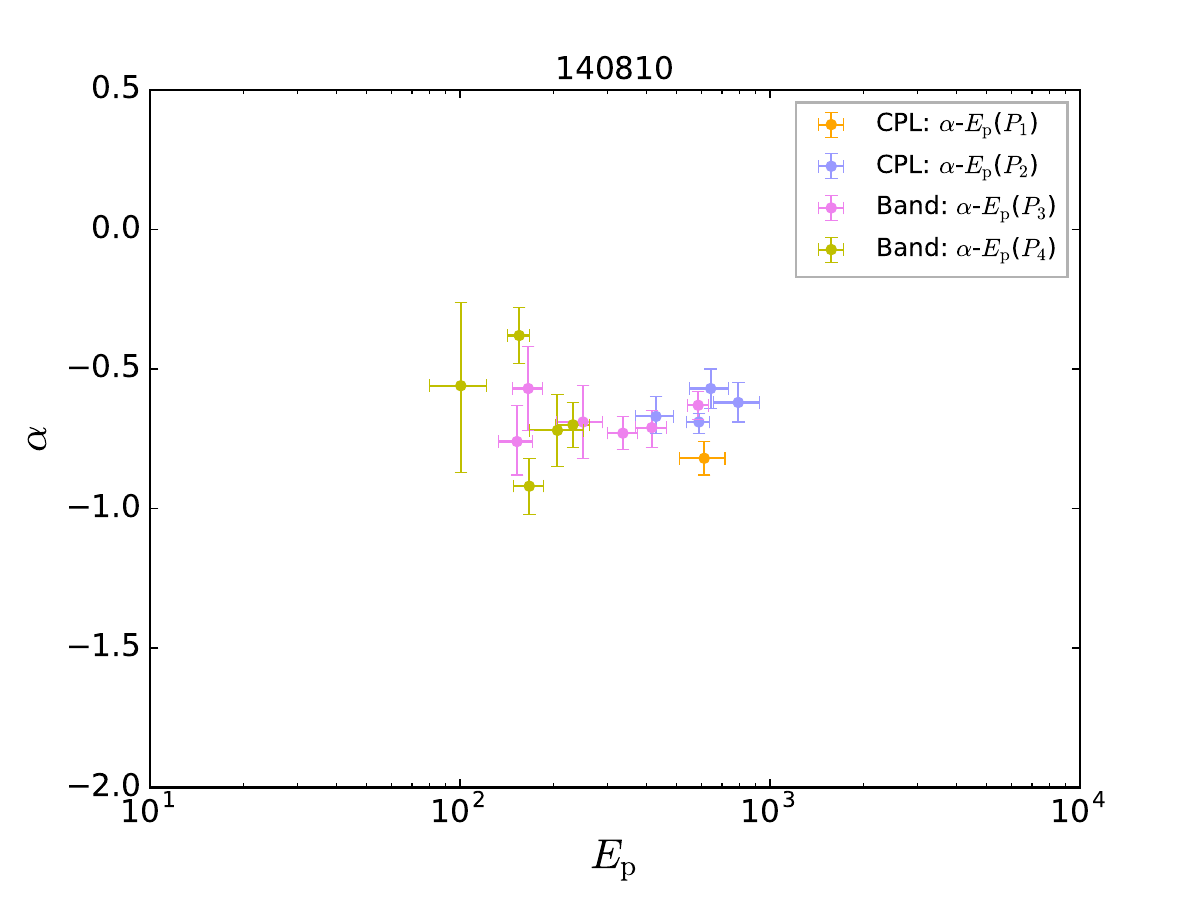}
\includegraphics[angle=0,scale=0.3]{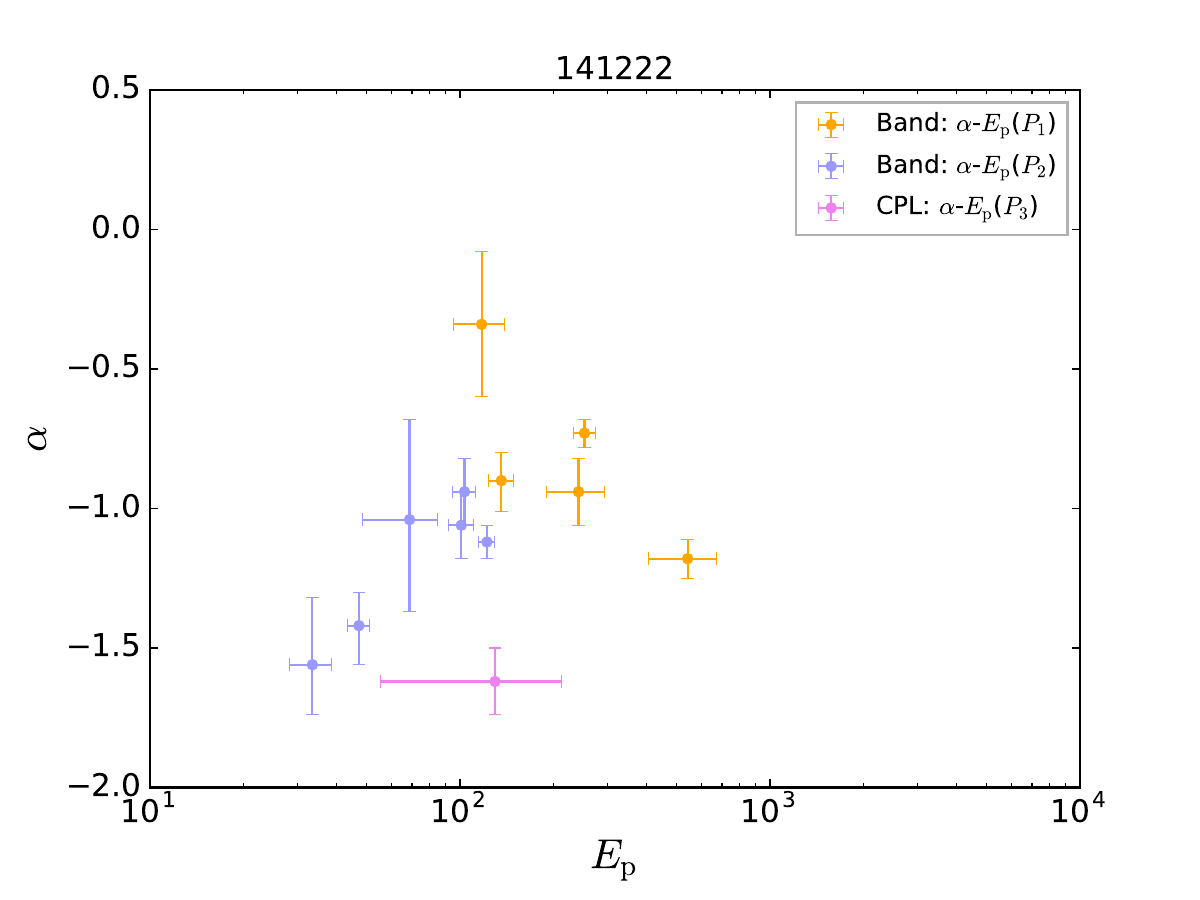}
\includegraphics[angle=0,scale=0.3]{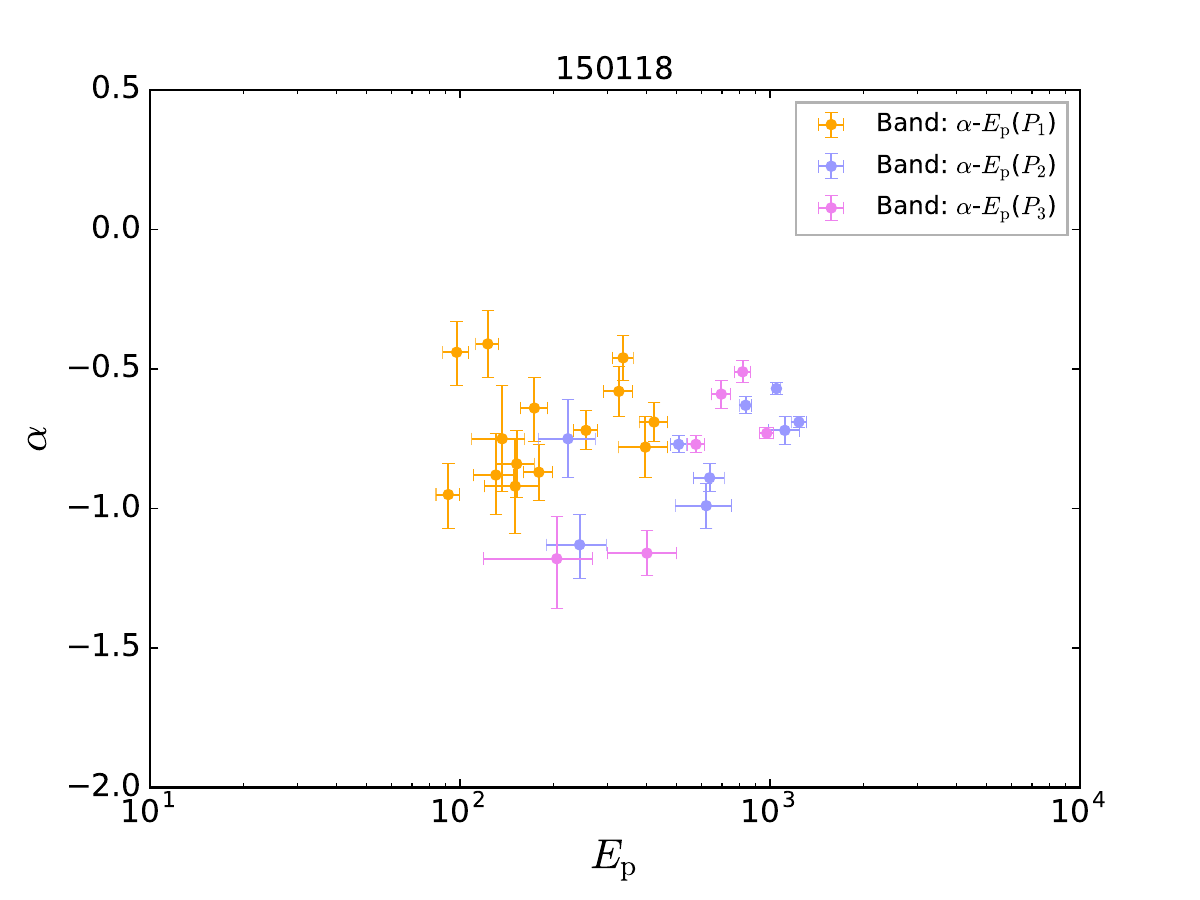}
\center{Fig. \ref{fig:EpAlpha_Best}--- Continued}
\end{figure*}
\begin{figure*}
\includegraphics[angle=0,scale=0.3]{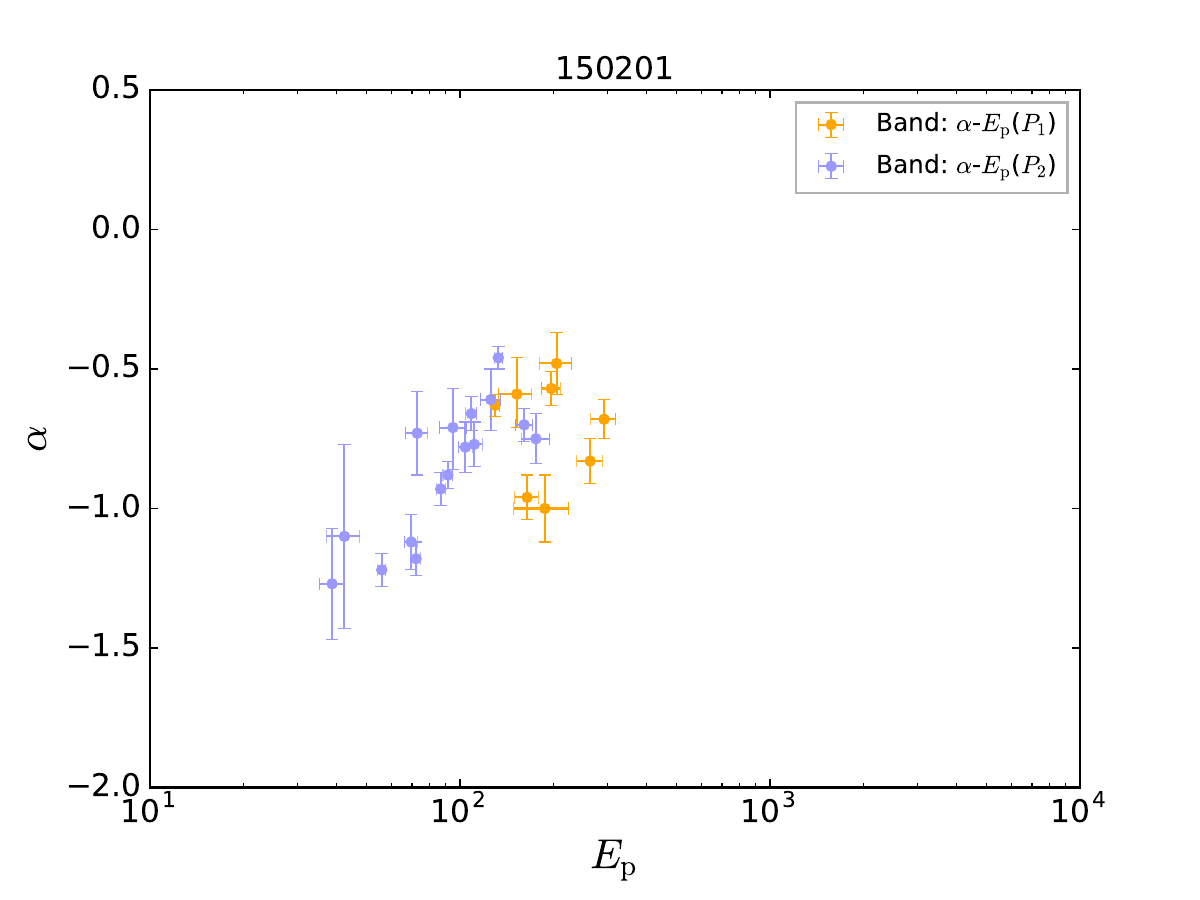}
\includegraphics[angle=0,scale=0.3]{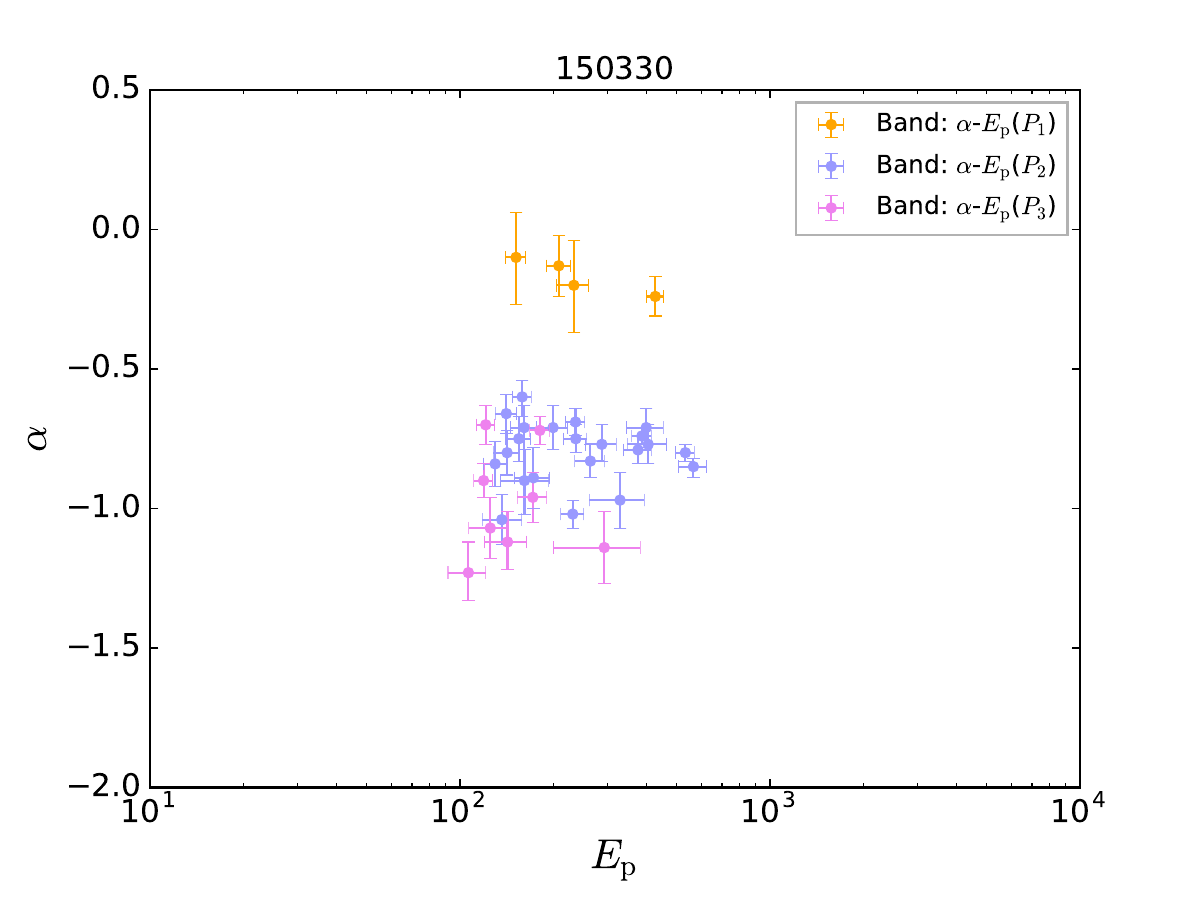}
\includegraphics[angle=0,scale=0.3]{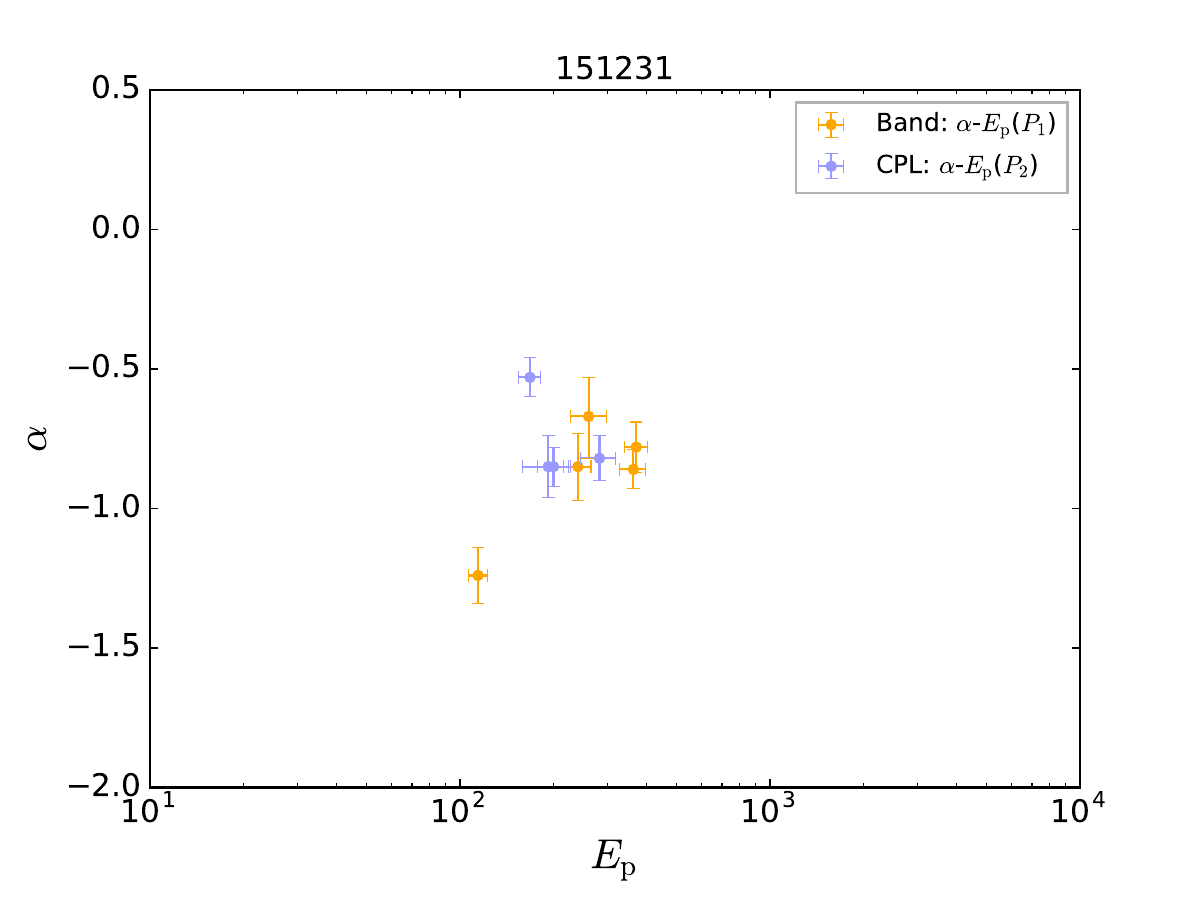}
\includegraphics[angle=0,scale=0.3]{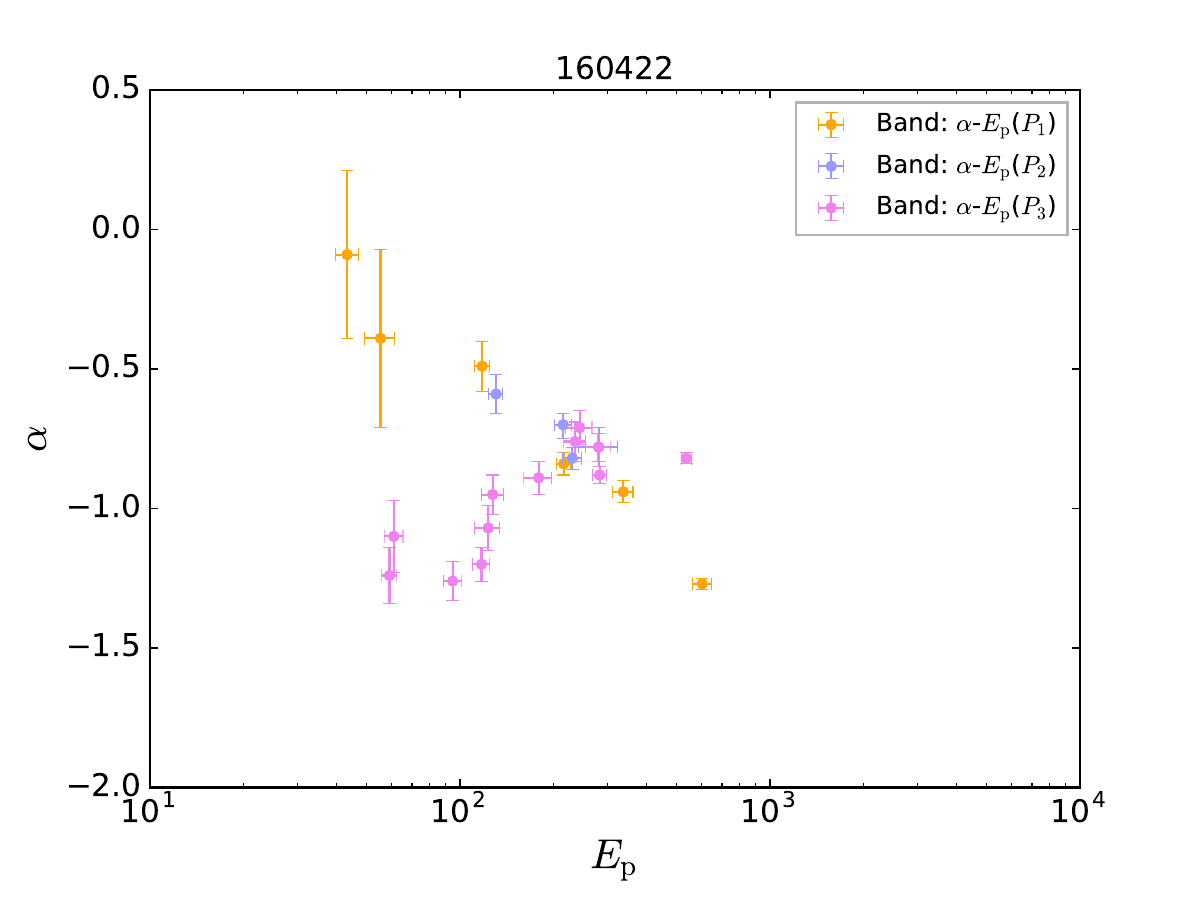}
\includegraphics[angle=0,scale=0.3]{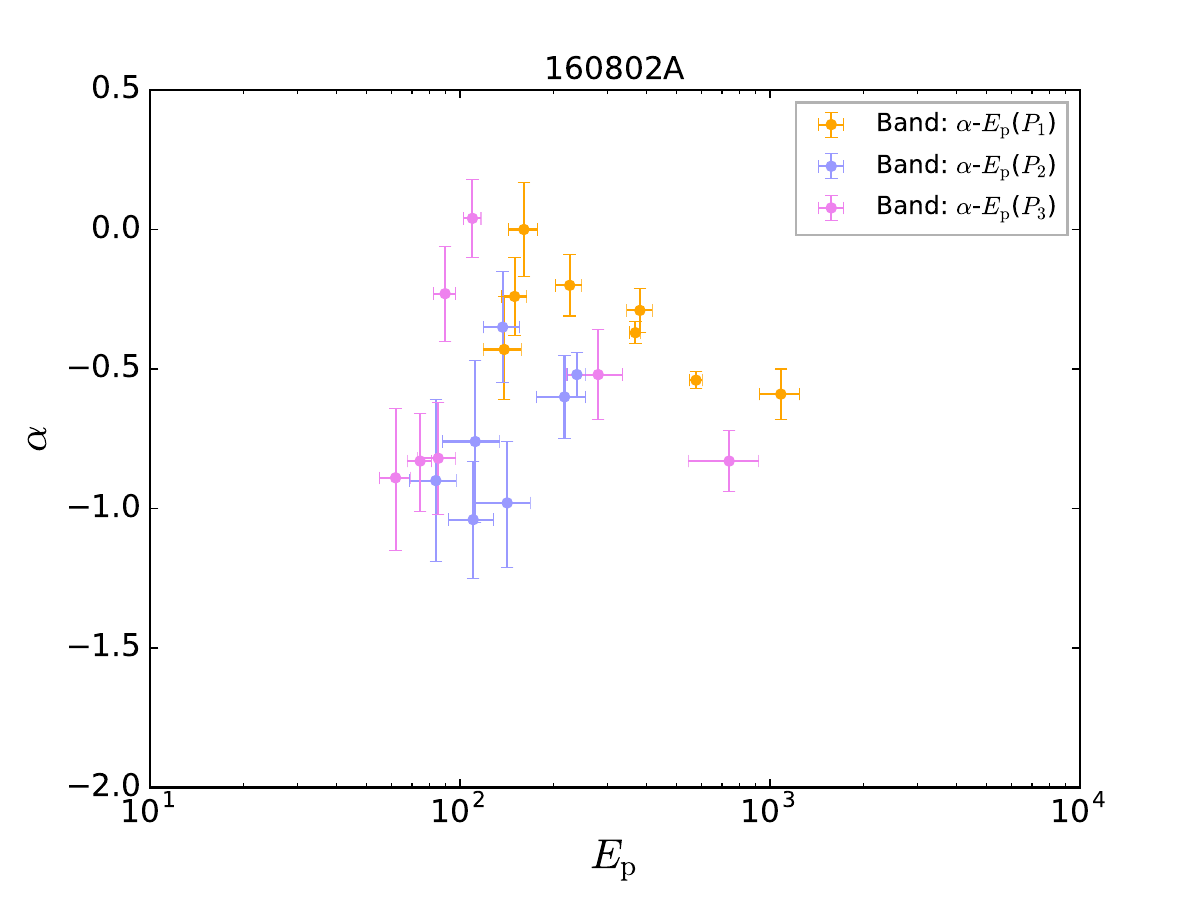}
\includegraphics[angle=0,scale=0.3]{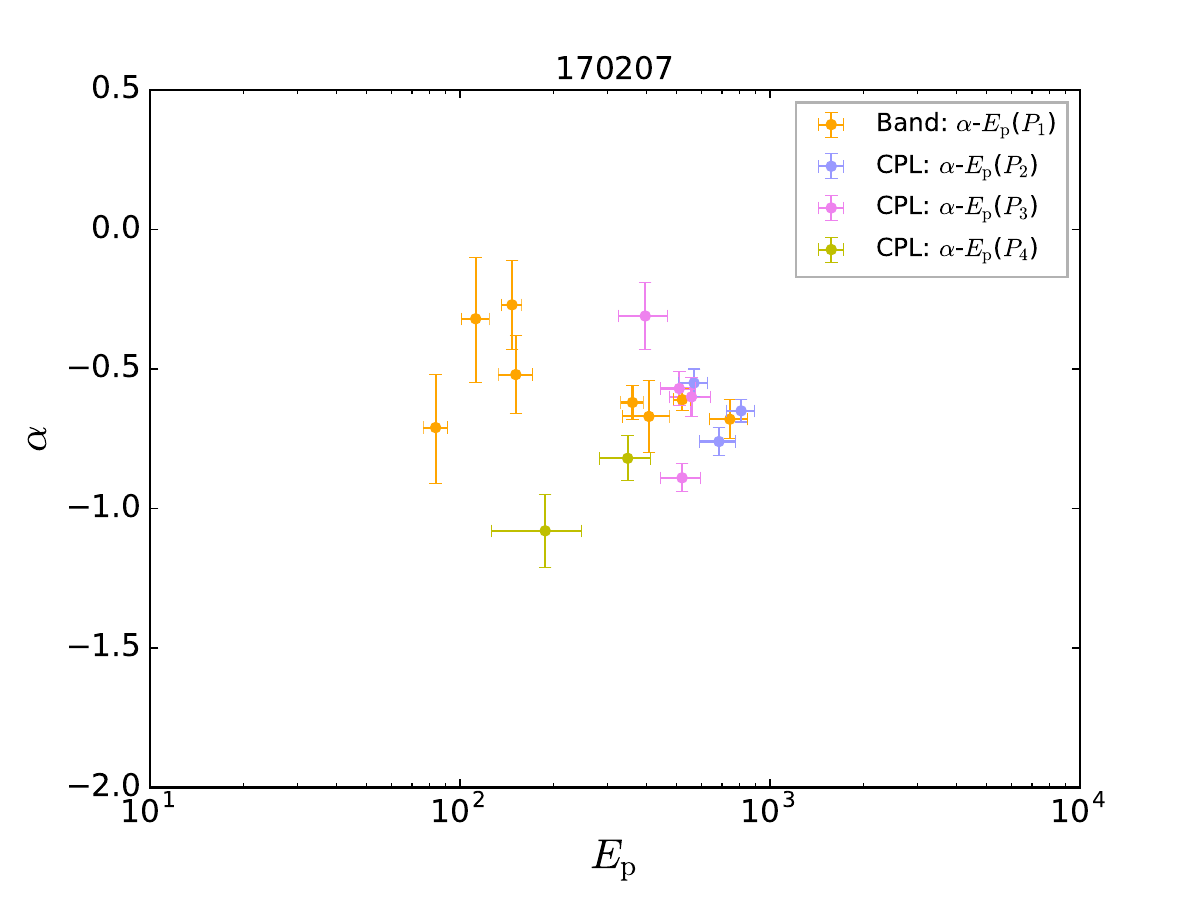}
\includegraphics[angle=0,scale=0.3]{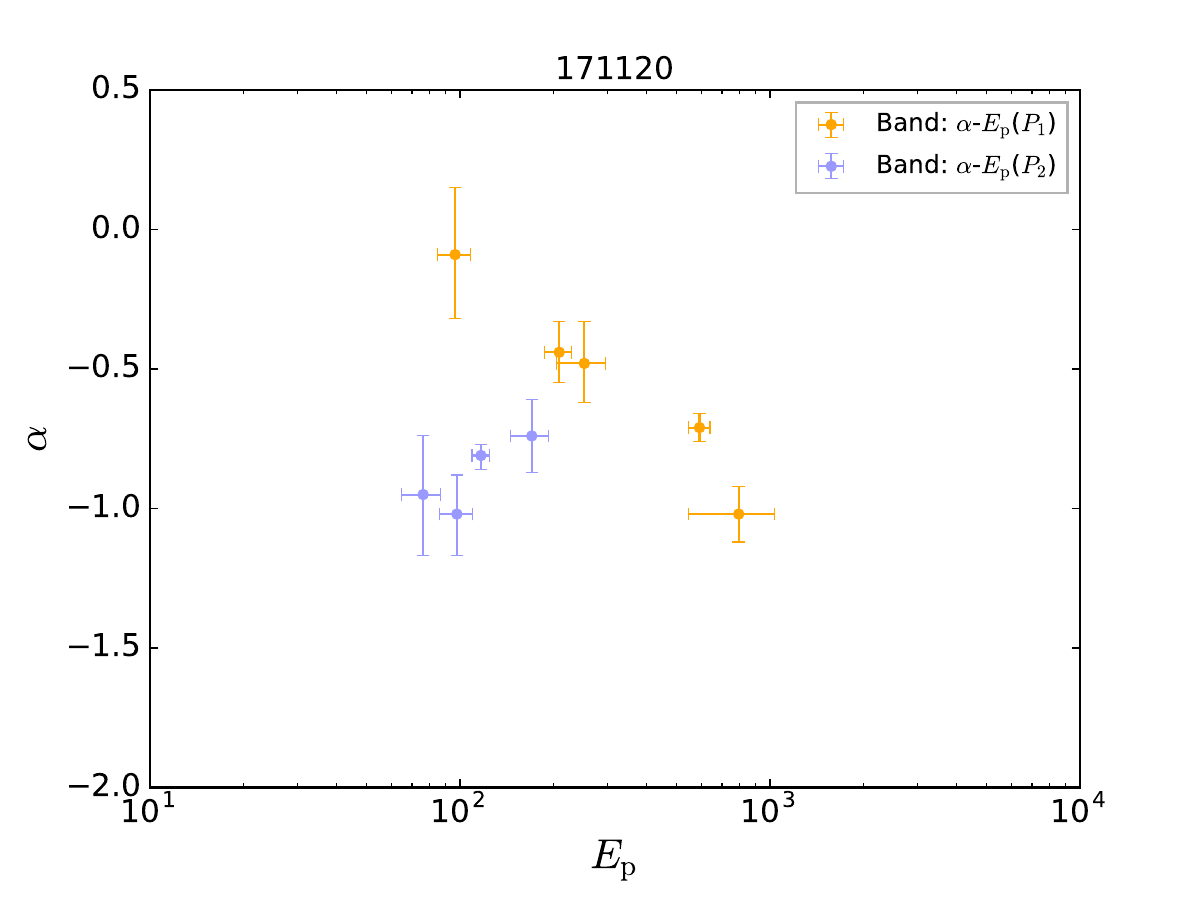}
\includegraphics[angle=0,scale=0.3]{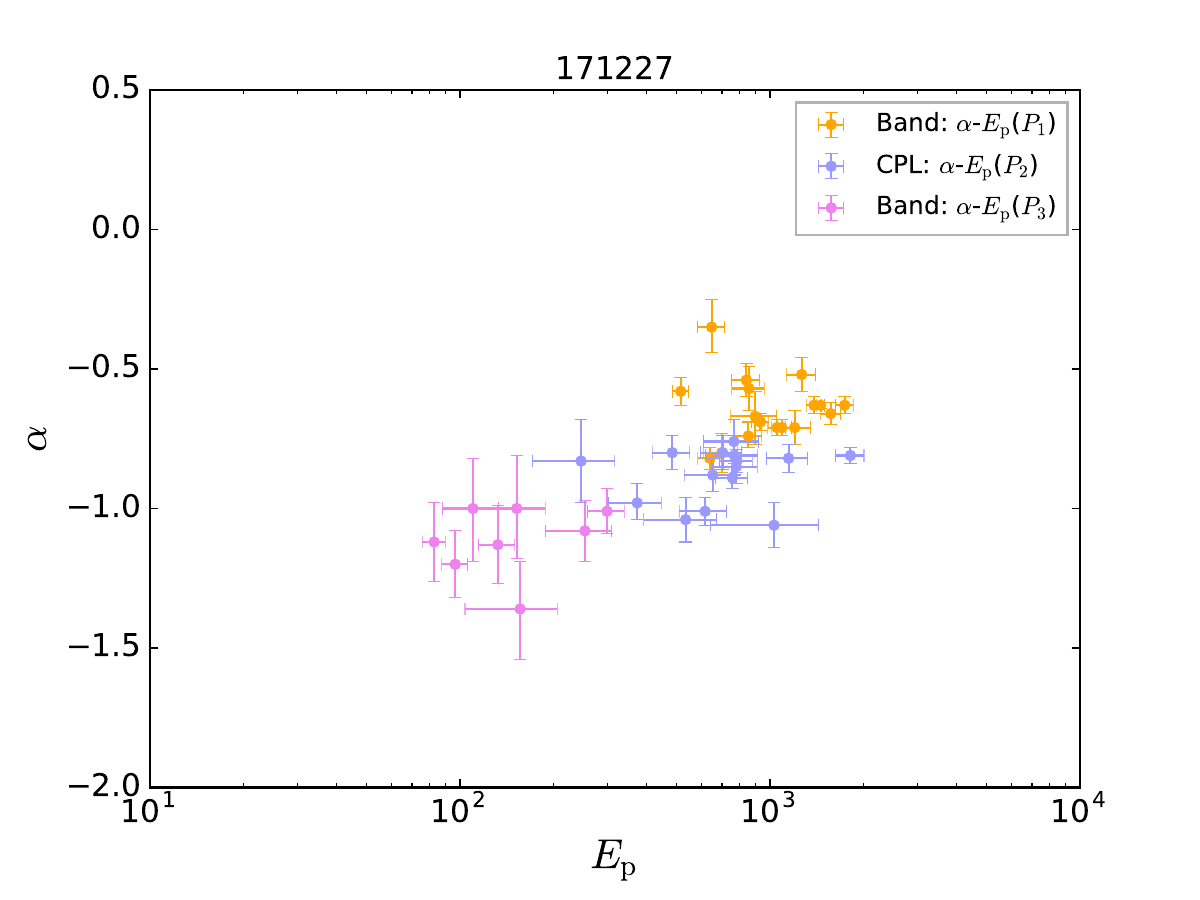}
\includegraphics[angle=0,scale=0.3]{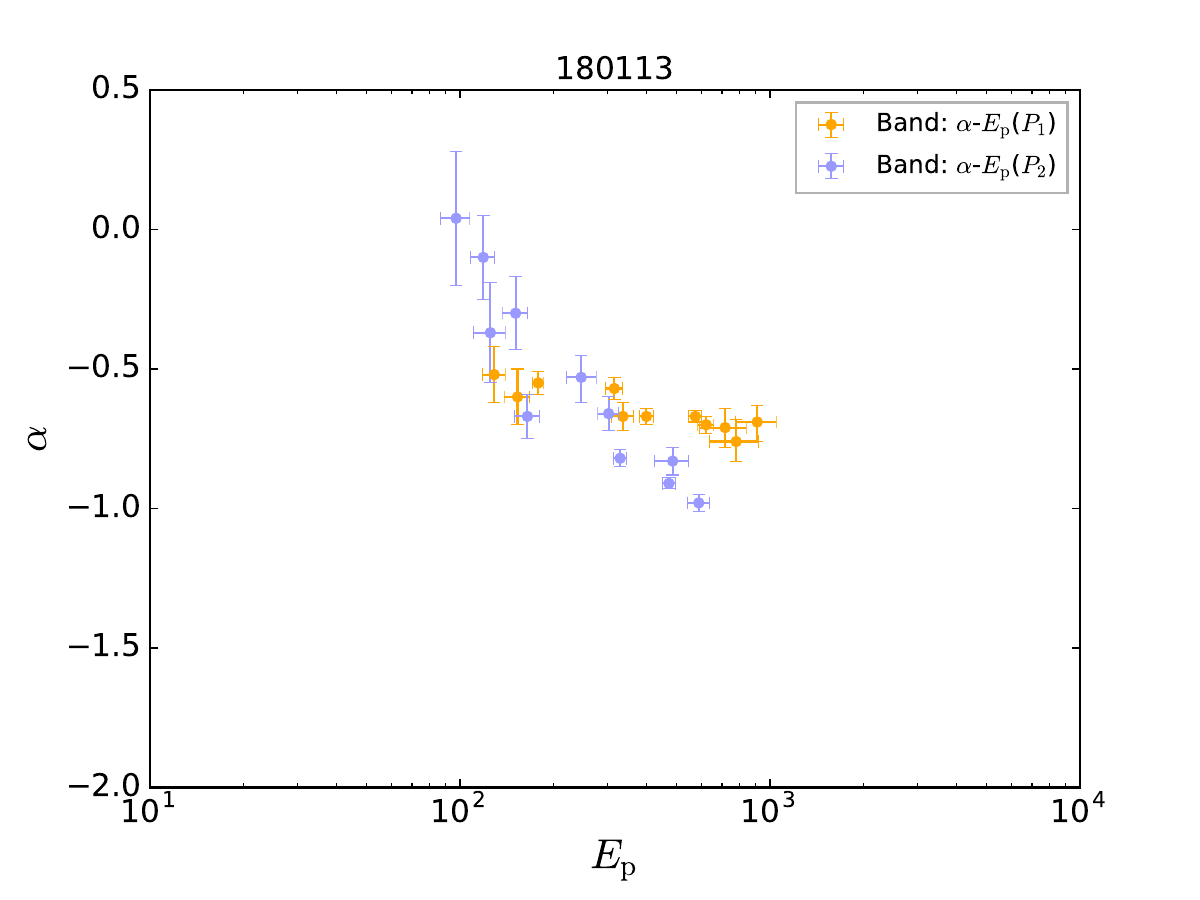}
\includegraphics[angle=0,scale=0.3]{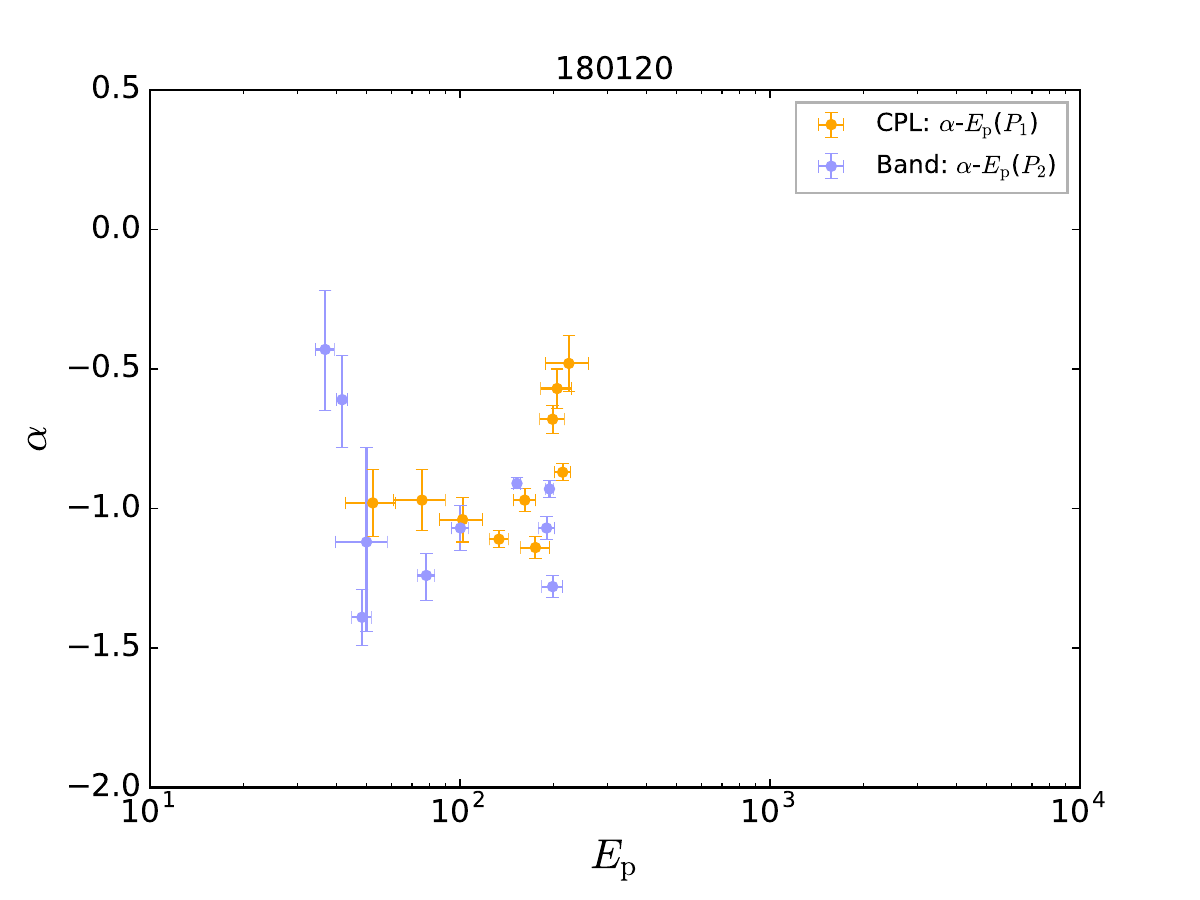}
\includegraphics[angle=0,scale=0.3]{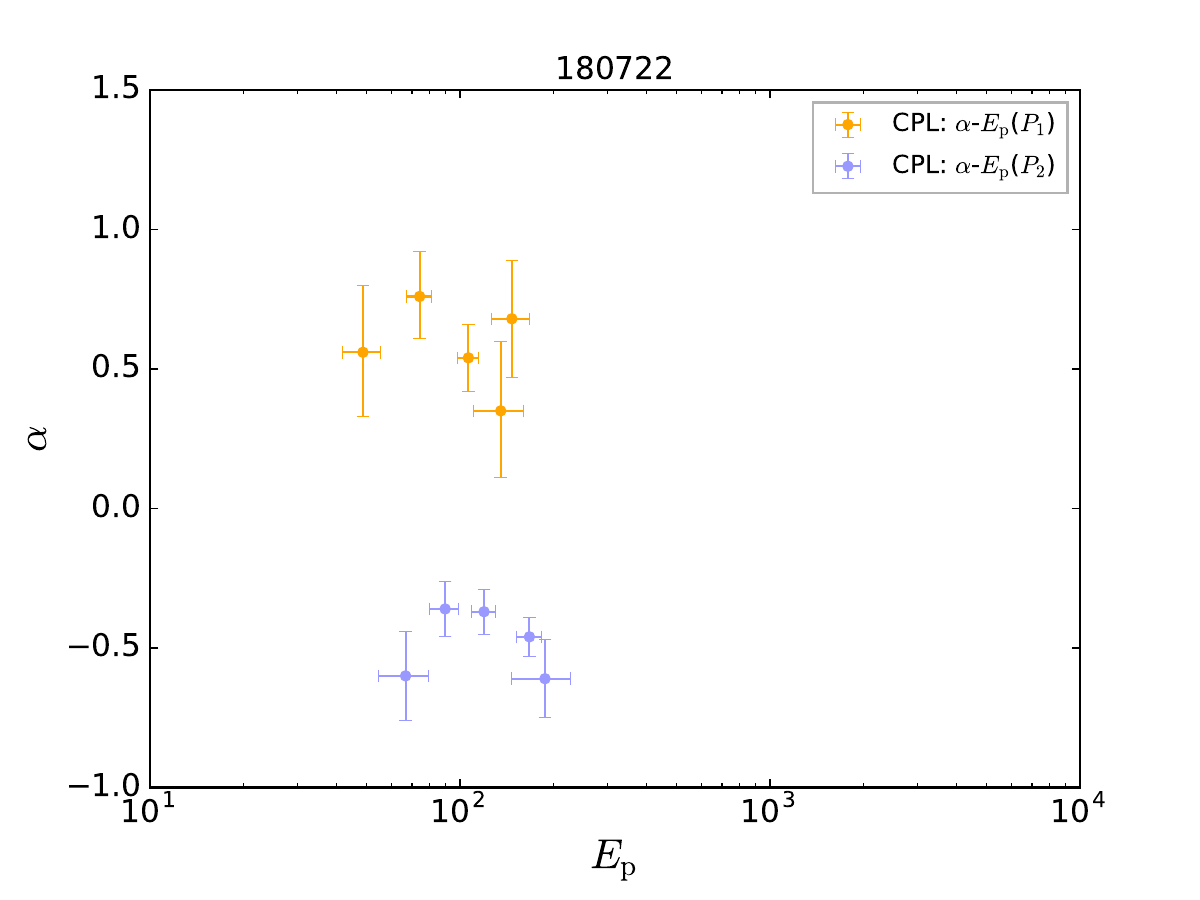}
\includegraphics[angle=0,scale=0.3]{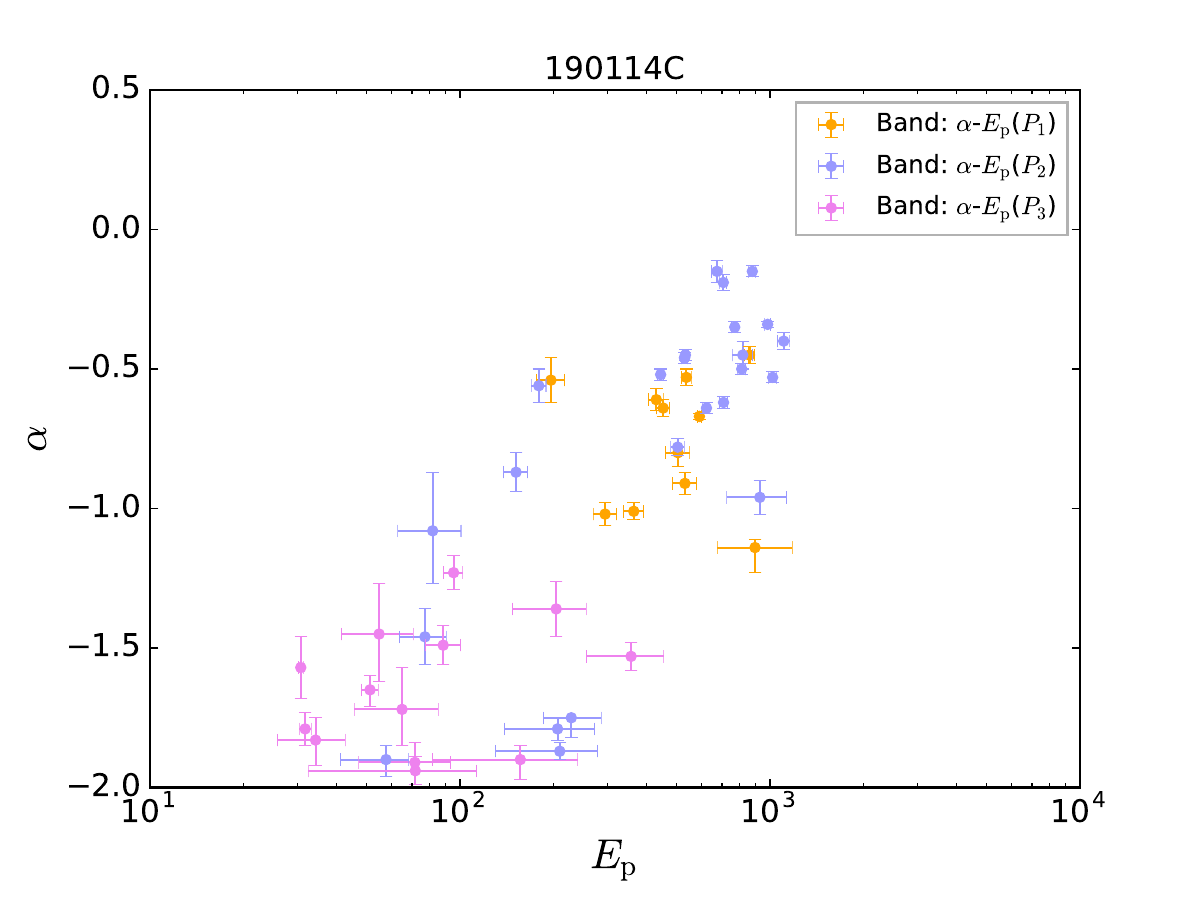}
\center{Fig. \ref{fig:EpAlpha_Best}--- Continued}
\end{figure*}

\clearpage
\subsection{Single Model-based Parameter Distributions in Different Pulses}

\begin{figure*}
\includegraphics[angle=0,scale=0.45]{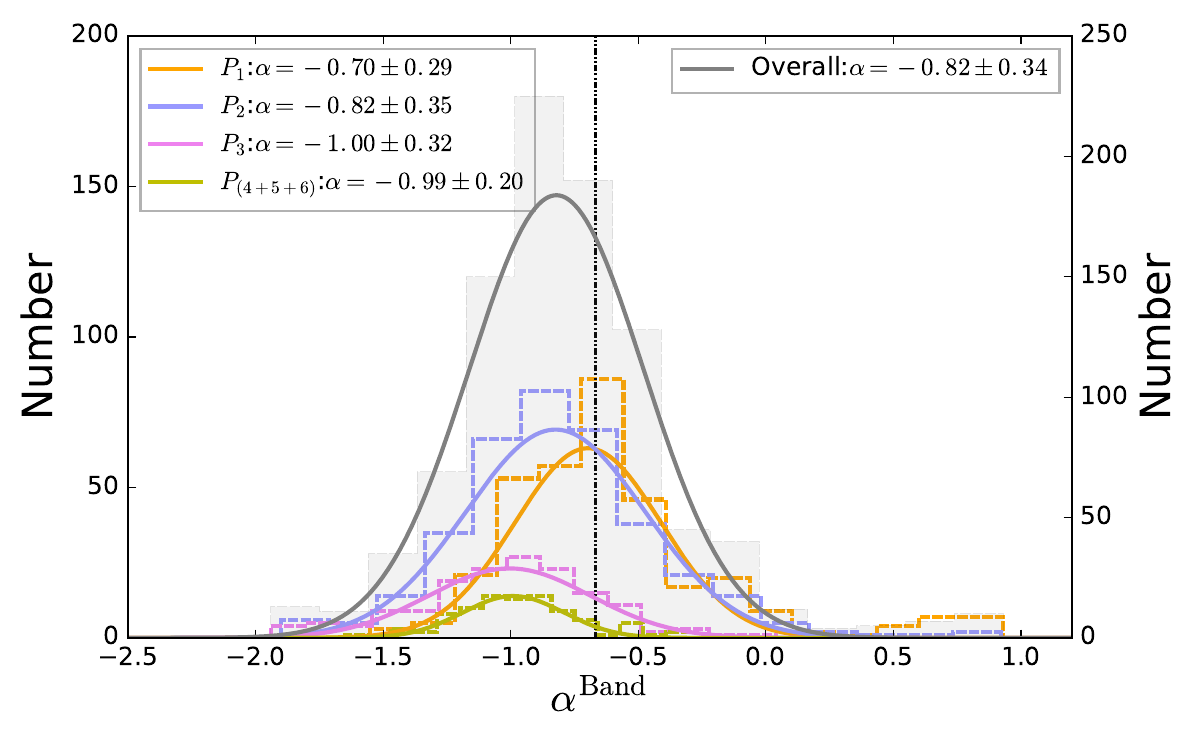}
\includegraphics[angle=0,scale=0.45]{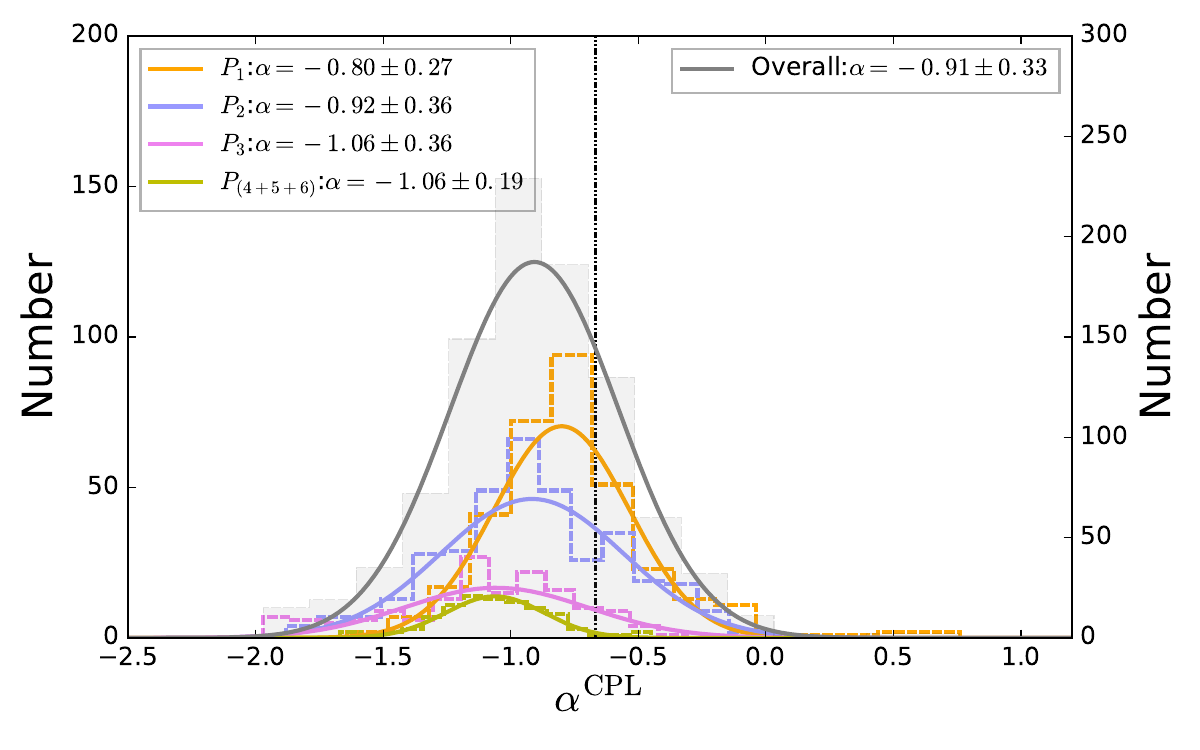}
\includegraphics[angle=0,scale=0.45]{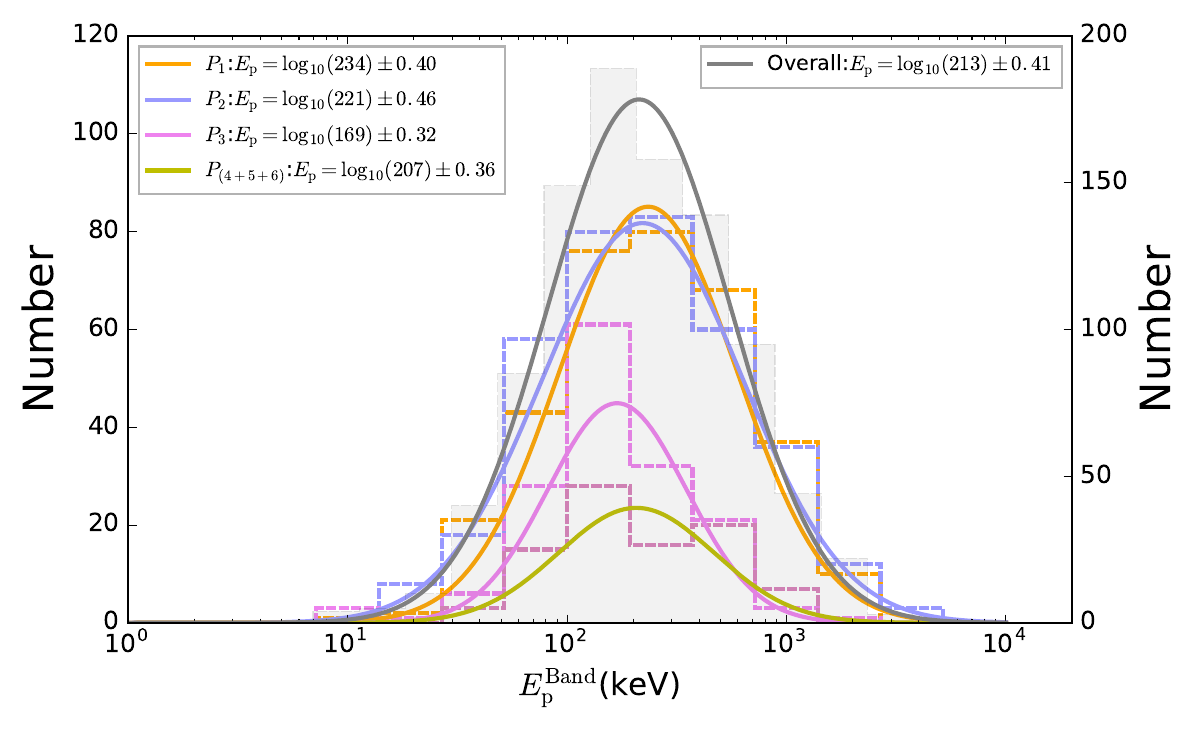}
\includegraphics[angle=0,scale=0.45]{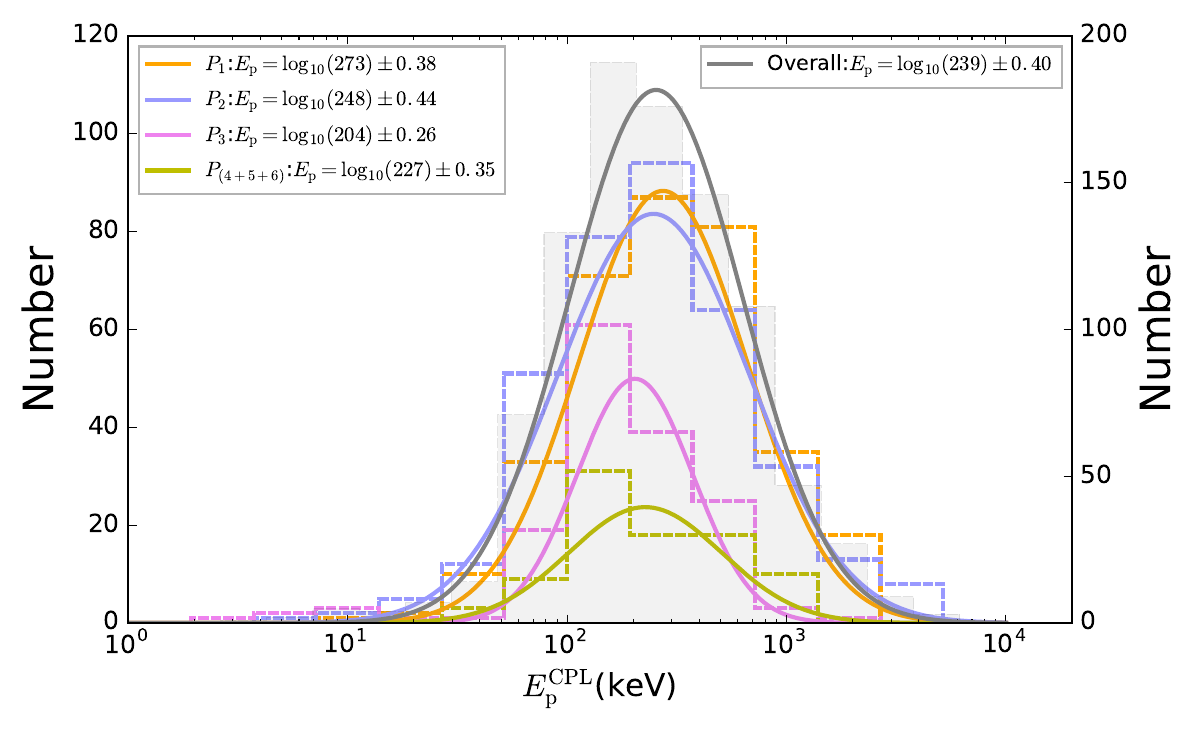}
\includegraphics[angle=0,scale=0.45]{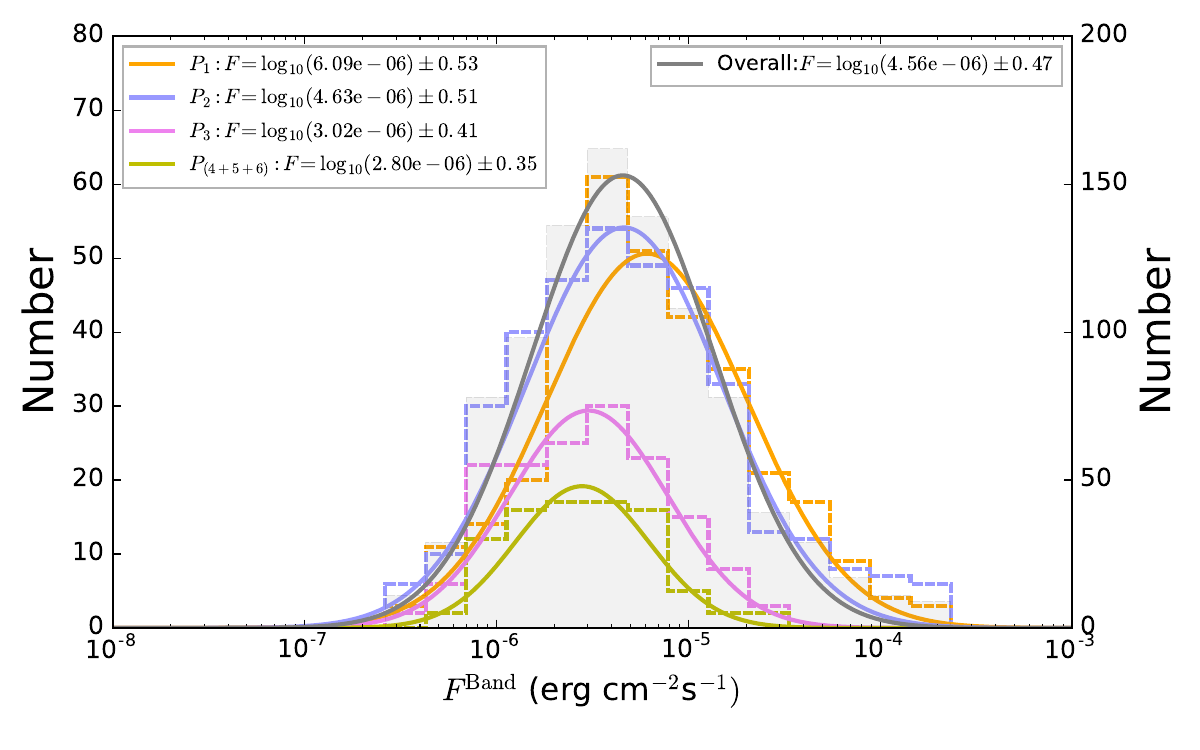}
\includegraphics[angle=0,scale=0.45]{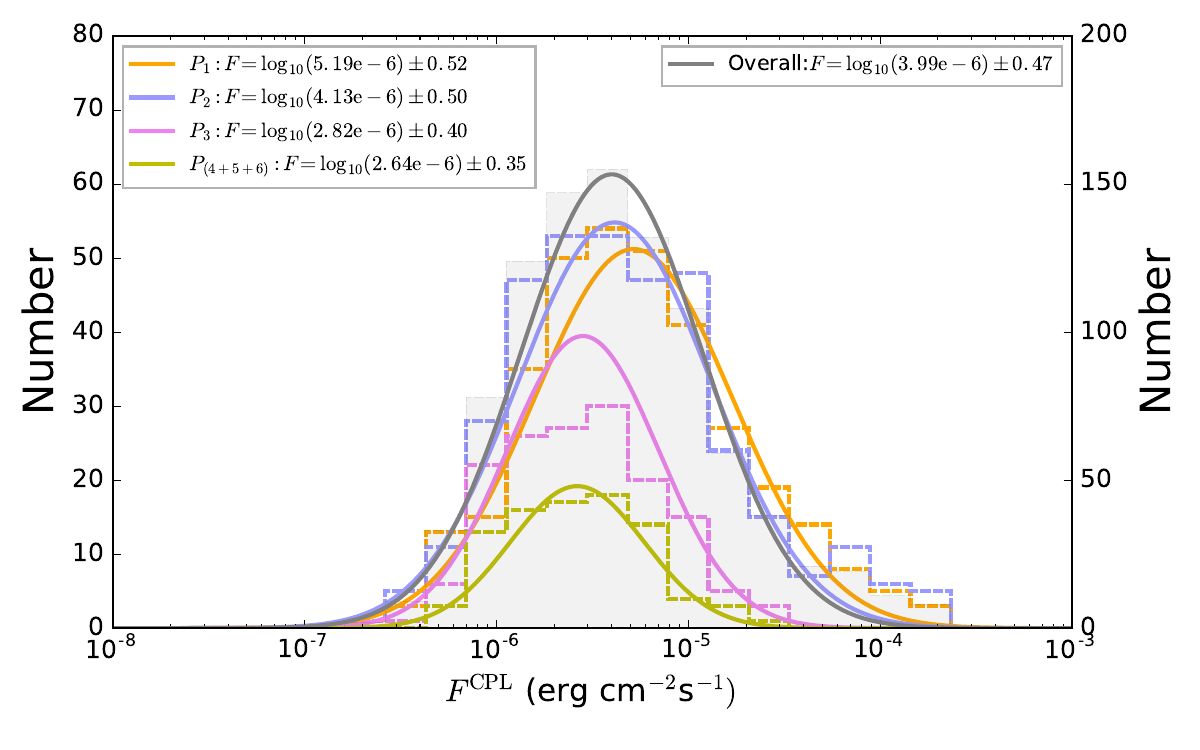}
\begin{center}
\includegraphics[angle=0,scale=0.45]{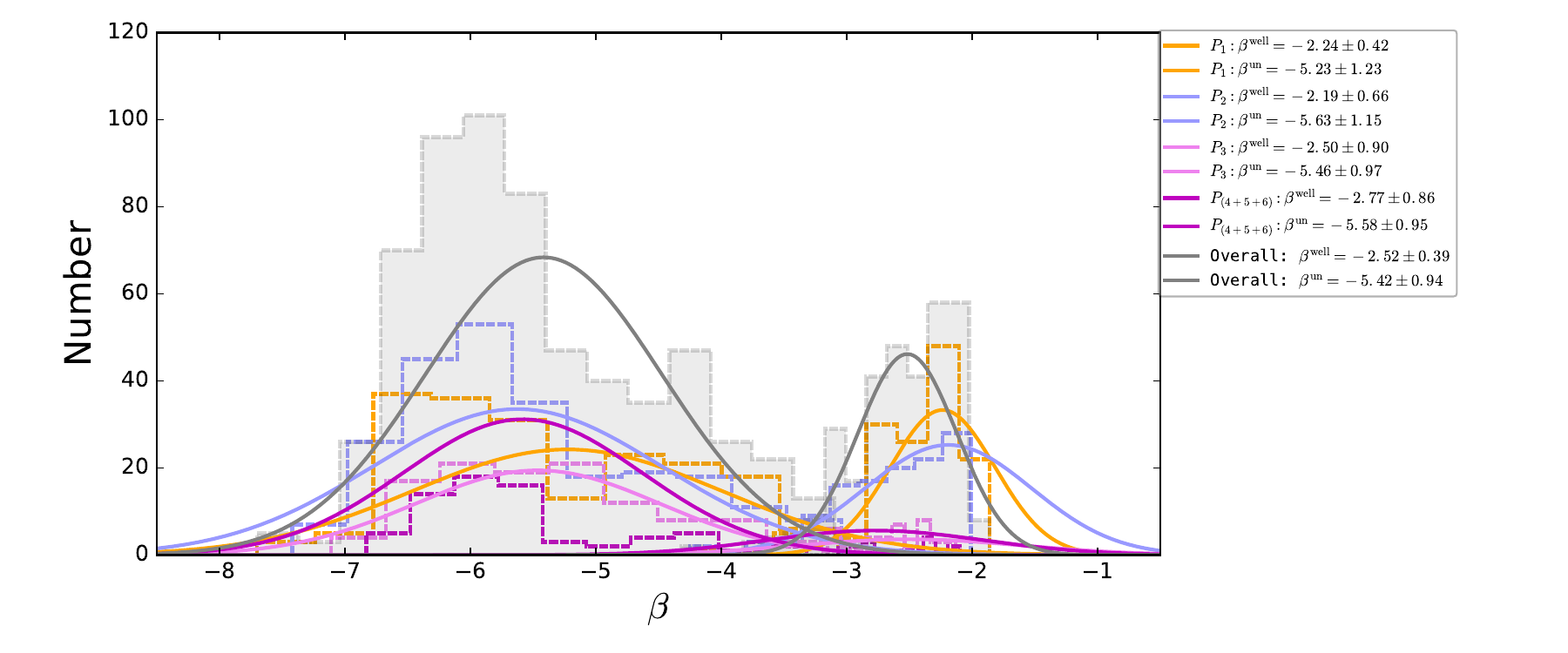}	
\end{center}
\caption{Same as Figure \ref{fig:distrubution_best_model}, but the analysis is based on the spectral parameters derived from single model (Band or CPL). Colors are the same as in Fig.\ref{fig:distrubution_best_model}. Left panels are for the Band model, and right panels are for the CPL model. The lower panel shows the distribution of $\beta$ for the Band model.}\label{fig:distribution_Band_CPL}
\end{figure*}

In Figure \ref{fig:distribution_Band_CPL}, we provide the results of the parameter distributions, separated by two empirical photon models (Band and CPL). The overall parameter distributions we studied include four parameters ($\alpha$, $\beta$, $E_{\rm p}$, and $F$) in the Band model and three parameters ($\alpha$, $E_{\rm p}$, and $F$) in the CPL model. The corresponding average values and standard deviations from the best Gaussian fit for each distribution is presented in Table \ref{table:distribution_CPL_Band}.

The main results of parameter distributions ($\alpha$, $E_{\rm p}$ and $F$) among pulses obtained from a single model (Fig.\ref{fig:distribution_Band_CPL}), either the Band model or the CPL model, are consistent with the finding in the best model (Fig.\ref{fig:distrubution_best_model}), as we discussed in the main text. For the $\beta$ distribution, we find a bimodal distribution for each time-series pulse sample, as well as the global sample, where the harder peak is at $\sim$ -2.3 and the softer peak is at $\sim$ -6.1. The two peaks are basically the same for all different time-series pulses. We also find that the $\beta$ indices are typically softer (about half of the Band spectra; the obtained $\beta$ indices cannot be well converged) than some previous catalogs, whose analysis is based on the frequency analysis method, but consistent with the results found in \cite{Yu2019}, who also adopted a fully Bayesian method but for a single-pulse sample.

\clearpage

\clearpage
\begin{deluxetable}{cccccccc}
%\rotate
\tablewidth{0pt}
\tabletypesize{\scriptsize}
\tablecaption{Results of the Average and Deviation Values of the Parameter Distribution}
\tablehead{
\colhead{Pulse}
&\colhead{Model}
&\colhead{Spectra}
&\colhead{$\alpha$}
&\colhead{$E_{\rm p}$}
&\colhead{$F$}
&\colhead{$\beta$}
&\colhead{$\beta$}\\
(Group)&(Selected)&(Number)&&(keV)&(erg s$^{-1}$ cm$^{-2}$)&(well-converged)&(un-converged)
}
\colnumbers
\startdata
\hline
$P_{1}$&CPL&338&-0.80$\pm$0.27&$\rm log_{10}$(273)$\pm$0.38&$\rm log_{10}$(5.19e-6)$\pm$0.52&\nodata&\nodata\\
$P_{2}$&CPL&361&-0.92$\pm$0.36&$\rm log_{10}$(248)$\pm$0.44&$\rm log_{10}$(4.13e-6)$\pm$0.50&\nodata&\nodata\\
$P_{3}$&CPL&156&-1.06$\pm$0.36&$\rm log_{10}$(204)$\pm$0.26&$\rm log_{10}$(2.82e-6)$\pm$0.40&\nodata&\nodata\\
$P_{(4+5+6)}$&CPL&89&-1.06$\pm$0.19&$\rm log_{10}$(218)$\pm$0.31&$\rm log_{10}$(2.98e-6)$\pm$0.38&\nodata&\nodata\\
Overall&CPL&944&-0.91$\pm$0.33&$\rm log_{10}$(239)$\pm$0.40&$\rm log_{10}$(3.99e-6)$\pm$0.47&\nodata&\nodata\\
\hline
\hline
$P_{1}$&Band&338&-0.70$\pm$0.29&$\rm log_{10}$(234)$\pm$0.40&$\rm log_{10}$(6.09e-6)$\pm$0.53&-2.24$\pm$0.42&-5.23$\pm$1.23\\
$P_{2}$&Band&361&-0.82$\pm$0.35&$\rm log_{10}$(221)$\pm$0.46&$\rm log_{10}$(4.63e-6)$\pm$0.51&-2.19$\pm$0.66&-5.63$\pm$1.15\\
$P_{3}$&Band&156&-1.00$\pm$0.32&$\rm log_{10}$(169)$\pm$0.32&$\rm log_{10}$(3.02e-6)$\pm$0.41&-2.50$\pm$0.90&-5.46$\pm$0.97\\
$P_{(4+5+6)}$&Band&89&-0.99$\pm$0.20&$\rm log_{10}$(207)$\pm$0.36&$\rm log_{10}$(2.98e-6)$\pm$0.39&-2.77$\pm$0.86&-5.58$\pm$0.95\\
Overall&Band&944&-0.82$\pm$0.34&$\rm log_{10}$(213)$\pm$0.41&$\rm log_{10}$(4.56e-6)$\pm$0.47&-2.52$\pm$0.39&-5.42$\pm$0.94\\
\enddata 
\vspace{3mm}
\tablecomments{Same as Table \ref{table:distribution} but the results are based on single empirical photon model.} 
\end{deluxetable}\label{table:distribution_CPL_Band}

\subsection{Single Model-based Parameter Relations in Different Pulses}

\begin{figure*}
\includegraphics[angle=0,scale=0.45]{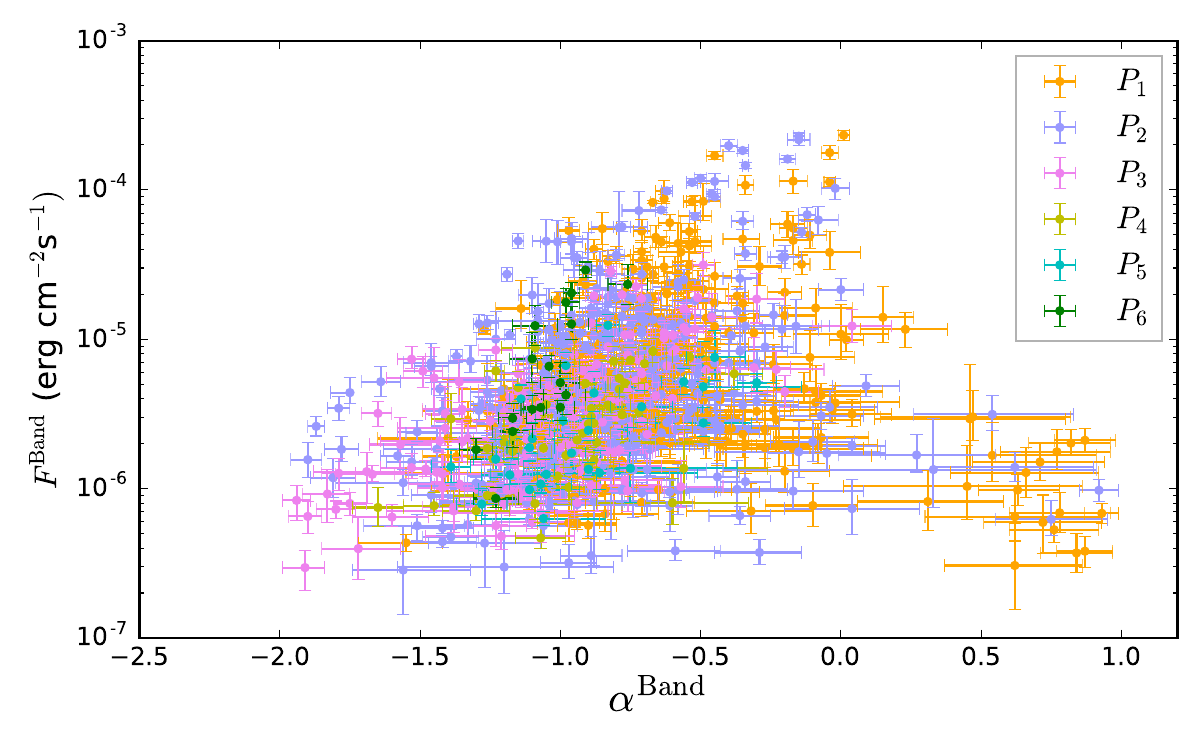}
\includegraphics[angle=0,scale=0.45]{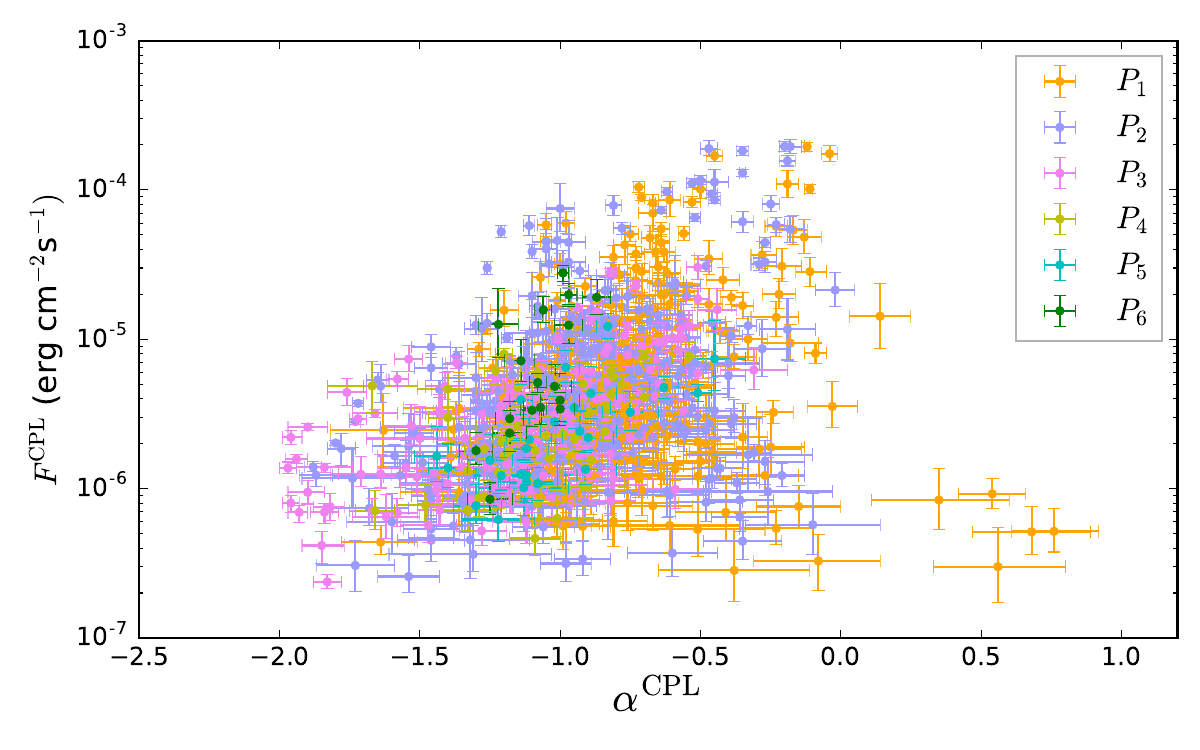}
\includegraphics[angle=0,scale=0.45]{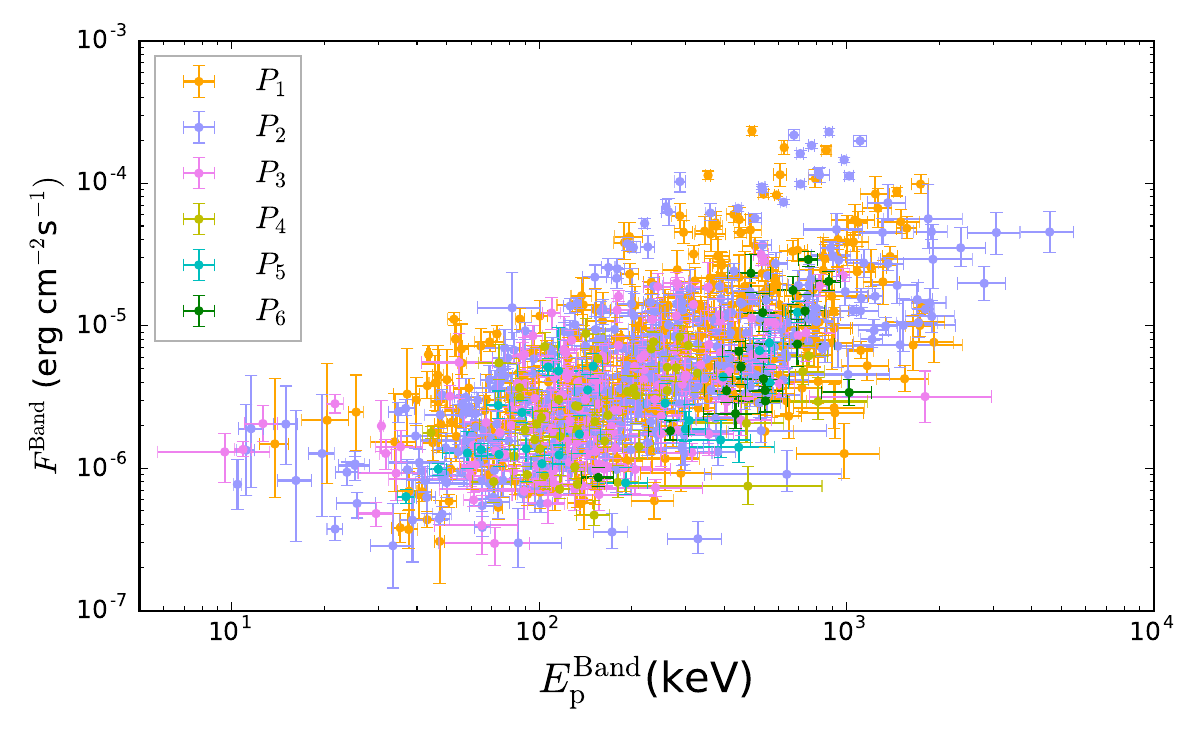}
\includegraphics[angle=0,scale=0.45]{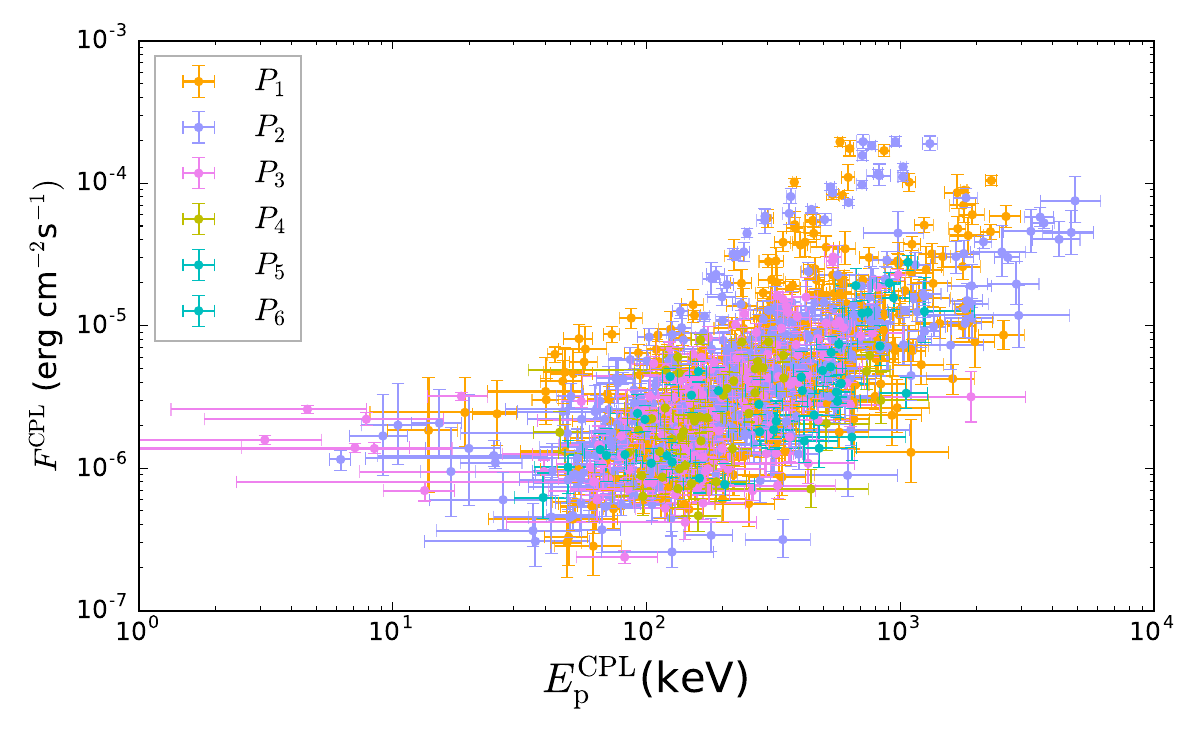}
\includegraphics[angle=0,scale=0.45]{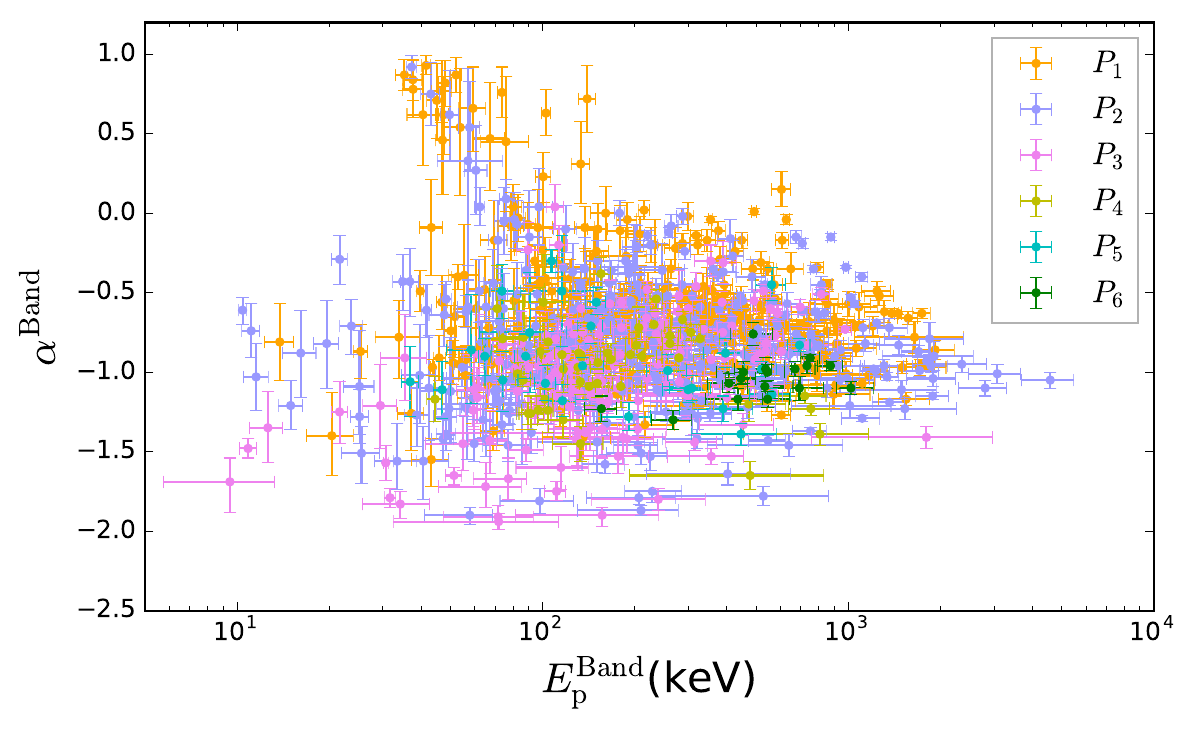}
\includegraphics[angle=0,scale=0.45]{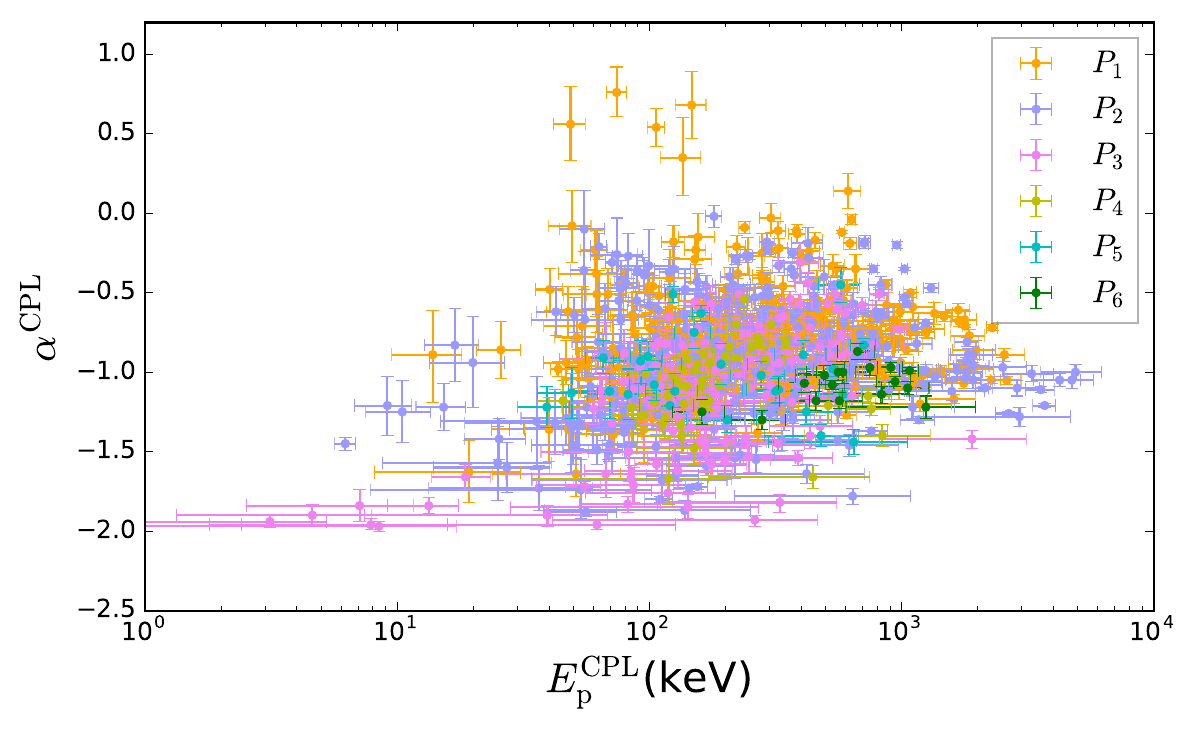}
\begin{center}
\includegraphics[angle=0,scale=0.45]{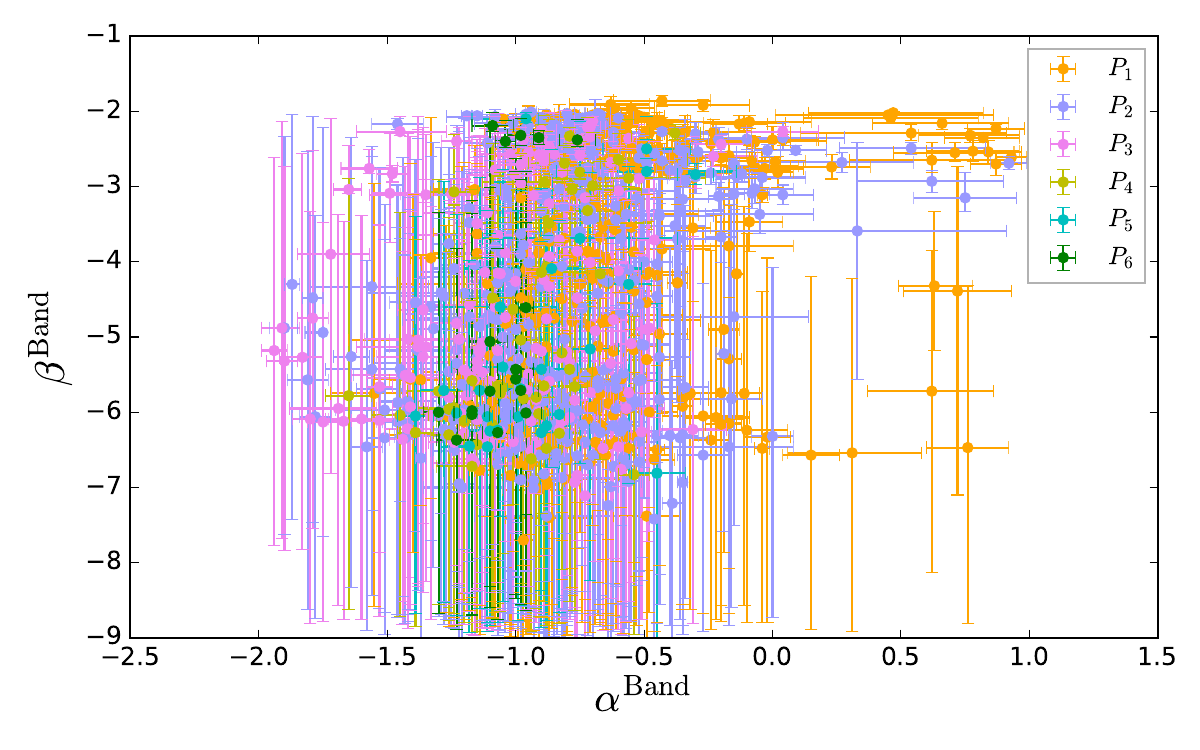}	
\end{center}
\caption{Pulsewise global relations of the fitted parameters of the Band and CPL models with statistical significance $S \geq 20$. The colors are the same as in Figure \ref{fig:distrubution_best_model}. Left panels are for the Band model, right panels are for the CPL model, and the bottom panel is only for the Band model.}\label{fig:globalcorrelation}
\end{figure*}

We show global pulsewise parameter relations for our full sample in Figure \ref{fig:globalcorrelation}. We present each parameter relation based on the two different photon models (the Band and CPL): $F$-$\alpha$ relation based on the Band model (first left panel), $F$-$\alpha$ relation based on the CPL model (first right panel), $F$-$E_{\rm p}$ relation based on the Band model (second left panel), $F$-$E_{\rm p}$ relation based on the CPL model (second right panel), $\alpha$-$E_{\rm p}$ relation based on the Band model (third left panel), $\alpha$-$E_{\rm p}$ relation based on the CPL model (third right panel), and $\alpha$-$\beta$ relation based on the Band model (bottom panel). 

We find that there is no significant difference in the global parameter relations between the Band and CPL models. In addition, the $F$-$E_{\rm p}$ Golenetskii relation based on the Band model shows a stronger monotonic positive correlation than the CPL model. For the $F$-$\alpha$ relation, a cluster of data points that significantly deviate from the peak of the distribution (probably mainly contributed by the Type 2 in the individual parameter relations) mainly come from the early ($P_{1}$) pulse. In short, the global $F$-$\alpha$ relation shows a monotone positive correlation following a break behavior. Such a break may be originated from thermal emission. The global $F$-$E_{\rm p}$ relation displays the type 1p behavior, while the global $\alpha$-$E_{\rm p}$ shows an anticorrelation type 1n behavior. Finally, we do not find a clear trend in the global $\alpha-\beta$ relation.

\clearpage

\clearpage
%\vspace{10mm}
\section{Additional Tables \label{sec:tab}}

The spectral parameters for each time bin are obtained by applying both the Band and CPL models to fit all 944 spectra. The results of the time-resolved spectral fits for each burst in our global sample, which include the start and stop times of the BBlocks (column 1), the statistical significance $S$ (column 2); the best-fitted parameters of the CPL model (columns 3-7), including the normalization $K$, the low-energy power-law photon spectral index $\alpha$, the break energy $E_{\rm c}$, peak energy\footnote{We note that in a few cases, the value of $E_{\rm p}$ is close to the lower limit of the detector range. Since the effective area decreases rapidly at the edges of the detector energy range and the determination of $\alpha$ suffers from the lack of dynamical range below the peak, the time bins with $E_{\rm p}< 20$ keV should be largely ignored \citep[see, e.g., ][]{Ravasio2019}.} $E_{\rm p}$, and the $\nu F_{\nu}$ flux $F$; and the best-fit parameters of the Band model (columns 8-12), containing the normalization, the low-energy power-law photon spectral index $\alpha$, the high-energy power-law photon spectral index $\beta$, the peak energy $E_{\rm p}$, and the $\nu F_{\nu}$ flux $F$; the $\Delta \rm DIC$ (column 13), the $p^{\rm CPL}_{\rm DIC}$ based on the CPL model fitting (column 14), and the $p^{\rm Band}_{\rm DIC}$ based on the Band model fitting (column 15), are listed in Tables \ref{table:081009140}-\ref{table:190114873} in Appendix. For the definition of $\Delta \rm DIC$, as well as the discussion for $p_{\rm DIC}$, please see \S \ref{sec:bestmodel} for details.

\clearpage
\begin{deluxetable}{cc|ccccc|ccccc|ccc}
\centering
\rotate
\tabletypesize{\scriptsize}
\tablecaption{Time-resolved Spectral Fit Results of GRB 081009140 \label{table:081009140}}
\tablehead{
\specialrule{0em}{3pt}{3pt}
\multirow{3}{*}{$t_{\rm start}$$\sim$$t_{\rm stop}$} 
&\multirow{3}{*}{$S$} 
&\multicolumn{5}{c}{Cutoff Power-Law Fitting}
&\multicolumn{5}{c}{Band Fitting}
&\multicolumn{3}{c}{Difference}\\
\cmidrule(lr){3-7}  \cmidrule(lr){8-12} \cmidrule(lr){13-15}
&
&\colhead{$K$}
&\colhead{$\alpha$}
&\colhead{$E_{\rm c}$}
&\colhead{$E_{\rm p}$}
&\colhead{$F$}
&\colhead{$K$}
&\colhead{$\alpha$}
&\colhead{$\beta$}
&\colhead{$E_{\rm p}$}
&\colhead{$F$}
&\colhead{$\Delta$DIC}
&\colhead{$p^{\rm CPL}_{\rm DIC}$}
&\colhead{$p^{\rm Band}_{\rm DIC}$}\\
\hline
}
\colnumbers
\startdata
\hline
Pulse 1\\
\hline
0.000$\sim$0.368&21.46&0.08$^{+0.03}_{-0.02}$&-1.36$^{+0.17}_{-0.15}$&75$^{+17}_{-18}$&48$^{+17}_{-16}$&1.31$^{+0.71}_{-0.41}$$\times$10$^{-6}$&0.38$^{+0.19}_{-0.21}$&-0.78$^{+0.23}_{-0.26}$&-2.72$^{+0.46}_{-0.02}$&34$^{+6}_{-6}$&1.53$^{+1.81}_{-0.85}$$\times$10$^{-6}$&-13.5&0.8&-5.4\\
0.368$\sim$1.114&47.31&0.12$^{+0.05}_{-0.05}$&-1.63$^{+0.20}_{-0.19}$&52$^{+14}_{-14}$&19$^{+12}_{-11}$&2.46$^{+1.87}_{-1.04}$$\times$10$^{-6}$&0.23$^{+0.14}_{-0.12}$&-1.40$^{+0.27}_{-0.25}$&-5.04$^{+1.93}_{-2.83}$&20$^{+4}_{-4}$&2.17$^{+3.21}_{-1.39}$$\times$10$^{-6}$&-11.9&-16.9&-28.5\\
1.114$\sim$1.358&39.81&0.24$^{+0.09}_{-0.09}$&-1.36$^{+0.20}_{-0.20}$&62$^{+16}_{-16}$&40$^{+16}_{-16}$&3.45$^{+2.55}_{-1.42}$$\times$10$^{-6}$&0.32$^{+0.10}_{-0.15}$&-1.26$^{+0.20}_{-0.23}$&-6.30$^{+2.53}_{-2.55}$&37$^{+4}_{-4}$&3.31$^{+3.59}_{-1.86}$$\times$10$^{-6}$&-7.4&-7.0&-15.0\\
1.358$\sim$1.662&57.46&0.66$^{+0.17}_{-0.18}$&-0.94$^{+0.14}_{-0.14}$&44$^{+6}_{-6}$&47$^{+9}_{-9}$&4.06$^{+1.59}_{-1.21}$$\times$10$^{-6}$&0.70$^{+0.19}_{-0.20}$&-0.91$^{+0.14}_{-0.14}$&-7.03$^{+1.96}_{-1.99}$&46$^{+2}_{-2}$&3.95$^{+2.24}_{-1.41}$$\times$10$^{-6}$&1.8&-0.5&0.9\\
1.662$\sim$1.895&58.56&1.53$^{+0.46}_{-0.52}$&-0.62$^{+0.16}_{-0.17}$&35$^{+4}_{-4}$&48$^{+8}_{-9}$&4.71$^{+2.35}_{-1.65}$$\times$10$^{-6}$&1.71$^{+0.56}_{-0.59}$&-0.56$^{+0.17}_{-0.17}$&-6.48$^{+2.07}_{-2.29}$&47$^{+2}_{-2}$&4.49$^{+2.77}_{-1.76}$$\times$10$^{-6}$&3.7&-6.1&-2.4\\
1.895$\sim$2.042&58.16&0.74$^{+0.20}_{-0.20}$&-1.04$^{+0.14}_{-0.14}$&60$^{+9}_{-10}$&57$^{+12}_{-12}$&6.84$^{+3.00}_{-1.98}$$\times$10$^{-6}$&0.77$^{+0.20}_{-0.20}$&-1.02$^{+0.14}_{-0.14}$&-6.84$^{+2.10}_{-2.14}$&56$^{+3}_{-3}$&6.85$^{+3.40}_{-2.31}$$\times$10$^{-6}$&2.0&-0.3&1.6\\
\enddata
\vspace{3mm}
Note---Column (1) list the start and stop times of the BBlocks time bins; Column (2) lists the significance $S$ of the time bin; Column (3)-(6) list the best-fit parameters for the CPL model: normalisation $K$, the low-energy power-law index $\alpha$, and the cut-off energy $E_{\rm c}$, and the peak energy $E_{\rm p}$=(2+$\alpha$)$E_{\rm c}$ in the $N^{2}N(E)$ or $\nu F_{\nu}$ spectrum; Column (7) lists the derived CPL ($\nu F_{\nu}$) energy flux $F$; Column (8)-(11) lists the best-fit parameters for the Band model: normalisation $K$, the low-energy power-law index $\alpha$, the high-energy power-law index $\beta$, and the peak energy $E_{\rm p}$; Column (12) lists the Band ($\nu F_{\nu}$) energy flux $F$; Column (13) lists the difference between the Deviance Information Criterion (DIC) for the CPL and the Band model, $\Delta$DIC=DIC$_{\rm Band}$-DIC$_{\rm CPL}$, Columns (14) and (15) list the effective number of parameters ($p_{\rm DIC}$) for the CPL and Band model, respectively. \end{deluxetable}

\clearpage
\begin{deluxetable}{cc|ccccc|ccccc|ccc}
\centering
\rotate
\tabletypesize{\scriptsize}
\tablecaption{Time-resolved Spectral Fit Results of GRB 081215784 \label{table:081215784}}
\tablehead{
\specialrule{0em}{3pt}{3pt}
\multirow{3}{*}{$t_{\rm start}$$\sim$$t_{\rm stop}$} 
&\multirow{3}{*}{$S$} 
&\multicolumn{5}{c}{Cutoff Power-Law Fitting}
&\multicolumn{5}{c}{Band Fitting}
&\multicolumn{3}{c}{Difference}\\
\cmidrule(lr){3-7}  \cmidrule(lr){8-12} \cmidrule(lr){13-15}
&
&\colhead{$K$}
&\colhead{$\alpha$}
&\colhead{$E_{\rm c}$}
&\colhead{$E_{\rm p}$}
&\colhead{$F$}
&\colhead{$K$}
&\colhead{$\alpha$}
&\colhead{$\beta$}
&\colhead{$E_{\rm p}$}
&\colhead{$F$}
&\colhead{$\Delta$DIC}
&\colhead{$p^{\rm CPL}_{\rm DIC}$}
&\colhead{$p^{\rm Band}_{\rm DIC}$}\\
\hline
}
\colnumbers
\startdata
\hline
Pulse 1\\
\hline
1.064$\sim$1.217&25.29&0.27$^{+0.04}_{-0.04}$&-0.63$^{+0.07}_{-0.07}$&984$^{+168}_{-165}$&1348$^{+240}_{-236}$&19.77$^{+10.09}_{-6.27}$$\times$10$^{-6}$&0.26$^{+0.04}_{-0.04}$&-0.62$^{+0.07}_{-0.07}$&-6.04$^{+2.56}_{-2.66}$&1313$^{+176}_{-181}$&20.28$^{+8.23}_{-6.25}$$\times$10$^{-6}$&-0.4&2.3&2.3\\
1.217$\sim$1.333&35.25&0.62$^{+0.07}_{-0.07}$&-0.77$^{+0.04}_{-0.04}$&1507$^{+226}_{-227}$&1854$^{+285}_{-286}$&42.89$^{+14.68}_{-11.40}$$\times$10$^{-6}$&0.45$^{+0.07}_{-0.07}$&-0.57$^{+0.08}_{-0.08}$&-2.21$^{+0.15}_{-0.14}$&1002$^{+181}_{-181}$&38.44$^{+20.81}_{-11.97}$$\times$10$^{-6}$&-20.8&2.7&2.3\\
1.333$\sim$1.409&41.78&0.77$^{+0.08}_{-0.08}$&-0.61$^{+0.04}_{-0.04}$&1208$^{+130}_{-130}$&1679$^{+187}_{-187}$&85.70$^{+28.88}_{-19.46}$$\times$10$^{-6}$&0.64$^{+0.08}_{-0.08}$&-0.49$^{+0.06}_{-0.06}$&-2.62$^{+0.22}_{-0.20}$&1241$^{+136}_{-135}$&83.91$^{+26.67}_{-21.10}$$\times$10$^{-6}$&-16.4&2.8&3.6\\
1.409$\sim$1.597&100.52&1.15$^{+0.06}_{-0.06}$&-0.50$^{+0.02}_{-0.03}$&723$^{+40}_{-41}$&1085$^{+62}_{-65}$&101.40$^{+15.07}_{-14.34}$$\times$10$^{-6}$&0.90$^{+0.06}_{-0.06}$&-0.34$^{+0.03}_{-0.03}$&-2.48$^{+0.08}_{-0.08}$&792$^{+38}_{-37}$&107.50$^{+17.10}_{-13.71}$$\times$10$^{-6}$&-124.8&2.9&3.9\\
1.597$\sim$1.699&53.58&0.89$^{+0.09}_{-0.09}$&-0.47$^{+0.05}_{-0.05}$&397$^{+38}_{-38}$&607$^{+61}_{-62}$&34.62$^{+11.37}_{-7.64}$$\times$10$^{-6}$&0.77$^{+0.08}_{-0.08}$&-0.35$^{+0.06}_{-0.07}$&-2.30$^{+0.12}_{-0.12}$&487$^{+41}_{-40}$&46.90$^{+13.27}_{-10.59}$$\times$10$^{-6}$&-35.0&2.6&3.7\\
1.699$\sim$1.888&55.12&0.90$^{+0.07}_{-0.07}$&-0.61$^{+0.05}_{-0.05}$&343$^{+31}_{-31}$&477$^{+47}_{-47}$&17.22$^{+4.39}_{-3.39}$$\times$10$^{-6}$&0.82$^{+0.07}_{-0.07}$&-0.51$^{+0.06}_{-0.07}$&-2.41$^{+0.17}_{-0.15}$&394$^{+37}_{-37}$&22.15$^{+6.87}_{-4.74}$$\times$10$^{-6}$&-15.8&2.8&3.6\\
\enddata
\vspace{3mm}
Note---All columns are the same as Table \ref{table:081009140} but for GRB 081215784.\\ 
\end{deluxetable}

\clearpage
\begin{deluxetable}{cc|ccccc|ccccc|ccc}
\centering
\rotate
\tabletypesize{\scriptsize}
\tablecaption{Time-resolved Spectral Fit Results of GRB 090131090 \label{table:090131090}}
\tablehead{
\specialrule{0em}{3pt}{3pt}
\multirow{3}{*}{$t_{\rm start}$$\sim$$t_{\rm stop}$} 
&\multirow{3}{*}{$S$} 
&\multicolumn{5}{c}{Cutoff Power-Law Fitting}
&\multicolumn{5}{c}{Band Fitting}
&\multicolumn{3}{c}{Difference}\\
\cmidrule(lr){3-7}  \cmidrule(lr){8-12} \cmidrule(lr){13-15}
&
&\colhead{$K$}
&\colhead{$\alpha$}
&\colhead{$E_{\rm c}$}
&\colhead{$E_{\rm p}$}
&\colhead{$F$}
&\colhead{$K$}
&\colhead{$\alpha$}
&\colhead{$\beta$}
&\colhead{$E_{\rm p}$}
&\colhead{$F$}
&\colhead{$\Delta$DIC}
&\colhead{$p^{\rm CPL}_{\rm DIC}$}
&\colhead{$p^{\rm Band}_{\rm DIC}$}\\
\hline
}
\colnumbers
\startdata
\hline
Pulse 1\\
\hline
2.963$\sim$3.560&48.79&1.56$^{+0.10}_{-0.10}$&-0.96$^{+0.09}_{-0.09}$&80$^{+10}_{-10}$&84$^{+13}_{-13}$&2.10$^{+0.52}_{-0.39}$$\times$10$^{-6}$&1.54$^{+0.10}_{-0.10}$&-0.92$^{+0.11}_{-0.11}$&-5.96$^{+2.62}_{-2.65}$&81$^{+5}_{-5}$&2.11$^{+0.30}_{-0.24}$$\times$10$^{-6}$&-0.1&2.6&2.7\\
3.560$\sim$3.900&24.03&1.15$^{+0.13}_{-0.13}$&-0.71$^{+0.25}_{-0.25}$&42$^{+9}_{-9}$&54$^{+16}_{-16}$&0.98$^{+0.59}_{-0.32}$$\times$10$^{-6}$&1.12$^{+0.13}_{-0.13}$&-0.65$^{+0.26}_{-0.26}$&-6.57$^{+2.33}_{-2.32}$&52$^{+4}_{-4}$&0.98$^{+0.19}_{-0.16}$$\times$10$^{-6}$&1.8&1.3&3.1\\
\hline
Pulse 2\\
\hline
6.140$\sim$6.715&46.22&1.61$^{+0.10}_{-0.11}$&-1.22$^{+0.10}_{-0.10}$&123$^{+24}_{-25}$&96$^{+23}_{-23}$&2.03$^{+0.64}_{-0.37}$$\times$10$^{-6}$&1.51$^{+0.10}_{-0.10}$&-0.62$^{+0.10}_{-0.10}$&-2.28$^{+0.09}_{-0.08}$&56$^{+4}_{-4}$&2.77$^{+0.50}_{-0.45}$$\times$10$^{-6}$&-34.1&2.0&3.1\\
6.715$\sim$6.809&25.06&2.62$^{+0.33}_{-0.33}$&-0.67$^{+0.33}_{-0.33}$&42$^{+12}_{-12}$&55$^{+21}_{-21}$&2.20$^{+2.31}_{-0.88}$$\times$10$^{-6}$&2.74$^{+0.31}_{-0.32}$&-0.64$^{+0.19}_{-0.20}$&-4.26$^{+1.50}_{-2.50}$&48$^{+5}_{-5}$&2.37$^{+0.66}_{-0.48}$$\times$10$^{-6}$&-2.5&-1.9&-0.0\\
6.809$\sim$7.232&70.73&3.80$^{+0.18}_{-0.18}$&-0.64$^{+0.11}_{-0.11}$&37$^{+3}_{-3}$&50$^{+6}_{-6}$&3.19$^{+0.71}_{-0.53}$$\times$10$^{-6}$&3.82$^{+0.16}_{-0.16}$&-0.54$^{+0.11}_{-0.11}$&-4.25$^{+0.91}_{-0.77}$&48$^{+2}_{-2}$&3.27$^{+0.30}_{-0.26}$$\times$10$^{-6}$&-7.9&2.7&0.7\\
7.232$\sim$7.495&41.35&3.39$^{+0.23}_{-0.23}$&-1.29$^{+0.17}_{-0.17}$&66$^{+15}_{-16}$&47$^{+16}_{-16}$&2.49$^{+0.75}_{-0.51}$$\times$10$^{-6}$&3.58$^{+0.32}_{-0.32}$&-0.43$^{+0.17}_{-0.18}$&-2.66$^{+0.14}_{-0.13}$&35$^{+2}_{-2}$&2.48$^{+0.42}_{-0.38}$$\times$10$^{-6}$&-20.2&1.1&3.3\\
\enddata
\vspace{3mm}
Note---All columns are the same as Table \ref{table:081009140} but for GRB 090131090.\\ 
\end{deluxetable}

\clearpage
\begin{deluxetable}{cc|ccccc|ccccc|ccc}
\centering
\rotate
\tabletypesize{\scriptsize}
\tablecaption{Time-resolved Spectral Fit Results of GRB 090618353 \label{table:090618353}}
\tablehead{
\specialrule{0em}{3pt}{3pt}
\multirow{3}{*}{$t_{\rm start}$$\sim$$t_{\rm stop}$} 
&\multirow{3}{*}{$S$} 
&\multicolumn{5}{c}{Cutoff Power-Law Fitting}
&\multicolumn{5}{c}{Band Fitting}
&\multicolumn{3}{c}{Difference}\\
\cmidrule(lr){3-7}  \cmidrule(lr){8-12} \cmidrule(lr){13-15}
&
&\colhead{$K$}
&\colhead{$\alpha$}
&\colhead{$E_{\rm c}$}
&\colhead{$E_{\rm p}$}
&\colhead{$F$}
&\colhead{$K$}
&\colhead{$\alpha$}
&\colhead{$\beta$}
&\colhead{$E_{\rm p}$}
&\colhead{$F$}
&\colhead{$\Delta$DIC}
&\colhead{$p^{\rm CPL}_{\rm DIC}$}
&\colhead{$p^{\rm Band}_{\rm DIC}$}\\
\hline
}
\colnumbers
\startdata
\hline
Pulse 1\\
\hline
1.511$\sim$21.588&54.02&0.28$^{+0.02}_{-0.02}$&-0.89$^{+0.05}_{-0.05}$&208$^{+20}_{-20}$&231$^{+25}_{-25}$&1.23$^{+0.29}_{-0.21}$$\times$10$^{-6}$&0.21$^{+0.02}_{-0.02}$&-0.43$^{+0.11}_{-0.11}$&-2.14$^{+0.06}_{-0.06}$&138$^{+10}_{-10}$&1.95$^{+0.50}_{-0.41}$$\times$10$^{-6}$&-23.1&2.7&3.3\\
21.588$\sim$28.423&30.65&0.35$^{+0.04}_{-0.04}$&-0.98$^{+0.13}_{-0.13}$&112$^{+20}_{-21}$&114$^{+25}_{-26}$&0.61$^{+0.26}_{-0.17}$$\times$10$^{-6}$&0.33$^{+0.04}_{-0.04}$&-0.84$^{+0.19}_{-0.18}$&-3.42$^{+0.90}_{-1.23}$&101$^{+12}_{-13}$&0.68$^{+0.24}_{-0.16}$$\times$10$^{-6}$&-1.5&1.8&1.9\\
\hline
Pulse 2\\
\hline
49.994$\sim$56.456&52.09&0.68$^{+0.04}_{-0.04}$&-1.25$^{+0.06}_{-0.06}$&220$^{+34}_{-34}$&165$^{+29}_{-29}$&1.31$^{+0.29}_{-0.22}$$\times$10$^{-6}$&0.65$^{+0.04}_{-0.04}$&-1.18$^{+0.10}_{-0.10}$&-3.29$^{+0.94}_{-1.22}$&147$^{+21}_{-24}$&1.47$^{+0.41}_{-0.32}$$\times$10$^{-6}$&-2.2&2.5&1.5\\
56.456$\sim$61.167&68.70&0.79$^{+0.04}_{-0.04}$&-1.07$^{+0.05}_{-0.05}$&210$^{+21}_{-22}$&195$^{+22}_{-23}$&2.19$^{+0.39}_{-0.29}$$\times$10$^{-6}$&0.70$^{+0.05}_{-0.05}$&-0.83$^{+0.10}_{-0.10}$&-2.34$^{+0.14}_{-0.07}$&138$^{+14}_{-14}$&2.98$^{+0.69}_{-0.58}$$\times$10$^{-6}$&-11.7&2.7&3.3\\
61.167$\sim$61.714&33.28&0.94$^{+0.11}_{-0.11}$&-1.10$^{+0.09}_{-0.09}$&330$^{+81}_{-82}$&297$^{+78}_{-80}$&3.55$^{+1.87}_{-1.05}$$\times$10$^{-6}$&0.92$^{+0.11}_{-0.11}$&-1.07$^{+0.10}_{-0.10}$&-4.28$^{+1.31}_{-1.32}$&278$^{+47}_{-47}$&3.64$^{+1.28}_{-0.83}$$\times$10$^{-6}$&0.8&1.9&2.6\\
61.714$\sim$62.186&43.95&1.37$^{+0.13}_{-0.13}$&-1.13$^{+0.07}_{-0.07}$&441$^{+95}_{-98}$&384$^{+88}_{-91}$&6.43$^{+2.57}_{-1.72}$$\times$10$^{-6}$&1.18$^{+0.18}_{-0.18}$&-0.89$^{+0.21}_{-0.20}$&-2.37$^{+0.39}_{-0.35}$&239$^{+89}_{-81}$&8.42$^{+6.54}_{-3.59}$$\times$10$^{-6}$&-15.3&2.1&-6.5\\
\enddata
\vspace{3mm}
Note---All columns are the same as Table \ref{table:081009140} but for GRB 090618353.\\ 
\end{deluxetable}

\clearpage
\begin{deluxetable}{cc|ccccc|ccccc|ccc}
\centering
\rotate
\tabletypesize{\scriptsize}
\tablecaption{Time-resolved Spectral Fit Results of GRB 091127976 \label{table:091127976}}
\tablehead{
\specialrule{0em}{3pt}{3pt}
\multirow{3}{*}{$t_{\rm start}$$\sim$$t_{\rm stop}$} 
&\multirow{3}{*}{$S$} 
&\multicolumn{5}{c}{Cutoff Power-Law Fitting}
&\multicolumn{5}{c}{Band Fitting}
&\multicolumn{3}{c}{Difference}\\
\cmidrule(lr){3-7}  \cmidrule(lr){8-12} \cmidrule(lr){13-15}
&
&\colhead{$K$}
&\colhead{$\alpha$}
&\colhead{$E_{\rm c}$}
&\colhead{$E_{\rm p}$}
&\colhead{$F$}
&\colhead{$K$}
&\colhead{$\alpha$}
&\colhead{$\beta$}
&\colhead{$E_{\rm p}$}
&\colhead{$F$}
&\colhead{$\Delta$DIC}
&\colhead{$p^{\rm CPL}_{\rm DIC}$}
&\colhead{$p^{\rm Band}_{\rm DIC}$}\\
\hline
}
\colnumbers
\startdata
\hline
Pulse 1\\
\hline
0.073$\sim$0.207&32.13&1.98$^{+0.16}_{-0.16}$&-1.10$^{+0.09}_{-0.09}$&174$^{+34}_{-35}$&157$^{+35}_{-35}$&4.20$^{+1.49}_{-1.01}$$\times$10$^{-6}$&1.88$^{+0.19}_{-0.19}$&-0.91$^{+0.19}_{-0.23}$&-3.88$^{+1.71}_{-3.02}$&124$^{+31}_{-32}$&5.03$^{+3.23}_{-1.76}$$\times$10$^{-6}$&-11.6&2.2&-6.2\\
0.207$\sim$0.282&32.79&3.30$^{+0.28}_{-0.28}$&-1.17$^{+0.11}_{-0.11}$&146$^{+32}_{-33}$&121$^{+31}_{-32}$&5.26$^{+2.12}_{-1.25}$$\times$10$^{-6}$&2.94$^{+0.26}_{-0.27}$&-0.48$^{+0.21}_{-0.21}$&-2.28$^{+0.13}_{-0.10}$&65$^{+8}_{-8}$&7.17$^{+2.40}_{-1.76}$$\times$10$^{-6}$&-15.4&2.0&2.7\\
0.282$\sim$0.526&82.83&6.15$^{+0.20}_{-0.20}$&-1.29$^{+0.04}_{-0.04}$&156$^{+15}_{-16}$&111$^{+13}_{-13}$&8.59$^{+0.99}_{-0.86}$$\times$10$^{-6}$&5.54$^{+0.20}_{-0.21}$&-0.40$^{+0.08}_{-0.08}$&-2.22$^{+0.04}_{-0.04}$&53$^{+2}_{-2}$&11.08$^{+1.31}_{-1.10}$$\times$10$^{-6}$&-86.8&2.8&3.1\\
0.526$\sim$0.782&72.41&5.17$^{+0.19}_{-0.19}$&-1.24$^{+0.05}_{-0.05}$&122$^{+12}_{-12}$&93$^{+11}_{-11}$&6.44$^{+0.88}_{-0.69}$$\times$10$^{-6}$&4.91$^{+0.20}_{-0.20}$&-0.94$^{+0.14}_{-0.14}$&-2.50$^{+0.13}_{-0.13}$&69$^{+7}_{-7}$&7.68$^{+1.33}_{-1.07}$$\times$10$^{-6}$&-17.3&2.8&3.3\\
0.782$\sim$0.886&36.25&3.75$^{+0.26}_{-0.26}$&-1.30$^{+0.10}_{-0.10}$&134$^{+26}_{-27}$&94$^{+23}_{-23}$&4.50$^{+1.29}_{-0.80}$$\times$10$^{-6}$&3.66$^{+0.27}_{-0.28}$&-1.20$^{+0.14}_{-0.16}$&-5.47$^{+2.79}_{-2.94}$&86$^{+12}_{-11}$&4.65$^{+1.19}_{-0.75}$$\times$10$^{-6}$&-1.2&2.4&1.6\\
0.886$\sim$1.223&50.75&3.39$^{+0.14}_{-0.15}$&-1.37$^{+0.08}_{-0.08}$&111$^{+17}_{-17}$&70$^{+14}_{-14}$&3.30$^{+0.53}_{-0.40}$$\times$10$^{-6}$&3.31$^{+0.15}_{-0.15}$&-1.16$^{+0.16}_{-0.16}$&-3.04$^{+0.58}_{-0.03}$&59$^{+7}_{-6}$&3.65$^{+0.63}_{-0.51}$$\times$10$^{-6}$&-12.9&2.5&-1.4\\
\hline
Pulse 2\\
\enddata
\vspace{3mm}
Note---All columns are the same as Table \ref{table:081009140} but for GRB 091127976.\\
\end{deluxetable}

\clearpage
\begin{deluxetable}{cc|ccccc|ccccc|ccc}
\centering
\rotate
\tabletypesize{\scriptsize}
\tablecaption{Time-resolved Spectral Fit Results of GRB 100719989\label{table:100719989}}
\tablehead{
\specialrule{0em}{3pt}{3pt}
\multirow{3}{*}{$t_{\rm start}$$\sim$$t_{\rm stop}$} 
&\multirow{3}{*}{$S$} 
&\multicolumn{5}{c}{Cutoff Power-Law Fitting}
&\multicolumn{5}{c}{Band Fitting}
&\multicolumn{3}{c}{Difference}\\
\cmidrule(lr){3-7}  \cmidrule(lr){8-12} \cmidrule(lr){13-15}
&
&\colhead{$K$}
&\colhead{$\alpha$}
&\colhead{$E_{\rm c}$}
&\colhead{$E_{\rm p}$}
&\colhead{$F$}
&\colhead{$K$}
&\colhead{$\alpha$}
&\colhead{$\beta$}
&\colhead{$E_{\rm p}$}
&\colhead{$F$}
&\colhead{$\Delta$DIC}
&\colhead{$p^{\rm CPL}_{\rm DIC}$}
&\colhead{$p^{\rm Band}_{\rm DIC}$}\\
\hline
}
\colnumbers
\startdata
\hline
Pulse 1\\
\hline
1.596$\sim$2.044&33.85&0.22$^{+0.03}_{-0.03}$&-0.33$^{+0.07}_{-0.07}$&320$^{+32}_{-32}$&534$^{+57}_{-57}$&10.01$^{+4.13}_{-2.80}$$\times$10$^{-6}$&0.22$^{+0.03}_{-0.03}$&-0.31$^{+0.07}_{-0.07}$&-3.55$^{+0.79}_{-0.98}$&517$^{+37}_{-37}$&11.03$^{+3.89}_{-2.53}$$\times$10$^{-6}$&-2.4&2.4&2.8\\
2.044$\sim$2.366&44.23&0.45$^{+0.05}_{-0.05}$&-0.47$^{+0.05}_{-0.05}$&385$^{+34}_{-34}$&589$^{+56}_{-56}$&17.11$^{+4.64}_{-3.99}$$\times$10$^{-6}$&0.44$^{+0.05}_{-0.05}$&-0.45$^{+0.06}_{-0.06}$&-4.18$^{+1.04}_{-1.20}$&573$^{+37}_{-37}$&17.53$^{+4.47}_{-3.32}$$\times$10$^{-6}$&-0.8&2.7&2.8\\
2.366$\sim$2.681&62.78&1.18$^{+0.08}_{-0.08}$&-0.73$^{+0.03}_{-0.03}$&594$^{+49}_{-47}$&755$^{+64}_{-63}$&30.05$^{+5.35}_{-4.76}$$\times$10$^{-6}$&1.14$^{+0.08}_{-0.08}$&-0.71$^{+0.04}_{-0.04}$&-2.86$^{+0.30}_{-0.26}$&693$^{+53}_{-52}$&33.79$^{+7.00}_{-5.51}$$\times$10$^{-6}$&-8.9&2.9&3.8\\
2.681$\sim$2.841&36.98&1.00$^{+0.11}_{-0.11}$&-0.71$^{+0.06}_{-0.06}$&384$^{+48}_{-49}$&495$^{+66}_{-67}$&15.37$^{+5.17}_{-3.67}$$\times$10$^{-6}$&0.99$^{+0.11}_{-0.11}$&-0.70$^{+0.06}_{-0.06}$&-5.47$^{+1.83}_{-1.75}$&483$^{+43}_{-43}$&15.62$^{+4.27}_{-2.97}$$\times$10$^{-6}$&0.3&2.7&2.9\\
\hline
Pulse 2\\
\hline
2.841$\sim$3.725&55.65&0.59$^{+0.04}_{-0.04}$&-0.61$^{+0.04}_{-0.04}$&242$^{+17}_{-17}$&336$^{+26}_{-26}$&7.10$^{+1.43}_{-1.18}$$\times$10$^{-6}$&0.59$^{+0.04}_{-0.04}$&-0.59$^{+0.05}_{-0.05}$&-4.24$^{+1.23}_{-1.39}$&327$^{+19}_{-18}$&7.43$^{+1.56}_{-1.10}$$\times$10$^{-6}$&-0.2&2.9&3.1\\
3.725$\sim$4.216&54.30&0.77$^{+0.06}_{-0.06}$&-0.52$^{+0.05}_{-0.05}$&190$^{+14}_{-14}$&281$^{+23}_{-23}$&8.50$^{+1.87}_{-1.56}$$\times$10$^{-6}$&0.60$^{+0.07}_{-0.08}$&-0.21$^{+0.13}_{-0.13}$&-2.37$^{+0.15}_{-0.14}$&203$^{+21}_{-21}$&11.66$^{+5.25}_{-3.30}$$\times$10$^{-6}$&-14.7&2.8&2.3\\
\enddata
\vspace{3mm}
Note---All columns are the same as Table \ref{table:081009140} but for GRB 100719989.\\ 
\end{deluxetable}

\clearpage
\begin{deluxetable}{cc|ccccc|ccccc|ccc}
\centering
\rotate
\tabletypesize{\scriptsize}
\tablecaption{Time-resolved Spectral Fit Results of GRB 100826957 \label{table:100826957}}
\tablehead{
\specialrule{0em}{3pt}{3pt}
\multirow{3}{*}{$t_{\rm start}$$\sim$$t_{\rm stop}$} 
&\multirow{3}{*}{$S$} 
&\multicolumn{5}{c}{Cutoff Power-Law Fitting}
&\multicolumn{5}{c}{Band Fitting}
&\multicolumn{3}{c}{Difference}\\
\cmidrule(lr){3-7}  \cmidrule(lr){8-12} \cmidrule(lr){13-15}
&
&\colhead{$K$}
&\colhead{$\alpha$}
&\colhead{$E_{\rm c}$}
&\colhead{$E_{\rm p}$}
&\colhead{$F$}
&\colhead{$K$}
&\colhead{$\alpha$}
&\colhead{$\beta$}
&\colhead{$E_{\rm p}$}
&\colhead{$F$}
&\colhead{$\Delta$DIC}
&\colhead{$p^{\rm CPL}_{\rm DIC}$}
&\colhead{$p^{\rm Band}_{\rm DIC}$}\\
\hline
}
\colnumbers
\startdata
\hline
Pulse 1\\
\hline
2.350$\sim$10.360&33.80&0.14$^{+0.01}_{-0.01}$&-0.72$^{+0.06}_{-0.06}$&275$^{+34}_{-34}$&352$^{+47}_{-47}$&1.32$^{+0.43}_{-0.28}$$\times$10$^{-6}$&0.10$^{+0.01}_{-0.01}$&-0.27$^{+0.18}_{-0.18}$&-1.92$^{+0.08}_{-0.08}$&188$^{+32}_{-32}$&2.50$^{+1.74}_{-0.90}$$\times$10$^{-6}$&-39.7&2.5&-3.9\\
10.360$\sim$13.706&41.03&0.22$^{+0.02}_{-0.02}$&-0.76$^{+0.04}_{-0.04}$&412$^{+48}_{-49}$&511$^{+61}_{-63}$&3.08$^{+0.76}_{-0.57}$$\times$10$^{-6}$&0.19$^{+0.02}_{-0.02}$&-0.59$^{+0.08}_{-0.08}$&-2.08$^{+0.11}_{-0.11}$&345$^{+43}_{-43}$&4.60$^{+1.52}_{-1.13}$$\times$10$^{-6}$&-20.5&2.7&3.1\\
13.706$\sim$15.608&54.28&0.33$^{+0.02}_{-0.02}$&-0.68$^{+0.04}_{-0.04}$&346$^{+29}_{-29}$&456$^{+41}_{-41}$&5.08$^{+1.03}_{-0.84}$$\times$10$^{-6}$&0.29$^{+0.02}_{-0.02}$&-0.54$^{+0.06}_{-0.06}$&-2.02$^{+0.07}_{-0.07}$&334$^{+30}_{-30}$&8.40$^{+2.15}_{-1.65}$$\times$10$^{-6}$&-35.4&2.9&3.6\\
15.608$\sim$18.208&75.38&0.41$^{+0.02}_{-0.02}$&-0.70$^{+0.03}_{-0.03}$&379$^{+24}_{-24}$&493$^{+33}_{-33}$&6.55$^{+0.95}_{-0.78}$$\times$10$^{-6}$&0.39$^{+0.02}_{-0.02}$&-0.66$^{+0.04}_{-0.03}$&-2.30$^{+0.11}_{-0.11}$&432$^{+28}_{-29}$&9.20$^{+1.47}_{-1.27}$$\times$10$^{-6}$&-23.2&3.0&3.9\\
18.208$\sim$22.288&112.66&0.50$^{+0.01}_{-0.01}$&-0.69$^{+0.02}_{-0.02}$&440$^{+18}_{-18}$&576$^{+26}_{-26}$&10.11$^{+0.95}_{-0.90}$$\times$10$^{-6}$&0.45$^{+0.02}_{-0.02}$&-0.60$^{+0.02}_{-0.02}$&-2.19$^{+0.05}_{-0.05}$&459$^{+20}_{-20}$&14.15$^{+1.43}_{-1.29}$$\times$10$^{-6}$&-107.6&2.9&4.0\\
22.288$\sim$22.536&38.51&0.61$^{+0.07}_{-0.06}$&-0.72$^{+0.07}_{-0.07}$&547$^{+100}_{-103}$&700$^{+133}_{-137}$&14.13$^{+5.80}_{-4.32}$$\times$10$^{-6}$&0.51$^{+0.06}_{-0.06}$&-0.55$^{+0.09}_{-0.09}$&-1.96$^{+0.09}_{-0.09}$&447$^{+70}_{-70}$&21.76$^{+9.52}_{-6.41}$$\times$10$^{-6}$&-34.8&2.1&3.0\\
\enddata
\vspace{3mm}
Note---All columns are the same as Table \ref{table:081009140} but for GRB 100826957.\\ 
\end{deluxetable}

\clearpage
\begin{longrotatetable}
\setlength{\tabcolsep}{0.15em}
\begin{deluxetable}{cc|ccccc|ccccc|ccc}
\centering
\rotate
\tabletypesize{\scriptsize}
\tablecaption{Time-resolved Spectral Fit Results of GRB 101014175 \label{table:101014175}}
\tablehead{
\specialrule{0em}{3pt}{3pt}
\multirow{3}{*}{$t_{\rm start}$$\sim$$t_{\rm stop}$} 
&\multirow{3}{*}{$S$} 
&\multicolumn{5}{c}{Cutoff Power-Law Fitting}
&\multicolumn{5}{c}{Band Fitting}
&\multicolumn{3}{c}{Difference}\\
\cmidrule(lr){3-7}  \cmidrule(lr){8-12} \cmidrule(lr){13-15}
&
&\colhead{$K$}
&\colhead{$\alpha$}
&\colhead{$E_{\rm c}$}
&\colhead{$E_{\rm p}$}
&\colhead{$F$}
&\colhead{$K$}
&\colhead{$\alpha$}
&\colhead{$\beta$}
&\colhead{$E_{\rm p}$}
&\colhead{$F$}
&\colhead{$\Delta$DIC}
&\colhead{$p^{\rm CPL}_{\rm DIC}$}
&\colhead{$p^{\rm Band}_{\rm DIC}$}\\
\hline
}
\colnumbers
\startdata
\hline
Pulse 1\\
\hline
0.322$\sim$0.565&31.24&0.59$^{+0.07}_{-0.07}$&-0.83$^{+0.08}_{-0.08}$&301$^{+57}_{-58}$&352$^{+71}_{-72}$&4.51$^{+2.13}_{-1.31}$$\times$10$^{-6}$&0.58$^{+0.07}_{-0.07}$&-0.81$^{+0.09}_{-0.09}$&-3.69$^{+1.11}_{-1.27}$&334$^{+47}_{-47}$&5.01$^{+1.95}_{-1.31}$$\times$10$^{-6}$&-0.1&2.3&2.8\\
0.565$\sim$0.842&45.15&0.61$^{+0.06}_{-0.06}$&-0.55$^{+0.06}_{-0.06}$&213$^{+21}_{-22}$&309$^{+33}_{-34}$&7.09$^{+2.14}_{-1.54}$$\times$10$^{-6}$&0.60$^{+0.06}_{-0.06}$&-0.54$^{+0.07}_{-0.07}$&-4.39$^{+1.19}_{-1.23}$&301$^{+22}_{-21}$&7.42$^{+1.67}_{-1.43}$$\times$10$^{-6}$&0.2&2.7&2.9\\
0.842$\sim$1.171&70.30&0.80$^{+0.05}_{-0.06}$&-0.43$^{+0.05}_{-0.05}$&186$^{+12}_{-12}$&293$^{+21}_{-21}$&11.34$^{+2.26}_{-1.95}$$\times$10$^{-6}$&0.75$^{+0.06}_{-0.06}$&-0.35$^{+0.06}_{-0.06}$&-2.78$^{+0.24}_{-0.22}$&263$^{+16}_{-16}$&13.80$^{+2.97}_{-2.71}$$\times$10$^{-6}$&-11.9&2.9&3.8\\
1.171$\sim$1.596&91.61&1.17$^{+0.06}_{-0.06}$&-0.56$^{+0.03}_{-0.03}$&214$^{+12}_{-12}$&308$^{+18}_{-18}$&13.37$^{+2.11}_{-1.60}$$\times$10$^{-6}$&1.14$^{+0.06}_{-0.06}$&-0.54$^{+0.04}_{-0.04}$&-3.87$^{+0.86}_{-1.11}$&296$^{+14}_{-14}$&14.20$^{+2.28}_{-1.92}$$\times$10$^{-6}$&-2.3&2.9&2.5\\
1.596$\sim$1.756&41.54&1.21$^{+0.11}_{-0.11}$&-0.72$^{+0.08}_{-0.08}$&159$^{+21}_{-21}$&204$^{+30}_{-30}$&5.89$^{+2.00}_{-1.39}$$\times$10$^{-6}$&1.20$^{+0.11}_{-0.12}$&-0.71$^{+0.08}_{-0.08}$&-5.60$^{+1.63}_{-1.60}$&202$^{+16}_{-16}$&6.06$^{+1.28}_{-1.06}$$\times$10$^{-6}$&0.6&2.7&3.0\\
1.756$\sim$1.957&36.67&0.99$^{+0.10}_{-0.10}$&-0.79$^{+0.09}_{-0.09}$&183$^{+28}_{-28}$&221$^{+38}_{-38}$&4.80$^{+1.89}_{-1.21}$$\times$10$^{-6}$&0.96$^{+0.10}_{-0.10}$&-0.73$^{+0.10}_{-0.10}$&-3.13$^{+0.75}_{-0.97}$&202$^{+24}_{-24}$&5.65$^{+2.31}_{-1.39}$$\times$10$^{-6}$&-2.8&2.5&2.8\\
\enddata 
\vspace{3mm}
Note---All columns are the same as Table \ref{table:081009140} but for GRB 101014175.\\ 
\end{deluxetable}
\end{longrotatetable}

\clearpage
\begin{deluxetable}{cc|ccccc|ccccc|ccc}
\centering
\rotate
\tabletypesize{\scriptsize}
\tablecaption{Time-resolved Spectral Fit Results of GRB 110301214 \label{table:110301214}}
\tablehead{
\specialrule{0em}{3pt}{3pt}
\multirow{3}{*}{$t_{\rm start}$$\sim$$t_{\rm stop}$} 
&\multirow{3}{*}{$S$} 
&\multicolumn{5}{c}{Cutoff Power-Law Fitting}
&\multicolumn{5}{c}{Band Fitting}
&\multicolumn{3}{c}{Difference}\\
\cmidrule(lr){3-7}  \cmidrule(lr){8-12} \cmidrule(lr){13-15}
&
&\colhead{$K$}
&\colhead{$\alpha$}
&\colhead{$E_{\rm c}$}
&\colhead{$E_{\rm p}$}
&\colhead{$F$}
&\colhead{$K$}
&\colhead{$\alpha$}
&\colhead{$\beta$}
&\colhead{$E_{\rm p}$}
&\colhead{$F$}
&\colhead{$\Delta$DIC}
&\colhead{$p^{\rm CPL}_{\rm DIC}$}
&\colhead{$p^{\rm Band}_{\rm DIC}$}\\
\hline
}
\colnumbers
\startdata
\hline
Pulse 1\\
\hline
0.190$\sim$0.963&78.55&1.38$^{+0.06}_{-0.06}$&-0.82$^{+0.05}_{-0.05}$&135$^{+11}_{-10}$&159$^{+14}_{-14}$&4.36$^{+0.76}_{-0.56}$$\times$10$^{-6}$&1.26$^{+0.07}_{-0.07}$&-0.60$^{+0.09}_{-0.09}$&-2.41$^{+0.12}_{-0.11}$&124$^{+9}_{-9}$&5.90$^{+1.31}_{-0.93}$$\times$10$^{-6}$&-27.2&2.8&3.9\\
0.963$\sim$1.686&102.59&1.89$^{+0.08}_{-0.08}$&-0.52$^{+0.05}_{-0.05}$&74$^{+3}_{-3}$&109$^{+6}_{-6}$&5.15$^{+0.68}_{-0.57}$$\times$10$^{-6}$&1.87$^{+0.08}_{-0.08}$&-0.49$^{+0.05}_{-0.05}$&-4.54$^{+0.97}_{-1.21}$&107$^{+3}_{-3}$&5.28$^{+0.51}_{-0.46}$$\times$10$^{-6}$&-1.5&3.0&2.8\\
1.686$\sim$1.974&81.20&2.97$^{+0.15}_{-0.15}$&-0.68$^{+0.06}_{-0.06}$&83$^{+6}_{-6}$&109$^{+9}_{-9}$&6.79$^{+1.14}_{-0.94}$$\times$10$^{-6}$&2.92$^{+0.15}_{-0.15}$&-0.64$^{+0.07}_{-0.07}$&-4.16$^{+0.85}_{-1.13}$&106$^{+4}_{-4}$&7.04$^{+0.84}_{-0.72}$$\times$10$^{-6}$&-1.9&2.9&2.8\\
1.974$\sim$2.688&159.10&4.50$^{+0.12}_{-0.11}$&-0.94$^{+0.03}_{-0.03}$&146$^{+6}_{-6}$&155$^{+8}_{-8}$&11.84$^{+0.96}_{-0.87}$$\times$10$^{-6}$&4.27$^{+0.12}_{-0.12}$&-0.83$^{+0.04}_{-0.04}$&-2.73$^{+0.11}_{-0.10}$&133$^{+5}_{-5}$&13.60$^{+1.20}_{-1.02}$$\times$10$^{-6}$&-26.9&3.0&4.0\\
2.688$\sim$3.128&103.34&3.69$^{+0.14}_{-0.14}$&-1.00$^{+0.04}_{-0.04}$&139$^{+10}_{-10}$&139$^{+11}_{-11}$&8.14$^{+0.97}_{-0.87}$$\times$10$^{-6}$&3.64$^{+0.14}_{-0.14}$&-0.97$^{+0.05}_{-0.05}$&-3.88$^{+0.88}_{-1.12}$&134$^{+6}_{-6}$&8.43$^{+0.94}_{-0.73}$$\times$10$^{-6}$&-1.7&2.9&2.7\\
3.128$\sim$3.725&97.15&3.26$^{+0.12}_{-0.12}$&-1.04$^{+0.04}_{-0.04}$&116$^{+8}_{-8}$&112$^{+9}_{-9}$&5.56$^{+0.72}_{-0.59}$$\times$10$^{-6}$&3.10$^{+0.13}_{-0.13}$&-0.88$^{+0.08}_{-0.08}$&-2.76$^{+0.19}_{-0.15}$&95$^{+6}_{-6}$&6.40$^{+0.84}_{-0.81}$$\times$10$^{-6}$&-10.8&2.9&3.8\\
\hline
Pulse 2\\
\enddata
\vspace{3mm}
Note---All columns are the same as Table \ref{table:081009140} but for GRB 110301214.\\ 
\end{deluxetable}

\clearpage
\begin{deluxetable}{cc|ccccc|ccccc|ccc}
\centering
\rotate
\tabletypesize{\scriptsize}
\tablecaption{Time-resolved Spectral Fit Results of GRB 110625881 \label{table:110625881}}
\tablehead{
\specialrule{0em}{3pt}{3pt}
\multirow{3}{*}{$t_{\rm start}$$\sim$$t_{\rm stop}$} 
&\multirow{3}{*}{$S$} 
&\multicolumn{5}{c}{Cutoff Power-Law Fitting}
&\multicolumn{5}{c}{Band Fitting}
&\multicolumn{3}{c}{Difference}\\
\cmidrule(lr){3-7}  \cmidrule(lr){8-12} \cmidrule(lr){13-15}
&
&\colhead{$K$}
&\colhead{$\alpha$}
&\colhead{$E_{\rm c}$}
&\colhead{$E_{\rm p}$}
&\colhead{$F$}
&\colhead{$K$}
&\colhead{$\alpha$}
&\colhead{$\beta$}
&\colhead{$E_{\rm p}$}
&\colhead{$F$}
&\colhead{$\Delta$DIC}
&\colhead{$p^{\rm CPL}_{\rm DIC}$}
&\colhead{$p^{\rm Band}_{\rm DIC}$}\\
\hline
}
\colnumbers
\startdata
\hline
Pulse 1\\
\hline
-0.917$\sim$3.056&20.65&0.15$^{+0.02}_{-0.02}$&-0.99$^{+0.10}_{-0.10}$&250$^{+61}_{-61}$&253$^{+66}_{-67}$&0.56$^{+0.30}_{-0.17}$$\times$10$^{-6}$&0.14$^{+0.02}_{-0.02}$&-0.97$^{+0.11}_{-0.11}$&-6.16$^{+2.58}_{-2.59}$&237$^{+37}_{-38}$&0.59$^{+0.20}_{-0.15}$$\times$10$^{-6}$&1.1&1.4&2.4\\
3.056$\sim$4.045&26.24&0.11$^{+0.02}_{-0.02}$&-0.25$^{+0.12}_{-0.12}$&160$^{+22}_{-21}$&280$^{+43}_{-42}$&1.90$^{+1.24}_{-0.72}$$\times$10$^{-6}$&0.10$^{+0.02}_{-0.02}$&-0.22$^{+0.13}_{-0.12}$&-6.07$^{+2.68}_{-2.69}$&272$^{+21}_{-21}$&1.97$^{+0.99}_{-0.56}$$\times$10$^{-6}$&0.8&1.0&2.0\\
4.045$\sim$4.967&36.11&0.11$^{+0.02}_{-0.02}$&-0.03$^{+0.09}_{-0.09}$&154$^{+13}_{-13}$&303$^{+29}_{-29}$&3.55$^{+1.70}_{-1.00}$$\times$10$^{-6}$&0.11$^{+0.02}_{-0.02}$&-0.02$^{+0.09}_{-0.09}$&-6.33$^{+2.38}_{-2.46}$&298$^{+14}_{-14}$&3.75$^{+1.17}_{-0.90}$$\times$10$^{-6}$&0.8&1.9&2.6\\
\hline
Pulse 2\\
\hline
10.229$\sim$10.436&23.41&0.21$^{+0.04}_{-0.04}$&-0.40$^{+0.11}_{-0.11}$&308$^{+52}_{-53}$&493$^{+90}_{-91}$&6.88$^{+5.01}_{-2.78}$$\times$10$^{-6}$&0.21$^{+0.04}_{-0.04}$&-0.40$^{+0.11}_{-0.10}$&-6.31$^{+2.50}_{-2.52}$&484$^{+52}_{-52}$&6.96$^{+3.37}_{-2.15}$$\times$10$^{-6}$&0.8&1.2&2.1\\
10.436$\sim$10.553&28.72&0.27$^{+0.05}_{-0.05}$&-0.19$^{+0.10}_{-0.10}$&235$^{+30}_{-30}$&426$^{+59}_{-60}$&11.65$^{+7.04}_{-4.39}$$\times$10$^{-6}$&0.26$^{+0.05}_{-0.05}$&-0.16$^{+0.12}_{-0.12}$&-5.82$^{+2.58}_{-2.78}$&412$^{+37}_{-37}$&12.23$^{+6.16}_{-4.15}$$\times$10$^{-6}$&-0.2&1.6&1.8\\
10.553$\sim$11.203&88.13&0.60$^{+0.03}_{-0.03}$&-0.33$^{+0.03}_{-0.04}$&195$^{+9}_{-9}$&325$^{+16}_{-17}$&12.33$^{+1.79}_{-1.64}$$\times$10$^{-6}$&0.55$^{+0.04}_{-0.04}$&-0.24$^{+0.05}_{-0.05}$&-2.82$^{+0.19}_{-0.15}$&292$^{+12}_{-12}$&14.56$^{+2.69}_{-2.17}$$\times$10$^{-6}$&-13.9&2.9&3.9\\
\enddata
\vspace{3mm}
Note---All columns are the same as Table \ref{table:081009140} but for GRB 110625881.\\ 
\end{deluxetable}

\clearpage
\begin{deluxetable}{cc|ccccc|ccccc|ccc}
\centering
\rotate
\tabletypesize{\scriptsize}
\tablecaption{Time-resolved Spectral Fit Results of GRB 120129580 \label{table:120129580}}
\tablehead{
\specialrule{0em}{3pt}{3pt}
\multirow{3}{*}{$t_{\rm start}$$\sim$$t_{\rm stop}$} 
&\multirow{3}{*}{$S$} 
&\multicolumn{5}{c}{Cutoff Power-Law Fitting}
&\multicolumn{5}{c}{Band Fitting}
&\multicolumn{3}{c}{Difference}\\
\cmidrule(lr){3-7}  \cmidrule(lr){8-12} \cmidrule(lr){13-15}
&
&\colhead{$K$}
&\colhead{$\alpha$}
&\colhead{$E_{\rm c}$}
&\colhead{$E_{\rm p}$}
&\colhead{$F$}
&\colhead{$K$}
&\colhead{$\alpha$}
&\colhead{$\beta$}
&\colhead{$E_{\rm p}$}
&\colhead{$F$}
&\colhead{$\Delta$DIC}
&\colhead{$p^{\rm CPL}_{\rm DIC}$}
&\colhead{$p^{\rm Band}_{\rm DIC}$}\\
\hline
}
\colnumbers
\startdata
\hline
Pulse 1\\
\hline
0.066$\sim$0.676&46.41&1.02$^{+0.09}_{-0.09}$&-0.75$^{+0.11}_{-0.11}$&96$^{+14}_{-14}$&120$^{+20}_{-20}$&2.44$^{+0.90}_{-0.59}$$\times$10$^{-6}$&0.97$^{+0.08}_{-0.08}$&-0.47$^{+0.13}_{-0.13}$&-2.55$^{+0.31}_{-0.10}$&90$^{+10}_{-10}$&3.35$^{+1.04}_{-0.79}$$\times$10$^{-6}$&-12.2&2.5&2.0\\
0.676$\sim$0.812&30.71&1.60$^{+0.24}_{-0.24}$&-0.84$^{+0.19}_{-0.19}$&113$^{+30}_{-32}$&131$^{+41}_{-43}$&3.56$^{+3.15}_{-1.36}$$\times$10$^{-6}$&1.52$^{+0.21}_{-0.21}$&-0.68$^{+0.23}_{-0.21}$&-4.16$^{+1.85}_{-3.15}$&107$^{+18}_{-19}$&4.03$^{+2.49}_{-1.28}$$\times$10$^{-6}$&-3.5&0.3&0.3\\
0.812$\sim$1.070&51.73&1.89$^{+0.17}_{-0.17}$&-0.74$^{+0.10}_{-0.10}$&103$^{+14}_{-14}$&130$^{+20}_{-20}$&5.04$^{+1.80}_{-1.25}$$\times$10$^{-6}$&1.77$^{+0.14}_{-0.15}$&-0.45$^{+0.12}_{-0.12}$&-2.56$^{+0.28}_{-0.08}$&96$^{+9}_{-10}$&6.80$^{+1.97}_{-1.61}$$\times$10$^{-6}$&-11.2&2.6&2.1\\
1.070$\sim$1.209&68.40&1.98$^{+0.18}_{-0.18}$&-0.60$^{+0.06}_{-0.06}$&222$^{+23}_{-23}$&311$^{+35}_{-35}$&21.04$^{+5.93}_{-4.94}$$\times$10$^{-6}$&1.89$^{+0.19}_{-0.19}$&-0.55$^{+0.08}_{-0.08}$&-3.54$^{+1.15}_{-1.36}$&281$^{+28}_{-29}$&24.66$^{+9.16}_{-5.90}$$\times$10$^{-6}$&-7.3&2.7&0.1\\
1.209$\sim$1.310&76.56&2.72$^{+0.23}_{-0.24}$&-0.66$^{+0.05}_{-0.05}$&315$^{+33}_{-33}$&422$^{+47}_{-47}$&38.49$^{+11.21}_{-7.94}$$\times$10$^{-6}$&2.51$^{+0.26}_{-0.26}$&-0.58$^{+0.08}_{-0.08}$&-3.23$^{+0.78}_{-0.25}$&361$^{+43}_{-40}$&43.84$^{+16.15}_{-10.55}$$\times$10$^{-6}$&-9.8&2.7&-0.2\\
1.310$\sim$1.665&170.07&3.16$^{+0.12}_{-0.13}$&-0.56$^{+0.02}_{-0.02}$&267$^{+10}_{-10}$&384$^{+16}_{-16}$&51.08$^{+5.86}_{-5.10}$$\times$10$^{-6}$&3.12$^{+0.14}_{-0.14}$&-0.54$^{+0.03}_{-0.03}$&-5.17$^{+2.00}_{-2.83}$&375$^{+15}_{-16}$&52.65$^{+6.96}_{-5.67}$$\times$10$^{-6}$&-3.4&2.9&0.9\\
\enddata
\vspace{3mm}
Note---All columns are the same as Table \ref{table:081009140} but for GRB 120129580.\\ 
\end{deluxetable}

\clearpage
\begin{deluxetable}{cc|ccccc|ccccc|ccc}
\centering
\rotate
\tabletypesize{\scriptsize}
\tablecaption{Time-resolved Spectral Fit Results of GRB 120328268 \label{table:120328268}}
\tablehead{
\specialrule{0em}{3pt}{3pt}
\multirow{3}{*}{$t_{\rm start}$$\sim$$t_{\rm stop}$} 
&\multirow{3}{*}{$S$} 
&\multicolumn{5}{c}{Cutoff Power-Law Fitting}
&\multicolumn{5}{c}{Band Fitting}
&\multicolumn{3}{c}{Difference}\\
\cmidrule(lr){3-7}  \cmidrule(lr){8-12} \cmidrule(lr){13-15}
&
&\colhead{$K$}
&\colhead{$\alpha$}
&\colhead{$E_{\rm c}$}
&\colhead{$E_{\rm p}$}
&\colhead{$F$}
&\colhead{$K$}
&\colhead{$\alpha$}
&\colhead{$\beta$}
&\colhead{$E_{\rm p}$}
&\colhead{$F$}
&\colhead{$\Delta$DIC}
&\colhead{$p^{\rm CPL}_{\rm DIC}$}
&\colhead{$p^{\rm Band}_{\rm DIC}$}\\
\hline
}
\colnumbers
\startdata
\hline
Pulse 1\\
\hline
2.602$\sim$4.487&29.97&0.15$^{+0.01}_{-0.01}$&-0.74$^{+0.06}_{-0.06}$&511$^{+96}_{-97}$&644$^{+125}_{-126}$&2.97$^{+1.36}_{-0.86}$$\times$10$^{-6}$&0.14$^{+0.02}_{-0.02}$&-0.70$^{+0.08}_{-0.08}$&-4.22$^{+2.06}_{-3.29}$&573$^{+103}_{-108}$&3.29$^{+1.76}_{-1.04}$$\times$10$^{-6}$&-5.4&2.2&-0.3\\
4.487$\sim$4.817&28.87&0.36$^{+0.04}_{-0.04}$&-0.76$^{+0.08}_{-0.07}$&541$^{+124}_{-130}$&671$^{+160}_{-165}$&6.95$^{+3.83}_{-2.31}$$\times$10$^{-6}$&0.34$^{+0.04}_{-0.04}$&-0.73$^{+0.09}_{-0.09}$&-5.59$^{+2.81}_{-2.90}$&616$^{+123}_{-124}$&7.14$^{+3.53}_{-2.17}$$\times$10$^{-6}$&-0.5&1.8&1.8\\
4.817$\sim$5.993&69.84&0.46$^{+0.02}_{-0.02}$&-0.69$^{+0.03}_{-0.03}$&453$^{+38}_{-38}$&593$^{+52}_{-52}$&9.78$^{+1.79}_{-1.50}$$\times$10$^{-6}$&0.40$^{+0.02}_{-0.02}$&-0.55$^{+0.05}_{-0.05}$&-2.11$^{+0.08}_{-0.08}$&424$^{+38}_{-37}$&13.96$^{+2.97}_{-2.64}$$\times$10$^{-6}$&-30.0&2.8&3.2\\
5.993$\sim$8.058&106.36&0.71$^{+0.02}_{-0.02}$&-0.65$^{+0.03}_{-0.03}$&212$^{+10}_{-10}$&286$^{+15}_{-15}$&6.19$^{+0.69}_{-0.62}$$\times$10$^{-6}$&0.63$^{+0.02}_{-0.02}$&-0.47$^{+0.04}_{-0.04}$&-2.13$^{+0.06}_{-0.06}$&211$^{+10}_{-10}$&10.02$^{+1.32}_{-1.11}$$\times$10$^{-6}$&-75.0&2.9&3.9\\
8.058$\sim$8.770&47.75&0.71$^{+0.04}_{-0.04}$&-0.88$^{+0.05}_{-0.05}$&239$^{+31}_{-31}$&268$^{+37}_{-37}$&3.74$^{+0.99}_{-0.72}$$\times$10$^{-6}$&0.68$^{+0.05}_{-0.05}$&-0.80$^{+0.09}_{-0.09}$&-3.28$^{+1.17}_{-1.67}$&228$^{+36}_{-36}$&4.89$^{+2.08}_{-1.56}$$\times$10$^{-6}$&-9.3&2.6&-1.8\\
8.770$\sim$10.470&61.39&0.69$^{+0.03}_{-0.03}$&-0.95$^{+0.05}_{-0.05}$&202$^{+21}_{-21}$&212$^{+24}_{-24}$&2.48$^{+0.46}_{-0.34}$$\times$10$^{-6}$&0.62$^{+0.03}_{-0.03}$&-0.70$^{+0.09}_{-0.09}$&-2.05$^{+0.04}_{-0.04}$&139$^{+13}_{-13}$&4.30$^{+0.98}_{-0.69}$$\times$10$^{-6}$&-30.5&2.8&2.7\\
\enddata
\vspace{3mm}
Note---All columns are the same as Table \ref{table:081009140} but for GRB 120328268.\\ 
\end{deluxetable}

\clearpage
\begin{deluxetable}{cc|ccccc|ccccc|ccc}
\centering
\rotate
\tabletypesize{\scriptsize}
\tablecaption{Time-resolved Spectral Fit Results of GRB 120711115 \label{table:120711115}}
\tablehead{
\specialrule{0em}{3pt}{3pt}
\multirow{3}{*}{$t_{\rm start}$$\sim$$t_{\rm stop}$} 
&\multirow{3}{*}{$S$} 
&\multicolumn{5}{c}{Cutoff Power-Law Fitting}
&\multicolumn{5}{c}{Band Fitting}
&\multicolumn{3}{c}{Difference}\\
\cmidrule(lr){3-7}  \cmidrule(lr){8-12} \cmidrule(lr){13-15}
&
&\colhead{$K$}
&\colhead{$\alpha$}
&\colhead{$E_{\rm c}$}
&\colhead{$E_{\rm p}$}
&\colhead{$F$}
&\colhead{$K$}
&\colhead{$\alpha$}
&\colhead{$\beta$}
&\colhead{$E_{\rm p}$}
&\colhead{$F$}
&\colhead{$\Delta$DIC}
&\colhead{$p^{\rm CPL}_{\rm DIC}$}
&\colhead{$p^{\rm Band}_{\rm DIC}$}\\
\hline
}
\colnumbers
\startdata
\hline
Pulse 1\\
\hline
63.488$\sim$64.961&20.65&0.15$^{+0.02}_{-0.02}$&-0.86$^{+0.06}_{-0.06}$&1729$^{+324}_{-321}$&1971$^{+383}_{-381}$&7.71$^{+3.67}_{-2.65}$$\times$10$^{-6}$&0.15$^{+0.02}_{-0.02}$&-0.86$^{+0.06}_{-0.06}$&-6.45$^{+2.44}_{-2.42}$&1924$^{+297}_{-305}$&7.63$^{+2.98}_{-2.10}$$\times$10$^{-6}$&-0.3&2.5&2.5\\
64.961$\sim$67.130&37.39&0.30$^{+0.02}_{-0.02}$&-0.95$^{+0.03}_{-0.03}$&1679$^{+213}_{-212}$&1763$^{+229}_{-228}$&9.92$^{+2.29}_{-1.86}$$\times$10$^{-6}$&0.30$^{+0.02}_{-0.02}$&-0.95$^{+0.03}_{-0.03}$&-6.56$^{+2.36}_{-2.39}$&1732$^{+183}_{-182}$&10.01$^{+2.03}_{-1.62}$$\times$10$^{-6}$&0.2&2.9&3.0\\
67.130$\sim$71.711&67.94&0.42$^{+0.02}_{-0.02}$&-0.96$^{+0.02}_{-0.02}$&1702$^{+115}_{-117}$&1770$^{+124}_{-126}$&13.40$^{+1.82}_{-1.41}$$\times$10$^{-6}$&0.42$^{+0.02}_{-0.02}$&-0.96$^{+0.02}_{-0.02}$&-6.64$^{+2.27}_{-2.28}$&1757$^{+99}_{-100}$&13.36$^{+1.36}_{-1.22}$$\times$10$^{-6}$&-0.0&3.0&3.0\\
71.711$\sim$75.400&50.21&0.47$^{+0.03}_{-0.03}$&-1.07$^{+0.03}_{-0.03}$&1204$^{+147}_{-145}$&1119$^{+142}_{-140}$&6.68$^{+1.34}_{-1.08}$$\times$10$^{-6}$&0.47$^{+0.03}_{-0.03}$&-1.07$^{+0.03}_{-0.03}$&-6.69$^{+2.21}_{-2.28}$&1111$^{+109}_{-109}$&6.69$^{+1.04}_{-0.83}$$\times$10$^{-6}$&0.4&2.9&3.1\\
75.400$\sim$75.693&21.66&0.57$^{+0.09}_{-0.09}$&-1.00$^{+0.08}_{-0.08}$&816$^{+210}_{-214}$&816$^{+220}_{-224}$&7.00$^{+4.15}_{-2.43}$$\times$10$^{-6}$&0.57$^{+0.09}_{-0.09}$&-0.99$^{+0.08}_{-0.08}$&-6.68$^{+2.35}_{-2.31}$&786$^{+152}_{-154}$&7.25$^{+3.23}_{-2.13}$$\times$10$^{-6}$&0.6&2.1&2.7\\
75.693$\sim$81.191&35.69&0.31$^{+0.02}_{-0.02}$&-1.17$^{+0.03}_{-0.03}$&1941$^{+411}_{-409}$&1611$^{+346}_{-345}$&4.24$^{+1.13}_{-0.93}$$\times$10$^{-6}$&0.31$^{+0.02}_{-0.02}$&-1.17$^{+0.04}_{-0.04}$&-5.64$^{+2.64}_{-2.82}$&1545$^{+297}_{-289}$&4.23$^{+1.00}_{-0.79}$$\times$10$^{-6}$&-0.7&2.7&2.5\\
\enddata
\vspace{3mm}
Note---All columns are the same as Table \ref{table:081009140} but for GRB 120711115.\\ 
\end{deluxetable}

\clearpage
\begin{longrotatetable}
\setlength{\tabcolsep}{0.15em}
\begin{deluxetable}{cc|ccccc|ccccc|ccc}
\centering
\rotate
\tabletypesize{\scriptsize}
\tablecaption{Time-resolved Spectral Fit Results of GRB 120728484 \label{table:120728484}}
\tablehead{
\specialrule{0em}{3pt}{3pt}
\multirow{3}{*}{$t_{\rm start}$$\sim$$t_{\rm stop}$} 
&\multirow{3}{*}{$S$} 
&\multicolumn{5}{c}{Cutoff Power-Law Fitting}
&\multicolumn{5}{c}{Band Fitting}
&\multicolumn{3}{c}{Difference}\\
\cmidrule(lr){3-7}  \cmidrule(lr){8-12} \cmidrule(lr){13-15}
&
&\colhead{$K$}
&\colhead{$\alpha$}
&\colhead{$E_{\rm c}$}
&\colhead{$E_{\rm p}$}
&\colhead{$F$}
&\colhead{$K$}
&\colhead{$\alpha$}
&\colhead{$\beta$}
&\colhead{$E_{\rm p}$}
&\colhead{$F$}
&\colhead{$\Delta$DIC}
&\colhead{$p^{\rm CPL}_{\rm DIC}$}
&\colhead{$p^{\rm Band}_{\rm DIC}$}\\
\hline
}
\colnumbers
\startdata
\hline
Pulse 1\\
\hline
-0.177$\sim$3.764&22.44&0.15$^{+0.02}_{-0.02}$&-0.81$^{+0.12}_{-0.12}$&168$^{+36}_{-38}$&200$^{+48}_{-49}$&0.60$^{+0.34}_{-0.19}$$\times$10$^{-6}$&0.11$^{+0.02}_{-0.02}$&0.45$^{+0.41}_{-0.44}$&-2.05$^{+0.04}_{-0.04}$&76$^{+14}_{-13}$&1.04$^{+0.90}_{-0.41}$$\times$10$^{-6}$&-25.3&1.5&-5.5\\
10.358$\sim$12.621&29.11&0.29$^{+0.03}_{-0.03}$&-0.67$^{+0.14}_{-0.14}$&92$^{+17}_{-17}$&123$^{+26}_{-26}$&0.77$^{+0.41}_{-0.24}$$\times$10$^{-6}$&0.24$^{+0.03}_{-0.03}$&0.66$^{+0.26}_{-0.27}$&-2.16$^{+0.08}_{-0.08}$&59$^{+6}_{-6}$&1.28$^{+0.61}_{-0.41}$$\times$10$^{-6}$&-29.3&1.6&1.9\\
12.621$\sim$15.450&52.23&0.52$^{+0.03}_{-0.03}$&-0.93$^{+0.06}_{-0.06}$&177$^{+22}_{-22}$&189$^{+26}_{-26}$&1.72$^{+0.44}_{-0.33}$$\times$10$^{-6}$&0.36$^{+0.03}_{-0.03}$&0.47$^{+0.35}_{-0.33}$&-2.02$^{+0.02}_{-0.02}$&67$^{+8}_{-8}$&3.02$^{+1.49}_{-0.92}$$\times$10$^{-6}$&-69.6&2.6&-3.6\\
15.450$\sim$16.538&43.39&0.63$^{+0.06}_{-0.06}$&-0.83$^{+0.08}_{-0.08}$&142$^{+19}_{-19}$&167$^{+25}_{-25}$&2.02$^{+0.63}_{-0.45}$$\times$10$^{-6}$&0.50$^{+0.06}_{-0.06}$&-0.23$^{+0.30}_{-0.28}$&-2.28$^{+0.17}_{-0.12}$&100$^{+17}_{-19}$&2.90$^{+2.37}_{-1.11}$$\times$10$^{-6}$&-15.2&2.5&-3.1\\
16.538$\sim$17.118&39.84&0.92$^{+0.08}_{-0.08}$&-0.95$^{+0.08}_{-0.08}$&185$^{+28}_{-29}$&194$^{+33}_{-34}$&2.96$^{+0.97}_{-0.66}$$\times$10$^{-6}$&0.89$^{+0.10}_{-0.10}$&-0.89$^{+0.08}_{-0.12}$&-6.00$^{+2.78}_{-2.66}$&181$^{+24}_{-18}$&3.00$^{+1.17}_{-0.76}$$\times$10$^{-6}$&-0.4&2.5&2.0\\
17.118$\sim$20.689&60.23&0.60$^{+0.03}_{-0.03}$&-0.59$^{+0.07}_{-0.07}$&71$^{+5}_{-5}$&101$^{+9}_{-9}$&1.35$^{+0.29}_{-0.22}$$\times$10$^{-6}$&0.52$^{+0.04}_{-0.04}$&0.87$^{+0.11}_{-0.11}$&-2.23$^{+0.05}_{-0.05}$&52$^{+2}_{-2}$&2.11$^{+0.41}_{-0.36}$$\times$10$^{-6}$&-69.9&2.8&2.9\\
\enddata
\vspace{3mm}
Note---All columns are the same as Table \ref{table:081009140} but for GRB 120728484.\\ 
\end{deluxetable}%
\end{longrotatetable}

\end{document}